\RequirePackage{silence} 
\documentclass[ twoside,openright,titlepage,numbers=noenddot,
                headinclude,headings=optiontoheadandtoc,
                footinclude,cleardoublepage=empty,abstract=on,
                BCOR=5mm,paper=a4,fontsize=11pt,
                ]{scrreprt}
\usepackage{ragged2e}
\usepackage{subcaption}
\usepackage{tensor}
\usepackage{physics}
\usepackage{mathtools}

\newcommand{\sVert}[1][0]{%
  \ifcase#1\relax
  \rvert\or\bigr|\or\Bigr|\or\biggr|\or\Biggr
  \fi
}

\PassOptionsToPackage{utf8}{inputenc}
  \usepackage{inputenc}

\PassOptionsToPackage{T1}{fontenc} 
  \usepackage{fontenc}

\PassOptionsToPackage{
  drafting=false,    
  tocaligned=false, 
  dottedtoc=true,  
  eulerchapternumbers=true, 
  linedheaders=true,       
  floatperchapter=true,     
  eulermath=false,  
  beramono=true,    
  palatino=true,    
  style=classicthesis 
}{classicthesis}

\newcommand{\myTitle}{Thermodynamics and statistical mechanical ensembles 
of black holes and self-gravitating matter\xspace}
\newcommand{\mySubtitle}{\xspace}

\newcommand{\myName}{Tiago Vasques Fernandes\xspace}

\newcommand{\myFaculty}{Put data here\xspace}

\newcommand{\myUni}{Instituto Superior Técnico\xspace}

\newcommand{\myVersion}{draft v1}

\providecommand{\mLyX}{L\kern-.1667em\lower.25em\hbox{Y}\kern-.125emX\@}

\newcommand{\tf}[1]{{\textcolor{cyan}{\textbf{[Tiago: #1]}} }}

\PassOptionsToPackage{american}{babel} 
    \usepackage{babel}

\usepackage{csquotes}
\PassOptionsToPackage{%
  backend=bibtex8,bibencoding=ascii,%
  language=auto,%
  style=phys,%
  sorting=none, 
  maxbibnames=10, 
  eprint=true,%
  biblabel=brackets,%
  pageranges=false,%
  hyperref=true,%
  maxcitenames=10,%
  natbib=true 
}{biblatex}
    \usepackage{biblatex}

\PassOptionsToPackage{fleqn}{amsmath}       
  \usepackage{amsmath,amssymb,lmodern,amsthm}

\usepackage{graphicx,psfrag,mathrsfs} %
\usepackage{adjustbox}
\usepackage{csquotes}
\usepackage{multirow}
\usepackage{bold-extra}
\usepackage{bm}
\usepackage{scrhack} 
\usepackage{xspace} 
\PassOptionsToPackage{printonlyused,smaller}{acronym}
  \usepackage{acronym} 

\usepackage{tabularx} 
\usepackage{subcaption}

\usepackage{listings}
\usepackage{classicthesis}

\hypersetup{%
  colorlinks=true, linktocpage=true, pdfstartpage=3, pdfstartview=FitV,%
  breaklinks=true, pageanchor=true,%
  pdfpagemode=UseNone, %
  plainpages=false, bookmarksnumbered, bookmarksopen=true, bookmarksopenlevel=1,%
  hypertexnames=true, pdfhighlight=/O,
  urlcolor=CTurl, linkcolor=CTlink, citecolor=CTcitation, 
  pdftitle={\myTitle},%
  pdfauthor={\textcopyright\ \myName, \myUni, \myFaculty},%
  pdfsubject={},%
  pdfkeywords={},%
  pdfcreator={pdfLaTeX},%
  pdfproducer={LaTeX with hyperref and classicthesis}%
}

\makeatletter
\@ifpackageloaded{babel}%
  {%
    \addto\extrasamerican{%
    }%
    }{\relax}
\makeatother

\def\be{\begin{equation}}
\def\ee{\end{equation}}
\def\beq{\begin{eqnarray}}
\def\eeq{\end{eqnarray}}

\newpairofpagestyles[scrheadings]{userightbotmark}{}
\begin{document}
\frenchspacing
\raggedbottom
\selectlanguage{american} 
\pagenumbering{roman}
\pagestyle{plain}
\thispagestyle{empty}
\pdfbookmark[1]{Title}{title}
%
%
%

\begin{flushleft} ~\\ \vspace{-30mm} \hspace{-12mm}  \includegraphics[width=8cm]{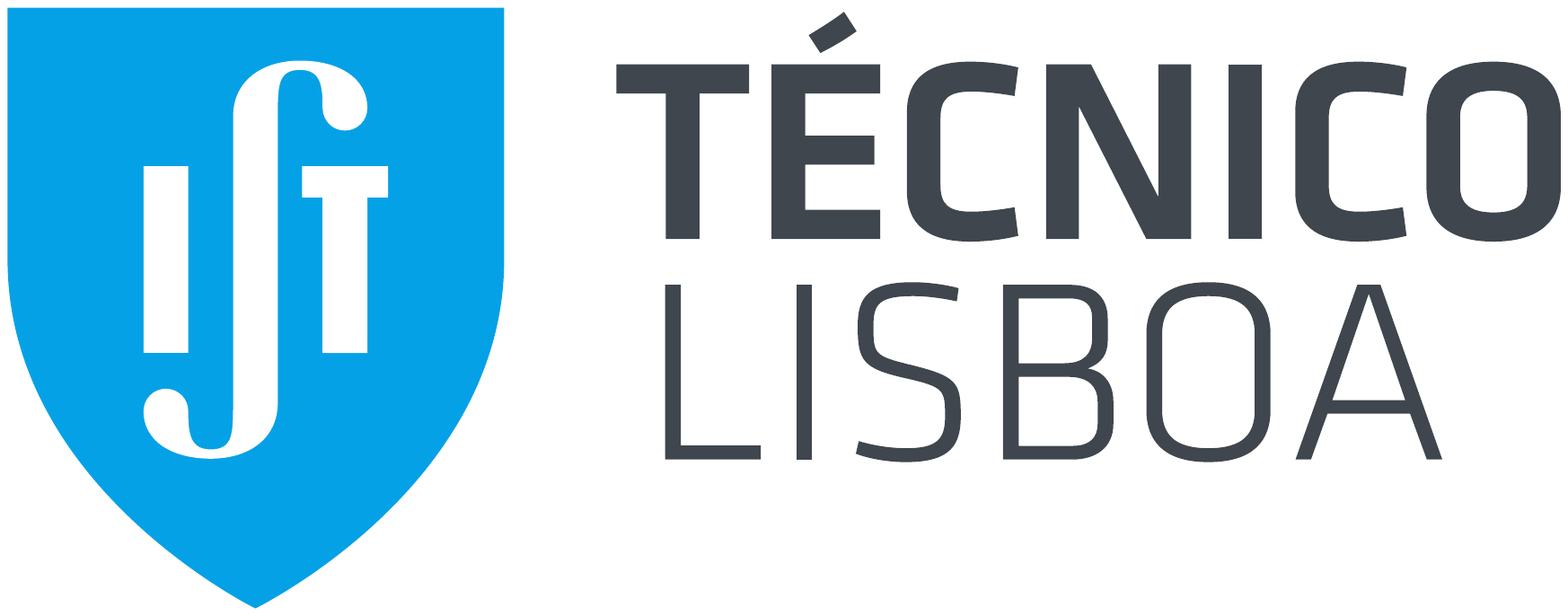} 
	
	\centering
	\LARGE \textbf{\spacedallcaps{UNIVERSIDADE DE LISBOA \\ INSTITUTO SUPERIOR TÉCNICO}}
	\\ \vspace{33mm}

	
	\centering
	\LARGE \spacedallcaps{\myTitle}
	\\ \vspace{5mm}
	\LARGE \spacedlowsmallcaps{\myName} 
	\vspace{2cm}
	
	\begin{flushleft}
			\hspace{-2mm} \large \spacedlowsmallcaps{\textbf{\uppercase{S}upervisor}}: \large Doctor José Pizarro de Sande e Lemos\\
			\vspace{3mm}
			\hspace{-2mm} \large \spacedlowsmallcaps{\textbf{\uppercase{C}o-\uppercase{S}upervisor}}: \large Doctor Vítor Manuel dos Santos Cardoso\\
	\end{flushleft}
	\vspace{10mm}
	\centering
	\small \spacedlowsmallcaps{ \uppercase{T}hesis approved in public session to obtain the \uppercase{P}h\uppercase{D} \uppercase{D}egree in:}\\
	\small \spacedallcaps{\textbf{\uppercase{P}hysics}}\\
	\vspace{3mm}
	\small \spacedlowsmallcaps{\uppercase{J}ury final classification:}
	\small \spacedallcaps{\textbf{\uppercase{P}ass with \uppercase{D}istinction and \uppercase{H}onour}}

	\vspace{5mm}
	
	\Large \textbf{2025} \\
	\let\thepage\relax
\end{flushleft}
\pagebreak

\thispagestyle{empty}
\pdfbookmark[1]{Title}{title}
%
%
%

\begin{titlepage}
\begin{flushleft} ~\\ \vspace{-30mm} \hspace{-12mm}  \includegraphics[width=8cm]{gfx/IST_A_CMYK_POS} 
	
	\centering
	\LARGE \textbf{\spacedallcaps{UNIVERSIDADE DE LISBOA \\ INSTITUTO SUPERIOR TÉCNICO}}
	\\ \vspace{10mm}

	
	\centering
	\Large \spacedallcaps{\myTitle}
	\\ \vspace{3mm}
	\Large \spacedlowsmallcaps{\myName} 
	\vspace{0.1cm}
	
	\begin{flushleft}
			\hspace{-2mm} \large \spacedlowsmallcaps{\textbf{\uppercase{S}upervisor}}: \large Doctor José Pizarro de Sande e Lemos\\
			\vspace{3mm}
			\hspace{-2mm} \large \spacedlowsmallcaps{\textbf{\uppercase{C}o-\uppercase{S}upervisor}}: \large Doctor Vítor Manuel dos Santos Cardoso\\
	\end{flushleft}
	\vspace{5mm}
	\centering
	\small \spacedlowsmallcaps{Thesis approved in public session to obtain the PhD degree in:}\\
	\small \spacedallcaps{\textbf{\uppercase{P}hysics}}\\
	\vspace{1mm}
	\small  \spacedlowsmallcaps{Jury final classification:}
	\small  \spacedallcaps{\textbf{\uppercase{P}ass with \uppercase{D}istinction and \uppercase{H}onour}}
	\vspace{5mm}
	
	\normalsize \spacedallcaps{\textbf{Jury}}
	
	\vspace{1mm}
	
	\begin{flushleft}
			\normalsize \spacedlowsmallcaps{Chairperson}: 
			\vspace{1mm}  
			\normalsize Doctor Ilídio Pereira Lopes\small, Instituto Superior Técnico, Universidade de Lisboa\\
			\vspace{2mm}
			\normalsize \spacedlowsmallcaps{Members of the Committee} :  \\
			\vspace{1mm}
			\normalsize Doctor Antonino Flachi\small, Faculty of Business and Commerce, Keio University, Japão\\ \vspace{1.5mm}
			\normalsize Doctor Christian Gerhard B\"ohmer\small, Department of Mathematics, Faculty of Mathematical \& Physical Sciences, 
			University College London, Reino Unido\\ \vspace{1.5mm}
			\normalsize Doctor José Pizarro de Sande e Lemos\small, Instituto Superior Técnico, Universidade de Lisboa\\ \vspace{1.5mm} 
			\normalsize Doctor Filipe Artur Pacheco Neves Carteado Mena\small, Instituto Superior Técnico, Universidade de Lisboa\\ \vspace{1.5mm}
			\normalsize Doctor José Pedro Oliveira Mimoso\small, Faculdade de Ciências, Universidade de Lisboa\\ \vspace{1.5mm} 
		\end{flushleft}
			
	\vspace{3mm}
	
	\Large \textbf{2025} \\
	\let\thepage\relax
\end{flushleft}
\pagebreak
\end{titlepage}
\begin{titlepage}
    \begin{center}
        \large

        \hfill

        \vfill

        \begingroup
            \LARGE \color{CTtitle}\spacedallcaps{\myTitle} \\  \bigskip
            \Large \mySubtitle
        \endgroup
        
        \vfill

        \Large \spacedlowsmallcaps{\myName}

        \vfill

        \spacedlowsmallcaps{2025}

    \end{center}
\end{titlepage}

\thispagestyle{empty}

\hfill

\vfill

\noindent\myName, \textit{\myTitle}, 
\textcopyright\ 2025.

%
%
%
%
%

\cleardoublepage
\thispagestyle{empty}
\phantomsection
\pdfbookmark[1]{Dedication}{Dedication}

\vspace*{1cm}
\vspace{1cm}


\begin{center}
    {À memória da minha avó, Ana...} \\ \smallskip
	{Dedicado ao meu avô, Albino.}
\end{center}

\cleardoublepage


\pdfbookmark[1]{Resumo}{Resumo}
\chapter*{Resumo}

Buracos negros existem por todo o nosso universo, e possuem uma 
larga variedade de massas. Até ao momento, os buracos negros têm sido 
usados para testar a relatividade geral em escalas astrofísicas, 
mas também poderão dar no futuro informação sobre a gravidade em escalas 
microscópicas. Os buracos negros parecem ter propriedades termodinâmicas 
como a entropy de Bekenstein e Hawking, que são mais relevantes quando se 
consideram buracos negros do tamanho de alguns centímetros ou mais pequenos ainda. 
Como a entropia está relacionada com os micro-estados de um sistema em mecânica 
estatística, isto levanta certas questões: o que dá origem à entropia 
de um buraco negro? Poderá esta origem ser explicada por uma descrição 
quântica da gravidade? Para compreender estas questões, a conexão entre 
a termodinâmica e a gravidade tem de ser explorada. 

Nesta tese de doutoramento, pretendemos compreender esta conexão usando duas 
descrições que levam à termodinâmica de espaços-tempos curvos. Começamos por impôr 
a primeira lei da termodinâmica a uma camada fina auto-gravitante com carga elétrica 
em dimensões arbitrárias. A camada fina com carga pode assumir uma equação de estado 
fundamental para a pressão, que é obtida apenas pela relatividade geral. Uma equação 
de estado para a temperatura da camada é escolhida para permitir o estudo do limite 
de buraco negro e a consequente recuperação da termodinâmica de buracos negros. 

Para além disso, usamos a abordagem da gravidade quântica através do integral de 
caminho Euclideano para construir \textit{ensembles} estatísticos de espaços-tempos 
com buracos negros e com matéria auto-gravitante, com o objetivo de estudar 
semiclassicamente as possíveis transições de fase entre matéria quente e buracos negros. 
Mostramos também a capacidade que o formalismo
tem para descrever as propriedades termodinâmicas de espaços-tempos curvos. 
Especificamente, estudamos os \textit{ensembles} canónicos e grão-canónicos 
de buracos negros com carga elétrica, dentro de uma cavidade com raio finito 
ou infinito. Adicionalmente, construimos \textit{ensembles} de camadas finas 
em anti-de Sitter e em espaços assintoticamente planos, para compreender 
as características termodinâmicas de camadas finas e as possíveis transições 
de fase para configurações de buracos negros.

\vspace{0.7cm}
\noindent \spacedlowsmallcaps{Palavras-chave: termodinâmica, \textit{ensembles} estatísticos, 
relatividade geral, buracos negros, matéria quente} . 

\vfill

\newpage

\chapter*{Abstract}
\pdfbookmark[1]{Abstract}{Abstract}

Black holes exist all over our Universe, possessing a very wide range of masses. 
At the moment, they serve as a probe to test general relativity at astrophysical scales, 
but in the future they may also give us information about gravity at the microscale.
Black holes seem to have thermodynamic properties, such as
the Bekenstein-Hawking entropy, which are important when considering 
black holes with size of a few centimeters or smaller. Since entropy in statistical mechanics is 
related to the number of possible microstates of a system, several questions arise: 
what gives rise to the black hole entropy? Can it be explained by a quantum description 
of gravity? In order to further study these questions, the 
connection between thermodynamics and gravity must be explored at the microscale. 

In this doctoral thesis, we aim to understand this connection using two descriptions 
that yield the thermodynamics of curved spacetimes. We start by imposing 
the first law of thermodynamics to a charged self-gravitating matter 
thin shell in higher dimensions. The fundamental pressure equation of state 
can be used for the shell, which is given solely by general relativity. 
An equation of state for temperature of the shell is also chosen, so that it 
allows the study of the black hole limit and the recovery of black hole thermodynamics. 

Furthermore, we use the Euclidean path integral 
approach to quantum gravity to construct statistical ensembles of 
black hole spacetimes and self-gravitating matter, in order to study semiclassically 
the possible phase transitions between hot matter and black holes. We also show 
the power of the formalism in obtaining the thermodynamic properties of 
curved spacetimes. Namely, we study the canonical and grand canonical ensemble of 
charged black holes inside a cavity, which may have a finite or infinite radius.
We also construct ensembles of a self-gravitating matter thin shell, 
both in anti-de Sitter and in asymptotically flat spaces, in order to understand 
the thermodynamic features of the shell and the possible phase transitions to 
black hole configurations.

\vspace{0.7cm}
\noindent \spacedlowsmallcaps{Keywords: thermodynamics, statistical ensembles, general relativity, black holes, 
hot matter} .


\vfill

\cleardoublepage
\pdfbookmark[1]{Preface}{preface}
\chapter*{Preface}

During the four years and five months of the Doctoral Programme, the 
official research that led to the development of this 
thesis has been conducted at Centro de Astrofísica e 
Gravitação (CENTRA), in the Physics Department at 
Instituto Superior Técnico (IST) - University of Lisbon. 
The research done during the Doctoral Programme 
was financially supported by Fundação 
para a Ciência e Tecnologia - FCT through the 
project UIDB/00099/2020, with the grant FCT no. RD0970, and 
through the project UIDB/00099/2025, with the grant FCT no. RD1415. 
Furthermore, I acknowledge the support from the
project 2024.04456.CERN. I declare that this thesis and its content 
has not been submitted for a degree, diploma or other qualification 
at any other university and it has been made specifically to obtain the 
PhD in Physics at IST.

The research developed in Chapters \ref{ch:chargedselfgravitating}, \ref{ch:grandcanonicalblackhole}, 
\ref{ch:Daviesblackhole} and \ref{ch:canonicalblackhole} was done in collaboration 
with José Sande Lemos. 
The research developed in Chapter \ref{ch:adslimits}
was done in collaboration with Francisco J. Gandum,
José Sande Lemos and Oleg Zaslavskii.
The research developed in Chapter \ref{ch:thinshellAdS}
was done in collaboration with Francisco J. Gandum and 
José Sande Lemos.  Finally, 
the research developed in Chapter \ref{ch:Chatelier}
was done in collaboration with
José Sande Lemos and Oleg Zaslavskii. 
Chapters~\ref{ch:chargedselfgravitating}, \ref{ch:grandcanonicalblackhole} and 
\ref{ch:canonicalblackhole} have been published, while
Chapters~\ref{ch:Daviesblackhole} has been accepted but not yet published 
and \ref{ch:thinshellAdS} has been submitted to a journal, 
with the manuscripts being available in the arXiv. 
The Chapters~\ref{ch:adslimits} and \ref{ch:Chatelier} are in preparation.
In sum, this doctoral thesis is mainly based on the following works:\\

{\noindent\cite{Fernandes:2022gjd} \fullcite{Fernandes:2022gjd} (Chapter \ref{ch:chargedselfgravitating});}\\

{\noindent\cite{Fernandes:2023byx} \fullcite{Fernandes:2023byx} (Chapter \ref{ch:grandcanonicalblackhole});}\\

{\noindent\cite{Tiago2024a} \fullcite{Tiago2024a} (Chapter \ref{ch:Daviesblackhole});}\\

\pagebreak
{\noindent\cite{Tiago2024b} \fullcite{Tiago2024b} (Chapter \ref{ch:canonicalblackhole});}\\

{\noindent\cite{Tiago2024bl} \fullcite{Tiago2024bl} (Chapter \ref{ch:adslimits});}\\

{\noindent\cite{Tiago2024bk} \fullcite{Tiago2024bk} (Chapter \ref{ch:thinshellAdS});}\\

{\noindent\cite{Tiago2024f} \fullcite{Tiago2024f} (Chapter \ref{ch:Chatelier}).}\\

There are also the following works which were started recently and came out of the ideas of this PhD thesis:\\

{\noindent\cite{Tiago2025fg} \fullcite{Tiago2025fg};}\\

{\noindent\cite{Tiago2025fw} \fullcite{Tiago2025fw}.}\\

During the years of my doctoral program, I also coauthored other works which are not discussed in this thesis. 
Two of these works were done in collaboration with David Hilditch, 
Vítor Cardoso and José Sande Lemos, published in Phys. Rev. D. Two of the works have been done in collaboration 
with David Lopes and José Sande Lemos, as part of David Lopes' master thesis, with one published in Phys. Rev. D and the 
other to be submitted. The remaining work was done in collaboration with Julian Barragán Amado and David Lopes, 
and it has been submitted to JHEP, with the manuscript being available in the arXiv. These 
works are the following:\\

{\noindent\cite{Fernandes:2021qvr} \fullcite{Fernandes:2021qvr}.}\\

{\noindent\cite{Fernandes:2022con} \fullcite{Fernandes:2022con}.}\\

{\noindent\cite{Lopes:2024ofy} \fullcite{Lopes:2024ofy}.}\\

{\noindent\cite{Tiago2025zzz} \fullcite{Tiago2025zzz}.}\\

{\noindent\cite{BarraganAmado:2025pfy} \fullcite{BarraganAmado:2025pfy}.}\\

\cleardoublepage
\pdfbookmark[1]{Acknowledgments}{acknowledgments}

\bigskip

\begingroup
\let\clearpage\relax
\let\cleardoublepage\relax
\let\cleardoublepage\relax
\chapter*{Agradecimentos}

Chegou o momento para agradecer a quem me apoiou e me acompanhou 
neste trecho da minha vida que culminou no meu doutoramento. 
É verdade que sou reservado, até demasiado às vezes, 
especialmente no momento de exprimir os meus desejos e sentimentos.
No entanto, vou tentar aqui deixar os meus agradecimentos 
da melhor forma que consiga.

O meu estudo e trabalho durante 
o doutoramento, que faz parte desta tese, não podia ter sido feito 
sem o apoio e os conselhos do meu orientador de doutoramento José Sande e Lemos. 
Devo dizer que é uma experiência bastante interessante ter o José 
como orientador. Ele proporcionou-me uma visão 
sobre o que é um artigo científico e a sua estrutura, e também de como 
se faz física. Ele sempre foi muito presente e eu pude sempre contar com 
ele sobre questões do doutoramento. Agradeço também pelas conversas 
que tivemos, especificamente sobre física e história da física, 
sobre o futebol e o Benfica, sobre ténis e sobre snooker. 

Agradeço aos 
meus colaboradores Oleg Zaslavskii, David Lopes, Francisco Gandum 
e Julián Barragán pelo trabalho que fizemos em conjunto em certos 
artigos, juntamente com conversas interessantes sobre variados temas.

Agradeço 
também a todos os membros do CENTRA pela camaradagem e conversas 
que tivemos, tornando a experiência do doutoramento mais fácil.
Em particular, agradeço aos meus colegas de doutoramento  
Christian, Diogo, Krinio, Arianna, Jorge, João e Zhen, que me 
fizeram sentir que não estava sozinho nesta caminhada. Devo agradecer 
adicionalmente a alguns membros do CENTRA: Alex, Edgar, Valentin, Richard, Nicolás, 
Joan, Matteo e Hannes. Em especial, agradeço ao David por ser outra referência 
minha no centro. E claro, não podia faltar um agradecimento a João Dinis 
pelas discussões sobre física e \textit{metal}. Deveras, ouvir \textit{metal} deu-me forças 
para desenvolver a tese.

Devo dar um agradecimento aos meus amigos que tornaram especial 
e mais divertido este percurso académico e de vida. Agradeço ao Filipe, 
Gonçalo, Rafael, Carlos, Rodolfo, Miguel e Henrique. Em particular, agradeço 
ao Filipe e ao Gonçalo pelas sessões longas de jogo. Também 
agradeço ao Rafael e Henrique pelas conversas sobre a vida académica.

O doutoramento não é um período fácil, longe disso. É preciso ter uma 
capacidade de gestão e controlo emocional para manter 
uma rotina de trabalho e de estudo. Mas isto não se aplica só 
no doutoramento mas sim a muitas vertentes da vida. Eu não poderia 
ter mantido esta capacidade sem o grande apoio da Carla. Foram todos 
os momentos que passamos juntos que me ajudaram a ser uma pessoa melhor 
e que me ajudaram a ter felicidade, neste meu barco que navega no rio da 
vida. Por isso, agradeço-te do fundo do meu coração, Carla. 

Neste seguimento, também não poderia deixar de agradecer à Mariana e ao 
Carlos por me fazer sentir parte da família.

Para terminar, toda a minha vida não poderia ser como era sem o apoio incondicional 
da minha família, em especial dos meus avós maternos Ana e Albino. 
Eles representam os pilares da minha vida, pois sem eles nem sei 
onde estaria. E é pelo apoio ao longa da minha vida desde a primeira instância 
que dedico esta tese a eles. Agradeço também ao meu tio Pedro, o meu 
padrinho, e à minha tia Lina pelo apoio adicional que me deram.

\endgroup

\cleardoublepage
\pagestyle{scrheadings}
\pdfbookmark[1]{\contentsname}{tableofcontents}
\setcounter{tocdepth}{2} 
\setcounter{secnumdepth}{3} 
\manualmark
\markboth{\spacedlowsmallcaps{\contentsname}}{\spacedlowsmallcaps{\contentsname}}
\tableofcontents
\automark[section]{chapter}
\renewcommand{\chaptermark}[1]{\markboth{\spacedlowsmallcaps{#1}}{\spacedlowsmallcaps{#1}}}
\renewcommand{\sectionmark}[1]{\markright{\textsc{\thesection}\enspace\spacedlowsmallcaps{#1}}}
\clearpage
\begingroup
    \let\clearpage\relax
    \let\cleardoublepage\relax
    \pdfbookmark[1]{\listfigurename}{lof}
    \listoffigures

    \vspace{8ex}


    \vspace{8ex}

    \newpage
    \pdfbookmark[1]{Conventions and Units}{conventions}
    \markboth{\spacedlowsmallcaps{Conventions and Units}}{\spacedlowsmallcaps{Conventions and Units}}
	\chapter*{Conventions, Notation and Units}
	In this thesis, the conventions of Refs.~\cite{Misner:1974qy,Wald:1984rg} are followed.
	There are four constants throughout the thesis that establish the units, the gravitational 
    constant $G$ in arbitrary dimensions $d$, 
    the speed of light $c$, the Planck constant $\hbar$ and the Boltzmann constant. 
    Unless stated otherwise, I use Planck units $G=\hbar=c=1$ and the Boltzmann constant is set to $1$ as well. 
    In some places, the Planck scale or the gravitational constant $l_{\mathrm{p}} = G^{\frac{1}{d-2}}$ is kept.
    Lorentzian spacetimes have the most positive metric signature, while Euclidean spaces 
    have the typical positive metric signature. \\
	
	\resizebox{\textwidth}{!}{\begin{tabular}{lll}
		&$\alpha,\beta,\gamma, ...$          & Lorentzian spacetime \\
                                         &   & and Riemannian space indices; \\
        &$a,b,c,d, ...,h$                    & timelike hypersurface indices in Lorentzian spacetime\\
                                         &   & and indices of a hypersurface parametrized \\
                                         &   & by imaginary time in Riemannian space;\\
		&$i,j, k, ...$      & spacelike hypersurface indices in Lorentzian spacetime \\
                         &   & and indices of a hypersurface of constant imaginary time \\
                         &   & in Riemannian space;\\
        &$A,B, C, ...$     & $d-2$-surface indices in Lorentzian spacetime \\
                         &   & and in Euclidean space;\\
		&$V_\alpha W^\alpha \equiv \sum_{\alpha=0}^{3} V_\alpha W^\alpha$ & Einstein's notation;\\
		&$T_{(\alpha_1\,...\,\alpha_l)}\equiv= \frac{1}{l!}\sum_{\,\sigma} T_{\alpha_{\sigma(1)}\,...\,\alpha_{\sigma(l)}}$ 
        & symmet. over all permutat.~$\sigma$;\\
		&$T_{[\alpha_1\,...\,\alpha_l]}\equiv \frac{1}{l!}\sum_{\,\sigma} 
        \epsilon_\sigma T_{\alpha_{\sigma(1)}\,...\,\alpha_{\sigma(l)}}$ & anti-symmet. over all permutat.~$\sigma$;\\
		&$g_{\alpha \beta}$       & curved Lorentzian/Riemannian metric;  \\
		&$(\,\cdot\,)_{,\alpha}=\partial_\alpha(\cdot)=\frac{\partial}{\partial x^\alpha}(\cdot)$ &  coord. derivative; \\
		&$(\,\cdot\,)_{;\alpha}= \nabla_\alpha(\cdot)$  & Levi-Civita derivative; \\
		&$\Box(\cdot) \equiv \nabla_\alpha \nabla^\alpha(\cdot)$ &  Levi-Civita d'Alembertian. \\
	\end{tabular}}
	


\endgroup

\cleardoublepage
\pagestyle{scrheadings}
\pagenumbering{arabic}
\cleardoublepage
\chapter{Introduction}\label{ch:introduction}

\section{Classical black holes in general relativity}

The theory of general relativity, with its definite formulation 
in~\cite{Einstein:1916}
by Einstein, has withstood for now as the theory that describes gravity 
at large length scales. 
Gravity in this theory is described as the link 
between the curvature of spacetime and the presence of matter, through 
the Einstein equations. As a consequence, the presence of matter curves 
spacetime in a nontrivial way, giving origin to a number of effects such 
as the precession of orbits, the deflection of light near massive objects, 
the gravitational redshift of light's frequency and
the time delay of light as it travels near a massive object. 
Surprisingly, right at its conception, general relativity 
was able to explain the perihelion motion of Mercury~\cite{Einstein:1915}, 
which could not be explained by Newton's gravity. Moreover, the deflection 
of light by the Sun was observed by Eddington~\cite{Eddington:1920}. 
Further tests were made, with 
the measurement of the gravitational redshift~\cite{Popper:1954} and the 
measurement of the gravitational time delay~\cite{Shapiro:1964, Shapiro:1965}, 
or Shapiro delay, which agreed with general relativity. Another characteristic 
of the theory is the existence of gravitational waves, ripples of 
spacetime originated from the motion of two massive compact objects. These were
eventually measured indirectly by Hulse and Taylor~\cite{Hulse:1975,Taylor:1981} 
and first directly measured by LIGO~\cite{LIGOScientific:2016aoc}. The 
present evidence continues to strengthen the position of general relativity 
as the theory of gravitation, at least at large length scales.

Although these effects are quite important, 
one of the most important predictions of general 
relativity is the existence of black holes. These objects are defined, in 
a general sense, as regions of spacetime from which light cannot escape. The 
first solution which describes such an object was given by 
Schwarzschild~\cite{Schwarzschild:1916} and additionally with electric charge 
in~\cite{Reissner:1916,Nordstrom:1918}, although initially these metrics 
were only thought to describe the exterior region of 
spherically symmetric self-gravitating objects. However, it was only with the work by 
Oppenheimer and Snyder~\cite{Oppenheimer:1939}, where they studied the gravitational 
collapse of dust in spherical symmetry, that black holes were put in 
the spotlight as remnants of gravitational collapse. 
Penrose~\cite{Penrose:1965} showed that gravitational collapse in general settings 
would occur, giving birth to black holes and to the occurrence of singularities. 
In general relativity, these black holes can be described by the Kerr-Newman
family~\cite{Kerr:1963,Newman:1965}, which extends the Schwarzschild 
and Reissner-Nordstr\"om solutions to include rotation. The fact that black holes 
can only be described by three parameters has been a quite enticing feature of 
general relativity and the possibility of probing regions of strong gravitational 
field with current technology could give us more insights in the validity of 
general relativity. Based on recent observations, it is most certain that 
black holes exist and that they populate our whole universe. 
The first detection of gravitational waves~\cite{LIGOScientific:2016aoc} 
not only demonstrated the existence of these objects but also initiated 
an era of probing the strong field regime of general relativity with 
gravitational waves. Moreover, the observations of
the center of the M87 galaxy~\cite{Akiyama:2019}, 
and the center of the Milky Way~\cite{Ghez:2008, Gillessen:2009,Akiyama:2022}, 
are in complete agreement with the existence of a supermassive black hole 
at the center of these galaxies and also with the predictions of general relativity.

At the theoretical level, the analysis of global causality provided by Penrose and Hawking, 
which were needed for the 
development of the singularity theorems~\cite{Penrose:1965, Hawking:1970zqf}, allowed for 
a better understanding of the properties of a black hole and its boundary, the event horizon.
It was realized by Hawking~\cite{Hawking:1971vc} that under the weak cosmic censorship and 
under the weak energy condition, the area of the event horizon always increases. Furthermore, 
using the Kerr solution, Smarr~\cite{Smarr:1973} showed that the mass of the black hole is 
related to the area of the event horizon and its angular momentum. These developments led to 
the establishment of the four laws of black hole mechanics~\cite{Bardeen:1973gs} for black 
holes in stationary spacetimes. While these revealed crucial properties of 
classical black holes, the most impact was felt in black hole thermodynamics.

Astrophysical black holes, which we discussed above, can have a mass ranging from a few solar masses 
up to millions of solar masses. There may be however another class of black holes, with masses much smaller than 
one solar mass~\cite{Hawking:1971ei} but larger than the Planck mass. 
These are micro black holes or quantum black holes, where 
their thermodynamic properties have the most importance. Theoretically, 
they may be formed by extreme heat, by a collision of two particles or even by overdensities in 
the early universe. Micro black holes are of great importance to probe the gravitational 
interaction at quantum scales and, in the particular context of this thesis, they are
an important avenue to study the link between gravity and thermodynamics.

\section{Thermodynamic properties of black holes}

Influenced by Wheeler's thought experiments regarding matter entropy and black holes,
Bekenstein~\cite{Bekenstein:1972,Bekenstein:1972b} proposed that black holes 
should have an entropy proportional to its event horizon area and generalized 
the second law of thermodynamics to include black hole entropy. The argument 
towards such proposal was based on the lowest integer power of the horizon 
radius that allowed for the entropy to always increase, following from the second 
law of black hole mechanics given by Hawking~\cite{Hawking:1971vc}. Bekenstein's 
proposal was seen with skepticism by Hawking, which in~\cite{Bardeen:1973gs} 
points out that classical black holes do not radiate and so the only connection between 
thermodynamics and the four laws of black hole mechanics was purely an analogy.

It was around this time that quantum field theory in curved spacetime was being 
explored. The notion of particle states provided by the Fock space of a field in 
a curved spacetime depends on the choice of a basis of positive frequencies 
for the field, i.e. it depends on the observer. However, such 
dependence seems to be akin to the choice of coordinates of a manifold in general relativity.
Hawking~\cite{Hawking:1974rv, Hawking:1975vcx} used the treatment of quantum fields in a 
collapsing spacetime that would asymptotically tend to a stationary 
black hole spacetime. Using the geometric optics approximation, he showed that 
black holes radiate neutral quanta with a spectrum similar to a black body 
at Hawking temperature $T = \frac{\hbar \kappa}{2\pi k}$, where $\hbar$ is the 
Planck constant, $k$ is the Boltzmann constant and $\kappa$ is the 
surface gravity of the event horizon. Boulware~\cite{Boulware:1975}, 
upon the work by Hawking, showed that the vacuum 
prescribed by choosing positive frequencies via the timelike Killing field led 
to no radiation for an eternal Schwarzschild black hole and that Hawking radiation was mostly 
due to the existence of gravitational collapse. The picture of black hole 
evaporation for an eternal Schwarzschild black hole was then developed by Unruh~\cite{Unruh:1976},
by choosing a different vacuum at the past horizon, 
as the one that could mimic gravitational collapse. However, both Boulware and 
Unruh vacuum states are not well-defined in the maximally extended Schwarzschild 
spacetime. Hartle and Hawking~\cite{Hartle:1976} obtained the unique 
vacuum state that is well-defined in the whole maximally extended Schwarzschild 
spacetime through Euclidean path integral techniques. An observer following 
the orbits of the timelike Killing vector in the exterior region of Schwarzschild 
will observe the Hartle-Hawking state as a thermal state. By consequence, a 
Schwarzschild black hole can be in thermal equilibrium~\cite{Gibbons:1976es} 
with a heat bath described by the Hartle-Hawking state, revealing the thermodynamic 
nature of black holes. Since these developments, quantum field theory in curved spacetime 
has established itself as a promising area of research, with the study of 
vacuum states in black hole spacetimes being extended to several cases, 
e.g. see~\cite{Hewitt:2014vwy}.

These developments established the first 
strong link between thermodynamics and black hole mechanics that was missing.
Black holes indeed radiate at the Hawking temperature and possess the 
Bekenstein-Hawking entropy $S= \frac{A_+}{4 A_p}$, where $A_+$ is 
the event horizon area and $A_p$ is the 
Planck area. The four laws of black hole mechanics in~\cite{Bardeen:1973gs} 
correspond to the laws of black hole thermodynamics. These laws were used in 
tandem with the Hawking temperature and the Bekenstein-Hawking temperature 
to describe the thermodynamic theory of black holes, by Davies~\cite{Davies:1977bgr}.
Since then, the first law of thermodynamics has been used to study black hole 
solutions and their stability~\cite{Katz:1993up, Parentani:1994wr, Gibbons:2004ai, 
Myung:2008ze, Maeda:2009uy, La:2010bx, Hendi:2014mna, Clement:2019ghi, 
Hajian:2021hje, Jiang:2021pna, Rodriguez:2021hks, Murk:2023rwl}.
However, it must be noted that such prescription is heuristic and so 
a fundamental formalism is required 
to describe the thermodynamics in black hole spacetimes.

\section{Thermodynamics and statistical ensembles in curved spacetimes}

From the existence of a thermal state describing a heat bath, 
one can extract thermodynamic properties of a curved spacetime. But the realization of 
temperature in curved spacetimes predates the substantial works on black hole thermodynamics.
To tackle this issue, Tolman and Ehrenfest~\cite{Tolman:1930} considered a system in thermal 
equilibrium with an external heat reservoir, but with the space between the system and 
the heat reservoir filled with electromagnetic radiation. The electromagnetic field 
was then averaged to describe black body radiation in the form of a perfect fluid with 
a radiative equation of state. From the equilibrium equations, Tolman and Ehrenfest 
showed that the local temperature is given by the redshifted 
temperature of the heat reservoir. This means that thermodynamic equilibrium in a 
gravitating system is not described by a constant local temperature but rather must 
be described by the redshift factor. In some sense, temperature follows the same
behavior as the frequency of photons in a stationary curved spacetime. This evidently distorts 
the notions of intensive and extensive variables in thermodynamics envolving curved spacetimes.

To understand further thermodynamics in curved spacetimes and its properties, 
one needs to have a more fundamental approach. A formal way of formulating 
thermodynamics is by using the tools of statistical mechanics. 
However, for that construction, one needs to have 
a microscopic description of gravity. For quantum systems without gravity, the 
microscopic description is given by quantum field theory and one can build the 
partition function of statistical ensembles using the statistical path integral
~\cite{Weinberg:1974hy, Bernard:1974bq}. 
The time parameter associated to the evolution of states is extended to the complex plane and 
one then works with an imaginary time, which has a period equal to the inverse 
temperature of the ensemble. In this framework, quantum field theory is transformed into 
thermal quantum field theory. Gibbons and Hawking~\cite{Gibbons:1977} extended this 
formalism to curved spacetimes, where the partition function is given by 
the Euclidean path integral approach to quantum gravity. Using the semiclassical 
approximation, they obtained the grand canonical ensembles of Kerr-Newman and 
Reissner-Nordstr\"om spaces and also the canonical ensemble of Schwarzschild space, 
in four dimensions. For the specific case of Schwarzschild, the heat capacity of 
the black hole is negative, which makes the canonical ensemble unstable and the 
semiclassical approximation not valid. York~\cite{York:1986} understood, motivated by 
the results of Page and Hawking~\cite{Page:1982dh}, that introducing a cavity at 
finite radius makes the canonical ensemble of a black hole stable and the semiclassical 
approximation valid in a specific range of the parameter space. Specifically, the introduction 
of the cavity gives rise to an additional black hole solution in thermal equilibrium 
with the cavity with positive heat capacity. This is the York 
formalism for the construction of canonical and grand canonical ensembles in 
curved spacetimes. Even though one is making a semiclassical approximation, the formalism 
allows the study of phase transitions between stable configurations, namely between 
matter at finite temperature and black hole configurations, which arise purely from 
quantum effects. Note that the configurations we are discussing must be microscopic so that 
semiclassical effects come into play, but they must also be far away from the Planck scale 
so that additional quantum effects can be ignored. This motivates the exploration of the 
York formalism to understand the 
implications of the Euclidean path integral approach to quantum gravity in thermodynamic 
and microscopic systems.

\section{Higher dimensional curved spacetimes}

From our sensorial perspective, we perceive the universe as being four dimensional. 
As we are observers, our spatial awareness tells us that there are three dimensions of 
space and we move in worldlines with a dedicated clock, i.e. one dimension of time. 
Up to now, it seems that all events that we observe are consistent 
with a four dimensional universe. However, the interest in higher dimensions increased 
mainly by the prospects of unifying the fundamental forces of the Universe. The original 
work by Kaluza~\cite{Kaluza:1921tu} and Klein~\cite{Klein:1926tv} tried to unify 
electromagnetism with gravity by considering a five dimensional spacetime with a 
cilindrical coordinate, with the metric being a solution of the five dimensional 
Einstein field equations. The metric did not depend on this cilindrical coordinate, 
and its effects on the matter fields could be minimized by having a scale much smaller than the 
Planck scale associated to the length of the extra dimension, i.e. one has a compactified 
dimension. This approach was expanded in order to include the forces of the 
standard model by considering supersymmetry~\cite{Freedman:1976xh}, which is 
named as Kaluza-Klein supergravity theory. It was shown that this could be 
done with an eleven dimensional supergravity theory~\cite{Witten:1981me}. However, 
many difficulties arose, e.g. the difficulty in including chirality for fermion fields 
and the presence of anomalies when quantizing the theory. The unification attempts were 
later focused on superstring theories, which had supergravity as their low energy limit, 
and ultimately M-theory, to tackle these difficulties. Still, 
there is an apparent absence of clear physical predictions, apart from supersymmetry,
that can be extracted from these unified theories. Additionally, from the observation of 
the gravitational wave signal and electromagnetic counterpart of GW170817~\cite{LIGOScientific:2017vwq,Pardo:2018ipy}, 
an event classified as a neutron star binary merger, there are constraints on the dimensions 
of the Universe that are compatible with four non-compactified dimensions.

Apart from unification theories, there has been a large interest in higher dimensional 
spacetimes due to the AdS/CFT conjecture~\cite{Maldacena:1997re}. 
This conjecture is a correspondence between a string theory in $(d+1)$
dimensions and a quantum conformal field theory without gravity in $d$ dimensions.
The main advantages of the conjecture is that one has a strong/weak coupling duality, 
i.e. string theory in the weak coupling regime is dual to a conformal field theory in the 
strong coupling regime. This motivated the study of strings and branes in higher dimensional 
AdS spacetimes, which can be done using higher dimensional general relativity, 
in order to obtain the properties of non-perturbative conformal fluids, 
with many applications to condensed matter and 
particle physics.

\section{Outline of the thesis}

In this thesis, we focus on the study of thermodynamic self-gravitating systems 
with the objective to further understand the interplay between 
gravity and thermodynamics. We work with microscopic systems much larger than the Planck length, 
where thermodynamics is important and semiclassical effects are present. 
This thesis consists on two main Parts.
In each Part, we choose a specific formulation to describe thermodynamics in curved spacetime. 
Moreover, each Part of the thesis is self-contained.

In the first Part, we formulate the thermodynamics for self-gravitating matter 
by imposing the first law of thermodynamics. Namely, in Chapter~\ref{ch:chargedselfgravitating},
we impose the first law of thermodynamics to a self-gravitating matter thin shell 
with electric charge, in arbitrary dimensions. The purpose is 
to understand the possibility and the implications of imposing 
the Martinez pressure equation of 
state, which arises naturally from the Einstein equations. Indeed, the Martinez pressure 
equation of state can in fact be imposed, and we obtain the entropy of the shell 
in terms solely of its gravitational radius and its Cauchy radius, related to the electric 
charge. We impose further equations of state that have a black hole-like behaviour, 
allowing the recovery of black hole thermodynamics from 
the thin shell in the black hole limit. The intrinsic thermodynamic stability of the 
thin shell is then analyzed, showing that the case of a thin shell with black hole 
equations of state in the black hole limit is marginally stable.

In the second Part, we construct statistical ensembles 
of curved spacetimes including matter in order to obtain their thermodynamic properties. 
To obtain the partition 
function of a statistical ensemble, we use the Euclidean path integral approach to quantum gravity, 
which gives a microscopic description of gravity, 
in the zeroth order of the saddle point approximation, i.e. 
the zero loop approximation. The state of the art and the formalism restricted to spherically symmetric 
metrics are explained in Chapter~\ref{ch:Euclideanpathintegral}, which is taken as a 
reference in the rest of the Chapters. Apart from this relationship between 
Chapter~\ref{ch:Euclideanpathintegral} and the rest of the Chapters, 
the remaining content of the Chapters is self-contained. Throughout the second 
Part, we apply this formalism to various cases involving a gauge field and matter 
in order to understand the intricacies of the formalism and the phase diagrams of the 
configurations considered. In Chapter~\ref{ch:grandcanonicalblackhole}, we consider 
the grand canonical ensemble of 
a Reissner-Nordstr\"om black hole inside a cavity, in arbitrary 
dimensions. In Chapter~\ref{ch:Daviesblackhole}, we consider the canonical ensemble 
of a Reissner-Nordstr\"om black hole with cavity at infinity, in arbitrary 
dimensions, where we establish a link between the Euclidean path integral approach 
to quantum gravity and the strategy of imposing the first law of thermodynamics. 
In Chapter~\ref{ch:canonicalblackhole}, we consider the canonical ensemble 
of a Reissner-Nordstr\"om black hole inside a cavity, in arbitrary 
dimensions. Note that for Chapters~\ref{ch:grandcanonicalblackhole},
~\ref{ch:Daviesblackhole}, and~\ref{ch:canonicalblackhole}, we obtain 
the phase diagrams between black hole configurations and hot flat space, 
where the models of hot flat space, i.e. flat space at a 
certain temperature, are considered for fixed electric potential and 
for fixed electric charge, which is a novelty. In Chapter~\ref{ch:adslimits}, 
we analyze the limits of the canonical ensemble 
of a black hole in AdS inside a cavity, which unify the
black hole solutions existing in the literature.
In Chapter~\ref{ch:thinshellAdS}, we consider
the canonical ensemble of a self-gravitating matter thin shell in anti-de Sitter (AdS), 
showing that for a particular equation of state, it mimics 
hot thermal AdS, i.e. pure AdS space with a thermal graviton gas, for a wide range 
of temperatures. In Chapter~\ref{ch:Chatelier}, 
we build the grand canonical ensemble of a matter thin shell with chemical potential 
involving a black hole, all surrounded by a cavity, showing the 
power of the formalism. We further show certain connections between the validity of the 
zero loop approximation, mechanical stability and thermodynamic stability. 

Finally in Chapter~\ref{ch:Conclusion}, we present some conclusion remarks 
regarding the main results of the thesis, including caveats of the study and 
possible future work.

\cleardoublepage
\part{Thermodynamics of self-gravitating matter using the first law}\label{pt:firstlaw}
\chapter{Electrically charged spherical matter shells in higher dimensions}
\label{ch:chargedselfgravitating}

\section{Introduction
\label{sec:introthinshell}}

The study of self-gravitating matter is fundamental to understand the 
effects of general relativity and ultimately to describe the 
astrophysical objects of the universe. While matter is typically distributed 
through the three dimensions of space, self-gravitating thin shells 
in general relativity~\cite{Israel:1966zz} have proven to be of great 
significance towards 
the understanding of the interaction between gravitational and matter 
fields. Namely, the dynamics of thin shells in Schwarzschild 
and Reissner-Nordstr\"om spacetimes, together with generalizations to 
higher dimensions~\cite{Israel:1967zz, Kuchar:1968, Kijowski:2005if, 
Dias:2006wu, Gao:2008jy, Eiroa:2012nd}, are able to capture in detail
the main features of gravitational collapse and the 
corresponding black hole formation.
Related to gravitational collapse 
and the stability of self-gravitating matter, 
the maximum compactness of stars has been studied through 
neutral and electrically charged thin shells~\cite{Andreasson:2008xw}.
Moreover, the landscape of spacetimes that can be constructed using 
thin shells is vast and uncovers the possible exotic configurations provided 
by general relativity such as wormhole spacetimes~\cite{Dias:2010uh, 
Bejarano:2011yz}, 
bubble universes~\cite{Lemos:2022lpu}, tension shells and 
stars~\cite{Katz:1991},
and regular black holes~\cite{Lemos:2011dq, Uchikata:2012zs, 
Brandenberger:2021ken}, with the case of an electrically charged shell 
with two different normal orientations being able to describe 
even more objects~\cite{Lemos:2021jtm}.

The interest in self-gravitating matter thin shells also extends to the 
treatment of thermodynamics within general relativity, as shells 
can be used to understand the thermodynamics of matter in a gravitational 
field and even the thermodynamics of the gravitational field itself. 
While one has the freedom to choose equations of state for the shell, there 
is a particular pressure equation of state that can be provided by general 
relativity in the case of a static spherical thin shell with Minkowski in 
its interior. This can be called as a fundamental pressure 
equation of state and allows the radius of the shell to be arbitrary. 
By imposing the first law of thermodynamics to the shell, one can restrict 
the expression for the temperature equation of state using the 
integrability conditions, leaving some freedom for its choice that can be 
motivated by a fundamental description of matter or one can always 
make a reasonable guess based on the behaviour that one wants to imprint to 
the shell. This treatment was first performed for a 
static shell with an exterior Schwarzschild spacetime 
in~\cite{Martinez:1996ni, PerezBergliaffa:2020zzv} and 
extended to higher dimensions in~\cite{Andre:2019zzo}. The inclusion 
of electric charge in the case of four dimensions has also been treated
in~\cite{Lemos:2015gna, Bergliaffa:2022}, with the extremal case 
being analyzed in~\cite{Lemos:2015ama, Lemos:2016pyc}, where
one also has a fundamental pressure equation of state and 
the exterior region is now Reissner-Nordstr\"om spacetime. As 
we shall see in the next chapters, the thermodynamics
of these matter thin shells has some relation with the 
thermodynamics of black holes and their statistical ensembles inside 
a cavity\cite{York:1986, Zaslavskii:1991, 
Andre:2020czm, Andre:2021ctu}. 
As the boundary of such cavity is given by a non-massive 
spherically symmetric shell, the thermodynamic variables 
of the two systems can be similar.

The thermodynamics of matter thin shells can also yield 
black hole thermodynamics using the quasiblack hole 
approach~\cite{Lemos:2007yh, Lemos:2009uk, Lemos:2010kw, 
Lemos:2020ooh}. This approach avoids the setting of the 
specific equations of state, as one keeps the shell's 
gravitational radius fixed and one changes its proper mass and 
radius, so that the configuration is maintained near the black hole 
threshold. By using the integration of the first law 
over these configurations, one retrieves the black hole properties 
in a model independent way.

Therefore, the study of thin shells, together with black holes and 
quasiblack holes, is of great importance in the understanding of 
thermodynamics of spacetimes. Following these themes, in this chapter, 
we analyze the entropy and the thermodynamic stability of 
a static electrically charged spherical thin shell in $d$ dimensions, 
with the fundamental pressure equation of state, and 
we also study the black hole limit, extending the analysis 
of~\cite{Andre:2019zzo} to the electrically charged case and 
of~\cite{Lemos:2015gna} to higher dimensions. We impose 
an equation of state for the temperature and the electric potential 
so that the entropy of the shell is described by a power law in the 
gravitational radius, and find that intrinsic stability for the possible 
fluctuations of the shell requires positive heat capacity, positive 
isothermal compressibility and positive isothermal electric susceptibility.
By performing the black hole limit, we find the black hole thermodynamic 
properties, such as the Smarr formula and thermodynamic stability. 
The intrinsic stability analysis followed here is provided in~\cite{Callen}.

The work presented in this chapter is mainly based 
on~\cite{Fernandes:2022gjd}. The chapter is organized as follows. 
In Sec.~\ref{sec:thinshellformalism}, we construct the spacetime 
solution containing an electrically charged matter thin shell 
using the thin shell formalism. In Sec.~\ref{sec:firstlaw}, we 
apply the first law of thermodynamics of the shell together with the 
fundamental pressure equation of state to obtain the entropy of the 
shell for two specific equations of state and further analyze the black hole 
limit. In Sec.~\ref{sec:intrinsicstabshell}, we analyze the intrinsic 
stability of the shell with the specific equations of state, including the 
case of the black hole limit. In Sec.~\ref{sec:intrinsicstablab}, we treat 
the intrinsic stability in terms of physical quantities, namely the 
heat capacity, the isothermal compressibility and the isothermal electric 
susceptibility. In Sec.~\ref{sec:conc1stlaw}, we conclude.

\section{Electrically 
charged matter thin shell
in higher dimensions\label{sec:thinshellformalism}}

\subsection{Formalism}

In order to construct the spacetime with an electrically matter thin shell, 
we first consider a curved spacetime 
containing a Maxwell field and additional matter. 
The metric is described by the Einstein 
equations 
\begin{align}
    G_{\alpha \beta}=8\pi G T_{\alpha \beta}\,,
\label{ch2eq:einstein}
\end{align}
where $G_{\alpha \beta}$ is the Einstein tensor given in therms of the 
metric $g_{\alpha \beta}$ and its derivatives, $T_{\alpha \beta}$ is 
the stress-energy tensor, while
the Maxwell field is described 
by the Maxwell equations
\begin{align}
    \nabla_b\hskip0.04cm{F^{ab}}= J^{a}\,,
    \label{ch2eq:max}
\end{align}
where $F_{\alpha \beta}$ is the Maxwell tensor obeying the 
internal equations $\nabla_{[\alpha} F_{\beta \gamma]}$,  
with the Maxwell tensor being described by the vector 
potential $A_\alpha$ as $F_{\alpha \beta} = \nabla_{[\alpha} A_{\beta]}$, 
and $J^a$ is the electric current. 
For the case of an electrovacuum 
spacetime, the stress-energy tensor $T_{\alpha \beta}$ is 
\begin{align}
    T_{\alpha \beta}=\varepsilon
\left(
{F_\alpha}^\gamma F_{\beta \gamma}-\frac14 g_{\alpha \beta}F^{\gamma \nu} 
F_{\gamma \nu}
\right)
\,,
\label{ch2eq:stressenergytensormaxwell}
\end{align}
where we define the parameter $\varepsilon$ 
as $\varepsilon = \epsilon \frac{(d-3)}{\Omega}$, 
the parameter $\epsilon$ being the electromagnetic coupling constant, 
and the area of a $d-2$ unit sphere is $\Omega = \frac{2\pi^{\frac{d-1}{2}}}
{\Gamma\left[\frac{d-1}{2}\right]}$.

As we have an electrically charged thin shell, the spacetime is 
divided into two bulk regions, the interior region $\mathcal{V}_1$
and the exterior region $\mathcal{V}_2$, both obeying to 
Eqs.~\eqref{ch2eq:einstein} and~\eqref{ch2eq:max}, with the 
stress energy tensor given by Eq.~\eqref{ch2eq:stressenergytensormaxwell}
and zero electric current. Moreover, there is a boundary 
timelike hypersurface, $\Sigma$, 
corresponding to the thin shell, between the two regions. 
In order to match the two regions at the thin shell, one requires 
the fulfillment of appropriate junction conditions, according 
to \cite{Israel:1966zz} in general relativity.

For the interior region $\mathcal{V}_1$, the coordinates 
associated to this region are $x_1^{\alpha}$ and the metric is 
$g\indices{_1_{\alpha \beta}}$, with the 
corresponding covariant derivative $\nabla\indices{_1_\alpha}$. 
The covector normal to the 
thin shell in this region is $n\indices{_1_{\alpha}}$, 
and so one can build tangent vectors 
$(e_1)\indices{^\alpha_a} = \frac{\partial x_1^\alpha}{\partial y^a}$, 
where $y^a$ are the associated coordinates to the thin shell. The 
vector potential in the interior region is $A\indices{_1_\alpha}$
with the field strength $F\indices{_1_{\alpha\beta}} = 
\nabla\indices{_1_{[\alpha}} A\indices{_1_{\beta]}}$.
In the same way, for the exterior region $\mathcal{V}_2$, 
we have the same definitions but with the subscript $2$, i.e. 
the coordinates $x_2^\alpha$, metric $g\indices{_2_{\alpha \beta}}$,
covariant derivative $\nabla\indices{_2_\alpha}$, 
the normal covector $n\indices{_2_\alpha}$, 
the tangent vectors on the hypersurface 
$(e_2)\indices{^\alpha_a}$, the vector potential 
$A\indices{_2_\alpha}$ and strength field tensor 
$F\indices{_2_{\alpha \beta}}$.

The boundary timelike hypersurface $\Sigma$, with coordinates 
$y^a$, is the thin shell and it is shared by the two regions 
$\mathcal{V}_1$ and $\mathcal{V}_2$. One can perform the pull-back 
of the tensorial quantities living in the product of 
cotangent spaces of both regions to define these quantities at 
the hypersurface. The junction conditions then yield the 
relation between the quantities evaluated in both regions. 
For the interior region, the pull-back of the metric 
is defined as $ 
g\indices{_1_{\alpha \beta}} (e_1)\indices{^{\alpha}_{a}}(e_1)
\indices{^{\beta}_{b}} 
\equiv h\indices{_i_{ab}}$, and the pull-back of 
$\nabla_{1\alpha} n_{1\beta}$ is
$\nabla\indices{_1_{\alpha}}n\indices{_1_{\beta}}
(e_1)^{\alpha}_{a} (e_1)^{\beta}_{b} = 
\nabla\indices{_1_{a}} n\indices{_1_{b}} \equiv K\indices{_1_{a b}}$,
where $K\indices{_1_{a b}}$ is the extrinsic curvature 
of the hypersurface measured in the interior region. 
Additionally, 
in the interior region, the pull-back of the vector potential 
is $A\indices{_1_\alpha}(e_1)\indices{^\alpha_a}\equiv A_{1a}$, 
and the strength field tensor can be decomposed in two parts, 
the complete pull-back 
$F_{1 \alpha \beta} (e_1)^{\alpha}_{a}(e_1)^{\beta}_{b}
\equiv F_{1ab}$ and the pull-back of $F_{1 \alpha \beta} n^{\beta}$ 
as $F_{1 \alpha \beta} n^{\beta} (e_1)^{\alpha}_{a} \equiv F_{1 a}$.
For the exterior region, the same definitions and 
pullbacks apply but where subscript $1$ is 
replaced by the subscript $2$.

The junction conditions to relate the above quantities in both 
regions can be obtained by assigning a normal
geodesic coordinate common to both regions in the neighbourhood 
of the hypersurface and the conditions for the metric are obtained by 
imposing regularity of the Levi-Civita connection and 
imposing the Einstein equations, considering only the Dirac delta terms. 
This can also be achieved by variational principle using the Einstein-Hilbert 
action together with the Gibbons-Hawking-York boundary terms at the 
hypersurface. From existence of the common normal coordinate, 
one has that the normal covectors in both regions must be the same 
at the hypersurface, but it does not apply to its derivatives. 
From the regularity of Levi-Civita connection, the 
junction condition reads
\begin{align}
    [h_{ab}] = 0\,, 
\label{ch2eq:Curvaturecondition1}
\end{align}
where $[h_{ab}]$ means $[h_{ab}]=h_{2ab}-
h_{2ab}$ and the same applies for other quantities.  The 
junction condition in Eq~\eqref{ch2eq:Curvaturecondition1} 
means that the induced metric at the hypersurface must be 
recovered from both sides, i.e. $h_{ab}=h_{1ab}=
h_{2\alpha\beta}$, and this establishes the relation between 
the coordinates chosen in both regions. From the Einstein 
equations, the junction condition reads
\begin{align}
    -\Big([K_{ab}] -
        [K]h_{ab} \Big)=
        {8\pi G} S_{ab}  \,,
        \label{ch2eq:Curvaturecondition}
\end{align}
where $K$ is defined as the trace of the extrinsic curvature 
$K_{a b}$,
and $S_{ab}$ is defined as the stress-energy tensor for
matter in the shell. For the stress-energy tensor, it is 
assumed that the matter is described by a perfect fluid, i.e. 
\begin{equation}
    S_{a b} = (\sigma + p)u_a u_b +
    p h_{a b}\,,\label{ch2eq:Sab}
\end{equation} 
where $\sigma$ is the matter density, $p$ is the matter pressure 
and $u_a$ is the velocity of the fluid on the boundary.

One also has junction conditions for the vector potential 
and the strength field tensor, as they can be extracted by 
imposing the continuity of the vector potential and 
the Maxwell equations. Indeed, one has
\begin{align}
    &[A_a] = 0\,,
    \label{ch2eq:Vectorcondition1}\\
    &[F_{a b}] = 0\,\,,\,\,\,
    [F_{a}] = j_a\,,
    \label{ch2eq:Vectorcondition2}
\end{align}
where $F_{a}$ is the pull-back of $F_{\alpha \beta}n^{\beta}$
for each region at the hypersurface, and 
$j_a$ is the electric current given by
\begin{align}
    j_a=
    {\zeta}
    \sigma_e u_a,
        \label{ch2eq:ja}
\end{align}
with $\sigma_e$ being the electric charge density, 
$\zeta$ defined as $\zeta=\frac{\Omega }{\epsilon_q}$,
and $\epsilon_q$ being the electric permittivity.

The formalism can now be applied to a $d$-dimentional 
spacetime with a Minkowski interior
in region $\mathcal{V}_1$, an electrically charged shell 
at the hypersurface and a Reissner-Nordstr\"om-Tangherlini 
exterior in region $\mathcal{V}_2$. The electromagnetic 
coupling $\epsilon$ appearing in 
Eq.~\eqref{ch2eq:stressenergytensormaxwell}
and the electric permittivity $\epsilon_q$ appearing in 
Eq.~\eqref{ch2eq:ja} are set to unity as a choice of 
convention, i.e. $\epsilon=1$ and $\epsilon_q=1$. However, 
the next subsection, the metric and vector potential dependence 
on these parameters are shown. The reason for 
not setting them at the start is to show the possible conventions 
in the literature and make the conversions easier.

\subsection{The solution of a spacetime with an electrically charged 
thin shell}

As the interior region is a vacuum $d$-dimensional spherically symmetric 
Minkowski spacetime, the line element for the metric in region 
$\mathcal{V}_1$ is
\begin{align}
    ds_1^2 = - dt_1^2 + dr^2 + r^2 d\Omega^2\,,
    \quad 0\leq r \leq R_1\,,
    \label{ch2eq:in1}
\end{align}
where spherical coordinates have been adopted, i.e. 
$x_1^\alpha = (t_1, r_1, \theta_1^A)$ together with 
$r\equiv r_1$, $d\Omega^2$ is the line element of 
a $(d-2)$ unit sphere, and $R_1$ is the radius of the 
shell at region $\mathcal{V}_1$. Regarding the vector potential 
$A_{1\alpha}$, we have
\begin{align}\label{ch2eq:covi}
    A_{1t_1} = A_1\,,
\end{align}
where $A_1$ is a constant, with the other remaining 
components being zero.

The exterior region $\mathcal{V}_2$ is described by the 
$d$-dimensional Reissner-Nordstr\"om spacetime, 
also known as Reissner-Nordstr\"om-Tangherlini spacetime, 
with the line element
\begin{align}
    ds_2^2 = - f(r)\, dt_2^2 + f(r)^{-1} dr^2 +
    r^2 d\Omega^2\,,
    \quad R_2\leq r\leq \infty\,,
    \label{ch2eq:ex1}
\end{align}
where Schwarzschild-like coordinates 
$x_2^{\alpha} = (t_2,r_2,\theta_2^A)$ have been adopted, 
$r\equiv r_2$ has been used, $R_2$ is the radius of the 
thin shell at the region $\mathcal{V}_2$, and the 
function $f(r)$ is
\begin{align}
    f(r) = 1 - \frac{2\mu m}{r^{d-3}} +
    \frac{q Q^2}{r^{2(d-3)}}\,,
\label{eqch2:fofr1}
\end{align}
with $m$ being the ADM mass, $Q$ being the 
total electric charge and
\begin{align}
    \mu = \frac{8\pi G}{(d-2)\Omega}\,,
\quad\quad \lambda = \frac{8\pi G
}{(d-2)\Omega}\,.
\label{eqch2:muq}
\end{align} 
It must be noted that the choice of $\epsilon$ and 
$\epsilon_q$ has been made to be unity, otherwise 
the quantity $\lambda$ would be given by 
$\lambda = \frac{8\pi G \epsilon}{(d-2)\Omega{\epsilon_q^2}}$.
Also, the definition of electric charge used here is 
$\frac{1}{2}\int F^{\alpha \beta}dS_{\alpha \beta} = 
\frac{\Omega Q}{\epsilon_q}$, where $dS_{\alpha \beta}$ 
is the surface element.

The exterior Reissner-Nordstr\"om metric has its 
gravitational radius, $r_+$, and Cauchy radius, $r_-$, 
given by the parameters $m$ and $Q$ in the following way
\begin{align}
    r_{\pm}^{d-3} = \mu m \pm \sqrt{\mu^2 m^2 - \lambda Q^2}
    \,,\label{ch2eq:rpm}
\end{align}
Since in this case, only the exterior region is 
Reissner-Nordstr\"om spacetime, the gravitational radius 
and the Cauchy radius are not horizon radii if the shell radius 
is larger than the gravitational radius. The extremal case is 
defined by the relation $r_+=r_-$, which in terms of the mass 
and charge is $\mu m = \sqrt{\lambda} Q$. This last relation 
for the choice of $\epsilon = \epsilon_q=1$ is 
$\sqrt{\mu} m = \sqrt{\mu} Q$. The area associated to 
the gravitational radius, $A_+$, is an important quantity 
that is considered in the analysis, and it is given by
\begin{align}
    A_+ = \Omega r_+^{d-2}\,.
    \label{ch2eq:A+}
\end{align}
It is also helpful to invert Eq.~\eqref{ch2eq:rpm}, to obtain the 
mass and the electric charge in terms of the horizon and Cauchy 
radii as 
\begin{align}
    &m = \frac{1}{2\mu}(r_+^{d-3} + r_-^{d-3})
    \,,\quad\quad
    Q = \frac{(r_+ r_-)^{\frac{d-3}{2}}}{\sqrt{\lambda}}\,.
\label{ch2eq:masscharge}
\end{align}
With the choice of $\epsilon=\epsilon_q = 1$, the parameter
$\lambda$ can be replaced by $\mu$. Nevertheless, the parameter $\lambda$ 
is kept throughout the chapter whenever the coefficient is associated to the 
electric charge $Q$. The function $f(r)$ can then be 
rewritten in terms of $r_+$ and $r_-$ as 
\begin{align}
    f(r) = 
\left(1-\left(\frac{r_+}{r}\right)^{d-3}\right)
    \left(1-\left(\frac{r_-}{r}\right)^{d-3}\right)
\,.
\label{ch2eq:fofr2}
\end{align}
For the exterior region as well, the vector potential that solves the 
vacuum Maxwell equations with the presence of a total electric charge is
\begin{align}
    A_{2t_2} = - \frac{Q}{ (d-3) r^{d-3}} \,,
    \label{ch2eq:cove}
\end{align}
where the constant of integration has been set to zero, 
as one can always make a gauge choice, and the other remaining 
components vanish. The strength field tensor has a non-zero 
component $F_2^{t_2 r} = \frac{Q}{r^{d-2}}$, which can be understood 
as the electric field with respect to a stationary observer. 
Notice that $\epsilon_q=1$ here, otherwise one would have 
$A_{2t_2} = - \frac{Q}{(d-3)\epsilon_q} r^{d-3}$ and 
$F_2^{t_2 r} = \frac{Q}{\epsilon_q r_2^{d-2}}$.

For the boundary hypersurface, describing the history of a thin 
shell, we assume spherical symmetry and so the induced metric 
$h_{ab}$ or the line element $ds_\Sigma^2 = h_{ab}dy^a dy^b$ 
is given by
\begin{align}
    ds_\Sigma^2 = -d\tau^2 + R(\tau)^2 d\Omega^2\,,
    \label{ch2eq:sigma1}
\end{align}
where the coordinate system $y^a=(\tau,\theta^A)$ has been adopted, 
with the coordinate $\tau$ being the proper time of the shell, and 
$R(\tau)$ being the radius of the shell. The vector potential 
must be constant along the shell due to spherical symmetry, up to 
gauge transformations, as
\begin{align}
    A_{\Sigma \tau} = A_\Sigma\,,
    \label{ch2eq:bsigma}
\end{align}
with $A_\Sigma$ being a constant and with the remaining components 
being zero.

With the two bulk regions and the hypersurface described, one now can 
proceed with the junction conditions. The pull-back of the metric 
in the interior region $\mathcal{V}_1$ evaluated at the hypersurface 
$\Sigma$ is
\begin{align}
    ds_{1,\Sigma}^2 = \left(-  \Dot{t}_1^2 +
    \Dot{R}_1^2\right) d\tau^2 +
   R_1(\tau)^2 d\Omega^2\,,
   \label{ch2eq:metricinpullb}
\end{align}
where the hypersurface $\Sigma$ is described by 
$r = R_1(t_1)$, $R_1(t_1)$ being the radius of the 
shell in function of the time coordinate in $\mathcal{V}_1$, 
and where $\Dot{\,\,} = \frac{d}{d\tau}$. The pull-back of the 
metric in the exterior region $\mathcal{V}_2$ evaluated at the 
hypersurface is
\begin{align}
    ds_{2,\Sigma}^2 =& \Big[- f(R_2(\tau)) \Dot{t}_2^2 +
    f(R_2(\tau))^{-1} \Dot{R}_2^2\Big] d\tau^2 + 
    R_2(\tau)^2 d\Omega^2\,,
    \label{ch2eq:metricout}
\end{align}
here the hypersurface $\Sigma$ is described by 
$r = R_2(t_2)$, $R_2(t_2)$ being the radius of the 
shell in function of the time coordinate in $\mathcal{V}_2$.
The first junction condition, Eq.~\eqref{ch2eq:Curvaturecondition1} 
now states that the induced 
metrics in Eqs.~\eqref{ch2eq:metricinpullb} 
and~\eqref{ch2eq:metricout} must correspond to the same physical 
induced metric and additionally must correspond to 
Eq.~\eqref{ch2eq:sigma1}. Since the three metrics are in the 
same coordinate system, the first junction condition yields
\begin{align}
    &R_2(\tau) = R_1(\tau) = R(\tau)\,,
    \label{ch2eq:conditioncontinuitymetric}\\
    & - \Dot{t}_1^2 + \Dot{R}^2 =-
    f(R) \Dot{t}_2^2 + f(R)^{-1} \Dot{R}^2
    =-1\,.\label{ch2eq:conditioncontinuitymetric2}
\end{align}
It must be noted that condition 
Eq.~\eqref{ch2eq:conditioncontinuitymetric} 
motivates the usage of the same coordinate $r$ 
for the interior and exterior region, as it was done 
apriori. 
The area of the shell can then be defined as 
\begin{align}
    A = \Omega R^{d-2}\,.
    \label{ch2eq:A}
\end{align}

Moving to the second junction condition, the normal covector 
must be specified for each region in order to compute the 
extrinsic curvature $K_{ab}$. For the interior, the normal covector 
can be deduced from the hypersurface equation $r=R_1(t_1)$ as
\begin{align}
n_{1\alpha}dx^\alpha = \left(1 -
\left(\frac{dR}{dt_1}\right)^2\right)^{-\frac12}
\left( - \frac{d R}{t_1} dt_1+dr \right)\,.
\end{align}
It is useful to write 
the components of the normal covector in terms of the 
coordinate $\tau$ at the hypersurface. Using the first junction condition 
in Eq.~\eqref{ch2eq:conditioncontinuitymetric2}, one has
\begin{align} 
&\eval{
\left(1 -
\left(\frac{dR}{dt_1}\right)^2\right)^{-\frac12}
}_\Sigma = \sqrt{1 +
    \Dot{R}^2}\,\,,\\
&
\eval{\frac{dR}{t_1}}_\Sigma = \frac{ \Dot{R}}{\sqrt{1
    + \Dot{R}^2}}\,\,, 
\end{align}    
so that
\begin{align}
n_{1\alpha}|_\Sigma = \left(-\Dot{R},\sqrt{1+ \Dot{R}^2},0,0\right)\,.
\end{align}
For the exterior region, the normal covector is given by
\begin{align}
n_{2\alpha}dx^{\alpha} =\
\sqrt{f(r_2)}\left(
f(r_2)^2 -
\left(\frac{dR}{dt_2}\right)^2\right)^{-\frac12}
\left(- \frac{dR}{dt_2} dt_2+ dr \right)\,.
\end{align}
Written in terms of the coordinate $\tau$, and using 
\begin{align}
&\sqrt{f(r_2)}\eval{
\left(
f(r_2)^2 -
\left(\frac{dR}{dt_2}\right)^2\right)^{-\frac12}
}_\Sigma = \frac{\sqrt{f(R) +
    \Dot{R}^2}}{f(R)}\,\,,\\
&\eval{\frac{dR}{dt_2}}_\Sigma = \frac{f(R) \Dot{R}}{\sqrt{f(R)
    + \Dot{R}^2}}\,,
\end{align}
the normal covector at the hypersurface is
\begin{align}
    n_{2\alpha}|_\Sigma = \left(-\Dot{R},  \frac{\sqrt{f(R) +
    \Dot{R}^2}}{f(R)},0,0\right)\,.
\end{align}
The extrinsic curvature can now be calculated, 
giving for the interior region 
\begin{align}
    K\indices{_1^{\tau}_{\tau}} =
    \frac{\Ddot{R}}{\sqrt{1 + \Dot{R}^2}}
\,,\quad
K\indices{_1^{\theta^A}_{\theta^A}}
    = \frac{\sqrt{1 + \Dot{R}^2}}{R}\,,
\label{ch2eq:ExtrinsicCurv}
\end{align}
and for the exterior region 
\begin{align}
    K\indices{_2^{\tau}_{\tau}} = \frac{\Ddot{R} +
        \frac{\partial_{R} f(R)}{2}}{\sqrt{f(R) + \Dot{R}^2}}
    \,,\quad
    K\indices{_2^{\theta^A}_{\theta^A}} = \frac{\sqrt{f(R) +
        \Dot{R}^2}}{R}\,,
    \label{ch2eq:ExtrinsicCurv2}
\end{align}
where the indices $A$ in this case are not to be summed over,
The shell is assumed to be static and in equilibrium, meaning 
$\Dot{R}= 0$ and $\Ddot{R} = 0$, together with $u^\alpha = (1,0,0)$, 
the velocity of the shell. From the second junction condition, 
i.e. Eqs.~\eqref{ch2eq:Curvaturecondition} and~\eqref{ch2eq:Sab}, 
together with the expressions for the extrinsic curvature 
in Eqs.~\eqref{ch2eq:ExtrinsicCurv}-\eqref{ch2eq:ExtrinsicCurv2}, 
the energy density and the pressure can be obtained in terms of the 
gravitational and Cauchy radii of the exterior region, as 
\begin{align}
    &\sigma = \frac{1 - k}{\mu\Omega R}\,,\label{ch2eq:sig}\\
    &p = \frac{1}{2\mu\Omega R^{2d-5}k} \frac{d-3}{d-2}
    \Bigg[(1-k)^2 R^{2(d-3)}-\lambda
    Q^2\Bigg]\,,
    \label{ch2eq:p}
\end{align}
where $k$ is defined as the redshift function evaluated at 
the shell radius $R$, as
\begin{align}
    k = \sqrt{f(R)}
\,.\label{ch2eq:reds}
\end{align}
As a reminder, the parameter $\lambda$ could be renamed 
as $\mu$ in Eq.~\eqref{ch2eq:p} due to the choice in 
Eq.~\eqref{eqch2:muq}. Knowing the energy density, the 
rest mass of the shell can be defined by 
\begin{align}
    M =  \Omega R^{d-2} \sigma = \frac{R^{d-3}}{\mu}
    \left(1-k\right)\,,\label{ch2eq:M}
\end{align}
This 
relation can be inverted to get the ADM mass in terms 
of the rest mass of the shell and the electric charge as 
\begin{align}
    m = M - \frac{\mu M^2}{2 R^{d-3}} +
    \frac{Q^2}{2
    R^{d-3}}\,,\label{ch2eq:m}
\end{align}
where it was used $k(M,R,Q) =\sqrt{ 1 - \frac{2\mu m}{R^{d-3}} +
\frac{\lambda Q^2}{R^{2(d-3)}}}$.
The expression in Eq.~\eqref{ch2eq:m} can be interpreted 
as the total energy of the self-gravitating shell being 
given by the rest mass plus the gravitational potential energy 
and the electric potential energy. Written with generic $\epsilon$ 
and $\epsilon_q$, Eq.~\eqref{ch2eq:m} is 
$m = M - \frac{\mu M^2}{2 R^{d-3}} +
\frac{{\epsilon}Q^2}{2{\epsilon_q^2}
R^{d-3}}$.

With respect to the junction conditions of the Maxwell 
vector potential, the first junction condition in 
Eq.~\eqref{ch2eq:Vectorcondition1}, which is $[A_\tau] = 0$,
together with the expressions of the vector potential 
in Eqs.~\eqref{ch2eq:covi} and~\eqref{ch2eq:cove},
determines the constant
\begin{align}
    A_i=-\frac{Q}{(d-3) R^{d-3}k} \,,
 \label{Aijunc}
\end{align}
which can be written as $A_i=-\frac{Q}{(d-3){\epsilon_q} R^{d-3}k}$
for generic $\epsilon$ and $\epsilon_q$. 
The relevant junction condition for the strength field tensor 
in Eq.~\eqref{ch2eq:Vectorcondition2} is 
$[F_{\alpha}] = j_\alpha$, 
which upon using
Eq.~\eqref{ch2eq:ja}
becomes 
$-(F_2)_{tr}k\Dot{t}_2 =\zeta \sigma_e$, 
with 
$\zeta = \frac{\Omega}{\epsilon_q}$, see
Eq.~\eqref{ch2eq:ja}, $\epsilon_q=1$ and
$\zeta=\Omega$.
The junction condition implies,
with Eqs.~\eqref{ch2eq:covi}
and~\eqref{ch2eq:cove}, that
\begin{align}
Q = R^{d-2} \sigma_e\,.
\end{align}
This relates the total
electric charge with its corresponding charge
density.

\section{Thermodynamics of the electrically charged thin shell
from the first law\label{sec:firstlaw}}\sectionmark{
    Thermodynamics of the shell from the first law
    }\thispagestyle{userightbotmark}

\subsection{The entropy of the shell from the first law of thermodynamics}

With the electrically charged matter thin shell spacetime constructed, 
one way to study its thermodynamics is by imposing the first law 
of thermodynamics. By matching the internal energy of the shell 
to its rest mass, the first law for the shell can be written 
in a way to determine the differential of the shell entropy $S$ as 
\begin{align}
    dS = \beta dM + \beta p dA - \beta \Phi dQ\,,
    \label{ch2eq:1stlaw2}
\end{align}
where 
\begin{align}
    \beta=\frac1T \,,
\label{ch2eq:inverseT}
\end{align}
is the inverse of the local temperature, $T$, of the shell, 
$M$ is the rest mass of the shell determined in the previous section, 
$p$ is the pressure of the shell determined in the previous section, 
$A$ is the area of the shell, $Q$ is the electric charge 
of the shell and $\Phi$ is the thermodynamic electric potential.
The main objective is to compute the entropy of the shell by integrating 
Eq.~\eqref{ch2eq:1stlaw2}. In order to proceed, one must provide 
an equation of state for the three quantities
\begin{align}
    &\beta = \beta(M,A,Q)\,\,,\label{ch2eq:es1}\\
    &p= p(M,A,Q)\,\,,\label{ch2eq:es2}\\
    &\Phi = \Phi(M,A,Q)\,\,,\label{ch2eq:es3}
\end{align}
in function of the rest mass $M$, the area of the shell $A$ and 
the electric charge $Q$. We now choose the pressure equation of state  
as the one in Eq.~\eqref{ch2eq:p}, stemming from the junction 
conditions, with the function $k$ being written in terms of 
$M$, $A$ and $Q$, and additionally $R$ being written in terms of 
$A$. This is referred to the fundamental pressure equation of state 
coming from the Einstein equations, and we shall see its consequences. 
The choice of the remaining equations of 
state $\beta(M,A,Q)$ and $\Phi(M,A,Q)$ is not completely free as 
the functions $\beta$ and $\Phi$ must satisfy integrability conditions. These conditions 
must be first analyzed before the equations of state are chosen.

The integrability conditions are related to the fact that $S$ is 
a function or scalar, depending on the thermodynamic parameters 
$(M,A,Q)$. So, its differential must be exact by definition, which 
is translated to the Hessian matrix of $S$ being a symmetric matrix.
From the first law in Eq.~\eqref{ch2eq:1stlaw2}, the first derivatives 
of the entropy are 
\begin{align}
    &\Big(\pdv{S}{M}\Big)_{A,Q} = \beta\,,\\
    &\Big(\pdv{S}{A}\Big)_{M,Q} = \beta p\,,\notag\\
    &\Big(\pdv{S}{Q}\Big)_{M,A} = - \beta \Phi\,,
\end{align}
and so the integrability conditions read
\begin{align}
    &\Big(\pdv{\beta}{A}\Big)_{M,Q} =
    \Big(\pdv{\beta p}{M}\Big)_{A,Q}\,,\notag\\
    &\Big(\pdv{\beta}{Q}\Big)_{M,A} = -
    \Big(\pdv{\beta \Phi}{M}\Big)_{A,Q}\,,\notag\\
    &\Big(\pdv{\beta p}{Q}\Big)_{M,A} = -
    \Big(\pdv{\beta \Phi}{A}\Big)_{M,Q}\,.
    \label{ch2eq:ExactConditions}
\end{align}

The idea is to use the fundamental pressure equation of state to 
determine the restrictions to the expression of the inverse 
temperature and to the thermodynamic electric potential. 
In order to simplify this task, it is useful to proceed with 
a parameter transformation from ($M,A,Q$) or ($M,R,Q$) 
to the parameters ($r_+,r_-,R$), which can be done using 
Eq.~\eqref{ch2eq:masscharge} and Eq.~\eqref{ch2eq:A}. For completeness, 
the redshift function in these parameters is 
\begin{align}
    k(r_+,r_-,R) =
\sqrt{\left(1-\left(\frac{r_+}{R}\right)^{d-3}\right)
    \left(1-\left(\frac{r_-}{R}\right)^{d-3}\right)}\,\,.
    \label{ch2eq:redshift}
\end{align} 
By using the chain rule, one can transform the differential of 
the entropy to be dependent on the parameters ($r_+,r_-,R$)
with the derivatives of the entropy being 
\begin{align}
    \left(\pdv{S}{r_\pm}\right)_{r_\mp,R} = \beta
    \left(\pdv{M}{r_\pm}\right)_{r_\mp,R} - \beta \Phi
    \left(\pdv{Q}{r_\pm}\right)_{r_\mp,R}.\label{ch2eq:derSderrpm}
\end{align}
and
\begin{align}
    &\left(\pdv{S}{R}\right)_{r_+,r_-} = \beta
    \left(\pdv{M}{R}\right)_{r_+,r_-} + \beta p
    \left(\pdv{A}{R}\right)_{r_-,r_+} = 0\,,\label{ch2eq:derSderR}
\end{align} 
where it was used
\begin{align}
    &\left(\pdv{M}{R}\right)_{r_+,r_-} = - p
    \left(\pdv{A}{R}\right)_{r_+,r_-}\,\,,
\end{align}
from Eqs.~\eqref{ch2eq:p},~\eqref{ch2eq:M} and~\eqref{ch2eq:masscharge}.
As it can be seen from Eq.~\eqref{ch2eq:derSderR}, 
the consequence of having 
the fundamental pressure equation of state is that the 
entropy, supposedly a function $S=S(r_+,r_-,R)$, does not 
depend on the radius of the shell $R$, it only depends on 
$r_+$ and $r_-$, i.e.
\begin{align}
    S=S(r_+,r_-)\,.
\label{entropysimplified}
\end{align}

The integrability conditions can be used to further restrict 
the remaining derivatives of the entropy in the parameters $(r_+,r_-,R)$, 
by finding the expressions for $\beta$ and $\Phi$. 

For $\beta$, one can compute the 
derivative of $\beta$ using the chain rule as
\begin{align}
    \left(\pdv{\beta}{R}\right)_{r_+,r_-} =
    \left(\pdv{\beta}{M}\right)_{A,Q}
    \left(\pdv{M}{R}\right)_{r_+,r_-}
    +\left(\pdv{\beta}{A}\right)_{M,Q}
    \left(\pdv{A}{R}\right)_{r_+,r_-}\,\,.
\end{align}
Using the first integrability condition in 
Eq.~\eqref{ch2eq:ExactConditions}, together with 
\begin{align}
    \left(\pdv{p}{M}\right)_{A,Q} =
    \frac{1}{(d-2)\Omega k
        R^{(d-3)}}\left(\pdv{k}{R}\right)_{r_+,r_-}\,\,,
        \label{ch2eq:dpdM}
\end{align}
one obtains a differential equation for $\beta$ 
\begin{align}
    \left(\pdv{\beta}{R}\right)_{r_+,r_-} = \frac{\beta}{k}
\left(\pdv{k}{R}\right)_{r_+,r_-}\,\,,
\end{align}
which can be integrated to give
\begin{align}
    \beta(r_+,r_-,R) = b(r_+,r_-) k\,,
\label{ch2eq:beta}
\end{align}
where $b(r_+,r_-)$ is a reduced equation of state and 
only depends on the nature of matter in the shell.
The meaning of $b(r_+,r_-)$ is made clear when the 
limit of $R\rightarrow++\infty$ is made in the expression of 
$\beta$, i.e. $\beta(r_+,r_-,+\infty) = b(r_+,r_-)$, where 
we emphasize that $k$ is given by Eq.~\eqref{ch2eq:redshift}
and becomes unity in the limit $R\rightarrow +\infty$. Therefore,
$b(r_+,r_-)$ is the inverse temperature of the shell measured 
by a stationary observer at infinity, while expression 
Eq.~\eqref{ch2eq:beta} describes the Tolman's formula for the 
temperature. We must stress that this is indeed a consequence 
of the choice of the fundamental pressure equation of state. 

For $\Phi$, one can use the chain rule once again to obtain the 
derivative
\begin{align}
    \left(\pdv{\Phi}{R}\right)_{r_+,r_-} =
\left(\pdv{A}{R}\right)_{r_+,r_-}\left(\pdv{\Phi}{A}\right)_{Q,M}
- p
\left(\pdv{A}{R}\right)_{r_+,r_-}\left(\pdv{\Phi}{M}\right)_{A,Q}\,\,,
\end{align}
using Eq.~\eqref{ch2eq:dpdM}. The integrability conditions 
in Eq.~\eqref{ch2eq:ExactConditions} can 
be rearranged to transform the last equation into 
\begin{align}
    \left(\pdv{\Phi}{R}\right)_{r_+,r_-} =
    -\left(\pdv{A}{R}\right)_{r_+,r_-}\left(\pdv{p}{Q}\right)_{M,A}
    - \Phi
    \left(\pdv{A}{R}\right)_{r_+,r_-}\left(\pdv{p}{M}\right)_{A,Q}\,\,.
\end{align}
Using the fundamental pressure equation of state, with its derivatives 
\begin{align}
    &\left(\pdv{p}{M}\right)_{A,Q} = \frac{1}{(d-2)\Omega k
    R^{(d-3)}}\left(\pdv{k}{R}\right)_{r_+,r_-}\,\,,\\
    &\left(\pdv{p}{Q}\right)_{M,A} =- \frac{Q(d-3)}{(d-2)\Omega k
    R^{2d-5}}\,\,,
\end{align}
the differential equation for the electric potential $\Phi$ is 
obtained as
\begin{align}
    \left(\pdv{k\Phi}{R}\right)_{r_+,r_-} = \frac{(d-3)Q}{R^{d-2}}\,\,,
\end{align}
which can be readily integrated into 
\begin{align}
    &\Phi(r_+,r_-,R) = \frac{Q}{k}
\Big[ c(r_+,r_-) -
    \frac{1}{R^{d-3}}
    \Big]\,,\label{ch2eq:Phi}
\end{align}
where $c(r_+,r_-)$ is a reduced equation of state like $b(r_+,r_-)$, 
and it depends on the nature of matter in the shell. Performing the 
limit $R\rightarrow +\infty$, one can see that 
$c(r_+,r_-) = \frac{\Phi(r_+,r_-,\infty)}{Q}$ and so $c(r_+,r_-)$ is the 
electric potential per charge 
measured by a stationary observer at infinity.

With the expressions for $\beta$ in Eq.~\eqref{ch2eq:beta} and 
for $\Phi$ in Eq.~\eqref{ch2eq:Phi}, one can obtain the derivatives 
of the entropy in Eq.~\eqref{ch2eq:derSderrpm} in terms of the reduced 
equations of state. The differential of the entropy $dS$ in the 
parameters $(r_+,r_-,R)$ becomes 
\begin{align}
    dS =& \frac{{(d-3)}b(r_+,r_-)}{2\mu}\Big[ \Big(1 - r_-^{d-3}
    c(r_+,r_-)\Big)r_+^{d-4}dr_+ \notag\\&+ \Big(1-
    r_+^{d-3}c(r_+,r_-)\Big)r_-^{d-4}dr_-\Big]\,.
    \label{ch2eq:entropydiff}
\end{align}
It must be noted that there are still integrability conditions that 
must be satisfied between $b(r_+,r_-)$ and $c(r_+,r_-)$ to ensure 
that the differential is exact, yielding 
\begin{align}
    &\pdv{b}{r_-}\,(1- c\, r_-^{d-3})r_+^{d-4} - \pdv{b}{r_+}\,(1-c
\,r_+^{d-3})r_-^{d-4} \notag\\&= \pdv{c}{r_-}b r_-^{d-3}r_+^{d-4} -
\pdv{c}{r_+}b r_+^{d-3}r_-^{d-4}\,.
\label{ch2eq:inteC}
\end{align}
The consequence of having the fundamental pressure equation of state 
is that the entropy $S(r_+,r_-)$ depends on two reduced equations of 
state $b(r_+,r_-)$ and $c(r_+,r_-)$, related by Eq.~\eqref{ch2eq:inteC}.
These two functions cannot be further specified by general relativity 
or the first law of thermodynamics, and so their expression must be 
chosen depending on the class of matter that it is of interest.

It may also be interesting to rewrite the entropy in terms of the 
ADM mass, i.e. $S(m,Q)$, with its differential being given by 
$dS = \frac{b}{2\mu}
d(r_+^{d-3}+r_-^{d-3})
-\frac{b}{2\mu}c\,d[(r_+ r_-)^{d-3}]$, or in a cleaner way, 
$dS =bdm- b\phi dQ$, where $\phi=Q\,c$. In this case, the 
equations of state are a function of $m$ and $Q$ as 
$b=b(m,Q)$ and $c(m,Q)$. Notice that we are using the convention 
$\lambda = \mu$, and we write $Q$ here
as the modulus of the electric charge. This further stresses 
the meaning of $b$ and $c$ as the inverse temperature and 
the electric potential per charge at infinity.

\subsection{The entropy of the shell for a specific choice of equations of 
state}

We are now going to choose the two reduced equations of state
for $b(r_+,r_-)$ and $c(r_+,r_-)$. The choice for  
the reduced equation of state for the inverse temperature of the 
shell is
\begin{align}
    b(r_+,r_-) = \frac{a \gamma \Omega^{a-1}}{d-3}
      \frac{r_+^{a(d-2)}}{r_+^{d-3}-r_-^{d-3}}\,\,,
      \label{ch2eq:tempeqstate}
\end{align}
where $a$ is a free exponent and $\gamma$ is a free parameter. 
The equation of state is only valid for $r_- \leq r_+$, with 
$r_+$ and $r_-$ assume real values from Eq.~\eqref{ch2eq:rpm}.
The shell for this choice of equation of state can be either 
undercharged or extremely charge but not overcharged.

From the integrability 
condition Eq.~\eqref{ch2eq:inteC} with the choice of 
the reduced equation of state in Eq.~\eqref{ch2eq:tempeqstate},
one possible solution that we choose for $c(r_+,r_-)$ is 
\begin{align}
    c(r_+,r_-) = \frac{1}{r_+^{d-3}}\,,
\label{ch2eq:electeqstate}    
\end{align}
yielding the typical Reissner-Nordstr\"om equation of state for
the electric potential.

We should give a comment regarding the constants appearing in 
the equation of state $b(r_+,r_-)$ in 
Eq.~\eqref{ch2eq:tempeqstate}. One has two parameters $a$ and 
$\gamma$. The power law exponent $a$ is adimensional and
it is the most relevant in the analysis. The constant $\gamma$ 
should be determined by the features of matter, including quantum 
effects, and so it is expected that depends on the Planck constant 
and the Boltzmann constant, which we set to unity here. Moreover, 
$\gamma$ must have the units of length to the power $(d-2)(1-a)$. 
Regarding the equation of state for the electric potential, it does 
not depend on any new free parameter. We also note that both 
equations of state depend on the dimensions $d$, and one can treat 
$d$ as a free parameter, as long as it is a finite positive integer and 
$d>3$, which is case of interest here. 
There may be some interest in performing the limit of 
infinite $d$, but that depends on the way the limit is taken. We do not 
pursue that limit here.

With the reduced equations of state Eqs.~\eqref{ch2eq:tempeqstate}
and~\eqref{ch2eq:electeqstate}, the differential of the entropy 
given in Eq.~\eqref{ch2eq:entropydiff} can be integrated, yielding 
\begin{align}
    S = \    \frac{\gamma}{16\pi G}A_+^a\,,
    \label{ch2eq:suggestedentropy}
\end{align}
and so the entropy of the shell, being dimensionless in our convention, is
proportional to a power of the gravitational area $A_+$. Indeed, 
the specific choice of equations of state make the entropy $S$ 
only dependent on the gravitational radius $r_+$ only as 
$S=S(r_+)$. It is also convenient to restrict the parameter $a$ 
to $a>0$, so that the entropy does not diverge when 
$r_+ = 0$. An additional note is that this is still 
the entropy of the shell and $r_+$ is the gravitational radius 
of the shell and not the horizon radius of a black hole.

We now explain the motivation for the choice of the reduced equations of 
state in Eqs.~\eqref{ch2eq:tempeqstate}
and~\eqref{ch2eq:electeqstate}. First, 
power laws in thermodynamics and statistical mechanics 
emerge ubiquitously, therefore it is natural 
for $b(r_+,r_-)$ and $c(r_+,r_-)$ to be described 
by power laws in $r_+$ and $r_-$, these parameters being dependent 
on the rest mass $M$ and the charge $Q$. Second, it is of interest 
to assign black hole like behaviour to the shell, so that 
it is possible to perform the black hole limit $R=r_+$ and 
study its implications. Moreover, if one sets $a=1$,
the inverse temperature has the same dependence as the 
Hawking temperature of a black hole, while the electric 
potential is identical as the one from a black hole. 
Consequently, the entropy of the shell in the case $a=1$ 
is $S =\frac{\gamma}{16\pi G}A_+$, which has the same functional 
dependence as the Bekenstein-Hawking black hole entropy. 
However, one indeed could choose other power laws for the 
equation of state and they could possibly have the same 
black hole features for appropriate choice of the exponents. 
Another possibility for the equations of state that could be 
worth exploring and that generalizes the 
choice in \cite{Andre:2019zzo} for higher dimensions 
is choosing power laws in the ADM mass and 
the charge, i.e. $b(r_+,r_-) = a (r_+^{d-3} + r_-^{d-3})^\alpha$
and $c(r_+,r_-) = \frac{f(r_+r_-)}{(r_+^{d-3} +
r_-^{d-3})^\alpha}$, which obey the integrability conditions.
However, we did not explored such choices here.

The expression for the entropy in Eq.~\eqref{ch2eq:suggestedentropy}
brings an important point, the entropy is the same if $r_+$ 
is fixed for any radius $R$ of the shell. If one imagines 
the process of increasing the radius $R$ of the shell with fixed 
$r_+$ and fixed $r_-$, this implies that electric charge $Q$ 
and the entropy are constant, while the area of the shell increases 
and the internal energy of the shell decreases. 
In the limit $R\rightarrow+\infty$,
the internal energy assumes the value of the ADM mass.
And so this is an isentropic process.

To end this subsection, one can also obtain an integrated version of the 
first law of thermodynamics to the shell, knowing the entropy with 
Eq.~\eqref{ch2eq:suggestedentropy}, the internal energy or rest mass 
$M$ with Eq.~\eqref{ch2eq:M}, the electric charge with 
Eq.~\eqref{ch2eq:masscharge} and the area of the shell with 
Eq.~\eqref{ch2eq:A}. The energy of the shell can be written in terms 
of the entropy, area and the charge of the shell as
\begin{align}
    M(S,A,Q) = \frac{1}{\mu}\left(\frac{A}{\Omega}
    \right)^{\frac{d-3}{d-2}} \left[1 -
    \hspace{-2mm}\sqrt{\left(1 - \left(\frac{16\pi G
    S}{\gamma A^a}\right)^{\frac{d-3}{a(d-2)}}\right) \left(1 -
    \frac{q Q^2 \Omega^{2\frac{d-3}{d-2}}}{
\left(\frac{16\pi G
S A^a}{\gamma}\right)^{\frac{d-3}{a(d-2)}}}\right)}\hspace{1mm}\right].
\label{ch2eq:massSAQ}
\end{align}
The function $M(S,A,Q)$ has the scaling property 
\begin{align}
M\left(\nu S^{\frac{1}{a}}, \nu A,
\nu Q^{\frac{d-2}{d-3}}\right) = \nu^{
\frac{d-3}{d-2}} M
\left(S^{\frac{1}{a}}, A, Q^{\frac{d-2}{d-3}}\right)\,\,,
\end{align} 
and since the first law of thermodynamics states 
$dM = TdS - pdA + \Phi dQ$, one can use the Euler 
relation theorem for homogeneous functions to get 
\begin{align}
    \frac{d-3}{d-2}M = a T S - p A +
    \frac{d-3}{d-2} \Phi Q\,\,.\label{ch2eq:EulerRel}
\end{align}
This shows that the choice of equations of state alter 
the homogeneity of the variables compared to what is typical 
in homogeneous thermodynamic systems.

\subsection{The case of a shell with black hole features and the black hole 
limit}

In this subsection, we focus on the shell with black hole 
features. This can be done by setting $a=1$ in 
Eqs.~\eqref{ch2eq:tempeqstate}-\eqref{ch2eq:suggestedentropy}. The resulting
temperature of the shell for this case $T_0 = \frac{1}{b|_{a=0}}$ is
$T_0 = \frac{ d-3}{\gamma } \frac{r_+^{d-3}-r_-^{d-3}}
    {r_+^{d-2}}$,
which translates into the Hawking temperature of the shell if additionally 
one considers $\gamma=4\pi$. The reduced equation of state for the electric 
potential is still described by $c(r_+,r_-) = \frac{1}{r_+^{d-3}}$, which 
corresponds to the black hole electric potential. Therefore, the shell 
with black hole features has the reduced equations of state
\begin{align}
b_+(r_+,r_-) = \frac{4\pi}{d-3}
\frac{r_+^{d-2}}{r_+^{d-3}-r_-^{d-3}} \,\,  , \,\,
c_+(r_+,r_-)  
=  \frac{1}{r_+^{d-3}}\,\,,
    \label{ch2eq:tempelectshellbhfinal}
\end{align}
where the subscript $+$ indicates thermodynamic quantities 
characteristic of black holes. 
The entropy of the shell becomes
\begin{align}
    S_+ = \frac{1}{4}\frac{A_+}{A_{\mathrm{p}}}\,,
    \label{ch2eq:entropyshellfinal}
\end{align}
with $A_{\mathrm{p}} = l_{\mathrm{p}}^{d-2} = G$ being the 
Planck area. This means the shell with black hole features, 
i.e. $a=1$ and $\gamma=4\pi$, has precisely the 
Bekenstein-Hawking entropy. While the spacetime does not describe 
a black hole spacetime but rather a shell spacetime, we have that 
the shell mimics thermodynamically the black hole 
spacetime with horizon radius equal to the gravitational radius of 
the shell, $r_+$. Even more, this is independent of the radius of the 
shell $R$, with the condition $R\geq r_+$, since the entropy remains 
the same as the Bekenstein-Hawking entropy.

In order to get a shell which not only has black hole
features but it is almost a black hole, meaning 
a quasiblack hole, we must take the limit of 
the shell radius $R\rightarrow r_+$. This can only be done 
through a sequence of quasistatic thermodynamic equilibrium configurations
if the temperature shell is precisely the Hawking temperature, 
$T_+(r_+,r_-) = \frac{ d-3}{4\pi}
\frac{r_+^{d-3}-r_-^{d-3}} {r_+^{d-2}}$, with the entropy of the shell 
being the Bekenstein-Hawking entropy, in 
Eq.~\eqref{ch2eq:entropyshellfinal}. This is to avoid divergences from the 
backreaction of the matter quantum fluctuations.
When the shell is placed at its 
own gravitational radius, the shell spacetime describes a quasiblack hole 
state, with the gravitational radius being a quasihorizon radius. 
The pressure of the shell is very large as one approaches 
the radius of the shell to its gravitational radius, becoming 
divergent the limit, see 
Eq.~\eqref{ch2eq:p}. In some sense, this means the shell must 
have all its degrees of freedom excited in this limit to maintain 
equilibrium and to make the entropy maximal. As such, the limit 
$R\rightarrow r_+$ must be envisioned as $R$ being very close 
to $r_+$ but not at exactly 
$r_+$. While the shell formalism indeed provides 
the black hole features in the appropriate limit, the 
quasiblack hole formalism, having some correspondence 
with the membrane paradigm formalism, can deal with generic matter systems 
on the verge of becoming a black hole and provides also 
the thermodynamic properties of black holes~\cite{Lemos:2009uk, Lemos:2010kw,
Lemos:2020ooh}.

The extremal case of the charged shell deserves a complete study but we give
some highlights here 
in connection to the 
extremal Reissner-Nordstr\"om black hole in $d$ dimensions.   
The extremal Reissner-Nordstr\"om spacetime
satisfies the relation
$r_+ = r_-$.
For a reduced equation of state of the form given in 
Eq.~\eqref{ch2eq:tempeqstate},
we have
that the extremal charged shell case
has zero temperature,
whereas the reduced electric potential
is still given by Eq.~\eqref{ch2eq:electeqstate},
thus both are well-defined in the
extremal case. The expression for the entropy of the shell, 
however, requires some subtleties and depends on the order 
of the limits performed. The first case is 
when we start from a nonextremal shell 
with $R>r_+$ and we take the limit $r_+=r_-$. The entropy of the 
shell is then Eq.~\eqref{ch2eq:suggestedentropy}, by continuity. 
The second case is when we consider from the start an extremal shell. 
Then, the 
entropy is some function of the gravitational radius $S(A_+)$, 
and we are free to choose it, see \cite{Lemos:2015ama}. Proceeding 
with the black hole limit in the extremal case, the first case gives the 
Bekenstein-Hawking entropy $S_+ = \frac{1}{4}\frac{A_+}{A_p}$ 
and the second case gives an entropy 
determined by an unspecified function of the gravitational radius. 
There is an additional third case in the black hole limit, which is 
when we start from an undercharged shell with $R>r_+$ and then 
we bring simultaneously the shell to the brink of extremality while 
approaching it to its gravitational radius \cite{Lemos:2016pyc}. But in 
the third case, the entropy of the shell also becomes 
the Bekenstein-Hawking entropy. Connecting these cases to 
the properties of the black hole, as one makes the black hole limit 
of the shell, the entropy of the extremal black hole depends on its 
past, or better, on how it was formed, see~\cite{Lemos:2010kw}.

Another property that we can obtain from the black hole limit of the 
charged thin shell is the Smarr formula for the charged black hole, 
starting from the integrated first law formula, Eq.~\eqref{ch2eq:EulerRel}.
Setting $a=1$ in the reduced equations of state, the 
integrated first law formula is 
$\frac{d-3}{d-2}k M  = T_+ S_+ -k p  A + \frac{d-3}{d-2}k \Phi_+  Q$, 
where the factor $k$ was multiplied and $\Phi_+$ is $\Phi$ defined
in Eq.~\eqref{ch2eq:Phi} with black hole characteristics, i.e.,
$\Phi_+(r_+,r_-,R) = Q \frac{ r_+^{-(d-3)} -
    R^{-(d-3)}}{k}$.
We now must take the black hole limit with care, as $R=r_+$ means that 
$k=0$, and so $kM =0$ and $k\Phi_+=0$. The remaining equation is 
$T_+ S_+ - k p A = 0$. Now, $k p A = \frac{1}{2\mu}\frac{d-3}{d-2}
(r_+^{d-3} + r_-^{d-3})
-\frac{1}{\mu}\frac{d-3}{d-2} r_-^{d-3}$ in the black hole limit. 
This can be simplified using Eq.~\eqref{ch2eq:masscharge}, 
as the first term becomes the ADM mass
$\frac{1}{2\mu}\left(r_+^{d-3} + r_-^{d-3}\right)
=m$, while second term becomes $\frac{d-3}{d-2}
\phi_+ Q$, where
the black hole potential $\phi_+$ is naturally
defined as $\phi_+=Qr_+^{-(d-3)}$. We then recover the Smarr formula 
in the black hole limit as 
\begin{align}
    m = \frac{d-2}{d-3}T_+ S_+ + \phi_+ Q\,\,.
\label{ch2eq:smarreissnernordstromtangherlini}
\end{align}
For the extremal case, $r_+=r_-$, the Smarr formula in 
Eq.~\eqref{ch2eq:smarreissnernordstromtangherlini} 
can also be applied, with
$T_+=0$ and 
$\phi_+ Q= \frac{Q}{\sqrt \mu}$, where 
Eq.~\eqref{ch2eq:masscharge} has been used,
and the equality $\mu=\lambda$ in our convention of units
has been applied. For the case $d=4$, the original Smarr formula 
is recovered, together with the extremal case.

\section{Intrinsic thermodynamic stability for the charged thin shell
\label{sec:intrinsicstabshell}}

\subsection{Stability conditions for the charged thin shell}

With the thermodynamic equilibrium of the shell described, we must analyze the intrinsic 
thermodynamic stability of the shell. We perform the analysis according to an 
extended formalism in Callen's book\cite{Callen}. 

If we consider a system in thermodynamic equilibrium, the system 
is always susceptible to fluctuations. Let's consider an isolated 
system with entropy $S$, which can always be split into two equal 
subsystems. For perturbations within the system, there can be exchanges in the 
thermodynamic variables $(M,A,Q)$ between the two subsystems. These 
fluctuations can lead the system slightly away from the initial 
equilibrium and the system's entropy becomes $S+\Delta S$, which can 
be assumed to be described by the sum of the entropies of the two 
subsystems. Using the second law 
of thermodynamics, systems tend always to be at the configuration that 
maximizes the entropy and so the system returns to the initial equilibrium 
if $\Delta S < 0$, i.e. the system is stable. Otherwise, the system 
departs from the initial equilibrium configuration, developing 
inhomogeneities, and so the system is unstable. Thus, for small fluctuations, 
the conditions of intrinsic stability are such that $d\bar{S}=0$ and 
$d^2\bar{S} < 0$. Note that these conditions are applied to a 
generalized entropy $\bar{S}$, which covers configurations not in equilibrium.
An example of such generalized entropy is precisely the one of 
\cite{Callen}, where $2 \bar{S} = 
S(M+\delta M,A,Q) + S(M - \delta M,A,Q)$ for small mass fluctuations only, 
with $S(M,A,Q)$ being precisely the entropy of the 
configuration in equilibrium, $\delta M$ being the 
variable establishing the non-equilibrium configuration. The condition 
$d\bar{S} = 0$ is satisfied if $\delta M = 0$, indeed $S(M,A,Q)$ is the 
entropy of the configuration in equilibrium. The condition $d^2\bar{S}< 0$
evaluated at $\delta M = 0$ translates into the condition that the 
hessian of $S(M,A,Q)$ must be negative definite. 
Note that for the case of the shell, we must calculate the 
second derivatives in order to the parameters $(M,A,Q)$ and not to the parameters $(r_+,r_-,R)$. 
First, they are not equivalent as hessians of scalars are not tensors. 
Second, the independent thermodynamic parameters that can be directly exchanged 
by the subsystems are the quantities $(M,A,Q)$.

The stability conditions can be written in terms of the hessian components 
of the entropy, i.e. the second derivatives $S_{h_i h_j} = \frac{\partial^2
S}{\partial h_i \partial h_j}$, where $h_1 = M$, $h_2=A$ and $h_3=Q$. 
The negative definite condition is equivalent to the condition that 
the eigenvalues of the hessian are negative. 
Since the hessian is a $3\times3$ matrix, the expression of the eigenvalues 
is not trivial but one can get the sufficient conditions through the 
Sylvester's criterion by looking at the leading principal 
minors of the matrix, which are related to 
the pivots when one does Gauss 
elimination in the hessian to have a matrix in row-echelon form. 
The full conditions are
\begin{align}
    & S_{MM} \leq 0 \,,\,\,S_{AA} \leq 0 \,,\,\,S_{QQ} \leq 0
    \,,\notag\\
    & S_{MM}S_{AA} - S_{MA}^2 \geq 0 \,,\notag\\& S_{MM}S_{QQ} -
    S_{MQ}^2 \geq 0 \,,\notag\\& S_{QQ}S_{AA} - S_{QA}^2 \geq
    0\,,\notag\\
    & (S_{MM}S_{AQ} - S_{MA}S_{MQ})^2 - (S_{AA}S_{MM} -
    S_{AM}^2)(S_{QQ}S_{MM} - S_{QM}^2 )\leq 0\,, \mathrm{if} \,
    S_{MM} \neq 0\,\,,\notag\\
    &- S_{MA}^2 S_{QQ} 
    + 2 S_{MQ} S_{MA} S_{AQ} - S_{AA} S_{MQ}^2 \leq 0\,\,, 
    \mathrm{if} \, S_{MM} = 0\,\,,
    \label{ch2eq:StabCond}
\end{align}
where the conditions were extended to the marginal case, seminegative 
definiteness. We must make a correction to 
\cite{Fernandes:2021qvr} here. While the negative definiteness follows 
sufficiently from only the leading principal minors or the pivots,
the condition of semi-negative definiteness does not follow by 
simply including the equal case, with the failing 
cases being when one has vanishing leading principal minors. 
One instead must look at all the principal minors of the matrix. 
It turns out that for the case of the shell, we verified that this 
does not make any difference and the analysis of \cite{Fernandes:2021qvr} follows.

For convenience, we present the entropy and its first derivatives 
here for the chosen equations of state. 
The entropy of the shell is 
\begin{align}
    S(M,A,Q) =  \frac{\gamma}{16\pi G}A_+^a\,,
    \label{ch2eq:suggestedentropystability}
\end{align}
where $A_+ = \Omega r_+^{d-2}$ and $r_+$ and $r_-$ are functions of $(M,A,Q)$
from Eqs.~\eqref{ch2eq:rpm},~\eqref{ch2eq:m}, and~\eqref{ch2eq:A}.
From Eq.~\eqref{ch2eq:beta} and the specific choice 
of the reduced equation of state, the inverse temperature is given by 
\begin{align}
    \beta(M,A,Q) = \frac{a \gamma \Omega^{a-1}}{d-3}
      \frac{r_+^{a(d-2)}}{r_+^{d-3}-r_-^{d-3}}k\,.
      \label{ch2eq:tempeqstatestability}
\end{align}
The pressure is given by the fundamental pressure equation of state 
as 
\begin{align}
    p(M,A,Q)  = 
\frac{1}{2\mu\Omega } \frac{d-3}{d-2}
    \Bigg[(1-k)^2 R^{2(d-3)}-\lambda
    Q^2\Bigg]\frac1{ R^{2d-5}k},
    \label{ch2eq:pstability}
\end{align}
Finally, the electric potential in Eq.~\eqref{ch2eq:Phi} with the 
choice of the reduced equation of state in Eq.~\eqref{ch2eq:electeqstate}
is given by 
\begin{align}
    &\Phi(M,A,Q)  = Q\left(
\frac{1}{r_+^{d-3}} -\frac{1}{R^{d-3}}
    \right)\frac1k\,.
    \label{ch2eq:Phistability}
\end{align}

The first derivatives of the entropy follow easily from
the first law given in Eq.~\eqref{ch2eq:1stlaw2} together with
Eqs.~\eqref{ch2eq:tempeqstatestability}-\eqref{ch2eq:Phistability}, yielding
\begin{align}
    &S_M = \frac{a \gamma \Omega^{a-1}}{d-3}
    \frac{r_+^{a(d-2)}}{r_+^{d-3} - r_-^{d-3}}k\,,\notag\\
    &S_A = \frac{a \gamma \Omega^{a-2}r_+^{a(d-2)}\Big[(1-k)^2
    R^{2(d-3)}-\lambda Q^2 \Big]}{2\mu(d-2)
    R^{2d-5}(r_+^{d-3}-r_-^{d-3})}\,,\notag\\
    &S_Q = -\frac{a \gamma \Omega^{a-1} Q}{{(d-3)}}\left(
    \frac{r_+^{3-d} - R^{3-d}}{r_+^{d-3} -
    r_-^{d-3}}\right) r_+^{a(d-2)}\,.
\end{align}
To compute the second derivatives of the entropy, it is useful to 
present the derivatives of $r_\pm$ with respect to the thermodynamic
variables as 
\begin{align}
    &\pdv{r_{\pm}}{M} = \pm2\mu\frac{r_\pm\, k}{(d-3)(r_+^{d-3} -
    r_-^{d-3})}\,\,,\\
    &\pdv{r_{\pm}}{R} = \pm \mu\frac{r_\pm}{r_+^{d-3} -
    r_-^{d-3}}\frac{\mu M^2 -
    Q^2}{R^{d-2}}\,\,,\\
    &\pdv{r_{\pm}}{Q} = \mp\frac{2 \lambda Q r_\pm
    \left( r_\pm^{3-d} - R^{3-d}\right)
    }{(d-3) (r_+^{d-3} -
    r_-^{d-3})}\,\,.
\end{align} 
The components of the hessian of the entropy are 
\begin{align}
    &S_{MM} = \frac{a \gamma \Omega^{a-2} 8\pi G
    r_+^{a(d-2)}}{(d-3)(d-2)(r_+^{d-3} - r^{d-3}_-)
    R^{d-3}}\,S_1 \,,\notag\\
&S_{AA} = \frac{a \gamma \Omega^{a-3}r_+^{a(d-2)}}{
         2\mu(d-2)^2(r_+^{d-3}-r_-^{d-3})R^{d-1}}\,S_2\,,\notag\\
&S_{QQ} = \frac{a\gamma
    \Omega^{a-1}r_+^{a(d-2)}(1-x)}{{(d-3)}(r_+^{d-3}-r_-^{d-3})
    r_+^{d-3}}\,S_3\,,\notag\\
    &S_{MA} = \frac{a\gamma
    \Omega^{a-2}r_+^{a(d-2)}}{(d-2)(r_+^{d-3} -
    r_-^{d-3})R^{d-2}}\,S_{12}\,,\notag\\
&S_{MQ} = - \frac{2\mu a \gamma \Omega^{a-1} r_+^{a(d-2)} Q k
    }{(d-3)^{2} (r_+^{d-3} -
    r_-^{d-3})r_+^{2d-6}}\,S_{13}\,,\notag\\
&S_{AQ} = -\frac{a\gamma \Omega^{a-2}r_+^{a(d-2)}Q
    }{{(d-2)}(r_+^{d-3}-r_-^{d-3})r_+^{d-3}R^{d-2}}\,S_{23}\,,
    \label{ch2eq:secondderivatives}
\end{align}
with 
\begin{align}
        &S_1 = \frac{2 k^2 \mathcal{G}}{(d-3)x} - 1\,\,\,,\,\,\,
        S_2 = \mathcal{F}\left[\frac{\mathcal{F} \mathcal{G}}{x} - 2d +
        5\right]\,,\notag\\
        &S_3 = -1 + \frac{2y}{d-3}\left[\mathcal{G}(1-x) -
        \frac{2(d-3)}{1-y}\right]\,,\notag\\
        &S_{12} = 1 - k + \frac{k \mathcal{G}}{x(d-3)}\mathcal{F}
        \,\,\,,\,\,\,
        S_{13} = \mathcal{G}(1-x) - \frac{(d-3)}{1-y}\,,\notag\\
        &S_{23} = x + \frac{\mathcal{F}}{x(d-3)}\Bigg[ \mathcal{G}(1-x) -
        \frac{(d-3)}{1-y}\Bigg]\,,
        \label{ch2eq:Sfunctions}
\end{align}
The auxiliary functions $\mathcal{G}$,
$\mathcal{F}$ and $k$ are given by 
\begin{align}
    &\mathcal{G} = \frac{1}{1-y}\Bigg[ a(d-2)-(d-3)\frac{1+y}{1-y}
        \Bigg]\,\,,\,\,
        \mathcal{F} = 2 - 2k{-} x(1-y)\,\,,\notag\\
        &k = \sqrt{(1-x)(1-xy)}\,,
        \label{ch2eq:Sfunctionsauxiliary}
\end{align}
and the parameters $x$ and $y$ are defined as 
\begin{align}
    x=\frac{r_+^{d-3}}{R^{d-3}}\,,\quad\quad
    y=\frac{r_-^{d-3}}{r_+^{d-3}}\,.
    \label{ch2eq:xy}
\end{align}
Note that the definition of $k$ is the same as above, but given 
in terms of $x$ and $y$.

The set of inequalities in Eq.~\eqref{ch2eq:StabCond} with the
entropy equation given in Eq.~\eqref{ch2eq:suggestedentropystability}
together with the
equations
of state given in
Eqs.~\eqref{ch2eq:tempeqstatestability}-\eqref{ch2eq:Phistability}
can be written as conditions in
terms of the functions given in Eq.~\eqref{ch2eq:Sfunctions}. The
conditions restrict the parameter space described by the
points $(d,a,x,y)$ for stable configurations. We constrain the parameter 
space to the region 
\begin{align}
d\geq 4\,,\quad
a > 0\,,\quad0<x<1\,,\quad 0<y<1\,.
\end{align}
We constrain the dimension $d$ as $d \geq 4$, since for lower $d$ there is
no proper Reissner-Nordstr\"om solution. We constrain
the parameter $a$ as $a>0$, because  
in the no black hole limit, $A_+= 0$,
and the entropy expression cannot diverge, see
Eq.~\eqref{ch2eq:suggestedentropystability}. We constrain also the 
parameter $x$ as
$0<x<1$, because the shell has to be in the limits
between no shell, $x=0$, and the black hole state,
$x=1$. Finally, we constrain the parameter $y$ as
$0<y<1$, due to the validity of the equations of state.
Overcharged shell, with $y>1$, are not treated here since
the equations of state,
Eqs.~\eqref{ch2eq:tempeqstatestability}-\eqref{ch2eq:Phistability},
do not apply to overcharged shells.

In what follows, we present the analysis of the 
stability conditions for each possible combination of fluctuations,
accompanied by plots to further understand these conditions.

\subsection{Stability of the shell for mass fluctuations only
\label{massfluconly}}

\begin{figure}[h]
    \vskip -0.0cm
    \centering
        \begin{subfigure}[h]{0.45\columnwidth}
        \centering
        \includegraphics[width=\columnwidth]
        {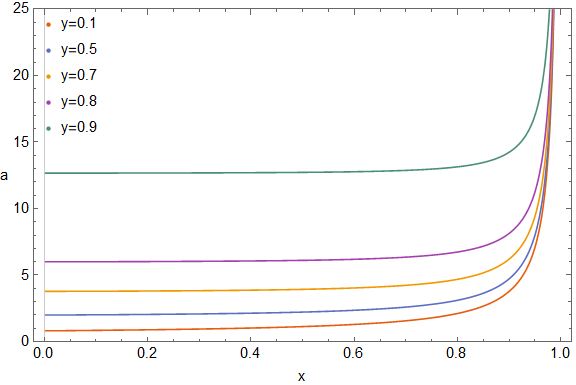}
        \caption{$y$ fixed}
        \label{fig:S1y}
        \end{subfigure}%
        \begin{subfigure}[h]{0.45\columnwidth} 
        \centering
        \includegraphics[width=\columnwidth]
        {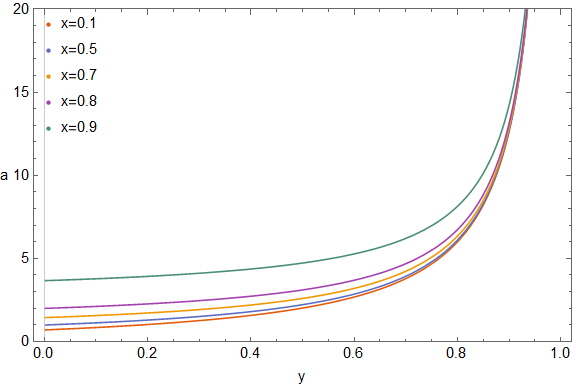}
        \caption{$x$ fixed}
        \end{subfigure}
        \caption{\small{Region of thermodynamic stability of the 
        shell for mass fluctuations only for $d=5$ 
        with the curves of marginal stability 
        $a(x,y)$ plotted in function $x= r_+^{2}/R^{2}$, and
        $y=r_-^{2}/r_+^{2}$: (a) 
        certain values of $y$; 
        (b) certain values of $y$;. 
        The regions below the curves 
        describe the stable configurations.}}
    \label{fig:S1}
    \end{figure}

\begin{figure}[h]
    \begin{subfigure}[h]{0.45\columnwidth}
    \centering
    \includegraphics[width=\columnwidth]
    {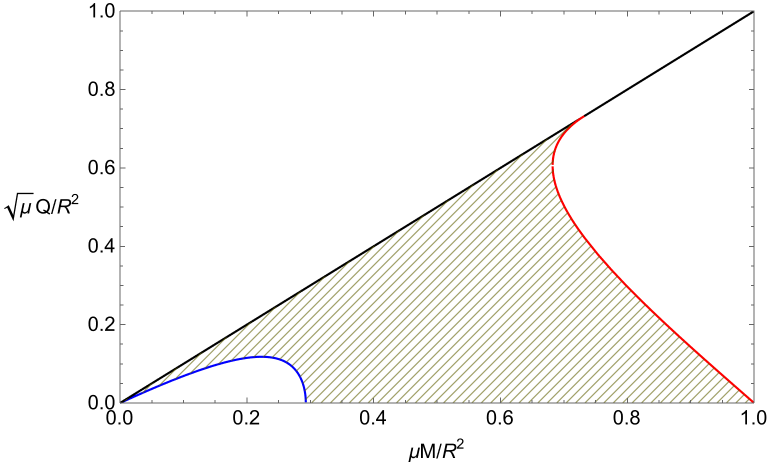}
    \caption{$d=5$ in $M$ and $Q$}
    \end{subfigure}\,\,
    \begin{subfigure}[h]{0.45\columnwidth}
            \centering
            \includegraphics[width=0.95\columnwidth]
            {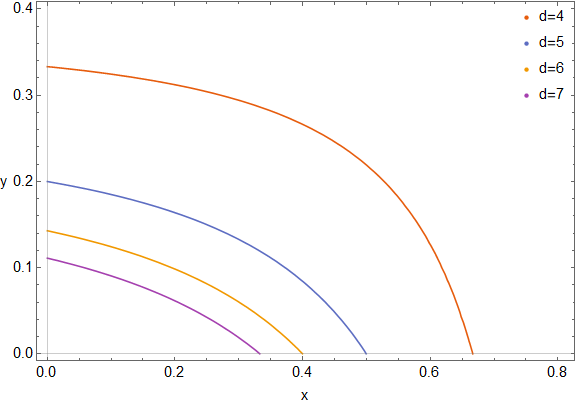}
            \label{fig:S1a1}
            \caption{in $x$ and $y$}
    \end{subfigure}
    \caption{\small{Region of thermodynamic stability of the shell for mass
    fluctuations only and $a=1$: (a) in stripes in terms of 
    $\mu M/R^2$ and $\sqrt{\mu}Q/R^2$; (b) above the curves for different $d$ in 
    terms of $x=r_+^{d-3}/R^{d-3}$ and $y=r_-^{d-3}/r_+^{d-3}$.}}
    \label{fig:S1a1MQheat}
\end{figure}

Considering a shell with only mass fluctuations, 
the stability condition is
given by $S_{MM} \leq 0$, see Eq.~\eqref{ch2eq:StabCond}.  For the
chosen equations of state, and with the help of
Eq.~\eqref{ch2eq:secondderivatives}, the condition $S_{MM} \leq 0$ can be
written as
\begin{align}
    S_1 \leq 0\,.
    \label{ch2eq:S1cond}
\end{align}
This inequality can be 
simplified using Eq.~\eqref{ch2eq:Sfunctions} to the condition
\begin{align}
    a \leq \frac{x(d-3)(1-y)}{2(d-2)k^2} +
    \frac{(d-3)}{(d-2)}\frac{(1+y)}{(1-y)}\,,\label{ch2eq:SMMa}
\end{align}
where Eqs.~\eqref{ch2eq:Sfunctionsauxiliary} and \eqref{ch2eq:xy} were
used. We should give some comments about this condition. 
The right-hand side of Eq.~\eqref{ch2eq:xy} tends to infinity at
the points $x=1$ or $y=1$. Moreover, it has its minimum value at $(x,y) =
(0,0)$, corresponding to $a = \frac{d-3}{d-2}$. We plot the 
right-hand side for $d=5$ in 
Figs.~\ref{fig:S1}(a) and~\ref{fig:S1}(b), where 
the region below the curves is stable. The curves increase overall with $d$. 
The case $a=1$ has some interest 
as it represents a shell with thermodynamic
black hole features, we plot this case in Figs.~\ref{fig:S1a1MQheat}(a) and 
~\ref{fig:S1a1MQheat}(b). 
For the uncharged
case $y=0$, there is thermodynamic stability for
$\frac{2}{d-1}<x<1$, in agreement with \cite{Lemos:2009uk}. Increasing
the value of $y$ also increases the range of $x$ for thermodynamic
stable configurations, meaning that a shell with more electric charge
may have higher radius $R$ and remain stable. Thermodynamic stability is
guaranteed in the full range of $x$ if $y\geq\frac{1}{2d-5}$, see also
Fig.~\ref{fig:S1a1MQheat}(b) for this case.
It is also interesting to see the stability with respect to the
variables $\frac{\mu M}{R^{d-3}}$ and $\frac{\sqrt{\mu}Q}{R^{d-3}}$, shown 
in Fig.~\ref{fig:S1a1MQheat}(a), which follow from the analysis above by 
a transformation of variables.

    \clearpage

\subsection{Stability of the shell for area fluctuations only
\label{areafluconly}}

\begin{figure}[h]
    \vskip 0.2cm
        \centering
        \begin{subfigure}[h]{0.45\columnwidth}
        \centering
        \includegraphics[width=\columnwidth]
        {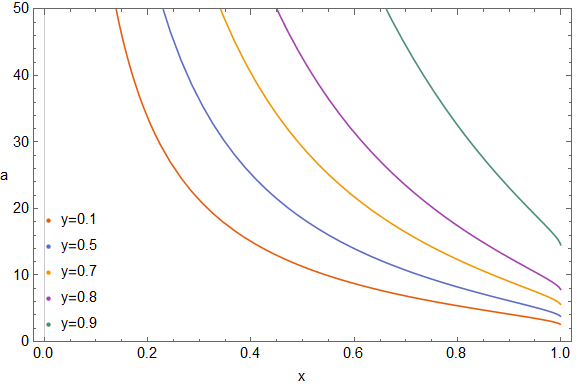}
        \caption{$y$ fixed}
        \label{fig:S2y}
        \end{subfigure}
        \begin{subfigure}[h]{0.45\columnwidth}
        \centering
        \includegraphics[width=\columnwidth]
        {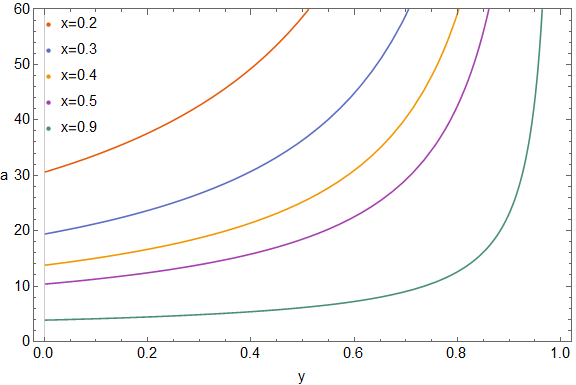}
        \caption{$x$ fixed}
        \label{fig:S2x}
        \end{subfigure}
        \caption{\small{Region of thermodynamic stability of the 
        shell for area fluctuations only for $d=5$ 
        with the curves of marginal stability 
        $a(x,y)$ plotted in function $x=r_+^{2}/R^{2}$, and
        $y=r_-^{2}/r_+^{2}$: (a) 
        certain values of $y$; 
        (b) certain values of $x$. 
        The regions below the curves describe 
        the stable configurations.}}
    \label{fig:S2}
    \end{figure}

Considering a shell with only area fluctuations, the stability
condition is given by
$S_{AA} \leq 0$, see Eq.~\eqref{ch2eq:StabCond}.
Using 
Eq.~\eqref{ch2eq:secondderivatives},
we have that $S_{AA} \leq 0$ can be written as
\begin{align}
    S_2\leq 0\,.\label{ch2eq:S2cond}
\end{align}
Now Eq.~\eqref{ch2eq:Sfunctions} can also be used to 
rearrange this inequality into
\begin{align}
    a \leq \frac{(2d-5)x (1-y)}{(d-2)\mathcal{F}} +
    \frac{(d-3)}{(d-2)}\frac{(1+y)}{(1-y)}\,,
    \label{ch2eq:SAAa}
\end{align}
where Eqs.~\eqref{ch2eq:Sfunctionsauxiliary} and \eqref{ch2eq:xy} have been
used. We also used that the factor $\mathcal{F}$
is always positive for $0<x<1$ and $0<y<1$, being proportional to
$M-m$. Regarding some properties of the condition, the right-hand side
of Eq.~\eqref{ch2eq:SAAa}
has the minimum at $(x=1,y=0)$, with the value $a
= 3- \frac{2}{d-2}$. Moreover, the function increases in the direction of
$x\rightarrow 0$ or $y\rightarrow 1$, where it tends to infinity.
The right-hand side is plotted for $d=5$ 
in Figs.~\ref{fig:S2}(a) and~\ref{fig:S2}(b). The curves increase with $d$.
The case of the shell with $a=1$, which has thermodynamic black hole
features, is always below the surface of
marginal stability, therefore it is stable
to area perturbations only.

\clearpage

\subsection{Stability of the shell for charge fluctuations only
\label{chargeofluconly}}

\begin{figure}[h]
    \vskip 2.5cm
        \centering
        \begin{subfigure}[h]{0.45\columnwidth}
        \centering
        \includegraphics[width=\columnwidth]
        {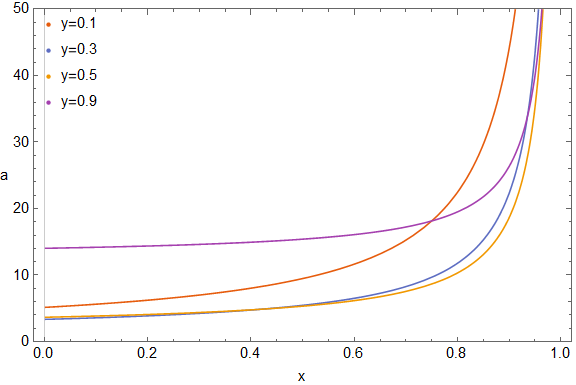}
        \caption{$y$ fixed}
        \label{fig:S3y}
        \end{subfigure}
        \begin{subfigure}[h]{0.45\textwidth}
        \centering
        \includegraphics[width=\textwidth]
        {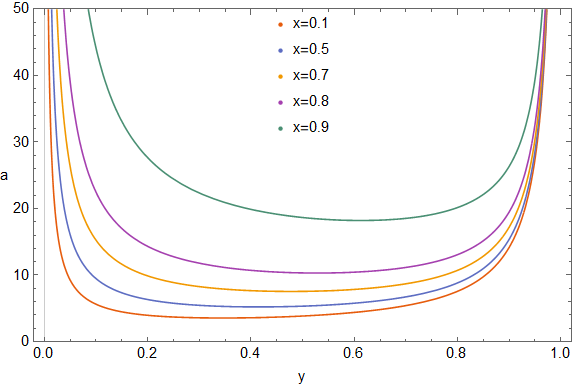}
        \caption{$x$ fixed}
        \label{fig:S3x}
        \end{subfigure}
        \caption{\small{Region of thermodynamic stability of the 
        shell for charge fluctuations only for $d=5$ 
        with the curves of marginal stability 
        $a(x,y)$ plotted in function $x=r_+^{2}/R^{2}$, and
        $y=r_-^{2}/r_+^{2}$: (a) 
        certain values of $y$; 
        (b) certain values of $x$. 
        The regions below the curves 
        describe the stable configurations.}}
        \label{fig:S3}
    \end{figure}

For a shell with only electric charge
fluctuations, the stability
condition is given by
$S_{QQ} \leq 0$, see Eq.~\eqref{ch2eq:StabCond}.
With the help of
Eq.~\eqref{ch2eq:secondderivatives},
the condition $S_{QQ} \leq 0$ can be written as
\begin{align}
    S_3\leq 0\,.
    \label{ch2eq:S3cond}
\end{align}
Using Eq.~\eqref{ch2eq:Sfunctions}, this inequality can be simplified into
\begin{align}
    &a\leq \frac{(d-3)(1-y)}{2(d-2)y(1-x)} +
    \frac{2(d-3)}{(d-2)(1-x)} \nonumber\\&+
    \frac{(d-3)}{(d-2)}\frac{(1+y)}{(1-y)}\,,
    \label{ch2eq:SQQa}
\end{align} 
where Eqs.~\eqref{ch2eq:Sfunctionsauxiliary} and \eqref{ch2eq:xy} have been
used. The right-hand side of Eq.~\eqref{ch2eq:SQQa}
depicts a concave surface, faced to
$a\rightarrow +\infty$. In the restricted parameter space, the minimum
is at $\left(x=0,y=\frac{1}{3}\right)$, where its value is $a =
5\frac{d-3}{d-2}$. The right-hand side diverges to 
infinity at the axes $x=1$,
$y=0$ and $y=1$.
The right-hand side is plotted in Figs.~\ref{fig:S3}(a) and~\ref{fig:S3}(b) 
for $d=5$.
For increasing $d$, the curves increase overall.
The shell with thermodynamic black hole
features, described by $a=1$, finds itself always below the surface of
marginal stability, therefore it is stable
to charge perturbations only.

\clearpage

\subsection{Stability of the shell for mass and area fluctuations together
\label{massandareafluctogether}}

\begin{figure}[h]
    \vskip 2.0cm
        \centering
        \begin{subfigure}[b]{0.45\columnwidth}
        \centering
        \includegraphics[width=\columnwidth]
        {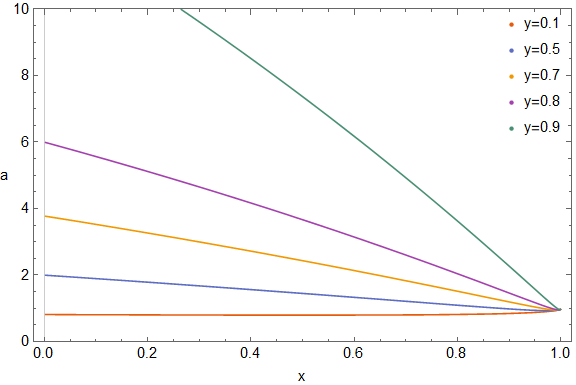}
        \caption{$d=5$, $y$ fixed}
        \label{fig:S4y}
        \end{subfigure}%
        \begin{subfigure}[b]{0.45\textwidth}
        \centering
        \includegraphics[width=\textwidth]
        {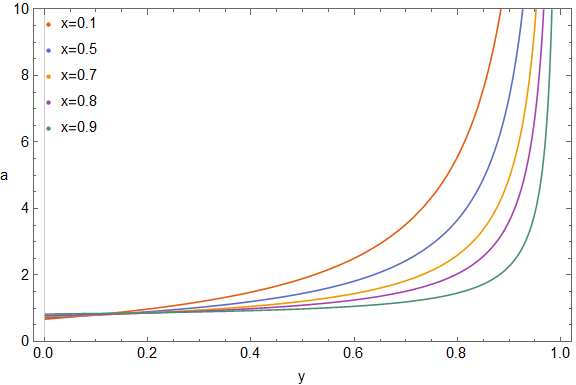}
        \caption{$d=5$, $x$ fixed}
        \label{fig:S4x}
        \end{subfigure}%
        \caption{\small{Region of thermodynamic stability of the 
        shell for mass and area fluctuations together for $d=5$ 
        with the curves of marginal stability 
        $a(x,y)$ plotted in function $x=r_+^{2}/R^{2}$, and
        $y=r_-^{2}/r_+^{2}$: (a) 
        certain values of $y$; 
        (b) certain values of $x$. 
        The regions below the curves describe 
        the stable configurations.}}
        \label{fig:S4}
    \end{figure}

\begin{figure}[h]
        \centering
        \includegraphics[width=0.4\columnwidth]
        {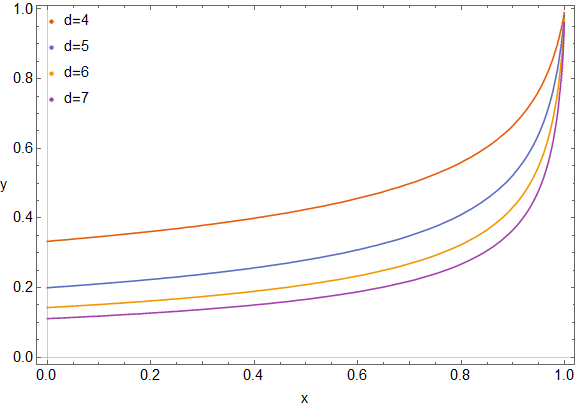}
        \caption{Region of thermodynamic stability for mass and area 
        fluctuations and for $a=1$, for different values of $d$, in terms 
        of $x=r_+^{d-3}/R^{d-3}$ and $y=r_-^{d-3}/r_+^{d-3}$. Region below 
        curves describes stability.}
        \label{fig:S4a1}
\end{figure}

Considering now a shell with mass and area fluctuations, the stability
conditions including the marginal condition is given by
$S_{MM} \leq 0$, $S_{AA} \leq 0$,
and  $S_{MM}S_{AA} - S_{MA}^2 \geq 0$, see Eq.~\eqref{ch2eq:StabCond}.
Without the marginal condition, it suffices to consider $S_{MM} < 0$
and $S_{MM}S_{AA} - S_{MA}^2 > 0$. However, the condition $S_{MM}S_{AA} -
S_{MA}^2 \geq 0$ for the case of the shell is the strongest even including 
the marginal case.
With
Eq.~\eqref{ch2eq:secondderivatives}, we have that
$S_{MM}S_{AA} - S_{MA}^2 \geq 0$
can be written as 
\begin{align}
    S_4 = -\frac{1}{2(d-3)}S_1 S_2 + S_{12}^2 \leq 0\,.
    \label{ch2eq:S4cond}
\end{align}
From Eq.~\eqref{ch2eq:Sfunctions}, this inequality can be simplified into
\begin{align}
    &a \leq \frac{(1-y)x \Big((d-\frac{5}{2})\mathcal{F} -
    (d-3)(1-k)^2\Big)}{(d-2)\mathcal{F}\Big(\frac{k^2}{d-3} +
    2 k + \frac{\mathcal{F}}{2}\Big)} +
    \frac{(d-3)}{(d-2)}\frac{(1+y)}{(1-y)}\,,
    \label{ch2eq:SMMAAa}
\end{align}
where Eqs.~\eqref{ch2eq:Sfunctionsauxiliary} and \eqref{ch2eq:xy} have been
used.
The right-hand side of
Eq.~\eqref{ch2eq:SMMAAa} assumes the minimum value at $x=1$, 
where $a = 1$. From $x=1$
towards $x=0$, the function bends towards $a
=\frac{d-3}{d-2}\frac{1+y}{1-y}$. At $y=1$, the right-hand side
tends to infinity. The right-hand side is plotted in
Figs.~\ref{fig:S4}(a) and~\ref{fig:S4}(b) for $d=5$. For higher $d$, the 
curves increase overall. The case $a=1$ of the shell, having thermodynamic
black hole features, has the property that
increasing the value of $y$ decreases the range of $x$ for
thermodynamic stable configurations, meaning, if the shell has more
electric charge than it needs to have lower $R$ for stability, see also
Fig.~\ref{fig:S4a1} for this $a=1$ case.

\subsection{Stability of the shell for mass and charge fluctuations together
\label{massandchargefluctogether}}

\begin{figure}[h]
    \vskip 2.2cm
        \centering
        \begin{subfigure}[b]{0.45\columnwidth}
        \centering
        \includegraphics[width=\columnwidth]
        {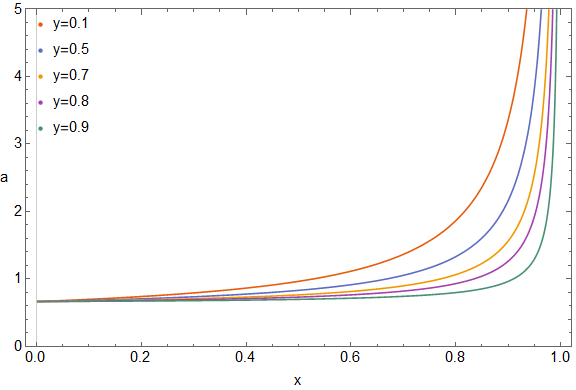}
        \caption{$y$ fixed}
        \label{fig:S5y}
        \end{subfigure}%
        \begin{subfigure}[b]{0.45\columnwidth}
        \centering
        \includegraphics[width=\columnwidth]
        {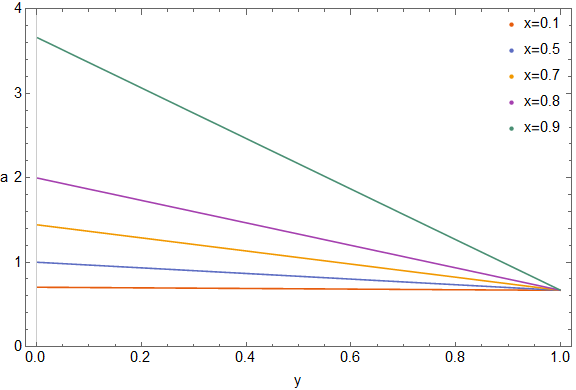}
        \caption{$x$ fixed}
        \label{fig:S5x}
        \end{subfigure}%
        \caption{\small{Region of thermodynamic stability of the 
        shell for mass and charge fluctuations together for $d=5$ 
        with the curves of marginal stability 
        $a(x,y)$ plotted in function $x=r_+^{2}/R^{2}$, and
        $y=r_-^{2}/r_+^{2}$: (a) 
        certain values of $y$; 
        (b) certain values of $x$. 
        The regions below the curves describe 
        the stable configurations.}}
        \label{fig:S5}
    \end{figure}

\begin{figure}[h]
        \centering
        \includegraphics[width=0.5\columnwidth]
        {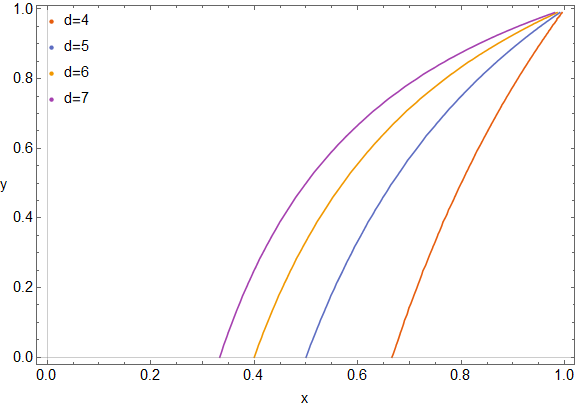}
        \caption{Region of thermodynamic stability for mass and area 
        fluctuations and for $a=1$, for different values of $d$, in terms 
        of $x=r_+^{d-3}/R^{d-3}$ and $y=r_-^{d-3}/r_+^{d-3}$. Region below 
        curves describes stability.}
        \label{fig:S5a1}
\end{figure}

We consider now the shell with mass and charge fluctuations. 
The stability conditions including the marginal case are given by
$S_{MM} \leq 0$, $S_{QQ}\leq 0$,
and  $S_{MM}S_{QQ} - S_{MQ}^2 \geq 0$, see Eq.~\eqref{ch2eq:StabCond}.
Without considering the marginal case, the sufficient
conditions are $S_{MM}< 0$ and $S_{MM}S_{QQ} -
S_{MQ}^2 > 0$. However, for the case of the shell, 
the condition $S_{MM}S_{QQ} - S_{MQ}^2 \geq 0$ is the strongest.
Using 
Eq.~\eqref{ch2eq:secondderivatives}, we have that
$S_{MM}S_{QQ} - S_{MQ}^2 \geq 0$
can be written as 
\begin{align}
    S_5 = - x(1-x)S_1S_3 +
    \frac{4 y k^2}{(d-3)^2}S_{13}^2\leq 0\,.
    \label{ch2eq:S5cond}
\end{align}
Using Eq.~\eqref{ch2eq:Sfunctions}, the inequality above can be rearranged
as
\begin{align}
   a \leq \frac{(d-3)}{2(d-2)}\,
    \frac{2 - x(1+y)}{1-x}\,,
    \label{ch2eq:SMMQQa}
\end{align}
where Eqs.~\eqref{ch2eq:Sfunctionsauxiliary} and \eqref{ch2eq:xy} have been
used. Some properties of right-hand side follow. At $x=0$
or $y=1$, the right-hand side takes the value 
$a= \frac{d-3}{d-2}$. The function
diverges to infinity at $x=1$. The function then bends from
a constant value
to $a = \frac{d-3}{2(d-2)}\,\frac{2-x}{1-x}$, going from $y=1$ to
$y=0$.
We present the plot of the right-hand side for $d=5$ in 
Figs.~\ref{fig:S5}(a) and~\ref{fig:S5}(b). The curves further increase with 
$d$. 
For the case with $a=1$, increasing the value
of $y$ decreases the range of $x$ for thermodynamic
stable configurations, see also Fig.~\ref{fig:S5a1} for
this $a=1$ case.

\subsection{Stability of the shell for area and charge fluctuations together
\label{areaandchargefluctogether}}

\begin{figure}[h]
    \vskip 2.0cm
        \centering
        \begin{subfigure}[b]{0.45\columnwidth}
        \centering
        \includegraphics[width=\columnwidth]
        {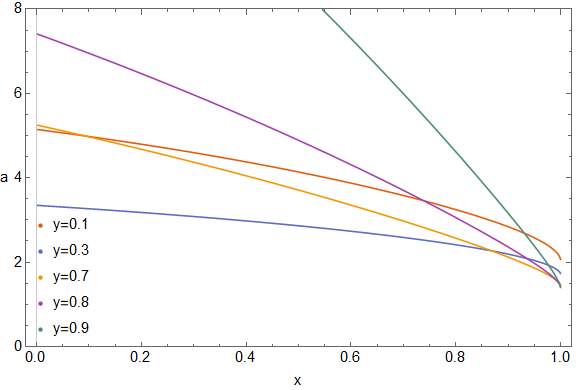}
        \caption{$y$ fixed}
        \label{fig:S6y}
        \end{subfigure}
        \begin{subfigure}[b]{0.45\columnwidth}
        \centering
        \includegraphics[width=\columnwidth]
        {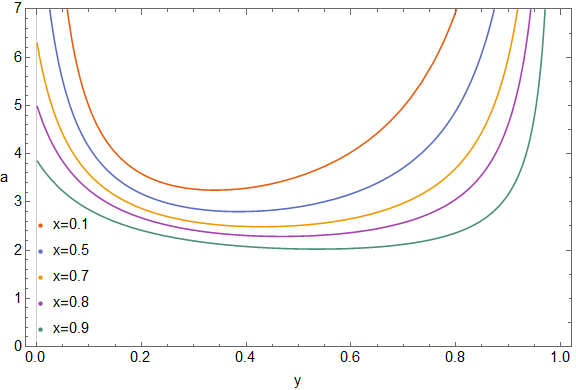}
        \caption{$x$ fixed}
        \label{fig:S6x}
        \end{subfigure}
        \caption{\small{Region of thermodynamic stability of the 
        shell for area and charge fluctuations together for $d=5$ 
        with the curves of marginal stability 
        $a(x,y)$ plotted in function $x=r_+^{2}/R^{2}$, and
        $y=r_-^{2}/r_+^{2}$: (a) 
        certain values of $y$; 
        (b) certain values of $x$. 
        The regions below the curves describe 
        the stable configurations.}}
        \label{fig:S6}
    \end{figure}

Regarding the case of a shell with area and charge fluctuations, 
the stability
conditions including the marginal case 
are given by $S_{AA} \leq 0$, $S_{QQ} \leq 0$, and
$S_{AA}S_{QQ} - S_{AQ}^2 \geq 0$, see Eq.~\eqref{ch2eq:StabCond}.
Without the marginal case, the sufficient
conditions can be chosen to be $S_{AA}< 0$ and $S_{AA}S_{QQ} -
S_{AQ}^2 > 0$. For the specific case of the shell, it turns out that 
$S_{AA}S_{QQ} -
S_{AQ}^2 \geq 0$ is sufficient.
Using
Eq.~\eqref{ch2eq:secondderivatives}, we have that
$S_{AA}S_{QQ} - S_{AQ}^2 \geq 0$
can be written as 
\begin{align}
    S_6 = -\frac{(1-x)}{2(d-3)}S_2S_3 + xy S_{23}\leq 0\,,
    \label{ch2eq:S6cond}
\end{align}
which can be simplified using Eq.~\eqref{ch2eq:Sfunctions} into
\begin{align}
&a \leq \frac{\frac{(1-x)\mathcal{F}(2d-5)}{2(d-3)}(1+3y) -
x^3 y(1-y) + 2 \mathcal{F}xy -
\frac{y \mathcal{F}^2}{x(1-y)}}{(d-2)(1-x)
\Big(\frac{\mathcal{F}^2}{2 x (d-3)} +
\frac{2d -5}{(d-3)^2}y(1-x)\mathcal{F} +
\frac{2 \mathcal{F} x y}{(d-3)} \Big)} +
\frac{(d-3)}{(d-2)}\,\frac{1+y}{1-y}\,.
\label{ch2eq:SAAQQa}
\end{align}
where Eqs.~\eqref{ch2eq:Sfunctionsauxiliary} and \eqref{ch2eq:xy} have been
used.
At $y=0$, the right-hand side function intersects
$S_2$. It then grows without bound at $(x=0,y=0)$ or
$y=1$. In the limit of $x\rightarrow 1$, the right-hand side approaches the
value of $a = \frac{8 + 6y - 3d (1+y)}{(d-2)(1+3y)}$. At $x=0$, the
right-hand side function approaches $S_3$ from below. The 
right-hand side function is plotted for $d=5$ in
Figs.~\ref{fig:S6}(a) and~\ref{fig:S6}(b). The curves increase with $d$. 
The case with $a=1$
is always below the surface of marginal stability, therefore
it is stable to area and charge perturbations.

\vspace{-5mm}

\clearpage

\begin{figure}[h]
    \vskip 2.5cm
        \centering
        \begin{subfigure}[b]{0.45\columnwidth}
        \centering
        \includegraphics[width=\columnwidth]
        {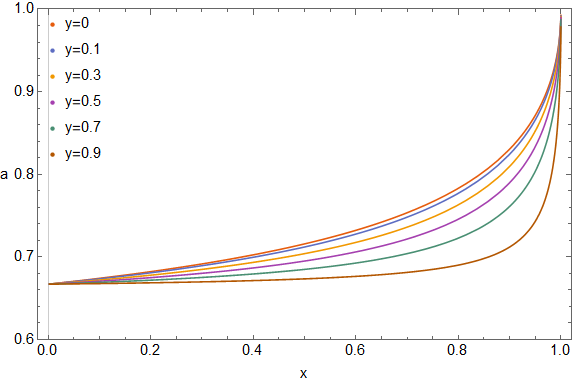}
        \caption{$y$ fixed}
        \label{fig:S7y}
        \end{subfigure}
        \begin{subfigure}[b]{0.45\columnwidth}
        \centering
        \includegraphics[width=\columnwidth]
        {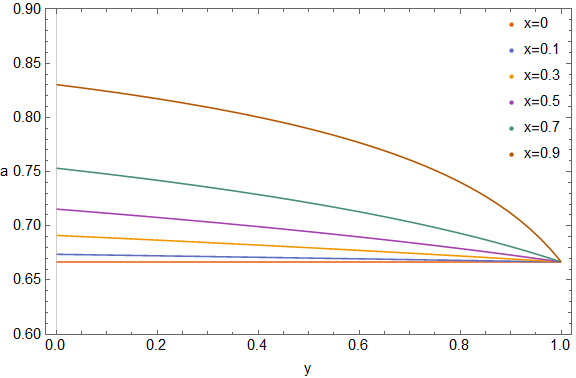}
        \caption{$x$ fixed}
        \label{fig:S7x}
        \end{subfigure}
        \caption{\small{Region of thermodynamic stability of the 
        shell for mass, area and charge fluctuations together for $d=5$ 
        with the curves of marginal stability 
        $a(x,y)$ plotted in function $x=r_+^{2}/R^{2}$, and
        $y=r_-^{2}/r_+^{2}$: (a) 
        certain values of $y$; 
        (b) certain values of $x$. 
        The regions below the curves describe 
        the stable configurations.}}
    \label{fig:S7}
    \end{figure}%

\subsection{Stability of the shell for mass, area and charge fluctuations
\label{massareaandchargeflucaltogether}}

For the shell with full perturbations, i.e.
mass, area, and charge fluctuations, the
stability conditions including the marginal case 
are given by all the inequalities in
Eq.~\eqref{ch2eq:StabCond}. Without the marginal case, the sufficient 
conditions are $S_{MM} < 0$, $S_{MM}S_{AA} -
S_{MA}^2 > 0$, and $ (S_{MM}S_{AQ} - S_{MA}S_{MQ})^2 -
(S_{AA}S_{MM} - S_{AM}^2)(S_{QQ}S_{MM} - S_{QM}^2 ) < 0$. 
However, for the case of the shell, it is sufficient to consider 
$ (S_{MM}S_{AQ} - S_{MA}S_{MQ})^2 -
(S_{AA}S_{MM} - S_{AM}^2)(S_{QQ}S_{MM} - S_{QM}^2 ) \leq 0$ for 
$S_{MM}\neq 0$ and $- S_{MA}^2 S_{QQ} 
    + 2 S_{MQ} S_{MA} S_{AQ} - S_{AA} S_{MQ}^2 \leq 0$ for $S_{MM} = 0$.
For $S_{MM}\neq0$, the condition simplifies to
\begin{align}
    S_7 = \left(x S_1 S_{23} -
    \frac{2 k}{d-3}S_{12}S_{13}\right)^2y -
    S_4 S_5 \leq 0\,,
    \label{ch2eq:S7cond}
\end{align}
while for $S_{MM}=0$, we must divide by $S_{MM}$ and make the limit 
$S_{MM} = 0$. In both cases, the inequality reduces to
\begin{align}
    a\leq \frac{d-3}{d-2}\left(
    \frac{4 - 4 k + x^2 (d(1-y)^2 +
    C)}{4 - 4k + x^2 d (1-y)^2 + x D}\right)
    \,,\label{ch2eq:SMMAAQQa}
\end{align}
where
$C = 2 x(1+y) (k-2) - 2 -2(y-4)y$,
and
$D = 4 k - 2y - 6 - x\left( 1 + y (3y-8)\right)$, and
Eqs.~\eqref{ch2eq:Sfunctionsauxiliary} and \eqref{ch2eq:xy} have been
used.
The right-hand side in the condition given in Eq.~\eqref{ch2eq:SMMAAQQa} 
has its lowest value of $a = \frac{d-3}{d-2}$ at $x=0$, 
for every $y$. The function then
increases towards $x=1$, with the limit $x=1$ giving $a=1$. 
At the limit of $y=1$, the right-hand side is given 
by the lowest value $a =
\frac{d-3}{d-2}$ for every $x$, except for the limit $x=1$ 
where it gives $a =1$.  
Therefore, the condition for stability in Eq.~\eqref{ch2eq:SMMAAQQa}
implies that every configuration with $a\leq \frac{d-3}{d-2}$ is
stable. For $\frac{d-3}{d-2} < a < 1$, the
stability region decreases with increasing $y$, being zero in the
limit of $y=1$. And so shells with more electric charge
have less configurations of stability.
The space of stable
configurations in the $a-d$ plane is similar to
the analysis made for the uncharged case in~\cite{Andre:2019zzo}.
The right-hand side is plotted for $d=5$ in
Figs.~\ref{fig:S7}(a) and~\ref{fig:S7}(b).
The curves increase with 
increasing $d$. The case of the shell with thermodynamic black hole features, 
with $a=1$, is always above the surface of
marginal stability, hence unstable, except for the points with $x = 1$
which lie on the limit of the surface, hence marginally stable. This means 
that, in
the black hole limit $a=1$ and $x=1$, the
configurations for every value of $y$ are marginally stable.

\subsection{Behaviour of the intrinsic stability with the parameter $a$: 
some comments}

We give here some additional comments on the analysis of the stability conditions above, 
namely on the stability of mass fluctuations only in terms of $M$ and 
$Q$, and also on the effect of the electric charge in the stability.

For mass fluctuations only, Sec.~\ref{massfluconly}, we barely mentioned the stability analysis 
in terms of $\frac{M}{R^{d-3}}$ and $\frac{Q}{R^{d-3}}$. However, 
there are still some interesting insights to be made.  
The condition given in Eq.~\eqref{ch2eq:SMMa} in terms 
of $\frac{M}{R^{d-3}}$ and $\frac{Q}{R^{d-3}}$
instead of $x=\frac{r_+^{d-3}}{R^{d-3}}$ and
$y=\frac{r_-^{d-3}}{r_+^{d-3}}$
becomes
\begin{align} &a \leq
    \frac{\mu}2\frac{(d-3)}{(d-2)}\frac{\left(\frac{Q^2}{R^{2(d-3)}} +
    \frac{\mu M^2}{R^{2(d-3)}} - \frac{2M}{R^{d-3}}\right)}{\left(1 -
    \frac{\mu M}{R^{d-3}}\right)^2 }
    \frac{\left(\frac{2\mu M}{R^{d-3}} + \frac{\mu (Q^2 -
     M^2 \mu^2)}{R^{2(d-3)}} -2\right)}
     {\sqrt{ \frac{\mu(\mu M^2 -
     Q^2)}{R^{2(d-3)}} \left(
     \left(2 - \frac{\mu M}{R^{d-3}}\right)^2 -\frac{\mu Q^2}
     {R^{2(d-3)}}\right) }}\,.
     \label{ch2eq:Sa1MQ}
\end{align}
The parameter space is restricted to the condition of subextremality, 
namely $\sqrt{\mu} M > Q$, and to the condition of no trapped surface, 
$\frac{r_+}{R}<1$. From the stability condition in Eq.~\eqref{ch2eq:Sa1MQ},
we find that the shell with small $\frac{M}{R^{d-3}}$ requires 
at least a minimum value of electric charge $\frac{Q}{R^{d-3}}$ to be stable.
When $\frac{M}{R^{d-3}}$ assumes the value corresponding to 
$x=\frac{2}{d-1}$, the minimum charge for stability becomes zero, or 
$y=0$. For higher mass $\frac{M}{R^{d-3}}$, the region of stable 
configurations is only constrained by the restrictions of the parameter space, 
meaning $\frac{\sqrt{\mu} M}{R^{d-3}} >
\frac{Q}{R^{d-3}}$ and $\frac{r_+}{R}<1$. The important point is that the 
condition in Eq.~\eqref{ch2eq:Sa1MQ} means that thermodynamic stability 
for small $\frac{M}{R^{d-3}}$ only happens for sufficiently large electric 
charge. Moreover, it is also interesting to consider the $y=0$ case. The 
shell is only stable for $x \geq \frac{2}{d-1}$, with equality being the 
marginally stable case. Since $x= \frac{2}{d-1}$ corresponds to the photonic 
orbit, the stable shell must always lay inside the photonic orbit, 
in agreement with \cite{Andre:2019zzo}. We must note that this behaviour 
is similar to black holes in the canonical ensemble
~\cite{York:1986, Zaslavskii:1991} and its generalization to 
higher dimensions~\cite{Andre:2020czm, Andre:2021ctu}, 
indeed there is a stable black hole 
solution which must be larger than $x=\frac{2}{d-1}$, where 
$x$ is the ratio between the horizon radius and the cavity radius. 
For larger values of $\frac{M}{R^{d-3}}$, it seems that increasing the 
electric charge does not change the stability of the shell, apart from the 
subextremality and no trapped surface conditions. This can be interpreted in 
terms of a thermal length scale, which is proportional to the reduced 
inverse temperature $b$. We have that, 
for small $\frac{M}{R^{d-3}}$ and $Q=0$, the thermodynamic unstable shells 
have radii higher than the photonic orbit. Since the thermal length $b$ 
is proportional to $M$ in the uncharged case, the thermal length is 
smaller or of the order of the radius of the shell, and so the shell 
loses energy and mass along these thermal lengths. The effect of loosing mass 
causes the thermal length $b$ to decrease and so we have a runaway process, 
the shell is unstable. For the case that charge $Q$ is increased, 
the thermal length $b$ gets also increased and so it happens that 
for sufficiently large charge, $b$ becomes greater than the radius of the 
shell, quenching the loss of energy. And so the shell becomes stable for 
charges larger than this minimum electric charge. For a shell close to 
extremality, the thermal length is proportional to $\frac{1}{\sqrt{M-Q}}$, 
which is divergent and so larger than $R$. For the exact value of 
$Q=0$ and $\frac{M}{R^{d-3}}$ corresponding to $x=\frac{2}{d-1}$, the 
shell is at the photonic orbit and it is marginally stable. Indeed, the 
thermal length is barely larger than the radius of the shell so that 
the shell is in the cusp of losing energy. For larger $\frac{M}{R^{d-3}}$
and $Q=0$, the shell resides inside the photonic orbit and the thermal 
length is larger enough to avoid the loss of energy, being thus stable. 
If we increase the charge, the thermal length increases even more and the 
shell remains stable. The discussion we presented here for generic
dimensions $d$ also applies to the $d=4$ electric charged case studied in
\cite{Lemos:2015gna} and is exemplified for $d=5$ in
Fig.~\ref{fig:S1a1MQheat}(a).

For the case of full fluctuations, i.e. mass, area and charge fluctuations,
shells with more electric charge have a lesser amount of stable 
configurations. This behaviour does differ from the case 
of mass fluctuations only, where more electric charge contributes to 
stability. But of course, the thermal length analysis is not enough to 
explain such stability since there are area and charge fluctuations to 
take into account. Stability is then more restrictive, meaning that 
configurations that are stable to mass fluctuations may not be 
stable to full fluctuations, while stable configurations to full 
perturbations must be stable to mass fluctuations only.

We must indicate another point regarding the stable values of $a$ in the 
case of one or two fluctuations together. Indeed, there are stable 
shell configurations with $a\geq 1$. And in turn, shells with higher 
$a$ for the same thermodynamic configurations have higher entropy as 
it goes with the power of $a$. For example, for area fluctuations only, 
the value of $a$ for marginal stability is $a= 3- \frac{2}{d-2}$ for 
$x=1$ and $y=0$, and then increases for other values. Since $a
= 3 - \frac{2}{d-2}$ is always larger than one, it may mean that 
a shell with lower $a$ suffers a transition to a shell with larger 
$a$, since the latter has more entropy. For that to happen, the 
matter of the shell would have to rearrange in order to change its 
equation of state. Note however that the stability analysis here is 
for fixed $a$, and so one would need to have a more fundamental description 
for the shell to understand if such transition is possible.

With the thermodynamic stability conditions worked out, we present the physical 
meaning of these conditions below.

\section{Intrinsic thermodynamic stability in terms of laboratory variables
\label{sec:intrinsicstablab}}

\subsection{The case for mass fluctuations only
\label{sec:labmassfluc}}

In thermodynamics, the stability conditions are linked to thermodynamic 
quantities that are measured in a laboratory. Here, we establish this link 
for the self-gravitating thin shell. A simple example is the one given by 
a shell with mass fluctuations only. The stability condition is tied 
to the heat capacity at constant area and charge, $C_{A,Q}$, defined 
as $C_{A,Q}^{-1} = \left(\frac{\partial T}{\partial M}\right)$, since 
$S_{MM} = -\beta^2 C_{A,Q}^{-1}$. For the shell to be stable in terms 
of mass fluctuations only, one has $S_{MM} \leq 0$, 
and so the shell must have a positive heat capacity. 
We extend this analysis below for mass and charge fluctuations together 
and for full fluctuations, as they are the most interesting cases in 
the context of the thesis.

\subsection{The case for mass and charge fluctuations
\label{sec:labmasschargefluc}}

We discuss here the stability conditions for mass and charge fluctuations in terms of 
laboratory variables. The interest in this case 
stems from the fact that, in canonical ensembles, the area is 
fixed, and this extends even to the case of black holes, where the 
area of the cavity is fixed. There are two variables that have an 
important role which are the two heat capacities $C_{A,Q}$ and 
$C_{A,\Phi}$, i.e. the heat capacity at constant area and charge, and 
the heat capacity at constant area and electric potential. There is 
an additional variable, the susceptibility at constant entropy and area 
$\chi_{S,A}$ which comes into play.

The idea is to write the second derivatives of the entropy in terms of 
the laboratory variables or thermodynamic coefficients. 
For that, we must start from the equations of state 
$T(M,A,Q)$, $p(M,A,Q)$ and $\Phi(M,A,Q)$, given in 
Eqs.~\eqref{ch2eq:tempeqstatestability}-\eqref{ch2eq:Phistability}, 
and rewrite them in terms of such variables. For mass and charge fluctuations,
we only need to consider $T(M,A,Q)$ and $\Phi(M,A,Q)$.

Regarding the equation of state for the temperature, 
$T(M,A,Q)$, it is convenient to define the laboratory quantities 
in terms of the derivatives of $S(T,A,Q)$. The heat capacity 
$C_{A,Q}$ is defined as $
\frac1T C_{A,Q}=(\frac{\partial S}{\partial T})_{A,Q}$, 
which is equivalent to the usual definition $C_{A,Q}^{-1} =
(\frac{\partial T}{\partial M})_{A,Q}$. The latent 
heat capacity at constant temperature and charge,
$\lambda_{T,Q}$, is defined by the derivative 
$\lambda_{T,Q}=(\frac{\partial
S}{\partial A} )_{T,Q}$. The latent heat capacity 
at constant temperature and area, $\lambda_{T,A}$, 
is defined by the derivative 
$\lambda_{T,A}=(\frac{\partial S}{\partial Q})_{T,A} $.
With these definitions, we can write the differential of $S(T,A,Q)$ 
and then invert the relation to get the differential of 
$T(S,A,Q)$. Using the first law as $TdS = dM+pdA-\Phi dQ$, 
the differential of $T(S,A,Q)$ can be transformed into 
the differential of $T(M,A,Q)$, yielding 
\begin{align}
    dT = \frac{1}{C_{A,Q}}dM -
    \frac{T\lambda_{T,Q} - p}{C_{A,Q}} dA-
    \frac{T \lambda_{T,A}+\Phi}{C_{A,Q}} dQ\,.
    \label{ch2eq:dTMassCharge}
\end{align}

Regarding the equation of state for the 
electric potential, $\Phi (M,A,Q)$, we can define the 
laboratory variables in terms of the derivatives of 
$\Phi(S,A,Q)$. The adiabatic electric susceptibility, 
$\chi_{S,A}$, is 
defined as $\frac{1}{\chi_{S,A}}= 
(\frac{\partial \Phi}{\partial Q} )_{S,A}$.
The electric pressure at constant entropy and charge, $P_{S,A}$, 
is defined as $P_{S,Q} =( \frac{\partial \Phi}{\partial
A})_{S,Q} $. The remaining derivative of 
$\Phi$ is given by the Maxwell relation 
$(\frac{\partial \Phi}{\partial S})_{A,Q}
= (\frac{\partial T}{\partial Q})_{S,A} = - \frac{T
\lambda_{T,A}}{C_{A,Q}}$. The differential of 
$\Phi(S,A,Q)$ can be written directly in terms of laboratory 
variables, and using the first law $TdS = dM+pdA-\Phi dQ$, 
we obtain the differential of $\Phi(M,A,Q)$ as
\begin{align}
    &d\Phi = - \frac{\lambda_{T,A}}{C_{A,Q}} dM +
\left(
P_{S,Q}
-
p\frac{\lambda_{T,A}}{C_{A,Q}}\right)dA
+
\left(\frac{1}{\chi_{S,A}} + \frac{\Phi
\lambda_{T,A}}{C_{A,Q}}\right)dQ 
\,.
\label{ch2eq:dphiMassCharge}
\end{align}
Additionally, it is important to define the heat capacity at 
constant area and electric potential as 
$C_{A,\Phi} = T (\frac{\partial
S}{\partial T})_{A,\Phi}$, which can be written 
as $ C_{A,\Phi}= C_{A,Q} (1 -
\frac{T \lambda_{T,A}^2}{C_{A,Q}}\chi_{S,A})^{
-1}$.

Returning to the stability conditions of a shell for mass and 
charge fluctuations, the relevant stability conditions are
$S_{MM}\leq 0$ and $S_{MM}S_{QQ} - S_{MQ}^2 \geq 0$, see 
Eq.~\eqref{ch2eq:StabCond}. The first condition is identical to 
the one of the mass fluctuations only, as we have
 $S_{MM} = -\beta^2 \frac1{C_{A,Q}}$ from 
 Eq.~\eqref{ch2eq:dTMassCharge}. 
 The second condition can be rewritten using 
 Eqs.~\eqref{ch2eq:dTMassCharge} and~\eqref{ch2eq:dphiMassCharge}, 
 together with the definition of the heat capacity $C_{A,\Phi}$ 
 to obtain $S_{MM}S_{QQ} - S_{MQ}^2 = \
\beta^2\frac{1}{C_{A,\Phi}\chi_{S,A}}$.
And so the thermodynamic stability for mass and charge fluctuations 
reduces to the conditions
\begin{align}\label{ch2eq:LabMQ}
    & C_{A,Q} \geq 0 \,,\nonumber\\
& C_{A,\Phi}
\,\,\chi_{S,A} \geq 0\,.
\end{align}

For the equations of state chosen, i.e. 
Eqs.~\eqref{ch2eq:tempeqstatestability}-\eqref{ch2eq:Phistability}
the adiabatic susceptibility is given by 
$\chi_{S,A}^{-1} = \Phi^2 \frac{\mu}{R^{d-3}\left(1-\mu
    \frac{M}{R^{d-3}}\right)} + \frac{\Phi}{Q}$, and so for the 
physical parameters of $(M,A,Q)$, $\chi_{S,A} \geq 0$. 
Therefore, the stability conditions become 
$C_{A,Q} \geq 0$ and $C_{A,\Phi} \geq 0$. For the case of the shell,
the condition $S_{MM}S_{QQ} - S_{MQ}^2 \geq 0$ is sufficient 
and so the condition for thermodynamic stability for mass and 
charge fluctuations is 
\begin{align}
    \label{ch2eq:SMMQQa2} 
C_{A,\Phi} \geq 0 \,.
\end{align}
Note that Eq.~\eqref{ch2eq:SMMQQa2} is equivalent to 
Eq.~\eqref{ch2eq:SMMQQa}, but it is important to stress that 
equality in Eq.~\eqref{ch2eq:SMMQQa} means that the heat capacity 
diverges to positive infinity.

\subsection{The case for mass, area and charge fluctuations
\label{sec:labmassareachargefluc}}

In this subsection, we treat the thermodynamic stability of a thin shell with 
full fluctuations, i.e. mass, area and charge fluctuations,
in terms of the laboratory variables. The analysis of the previous 
subsection highlighted the importance of the heat capacities 
$C_{A,Q}$ and $C_{A,\Phi}$ in the description of thermodynamic stability 
with fixed area. For the case of full fluctuations, there are other 
thermodynamic coefficients that play an important role, namely 
the expansion coefficient at constant temperature and electric charge, 
$\kappa_{T,Q}$, and the electric susceptibility at constant pressure 
and temperature $\chi_{p,T}$. Additionally, the heat capacity $C_{A,Q}$ 
also appears here.

As in the previous subsection, it is helpful to obtain the 
differential of the equations of state $T(M,A,Q)$, $p(M,A,Q)$ and 
$\Phi(M,A,Q)$ in terms of the laboratory variables or thermodynamic 
coefficients. These can then be related to the second order derivatives 
of the entropy since $dS = \beta dM + \beta p dA - \beta \Phi dQ$. 
To that effect, we define the laboratory variables considering the 
derivatives of $S(T,p,Q)$, $A(T,p,Q)$ and $\Phi(T,p,Q)$, as they simplify the 
considered stability conditions. Note that the three functions 
$S(T,p,Q)$, $A(T,p,Q)$ and $\Phi(T,p,Q)$ are precisely the derivatives 
of the Gibbs potential, i.e. $dG = - SdT + Adp + \Phi
dQ$. Starting with the equation of state $A(T,p,Q)$, the 
expansion coefficient $\alpha_{p,Q}$ is 
defined as $\alpha_{p,Q} = 
\frac1A (\frac{\partial A}{\partial T})_{p,Q} $, 
the isothermal compressibility $\kappa_{T,Q}$ is defined as 
$\kappa_{T,Q}=-\frac1A
( \frac{\partial A}{\partial p})_{T,Q}$, 
and the electric compressibility $\kappa_{p,T}$ is defined as 
$\kappa_{p,T}=-\frac1A
( \frac{\partial A}{\partial Q})_{T,p}$. 
For the equation of state $S(T,p,Q)$, the derivative 
$(\frac{\partial S}{\partial T})_{p,Q}$ can be written 
as $(\frac{\partial S}{\partial T})_{p,Q} =
    \frac{C_{A,Q}}{T} + A \frac{\alpha_{p,Q}^2}{\kappa_{T,Q}}$, 
while the derivative $(\frac{\partial S}{\partial
p})_{T,Q}$ can be calculated
using the Maxwell relation $(\frac{\partial S}{\partial
p})_{T,Q} = -(\frac{\partial A}{\partial T})_{T,Q}$ 
to get $(\frac{\partial S}{\partial T})_{p,Q} =
    \frac{C_{A,Q}}{T} + A \frac{\alpha_{p,Q}^2}{\kappa_{T,Q}}$,
$(\frac{\partial S}{\partial p})_{T,Q}$. For the remaining derivative,
the 
latent heat capacity $\lambda_{p,T}$ is defined as 
$\lambda_{p,T}=
(\frac{\partial S}{\partial Q})_{p,T}$. 
For the equation of state $\Phi(T,p,Q)$, two of its derivatives 
are given by the Maxwell relations as 
$(\frac{\partial \Phi}{\partial T})_{p,Q} =
-(\frac{\partial S}{\partial Q})_{p,T}=\lambda_{p,T}$
and $(\frac{\partial \Phi}{\partial p})_{T,Q}
= (\frac{\partial A}{\partial Q})_{p,T}
=-A\kappa_{p,T}$, while the isothermal electric susceptibility 
$\frac1{\chi_{p,T}}$ is defined as $\frac1{\chi_{p,T}}=
\left(\frac{\partial \Phi}{\partial Q}\right)_{p,T}$.

Having these definitions together with the differentials 
$dA(T,p,Q)$ and $dS(T,p,Q)$ in terms of laboratory variables, 
we can invert the relations to 
obtain $dT(S,A,Q)$ and $dp(S,A,Q)$. Using then the first law 
$TdS=dM+pdA-\Phi dQ$, the differentials 
$dT(M,A,Q)$ and $dp(M,A,Q)$ are obtained. Inserting these two 
differentials into the differential of $\Phi(T,p,Q)$, we obtain
the differential $d\Phi(M,A,Q)$. The differentials in terms of the 
laboratory variables and the thermodynamic variables $(M,A,Q)$ are 
\begin{align}
    &dT = \frac{dM}{C_{A,Q}} + \left(\frac{p}{C_{A,Q}} -
    T\frac{\alpha_{p,Q}}{C_{A,Q}\kappa_{T,Q}}\right)dA -
    \left(\frac{\Phi}{C_{A,Q}} + T\frac{\lambda_{p,T}}{C_{A,Q}} + A
    \frac{\alpha_{p,T} \kappa_{p,T}}{\kappa_{T,Q} C_{A,Q}}
    \right)dQ\,,\label{ch2eq:dTMAQ}\\
    &dp = \frac{\alpha_{p,Q}}{C_{A,Q}\kappa_{T,Q}} dM -
    \left[ \frac{1}{A \kappa_{T,Q}} - \frac{\alpha_{p,Q}}{C_{A,Q}
    \kappa_{T,Q}}\left( p - T \frac{\alpha_{p,Q}}{\kappa_{T,Q}}\right)
    \right] dA \nonumber\\& - \left(\frac{\kappa_{p,T}}{\kappa_{T,Q}}
    - \frac{\alpha_{p,Q}}{C_{A,Q} \kappa_{T,Q}}\mathcal{C}\right)dQ\,,
    \label{ch2eq:dpMAQ}\\
    &d\Phi = - \mathcal{B} dM +
    \left[\frac{\kappa_{p,T}}{\kappa_{T,Q}} - \left(p -
    T\frac{\alpha_{p,Q}}{\kappa_{T,Q}}\right)\mathcal{B}
    \right]dA + \left(\mathcal{B}\mathcal{C} +
    \frac{1}{\chi_{p,T}} + A\frac{\kappa_{p,T}^2}{\kappa_{T,Q}}
    \right)dQ\,,\label{ch2eq:dPhiMAQ}
\end{align}
where $\mathcal{B}$ is defined as
$\mathcal{B} = A\frac{\kappa_{p,T} \alpha_{p,Q}}{C_{A,Q}
\kappa_{T,Q}} + \frac{\lambda_{p,T}}{C_{A,Q}}$,
and $\mathcal{C}$ is defined as $\mathcal{C} =
T A \frac{\kappa_{p,T}\alpha_{p,Q}}{\kappa_{T,Q}} + T\lambda_{p,T} +
\Phi$. With Eqs.~\eqref{ch2eq:dTMAQ}-\eqref{ch2eq:dPhiMAQ}, the second 
derivatives of the entropy in terms of the laboratory variables can 
be found through the first law of thermodynamics.

The intrinsic thermodynamic stability of the shell for mass, area 
and charge fluctuations is by the relevant stability conditions 
$S_{MM}\leq 0 $,
$S_{MM}S_{AA} - S_{MA}^2 \geq 0$, and 
$(S_{MM}S_{AQ} - S_{MA}S_{MQ})^2
-(S_{AA}S_{MM} - S_{AM}^2)(S_{QQ}S_{MM} - S_{QM}^2 ) 
\leq 0$, taken from Eq.~\eqref{ch2eq:StabCond}. 
Now, in terms of laboratory variables, the first condition 
is given by $S_{MM} = -\beta^2\frac{1}{C_{A,Q}}$, 
the second condition is given by 
$S_{MM}S_{AA} - S_{MA}^2 = -\beta^3\frac{1}{A \kappa_{T,Q}
C_{A,Q}}$ and finally the third condition is 
given by $(S_{MM}S_{AQ} - S_{MA}S_{MQ})^2
-(S_{AA}S_{MM} - S_{AM}^2)(S_{QQ}S_{MM} - S_{QM}^2 )
- \beta^6\frac{1}{A C_{A,Q}^2 \kappa_{T,Q} \,\,\chi_{p,T}}$. 
It follows that the stability conditions for mass, area and charge 
fluctuations in terms of the laboratory variables reduce to
\begin{align}
    &C_{A,Q}\geq 0,\,\nonumber\\
    &\kappa_{T,Q}\geq 0\,,\nonumber\\
    & \chi_{p,T}\geq 0\,.\label{ch2eq:stabcondbetter}
\end{align}
Hence, all the three laboratory quantities have to be positive,
specifically, 
the heat capacity $C_{A,Q}$ related to the temperature 
equation of state, 
the isothermal compressibility $\kappa_{T,Q}$
related to the pressure equation of state, and the
isothermal electric
susceptibility $\chi_{p,T}$ related to the 
electric potential equation of state, have to be positive,
with the case of marginal stability
corresponding to these physical
variables going to infinity.

From the conditions in Eqs.~\eqref{ch2eq:stabcondbetter}, 
the sufficient stability condition for the case of the shell 
with the specific choice of equations of state is the last condition in 
Eq.~\eqref{ch2eq:stabcondbetter}, namely
\begin{align}
    \label{ch2eq:SMAQlab} 
    \chi_{p,T}\geq 0 \,.
\end{align}
In connection with Sec.~\ref{massareaandchargeflucaltogether}, 
the condition in Eq.~\eqref{ch2eq:SMAQlab} is equivalent to 
Eq.~\eqref{ch2eq:SMMAAQQa}, meaning that 
for $a<\frac{d-3}{d-2}$, the isothermal electric susceptibility 
is positive, and for $\frac{d-3}{d-2}<a<1$ it is positive 
depending on the values of $(r_+, r_-,R)$. For $a\geq 1$ and 
$r_+<R$, the susceptibility is negative. The shell with black hole 
features has to be treated carefully as it resides in the 
marginal surface. If the shell with $a=1$ and 
$r_+<R$ approaches its own gravitational radius, it is thermodynamically 
unstable as the susceptibility tends to $\chi_{p,T}
\rightarrow -\infty$. However, there can be a configuration 
with $R=r_+$ from the start that does not belong to this 
sequence of quasistatic configurations. The stability of the 
black hole limit depends on whether the exponent $a$ of the 
equation of state
approaches $a=1$ from below
or from above. If it is possible to have the exponent $a$ of 
the temperature equation of state to approach $a=1$ from 
below, the configuration with $R=r_+$ is marginally stable with 
$\chi_{p,T}\rightarrow +\infty$. Having such diverging susceptibility 
means that changes in the electric charge
of the configuration
don't have any impact on the electric potential.
For the case that the exponent $a$ approaches $a=1$ from above, 
the configuration with $R=r_+$ is unstable, and  $\chi_{p,T}
\rightarrow -\infty$.

\section{Conclusions\label{sec:conc1stlaw}}

In this chapter, we used the thin shell formalism to determine the 
mechanics of a static charged spherical thin shell in $d$ 
dimensions in general relativity. Furthermore, we studied the thermodynamics of 
the shell by imposing the first law of thermodynamics.
The use of the pressure equation of state as given by general relativity 
and the relation between the rest mass of the shell and the quasilocal 
energy give special thermodynamic properties to the shell, indicating 
a link between thermodynamics and general relativity. One of such 
remarkable 
thermodynamic properties is that the entropy of shell depends on 
$r_-$ and $r_+$ and not on the radius of the shell. Note that this 
property has also been found for other thin shell spacetimes.

In order to proceed with the thermodynamic analysis of the shell, we provided
two equations of state to the shell, one where the 
temperature is described by a power law in $r_+$ with exponent $a$, 
and another where the electric potential is described by the typical 
potential of the Reissner-Nordstr\"om spacetime. We were interested in these 
specific shells due to the possibility of performing the black hole limit, and also for 
having shells with thermodynamic black hole features.

We studied the thermodynamic intrinsic stability of the shell. A shell 
is stable if the hessian of the entropy is negative semidefinite, 
where the marginal case was included. We analyzed the stability for 
seven types of fluctuations. The most general case 
constitutes the one with mass, area and charge fluctuations, for which 
the shell is always stable in the case  $0<a\leq \frac{d-3}{d-2}$. 
For $\frac{d-3}{d-2}<a<1$, the stability depends on the mass 
and electric charge, while for $a\geq 1$ and $r_+<R$ the shell is unstable.
For the shell with $a=1$ at its own gravitational radius, there is marginal 
stability.

We have seen the thermodynamic intrinsic stability of the shell from the 
perspective of laboratory variables. For the generic type of 
fluctuations, stable shells have positive heat capacity, 
positive isothermal compressibility and positive 
isothermal electric susceptibility. We found, for the specific 
shells considered, that the positivity of the isothermal electric 
susceptibility is sufficient for the thermodynamic intrinsic stability 
of the shell. The marginal stability case corresponds to an infinite 
electric susceptibility, with its positivity depending on the way 
one approaches the marginal points. If the shell has negative 
susceptibility, there is a runaway process, making them depart from 
equilibrium towards a stable equilibrium configuration or even 
towards a breakdown of the shell.

In this chapter, we 
have derived some thermodynamic properties for electrically
charged spherical matter shells in higher dimensions, complementing a
set of works on the thermodynamics of thin shells. There is still 
more future work that has to be done in 
regarding the link between
thermodynamics and general relativity, hopefully contributing to the
understanding of black hole physics and with it to grasp
gravitation at the tiniest possible scales.

\cleardoublepage
\part{Statistical mechanical ensembles of black holes and matter 
using the Euclidean path integral approach}\label{pt:ensembless}
\chapter[Thermodynamics in curved spaces
through the Euclidean path integral]{Thermodynamics in curved spaces
through the Euclidean path integral approach}
\label{ch:Euclideanpathintegral}

\section{Thermodynamic black hole ensembles}

\subsection{The Gibbons-Hawking statistical path integral 
and York formalism}

The thermodynamics of stationary configurations involving 
gravity can be obtained from the construction of
statistical ensembles through the Euclidean path integral 
approach to quantum gravity. The approach is based on 
extending the statistical path integral to the gravitational 
sector, where one performs a map from the 
Lorentzian metric to a Riemannian or pseudo-Riemannian 
metric \cite{Hawking:1980gf, Brown:1994su}. 
This map is usually a Wick transformation $t \rightarrow -i \tau$, 
where $t$ is a Lorentzian time coordinate and $\tau$ is 
an imaginary time. The path integral of the Euclidean gravitational 
action is then performed 
over the possible 
metrics with fixed boundary conditions which are extracted from 
the configuration one wishes to study. In the canonical ensemble,
one fixes the inverse temperature given by the total imaginary proper 
time at the boundary. The boundary of the space then acts as 
a heat reservoir. The statistical path integral gives the 
partition function of the ensemble, which is associated to a 
thermodynamic potential. One can use this link to extract the thermodynamic
properties of the system.

The Euclidean path integral approach has several shortcomings. To start,
the map between Lorentzian spacetimes 
and Riemannian or quasi-Riemannian spaces is not well-defined in general.
Moreover, the Euclidean gravitational action can be unbounded from 
below, which makes the path integral ill-defined. It has been suggested that 
this last issue can be tackled by using conformal 
classes of metrics and change the contour of the 
integration~\cite{Gibbons:1978ac,Hawking:1978jz}. 
There is also a problem regarding the measure 
of the gravitational metric. For gauge fields, the measure is well understood, 
where the overcounting coming from the gauge freedom is removed using ghost contributions. 
For general metrics, it is not yet clear 
how to remove the overcounting from diffeomorphisms. Moreover, 
it is also not clear what is the relative measure between metrics 
of different topology. Nevertheless, the Euclidean path integral 
approach yields interesting results when the saddle point approximation 
is performed. This approximation or the zeroth order version of it, the 
zero loop approximation, avoids the shortcomings of the 
Euclidean path integral. It consists on expanding the Euclidean action 
over the paths that extremize the action. In the semiclassical limit, these
paths contribute the most to the partition function and have a 
correspondence to physical Lorentzian spacetimes. For the approximation 
to be valid, one must consider the stationary points that minimize
the action. This can be seen from the first order loop corrections, 
where the integrand can be put in terms of the eigenvalues of an 
operator. For the stationary points that minimize the action, 
the operator has positive eigenvalues, yielding a path integral 
with real values. Otherwise, the stationary 
point is called an instanton and the first loop 
corrections yield complex contributions to the partition function 
with the imaginary part indicating the 
decay probability of the instanton.

The construction of statistical ensembles through the 
Euclidean path integral approach to quantum gravity was 
first applied by Gibbons and Hawking~\cite{Gibbons:1977}. In the 
zero loop approximation, the grand canonical and canonical ensembles 
of Kerr-Newmann black hole spacetimes in four dimensions was considered with 
boundary at infinity. For the case of a Schwarzschild black hole as a 
stationary point of the Euclidean gravitational action, 
it was observed that the heat capacity of the black hole was negative.
In the canonical ensemble, this means that the Schwarzschild black hole 
is thermodynamically unstable and also deems the zero loop approximation 
invalid. 
It was further shown in~\cite{Gross:1982cv} that the Schwarzschild 
instanton was a saddle point of the Euclidean action and not a maximum, 
and its existence caused a global instability of flat space. The first 
loop corrections of the Schwarzschild black hole led to a complex contribution 
to the partition function due to a negative mode perturbation, which 
disappeared if the boundary was put at a finite radius~\cite{Allen:1984bp}. 
An additional analysis of the negative mode was done in~\cite{Gregory:2001bd}. 
Moreover, Hawking and Page~\cite{Page:1982dh} applied the 
same formalism to the Schwarzschild-anti de Sitter black hole 
in four dimensions, with boundary at infinity. In this case, 
two possible solutions for the black hole mass were found 
for a fixed temperature, with one having a positive heat capacity 
and thus being stable. The existence of the stable 
solution is related to the fact that anti-de Sitter space acts as 
a finite box. In order to cure the canonical ensemble of a Schwarzschild black hole, 
taking into consideration the works above,
York analyzed the Schwarzschild black hole inside a finite cavity~\cite{York:1986}.
Two solutions for the Schwarzschild radius were found, in analogy to the 
Schwarzschild-anti de Sitter case, with one having again a positive heat 
capacity and thus being stable. By putting the stationary configuration in 
a finite cavity, the zero loop approximation becomes valid. 
This is the 
York formalism in the construction of statistical ensembles of 
curved spaces. Moreover, York constructed a generalized free energy 
that allowed the study of phase transitions between hot flat space, 
i.e. vacuum flat space at a fixed temperature, and the stable Schwarzschild 
black hole. The motivation for the generalized free energy was given in~\cite{Whiting:1988qr},
where the generalized free energy is obtained from the reduced action, which is 
the Euclidean action with imposed constraint equations that partially 
extremize the action.

In this part of the thesis, we are interested in exploring the York and 
Gibbons-Hawking formalism to construct statistical ensembles 
of various spacetimes. We focus on charged black hole spacetimes and 
spacetimes involving self-gravitating matter thin shells. 
The objective is to further understand 
the phase diagrams when we include matter and gauge fields, 
and to uncover possible links between dynamics and thermodynamics.

\subsection{Application to different configurations}

The construction of statistical ensembles through the Euclidean path integral 
approach was further extended to other stationary spacetime configurations 
and different ensembles. Namely, the formulation of different ensembles 
with a gravitational action was done in~\cite{Brown:1989fa}, and 
specifically the microcanonical ensemble was formulated more explicitly
in~\cite{Brown:1992bq}. Regarding other stationary black hole spacetimes 
without matter, 
the ensemble and thermodynamics of a two-dimensional black hole in 
the Teitelboim-Jackiw theory was treated in~\cite{Lemos:1996bq}, 
the formalism was extended to anti-de Sitter black 
holes~\cite{Brown:1994gs, Akbar:2004ke}, and to 
de Sitter spaces~\cite{Ghezelbash:2001vs,
Miyashita:2021iru,Banihashemi:2022jys, Jacobson:2022jir,Lemos:2024sjs}. 
An important study 
of the canonical ensemble of five-dimensional 
and $d$-dimensional Schwarzschild black holes 
was done in~\cite{Andre:2020czm, Andre:2021ctu}, where a link was 
established between the Buchdahl bound~\cite{Buchdahl:1959zz, PoncedeLeon:2000pj} 
and the radius marking the phase transition from hot flat space 
to a black hole phase. It is important to note that the   
Buchdahl bound indicates the maximum bound for the compactness 
of fluid spheres above which the configuration is singular, when certain energy conditions are obeyed.
This bound has been generalized to charged configurations~\cite{Andreasson:2008xw}, 
for positive cosmological constant~\cite{Andreasson:2009pe} and for higher dimensions~\cite{Wright:2015dma}.
Therefore, the work in~\cite{Andre:2020czm, Andre:2021ctu} suggests the existence of 
a link between matter dynamics and black hole thermodynamics, which shall be explored in this thesis.

The formalism was extended as well to 
include the Maxwell vector potential, allowing the treatment of 
charged black holes. The grand canonical ensemble for Reissner-Nordstr\"om 
black holes in four dimensions was done in~\cite{Braden:1990hw} 
and its extension to anti-de Sitter in~\cite{Peca:1998cs}. The 
thermodynamics and the construction of the ensembles of Kerr-Newmann 
black holes through the York formalism was sketched in~\cite{Brown:1990fk}.
Moreover, the canonical ensemble of a Reissner-Nordstr\"om 
black hole in four dimensions was worked out in \cite{Carlip:2003ne, 
Lundgren:2006kt}, and the $d$ dimensional Reissner-Nordstr\"om-anti-de Sitter
was worked out in~\cite{Chamblin:1999hg}.
The inclusion of matter, namely 
of a spherical matter thin shell with a black hole inside 
was done in~\cite{Martinez:1989hn}, 
where it was shown the additivity of the matter and black hole entropies. 
A more thorough study of this case was done in~\cite{Lemos:2023yiz}. 
The canonical ensemble for arbitrary configurations of a self-gravitating system 
was studied in~\cite{ZASLAVSKII1991463}.

The analysis of ensembles of anti-de Sitter spaces has a deeper motivation 
related to superstring and supergravity theories, and gauge/gravity duality. 
In supergravity theories, one usually has a collection of branes living 
in a world space. Through the gauge/gravity duality, the low energy 
supergravity in the world space can have a correspondence to 
a strongly coupled field theory at the boundary. An example of a gauge/gravity 
duality is the AdS/CFT correspondence~\cite{Maldacena:1997re, 
BenettiGenolini:2023rkq}.
The important feature here of the gauge/gravity duality is that 
the thermodynamic properties of the branes carry 
over to the field theory. In that regard, the thermodynamics 
of black branes through statistical ensembles have been 
studied~\cite{Gubser:1998jb, 
Hawking:1999dp, Reall:2001ag,Lu:2010au,Lu:2010xt}, including 
electric charge as well. Another motivation for the study of such 
thermodynamic ensembles is the Gubser-Mitra conjecture~\cite{Gubser:2000mm}. 
This conjecture states that black branes are stable to linear 
perturbations if and only if they are thermodynamically stable. 
The linear instability of black branes is mainly driven by the 
Gregory-Laflame instability~\cite{Gregory:1993vy, Collingbourne:2020sfy} 
and it was shown to have a connection with the negative mode arising 
in the perturbation of the respective instantons for particular 
cases~\cite{Reall:2001ag, Miyamoto:2007mh}, thus complying with the 
conjecture.

\subsection{Physical scales and the applicability 
of the zero loop approximation}

It is important to state the applicability of the zero loop 
approximation to obtain the partition function of self-gravitating 
systems. The calculation of the one loop corrections can 
give us a hint, by evaluating when these corrections are 
negligible. Formally, the loop contributions can be computed by 
renormalization and regularization techniques, which 
for the case of statistical path integrals, the zeta 
regularization procedure~\cite{Hawking:1976ja} and the 
expansion of the heat kernel through DeWitt-Schwinger 
proper time~\cite{Schwinger:1951nm, DeWitt:1975ys} are the most 
utilized.
The one loop corrections arise 
in the form of logarithmic terms that are added to the 
thermodynamic potential associated to the ensemble, for example 
see~\cite{Fursaev:1994te,Solodukhin:1994yz,
Mann:1996bi,Sen:2012dw} where 
the one loop contributions for different fields have been computed
arising from higher order local and non-local curvature terms. 
There is also a computation of the one loop contributions arising from
thermodynamic fluctuations~\cite{Akbar:2003mv}, where the heat capacity 
plays an important role.

The scale controlling the one loop contributions is 
generally attributed to the Planck scale, $l_p = 1.6\times 10^{-35} m$. 
It is expected that one loop contributions are negligible 
for scales much larger than $l_p$. Another scale which is 
fixed in the canonical and grand canonical ensemble is the 
temperature and the radius of the cavity. It is then useful 
to understand what are the scales of 
interest~\cite{Lemos:1996jv, Wondrak:2017ttq} in the semiclassical 
regime where the zero loop approximation is still valid. 
We can work with the stable black hole of York~\cite{York:1986},
for which the zero loop approximation is valid. 
First, we require that the regime must be far 
from the Planck length. In the canonical ensemble, this means 
that the temperature of ensemble must be below 
Planck temperature $T_{\mathrm{p}} = 10^{32}\mathrm{K}$. 
For this temperature and higher, the stable black hole is close enough to 
the cavity such that the full quantum regime must be taken 
into account. Moreover, we want to have a cavity radius 
far from the Planck length but still small compared to 
SI units in order to probe the semiclassical regime, 
e.g. we can choose a cavity radius $R = 10^{20}l_{\mathrm{p}}$. 
For the parameters chosen, the stable black hole solution 
exists for temperatures higher than $T = 2\times 10^{11}\mathrm{K}$. 
However, one loop corrections seem quite relevant near the temperatures 
at which the large black hole solution starts to exist~\cite{Akbar:2003mv}.
Hence, the zero loop approximation is valid for the range of temperatures 
$2\times10^{11}\mathrm{K} \ll T \ll 10^{32}\mathrm{K}$, with the large black hole 
radius being of the order $r_+ = 6\times 10^{19} l_{\mathrm{p}}$. 
Indeed, these orders of magnitude imply a cavity with $R= 10^{-13}\mathrm{cm}$
and a black hole with mass $m=6\times10^{14}\mathrm{g}$, which means 
the system is microscopic, where semiclassical effects enter into 
play. We could also make the same analysis for the case 
of an infinite cavity, where we have to consider 
the Gibbons and Hawking black hole. For such a black hole, the 
zero loop approximation of the canonical ensemble is 
not valid but it is useful for the study of Hawking radiation~\cite{Nicolini:2011nz}.

The scale analysis above has the purpose of motivating 
the study of thermodynamic ensembles through the zero 
loop approximation in order to probe semiclassical effects. 
Such effects can lead to phase transitions between the stable 
black hole phase and hot space, which is our main object of 
study here. Although these phase transitions may occur for temperatures 
close to the starting point of existence of the stable 
black hole, where loop corrections may be relevant, 
we extrapolate the analysis of the zero loop approximation 
to this regime with the expectation that the qualitative 
behaviour remains the same.

\subsection{Outline}

The role of this chapter is to introduce the Euclidean 
path integral approach to quantum gravity and its 
application to the construction of the statistical 
path integral for curved spaces. Moreover, this 
chapter serves as a preparation for the remaining 
chapters of the second part of the thesis. 

In Sec.~\ref{sech2:PathIntegral}, we discuss the extension 
of the Euclidean path integral to obtain the 
partition function of curved spaces. In Sec.~\ref{sech2:symmetricmetric}, we work out
the restriction of the path integral to spherically 
symmetric metrics. In Sec.~\ref{sech2:regularitycond}, we explain
the regularity conditions for the spherically symmetric metrics 
that enter in the path integral. In Sec.~\ref{sech2:boundarycond}, we present
the boundary conditions that establish the data that is fixed 
in the path integral. In Sec.~\ref{sech2:actionsphericalsym}, we calculate
the gravitational action that enters in the path integral 
for spherically symmetric metrics, which 
is relevant for the upcoming chapters. In Sec.~\ref{sech3:thermo}, we connect
the statistical path integral to the relevant 
thermodynamic potential of an ensemble, from which we can 
derive the thermodynamic quantities of the ensemble.
Finally, 
in Sec.~\ref{sech2:summary}, we summarize the chapter.

\section{The Euclidean path integral approach\label{sech2:PathIntegral}}

In order to construct the generating function of the 
statistical ensemble, also called the partition function, 
we employ the Euclidean path integral approach. This 
approach is mainly used to obtain the partition function 
of systems with quantum fields, giving origin to the 
study of thermal field theory. Suppose that one has 
a quantum system being described by a quantum field 
$\hat{\psi}$, with its classical counterpart $\psi$, 
and with the associated Hamiltonian operator $H$. 
The ensemble of the system with fixed temperature and 
volume, i.e. the canonical ensemble, has the 
partition function given by $Z = 
\Tr\left[\mathrm{e}^{-\beta H}\right]$, where $\Tr$ 
is the trace of the operator over a basis of the Hilbert 
space, where the quantum field theory is modelled. This trace 
can be rewritten in terms of a path integral. In order to 
see this, one can use the formula of the Feynman path integral. 
Let the system be at the quantum state $\ket{\psi_1}$ at a 
time $t_1$, the amplitude for the system to be at a quantum 
state $\ket{\psi_2}$ at time $t_2$ is $\bra{\psi_2} 
\mathrm{e}^{-i(t_2-t_1)H} \ket{\psi_1}$, but in turn this 
can also be given by the Feynman path integral 
$\int D\psi\, \mathrm{e}^{i I_L}$, where $I_L$ is the 
Lorentzian action of the field $\psi$, and where the 
path integral has boundary conditions $\psi(t_1) = \psi_1$
and $\psi(t_2)=\psi_2$. One can make now the transformation 
$i (t-t_1) = \tau'$, where $\tau'$ is an imaginary time with period 
$\beta = i(t_2 - t_1)$, and also set $\psi_1=\psi_2$. 
Therefore, the partition function of the ensemble is given by 
$Z = \int D\psi\, \mathrm{e}^{-I}$, where $I$ is the Euclidean 
action and the integration is done with periodic boundary conditions 
with period $\beta$ for bosonic fields, and anti-periodic boundary conditions 
for fermionic fields, due to 
the properties of the commutator and anticommutator 
between the fields.

The idea of the Euclidean path integral approach to quantum gravity is to 
apply the aforementioned logic to the gravitational field. The first ingredient 
is the map between a $d$ dimensional Lorentzian spacetime $M_L$ 
and a $d$ dimensional Riemannian 
space $M$, through a Wick rotation $t\rightarrow - i \tau'$, where 
$t$ is a Lorentzian time coordinate and $\tau'$ is the imaginary 
time. Of course, such map is not covariant and may be ill-defined. 
Usually, this issue can be 
overlooked for static spacetimes, while for stationary spacetimes 
one must consider a map to a quasi-Riemannian space instead, which 
satisfies allowable conditions, see~\cite{Kontsevich:2021dmb, Witten:2021nzp}. 
The time coordinate chosen for the map is usually associated to 
a Killing vector, which is timelike in some region. For black hole 
spacetimes, the time coordinate chosen is associated to the
Killing vector that is timelike near the horizon and 
becomes null at the horizon. 
At the boundary of the spacetime, $\partial M_L$, 
one has the heat reservoir represented by 
a timelike hypersurface in the Lorentzian spacetime. This hypersurface 
can be brought to the Riemannian or quasi-Riemannian space through the 
map, obtaining 
a hypersurface $\partial M$.
Now, there must be an identification of points such that the imaginary 
time $\tau'$ is periodic with some constant period. It is better to perform 
a coordinate transformation $\tau'(\tau)$ and work with an 
imaginary time $\tau$ with period $2\pi$. This map allows the 
correspondence of a Riemannian or quasi-Riemannian space with the 
physical spacetime, and more importantly the physical boundary data of the 
heat reservoir can be established, namely its geometry 
and its quasilocal quantities, such as the energy and angular 
momentum. This data at the boundary of space is what we need to 
consider in the construction of an ensemble, while the specific geometry of the 
Riemannian or quasi-Riemannian space is not needed in principle but it 
plays a huge role as we shall see.

The partition function of an ensemble for a curved spacetime including 
matter fields, through the Euclidean path integral approach, is formally defined 
by 
\begin{align}\label{eqch3:pathint}
    Z = \int Dg_{\alpha \beta} D\psi \,\mathrm{e}^{-I[g_{\gamma \nu},\psi]}\,\,,
\end{align}
where $I$ is the Euclidean action,
$g_{\alpha \beta}$ is the 
Euclidean metric of the Riemannian space (not to be confused with the Euclidean flat metric), 
$\psi$ represents any kind 
of matter or gauge field, $Dg_{\alpha \beta}$ is the integration 
measure over the paths of $g_{\alpha \beta}$ and $D\psi$ is the 
integration measure over the path of $\psi$. The path integral 
is done over periodic $g_{\alpha \beta}$ and $\psi$, if bosonic.
In general, all paths of $g_{\alpha \beta}$ 
and $\psi$ may not have locally a physical 
correspondence, but one does need to give fixed data at the boundary 
of the Riemannian space, $\partial M$, where the heat reservoir sits, 
corresponding to the same data of the stationary Lorentzian 
spacetime one wants to study, through the Wick rotation mentioned above.
Namely for static configurations, one can use
the Dirichlet boundary conditions for the induced 
metric at the boundary, with fixed inverse temperature defined by 
$\beta = \int_0^{2\pi} b|_{\partial M} d\tau$, where $b = 1/\sqrt{g^{\tau\tau}}$, 
and with the remaining components describing the spatial geometry of 
the boundary in the Lorentzian static configuration. Moreover, 
one must also give data for the field $\psi$ at the boundary, depending 
on the type of ensemble one wants to consider. We explain the matter boundary 
conditions in the following 
chapters, according to the ensemble under study. With such boundary conditions, 
the partition function can then be determined. Note that the reason for 
associating the boundary conditions of the Riemannian space 
to the data of a Lorentzian configuration allows 
the Riemannian space obtained from the map of such configuration to be included 
in the sum of paths, with this Riemannian space extremizing the Euclidean action. 
This is how a physical meaning is given to the Euclidean path integral, since there 
is a correspondence to a physical spacetime.

We assume that the Euclidean action $I[g_{\alpha \beta},\psi]$ is given by the 
sum $I[g_{\alpha \beta},\psi]= I_g[g_{\alpha \beta}] 
+ I_m[g_{\alpha \beta},\psi]$, where
$I_g[g_{\alpha \beta}]$ is the gravitational Euclidean action 
given by the
Euclidean Einstein-Hilbert action with the Gibbons-Hawking-York 
boundary term, i.e.
\begin{align}\label{eqch3:gravaction1}
    I_g = -\frac{1}{16\pi l_p^{d-2}}\int_{M}(R - 2\Lambda)\sqrt{g}d^dx 
    - \frac{1}{8\pi l_p^{d-2}}\int_{\partial M}K \sqrt{\gamma} d^{d-1}x 
    - I_{\mathrm{ref}}\,\,,
\end{align} 
where $R$ is the Ricci scalar, $\Lambda$ is the cosmological constant, 
$g$ is the metric determinant, 
$K = {n^\alpha}_{;\alpha}$ is trace of the extrinsic curvature of $\partial M$, 
$n^\alpha$ is the unit normal vector to $\partial M$, 
$\gamma$ is the determinant of the induced metric $\gamma_{ab}$ 
of $\partial M$ and $I_{\mathrm{ref}}$ is the action of a reference metric 
to make $I_g$ finite. The action $I_m[g_{\alpha \beta},\psi]$ is the 
matter action which is specified in the following chapters depending 
on the case of study. One can obtain the Euclidean action $I_g$ from the 
Lorentzian action $I_L$ by performing the map referred above from a Lorentzian spacetime 
to a Riemannian space. Neglecting the boundary term on the 
spacelike hypersurfaces, the effect of the map can be seen by 
changing the volume elements as $\sqrt{-g_L}d^dx \rightarrow -i \sqrt{g}d^dx$ 
and $\sqrt{-\gamma_L}d^{d-1}x \rightarrow -i \sqrt{\gamma}d^{d-1}x$, 
as the integrands are left invariant, where $g_L$ is the determinant of the 
Lorentzian metric and $\gamma_L$ is the determinant of the induced Lorentzian 
metric. The Euclidean action is then defined 
with an overall minus sign so that $I_L \rightarrow i I_g$, which explains the minus 
sign in the Ricci term in Eq.~\eqref{eqch3:gravaction1}, as the Lorentzian 
action is defined with a positive sign in the Ricci term. The analysis 
of the gravitational Euclidean action is going to be split into two 
cases for spherically symmetric metrics: 
the zero cosmological constant case and the negative cosmological case.
For the negative cosmological case, the anti-de Sitter or AdS length is 
defined by $l^{2} = \frac{(d-1)(d-2)}{-2\Lambda}$.

\section{The class of spherically symmetric metrics\label{sech2:symmetricmetric}}

\subsection{Smooth metrics}

In this thesis, we focus on statistical ensembles of configurations 
with spherical symmetry. In order to avoid the repetition in the upcoming 
chapters, we analyze here in detail 
the Euclidean gravitational action in spherical symmetry.

To avoid the problems coming from a sum over topologies, 
we can restrict the paths in the path 
integral to metrics with spherical symmetry. In cases where the system 
is described by a finite cavity, this can be motivated by the fact that 
spherically symmetric metrics are expected to contribute the most 
to the path integral. The Euclidean metric for the Riemannian space $M$ can 
then be written as 
\begin{align}\label{eqch3:metric}
    ds^2 = b(u)^2 d\tau^2 + a(u)^2 du^2 + r(u)^2 d\Omega^2_{d-2}\,\,,
\end{align}
where $b(u)$, $a(u)$ and $r(u)$ are arbitrary functions of $u$, 
the coordinate $\tau$ is spanned by $\tau\in\,]0,2\pi[$, 
the coordinate $u$ is spanned by $u\in \,]0,1[$ and 
$d\Omega^2_{d-2}$ is the $(d-2)$--sphere metric in spherical 
coordinates $\theta^A$ with total area $\Omega_{d-2}=
\frac{2\pi^{\frac{d-1}{2}}}{\Gamma(\frac{d-1}{2})}$,
where $\Gamma$ is the gamma function.

The boundary of the space is described by the hypersurface 
$u=1$, which may be singular for the reservoir at infinity 
or smooth for the reservoir at a finite radius. 
It is then useful to analyze hypersurfaces of constant $u$, 
which have an induced metric 
\begin{align}\label{eqch3:uconstant}
    ds^2|_{u} = b(u)^2 d\tau^2 + r(u)^2 d\Omega^2_{d-2}\,\,. 
\end{align}
The dependence on $u$ is now going to be dropped for convenience, 
except for occasions where clarity demands it.
The extrinsic curvature $K_{ab}$ of the constant $u$ hypersurfaces 
can be calculated using the unit normal 
$n_\alpha dx^\alpha = a du$ as
\begin{align}\label{eqch3:extrinsiccurvy}
    K_{ab}dx^a dx^b = \frac{b' b}{a}d\tau^2 
    + \frac{r' r}{a}d\Omega^2_{d-2}\,\,,
\end{align} 
where a prime means the derivative over $u$, i.e. 
$b' = \frac{db}{du}$. The trace of the extrinsic curvature 
is given by
\begin{align}\label{eqch3:traceextrinsiccurvy}
    K = \frac{b'}{ab} + (d-2)\frac{r'}{a r}\,\,.
\end{align}

One can use the Cartan structure equations to determine the 
Ricci tensor, $R_{\alpha \beta}$, 
of the metric in Eq.~\eqref{eqch3:metric}, 
together with a differential relation between the components 
of $d\Omega^2_{d-2}$. The components of the 
Ricci tensor have the following expression 
\begin{align}\label{eqch3:Riccitensor}
    & R\indices{^\tau_\tau} = - \frac{1}{ab r^{d-2}}
    \left(\frac{b' r^{d-2}}{a}\right)'\,\,,\notag\\
    &  R\indices{^y_y} = R\indices{^\tau_\tau} 
    - \frac{d-2}{r a}\left( \frac{r'}{a}\right)' 
    + (d-2)\frac{b' r'}{a^2 b r}\,\,,\notag\\
    & R\indices{^{\theta^A}_{\theta^A}} = 
    - \frac{b' r'}{b a^2 r} + \frac{1}{ra}\left(\frac{r'}{a}\right)'
    - \frac{1}{r' r^{d-2}}\left[r^{d-3}\left(\left(\frac{r'}{a}\right)^2 
    - 1 \right) \right]'\,\,, 
\end{align}
where the indices $A$ are not being summed. The Ricci scalar, $R$, 
is then given by 
\begin{align}\label{eqch3:Ricciscalar}
    R =  - \frac{2}{ab r^{d-2}}
    \left(\frac{b' r^{d-2}}{a}\right)' - 2G\indices{^\tau_\tau}\,\,,
\end{align}
where $G\indices{^\tau_\tau}$ is the $\tau\tau$ component of the 
Einstein tensor, given by
\begin{align}\label{eqch3:Einsteintensor}
    G\indices{^\tau_\tau} = \frac{(d-2)}{2r' r^{d-2}} 
    \left[r^{d-3}\left(\left(\frac{r'}{a}\right)^2 
    - 1 \right) \right]'\,\,.
\end{align}

\subsection{$C^0$ metrics}

For the purpose of the thesis, we assume 
metrics with the form of Eq.~\eqref{eqch3:metric} to be smooth 
except when there is the presence of a spherical matter thin shell, 
described by the hypersurface $\mathcal{C}$. In such case, the 
hypersurface $\mathcal{C}$ separates the space $M$ into 
the inner region $M_1$ and the outer region $M_2$ with metrics
\begin{align}
    &ds^2_1 = b_1(u)^2\frac{b_2(u_\mathrm{m})^2}{b_1(u_\mathrm{m})^2} 
    d\tau^2 + a_1(u)^2 du^2 + r(u)^2 d\Omega^2_{d-2}\,\,,\label{eqch3:metricM1}\\
    &ds^2_2 = b_2(u)^2 d\tau^2 + a_2(u)^2 du^2 + r(u)^2 d\Omega^2_{d-2}\,\,,
    \label{eqch3:metricM2}
\end{align}
respectively, where $u_\mathrm{m}$ is the position label of the 
matter thin shell with the hypersurface $\mathcal{C}$ being 
described by $u=u_\mathrm{m}$, and, $b_1(u)$, $b_2(u)$, 
$a_1(u)$ and $a_2(u)$ are smooth arbitrary functions. The coordinate 
$u$ in $M_1$ has the range $u\in ]0,u_\mathrm{m}[$ and in $M_2$ it has 
the range $u\in ]u_\mathrm{m},u[$. The metrics in Eqs.~\eqref{eqch3:metricM1} 
and~\eqref{eqch3:metricM2} can be described by the metric in Eq.~\eqref{eqch3:metric}, 
with metric components 
\begin{align}\label{eqch3:shellmetric}
    & b(u) = \begin{cases}
        \frac{b_1(u) b_2(u_\mathrm{m})}{b_1(u_\mathrm{m})} & 0<u<u_\mathrm{m} \\
        b_2(u)  & u_\mathrm{m}\leq u<1\end{cases}\,\,,\notag\\
    & a(u) = \begin{cases}
        a_1(u) & 0<u<u_\mathrm{m} \\
        a_2(u) & u_\mathrm{m}\leq u<1
    \end{cases}\,\,,
\end{align}
While in this case $b(u)$ is continuous, $a(u)$ is not necessarily.
Using further a coordinate transformation to a geodesic coordinate 
$\rho = \int_{u_\mathrm{m}}^u a(s) ds$, 
one turns the discontinuity in $a$ into the property that $r(\rho)$ 
is $C^0$ in function of the geodesic coordinate. The metric 
with these properties is then $C^0$ with the form
\begin{align}
    ds^2 = b(\rho)^2 d\tau^2 + d\rho^2 + r(\rho)^2d\Omega^2_{d-2}\,\,,
\end{align} 
where
\begin{align}
    b(\rho) = b_1(u(\rho))\frac{b_2(u_\mathrm{m})}{b_1(u_\mathrm{m})}
    \theta[-\rho] + b_2(u(\rho)) \theta[\rho]\,\,, 
\end{align}
where $\theta[\rho]$ is the Heaviside function and the inverse 
of $\rho(u)$, i.e. $u(\rho)$ was used.

The matter shell and the boundary of the space are described by the hypersurfaces 
of constant $u$, i.e. $u=u_\mathrm{m}$ and $u=1$ respectively. 
As the previous case of smooth metrics, it 
is useful to analyze hypersurfaces of constant $u$, 
which have an induced metric as Eq.~\eqref{eqch3:uconstant}, 
or explicitly at the shell one has 
\begin{align}\label{eqch3:umconstant}
    ds^2|_{u=u_\mathrm{m}} = b_2(u_\mathrm{m})^2 d\tau^2 + r(u_\mathrm{m})^2 d\Omega^2_{d-2}\,\,,
\end{align}
and at the boundary one has
\begin{align}\label{eqch3:u1constant}
    ds^2|_{u\rightarrow 1} = b_2(u\rightarrow 1)^2 d\tau^2 + r(u\rightarrow 1)^2 d\Omega^2_{d-2}\,\,,
\end{align}
The extrinsic curvature $K_{ab}$ of the constant $u$ hypersurfaces 
is given by Eq.~\eqref{eqch3:extrinsiccurvy}, taking into account the 
metrics in Eqs.~\eqref{eqch3:metricM1} and~\eqref{eqch3:metricM2}. 
Explicitly at the shell, the extrinsic curvature suffers 
a jump in general, i.e. 
the extrinsic curvature 
computed in one side of the shell is not the same as the 
one computed at the other side. The jump is defined by 
square brackets on a tensor living in the hypersurface, 
e.g. $[K_{ab}] = K_{2ab} - K_{1ab}$, 
where $K_{2ab}$ is the extrinsic curvature evaluated 
at the side towards $u>u_\mathrm{m}$ and 
$K_{1ab}$ is the extrinsic curvature evaluated 
at the side towards $u<u_\mathrm{m}$. Namely, the extrinsic 
curvature at the shell from the side of $M_1$ is
\begin{align}\label{eqch3:extrinsiccurvyM1}
    K_{1ab}dx^a dx^b = \frac{b'_1 b_2^2}{a_1 b_1}d\tau^2 
    + \frac{r' r}{a_1}d\Omega^2_{d-2}\,\,,
\end{align} 
and from the side of $M_2$ is
\begin{align}\label{eqch3:extrinsiccurvyM2}
    K_{2ab}dx^a dx^b = \frac{b'_2 b_2}{a_2}d\tau^2 
    + \frac{r' r}{a_2}d\Omega^2_{d-2}\,\,,
\end{align} 
where the components are written here in terms of the 
coordinate $u$ and the prime means the derivative in $u$.
Moreover, the trace of the extrinsic curvature 
at each side is given by
\begin{align}\label{eqch3:traceextrinsiccurvyM1M2}
    &K_1 = \frac{b'_1}{a_1 b_1} + (d-2)\frac{r'}{a_1 r}\,\,,\notag\\
    &K_2 = \frac{b'_2}{a_2 b_2} + (d-2)\frac{r'}{a_2 r}\,\,.
\end{align}
At the boundary of space $u=1$, one has the extrinsic curvature 
$K_{2ab}$ and its trace $K_2$ with the same form of 
Eqs.~\eqref{eqch3:extrinsiccurvyM2} and~\eqref{eqch3:traceextrinsiccurvyM1M2} 
evaluated at $u=1$.

In the presence of a matter thin shell, the $C^0$ metric 
induces Dirac delta terms in the Ricci tensor. We are interested here 
on the Ricci scalar and the Einstein tensor in particular, 
which in this case have the expression 
\begin{align}\label{eqch3:expansion}
    & R = R_1 \theta[-\rho] + R_2 \theta[\rho] - 2[K]\delta[\rho] \,\,,\notag\\
    & G\indices{^\tau_\tau} = {G_1}\indices{^\tau_\tau}
    \theta[-\rho] + {G_2}\indices{^\tau_\tau} \theta[\rho] + ([K] 
    - [K\indices{^\tau_\tau}])\delta[\rho]\,\,,
\end{align} 
where $R_1$ is the Ricci scalar evaluated at $\rho <0$ or $u<u_\mathrm{m}$, $R_2$ is the Ricci 
scalar evaluated at $\rho >0$ or $u>u_\mathrm{m}$, ${G_1}\indices{^\tau_\tau}$ is the Einstein 
tensor component evaluated at $\rho < 0$ or $u<u_\mathrm{m}$, and 
${G_2}\indices{^\tau_\tau}$ is the Einstein 
tensor component evaluated at $\rho > 0$ or $u>u_\mathrm{m}$, with these 
quantities being given by 
\begin{align}
    &R_1 =  - \frac{2}{a_1 b_1 r^{d-2}}
    \left(\frac{b'_1 r^{d-2}}{a_1}\right)'\frac{b_2(u_\mathrm{m})^2}{b_1(u_\mathrm{m})^2} 
    - 2{G_1}\indices{^\tau_\tau}\,\,,\\
    &R_2 =  - \frac{2}{a_2 b_2 r^{d-2}}
    \left(\frac{b'_2 r^{d-2}}{a_2}\right)' - 2{G_2}\indices{^\tau_\tau}\,\,,\\
    &{G_1}\indices{^\tau_\tau} = \frac{(d-2)}{2 r' r^{d-2}} 
    \left[r^{d-3}\left(\left(\frac{r'}{a_1}\right)^2 
    - 1 \right) \right]'\,\,,\\
    &{G_2}\indices{^\tau_\tau} = \frac{(d-2)}{2r' r^{d-2}} 
    \left[r^{d-3}\left(\left(\frac{r'}{a_2}\right)^2 
    - 1 \right) \right]'\,\,,
\end{align}
written in terms of the coordinate $u$. 
Notice that the expansion in 
Eq.~\eqref{eqch3:expansion} can be computed by writing 
the Ricci scalar and the Einstein tensor in terms of the 
first and second derivatives of $b$ and $r$, use the chain rule 
$\frac{d}{d\rho} = \frac{1}{a(u)}\frac{d}{du}$, and then use 
the expansion in Heaviside functions, with the 
identity $\frac{d\theta[\rho]}{d\rho} = \delta(\rho)$. 
This is indeed the same procedure as the thin shell 
formalism~\cite{Israel:1966zz}, where the continuity of the metric 
is imposed as the first junction condition.
The expression for the Dirac delta term in the Einstein tensor is given by 
\begin{align}\label{eqch3:hamiltoniandelta}
    [K] - [K\indices{^\tau_\tau}] = 
    \frac{(d-2)}{r}\eval{\left(\frac{r'}{a_2} 
    - \frac{r'}{a_1}\right)}_{u=u_\mathrm{m}}\,\,,
\end{align}
written in terms of the coordinate $u$.
The expression in Eq.~\eqref{eqch3:hamiltoniandelta} 
is useful further on and constitutes one component of the 
gravitational part of the second junction condition.

We have to insert the metric, in Eq.~\eqref{eqch3:metric} for smooth metrics or 
in Eqs.~\eqref{eqch3:metricM1} and~\eqref{eqch3:metricM2} for $C^0$ metrics, 
in the path integral in Eq.~\eqref{eqch3:pathint}, together with the
boundary conditions at the hypersurface $u=1$ 
describing the boundary of space $\partial M$ and also 
the heat reservoir. These boundary conditions are fixed 
while performing the path integral. One then should sum over 
all the possible metrics on the path integral. 
In principle, the sum over the metrics can be decomposed
in terms of their topology class. In the case here treated, the 
topology class depends on a set of regularity conditions for the 
spherically symmetric metric at $u=0$.   
Below, we present the regularity and 
boundary conditions used in the thesis for spherically symmetric 
metrics.

\section{Metric regularity conditions\label{sech2:regularitycond}}

\subsection{Black hole-like conditions}

For spherically symmetric metrics, we need to impose regularity conditions 
at the center of the space or at its minimal 
surface, which for the metric in Eq.~\eqref{eqch3:metric} is 
situated at $u=0$. The black hole-like conditions correspond to 
the choice
\begin{align}\label{eqch3:bhcond1}
    & b(0) = 0\,\,,\,\, r(0) = r_+\,\,,
\end{align}
for the components of the metric, where $r_+$ is the 
horizon radius. 

This choice alone can induce possible 
divergences in the Ricci scalar and topological defects, 
which here they must be avoided by imposing conditions to the 
derivatives of the components of the metric. These conditions can 
be found by expanding the metric near $u=0$ as
\begin{align}\label{eqch3:metricnearhorizon}
    &ds^2=\left[\left(\frac{b'}{a}\right)^2\sVert[3]_{u=0} \varepsilon^2 
    + \left(\frac{b'}{a^2}\left(\frac{b'}{a}\right)'\right)\sVert[3]_{u=0} \varepsilon^3
    + \mathcal{O}(\varepsilon^4)\right] d\tau^2\notag\\
    & + d\varepsilon^2 + \left[r_+ + (r'a^{-1})\sVert[1]_{u=0} \varepsilon 
    + \mathcal{O}(\varepsilon^2)
    \right]^2d\Omega^2_{d-2}\,\,,
\end{align}
where $\epsilon = \int^{\delta}_0 a du$ for small $\delta$, assuming 
that $\int^{\delta}_0 a du$ is finite. Otherwise, 
$u=0$ should be understood as a boundary of the space.
The condition $b(0)=0$ means that a hypersurface with constant 
$u$, having topology $\mathbb{S}^1\times \mathbb{S}^{d-2}$, 
becomes topologically $\{u=0\}\times \mathbb{S}^{d-2}$ in the limit 
of $u=0$, i.e. 
a point times a $(d-2)$--sphere and the hypersurface volume 
becomes zero. This behaviour is precisely described 
by the metric in Eq.~\eqref{eqch3:metricnearhorizon}, namely, 
the $(\tau, \varepsilon)$ sector describes approximately 
the metric of a cone in general, with a possible conical singularity which 
introduces a topological defect in the Riemannian space. 
In order to avoid the existence of such singularity, we impose the regularity 
condition
\begin{align}\label{eqch3:bhcond2}
    \frac{b'}{a}\sVert[3]_{u=0}=1\,\,,
\end{align} 
and so the $(\tau,\varepsilon)$ sector of metric 
describes Euclidean flat space near $u=0$. The remaining conditions are 
found from avoiding the divergence of the Ricci scalar. The Ricci scalar 
near $u=0$ is given by 
\begin{align}\label{eqch3:riccibh}
    R = -\frac{2(d-2)}{\varepsilon r_+}\left(\frac{r'}{a}\right)\sVert[2]_{u=0}
    - \frac{2}{\varepsilon}\left(\frac{1}{a}\left(\frac{b'}{a}\right)'\right)\sVert[2]_{u=0} 
    + \mathcal{O}(1)\,\,,
\end{align}
and so the regularity conditions are 
\begin{align}\label{eqch3:bhcond3}
    & \left(\frac{r'}{a}\right)\sVert[2]_{u=0} = 0\,\,,\,\, 
    \left(\frac{1}{a}\left(\frac{b'}{a}\right)'\right)\sVert[2]_{u=0} = 0\,\,.
\end{align}
It is interesting to note that the first equation of Eq.~\eqref{eqch3:bhcond3}
is equivalent, in even dimensions, to the condition that the 
Riemannian space must have an Euler characteristic $\chi = 2$.

The conditions in Eqs.~\eqref{eqch3:bhcond1}--\eqref{eqch3:bhcond3} are precisely 
the conditions of the metric that one would obtain if the Wick transformation of 
a stationary black hole spacetime metric was performed. The $(d-2)$--surface at $u=0$ coincides 
with the bifurcate $(d-2)$--sphere of the horizon of the stationary black hole. The topology of the 
Riemannian space is $\mathbb{R}^2\times\mathbb{S}^{d-2}$ with these conditions.

\subsection{Flat conditions}

Other possible regularity conditions are the flat 
conditions, which are achieved by choosing
\begin{align}\label{eqch3:flatcond1}
    &b(0)\,\, \mathrm{finite}\,\mathrm{and}\,\mathrm{non}\,\mathrm{zero}\,\,,\,\, 
    r(0) = 0\,\,,
\end{align}
for the components of the metric. By expanding the 
metric near $u=0$ with the conditions in Eq.~\eqref{eqch3:flatcond1}, 
one has 
\begin{align}\label{eqch3:flatmetricnearzero}
    &ds^2 = \left(b(0) + \left(\frac{b'}{a}\right)\sVert[2]_{u=0} \varepsilon 
    + \mathcal{O}(\varepsilon^2) \right)d\tau^2 
    + d\varepsilon^2 \notag\\
    &+ \left[\left(\frac{r'}{a}\right)\sVert[2]_{u=0} 
    \varepsilon + \frac{1}{a}\left(\frac{r'}{a}\right)'\varepsilon^2 
    + \mathcal{O}(\varepsilon)\right]^2d\Omega^2_{d-2}\,\,,
\end{align}
where again $\varepsilon = \int_{0}^{\delta}a du$ for small $\delta$ and 
it is assumed that $\int_{0}^{\delta}a du$ is finite. The remaining 
regularity conditions must be extracted from the condition that 
the Ricci scalar is well-behaved at $u=0$. The Ricci scalar is
\begin{align}\label{eqch3:ricciflatnear}
    &R = - \frac{2(d-2)}{b\varepsilon}\left(\frac{b'}{a}\right)\sVert[3]_{u=0}
    - \frac{2(d-2)}{\varepsilon}\left(\frac{1}{r'}\left(\frac{r'}{a}\right)'\right)\sVert[3]_{u=0}
    \notag\\
    &- \frac{(d-2)(d-3)}{\varepsilon^2}\left[1 - \left(\frac{a}{r'}\right)^2 \right]\sVert[3]_{u=0}\,\,,
\end{align}
near $u=0$. Therefore, in order to avoid the divergence of the Ricci scalar,
the following regularity conditions
\begin{align}\label{eqch3:flatcond2}
    & \left(\frac{b'}{a}\right)\sVert[3]_{u=0} = 0\,\,,\,\,
    \left(\frac{r'}{a}\right)\sVert[3]_{u=0} = 1\,\,,\,\, 
    \left(\frac{1}{a}\left(\frac{r'}{a}\right)'\right)\sVert[3]_{u=0} = 0\,\,,
\end{align}
are necessary.

The regularity conditions in Eqs.~\eqref{eqch3:flatcond1} and~\eqref{eqch3:flatcond2}
are the conditions of the Riemannian metric if the Wick transformation was performed 
to a flat Lorentzian metric. The topology of the Riemannian space with these regularity 
conditions is $\mathbb{S}^1\times \mathbb{R}^{d-1}$.

\section{Metric boundary conditions\label{sech2:boundarycond}}

\subsection{Finite cavity}

The boundary conditions that we impose at the boundary of the 
Riemannian space for the metric are given here 
by the Dirichlet boundary conditions. In the case of 
the spherically symmetric metric in Eq.~\eqref{eqch3:metric},
the boundary of the space is positioned at $u=1$ with 
induced metric
\begin{align}\label{eqch3:inducedmetricu1}
    ds^2\sVert[1]_{u=1} = b(1)^2 d\tau^2 + r(1)^2 d\Omega_{d-2}^2\,\,.
\end{align}
According to the Dirichlet boundary conditions, for 
a finite boundary, we must fix
\begin{align}\label{eqch3:boundfinitecavity}
    &\beta = 2\pi b(1)\,\,,\,\,R = r(1)\,\,,
\end{align}
that is, we must fix the inverse temperature $\beta$ of the 
spherical shell at $u=1$, representing the heat reservoir, 
that is given as the total imaginary time length, and moreover 
we fix the radius $R$ of the shell. We therefore 
have a Riemannian space which represents a 
finite cavity, assuming the regularity conditions in the previous 
section.

\subsection{Infinite cavity: zero cosmological constant}

For the case where the boundary of the space is infinite, i.e. 
when $r(u)\sVert[1]_{u\rightarrow 1}$ is infinite, 
the boundary conditions of the Riemannian space are 
given according to the asymptotic behaviour of the metric when 
$u\rightarrow 1$.
The cases for zero and negative cosmological constant must be 
analyzed separately as the metric has different 
asymptotic behaviour.

When the cosmological constant is zero, the boundary conditions 
imposed are the same as the asymptotically flat spacetime 
conditions but translated to Riemannian space. In this sense, 
the behaviour of the metric components must be that
\begin{align}\label{eqch3:boundarycavityinfinite}
    &b(u)\sVert[1]_{u\rightarrow1} = \frac{\beta}{2\pi}\,\,,\,\,
    \frac{r'}{a}\sVert[2]_{u\rightarrow 1} = 1\,\,,
\end{align}
where $b(u)\sVert[1]_{u\rightarrow1}$ must be a fixed constant and 
it is given by $\beta$, the inverse temperature measured at infinity.

\subsection{Infinite cavity: negative cosmological constant}

When the cosmological constant is negative, for an infinite 
hypersurface $\partial M$, the boundary conditions 
imposed are the ones of asymptotically anti-de Sitter or AdS, but translated 
to Riemannian space. This amounts to the metric satisfying asymptotically the 
Euclidean Einstein equations with a negative cosmological 
constant, i.e. $R_{\alpha \beta} = - \frac{(d-1)}{l^2}g_{\alpha \beta}$ 
and fixing the remaining freedom in the metric. In order to 
put these conditions in terms of the components of the metric and 
their asymptotic behaviours, we must perform a conformal transformation 
in the metric $g_{\alpha \beta} = w^2 \bar{g}_{\alpha \beta}$, where 
$\bar{g}_{\alpha \beta}$ is the conformal metric and $w$ is the conformal 
factor. In order to have a nonsingular conformal metric, the 
conformal transformation must have a behaviour $w=\frac{c}{r(u)}$, where 
$c$ is a constant. We choose $c=1$. We further make the coordinate 
transformation $w = w(u)$ so that the conformal metric assumes 
the form in the neighbourhood $\mathcal{N}(\partial M)$ 
of the hypersurface $\partial M$ as
\begin{align}\label{eqch3:conformalmetric}
    d\bar{s}^2\sVert[1]_{\mathcal{N}(\partial M)} =  
    \frac{b(u)^2}{r(u)^2}d\tau^2 + \left(\frac{a(u) r(u)}{r'(u)}\right)^2 dw(u)^2
    + d\Omega^2_{d-2}\,\,.
\end{align}
The hypersurface $\partial M$ is then defined by $w=0$.
The asymptotic behaviour of the metric $g_{\alpha \beta}$ translates into 
conditions for the metric $\bar{g}_{\alpha \beta}$, 
see~\cite{Ashtekar:1984zz, Henneaux:1985tv}. In fact, the condition 
$R_{\alpha \beta} = - \frac{(d-1)}{l^2}g_{\alpha \beta}$ can be 
split into two conditions for the conformal metric, namely that the 
boundary $w=0$ is described by the metric $d\bar{s}^2|_{w=0} = d\bar{\tau}^2 
+ d\Omega^2_{d-2}$ and that $\bar{g}^{\alpha \beta}\nabla_{\alpha} w\nabla_{\beta}w 
= \frac{1}{l^2}$, where $\bar{\tau}$ is proportional to $\tau$ by some constant. 
And so the boundary conditions chosen for the metric elements are
\begin{align}\label{eqch3:adsboundary}
    & \frac{b(u)}{r(u)}\sVert[2]_{u\rightarrow1} = \frac{\bar{\beta}}{2\pi l}\,\,,\,\,
     \frac{a(u)r(u)}{r'(u)}\sVert[2]_{u\rightarrow1} = l\,\,,
\end{align}  
where $\bar{\beta}$ is defined as the inverse temperature measured at the 
conformal boundary of the asymptotically AdS
space.

\section{The gravitational path integral in spherical 
symmetry\label{sech2:actionsphericalsym}}

\subsection{Smooth metrics}

\subsubsection{General considerations}

We now proceed to reduce the path integral in Eq.~\eqref{eqch3:pathint}
to the case of spherically symmetric Riemannian spaces. For the 
case of zero cosmological constant, one has the partition function 
$Z = \int Dg_{\alpha \beta}D\psi \mathrm{e}^{-I_g[g_{\mu \nu}] 
- I_m[g_{\mu \nu},\psi]}$ with the gravitational action being 
given by 
\begin{align}\label{eqch3:actionzerocosmological}
    I_g = -\frac{1}{16\pi l_p^{d-2}}\int_{M}(R-2\Lambda)\sqrt{g}d^dx 
    - \frac{1}{8\pi l_p^{d-2}}\int_{\partial M}K \sqrt{\gamma} d^{d-1}x 
    - I_{\mathrm{ref}}\,\,.
\end{align}
For the case of a smooth metric, one can use the expression of the 
Ricci scalar in Eq.~\eqref{eqch3:Ricciscalar} and the expression of 
the trace of the extrinsic curvature in 
Eq.~\eqref{eqch3:traceextrinsiccurvy} to obtain
\begin{align}\label{eqch3:intRicci}
    & -\frac{1}{16\pi l_p^{d-2}}\int_{M} R \sqrt{g}d^dx = 
    -\frac{\Omega_{d-2}}{4l_p^{d-2}}\left(\frac{b' r^{d-2}}{a}\right)\sVert[3]_{u=0} 
    + \frac{\Omega_{d-2}}{4l_p^{d-2}}\left(\frac{b' r^{d-2}}{a}\right)\sVert[3]_{u\rightarrow1}
    \notag \\
    &+ \frac{1}{8\pi l_p^{d-2}}\int_{M} a b r^{d-2}G\indices{^\tau_\tau} d^dx\,\,,\notag\\
    & -\frac{1}{8\pi l_p^{d-2}}\int_{\partial M} K \sqrt{\gamma}d^{d-1}x 
    = -\frac{2\pi}{\mu}\left(\frac{b r'r^{d-3}}{a}\right)\sVert[3]_{u\rightarrow1}
    - \frac{\Omega_{d-2}}{4l_p^2}\left(\frac{b' r^{d-2}}{a}\right)\sVert[3]_{u\rightarrow 1}\,\,,
\end{align}
where $\sqrt{g} = a b r^{d-2}$, $\sqrt{\gamma} = b r^{d-2}$ and 
$\mu = \frac{8\pi l_p^{d-2}}{(d-2)\Omega_{d-2}}$. 
Putting together the expressions in Eqs.~\eqref{eqch3:intRicci} 
into Eq.~\eqref{eqch3:actionzerocosmological}, 
one finally has 
\begin{align}
    &I_g = 
    -\left(\frac{2\pi b r^{d-3}}{\mu}\left(\frac{r'}{a}\right)\right)\sVert[3]_{u\rightarrow1}
    - \frac{\Omega_{d-2}}{4l_p^{d-2}}\left(\frac{b' r^{d-2}}{a}\right)\sVert[3]_{u=0}\notag\\
    &+ \frac{1}{8\pi l_p^{d-2}}\int_{M} a b r^{d-2}(G\indices{^\tau_\tau} + \Lambda)d^dx
     - I_{\mathrm{ref}}\,\,.\label{eqch3:actionspherical1}
\end{align}
The form of the gravitational action in Eq.~\eqref{eqch3:actionspherical1}
assumes the form of the typical decomposition of the space in a foliation 
of hypersurfaces, with $G\indices{^\tau_\tau}$ term and the boundary term at 
$u\rightarrow 1$ being part of the Hamiltonian of the space. 
The remaining terms are determined by the regularity and 
boundary conditions, and also by the choice of the reference space. It is interesting 
to note that the term at $u=0$ seems to be topological, due to its link with the 
regularity conditions.

\subsubsection{Zero cosmological constant}

For the case of zero cosmological constant, the reference space is 
the Riemannian space obtained by performing the map to flat Lorentzian 
spacetime at the same temperature, which we call hot flat space, giving
\begin{align}
    ds^2_{\mathrm{flat}} = b(1)^2 d\tau^2 + dr^2 + r^2 d\Omega_{d-2}^2\,\,,
\end{align}
where the coordinate transformation $r=r(y)$ was performed and 
$r\in\, ]0, r(1)[$. It is important to distinguish this space from the 
flat Riemannian space since the former has topology $\mathbb{S}^1\times 
\mathbb{R}^{d-1}$ while the latter has topology $\mathbb{R}^d$. The 
action for hot flat space can be written as 
\begin{align}\label{eqch3:actionrefflat}
    I_{\mathrm{flat}} =  -\frac{1}{8\pi l_p^{d-2}}\int_{\partial M} K_{\mathrm{flat}}d^{d-1}x\,\,,
\end{align}  
or alternatively can be obtained from Eq.~\eqref{eqch3:actionspherical1}
by setting $\left(\frac{r'}{a}\right)_{\mathrm{flat}} = 1$, 
$\left(\frac{b'}{a}\right)_{\mathrm{flat}} = 0$,
$G\indices{^\tau_\tau} = 0$, with $\Lambda =0$ and flat regular conditions, 
yielding 
\begin{align}
    I_{\mathrm{flat}} = 
    -\frac{2\pi}{\mu}\left(b r^{d-3}\right)\sVert[3]_{u\rightarrow1}\,\,.
    \label{eqch3:actionflat2}
\end{align}
The gravitational action for a spherically smooth metric with zero cosmological 
constant is then 
\begin{align}
    &I_{g\mathrm{f}} = 
    \left(\frac{2\pi b r^{d-3}}{\mu}\left(1 - \frac{r'}{a}\right)\right)\sVert[3]_{u\rightarrow1}
    - \frac{\Omega_{d-2}}{4l_p^{d-2}}\left(\frac{b' r^{d-2}}{a}\right)\sVert[3]_{u=0}\notag\\
    &+ \frac{1}{8\pi l_p^{d-2}}\int_{M} a b r^{d-2}G\indices{^\tau_\tau}d^dx\,\,.
    \label{eqch3:actionspherical1flat}
\end{align}

\subsubsection{Negative cosmological constant}

For negative cosmological constant, the reference space chosen is the 
Riemannian space obtained from performing the map to AdS
spacetime, which we call hot AdS space, at the same temperature. 
The metric describing hot AdS space is
\begin{align}\label{eqch3:adsmetric}
    ds_{\mathrm{AdS}}^2 = b_{\mathrm{AdS}}(y(r))^2 d\tau^2 
    + \left(\frac{a}{r'}(r)\right)^2_{\mathrm{AdS}}
    dr^2 + r^2 d\Omega^2_{d-2}\,\,,
\end{align}
where $b_{\mathrm{AdS}}(1) = b(1)$, 
$\left(\frac{b'}{a}\right)_{\mathrm{AdS}}\sVert[2]_{u=0}=0$
and $\left(\frac{r'}{a}\right)_{\mathrm{AdS}}^2 = 1 + \frac{r^2}{l^2}$. 
The action for the hot AdS space is more contrived compared to hot flat space, 
since the Ricci scalar is $R = - \frac{d(d-1)}{l^2}$, and it is 
given by 
\begin{align}\label{eqch3:AdSaction}
    I_{\mathrm{AdS}} = \frac{(d-1)}{8\pi l^2 l_p^{d-2}}\int_M \sqrt{g}d^dx
    - \frac{1}{8\pi l_p^{d-2}}\int_{\partial M}K_{\mathrm{AdS}}\sqrt{\gamma}d^{d-1}x \,\,.
\end{align}
Alternatively, one can use also Eq.~\eqref{eqch3:actionspherical1} to evaluate 
the action of hot AdS space by using the expression of the components of the 
metric at the boundary, plus that $G\indices{^\tau_\tau} = - \Lambda$ and 
the flat regularity conditions, to obtain
\begin{align}\label{eqch3:AdSaction2}
    I_{\mathrm{AdS}} = 
    -\left(\frac{2\pi b r^{d-3}}{\mu}\left(\frac{r'}{a}\right)_{\mathrm{AdS}}
    \right)_{u\rightarrow1}\,\,.
\end{align}
The gravitational action for a smooth spherically symmetric metric 
with a negative cosmological constant is then
\begin{align}
    \label{eqch3:actionspherical2}
    &I_{g l} = 
    \left(\frac{2\pi b r^{d-3}}{\mu}\left(\left(\frac{r'}{a}\right)_{\mathrm{AdS}} 
    - \frac{r'}{a}\right)\right)\sVert[3]_{u\rightarrow1}
    - \frac{\Omega_{d-2}}{4l_p^{d-2}}\left(\frac{b' r^{d-2}}{a}\right)\sVert[3]_{u=0}\notag\\
    &+ \frac{1}{8\pi l_p^{d-2}}\int_{M} a b r^{d-2}\left(G\indices{^\tau_\tau} 
    - \frac{(d-1)(d-2)}{2l^2})\right)d^dx\,\,.
\end{align}

\subsection{$C^0$ metrics}

\subsubsection{General considerations}

For the case where the metric is $C^0$ with non-differentiability 
at the hypersurface $\mathcal{C}$, described by $u=u_\mathrm{m}$, the gravitational 
action can be given by 
\begin{align}\label{eqch3:actionzerocosmologicalsplit}
    &I_g = -\frac{1}{16\pi l_p^{d-2}}\int_{M_1}(R_1-2\Lambda)\sqrt{g}d^dx
    -\frac{1}{16\pi l_p^{d-2}}\int_{M_2}(R_2-2\Lambda)\sqrt{g}d^dx\notag \\
    &  + \frac{1}{8\pi l_p^{d-2}}\int_{\mathcal{C}} [K] \sqrt{\gamma} d^{d-1}x
    - \frac{1}{8\pi l_p^{d-2}}\int_{\partial M}K \sqrt{\gamma} d^{d-1}x 
    - I_{\mathrm{ref}}\,\,,
\end{align}
where Eq.~\eqref{eqch3:expansion} was used to decompose the Ricci 
scalar in terms of Heaviside functions and the Dirac delta. Another 
way of getting Eq.~\eqref{eqch3:actionzerocosmologicalsplit} is by 
summing the Einstein-Hilbert action with the Gibbons-Hawking-York 
boundary term in each region $M_1$ and $M_2$, noting that 
the difference on the trace of the extrinsic curvature is not zero. 
The integrals can be decomposed in terms of 
the spherically symmetric metric components as
\begin{align}
    \label{eqch3:intRicci2}
    & -\frac{1}{16\pi l_p^{d-2}}\int_{M_1} R \sqrt{g}d^dx = 
    -\frac{\Omega_{d-2}}{4l_p^{d-2}}\left(\frac{b'_1 b_2(u_\mathrm{m}) r^{d-2}}{a_1 b_1(u_\mathrm{m})}\right)\sVert[3]_{u=0} 
    \notag \\
    &+ \frac{\Omega_{d-2}}{4l_p^{d-2}}\left(\frac{b'_1 b_2(u_\mathrm{m}) r^{d-2}}{a_1 b_1(u_\mathrm{m})}\right)\sVert[3]_{u = u_\mathrm{m}}
    + \frac{1}{8\pi l_p^{d-2}}\int_{M_1} a_1 b_1 \frac{b_2(u_\mathrm{m})}{b_1(u_\mathrm{m})} 
    r^{d-2}G\indices{_1^\tau_\tau} d^dx\,\,,\notag\\
    & -\frac{1}{16\pi l_p^{d-2}}\int_{M_2} R_2 \sqrt{g}d^dx = 
    -\frac{\Omega_{d-2}}{4l_p^{d-2}}\left(\frac{b'_2 r^{d-2}}{a_2}\right)\sVert[3]_{u=u_\mathrm{m}} 
    + \frac{\Omega_{d-2}}{4l_p^{d-2}}\left(\frac{b'_2 r^{d-2}}{a_2}\right)\sVert[3]_{u\rightarrow1}
    \notag \\
    &+ \frac{1}{8\pi l_p^{d-2}}\int_{M_2} a_2 b_2 r^{d-2}G\indices{_2^\tau_\tau} d^dx\,\,,\notag\\
    & \frac{1}{8\pi l_p^{d-2}}\int_{\mathcal{C}} [K] \sqrt{\gamma}d^{d-1}x 
    = -\frac{1}{8\pi l_p^{d-2}}\int_{\mathcal{C}} ([K\indices{^\tau_\tau}] - [K])\sqrt{\gamma}d^{d-1}x\notag\\
    &+ \frac{\Omega_{d-2}}{4l_p^2}\left(\frac{b'_2 r^{d-2}}{a_2}\right)\sVert[3]_{u\rightarrow u_\mathrm{m}}
    - \frac{\Omega_{d-2}}{4l_p^2}\left(\frac{b'_1 b_2(u_\mathrm{m}) r^{d-2}}{a_1 b_1(u_\mathrm{m})}
    \right)\sVert[3]_{u\rightarrow u_\mathrm{m}}\,\,,\notag\\
    & -\frac{1}{8\pi l_p^{d-2}}\int_{\partial M} K \sqrt{\gamma}d^{d-1}x 
    = -\frac{2\pi}{\mu}\left(\frac{b_2 r'r^{d-3}}{a_2}\right)\sVert[3]_{u\rightarrow1}
    - \frac{\Omega_{d-2}}{4l_p^2}\left(\frac{b'_2 r^{d-2}}{a_2}\right)\sVert[3]_{u\rightarrow 1}\,\,.
\end{align}
Putting together the terms in Eq.~\eqref{eqch3:intRicci2}, the action for the spherically 
symmetric $C^0$ metric is
\begin{align}\label{eqch3:actionspherical3}
    &I_g = -\frac{2\pi}{\mu}\left(\frac{b_2 r'r^{d-3}}{a_2}\right)\sVert[3]_{u\rightarrow1}
    -\frac{\Omega_{d-2}}{4l_p^{d-2}}\left(\frac{b'_1 b_2(u_\mathrm{m}) r^{d-2}}{a_1 b_1(u_\mathrm{m})}\right)\sVert[3]_{u=0} \notag\\
    & + \frac{1}{8\pi l_p^{d-2}}\int_{M_1} a_1 b_1 \frac{b_2(u_\mathrm{m})}{b_1(u_\mathrm{m})} 
    r^{d-2}(G\indices{_1^\tau_\tau}+ \Lambda) d^dx
    + \frac{1}{8\pi l_p^{d-2}}\int_{M_2} a_2 b_2 r^{d-2}(G\indices{_2^\tau_\tau}+ \Lambda)d^dx\notag\\
    & - \frac{1}{8\pi l_p^{d-2}}\int_{\mathcal{C}} ([K\indices{^\tau_\tau}] - [K])\sqrt{\gamma}d^{d-1}x
    - I_{\mathrm{ref}}\,\,,
\end{align}
which is basically Eq.~\eqref{eqch3:actionspherical1} but with $b(u)$, $a(u)$ and 
$G\indices{^\tau_\tau}$ expanded in Heaviside functions and Dirac delta. 
The non-smoothness of the metric leads to the additional boundary term of the action 
at the hypersurface $\mathcal{C}$ compared to the action of smooth metrics. 
Indeed, this is expected as the term $[K\indices{^\tau_\tau}] - [K]$ represents the 
junction condition for the shell that comes from the Dirac delta term of $G\indices{^\tau_\tau}$.

\subsubsection{Zero cosmological constant}

The action for spherically symmetric $C^0$ metric with zero cosmological constant 
follows from the analysis with the smooth metric. The action for the reference 
space is $I_{\mathrm{flat}}$ from Eq.~\eqref{eqch3:actionrefflat}, which 
in terms of the metric components is 
\begin{align}
    I_{\mathrm{flat}} = 
    -\frac{2\pi}{\mu}\left(b_2 r^{d-3}\right)\sVert[3]_{u\rightarrow1}\,\,.
    \label{eqch3:actionflat2C0}
\end{align}
And so the action of a $C^0$ metric with a zero cosmological constant in Eq.~\eqref{eqch3:actionspherical3}
becomes
\begin{align}\label{eqch3:actionsphericalflat2}
    &I_{g\mathrm{f}} = 
    \left(\frac{2\pi b_2 r^{d-3}}{\mu}\left(1 - \frac{r'}{a_2}\right)\right)\sVert[3]_{u\rightarrow1}
    -\frac{\Omega_{d-2}}{4l_p^{d-2}}\left(\frac{b'_1 b_2(u_\mathrm{m}) r^{d-2}}{a_1 b_1(u_\mathrm{m})}\right)\sVert[3]_{u=0} \notag\\
    & + \frac{1}{8\pi l_p^{d-2}}\int_{M_1} a_1 b_1 \frac{b_2(u_\mathrm{m})}{b_1(u_\mathrm{m})} 
    r^{d-2}G\indices{_1^\tau_\tau} d^dx
    + \frac{1}{8\pi l_p^{d-2}}\int_{M_2} a_2 b_2 r^{d-2}G\indices{_2^\tau_\tau} d^dx\notag\\
    & - \frac{1}{8\pi l_p^{d-2}}\int_{\mathcal{C}} ([K\indices{^\tau_\tau}] - [K])\sqrt{\gamma}d^{d-1}x\,\,.
\end{align}

\subsubsection{Negative cosmological constant}

For the case of negative cosmological constant, the reference action 
is $I_\mathrm{AdS}$ from Eq.~\eqref{eqch3:AdSaction}, which in terms of 
the metric components is 
\begin{align}\label{eqch3:AdSaction2C0}
    I_{\mathrm{AdS}} = 
    -\left(\frac{2\pi b_2 r^{d-3}}{\mu}\left(\frac{r'}{a_2}\right)_{\mathrm{AdS}}
    \right)_{u\rightarrow1}\,\,.
\end{align}
Therefore, the action of a spherically symmetric $C^0$ metric 
with negative cosmological constant, in Eq.~\eqref{eqch3:actionspherical3}, 
becomes
\begin{align}\label{eqch3:actionnegativecosmological2}
    &I_{gl} = 
    \left(\frac{2\pi b_2 r^{d-3}}{\mu}\left(\left(\frac{r'}{a}\right)_{\mathrm{AdS}} 
    - \frac{r'}{a_2}\right)\right)\sVert[3]_{u\rightarrow1}
    -\frac{\Omega_{d-2}}{4l_p^{d-2}}\left(\frac{b'_1 b_2(u_\mathrm{m}) r^{d-2}}{a_1 b_1(u_\mathrm{m})}\right)\sVert[3]_{u=0} \notag\\
    & + \frac{1}{8\pi l_p^{d-2}}\int_{M_1} a_1 b_1 \frac{b_2(u_\mathrm{m})}{b_1(u_\mathrm{m})} 
    r^{d-2}\left(G\indices{_1^\tau_\tau} 
    - \frac{(d-1)(d-2)}{2l^2}\right) d^dx \notag\\
    &+ \frac{1}{8\pi l_p^{d-2}}\int_{M_2} a_2 b_2 r^{d-2}\left(G\indices{_2^\tau_\tau} 
    - \frac{(d-1)(d-2)}{2l^2}\right)d^dx\notag\\
    & - \frac{1}{8\pi l_p^{d-2}}\int_{\mathcal{C}} ([K\indices{^\tau_\tau}] - [K])\sqrt{\gamma}d^{d-1}x\,\,.
\end{align}

\section{The statistical path integral and its connection to thermodynamics
\label{sech3:thermo}}

With the shape of the action decomposed into the spherically symmetric 
metric components, the Euclidean path integral is composed by the sum 
over the possible paths of the metric components and matter fields as
\begin{align}\label{eqch3:sphericalpathintegral}
    Z = \int Db Da Dr D\psi\, \mathrm{e}^{-I_g - I_m}\,\,,
\end{align}
where $I_g$ can be given by $I_{g\mathrm{f}}$ in the case of 
zero cosmological constant or $I_{gl}$ for the case of negative 
cosmological constant, with the sum being made over components 
obeying the boundary conditions and over the possible 
regularity conditions.

As discussed above, even in this form, an expression for the 
path integral seems quite elusive. Typically, one performs the 
saddle point approximation to find the paths of the metric 
components $b$, $a$, $r$ and $\psi$, that minimize the 
full action. We are going to apply this approximation in the upcoming chapters 
for the particular cases of interest, namely for black hole spaces 
with a static electromagnetic field or with matter, and also for 
the case of a self-gravitating matter thin shell. In the saddle 
point approximation, one can consider only the zeroth order 
contribution which translates into a partition function 
$Z = \mathrm{e}^{-I_0}$, where $I_0$ is the action evaluated 
at a minimum path. This is the zero loop approximation.

Depending on the ensemble, the partition function is tied 
to a thermodynamic potential. This can be seen from the 
definition of the partition function and the possible mean 
thermodynamic values that one can obtain. For example, 
in the canonical ensemble, with the inverse temperature 
and area fixed, one can obtain the mean energy through 
\begin{align}
    E = - \frac{\partial \log(Z)}{\partial \beta}\,\,,
\end{align}
since $-\frac{\partial \log(Z)}{\partial \beta}$ can be 
formally written as $\sum_i E_i \mathrm{e}^{-\beta E_i}/Z$, 
where the $i$ subscript means with respect to each microstate.
Moreover, the entropy is defined as the Gibbs entropy with 
the formula $S = -\sum_i p_i \log(p_i)$, with $p_i = \mathrm{e}^{-\beta E_i}/Z$,
which can be written in terms of $\log(Z)$ as 
\begin{align}
    S = - \beta \frac{\partial \log(Z)}{\partial \beta} + \log(Z)\,\,.
\end{align}
Using the formula for the energy and the entropy, the partition 
function can be related to the free energy $F$ 
in the canonical ensemble as 
\begin{align}
    \beta F = -\log(Z)\,\,,
\end{align}
where the free energy is defined by the Legendre transform 
of the mean energy, $F= E - T S$. We can now establish the 
connection between the zero loop approximation and thermodynamics. 
Since $\log(Z) = - I_0$, then we have
\begin{align}
    F = T I_0\,\,, 
\end{align}
which means that the action evaluated at the minimizing paths 
translates into the free energy of the ensemble. Having the 
action in the zero loop approximation, we can obtain straightforwardly 
the free energy and the remaining thermodynamic quantities, 
i.e. the mean energy, the entropy and the 
pressure, by calculating the derivatives of the free energy.

A similar analysis holds for the case of the 
grand canonical ensemble, where the thermodynamic 
potential that has the connection with the partition 
function is the grand potential $W = -\log(Z)$, where 
we define $W = E - T S - \mu N$, with $\mu$ being 
a chemical potential and $N$ being a mean number. Hence, 
the grand potential is given by the action at the 
minimum path as $W = T I_0$, and we can obtain the 
thermodynamic quantities by calculating the derivatives 
of $W$ or $I_0$.

\section{Summary\label{sech2:summary}}

In this chapter, we gave an introduction to the construction of statistical 
ensembles of curved spacetimes, here with focus on the metric. 
In order to study the statistical ensemble of a configuration described 
by a stationary Lorentzian spacetime, with a timelike hypersurface as 
the boundary and describing the heat reservoir, we must perform a 
map of the spacetime to a Riemannian or pseudo-Riemannian metric. 
The shape of the Riemannian or pseudo-Riemannian 
metric has to be relaxed except for the 
fixed data at the boundary, which must be the same data of the 
configuration that we want to study. For spherically symmetric 
spaces, we impose Dirichlet boundary conditions that compose the fixed data
at the boundary space, which in this case is described by a spherical shell 
and the inverse temperature of the ensemble is fixed to be 
the total imaginary time length at the boundary. The partition function 
is then given by the Euclidean path integral over the Riemannian or 
pseudo-Riemannian metrics with fixed boundary data. For spherically 
symmetric metrics, we also need to sum over the discrete set of 
regularity conditions which are tied to the topology of the 
Riemannian space. 

Here, we restricted the shape of the metric to spherically symmetric metrics as the 
boundary data is given for a spherical shell. We explained the possible regularity 
and boundary conditions, which are going to be used 
in the next chapters. Moreover, we decomposed the gravitational action 
in terms of the spherically symmetric metric components, ready to be 
used for the analysis of specific configurations including matter. 
We performed the calculations in this chapter to avoid repetition 
in the following chapters.

Finally, we established the connection of the partition function through the 
Euclidean path integral with the thermodynamics of the ensemble. 
In order to obtain the partition function, we are going to employ the 
zero loop approximation in the upcoming chapters. 
The action evaluated at the minimizing paths gives the 
relevant thermodynamic potential of the ensemble and through 
its derivatives, we can obtain the thermodynamic properties of the 
system. This analysis is expanded further in the upcoming chapters.

\chapter{Grand canonical ensemble of a $d$-dimensional 
Reissner-Nordstr\"om black hole in a cavity}
\chaptermark{Grand canonical 
ensemble of $d$-dimensional RN black hole in a cavity}
\label{ch:grandcanonicalblackhole}

\section{Introduction}

As previously discussed in Chapter~\ref{ch:Euclideanpathintegral}, 
the Euclidean path integral approach~\cite{Gibbons:1977} allows the construction 
of statistical ensembles in curved spaces. Moreover, the use of the zero loop 
approximation allows the computation of the partition function in terms of the 
classical paths of the action. The approach was extended to the York 
formalism~\cite{York:1986, Whiting:1988qr},
where a finite cavity is introduced, allowing for stable equilibrium configurations. 
The formalism was used to construct the grand canonical ensemble of a 
charged black hole inside a cavity in four dimensions~\cite{Braden:1990hw, Peca:1998cs} and 
for black branes~\cite{Lu:2010au}. Also, York's analysis was extended to higher 
dimensions~\cite{Andre:2020czm, Andre:2021ctu}, where there was an emphasis on the 
connection between the statistical ensemble and matter dynamic stability 
in curved spacetime. Namely, the two black hole solutions of the ensemble 
bifurcate when the cavity is at the light ring radius, and also the stable 
black hole starts to be more favorable than hot flat space when its radius corresponds 
to the Buchdahl bound.

Motivated by these developments, in this chapter, we construct the 
grand canonical ensemble of a Reissner-Nordstr\"om black hole inside a 
cavity in higher dimensions using the Euclidean path integral 
approach to quantum gravity, with fixed temperature and fixed electric 
potential. We perform the zero loop approximation in steps. 
First, the Hamiltonian and Gauss constraints are imposed to find a reduced action 
and then the stationary points of the reduced action are found, corresponding to
black hole solutions of the ensemble. It is found that there are up to two solutions 
of the ensemble, with its qualitative behaviour being presented. The two solutions 
bifurcate at a certain ratio between horizon radius and cavity radius, 
which does not correspond to the light ring ratio. The main objective was to study 
the phase transitions and read off possible connections to matter dynamic stability.
We therefore analyzed the phase transitions 
between the black hole solutions and hot flat space, finding a first order phase 
transition. We found that 
the black hole solution is more favorable when its radius is slightly lower than the 
Buchdahl-Andr\'easson-Wright bound~\cite{Wright:2015dma}, 
which is related to the maximum compactness of a 
charged configuration obeying certain energy conditions. We present in detail the five dimensional case, $d=5$, where an analytic 
expression for the black hole solutions was found.

This chapter is organized as follows. In Sec.~\ref{sech4:pathintegral}, we consider the 
partition of the grand canonical ensemble for spherically symmetric 
metrics, obeying regularity and boundary conditions, namely the fixed inverse temperature 
is established as the total imaginary proper time at the boundary of the cavity and the 
radius of the cavity is fixed. In Sec.~\ref{sech4:zeroloop}, we perform the zero loop approximation, 
where we first impose the constraints to find the reduced action and we find the stationary 
points of the reduced action. In Sec.~\ref{sech4:thermo}, we obtain the thermodynamics 
of the system from the partition function in the zero loop approximation 
and we further analyze the possible phase transitions. 
In Sec.~\ref{sech4:zeroloop5d}, we present in detail the five dimensional case, $d=5$. 
In Sec.~\ref{sech4:radii}, we compare the bifurcation radius with the light ring radius, and
also the thermodynamic radius with the Buchdahl-Andreásson-Wright bound.
In Sec.~\ref{sech4:concl}, we conclude the chapter. We note that the work in this chapter 
is based on~\cite{Fernandes:2023byx}.

\section{The grand canonical ensemble of a charged black hole 
in the Euclidean path integral approach\label{sech4:pathintegral}}\sectionmark{
The grand canonical ensemble of a charged black hole
}\thispagestyle{userightbotmark}

\subsection{The partition function}

Through the Euclidean path integral approach, we can
construct the grand canonical ensemble of a charged black hole 
inside a finite cavity, in $d$ dimensions, by considering the
partition function 
\begin{align}\label{eqch4:partitionfunction}
    Z = \int Dg_{\alpha \beta}DA_{\gamma}\, 
    \mathrm{e}^{-I[g_{\mu\nu},A_{\sigma}]}\,\,,
\end{align}
with the Euclidean action 
\begin{align}
    I = - \int_{\mathcal{M}}\left(\frac{R}{16\pi l_p^{d-2}} -
   \frac{F_{ab}F^{ab}}{4}\right)\sqrt{g} d^dx -
   \frac{1}{8\pi l_p^{d-2}}\int_{\partial M} (K-K_0) \sqrt{\gamma}
   d^{d-1}x\,,
   \label{eqch4:EucAction1}
\end{align}
where $R$ is the Ricci scalar, $g$ is the determinant of the Euclidean 
metric $g_{\alpha \beta}$, $F_{\alpha \beta} = \partial_\alpha A_{\beta} 
- \partial_\beta A_{\alpha}$ is the strength field tensor of the 
Maxwell vector potential $A_\alpha$, $\gamma$ is the determinant of 
the induced metric $\gamma_{ab}$ of the hypersurface describing the 
boundary $\partial M$, $K = n^\alpha_{;\alpha}$ is the trace of the 
extrinsic curvature of the hypersurface with $n^\alpha$ being the 
outward unit normal to it, and $K_0$ is the 
extrinsic curvature of the boundary embedded in flat space, giving 
the action of hot flat space. The action in Eq.~\eqref{eqch4:EucAction1} 
can be split into $I = I_{g\mathrm{f}} + I_A$, where 
\begin{align}
    & I_{g\mathrm{f}} =  -\frac{1}{16\pi l_p^{d-2}} \int_{\mathcal{M}}R\sqrt{g}d^dx 
    -\frac{1}{8\pi l_p^{d-2}}\int_{\partial M} (K-K_0) \sqrt{\gamma}
    d^{d-1}x\,,\\
    & I_{A} = \int_M \frac{F_{ab}F^{ab}}{4}\sqrt{g} d^dx\,\,.
\end{align}
The path integral in the partition function in Eq.~\eqref{eqch4:partitionfunction} 
is performed along Riemannian metrics with fixed boundary conditions, which are 
periodic in the imaginary time. We make a reminder that among the possible 
paths are Riemannian metrics that correspond to physical static Lorentzian 
spacetimes by a Wick transformation in the imaginary time. For further details 
on the construction of the path integral, 
see Chapter~\ref{ch:Euclideanpathintegral}.

\subsection{Geometry and boundary conditions}

In this case, we consider the boundary of space to be described by a spherical shell 
with fixed temperature and electric potential. This means we are dealing with 
the grand canonical ensemble. Due to the spherical symmetry of the boundary, 
the metrics with spherical symmetry should contribute the most 
to the path integral. And so, we restrict the path integral 
to spherically symmetric metrics of the form 
\begin{align}\label{eqch4:metricspherical}
    ds^2 = b(u)^2 d\tau^2 + a(u)^2 du^2 + r(u)^2 d\Omega^2_{d-2}\,\,,
\end{align}
where $b(u)$, $a(u)$ and $r(u)$ are arbitrary smooth functions of $u$, 
the coordinates have the range $\tau \in \,]0,2\pi[$ and $u\in\,]0,1[$, 
and $d\Omega^2_{d-2}$ is the $(d-2)$--sphere line element. 

In principle, the path integral should include a sum over topologies of 
the Riemannian space with a metric of the form of Eq.~\eqref{eqch4:metricspherical}. 
The sum over topologies is related to the sum 
over metrics with different regularity conditions. Here, we choose the 
black hole-like regularity conditions, described by 
\begin{align}\label{eqch4:regularitymetric}
    & b(0) = 0\,\,,\notag\\
    & r(0) = r_+\,\,,\notag\\
    & \eval{(b'\alpha^{-1})}_{u=0} = 1\,\,,\notag\\
    &   \eval{\alpha^{-1}(b'\alpha^{-1})'}_{u=0} = 0\,\,,\notag\\
    & \eval{\left(\frac{r'}{\alpha}\right)}_{u=0} = 0\,\,,
\end{align} 
where $r_+$ is the horizon radius and a prime denotes the derivative 
of a function in $u$, e.g. $b'=\frac{db}{du}$.
The boundary conditions, as already stated, are such that the boundary 
is described by a spherical shell, in this case with the boundary at 
$u=1$, with induced metric
\begin{align}\label{eqch4:inducedmetricboundary}
    ds_{\partial M}^2 = b(1)^2 d\tau^2 + R^2 d\Omega^2_{d-2}\,,
\end{align}
where $R$ is the radius of the shell. As part of the boundary conditions, 
we fix the radius of the shell $R$ or equivalently its area 
defined by 
\begin{align}\label{eqch4:areaA}
    A= \Omega_{d-2} R^{d-2}\,\,,
\end{align}
where $\Omega_{d-2} = \frac{2 \pi^{\frac{d-1}{2}}}
{\Gamma\left(\frac{d-1}{2}\right)}$ is the area of the unit 
$(d-2)$--sphere, with $\Gamma$ being the gamma function. 
For $d=4$, we have $\Omega_{4} = 4\pi$ and, for $d=5$, 
we have $\Omega_5 = 2 \pi^2$. We also fix the inverse temperature 
$\beta$ at the boundary of space, which corresponds to the 
component $b(1)$ as 
\begin{align}
    & \beta = 2\pi b(1)\,\,,\label{eqch4:betadef}
\end{align}
where $\beta = 1/T$, with $T$ being the temperature of the 
heat reservoir. 

We also need to provide regularity and 
boundary conditions for the Maxwell field $A_\alpha$ 
according to the regularity conditions of the metric 
and to the ensemble in question. In spherical symmetry, 
and without considering magnetic monopoles, the strength 
field tensor is zero except for the component 
$F_{y\tau} = - F_{\tau y} = A_\tau'$, where we choose a gauge 
in which only the Maxwell component $A_\tau$ is 
non-zero. At $u=0$, we enforce the regularity condition 
\begin{align}
    & A(0) = 0\,\,,\label{eqch4:Ay0}
\end{align}
while at the boundary, $u=1$, we fix the electric potential 
given by 
\begin{align}
    \beta \phi = 2\pi i A_{\tau}(1)\,\,.\label{eqch4:phidef}
\end{align}
We note that the correspondence between the electric potential 
and the Maxwell field can be deduced by defining $\phi$ 
as the electric potential measured by a stationary observer 
in the physical Lorentzian spacetime and then use the Euclidean 
Maxwell field instead of the physical one.

\subsection{Action in spherical symmetry}

Having the expression of the metric with the regularity and the 
boundary conditions, we can now simplify the action in 
Eq.~\eqref{eqch4:EucAction1} by using the explicit form of the 
spherically symmetric metrics in Eq.~\eqref{eqch4:metricspherical}. 
By using the results of Chapter~\ref{ch:Euclideanpathintegral}, the 
gravitational action can be written as
\begin{align}\label{eqch4:actionspherical2}
    & I_{g\mathrm{f}} =
     \left(\frac{2\pi b r^{d-3}}{\mu}
     \left(1 - \frac{r'}{a}\right)\right)\sVert[3]_{u=1}
    - \frac{\Omega_{d-2}}{4l_p^{d-2}}\left(\frac{b' r^{d-2}}{a}\right)\sVert[3]_{u=0}\notag\\
    &+ \frac{1}{8\pi l_p^{d-2}}\int_{M} a b r^{d-2}G\indices{^\tau_\tau}d^dx\,\,,
\end{align}
where 
\begin{align}\label{eqch4:mu}
    \mu = \frac{8\pi l_p^{d-2}}{(d-2)\Omega_{d-2}}\,\,,
\end{align}
and the Einstein tensor component $G\indices{^\tau_\tau}$ is given by
\begin{align}
    G\indices{^\tau_\tau} = \frac{(d-2)}{2r'r^{d-2}}\left(r^{d-3}\left(\frac{r^{\prime2}}{a^2} 
    -1\right)\right)'\,\,.
\end{align}
Together with the regularity and boundary conditions in 
Eqs.~\eqref{eqch4:regularitymetric}--\eqref{eqch4:betadef}, the action in 
Eq.~\eqref{eqch4:actionspherical2} becomes
\begin{align}\label{eqch4:actionspherical3}
    & I_{g\mathrm{f}} =
     \left(\frac{\beta R^{d-3}}{\mu}
     \left(1 - \frac{r'}{a}\right)\right)\sVert[3]_{u=1}
    - \frac{\Omega_{d-2} r_+^{d-2}}{4l_p^{d-2}}
    &+ \frac{1}{8\pi l_p^{d-2}}\int_{M} a b r^{d-2}G\indices{^\tau_\tau}d^dx\,\,.
\end{align}
The action for the Maxwell field can also be simplified using that 
$F^{\alpha \beta}F_{\alpha \beta} = 2 F_{u\tau}F^{u\tau} = 2\frac{A_\tau'^2}{b^2a^2}$. 
It is more convenient to work with the integrand written as 
\begin{align}
    \frac{\sqrt{g}}{4}F_{\alpha \beta} F^{\alpha \beta} = 
    - \frac{1}{2}\frac{r^{d-2}A^{\prime2}_\tau}{b a} 
    + \left( \left(\frac{r^{d-2} A'_\tau}{b a}\right) A_\tau\right)'
    - \left(\frac{r^{d-2} A'_\tau}{b a}\right)' A_\tau\,\,,
\end{align}
and so the Maxwell action is given by
\begin{align}
    I_A = - \frac{1}{2}\int_M \frac{r^{d-2}A^{\prime2}_\tau}{b a}d^dx 
    - i \beta \phi \Omega_{d-2}\left(\frac{r^{d-2}A'_\tau }{b a}\right)\sVert[3]_{u=1} 
    - \int_{M} \left(\frac{r^{d-2} A'_\tau}{b a}\right)' A_\tau d^dx\,\,,
\end{align}
where the regularity and boundary conditions in Eqs.~\eqref{eqch4:Ay0} 
and~\eqref{eqch4:phidef} have been used. Finally, 
putting together both actions, we have that the full action $I$ is 
\begin{align}\label{eqch4:actionspherical5}
    &I =  \left(\frac{\beta R^{d-3}}{\mu}
    \left(1 - \frac{r'}{a}\right)\right)\sVert[3]_{u=1}
   - \frac{\Omega_{d-2} r_+^{d-2}}{4l_p^{d-2}} 
   - i \beta \phi \Omega_{d-2}\left(\frac{r^{d-2}A'_\tau }{b a}\right)\sVert[3]_{u=1}\notag\\
   &+ \frac{1}{8\pi l_p^{d-2}}\int_{M} a b r^{d-2}
   \left(G\indices{^\tau_\tau} - 4\pi l_p^{d-2}\frac{A^{\prime2}_\tau}{b^2 a^2}\right) d^dx 
   - \int_{M} \left(\frac{r^{d-2} A'_\tau}{b a}\right)' A_\tau d^dx\,\,,
\end{align}
which must be inserted in the Euclidean path integral 
\begin{align}
    Z = \int Db Da Dr DA_\tau \mathrm{e}^{-I}\,\,.
\end{align}

\section{Zero loop approximation and the black hole solutions\label{sech4:zeroloop}}

\subsection{The constrained path integral and reduced action}

As a step towards the zero loop approximation, we constrain the path integral 
along metrics that obey the constraint equations 
that partially minimize the action. The constraint equations are composed 
by the Hamiltonian constraint, momentum constraint and the Gauss constraint. 
The momentum constraint is satisfied apriori since the metric is static. 
The Hamiltonian 
constraint is given by $G\indices{^\tau_\tau} = 8\pi l_p^{d-2} T\indices{^\tau_\tau}$, 
where $T\indices{_\alpha_\beta} = F_{\alpha \mu}F_{\beta\nu}g^{\mu \nu} 
- \frac{1}{4}g_{\alpha \beta}F_{\nu \mu}F^{\nu \mu}$ 
is the stress energy tensor of the Maxwell field. This Einstein equation is precisely 
obtained by calculating the first order variation of the Euclidean action in the 
metric component $b(u)$. In terms of the components of the 
metric, the Hamiltonian constraint is given by
\begin{align}\label{eqch4:Hamiltonian}
    \frac{(d-2)}{2r' r^{d-2}}\left(r^{d-3}\left(\frac{r^{\prime2}}{a^2}-1\right)\right)'
    = 4\pi l_p^{d-2}\frac{A^{\prime2}_\tau}{a^2 b^2}\,\,.
\end{align}
The Gauss constraint 
is given by the Maxwell equation $\nabla_{u} F^{\tau u} = 0$, 
which in terms of the metric and Maxwell field components yields 
\begin{align}\label{eqch4:GaussConstraint}
    \left(\frac{r^{d-2}A'_\tau}{a b}\right)' = 0\,\,,
\end{align}
which can be obtained by performing the first order variation 
of the Euclidean action in the Maxwell component $A_\tau$.
The Gauss constraint can be first integrated to give 
\begin{align}\label{eqch4:charge}
\frac{r^{d-2}}{a b} A'_\tau  = -
i \frac{q}{\Omega_{d-2}}\,\,,
\end{align} 
where $q$ is the electric charge, having dimensions 
of length $L^{\frac{d}{2}-2}$, where $L$ is some unit of length. 
Plugging this into the Hamiltonian 
constraint, one can integrate Eq.~\eqref{eqch4:Hamiltonian} to obtain
\begin{align}
    \left(\frac{r'}{a}\right)^2 = f(r;r_+,q) = 1 - \frac{r_+^{d-3}}{r^{d-3}} 
    - \frac{\lambda q^2}{r^{2d-6}} +
\frac{\lambda q^2}{r^{2d-6}}\,,
\label{eqch4:alphavalue}
\end{align}
where we used the regularity condition $r(0) = r_+$,  
we defined $f(r;r_+,q)$, and $\lambda$ is given by
\begin{align}\label{eqch4:lambda}
    \lambda = \frac{8\pi l_p^{d-2}}{\Omega_{d-2}^2(d-2)(d-3)}\,\,.
\end{align}
The action in Eq.~\eqref{eqch4:actionspherical5} 
with the Hamiltonian and Gauss constraints imposed is called the reduced 
action and assumes a simple expression, since the bulk terms disappear, 
yielding 
\begin{align}
    I_*[\beta, \phi, R;r_+,q] =  
    \frac{R^{d-3}\beta}{\mu}\left(1-\sqrt{f[R;r_+,q]}\right) -
    q \beta \phi - \frac{\Omega_{d-2} r_+^{d-2}}{4 l_p^{d-2}}\,.
    \label{eqch4:reducedaction}
\end{align}
The reduced action $I_*$ becomes a functional of the parameters 
$r_+$ and $q$, and a function of the fixed parameters $\beta$, 
$\phi$ and $R$. With the constraints applied to the path integral in 
Eq.~\eqref{eqch4:partitionfunction}, 
the integral over $b$ can be neglected as the reduced action does not depend 
on $b$. Also, the integral over $A_\tau$ can also be neglected for the same reason.
The remaining integrals over $a$, $r$ and $A_\tau$ transform into 
integrals over $r_+$ and $q$. In order to see this, first one can perform 
an arbitrary coordinate transformation $r=r(y)$, which gives a metric 
only as functional of $b(y)$, $r_+$ and $q$. And so, the sum over configurations 
obeying the constraints must be done only in $r_+$ and $q$. Then, 
the constrained path integral 
can be written as 
\begin{align}
    Z[\beta,\phi,R] = \int D[r_+]D[q] \mathrm{e}^{-I_*[\beta,\phi,R;r_+,q]}\,.
\label{eqch4:ZasI}
\end{align}  
To proceed with the zero loop approximation, one must impose the 
remaining Einstein-Maxwell equations. These equations are equivalent to 
the conditions for the stationary points 
of the reduced action in the plane $r_+ \times q$. 
For the zero loop approximation to remain valid, the stationary points must be 
local minima of the action. The motivation of imposing the 
constraint equations to the path integral is that we are able from 
Eq.~\eqref{eqch4:ZasI} to verify the stability of the solutions 
given by the stationary point, at least along the hypersurface of 
metrics obeying the constraints, see~\cite{Whiting:1988qr}.

\subsection{Stationary points of the reduced action}

The partition function in Eq.~\eqref{eqch4:ZasI} describes the 
grand canonical ensemble of a charged black hole inside a cavity constrained 
to the hypersurface where the Hamiltonian and Gauss constraints 
are satisfied. Here, we are interested in performing the full zero loop 
approximation of the path integral to obtain the equilibrium 
solutions for the black hole. The solutions are described by the 
stationary points of the reduced action in Eq.~\eqref{eqch4:reducedaction}, 
which satisfy the conditions
\begin{align}
    &\frac{\partial I_*}{\partial r_+} = 0\,,\label{eqch4:derI1}\\
    &\frac{\partial I_*}{\partial q} = 0\,.\label{eqch4:derI2}
\end{align}
These two conditions can be written in terms of the fixed variables of the 
ensemble, $\beta$ and $\phi$, and the variables evaluated at the 
stationary points, $r_+$ and $q$, as
\begin{align}
    &\beta = \frac{4\pi}{(d-3)}\frac{r_+^{2d-5}}{r_+^{2d-6}-\lambda
q^2}\sqrt{f[R,r_+,q]}\,,\label{eqch4:betaexpress}\\
&\phi = \frac{q}{(d-3)\Omega_{d-2}
\sqrt{f[R,r_+,q]}}\left(\frac{1}{r_+^{d-3}}-
\frac{1}{R^{d-3}}\right)\,,
\label{eqch4:phiexpress}
\end{align}
respectively. In order to find the solutions of the 
ensemble, one must solve the inverse problem of 
the system in Eqs.~\eqref{eqch4:betaexpress} 
and~\eqref{eqch4:phiexpress} to have the 
functions $r_+=r_+(\beta,\phi,R)$ and $q=q(\beta,\phi,R)$.
The reduced action evaluated at the stationary points 
$r_+=r_+(\beta,\phi,R)$ and $q=q(\beta,\phi,R)$ is defined 
as
\begin{align}
    I_0[\beta,\phi,R]=
I_*[\beta,\phi,R;r_+[\beta,\phi,R],q[\beta,\phi,R]]
\,.
\label{eqch4:actionI0}
\end{align}
Using the expression of the reduced action in 
Eq.~\eqref{eqch4:reducedaction},
the action $I_0$ can be further written as 
\begin{align}
    &I_0[\beta,\phi,R] =\nonumber\\
    & \frac{R^{d-3}\beta}{\mu}\left(1- 
    \sqrt{f[R;r_+[\beta,\phi,R],q[\beta,\phi,R]]}\right)
    \nonumber
    \\
    & -q[\beta,\phi,R] \beta \phi -
    \frac{\Omega_{d-2} r_+^{d-2}[\beta,\phi,R]}{4l_p^{d-2}}\,.
    \label{eqch4:actionI0full}
\end{align} 
In the zero loop approximation, the partition function 
is given solely by the contribution of the stationary point, 
i.e.
\begin{align}
    Z[\beta, \phi, R] = \mathrm{e}^{-I_0[\beta,\phi,R]}\,,
\label{eqch4:partaction}
\end{align}
where $I_0[\beta,\phi,R]$ is taken from 
Eq.~\eqref{eqch4:actionI0full}.

We analyze now the solutions of the stationary conditions. 
It is useful to make the following definitions 
\begin{align}
    &\gamma =
\frac{16\pi^2 R^2}{(d-3)^2}
\frac{\Phi^2}{\beta^2(1-\Phi^2)^2}
\,,
\label{eqch4:definitions2}
\\
&\Phi = (d-3) \Omega_{d-2} \sqrt{\lambda} \phi\,,
\label{eqch4:newPhi}
\\
&x=\frac{r_+}{R}\,,
\label{eqch4:definitionsx}
\\
&y=\frac{\lambda q^2}{R^{2d-6}}\,.
\label{eqch4:definitions}
\end{align}
The parameter $\gamma$ takes the role of 
the temperature $T=\frac{1}{\beta}$, while 
$\Phi$ takes the role of $\phi$. The radius 
of the reservoir $R$ is taken to be a scale, 
that can be absorbed in the variables $r_+$ 
and $q$, having thus the horizon radius in units 
of $R$, yielding $x$, and the charge squared in units of $R$ multiplied 
by the ratio between the Planck length and $R$, yielding $y$.
The system of equations in Eqs.~\eqref{eqch4:betaexpress} and~\eqref{eqch4:phiexpress} 
can be inverted to give equations for $x$ and $y$.
Inverting Eq.~\eqref{eqch4:phiexpress}, 
i.e. $y=\frac{x^{2d-6}\Phi^2}{1-(1-\Phi^2)x^{d-3}}$, and 
substituting into 
Eq.~\eqref{eqch4:betaexpress}, one arrives to the 
equation
\begin{align}
    (1-\Phi^2)x^{d-1} - x^2 + \frac{\Phi^2}{\gamma} = 0\,.
\label{eqch4:extremaaction}
\end{align}
Now using Eq.~\eqref{eqch4:extremaaction} into 
$y=\frac{x^{2d-6}\Phi^2}{1-(1-\Phi^2)x^{d-3}}$, one gets 
the second equation as
\begin{align}
    &y = \gamma x^{2(d-2)}\,.
\label{eqch4:extrema2action}
\end{align}
Hence, the solutions to the stationary conditions 
are obtained by solving Eq.~\eqref{eqch4:extremaaction} 
for $x(\gamma, \Phi)$ and the value of $y(\gamma,\Phi)$ can then be 
read off from Eq.~\eqref{eqch4:extrema2action}.

The solutions for the horizon radius satisfying the polynomial 
equation Eq.~\eqref{eqch4:extremaaction} can only be found analytically 
for specific values of $d$. For example, analytical solutions can be found 
for $d=4$ as we have a third order polynomial equation for $x$~\cite{Braden:1990hw}, and 
also for $d=5$ as we have a second order polynomial equation for $x^2$. We analyze this last case 
below, separately. For generic $d$, however, it is not possible 
to find an analytic expression for $x$.

Notwithstanding, we can study the qualitative behaviour of Eq.~\eqref{eqch4:extremaaction} 
in terms of the dimension $d$ and the parameters $\gamma$, which represents the fixed temperature 
of the ensemble, and $\Phi$, which represents the fixed electric potential of the 
ensemble. We impose that the solutions for $x$ and $y$ must be physical. This is so 
since the Riemannian space that minimizes the action must be correspondent to 
a Lorentzian space through the inverse Wick rotation. Here, we assume that 
the black hole must lie inside the cavity and that it must be subextremal, i.e.
\begin{align}
   & 0\leq x<1\,,
\label{eqch4:insidecond}\\
& 0\leq \frac{y}{x^{2(d-3)}} < 1\,,\label{eqch4:subex}
\end{align}
respectively. One can use Eq.~\eqref{eqch4:extrema2action} 
to rewrite the last condition Eq.~\eqref{eqch4:subex} into 
\begin{align}
    \gamma x^2 < 1
\label{eqch4:Imposecond1}\,\,.
\end{align}
Moreover, using Eq.~\eqref{eqch4:extremaaction} in 
Eq.~\eqref{eqch4:Imposecond1}, one obtains a condition 
for $\Phi$ as
\begin{align}
    0 \leq \Phi^2 < 1\,.
\label{eqch4:Phi2cond}
\end{align}
Therefore, the conditions for physical solutions reduce to the 
restrictions of the parameters $\gamma$ and $\Phi$ through 
Eqs.~\eqref{eqch4:Imposecond1} and~\eqref{eqch4:Phi2cond}, respectively.

In order to study further Eq.~\eqref{eqch4:extremaaction}, it is useful 
to define 
\begin{align}
    h(x) = (1-\Phi^2)x^{d-1} - x^2 + \frac{\Phi^2}{\gamma}
\,.
\label{eqch4:hofx}
\end{align}
The values of the function $h(x)$ at the boundaries established 
by the condition Eq.~\eqref{eqch4:insidecond} are 
$h(0) = frac{\Phi^2}{\gamma} >0$ and $h(1) = h(0)(1-\gamma)$. 
The parameter $\gamma$, although being restricted by Eq.~\eqref{eqch4:Imposecond1}, 
can assume values higher than unity. Since $\gamma$ is proportional to 
the temperature, $\gamma$ can attain large values for large values of the 
temperature, for fixed $\phi$ or $\Phi$. We thus need to separate the 
analysis into three regions, $\gamma < 1$, $\gamma=1$, and $\gamma > 1$.

Starting with $\gamma < 1$, one has that $h(x)$ is positive at the 
boundaries $h(0)>0$ and $h(1)>0$. To further gain knowledge on the amount of zeros, 
one must compute the zeros of the first derivative $h'(x)$ and 
the sign of the second derivative $h''(x)$. The derivatives 
are given by
\begin{align}
    &h'(x) = x(d-1)(1-\Phi^2)\left(x^{d-3} -
\frac{2}{(d-1)(1-\Phi^2)}\right)\,\,,\\
& h''(x) = (d-1)(d-2)(1-\Phi^2)x^{d-3}-2\,\,.
\end{align} 
The derivative of $h(x)$ vanishes at a bifurcation point 
\begin{align}
    x_{\mathrm{bif}} =
\left(\frac{2}{(d-1)(1-\Phi^2)}\right)^{\frac{1}{d-3}}\,,
\label{eqch4:xmin}
\end{align}
i.e. $h'(x_{\mathrm{bif}}) = 0$, 
and the second derivative is positive there, i.e. $h''(x_{\mathrm{bif}})>0$.
So, the bifurcation point $x_{\mathrm{bif}}$ marks the location of the only minimum 
of $h(x)$. In the case of $\gamma < 1$, if the location of the minimum 
lies out of bounds, $x_{\mathrm{bif}}>1$, then it is certain that 
there are no zeros of $h(x)$ in the interval $0\leq x<1$. If the minimum 
lies in the interval $0<x_{\mathrm{bif}}<1$, then $h(x)$ may have one or two 
zeros in $0\leq x<1$. This happens for the range of 
$0 \leq\Phi^2 < \frac{d-3}{d-1}$. However, in order for the zeros to exist, 
one must require that $h(x_{\mathrm{bif}})<0$, which only happens when
\begin{align}
    \gamma\geq \gamma_{\mathrm{bif}}(\Phi,d)\equiv
\frac{(d-1)^{\frac{d-1}{d-3}}}{4^{\frac{1}{d-3}}(d-3)}\Phi^2
(1-\Phi^2)^{\frac{2}{d-3}} \,.
\label{eqch4:solExcondition}
\end{align}

Summarizing the results in the interval $\gamma < 1$ and 
$0 \leq\Phi^2 < \frac{d-3}{d-1}$, for
\begin{align}
    \gamma <
\gamma_{\mathrm{bif}}\,,
\label{eqch4:nosolutionscond}
\end{align} 
there are no solutions. For 
\begin{align}
    \gamma_{\mathrm{bif}}\leq\gamma <1\,,
\label{eqch4:twosolutionscond}
\end{align}
there are two solutions. The solutions 
can be denominated by $x_1$ and $x_2$, 
with $x_1 \leq x_2$. Moreover, one must 
have
\begin{align}
    x_1 \leq
x_{\mathrm{bif}}\leq x_2\,.
\label{eqch4:x1x2xmin}
\end{align}
When the equality is reached in Eq.~\eqref{eqch4:twosolutionscond}, 
i.e. $\gamma = \gamma_{\mathrm{bif}}$, the two solutions 
merge into $x_1 = x_2 = x_{\mathrm{bif}}$, hence the nomenclature of 
$x_{\mathrm{bif}}$ as the bifurcation point. For 
$\gamma < 1$ and $\frac{d-3}{d-1} \leq \Phi^2 < 1$, 
there are no solutions.

For $\gamma = 1$, the function $h(x)$ has always one zero at $x=1$.
We note that this point is a critical point since the 
derivatives of the action are not well-defined there. If 
$0 \leq \Phi^2 < \frac{d-3}{d-1}$, the other zero lies in the 
interval $0<x<1$ and so it corresponds to $x_1$ while $x_2 = 1$. 
For the case of equality $\Phi^2 = \frac{d-3}{d-1}$, one has 
$x_1 = x_2 = 1$. For $\frac{d-3}{d-1} < \Phi^2 < 1$, $x_1=1$ 
while $x_2$ lies out of bounds as $x_2 > 1$ and it is unphysical.

\begin{figure}[t]
    \centering
    \includegraphics[width=0.45\columnwidth]{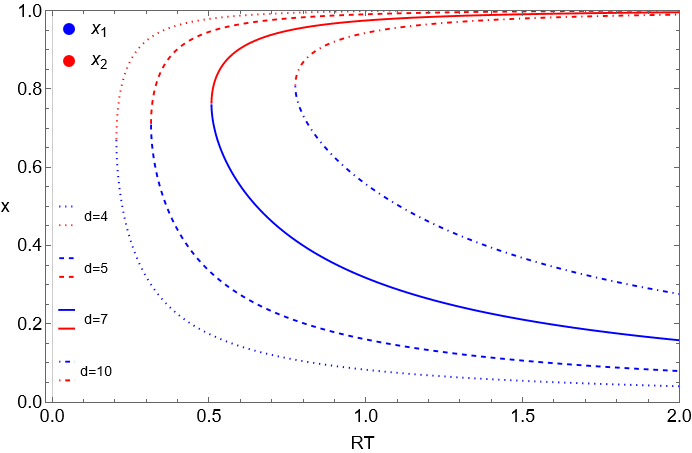}%
    \hskip 2mm %
    \includegraphics[width=0.45\columnwidth]{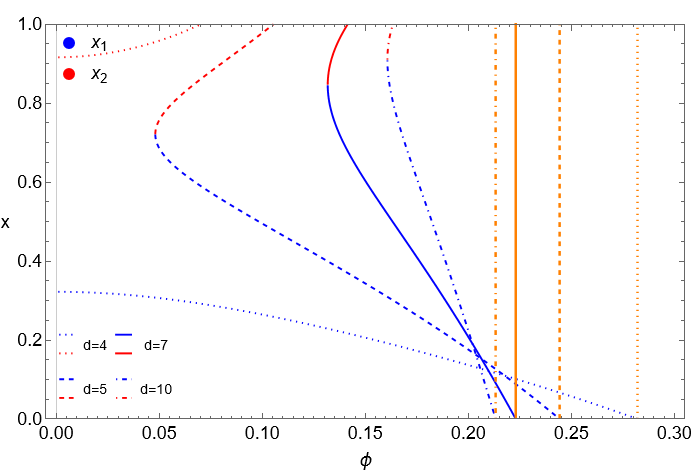}
    \caption{
    Left plot: Stationary points of the action are plotted, with $x_1$ in blue 
    and $x_2$ in red, in function of $RT$, for a constant $\phi = 0.02$ and 
    for four values of $d$: $d = 4$ in dotted lines, $d = 5$ in dashed lines, $d = 7$ in
    solid lines, and $d = 10$ in dot dashed lines.
    Right plot: Stationary points of the action are plotted, with $x_1$ in blue and
    $x_2$ in red, in function of $\phi$, for a constant $RT=0.3$, and the maximum
    value of $\phi$ (in orange) corresponding to $\Phi=1$, for four values
    of $d$: $d = 4$ in dotted lines, $d = 5$ in dashed lines, $d = 7$ in
    solid lines, and $d = 10$ in dot dashed lines. Here, $l_p$ is set to one.
    }
    \label{figch4:2solsphiRTconst}
    \end{figure}

For $\gamma > 1$, the function $h(x)$ has always one zero, $x_1$, 
in the interval $0 \leq x < 1$, while the remaining zero $x_2$ 
lies in the interval $x > 1$, all of this for $0 \leq \Phi^2 < 1$. 
The zero $x_1$ is physical while the zero $x_2$ is unphysical.

We must highlight some important comments on the behaviour of the solutions. 
From the definition of $\gamma$ in Eq.~\eqref{eqch4:definitions2}, 
one has that $\gamma \propto \frac{R^2}{\beta^2} \propto (RT)^2$. 
Considering the solution $x_2$ in function of the temperature 
$RT$, one has that $x_2$ exists with an image in the interval 
$0 < x_2 <1$ for a finite range of $T= \frac{1}{\beta}$ and 
for $\Phi^2 < \frac{d-3}{d-1}$. Specifically, for 
$RT = \frac{d-3}{4\pi \abs{\Phi}}(1- \Phi^2)$, i.e. $\gamma = 1$, 
one has $x_2=1$ and so for higher values of $RT$, the solution $x_2$
goes out of bounds and becomes unphysical. This behaviour is not 
present in the neutrally charged case, analyzed in~\cite{Andre:2021ctu}. 
The plot of the solutions $x_1$ and $x_2$ as functions of $RT$ for 
a constant $\phi = 0.02$ and for four values of $d$ is shown in 
the left part of Fig.~\ref{figch4:2solsphiRTconst}, with $l_p=1$. 
Also, the plot of the solutions $x_1$ and $x_2$ as functions of 
$\phi$ for a constant $RT= 0.3$ and for four values of $d$ is shown in 
the right part of 
Fig.~\ref{figch4:2solsphiRTconst}, with $l_p=1$. In these plots, we chose 
the quantity $\phi$ instead of $\Phi$ since we want to showcase 
the full dependence in the dimension $d$. Indeed, $\phi$ is the fixed 
independent parameter at the cavity while $\Phi$ is a quantity 
proportional to $d$.

\subsection{Beyond zero loop approximation and 
stability of the stationary points}\sectionmark{
Stability of stationary points
}\thispagestyle{userightbotmark}

With the stationary points analyzed, we proceed with the analysis 
of their stability. In order to probe the stability, we 
can go beyond the zero loop approximation of the reduced action. 
The reduced action is to be expanded around the stationary points 
up to second order terms, i.e. the reduced action 
can be written as
\begin{align}
    I_*[\beta,\phi, R; r_+,q] =
I_{0}[\beta,\phi, R]+
\sum_{ij}{I_{*0}}_{ij} \delta i \delta j
\,,
\label{eqch4:reducedactionperturbe}
\end{align}
where $I_{0}[\beta,\phi, R]$
is defined generically in Eq.~\eqref{eqch4:actionI0}
and specifically for the grand canonical ensemble 
of a charged black hole in 
Eq.~\eqref{eqch4:actionI0full},
and
${I_{*0}}_{ij}$ are 
the second derivatives of the reduced action
${I_{*}}_{ij} =
\frac{\partial^2 I_*}{\partial i \partial j}$ evaluated 
at an extremum of the action, with
$i$ and $j$ being either $r_+$ or $q$. The partition function 
with this expansion around a particular stationary point is given 
by 
\begin{align}
    Z[\beta, \phi, R] = {\mathrm{e}}^{-I_{0}[\beta,\phi, R]} \int D[\delta
q]D[\delta r_+] {\mathrm{e}}^{-\sum_{ij}{I_{*0}}_{ij} \delta i \delta j}\,.
\end{align}
In order to have a well-defined path integral in this approximation, 
which takes into account the static one loop corrections obeying the 
Hamiltonian and Gauss constraints, the exponent must be always negative, 
i.e. the stationary point must be a minimum of the reduced action.

To obtain the condition of stability in terms of the solutions 
of the ensemble, one must analyze the hessian of the reduced 
action, which has in this case the components
\begin{align}
    &{I_{*0}}_{r_+ r_+} = \frac{(d-2)\Omega_{d-2} R^{d-3} \beta}{16 \pi
   \sqrt{f} r_+^2 l_p^{d-2}} \mathcal{I}_{r_+r_+}\,,\label{eqch4:Irprp}\\
   &{I_{*0}}_{r_+ q} = \frac{(d-2)\Omega_{d-2} R^{d-3} \beta}{16 \pi \sqrt{f}
   r_+ q l_p^{d-2}} \mathcal{I}_{r_+ q}\,,\label{eqch4:Irpq}\\
   &{I_{*0}}_{qq} = \frac{(d-2) \Omega_{d-2} R^{d-3} \beta}{16 \pi \sqrt{f}
   q^2 l_p^2} \mathcal{I}_{q q}\,,\label{eqch4:Iqq}
\end{align}
with 
\begin{align}
    &\mathcal{I}_{r_+ r_+}{\hskip -2mm} = \frac{d-3}{f
    x^{2d-6}}\Big[\frac{d-3}{2}\left(x^{2d-6} - y\right)^2 {\hskip -2mm}-
    \left(x^{2d-6}-(2d-5)y\right)
    \left(1-x^{d-3}\right)\left(x^{d-3}{\hskip -2mm}-y\right)
    \Big]\,,\label{eqch4:Icalpp}\\
    &\mathcal{I}_{r_+ q} = -\frac{(d-3)}{x^{d-3}}\frac{\left(2 x^{d-3}
    - x^{2d-6}-y\right)}{x^{d-3} - y}y\,,\label{eqch4:Icalpq}\\
    &\mathcal{I}_{qq} = 2\frac{1-x^{d-3}}{x^{d-3} -
    y}y\,.\label{eqch4:Icalqq}
\end{align}
For the stationary point to be a minimum, the matrix ${I_{*0}}_{ij}$ 
must be positive definite. In turn, this condition can be translated 
into the principal minors being positive through the 
Sylvester criterion, i.e. 
\begin{align}
    &\mathcal{I}_{r_+ r_+} > 0\,,\label{eqch4:condstab1}\\
    &\mathcal{I}_{r_+ r_+}\mathcal{I}_{q q}-\mathcal{I}_{r_+
q}^2>0\,.
\label{eqch4:condstab2}
\end{align}
In this case, $\mathcal{I}_{qq}$ is always positive and so 
the last condition, Eq.~\eqref{eqch4:condstab2}, is 
sufficient to ensure positive definiteness. 
Using Eqs.~\eqref{eqch4:Icalpp}-\eqref{eqch4:Icalqq}
and~\eqref{eqch4:extrema2action}, the condition in
Eq.~\eqref{eqch4:condstab2} reduces to
\begin{align}
  - (d-3)\gamma x^{d-1}+  (d-1)x^{d-3}  - 2 > 0\,.
\label{eqch4:stabilityx}
\end{align}
This is the sufficient condition for a stationary point, given by
Eqs.~\eqref{eqch4:extremaaction} and \eqref{eqch4:extrema2action}, to be a
minimum of the action.

Since the stability condition in Eq.~\eqref{eqch4:stabilityx} is to 
be evaluated at the stationary points, one can use Eq.~\eqref{eqch4:extremaaction} 
to further simplify Eq.~\eqref{eqch4:stabilityx}. By rewriting 
Eq.~\eqref{eqch4:extremaaction} as
$\gamma = \frac{\Phi^2}{x^2 - (1-\Phi^2)x^{d-1}}$ and by 
substituting $\gamma$ in Eq.~\eqref{eqch4:stabilityx}, the stability condition 
simplifies into a factorized polynomial 
\begin{align}
    \frac{((d-1)(1- \Phi^2)x^{d-3} - 2) (1 - x^{d-3})}
    {1 - (1-\Phi^2)x^{d-3}}
    > 0\,\,\,.\label{eqch4:stabilitysimple}
\end{align}
Now, the physical range of solutions is $0\leq x^{d-3}<1$ 
which means the denominator is always greater than zero. Therefore, 
the condition can be reduced to $(d-1)(1- \Phi^2)x^{d-3} - 1 > 0$. 
The stationary point is thus stable if 
\begin{align}
   x >  x_{\mathrm{bif}} \,,\label{eqch4:criterionstability}
\end{align}
where
$x_{\mathrm{bif}} = \left(\frac{2}{(d-1)(1-\Phi^2)}
\right)^{\frac{1}{d-3}}$. Since $x_{\mathrm{bif}}$ 
marks the point of bifurcation of the two solutions 
$x_1$ and $x_2$, one has $x_1 < x_{\mathrm{bif}}<x_2$.
The bifurcation radius thus marks marginal stability. 
It must be noted that in the uncharged 
case, the bifurcation radius also marks marginal stability 
and it coincides with the photon sphere radius. It turns out 
that in the case of the grand canonical ensemble, this is not 
the case, see Sec.~\ref{sech4:radii}.

Coming back to the analysis of the stability, for $\gamma_{\mathrm{bif}}
<\gamma < 1$ and $\Phi^2 < \frac{d-3}{d-1}$, one has 
$x_1 < x_{\mathrm{bif}}<x_2$ and so $x_1$ is unstable, corresponding 
to a saddle point of the reduced action, while 
$x_2$ is stable, corresponding to a minimum of the reduced action. 
For the case $\frac{d-3}{d-1}\leq \Phi^2<1$, there are no solutions 
in the physical interval $0 < x < 1$. In case of the equality 
$\gamma = \gamma_{\mathrm{bif}}$, the solutions $x_1= x_2 = x_{\mathrm{bif}}$
coincide. While this may signal marginality on the stability, one 
should be more careful and analyze higher derivatives at this point. 
By inspection of the plots of the action $I(x,y)$, one finds that 
$x_{\mathrm{bif}}$ is a saddle point. 

For the case of $\gamma =1$, and with $\Phi^2 < \frac{d-3}{d-1}$, 
solution $x_1$ is unstable, while the solution $x_2$ resides at the 
boundary of the cavity, $x_2=1$. At $x_2=1$, the derivatives of the 
action are not well-defined and so the stability cannot be specified. 
For $\Phi^2 = \frac{d-3}{d-1}$, the two solutions $x_1$ and $x_2$ coincide 
and reside at the boundary of the cavity, sharing the properties of 
$x_2=1$. For $\frac{d-3}{d-1} < \Phi^2 < 1$, the solution 
$x_1=1$ resides at the cavity, so the stability cannot be specified, while 
$x_2$ lies outside the bounds of the cavity and it is unphysical.

Finally for $\gamma > 1$ and for $0 < \Phi^2 < 1$, the solution 
$x_1$ is the only physical solution and it is unstable.

\subsection{Most probable configurations of the ensemble}

With the stationary points obtained and their stability characterized, 
we are interested to see the configurations that are most favorable 
or most probable. From Eq.~\eqref{eqch4:ZasI}, it can be seen that the paths with 
the lowest $I_*$, or from
Eq.~\eqref{eqch4:partaction} the paths with the lowest $I_0$, are the ones that
contribute the most to the partition function, and so they correspond to the most
probable states. Here, we make a comparison between the critical points of the 
reduced action obtained above.

In the electrically uncharged case done in
\cite{York:1986} for $d=4$ and
in \cite{Andre:2020czm,Andre:2021ctu}
for generic $d$, the comparison between the
stable black hole solution and hot flat space was made regarding what was the most 
favorable state. The stable
black hole is a stationary point of the reduced action, and the hot flat
space solution is an extra stationary point existing in another topological 
sector. Hot flat space here is defined by the solution of the vacuum 
Einstein equations with topology $\mathbb{S}^1\times \mathbb{R}^{d-1}$, where 
the total imaginary time is the inverse temperature.
As already stated, the most probable state is
the one with the lowest value of the action.  In the case of no electric 
charge, the value of the action $I_0$ depends on $\beta$, while in the
case of hot flat space one has $I_{\mathrm{hot}\,\mathrm{flat}\,\mathrm{space}} = 0$. In
\cite{Andre:2020czm,Andre:2021ctu}, it was shown
for any dimension $d\geq4$ that the black hole is more favorable 
than hot flat space, $I_0 < I_{\mathrm{hot}\,\mathrm{flat}\,\mathrm{space}}$,
if $\beta$ is such that $\frac{r_+}{R} > \frac{r_+}{r_{\mathrm{Buch}}}$,
where $r_{\mathrm{Buch}}$ is the Buchdahl radius. Additionally, in
\cite{Andre:2020czm,Andre:2021ctu}, a comparison between the stable black
hole solution and quantum hot flat space was also done.

In the electrically charged case, we can also make a comparison of
the stable black hole with an equivalent of the uncharged hot
flat space. The electrically charged case is more
rich than the uncharged one since the dimensionality of the 
reduced action increases.  In the charged case, besides the
stationary point related to the stable black hole, there are two additional
critical points that are possible stable solutions of the ensemble.
One critical point
is $r_+ = 0$ and $q = 0$, corresponding
to a cavity without a black hole and without charge. Note however that 
this critical point must be seen as a limit, since the regularity conditions 
must be changed to cover this point. The configuration with $r_+=0$ and $q=0$ seems
unphysical for a fixed nonzero value of $\phi$. For this configuration,
there is a difference of electric potential, which in turn implies the
existence of an electric field and thus of an electric charge.  For
this reason, $q=0$ seems rather unphysical. Nevertheless, one must consider that
the path integral approach in the semiclassical approximation deals
intrinsically with quantum systems. And so, when one writes $q=0$, one
should mean $q$ of the order of the Planck charge, and a particle,
say, carrying such a charge should be envisaged as having the
dimensions of the order of or slightly larger than a Planck length. 
Therefore, one has to seek a corresponding action for such a particle 
in a reservoir of fixed $R$ and $\beta$. The other critical point is 
described by $r_+=R$
and $\sqrt{\lambda} q=R^{d-3}$, so that $r_+= (\sqrt{\lambda}
q)^\frac1{d-3}=R$.  This critical point actually corresponds to 
an extremal black hole with the horizon localized at the radius of the cavity,
and for this case the volume of the Riemannian space is zero, which
may require a different procedure to be analyzed.
Again, this critical point should be understood as a quantum system treated
semiclassically, and so one should think of a black hole almost at its
extremal state, failing to be extremal by a Planck charge and not
touching the reservoir at $R$ by a Planck length. 
The question of which configuration is the ground state 
is a pertinent one.

Starting with the first critical point $r_+ = 0$ and $q =
0$, as stated before, the derivative of the action in order to $q$ is not
well-defined. Nonetheless,
we argue that this critical point can be considered 
as a local minimum of the action in the physical domain but not in the 
typical sense, see
Sec.~\ref{sech4:gradient}. 
In order to find an equivalent of hot flat space for the charged 
case, we consider a hot sphere,
made of a perfect conductor material, with a certain radius $r_{\mathrm{hs}}$,
inside the reservoir at constant $\beta$ and $\phi$, and with its
center situated at the center of the reservoir. For simplicity, 
the gravitational interaction is neglected, i.e., the constant of gravitation is put to
zero. The action is then composed solely by the Maxwell term in
Eq.~\eqref{eqch4:EucAction1}. The hot sphere conductor 
depends on a fixed radius $r_{\mathrm{hs}}$, 
in the boundary conditions. Using the Gauss constraint,
the charge of the conducting sphere can be related to the
value of $\phi$, as $\phi =
\frac{q}{(d-3)\Omega_{d-2}}\left(\frac{1}{r_{\mathrm{hs}}^{d-3}}-
\frac{1}{R^{d-3}}\right)$, see also Eq.~\eqref{eqch4:phiexpress}.
The action for this cavity becomes $I=-\frac12 q\beta\phi$,
and one can use the relation between the electric charge and 
the electric potential to obtain the action of a hot spherical sphere 
in function of $\beta$, $\phi$, 
$r_{\mathrm{hs}}$ and $R$, yielding
\begin{align}
I_{\mathrm{hot}\,\mathrm{sphere}} = - \frac12\, \frac{(d-3)\Omega_{d-2}}
{\frac{1}{r_{\mathrm{hs}}^{d-3}}-
\frac{1}{R^{d-3}}}\beta\phi^2
\,.
\label{eqch4:Ihotflatsphere}
\end{align}
We can then make a comparison between the action of the conducting hot sphere
with radius $r_{\mathrm{hs}}$ given in
Eq.~\eqref{eqch4:Ihotflatsphere}, with the action of the stable configuration
of the charged black hole, which is Eq.~\eqref{eqch4:reducedaction} with
the $r_{+2}$ solution of Eq.~\eqref{eqch4:extremaaction}.  
Analyzing Eq.~\eqref{eqch4:Ihotflatsphere}, 
if $r_{\mathrm{hs}}$ is high, of the
order of $R$, then $I_{\mathrm{hot}\,\mathrm{sphere}}$ is large and
negative and so the hot flat sphere is
the most probable solution when
compared to the stable black hole $r_{+2}$. On the other hand, if
$r_{\mathrm{hs}}$ is small, as we expect to be when dealing with a case analogous
to hot flat space, then $I_{\mathrm{hot}\,\mathrm{sphere}}=0$ or close to zero.  
In the limit of $r_{\mathrm{hs}} = 0$, we can say that $I_{\mathrm{hot}\,
\mathrm{sphere}}$ is indeed $I_{\mathrm{hot}\,\mathrm{flat}\,\mathrm{space}}$ 
which is a configuration with zero action. The configuration of the hot conducting 
sphere shows clearly that the critical point $r_+=q=0$ is rather a limit of 
a very small electric charge at the center with a very small radius 
but such that $\phi$ is kept finite. The stable black hole has
a positive action
for low temperatures $T$, specifically, 
near the minimum temperature where the stable black
hole exists. Therefore, the very small charged sphere that
emulates hot flat space is more probable for a short interval of
low temperatures when compared with the stable black hole. For a long interval 
of temperatures,
the black hole is eventually more probable.
More specifically, when the solution of the stable black hole has
a horizon radius
\begin{align}
&\frac{r_{+2}^{d-3}}{R^{d-3}} \geq \frac{\mu m_{0}}{R^{d-3}} + \sqrt{\frac{\mu^2
m^2_0}{R^{2d-6}} - \frac{\lambda q^2}{R^{2d-6}}}\,,\notag\\
&\frac{\mu m_0}{R^{d-3}}= 
-\frac{4 (d-2)^2}{(d-1)^2(d-3)^2} + \frac{2(d-2)((d-2)^2
+ 1)}{(d-1)^2(d-3)^2}\sqrt{1 + \frac{(d-1)^2 (d-3)^2}{4(d-2)^2}
\frac{\lambda q^2}{R^{2d-6}}}\,\,.
\label{eqch4:conditionbhmoreprobable}
\end{align}
When Eq.~\eqref{eqch4:conditionbhmoreprobable}
is satisfied, then the action for the black hole $r_{+2}$
is negative
and the black hole is more
probable. When Eq.~\eqref{eqch4:conditionbhmoreprobable}
is not satisfied, the very small charged sphere
is more probable. This radius $R$ does not correspond 
to the Buchdahl-Andre\'asson-Wright radius~\cite{Wright:2015dma},
a radius that generalizes the Buchdahl bound
for $d$-dimensional
self-gravitating electric charged spheres. 
In fact, the horizon radius with
zero action is equal or lower than the Buchdahl-Andre\'asson-Wright radius
in the case of $d=4$,
with a difference up to $0.004$ in $\frac{\mu m}{R}$, and being equal
in the uncharged case and the extreme case $\sqrt{\lambda}q = R$.
So, the equality in the uncharged situation
of the minimum most probable
radius of a black hole in the canonical ensemble and
the Buchdahl radius does not seem to hold when other fields
are added. It is a very restricted equality
holding only in the pure gravitational situation.

Regarding the second critical point $r_+=R$ and
$\sqrt{\lambda}q=R^{d-3}$, it describes an extremal black hole with the
horizon localized at the radius of the cavity, bearing in mind that
the precise extremality and the precise location can fluctuate by
Planck order quantities. This is a critical point in the
sense that the gradient of the action is not defined, even as a limit. 
To analyze the limit, 
one can calculate the gradient of the reduced action in
Eq.~\eqref{eqch4:reducedaction} and perform the limit to $r_+=
(\sqrt{\lambda}q)^{\frac1{d-3}}=R$ along the curve $\frac{r_+}{R} = (1
- \epsilon)^{\frac1{d-3}}$ and $\frac{\sqrt{\lambda} q}{R^{d-3}} =
\sqrt{(1 - \eta \epsilon)}$, where $\eta$ is a positive constant and
$\epsilon$ parametrizes the curve.  The constant $\eta$ is restricted
here to the physical domain of the action, with the condition $\eta >
2$. After substituting the variables by the parameterization of the
curve in the expression of the gradient and performing the limit
$\epsilon \rightarrow 0^+$, one obtains
an expression 
that depends on the constant $\eta$.
Since the limit is different for different values of $\eta$, then the
gradient cannot be extended as a limit to that point,
but one can still analyze
the directional derivatives along the considered paths. The
directional derivatives along decreasing $\epsilon$, that go from lower
$r_+$ and $q$ towards  $r_+=
(\sqrt{\lambda}q)^{\frac1{d-3}}=R$, may be either positive, zero, or
negative, depending on the value of $\eta$.
So, the critical point does not resemble a local minimum restricted to a corner.
Particularly, there is a set of temperatures and electric potential
given by the condition $\gamma = 1$, where the stable black hole
solution tends to this extremal black hole. It can be seen
that for such values of temperature and electric potential, there is a
value of $\eta$ in which the limit of the gradient vanishes, but the
fact still remains that the gradient is undefined here, see
Sec.~\ref{sech4:gradient} for a detailed
analysis of the gradient at this critical point. 
It may be that this critical point 
smooths up by taking in consideration higher loops in the path
integral or a different theory of gravity.  The action for this
critical point can be evaluated from Eq.~\eqref{eqch4:reducedaction}, i.e.,
$I_{\mathrm{extreme}\,\mathrm{black}\,\mathrm{hole}} = \frac{
R^{d-3}\beta}{\mu}\Bigl(1- \Bigr.  \left.\sqrt{f(R,r_+,q)}\right) -q
\beta \phi - \frac{\Omega_{d-2} r_+^{d-2}}{4l_p^{d-2}}$, where 
$r_+$ and $q$ have extremal values, so that
$R=r_+$ and $f(R,r_+,q)=0$.  Then,
\begin{align}
I_{\mathrm{extreme}\,\mathrm{black}\,\mathrm{hole}} =& 
\frac{R^{d-3}\beta}{\mu}
-\frac{R^{d-3}}{\sqrt\lambda} \beta \phi -
\frac{\Omega_{d-2} R^{d-2}}{4l_p^{d-2}}\,.
\label{eqch4:Icoldcurvedspace}
\end{align}
Comparing the action of the critical point, 
$I_{\mathrm{extreme}\,\mathrm{black}\,\mathrm{hole}}$, 
it seems that the stable black hole is
always a more probable
configuration than the extreme black hole with horizon at the cavity.

\section{Thermodynamics of a charged black hole in higher dimensions 
inside a cavity\label{sech4:thermo}}\sectionmark{Thermodynamics of a RN 
black hole in $d$ dimensions 
inside a cavity}\thispagestyle{userightbotmark}

\subsection{Thermodynamic properties from the grand canonical ensemble}

Having the grand canonical ensemble constructed, with the partition 
function $Z$ obtained in the zero loop approximation, we 
can obtain the thermodynamic properties of the system 
by relating the partition function $Z$ to the thermodynamic grand potential $W$. 
The relation is established by 
\begin{align}
    Z(\beta,\phi,R) = \mathrm{e}^{-\beta W[\beta, \phi,R]}\,, 
    \label{eqch4:PartGrandCanonical} 
\end{align}
or $\beta W=-\ln Z$.

From the semiclassical zero loop approximation, 
the partition function is $Z = \mathrm{e}^{-I_0}$, 
hence one has the correspondence $\beta W[\beta, \phi,R] = 
I_0[\beta,\phi,R]$. Considering $\beta = \frac{1}{T}$, 
the grand potential is
\begin{align}
    W[T,\phi,A(R)] = T\,I_0[T,\phi,R]\,,
    \label{eqch4:grandpot}
\end{align}
where $A(R) = \Omega_{d-2}R^{d-2}$,
or written explicitly
\begin{align}
    W=& \frac{
    R^{d-3}}{\mu}
    \Bigl(1-\sqrt{f\left( R,T,\phi\right)}\Bigr) -
    T\frac{\Omega_{d-2} r_+^{d-2}(R,T,\phi)}{4l_p^{d-2}}-
    q(R,T,\phi) \phi \,,
    \label{eqch4:grandpotentialactionI0full}
\end{align}
where $f(R,T,\phi)$ is given by Eq.~\eqref{eqch4:alphavalue} and the 
solutions satisfy Eqs.~\eqref{eqch4:betaexpress} 
and~\eqref{eqch4:phiexpress}.

We must comment about the connection of the reduced action
to the grand potential in the zero loop approximation. The reduced action 
can be seen as a generalized grand potential where the relation between 
the first derivatives of the mean energy and the quantities fixed at the ensemble 
is relaxed. In fact, the minimization of the reduced action leads to the 
identification of the first derivatives of the mean energy to be the 
temperature $T$ fixed at the cavity and the electric potential $\phi$
fixed at the cavity. 

The grand canonical potential $W$, which is the potential 
directly related to the partition function of the grand canonical 
ensemble, is defined by the Legendre transformation of 
the mean energy $E$ as
\begin{align}
    W = E - TS-Q \phi\,,
\label{eqch4:grandpotthermodef}
\end{align}
where $E(S,Q,A)$. The thermodynamic quantities 
which are the entropy, the mean charge, and the thermodynamic 
pressure can be obtained by evaluating the derivatives 
of the grand potential, and, from Eq.~\eqref{eqch4:grandpotthermodef}, 
one can also extract the mean energy. The differential 
of the grand potential $W=W[T,\phi,A(R)]$ can be written as 
\begin{align}
    dW = -SdT -pdA-Qd\phi\,,
\label{eqch4:grandpotdiffform}
\end{align}
i.e. the first derivatives of the grand potential are 
$S=-\left(\frac{\partial W}{\partial T}\right)_{A,\phi}$,
$p\equiv-\left(\frac{\partial E}{\partial A}\right)_{S,Q}=
-\left(\frac{\partial W}{\partial A}\right)_{T,\phi}$,
and $Q=-\left(\frac{\partial W}{\partial \phi}\right)_{A,T}$. 
The subscript in this section means that the derivative 
is taken with the variables in subscript kept constant.

From Eq.~\eqref{eqch4:grandpotentialactionI0full}, the entropy 
can be computed through 
$S=-\left(\frac{\partial W}{\partial T}\right)_{A,\phi}$. 
It is useful to consider the grand potential with the 
dependence of the reduced action evaluated at the stationary 
points, meaning $W=W(T,\phi,A,r_+(T,\phi,R), q(T,\phi,R))$. 
Using the chain rule, one has 
\begin{align}
    S=\hskip-1mm - \hskip-1mm
\left(\frac{\partial W}{\partial T}\right)_{A,\phi}=
\hskip-1mm - \hskip-1mm
\left(\frac{\partial W}{\partial T}
\right)_{r_+,q,A,\phi}
\hskip-1mm - \hskip-1mm \left(\frac{\partial W}{\partial r_+}
\right)_{q,T,A,\phi}
\left(\frac{\partial r_+}{\partial T}
\right)
\hskip-1mm - \hskip-1mm
\left(\frac{\partial W}{\partial q}
\right)_{r_+,T,A,\phi}
\left(\frac{\partial q}{\partial T}
\right)\,.
\end{align}
Now, using the fact that the derivatives must be evaluated 
at the solutions $r_+(T,\phi,R)$ and
$q(T,\phi,R)$, one has that the partial derivatives 
with respect to $r_+$ and $q$ vanish. Hence, one only has to 
evaluate the partial derivative $\left(\frac{\partial W}{\partial T}
\right)_{r_+,q,A,\phi}$ to give
\begin{align}
    S = \frac{A_+}{4l_p^{d-2}}\,,
    \label{eqch4:entropyrp}
\end{align}
with $A_+$ is the area of the horizon given by
$A_+=\Omega_{d-2} r_+^{d-2}$. The entropy of the system 
is then the Bekenstein-Hawking entropy of the electrically charged
black hole.

In the
same way, one can calculate the electric charge
$Q$ to give the expression
$Q = - \left(\frac{\partial W}{\partial \phi}\right)_{T,A}
= - \left(\frac{\partial W}{\partial \phi}\right)_{r_+,q,T,A} 
$, yielding
\begin{align}
Q =  q\,,
\label{eqch4:meanQ}
\end{align}
so the thermodynamic value of the electric charge $Q$
is equal to the typical electric
charge $q$ of an electrically charged black hole.

The thermodynamic pressure is given by $p= - 
\left(\frac{\partial W}{\partial A}\right)_{T,\phi}$,
and so we obtain
\begin{align}
p= \frac{d-3}{16\pi l_p^{d-2} R \sqrt{f}}\left(\left(1 -
\sqrt{f}\right)^2 - \frac{\lambda q^2}{R^{2d-6}}\right)\,,
\label{eqch4:meanpressure}
\end{align}
which is the gravitational tangential pressure at
the heat reservoir of radius $R$.

The remaining quantity to be calculated is the mean energy, which 
can be achieved by
putting Eqs.~\eqref{eqch4:entropyrp}-\eqref{eqch4:meanpressure} into
Eq.~\eqref{eqch4:grandpotthermodef}. The mean energy is
\begin{align}
E = \frac{(d-2)\Omega_{d-2} R^{d-3}}{8\pi l_p^{d-2}} \left(1 -
\sqrt{f} \right)\,,
\label{eqch4:Energy}
\end{align}
which is the quasilocal energy evaluated at 
radius $R$.

One can  verify from the previous equations that 
the first law of thermodynamics is satisfied, i.e.
\begin{align}
TdS=   dE  +pdA - \phi dQ\,.
\label{1stlawusual}
\end{align}
In this spirit, one can rewrite the expression of the 
energy in function of $S$, $A$ and $Q$ as
\begin{align}
    E =&
    \frac{(d-2)A^{\frac{d-3}{d-2}}
    \Omega^{\frac{1}{d-2}}_{d-2}}{8\pi l_p^{d-2}}
    \left(1-\sqrt{\left(1-\left(\frac{4
    S}{A}\right)^{\frac{d-3}{d-2}}\right)\left(1-\frac{\lambda Q^2
    \Omega^{2\frac{d-3}{d-2}}_{d-2}}{(4SA)^{\frac{d-3}{d-2}}l_p^{d-3}}\right)}
    \right)\,.\label{eqch4:energyintermsof}
\end{align}
Using the Euler's homogeneous function theorem and the 
rescaling property $E\left(\nu S,\nu A,\nu Q^{\frac{d-2}{d-3}}\right) =
\nu^{\frac{d-3}{d-2}} E\left(S,A, Q^{\frac{d-2}{d-3}}\right)$ with 
$\nu$ being a constant, one can find an integrated version of the 
first law given by
\begin{align}
    E = \frac{d-2}{d-3}(TS - pA) + \phi
    Q\,,\label{eqch4:1stlawintegrated}
\end{align}
which is also called the Euler equation, in this case 
for system of a
$d$-dimensional electrically charged black holes in a
heat reservoir. By differentiating Eq.~\eqref{eqch4:1stlawintegrated} 
and considering
that $dE = TdS - pdA + \phi dQ$, one obtains a modified version 
of the Gibbs-Duhem relation 
\begin{align}
  TdS - pdA + (d-2)(SdT-Adp) + (d-3) Qd\phi = 0\,.
\label{eq:gibbs}
\end{align}
Note that this relation depends on the differential 
of $T$ and $S$, as well as $p$ and $A$ simultaneously. This is 
an indication of the lack of homogeneity of degree one of the system.

It is also interesting to consider the limit of infinite 
radius of the cavity. In fact, the integrated first law 
in Eq.~\eqref{eqch4:1stlawintegrated} becomes the Smarr 
formula for the charged black hole. To see this, one must 
consider the limit of infinite radius for the thermodynamic 
quantities. The temperature in Eq.~\eqref{eqch4:betaexpress} 
reduces for the infinite cavity to the Hawking temperature 
$T=T_{\mathrm{H}}=
\frac{d-3}{4\pi}\left(\frac{1}{r_+} 
- \frac{\lambda q^2}{r_+^{2d-5}} \right)$, while 
the electric potential in Eq.~\eqref{eqch4:phiexpress} reduces to the
electric potential of the Reissner-Nordstr\"om black hole 
$\phi = \phi_{\mathrm{H}} = \frac{q}{(d-3)\Omega_{d-2} r_+^{d-3}}$.
The quantity $p A$ with $p$ in
Eq.~\eqref{eqch4:meanpressure} being proportional to $\frac{1}{R^{d-3}}$
vanishes in the limit of infinite cavity. The mean energy reduces to 
the ADM mass $E=m$ determined by $m = \frac{1}{2\mu}\left(r_+^{d-3} 
+ \frac{\lambda q^2}{r_+^{d-2}}\right)$. With these ingredients, the 
integrated first law becomes 
\begin{align}
    m = \frac{d-2}{d-3} T_{\mathrm{H}}S + \phi_{\mathrm{H}}
    Q \,,\label{eqch4:smarr}
\end{align}
which is the Smarr formula. However, the Smarr formula is only valid 
for the solution that exists in the limit of infinite cavity, i.e. 
$r_{+1}$, where the zero loop approximation is not valid.

\subsection{Thermodynamic stability}
\label{secch4:Equilibria}

We now analyze the thermodynamic stability of the system in the zero loop 
approximation. We must note that thermodynamic 
stability in general has different connections to the 
stability of the stationary points of the reduced action. 
In this case, we show that they coincide. 

To understand the thermodynamic 
stability of an ensemble, we must consider the following.
In a thermodynamic system with fixed size, fixed temperature, and fixed
electric potential at a heat reservoir, there can be an exchange 
of energy, entropy, and electric
charge between the heat reservoir and the system.
In any thermodynamic process within the system,
the grand canonical potential $W$ tends to
decrease down to its minimum or stay at its minimum.  In particular,
a spontaneous process in the
grand canonical ensemble can never increase the grand canonical
potential $W$, otherwise it violates the second law of thermodynamics.

To see this, we must resort to the second law
of thermodynamics applied to the total structure.
A variation $dS$ in entropy
of the system plus a variation $dS_{\mathrm{reservoir}}$
in entropy of the reservoir add up to
a variation $dS_{\mathrm{total}}$ of the 
total entropy of the system plus reservoir, 
as $dS_{\mathrm{total}}=dS+dS_{\mathrm{reservoir}}$.
Now consider a perturbation in which the thermodynamic
system absorbs energy $dE$
and charge $dQ$ from the reservoir.
By conservation of energy and charge, the reservoir
has to absorb energy $-dE$ and charge $-dQ$.
The first law of thermodynamics states
that the change in entropy of the reservoir is
$TdS_{\mathrm{reservoir}}=
-dE+\phi dQ$, where
the reservoir's temperature and electric
potential are kept constant due to the quality of the reservoir.
The total change in entropy can be written then as 
$TdS_{\mathrm{total}}=TdS-dE+\phi dQ=-d(E-TS-\phi Q)=
-d{\bar W}$, where $T$ and $\phi$ are
constant since they are the reservoir values.
The potential ${\bar W}$ has been defined as
\begin{align}
{\bar W}[{\bar T},A,{\bar \phi}]\equiv E({\bar T},A,{\bar \phi})
- T S({\bar T},A,{\bar \phi}) - \phi Q({\bar T},A,{\bar \phi})\,,
\label{eqch4:Wperturb}
\end{align}
as the grand canonical potential related
to the nonequilibrium situation.
Due to the variation towards a 
nonequilibrium situation, the thermodynamic
system attains in general 
a new temperature
$\bar T$ and a new potential $\bar \phi$
different from $T$ and $\phi$ of the reservoir.
The energy $E$, the entropy $S$ and 
the charge $Q$ that arise in the variation
of the nonequilibrium situation have
the same functional form of ${\bar T}$, $A$, and 
${\bar \phi}$, as they had of $T$, $A$, and $\phi$
before the nonequilibrium process set in,
but ${\bar W}[{\bar T},A,{\bar \phi}]$
has a different functional dependence 
than the typical grand potential,
since $T$ and $\phi$
that appear in Eq.~\eqref{eqch4:Wperturb}
are quantities of the heat reservoir
fixed by assumption.
The area $A$ has been kept fixed in the system and 
reservoir. 
In brief, one has $TdS_{\mathrm{total}}=
-d{\bar W}$. Since
one must have $dS_{\mathrm{total}}\geq0$ by the second law of thermodynamics, 
one deduces that $d{\bar W}\leq0$. Any spontaneous process decreases
the grand canonical potential. For
a review of this discussion, see
Sec.~8.2 and the following sections of \cite{Reif:1965}.

The equilibrium situation is reached when
\begin{align}
{\bar T}=T\,,\quad\quad {\bar \phi}=\phi\,,
\label{eqch4:WS0equilibrium}
\end{align}
in which case ${\bar W}$ must reach a minimum. 
Therefore, to be stable, 
the hessian of the potential ${\bar W}$ must be positive 
definite, which can be summarized into the conditions
\begin{align}
\left(\frac{\partial^2 {\bar W}}{\partial {\bar T}^2}
\right)_{{\bar \phi},A} > 0\,,
\label{eqch4:WS2}
\end{align}
\begin{align}
\left(\frac{\partial^2 {\bar W}}{\partial {\bar T}^2}\right)_{{\bar \phi},A} 
\left(\frac{\partial^2 {\bar W}}{\partial {\bar \phi}^2}\right)_{{\bar T},A} 
-
\left(\frac{\partial^2 {\bar W}}{\partial {\bar T} \partial {\bar \phi}}
\right)_{A} ^2
> 0\,,\label{eqch4:WSQ}
\end{align}
\begin{align}
\left(\frac{\partial^2 {\bar W}}{\partial {\bar \phi}^2}
\right)_{{\bar T},A} > 0
\,,
\label{eqch4:WQ2}
\end{align}
where all the derivatives are to be
calculated at the solutions of the ensemble.
Only two conditions from
Eqs.~\eqref{eqch4:WS2}-\eqref{eqch4:WQ2} are sufficient, and so 
we choose
Eqs.~\eqref{eqch4:WS2} and~\eqref{eqch4:WSQ}.
From the expression of ${\bar W}$,
the second derivative in order to $T$ is
$\left(\frac{\partial^2 {\bar W}}{\partial {\bar T^2}}\right)_{A,
{\bar \phi}} =
\left(\frac{\partial  S}{\partial  T}\right)_{A,\phi}$,
where the bars have been dropped
on the right-hand side of the equality because $S$ has
the same functional form of ${\bar T}$, $A$, and 
${\bar \phi}$, as it has of $T$, $A$, and $\phi$,
and at equilibrium ${\bar T}=T$.
In the same way, one has 
$\left(\frac{\partial^2 {\bar W}}{\partial {\bar \phi}^2}
\right)_{{\bar T},A} =\left(\frac{\partial Q}{\partial
\phi}\right)_{T,A}$,
and
$\left(\frac{\partial^2 {\bar W}}{\partial {\bar T}
\partial {\bar \phi}}\right)_{{\bar T};{\bar \phi},A}
=\left(\frac{\partial Q}{\partial T}\right)_{A,\phi}=
\left(\frac{\partial S}{\partial \phi}\right)_{T,A}$.
The two
sufficient conditions,
Eqs.~\eqref{eqch4:WS2} and~\eqref{eqch4:WSQ},
can be written in terms of first derivatives of the 
entropy and charge as
\begin{align}
& \left(\frac{\partial S}{\partial T}\right)_{A,\phi} > 0\,,
\label{eqch4:WS2better}\\
& \left(\frac{\partial Q}{\partial\phi} \right)_{T,A}
\left(\frac{\partial S}{\partial T} \right)_{A,\phi} 
- \left(\frac{\partial S}{\partial \phi}\right)^2_{T,A} > 0\,
\label{eqch4:WSQbetter}
\end{align}
respectively.

We can now link the thermodynamic stability 
to thermodynamic coefficients. 
First, we can define the isochoric heat capacity 
at constant electric potential as
\begin{align}
C_{A,\phi} = 
T\left(\frac{\partial S}{\partial T}\right)_{A,\phi}\,.
\label{eqch4:CAphi}
\end{align}
Second, we can define
the adiabatic electric susceptibility as 
\begin{align}
    \chi_{S,A}
= \left(\frac{\partial
Q}{\partial \phi}\right)_{S,A}\,\,.
\end{align}
From a change of variables 
$Q(T,A,\phi)$ to $Q(T(S,A,\phi), A,\phi)$, where $T(S,A,\phi)$ is 
the inverse function of $S(T,A,\phi)$, one gets
\begin{align}
\chi_{S,A}=
\frac{\left(\frac{\partial Q}{\partial\phi} \right)_{T,A} 
\left(\frac{\partial S}{\partial T}\right)_{A,\phi}
- \left(\frac{\partial S}{\partial \phi}\right)^2_{T,A} }
{\left(\frac{\partial S}{\partial T}\right)_{A,\phi}}
\,,
\label{eqch4:chisa}
\end{align}
Hence, the two stability conditions,
Eqs.~\eqref{eqch4:WS2better} and~\eqref{eqch4:WSQbetter}, are now
\begin{align}
C_{A,\phi}> 0
\,,\label{eqch4:cphia}
\end{align}
\begin{align}
\chi_{S,A} C_{A,\phi}>0\,,\label{eqch4:cond2}
\end{align}
respectively.
The above analysis to obtain the stability conditions is equivalent 
to the requirement that the matrix of variances in the grand 
canonical ensemble is positive definite. The matrix of variances 
contains the variances $\Delta E^2$, $\Delta Q^2$ and the correlation 
$\Delta E \Delta Q$, where $E$ and $Q$ are the quantities 
that are exchanged with the heat reservoir. 
By working out the conditions of positive
definiteness through the Sylvester's criterion, 
one recovers also the conditions Eqs.~\eqref{eqch4:cphia} and~\eqref{eqch4:cond2}.

For the specific case of the electrically charged black hole
in a cavity, the susceptibility is
\begin{align}
    \chi_{S,A} = \frac{(d-3)\Omega_{d-2} r_+^{d-3}
  (1 - \frac{r_+^{d-3}}{R^{d-3}})}
  {(1 - (1 -
  \Phi^2)(\frac{r_+}{R})^{d-3})^{\frac{3}{2}}}\,\,.
\end{align}
The adiabatic susceptibility is then positive 
for all physical 
configurations of the charged black hole. 
Therefore, the two conditions for stability
can be reduced to a single one
given in Eq.~\eqref{eqch4:cphia},
$C_{A,\phi}>0$, which in terms of the 
ensemble quantities is
\begin{align}
  C_{A,\phi} = \frac{A (d-3)^2(d-2) x^{d-4} (1 -
  \Phi^2)^2}{32 l_p^{d-2} (\pi R T)^2 ((d-1)(1 - \Phi^2)x^{d-3} - 2)}> 0\,\,,
  \label{eqch4:heatcapacityPhibetter}
\end{align}
with the 
dependence on the variable $x = \frac{r_+}{R}$ 
being maintained for convenience. With
Eq.~\eqref{eqch4:heatcapacityPhibetter},
one recovers
Eq.~\eqref{eqch4:stabilitysimple} for thermodynamic stability.
This means that the stability of the stationary points, 
that point towards the validity of the zero loop approximation, 
coincides with thermodynamic stability, in this case. 
For the 
case of $\Phi^2 = 0$, $C_{A,\phi}$ becomes the heat capacity 
at constant area $C_{A}$ with the expression given in
\cite{Andre:2021ctu}. Moreover,  
the bifurcation radius, indicating marginal stability,
and the photon sphere radius are the same for $\Phi^2 = 0$.
We make a comparison between the bifurcation
radius and the photon sphere radius in Sec.~\ref{sech4:radii},
showing that these radii do not coincide. The connection 
displayed in the
uncharged case
does not seem to be generic, it seems to hold only in the pure
gravitational situation.

It is worth making a comparison of the thermodynamic analysis that we have done 
here with the case of a self-gravitating static electrically charged
thin shell in $d$-dimensions presented in Chapter~\ref{ch:chargedselfgravitating}, 
or in \cite{Fernandes:2022gjd}. 
It is quite remarkable that the thermodynamic pressure given
in Eq.~\eqref{eqch4:meanpressure} and the thermodynamic energy given in
Eq.~\eqref{eqch4:Energy} in the grand canonical ensemble have the same
expression as the matter pressure and the matter rest energy 
of the corresponding self-gravitating charged spherical shell
in equilibrium. Additionally, by choosing for the matter of the thin
shell the equations of state corresponding to the temperature and
electric potential of the black hole, the shell also has the
Bekenstein-Hawking entropy and its stability at constant area is given
by the same condition, i.e., positive heat capacity at constant electric
potential.

\subsection{Thermodynamic phases and
phase transitions}


In a thermodynamic system characterized by the grand canonical
ensemble, spontaneous processes always occur
in order to decrease $W$ to its lowest value. As shown above, this is 
a consequence of the second law of thermodynamics. 
The configuration we are studying here is
a black hole inside a reservoir characterized by a fixed area $A$, a
fixed temperature $T$, and a fixed electric potential $\phi$. So
thermodynamically, $W$ is the most suited thermodynamic potential 
to be analyzed.  
It is relevant to know
whether the stable black hole is the thermodynamic state with less
energy $W$, or if there is another state to which the black hole can make
a phase transition. Indeed, the stable black hole is a local minimum 
but may not be the global minimum of the potential.
Here, we can use the thermodynamic
language now, and so we can analyze phase transitions instead of 
quantum transitions as done previously in the analyses of probable configurations. 
But the results are the same, as here we use
$W$ instead of $I_0$, with the connection $TI_0=W$. 
We summarize the results using the grand canonical potential $W$.

In the uncharged case, one has $W_{\mathrm{hot}\,\mathrm{flat}\,\mathrm{space}} = 0$ 
and so the favorability of the black hole depends on whether
the black hole with horizon radius $r_{+2}$ has
a $W$ lower or greater than zero.
We found that the radius where the stable black hole with 
$r_{+2}$ yields $W=0$ when the cavity is at the Buchdahl radius, $r_{\mathrm{Buch}}$,
where the first order phase transition from hot flat space to black hole phase 
occurs.
For the ratio $\frac{r_{+2}}{R}$ higher than $\frac{r_{+2}}{r_{\mathrm{Buch}}}$, 
the black hole phase is favored. 
In the electrically charged case, the grand potential for 
hot flat space is $W_{\mathrm{hot}\,\mathrm{flat}\,\mathrm{space}}
= 0$, and corresponds to a cavity without a black hole and
without charge. We emulated hot flat space by a
very small electric hot sphere in flat space. Its grand potential 
is $W_{\mathrm{hot}\,\mathrm{sphere}} = T I_{\mathrm{hot}\,\mathrm{sphere}}$,
which tends to zero as the radius of the sphere tends to zero.
Essentially, in this setting,
the black hole phase is favored when its grand potential $W$ is less than zero.
We established that the ratio $\frac{r_{+2}}{R}$
which yields $W=0$ is not related to the 
Buchdahl-Andr\'easson-Wright ratio, a generalization for the
Buchdahl ratio to any higher dimension $d$
that includes electric charge, see
Sec.~\ref{sech4:radii}.
There is also a critical phase, the extreme black hole solution localized at the radius of the
cavity. We found that the stable black hole $r_{+2}$ has
always lower or equal $W$ than 
$W_{\mathrm{
extreme}\,\mathrm{black}\,\mathrm{hole}}$, and hence the stable black hole is always
more favorable than the extremal black hole with horizon at the
cavity.

\section{Zero loop approximation and thermodynamics 
for $d=5$\label{sech4:zeroloop5d}}

\subsection{Zero loop approximation}

\subsubsection{Reduced action, stationary solutions and stability conditions}

Here, we apply the whole formalism to the specific five dimensional case, $d=5$, 
since we want to stress some analytical results pertaining this case. 
The $d=4$ case
recovers the analysis of
\cite{Braden:1990hw}, see~\cite{Fernandes:2023byx}. In $d=5$, 
the reduced action in Eq.~\eqref{eqch4:reducedaction} can be rewritten as 
\begin{align}
I_* = \frac{3\pi}{4l_p^3}\beta R^{2}\left(1-\sqrt{f}\right) -
q \beta \phi - \frac{\pi^2 r_+^{3}}{2l_p^{3}}\,,
\label{eqch4:reducedaction5d}
\end{align}
where 
\begin{align}
f =
\left(1-\frac{r_+^{2}}{R^{2}}\right)\left(1-
\frac{1}{3\pi^3}
\frac{
q^2}{r_+^{2} R^{2}}\right).
\label{eqch4:f5d}
\end{align}
with $I_*=I_*[\beta,\phi, R; r_+,q]$
and $f=f[R;r_+,q]$. For $d=5$,
one has $\Omega=2\pi^2$, $\mu = \frac{4 l_p^3}{3\pi}$ and 
$\lambda = 
\frac{l_p^{3}}{3\pi^3}$.

The stationary solutions for $r_+$ that minimize the reduced action in 
Eq.~\eqref{eqch4:reducedaction5d} satisfy the 
relation in Eq.~\eqref{eqch4:extremaaction}, which for $d=5$ is 
\begin{align}
    (1-\Phi^2)\left(\frac{r_+}{R}\right)^4 -
    \left(\frac{r_+}{R}\right)^2 + (1 - \Phi^2)^2\frac{1}{(2\pi
    RT)^2}=0\,,\label{eqch4:extremaaction5d}
\end{align}
where $\gamma=(2\pi RT)^2\frac{\Phi^2}{(1-\Phi^2)^2}$. 
The electric charge is given in terms of the horizon radius 
$r_+$ by Eq.~\eqref{eqch4:extrema2action}, becoming 
\begin{align}
    \frac{l_p^{\frac{3}{2}}\abs{q}}{R^2}=2\sqrt{3}\,\pi^{\frac{5}{2}}\, 
    \frac{RT\abs{\Phi}}{1-\Phi^2}\left(\frac{r_+}{R}\right)^3\,.
    \label{eqch4:extrema2action5d}
\end{align}
The stationary condition in Eq.~\eqref{eqch4:extremaaction5d} 
must be solved to obtain the stationary point $r_+=r_+(T,\Phi,R)$, 
which in turn can be plugged in Eq.~\eqref{eqch4:extrema2action5d} 
to get $q=q(T,\Phi, R)$ of the stationary point. Since 
Eq.~\eqref{eqch4:extremaaction5d} is a quadratic equation for $r_+^2$, 
analytic expressions can be obtained for the two solutions 
$r_{+1}$ and $r_{+2}$. Namely, the solution $r_{+1}$ is given 
by 
\begin{align}
    \frac{r_{+1}}{R} = \frac{1}{\sqrt{2(1-\Phi^2)}}
    \left[1 - \sqrt{1 -
    \frac{(1 - \Phi^2)^3}{(\pi R T)^2}}
    \right]^{\frac{1}{2}}\,,
    \label{eqch4:solutionr+1}
    \end{align}
    \begin{align}
    \frac{l_p^{\frac{3}{2}}\abs{q_1}}{R^2}=
    \sqrt{\frac32}\, 
    \frac{\pi^{\frac{5}{2}} RT\Phi}{(1-\Phi^2)^{\frac52}}
    \left[1 - \sqrt{1 -
    \frac{(1 - \Phi^2)^3}{(\pi R T)^2}} \right]^{\frac{3}{2}}\,.
    \label{eqch4:solutionq1}
\end{align}
The solution $r_{+1}$ for the horizon radius of the charged black hole 
was designated $x_1$ in the analysis above for generic $d$. 
The second solution $r_{+2}$, designated as $x_2$ above, is given by 
\begin{align}
    \frac{r_{+2}}{R} = \frac{1}{\sqrt{2(1-\Phi^2)}}\left[1 + \sqrt{1 -
    \frac{(1 - \Phi^2)^3}{(\pi R T)^2}} \right]^{\frac{1}{2}}\,,
    \label{eqch4:solutionr+2}
    \end{align}
    \begin{align}
    \frac{l_p^{\frac{3}{2}}\abs{q_2}}{R^2}=
    \sqrt{\frac32}\, 
    \frac{\pi^{\frac{5}{2}} RT\Phi}{(1-\Phi^2)^{\frac52}}
    \left[1 + \sqrt{1 -
    \frac{(1 - \Phi^2)^3}{(\pi R T)^2}} \right]^{\frac{3}{2}}\,.
    \label{eqch4:solutionq2}
\end{align}
The analysis of the behaviour of the solutions can be easily understood 
compared to the $d$ generic case. For the two solutions to exist, one 
requires that Eq.~\eqref{eqch4:twosolutionscond}, which here reduces to 
\begin{align}
    0\leq (1 - \Phi^2)^3\leq (\pi R T)^2 <\infty\,,
    \label{eqch4:range}
\end{align}
in $d=5$ dimensions. For the uncharged case, i.e. $\Phi=0$, 
the condition of existence of both solutions Eq.~\eqref{eqch4:range}
reduces to the interval  $1\leq (\pi R T)^2 <\infty$, being 
precisely the interval of existence for the $d=5$ Schwarzschild-Tangherlini 
black hole solutions, see~\cite{Andre:2020czm}.

Before proceeding to a careful analysis of the stationary points,
it is useful to make an analysis of the limits. First, 
for very large $\pi RT$, or $(\pi R T)^2\to\infty$,
and constant $\Phi$, 
the solution $r_{+1}$ behaves as $\frac{r_{+1}}{R} \rightarrow
\frac{(1 - \Phi^2)}{2\pi R T}$,
and, since $\abs{\Phi}< 1$, the solution always exists.
For very large $\pi RT$, or $(\pi R T)^2\to\infty$,
and constant $\Phi$, 
the solution $r_{+2}$
behaves as $\frac{r_{+2}}{R}
\rightarrow \frac{1}{\sqrt{1 - \Phi^2}}$, which for values of $\Phi^2
< 1$, one has $r_{+2}> R$, so
the solution is
unphysical. This situation is different from the uncharged case,
where the solution with larger mass, $r_{+2}$, only meets the cavity at
infinite temperature, while in the charged case, the solution $r_{+2}$
meets the cavity at finite temperature, as already seen above in qualitative 
terms. Second, for $\Phi^2 \rightarrow 1$,
and constant $(\pi R T)^2$,
the solution $r_{+1}$ tends to $r_{+1}\rightarrow 0$.
For $\Phi^2 \rightarrow 1$,
and constant $(\pi R T)^2$,
the solution $r_{+2}$ tends to
$r_{+2}\rightarrow \infty$, which is unphysical.

\begin{figure}[t]
    \centering
    \includegraphics[width= 0.45\columnwidth]{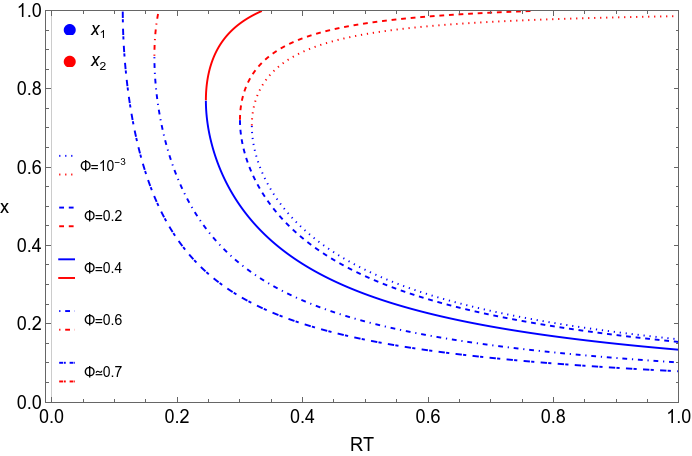}
    \includegraphics[width= 0.45\columnwidth]{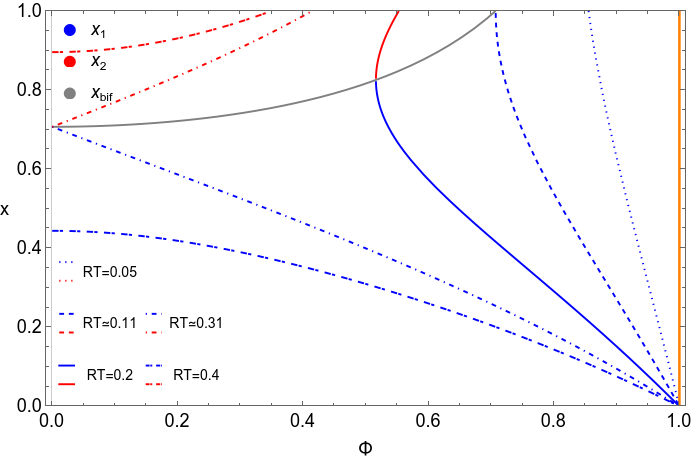}
    \caption{Left plot:
    Stationary points $\frac{r_{+1}}{R}=x_1$ (in blue) and
    $\frac{r_{+2}}{R}=x_2$ (in red) of the reduced action $I_*$ as a
    function of $RT$, for $d=5$ dimensions, and for five values of $\Phi$,
    namely, $\Phi = 0.001$ in dotted lines, $\Phi = 0.2$ in dashed lines,
    $\Phi = 0.4$ in solid lines, $\Phi = 0.6$ in dot dashed lines and
    $\Phi = \frac1{\sqrt{2}}=0.7$, the last equality
    is approximate, in dot
    double dashed lines.
    Right plot: Stationary points $\frac{r_{+1}}{R}=x_1$ (in blue) and
    $\frac{r_{+2}}{R}=x_2$ (in red) of the reduced action $I_*$ as a
    function of $\Phi$, for $d=5$ dimensions, and for five values of $RT$,
    namely, $RT = 0.05$ in dotted lines, $RT = \frac{1}{2\sqrt{2}\pi} =
    0.112$, the last equality is approximate, in dashed lines, $RT = 0.2$
    in solid lines, $RT = \frac1\pi= 0.318$, the last equality is
    approximate, in dot dashed lines, and $RT=0.4$ in dot double dashed
    lines. The gray line corresponds to the bifurcation points where
    the solutions $x_1$ and $x_2$ coincide.  The orange line corresponds
    to $\Phi = 1$, which is the maximum possible electric potential of the 
    ensemble.}
    \label{figch4:2solsd=5phi5piRT}
\end{figure}

\begin{figure}[t]
    \centering
    \includegraphics[width=0.45\columnwidth]{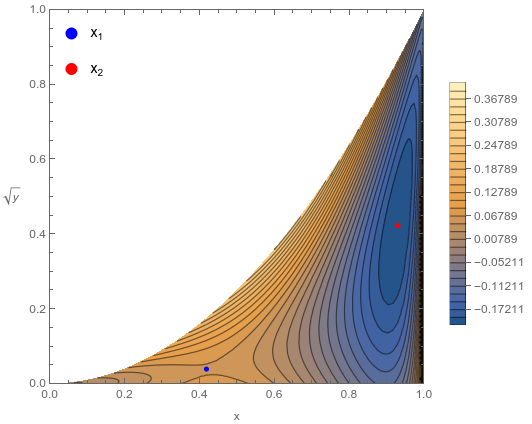}
    \caption{Contour plot of the reduced action $\frac{4 l_p^3 I_*}{3\pi R^{2}}$ in $d=5$ dimensions,
    in function of
    $\frac{r_{+}}{R}=x$ and $\frac{l_p^{\frac{3}{2}}\abs{q}}{\sqrt{3\pi^3}R^2}=\sqrt{y}$,
    for $\Phi = 0.2$ and $RT = 0.4$. The blue dot
    corresponds to $\frac{r_{+1}}{R}=x_1$ and it is a saddle point, 
    while the red dot corresponds to $\frac{r_{+2}}{R}=x_2$ and it is a minimum.}
    \label{figch4:ContourPlotRedact}
\end{figure}

\begin{figure}[t]
    \centering
    \includegraphics[width=0.45\columnwidth]{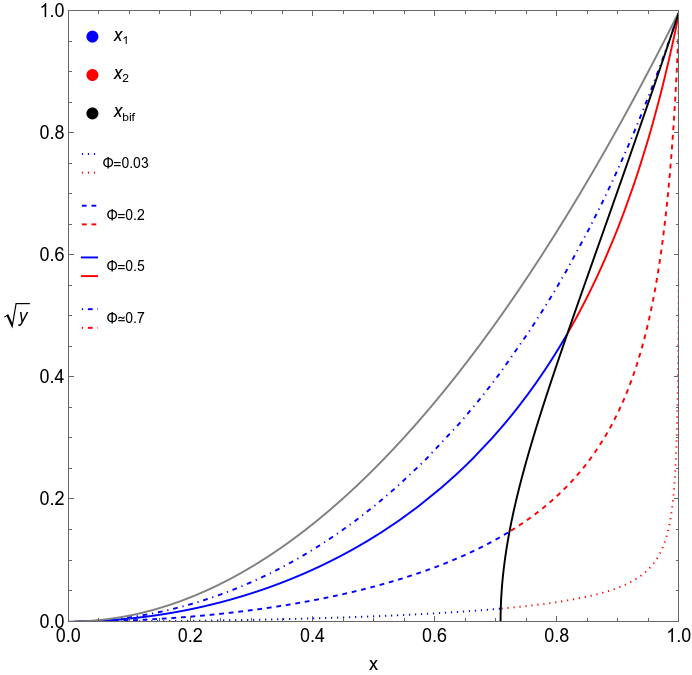}
    \includegraphics[width=0.45\columnwidth]{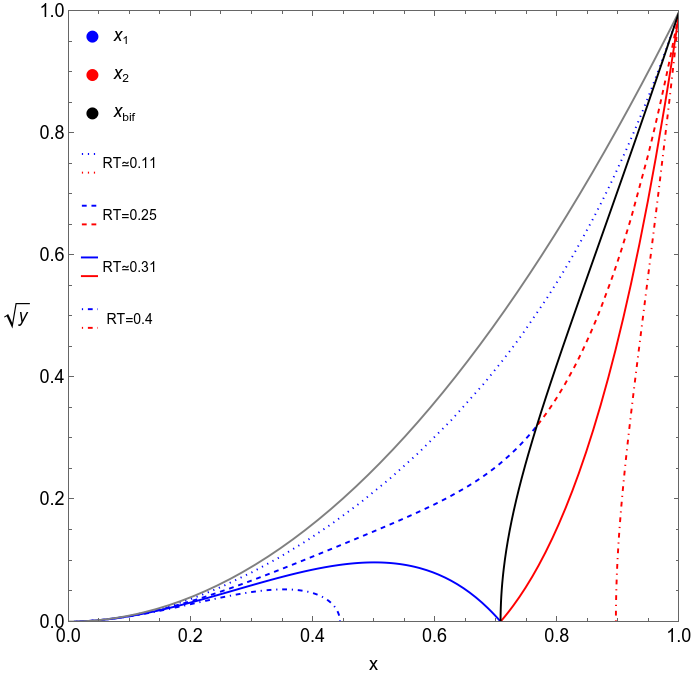}
    \caption{
    Left plot: Curves describing the path of the solutions $\frac{r_{+1}}{R}=x_1$ (in blue) and
    $\frac{r_{+2}}{R}=x_2$ (in red) in the $x\times \sqrt{y}$ plane, for
    $d=5$, with $x$ and $\sqrt{y}$ parametrized by $RT$, for $\Phi= 0.03$
    in dotted lines, $\Phi = 0.2$ in dashed lines, $\Phi = 0.5$ in solid
    lines and $\Phi = \frac1{\sqrt{2}}= 0.707$, the last equality being
    approximate, in dot dashed lines. The gray line corresponds to
    the class of extremal black holes inside the cavity, i.e., $\sqrt{y} = x^2$. 
    The black line corresponds to the bifurcation points $x$ and $\sqrt{y}$, where $x_1$ and $x_2$
    coincide.  Right plot: Curves describing the path of solutions $x_1$ (in blue) and $x_2$
    (in red) in the $x\times\sqrt{y}$ plane, for $d=5$, with $x$ and
    $\sqrt{y}$ parametrized by $\Phi$, for
    $RT=\frac{1}{2\sqrt{2}\pi}=0.11$, the last equality being approximate,
    in dotted line, $RT = 0.25$ in dashed lines, $RT = \frac1\pi=0.3$, the
    last equality being approximate, in solid lines, and $RT = 0.4$ in dot
    dashed lines.  The gray line corresponds to the condition of extremal
    black holes inside the cavity, i.e., $\sqrt{y} = x^2$. The black line corresponds to
    the bifurcation points $x$ and $\sqrt{y}$ where $x_1$ and $x_2$ coincide.
    }
    \label{figch4:ContourPlotRedactPathRTPhi}
\end{figure}

We now give a careful analysis of the stationary points described by the solution
$r_{+1}$ of Eqs.~\eqref{eqch4:solutionr+1}-\eqref{eqch4:solutionq1} and by the
solution $r_{+2}$ of Eqs.~\eqref{eqch4:solutionr+2}-\eqref{eqch4:solutionq2}. 
We show the plots of the solutions in
Figs.~\ref{figch4:2solsd=5phi5piRT}$-$\ref{figch4:ContourPlotRedactPathRTPhi},
which complement the behaviour expressed in Eqs.~\eqref{eqch4:solutionr+1}
--\eqref{eqch4:range}. In $d=5$, one has that
the value $\Phi^2=\frac12$ plays an important role in the
analysis. Thus, the analysis is divided into two parts,
namely, $0\leq\Phi^2 \leq \frac{1}{2}$
and $\frac{1}{2}<\Phi^2 <1$.

(i) For $0\leq\Phi^2 \leq \frac{1}{2}$, 
there are three different branches. 

\noindent
(a) for $0\leq (\pi R T)^2 <(1-\Phi^2)^3$, 
there are no stationary points or solutions for the charged black hole. 

\noindent
(b) For $(1-\Phi^2)^3 \leq (\pi R T)^2 \leq \frac{(1 - \Phi^2)^2}{4\Phi^2}$,
there are two black hole solutions and they lie inside
the cavity, i.e., $r_{+1}\leq R$ and $r_{+2}\leq R$.
In the case of the equality
$(\pi R T)^2 = \frac{(1 - \Phi^2)^2}{4\Phi^2}$, the
solution $r_{+1}$ obeys $r_{+1}<R$, and the solution $r_{+2}$
satisfies $r_{+2}=R$ with the charge $q_2$ obeying $l_p^{\frac{3}{2}}\abs{q_2}=
\sqrt{3 \pi^3}\,r_+^2$, which means that the $r_{+2}$
solution is an extremal electrically charged black hole.
The particular case $\Phi^2 = \frac{1}{2}$ and 
$(\pi R T)^2 = \frac{(1 - \Phi^2)^2}{4\Phi^2}$
yields that $(1-\Phi^2)^3=\frac{(1 -
\Phi^2)^2}{4\Phi^2}= (\pi R T)^2 = \frac18$, and 
the $r_{+1}$ and $r_{+2}$ solutions merge into one,
an extremal electrically charged black hole that
obeys $r_{+1}=r_{+2} = R$.

\noindent
(c) For $\frac{(1 - \Phi^2)^2}{4\Phi^2}<
(\pi R T)^2 <\infty$, 
the solution $r_{+1}$ has always $r_+<R$ and so it is physical. 
For $\Phi$ near zero,
$r_{+1}$ is small and as the
value of $(\pi R T)^2$
increases, $r_{+1}$ tends to zero.
For $\Phi$ near $\frac12$ approaching from below, 
$r_{+1}$ approaches $R$ from below
and as $(\pi R T)^2$
increases, $r_{+1}$ tends to zero.
Regarding the other solution $r_{+2}$,
it obeys $r_{+2} > R$, so it is unphysical.

\vskip 0.3cm
\noindent
(ii) For $\frac{1}{2}< \Phi^2 < 1$, there are three different branches.

\noindent
(a) For $0\leq (\pi R T)^2 <(1-\Phi^2)^3$, there are no black hole
solutions.

\noindent
(b)
For $(1-\Phi^2)^3 \leq (\pi R T)^2 \leq \frac{(1 -
\Phi^2)^2}{4\Phi^2}$,
 both solutions $r_{+1}$ and $r_{+2}$ exist but
lie outside the cavity and so they are unphysical.
This means that within this range there are no physical black hole
solutions.

\noindent
(c) For $\frac{(1 - \Phi^2)^2}{4\Phi^2}< (\pi R T)^2 <\infty$, the
solution $r_{+1}$ starts at $r_{+1} = R$ in the case of $(\pi R T)^2 =
\frac{(1 - \Phi^2)^2}{4\Phi^2}$ and then decreases toward zero as the
temperature increases.  On the other hand, the solution $r_{+2}$
remains outside the cavity, being thus unphysical.

In Fig.~\ref{figch4:ContourPlotRedact}, we present a contour plot of the reduced
action $I_*$,
for $RT = 0.5$ and $\Phi = 0.2$, as a function
of $\frac{r_{+}}{R}=x$ and
$\frac{l_p^{\frac{3}{2}}\abs{q}}{\sqrt{3\pi^3}R^2}=\sqrt{y}$. The 
reduced action is given by Eq.~\eqref{eqch4:reducedaction} with $d=5$.
The two stationary
points $\frac{r_{+1}}{R}=x_1$ and $\frac{r_{+2}}{R}=x_2$ are displayed 
as a blue dot and a red dot, respectively.
The contour plot allows for a visual identification of the nature of the 
stationary points, 
with $r_{+1}$ being a saddle point and $r_{+2}$ being a minimum.

It is interesting to see the effects of changing $T$ and $\Phi$ to the contour plot, 
namely to see the trajectories of the solutions $x_1$ and $x_2$.
In the left plot of Fig.~\ref{figch4:ContourPlotRedactPathRTPhi}, the migration path
of the two stationary points $\frac{r_{+1}}{R}=x_1$ and
$\frac{r_{+2}}{R}=x_2$ from a point in the central region where they
coincide (a bifurcation point) to the two points at the corners is shown as a function of
$RT$ for four different values of $\Phi$. The
gray line corresponds to the condition of extremal black holes inside a cavity,
namely,
$\sqrt{y} = x^2$, i.e., $\frac{l_p^\frac{3}{2}|q|}{\sqrt{3\pi^3}} = r_+^2$.  
The black line corresponds to the points $x$ and $\sqrt{y}$ where the solutions
$x_1$ and $x_2$ coincide, i.e. the class of bifurcation points. 
For the minimum possible temperature in each
case, the solutions start at the black line, and as one increases the
temperature, $x_1$ decreases toward the origin $x=\sqrt{y}=0$, where
$RT \rightarrow +\infty$, and $x_2$ increases toward $x=\sqrt{y}=1$,
where $RT \rightarrow \frac{(1-\Phi^2)^2}{4\Phi^2}$.
In the right plot of Fig.~\ref{figch4:ContourPlotRedactPathRTPhi}, the migration
path of the two stationary points $\frac{r_{+1}}{R}=x_1$ and
$\frac{r_{+2}}{R}=x_2$ from a point in the central region where they
coincide to the two points at the corners is shown as a function of
$\Phi$ for four different values of $RT$. In these plots, the
quantity $\Phi$ was chosen instead of $\phi$ so that the comparison
between the analytical study and the plots is straightforward, and also 
to avoid setting Planck units. Since $\Phi = \sqrt{\frac{16\pi}{3}}l_p^{\frac{3}{2}} \phi$, 
one has that $\Phi$ is fixed as a consequence of fixing $\phi$.
The gray line corresponds to the condition of
extremal black holes inside a cavity, namely $\sqrt{y} = x^2$, i.e.,
$\frac{l_p^\frac{3}{2}|q|}{\sqrt{3\pi^3}} = r_+^2$. 
The black line corresponds to the points $x$ and $\sqrt{y}$ where solutions $x_1$ and $x_2$
coincide, i.e. the bifurcation points. 
For minimum potential, the solutions either start from the
black line where the solutions coincide at the bifurcation points or start separated in
the $\sqrt{y}=0$ line. As one increases further the potential, $x_1$ tends to the
origin $x=\sqrt{y}=0$, where $\Phi \rightarrow 1$, and $x_2$ tends to
$x=\sqrt{y}=1$, where $\Phi \rightarrow \sqrt{(\pi RT)^2 +1}-\pi RT$.

Regarding stability, using Eq.~\eqref{eqch4:stabilitysimple}
with $x\equiv\frac{r_+}{R}$, one finds that
the solutions are stable if they obey
\begin{align}
\frac{\left(4(1 - \Phi^2)\left(\frac{r_+}{R}\right)^{2} - 2\right)
\left(1 - \left(\frac{r_+}{R}\right)^2\right)}
{(1 - (1 - \Phi^2) \left(\frac{r_+}{R}\right)^2)}>0\,\,,
\label{eqch4:stabilitysimple5d}
\end{align}
for $d=5$, where the physical range is $\frac{r_+}{R}<1$.
Hence, the solutions are stable if 
$r_+ >  r_{+ \mathrm{bif}}$,
where $r_{+\mathrm{bif}}
=\frac{R}{\sqrt{2(1-\Phi^2)}}$ is the
bifurcation radius from which the solutions $r_{+2}$ and
$r_{+1}$ bifurcate at $(\pi RT)^2 = (1 - \Phi^2)^3$. 
For $r_{+1}$, this condition means that for $(1-\Phi^2)^3 \leq (\pi R T)^2 \leq
\frac{(1 - \Phi^2)^2}{4\Phi^2}$, in the case $0\leq \Phi^2 \leq
\frac{1}{2}$, the solution does not obey the stability condition, and
so it is thermodynamically unstable, and in the case $\frac12< \Phi^2 <1$
the solution $r_{+1}$ does not physically exist as it lies outside the
cavity.  For $\frac{(1 -\Phi^2)^2}{4\Phi^2}< (\pi R T)^2 <\infty$ and
$0\leq \Phi^2 <1$, the solution $r_{+1}$ does not obey the stability
condition, and so it is thermodynamically unstable. Moreover, $r_{+1}$
corresponds to a saddle point of the action as seen from 
Fig.~\ref{figch4:ContourPlotRedact}.
For $r_{+2}$, this condition means that for $(1-\Phi^2)^3 \leq (\pi R T)^2 \leq
\frac{(1 - \Phi^2)^2}{4\Phi^2}$, in the case $0\leq \Phi^2 \leq
\frac{1}{2}$, the solution obeys the stability condition, therefore
for this range of parameters the solution is thermodynamically stable,
and it is also a
minimum of the action, as seen in Fig.~\ref{figch4:ContourPlotRedact}.
In the case $\frac12< \Phi^2 <1$, the solution $r_{+2}$
lies outside the cavity and it is not physical.  For $\frac{(1 -
\Phi^2)^2}{4\Phi^2}< (\pi R T)^2 <\infty$ and $0< \Phi^2 < 1$ the
solution $r_{+2}$ does not physically exist also, lying outside the cavity.

\subsubsection{Most favorable or probable configurations\label{sech4:favorabless}}

We study here the most probable configurations in the
case $d=5$. The analysis follows from the generic $d$ case, 
and additionally, we present the phase diagram for this case. Essentially, we 
perform the comparison 
between the
stable black hole and the charged equivalent hot flat space.
The reduced action has two stable stationary points, in particular,
the stationary point
$r_{+2}$ related to the stable black hole,
and the stationary point $r_+ = 0$
and $q = 0$, which corresponds to a cavity without a black hole and
without charge. The action also has 
a critical point corresponding to an extremal black hole with the 
radius of the cavity, $r_+=R$ and
${\frac{q}{\sqrt{3\pi^3}}} =R$.

In order to model the stationary point $r_+ = 0$ and $q = 0$, we have put forward 
a model described by a nongravitating perfect conductor hot sphere with radius
$r_{\mathrm{hs}}$, inside the reservoir at constant $\beta$ and $\phi$.
The electric potential for the case of the perfect conductor is $\phi =
\frac{q}{4\pi^2}\left(\frac{1}{r_{\mathrm{hs}}^2}- \frac{1}{R^2}\right)$, see also
Eq.~\eqref{eqch4:phiexpress}. The action for a hot sphere,
as a model of hot flat space, in five dimensions is then
\begin{align}
I_{\mathrm{hot}\,\mathrm{sphere}} = - \frac12\, \frac{4\pi^2}
{\frac{1}{r_{\mathrm{hs}}^2}-
\frac{1}{R^2}}\beta\phi^2
\,.
\label{eqch4:Ihotflatsphere5d}
\end{align}
We now compare the action of the conducting hot sphere given in
Eq.~\eqref{eqch4:Ihotflatsphere5d} with the action of the stable
configuration of the charged black hole given in 
Eq.~\eqref{eqch4:reducedaction5d} together
with Eqs.~\eqref{eqch4:solutionr+2} and
\eqref{eqch4:solutionq2}.  From
Eq.~\eqref{eqch4:Ihotflatsphere5d}, one can see that for small $r_{\mathrm{hs}}$, which is
the case analogous to hot flat space, the action is approximately zero, 
$I_{\mathrm{hot}\,\mathrm{sphere}}=0$, and so one can assign essentially 
$I_{\mathrm{hot}\,\mathrm{sphere}}
=I_{\mathrm{hot}\,\mathrm{flat}\,\mathrm{space}}$ in this case.

Regarding the stable black hole solution, it assumes a positive action
only in a small range of low temperatures, namely,
for temperatures near the minimum
temperature for which the stable black hole exists.
For higher temperatures, the action for the stable black hole solution
is negative. Hence, one finds
that the small charged sphere that emulates hot flat space is more
probable or favorable than the stable black hole solution 
for a small interval of temperatures. In fact, 
when the solution of the stable black hole obeys the condition for 
its horizon radius
\begin{align}
\frac{r_{+2}^2}{R^2}\geq \frac{4l_p^3 m}{3\pi} + \sqrt{\frac{16 l_p^6 m^2}{9 \pi^2} -
\frac{l_p^3 q^2}{3\pi^3}}\,,
\label{eqch4:bhmoreprobable5d}
\end{align}
with $\frac{4 l_p^3 m}{3\pi} = 
-\frac{9}{16} + \frac{15}{16}\sqrt{1 +
\frac{16 l_p^3}{27\pi^3}\frac{q^2}{R^4}}$,
the corresponding action is negative
and the black hole is more probable than the very small
charged sphere. The radii ratio in Eq.~\eqref{eqch4:bhmoreprobable5d} 
does not have a connection to the Buchdahl-Andr\'easson-Wright bound,
in contrast to the uncharged case, 
see also Sec.~\ref{sech4:radii}.

\begin{figure}[t]
    \centering
    \includegraphics[width=0.45\columnwidth]{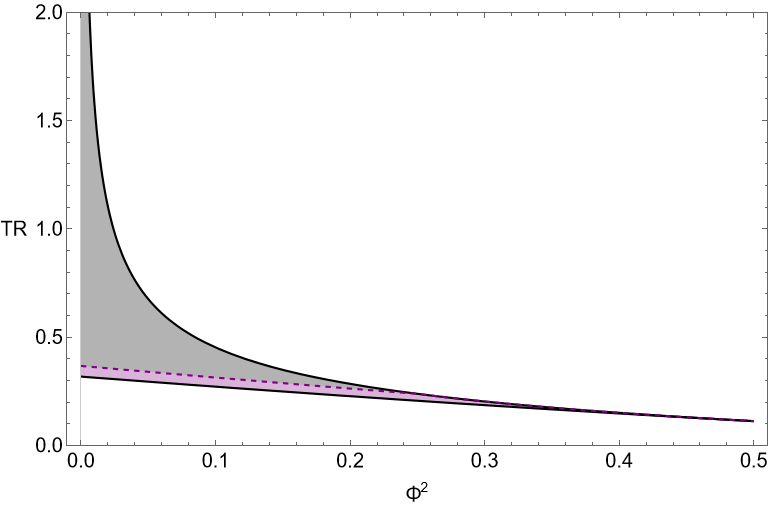}
    \includegraphics[width=0.45\columnwidth]{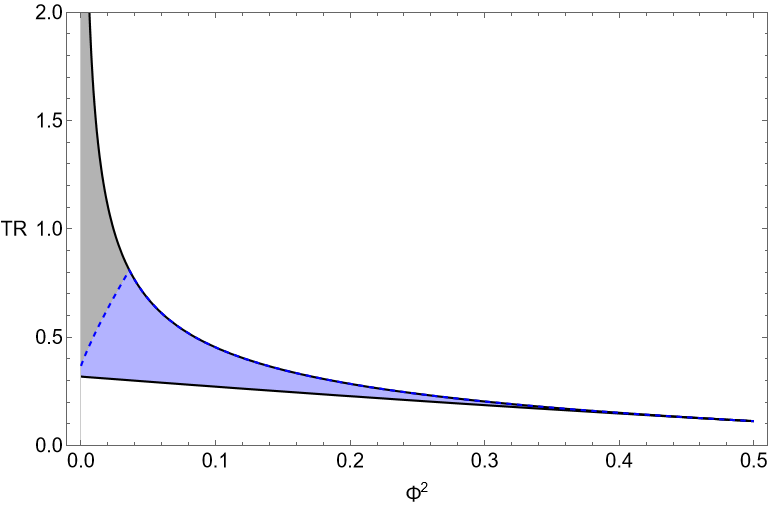}
    \caption{ Regions of favorability in five dimensions, $d=5$,
    between the stable black hole solution and the charged conducting
    sphere, in function of $RT$ and $\Phi$.  Left plot: $\frac{r_{\mathrm{hs}}}{R} \to
    0$.  The region in gray represents the points where the black hole
    solution is more favorable. The region in purple
    represents the points where the infinitesimal charged conducting
    sphere, emulating
    electrically charged hot flat space, is
    more favorable. The regions in
    white do not have a stable black hole solution, so
    presumably the most favorable state is hot flat space.
    Right plot:
    $\frac{r_{\mathrm{hs}}}{R} = 0.99$.  The region in gray represents the points
    where the black hole solution is more favorable. The region in blue
    represents the points where the charged conducting sphere is more
    favorable, with $\frac{r_{\mathrm{hs}}}{R} = 0.99$. The regions in white do not
    have a stable black hole solution, so
    presumably the most favorable state is hot flat space.
    }
    \label{figch4:condspheres0and0.99}
\end{figure}

The comparison between the hot flat sphere and the stable black hole is
shown in Fig.~\ref{figch4:condspheres0and0.99} for $d=5$ dimensions. 
In the two plots of the figure, the gray region represents the points
$(RT,\Phi)$ in which the stable black hole solution $r_{+2}$ is
more favorable or probable. The
regions in purple in the left plot of the figure,
and in blue in the right plot,
represent the points in which the charged conducting
sphere with radius $r_{\mathrm{hs}}$ is more
probable. The regions in white represent
points where there is no stable black hole solution, so
presumably the most favorable state is hot flat space. The upper white
region is quite different from the uncharged case, see~\cite{Andre:2021ctu},
because in the uncharged case the stable black hole solution exists
for temperatures up to infinite ones, whereas in the electrically
charged case, the stable black hole solution only exists within a
range of finite temperatures.
In the left plot of
Fig.~\ref{figch4:condspheres0and0.99},
one can see that for small values of $r_{\mathrm{hs}}$,
the larger the region gets where the
stable black hole solution is
more favorable over the conducting sphere,
until the point where one has a microscopic sphere. This
case of a microscopic electrically charged sphere is precisely the case that
emulates hot flat space.
In the right plot of
Fig.~\ref{figch4:condspheres0and0.99},
one can see that for large values of $r_{\mathrm{hs}}$,
the smaller the region gets where the
stable black hole solution is
more favorable, but this case is contrived,
as it does not emulate hot flat space.
Moreover, it must be stated that for
relatively small values of $r_{\mathrm{hs}}$, 
the region of favorability for the
electrically charged shell does not change much,
as even with $r_{\mathrm{hs}}
= 0.7$, the difference to $r_{\mathrm{hs}}=0$ is considerably small. Only
variations of $r_{\mathrm{hs}}$ close to $R$ induce substantial changes to the regions of
favorability.
We must give some comments regarding the uncharged case, in $d=5$, and the comparison
between the stable black hole solution and hot flat space in~\cite{Andre:2020czm}. 
It was shown that the black hole is favorable, $I_0 <
I_{\mathrm{hot}\,\mathrm{flat}\,\mathrm{space}}$, if $\beta$ is such 
that $\frac{r_+}{R} >
\frac{r_{+}}{r_\mathrm{Buch}}$, where $r_{\mathrm{Buch}}$ is the Buchdahl radius.
In the pure gravitational case, the radii ratio that establishes favorability 
agrees with the Buchdahl bound. However, here we have shown that this agreement 
does not seem to hold when other fields are present, since
the radii ratio for the electrically
charged black hole to be the dominant phase does not coincide
with the Buchdahl-Andr\'easson-Wright bound.

Finally, in $d=5$, one can make also a
comparison of the stable black hole $r_{+2}$ with the critical point
given by $r_+=R$ and
$\frac{l_p^\frac{3}{2}q}{\sqrt{3\pi^3}}=R$, which is an extremal black hole with
the horizon localized at the radius of the cavity, bearing in mind
that the precise extremality and the precise location can fluctuate by
Planck order quantities.  The gradient of the action is not defined at
this critical point but it may be smoothed up by taking in
consideration higher loops in the path integral or a different theory
of gravity, see Sec.~\ref{sech4:gradient} for more details on the critical point. 
The action for the extremal black hole at the cavity is 
Eq.~\eqref{eqch4:Icoldcurvedspace} in
the $d=5$ case, i.e.,
\begin{align}
I_{\mathrm{extreme}\,\mathrm{black}\,\mathrm{hole}} = \frac{3\pi R^2\beta}{4}
-\sqrt{3\pi^3}\,R^2 \beta \phi - \frac{\pi^2
R^3}{2}\,.
\label{eqch4:Icoldcurvedspace5d}
\end{align}
We found that, for every instance, the stable black hole is a more favorable
configuration than the extreme black hole with horizon at the cavity.

\subsection{Thermodynamics}

Here, we analyze the thermodynamics for the particular case 
of $d=5$ dimensions.
The grand potential $W$ has the dependence $W = W[T, \phi,
A]$, where $A$ is the surface area of the
$3$-sphere at the boundary $\partial M$.
The correspondence between thermodynamics and the action of the
system is given by Eq.~\eqref{eqch4:grandpot}. 
For $d=5$, one has
\begin{align}
W=&
\frac{3\pi}{4 l_p^3} R^{2}\left(1-
\sqrt{\left(1-\frac{r_+^{2}}{R^{2}}\right)\left(1-
\frac{1}{3\pi^3}
\frac{
q^2}{r_+^{2} R^{2}}\right)}\right)
\nonumber\\
&
- T\,\frac{\pi^2 r_+^{3}}{2 l_p^3}-
q \phi\,,
\label{eqch4:W5d}
\end{align}
The grand potential is defined by
the expression $W = E - ST- Q \phi$,
with $dW = -S dT - Q d\phi - p dA$
and with the first law of thermodynamics
$TdS=dE-\phi dQ+pdA$ holding, see Eq.~\eqref{eqch4:grandpot}.

The physical quantities of the system such as the entropy, electric
charge, surface pressure, thermodynamic energy, and area can be given
in this case, through the derivatives of the grand potential
The entropy can be directly obtained from Eq.~\eqref{eqch4:entropyrp} in
$d=5$ as
\begin{align}
S = \frac{A_+}{4l_p^3}\,,
\label{eqch4:entropyrp5d}
\end{align}
which is the Bekenstein-Hawking entropy of a black
hole, with $A_+=2\pi^2 r_+^3$.
The electric charge can be computed from Eq.~\eqref{eqch4:meanQ}, which in
$d=5$ it has the same appearance as in general $d$, i.e. $Q=q$.
The gravitational thermodynamic surface pressure at $R$ can be
calculated from Eq.~\eqref{eqch4:meanpressure} to yield
\begin{align}
 p = \frac{1}{8\pi R l_p^3 \sqrt{f}}\left(\left(1-
\sqrt{
\left(1-\frac{r_+^{2}}{R^{2}}\right) \left(1-\frac{
q^2}{3\pi^3r_+^{2} R^{2}}\right)}
\right)^2
-\frac{q^2}{3\pi^3R^{4}}\right)\,,
\end{align}
where $f$ is given
in Eq.~\eqref{eqch4:f5d}.
The tangential surface pressure $p$ acts along an
area $A$ that in $d=5$ is
$A=2\pi^2 R^3$.
Finally, the mean thermodynamic energy
can be taken from Eq.~\eqref{eqch4:Energy}
to the $d=5$ case and is given by
\begin{align}
E = \frac{3\pi R^{2}}{4l_p^3}
\left(1-
\sqrt{
\left(1-\frac{r_+^{2}}{R^{2}}\right) \left(1-\frac{
q^2}{3\pi^3r_+^{2} R^{2}}\right)}
\right)\,.
\label{eqch4:Energy5d}
\end{align}
This is again the same expression as the quasilocal energy evaluated at a
spherical shell of radius $R$ in $d=5$.

From Eq.~\eqref{eqch4:Energy5d}, one can write the mean energy
in terms of the
entropy $S$ of Eq.~\eqref{eqch4:entropyrp5d},
electric charge $Q$, and
surface area of the cavity $A$, 
as
\begin{align}
E =
\frac{3
(2\pi^2)^{\frac{1}{3}}  A^{\frac{2}{3}}  }{8\pi l_p^3} 
\left(1-\sqrt{\left(1-\left(\frac{4
S}{A}\right)^{\frac{2}{3}}\right)\left(1-\frac{ Q^2
(2\pi^2)^{\frac{4}{3}}}{3\pi^3(4SA)^{\frac{2}{3}}}\right)}
\right)\,.
\label{eqch4:energyintermsof5d}
\end{align}
Hence, one can use the Euler's homogeneous function
theorem considering that under rescaling of its arguments, the energy
has the property $E\left(\nu S,\nu A,\nu
Q^{\frac{3}{2}}\right) = \nu^{\frac{2}{3}} E\left(S,A,
Q^{\frac{3}{2}}\right)$. An integrated version of the
first law of thermodynamics can be obtained from the theorem,
Eq.~\eqref{eqch4:1stlawintegrated}, which in $d=5$ is
\begin{align}
\frac{2}{3}E = TS - pA + \frac{2}{3}\phi
Q\,.\label{eqch4:1stlawintegrated5d}
\end{align}
This is the Euler equation for
the system of a $d=5$
electrically charged black hole in a
heat reservoir.
By differentiating Eq.~\eqref{eqch4:1stlawintegrated5d} and considering
that $dE = TdS - pdA + \phi dQ$, the Gibbs-Duhem relation
\begin{align}
  TdS - pdA + 3(SdT-Adp) + 2 Qd\phi = 0\,.
\label{eq:gibbs5d}
\end{align}
is obtained
for the $d=5$
electrically charged black hole in a
heat reservoir.
By putting the reservoir at infinity, the integrated 
first law yields the Smarr formula in $d=5$ 
\begin{align}
m = \frac32 T_{\mathrm{H}}S + \phi_{\mathrm{H}}
Q\,,
\label{eq:smarr5d}
\end{align}
see Eq.~\eqref{eqch4:smarr}. We must note that the Smarr
formula is valid for the small black hole
solution only.

Regarding thermodynamic stability,
the heat capacity $C_{A,\phi} = 
T\left(\frac{\partial S}{\partial T}\right)_{A,\phi}$ controls the 
stability of the ensemble and it
is given by Eq.~\eqref{eqch4:heatcapacityPhibetter}. 
By setting $d=5$, the heat capacity is given by 
\begin{align}
C_{A,\phi}=
\frac{3 \left(\frac{r_+}{R}\right)\left(1 - \Phi^2\right)^2}
{8 \pi\,l_p^{3} T^2 \left(2(1 -
\Phi^2)\left(\frac{r_+}{R}\right)^{2} - 1\right)}
\,,
\label{eqch4:heatcapacityPhiA5d}
\end{align}
where we have that
$A$ is the area of the reservoir,
and
$x=\frac{r_+}{R}$. 
So $ C_{A,\phi} > 0$ is the same condition as the 
validity of the zero loop approximation 
Eq.~\eqref{eqch4:stabilitysimple5d}.

The most favorable thermodynamic configuration
is found from 
the state with the lowest value of the grand potential $W$, as we have
done previously for generic $d$.
Since $W=TI_0$, the analysis is practically the same if done
in $I_0$ or in $W$. The only difference is the 
perspective. In $I$, one talks about the
most probable state and about quantum transitions,
and when using $W$ one talks about
the most favorable state and thermodynamic phase transitions. 
See the analysis in the subsection above, Sec.~\ref{sech4:favorabless}.

\section{Thermodynamic radii and spacetime radii comparison
\label{sech4:radii}}

\subsection{Thermodynamic bifurcation radius and
the photon sphere radius comparison}

In the case
of the grand canonical ensemble of a $d$-dimensional
Reissner-Nordstr\"om black hole in a cavity,
we have seen in Eq.~\eqref{eqch4:xmin}
that the two
thermodynamic black hole solutions, represented by
$r_{+1}$ and $r_{+2}$,
bifurcate
from a horizon radius obeying
$\frac{r_{+\mathrm{bif}}}{R}=\frac{2^{\frac1{d-3}}}{\left({(d-1)}(1 -
\Phi^2)\right)^{\frac1{d-3}}}$, or in terms of $R$,
\begin{align}
R= 
\left(\frac{(d-1)}{2}(1 - \Phi^2)\right)^{\frac1{d-3}}r_+\,.
\label{eqch4:horizonbifurcradius}
\end{align}
We wave shown that a black hole
for which the horizon radius $r_+$
satisfies Eq.~\eqref{eqch4:horizonbifurcradius}
is marginally
stable to thermodynamic perturbations, and that black
holes with larger radius $r_+$
are thermodynamically stable.
Hence, the bifurcation radius is also
the marginal thermodynamic stable radius.

The photon sphere radius $R$ of a $d$-dimensional
Reissner-Nordstr\"om black hole
is given by
\begin{align}
R = \left(\frac{(d-1)}{2}
\left(1 + \frac{d-3}{d-2}\Phi^2\right)
\right)^{\frac1{d-3}}r_+
\,.
\label{eqch4:photonsphereradius}
\end{align}
At this radius, null geodesics and photons can have
circular trajectories.

From direct comparison
between Eqs.~\eqref{eqch4:horizonbifurcradius}
and \eqref{eqch4:photonsphereradius}, we observe that the two radii 
are distinct in any dimension
$d$, therefore in the grand canonical
ensemble of the Reissner-Nordstr\"om black hole
there is no connection between them.
Of course, when there is no charge or electric potential,
the two radii coincide as Eqs.~\eqref{eqch4:horizonbifurcradius}
and \eqref{eqch4:photonsphereradius} both yield
$R= 
\left(\frac{d-1}{2}\right)^{\frac1{d-3}}r_+$,
and so the radius of the
cavity at which
a stable black hole appears 
corresponds to the photon sphere radius, as seen in~\cite{Andre:2021ctu}.

\subsection{The marginal favorability radius and
the Buchdahl-Andr\'easson-Wright sphere radius comparison}
\label{sech4:bawradius}

In the case of grand canonical ensemble of a $d$-dimensional
Reissner-Nordstr\"om black hole in a cavity, the stable black hole 
solution has a
negative action $I_0$, see Eq.~\eqref{eqch4:actionI0full},
or equivalently, a negative
grand potential $W$, see Eq.~\eqref{eqch4:grandpotentialactionI0full},
if
\begin{align}
&\frac{\mu m}{R^{d-3}} \leq -\frac{4 (d-2)^2}{(d-1)^2(d-3)^2}
\nonumber \\&+ \frac{2(d-2)((d-2)^2 + 1)}{(d-1)^2(d-3)^2}\sqrt{1 
+ \frac{(d-1)^2 (d-3)^2}{4(d-2)^2}
\frac{\lambda q^2}{R^{2d-6}}}\,.
\label{eqch4:bhzeroaction}
\end{align}
The condition in Eq.~\eqref{eqch4:bhzeroaction} 
for $d=4$ is given by
$\frac{l_p^2 m}{R} \leq -\frac{16}{9} +
\frac{20}{9}\sqrt{1 + \frac{9}{16} \frac{
q^2}{4\pi R^{2}}}$.

The Buchdahl-Andr\'easson-Wright
bound is the minimum radius,
below which, an electrically charged
matter distribution obeying certain
conditions, in general relativity 
coupled to Maxwell electromagnetism in $d$ dimensions,
the spacetime is singular. 
The Buchdahl-Andr\'easson-Wright radius
was obtained in~\cite{Wright:2015dma} and can 
also be found from~\cite{Fernandes:2022gjd} 
by imposing that the trace of the stress-energy
tensor of the matter in the thin shell
is zero. The bound is given by
\begin{align}
&\frac{\mu m}{R^{d-3}} = \frac{d-2}{(d-1)^2} 
+ \frac{1}{d-1}\frac{\lambda q^2}{ R^{2d-6}} 
+ \frac{d-2}{(d-1)^2}\sqrt{1 + (d-1)(d-3)\frac{\lambda q^2}{R^{2d-6}}}\,.
\label{eqch4:buchandreassonwrightddimensions}
\end{align}
For $d=4$, this is
$\frac{l_p^2 m}{R} \leq \left( \frac{1}{3} + \sqrt{\frac{1}{9} +
\frac{1}{3}\frac{\lambda q^2}{R^2}}\right)^2$, i.e.,
$
\frac{m}{R} \leq
\frac{2}{9}+
\frac{1}{3}\frac{l_p^2 q^2}{4\pi R^2}+
       \frac23\sqrt{\frac{1}{9} +
\frac{1}{3}\frac{l_p^2 q^2}{4\pi R^2}}$.

Comparing Eqs.~\eqref{eqch4:bhzeroaction} and
\eqref{eqch4:buchandreassonwrightddimensions}, it can be seen that the marginal
favorability radius and the Buchdahl-Andr\'easson-Wright
radius are distinct for any dimension $d$, and so there is no
connection between them.
When there is no charge, $q=0$, and no electric potential, $\Phi=0$,
both radii are equal to Buchdahl radius, i.e. $\frac{r_+}{R}
\geq \left(\frac{4(d-2)}{(d-1)^2}\right)^\frac{1}{d-3}$. 
In this case, the stable solution has a negative
free energy if the radius of the black hole is larger than the
Buchdahl radius, see~\cite{Andre:2021ctu}.

\section{Gradient of the action for the two ill-behaved
critical points\label{sech4:gradient}}

We analyze the gradient of the action near the critical point
$r_+=q=0$, and also $r_+^{2d-6} = \lambda q^2 = R^{2d-6}$, to understand 
their metastability.
The gradient of the action in Eq.~\eqref{eqch4:reducedaction} can be written 
as
\begin{align}
&\frac{\mu}{R^{d-2}}\frac{\partial I_*}{\partial x} = 
\frac{(d-3)\beta}{2R x\sqrt{f}}\left[x^{d-3} -
\frac{y}{x^{d-3}}\right]
-2\pi x^{d-3}\,,
\label{eqappch4:graddaaction}
\\
&\frac{\mu}{R^{d-2}}\frac{\partial I_*}{\partial \sqrt{y}} = 
\frac{\beta \sqrt{y}}{R x^{d-3}\sqrt{f}}(1-x^{d-3}) -
\frac{\beta \Phi}{R}\,,
\label{eqappch4:graddaaction1}
\end{align}
where $x = \frac{r_+}{R}$, $y = \frac{\lambda q^2}{R^{2d-6}}$ and
$\Phi = (d-3)\Omega \sqrt{\lambda}\phi$.

First, we proceed with the analysis for hot flat space $r_+=q=0$. For that,
the limit of the gradient for $x=y=0$ is done along a family of curves $y
= (\eta)^{2} x^{2d-6}$, where $\eta$ is a positive constant of the
curve. One must consider $\eta < 1$ so that the curve
is inside the physical domain of the action and it covers
all the possible directions inside it. The gradient near $r_+=q=0$ is given by
\begin{align}
& \frac{\mu}{R^{d-2}} \frac{\partial I_*}{\partial x} =
\frac{(d-3)\beta 
x^{d-4}}{2 R}\left(1 - \eta^2\right)\,,\\
& \frac{\mu}{R^{d-2}}\frac{\partial I_*}{\partial \sqrt{y}} =
\frac{\beta}{R}\left(\eta - \Phi\right)\,.
\end{align}
The dependence in $x^{d-4}$ was left since it gives different
limits for the case of $d=4$ and $d>4$. Since the gradient depends on
$\eta$, then the limit of the gradient is not defined. Nevertheless, one can
calculate the directional derivative along the vector $v =
\frac{1}{\sqrt{1 + (d-3)^2\eta^2 x^{2d - 8}}} (1 ,
(d-3)\eta x^{d-4})^T$, which is given by
\begin{align}
D_v I_* =& \frac{(d-3)\beta x^{d-4}}{2R}\left(1 + \eta^2 -
2\eta\Phi \right)\,\,,
\end{align}
so for $d>4$, the directional derivative
vanishes. Additionally, the directional $(d-3)$th derivative is positive 
since $1 + \eta^2 -2\eta\Phi>0$ for $\eta < 1$ and $\Phi \leq 1$, and
so this can be considered as a minimum, although formally the partial
derivative in $\sqrt{y}$ is undefined. For $d=4$ case, the directional
derivative does not vanish, however, since $1 + \eta^2 -2\eta\Phi>0$ for 
$\eta < 1$ and $\Phi \leq 1$, one can observe
that the directional derivative is positive in the
physical domain. Hence, the action resembles a conical potential
well at the origin and so hot flat space can be considered as a
solution.

Now, we analyze the extremal black hole located at the cavity.
In order to study the gradient
in the critical point $x=1$ and $y=1$, one can calculate the gradient of
the action in this limit along the curve $x^{d-3}= 1 - \epsilon$ and
$y = 1 - \eta \epsilon$, where $\eta$ is a constant of the
curve and $\epsilon$ parametrizes the curve. The limit of $\epsilon
\rightarrow 0^+$ is then performed, giving the gradient
\begin{align}
&\frac{\mu}{R^{d-2}}\frac{\partial I_*}{\partial x} = 
\frac{(d-3)\beta}{2R\sqrt{\eta - 1}}(\eta-2) -
2\pi \,,
\label{eqch4:gradextreme1}\\
&\frac{\mu}{R^{d-2}}\frac{\partial I_*}{\partial \sqrt{y}} = 
\frac{\beta}{R\sqrt{\eta - 1}}-\frac{\beta \Phi}{R}\,,
\label{eqch4:gradextreme2}
\end{align}
where it is required that $\eta > 2$ so that the curve is done along configurations 
of subextremal black holes, coming from the condition $y < x^{2d-6}$.
Since there is a dependence on the curve one chooses to perform the limit,
the limit of the gradient at the extremal point is not defined.

Interestingly, for $\gamma = 1$, i.e.,
$\beta = \frac{4\pi}{d-3} \frac{\abs{\Phi}}{1 - \Phi^2}R$, the gradient
vanishes in the limit along a curve with $\frac{1}{\eta} = 1 +
\frac{1}{\Phi^2}$.  In fact, this set of temperatures corresponds to the
stable black hole solution hitting the extremal point $x=y=1$. 
But this only happens in one particular curve, the limit of the gradient is
still undefined.

Finally, one should consider the directional derivative along these 
curves, in the direction of smaller $\epsilon$. Indeed, the direction 
can be described by the vector $v = 
\frac{1}{\sqrt{1 + (d-3)^2\eta^2/4}}(1,
\frac{\eta (d-3)}{2})^T$,
and so the directional derivative gives
\begin{align}
&\frac{\mu}{R^{d-2}} D_v I_* =
\frac{\frac{\beta (d-3)}{2R}\left(2\sqrt{\eta -1} -
\eta \Phi\right)
 - 2\pi}{\sqrt{1 +
(d-3)^2\eta^2/4}} \,.
\label{eqch4:B8}
\end{align}
The directional derivative depends also on $\eta$ and it can be either
positive or negative. Particularly, for values of $\eta$ and $\Phi$
where $\gamma_{\mathrm{bif}}(\Phi,d)< \frac{4 (\eta -1)
\Phi^2}{(1-\Phi^2)^2}\left(1 -
\frac{\eta}{2\sqrt{\eta-1}}\Phi\right)^2$, the directional derivative
in Eq.~\eqref{eqch4:B8} can be positive in a region $\gamma_{\mathrm{
bif}}(\Phi,d)<\gamma < \frac{4 (\eta -1) \Phi^2}{(1-\Phi^2)^2}\left(1 -
\frac{\eta}{2\sqrt{\eta-1}}\Phi\right)^2$, with $\gamma_{\mathrm{bif}}$
given in Eq.~\eqref{eqch4:solExcondition}. Hence, the action near
this critical point does not resemble a potential well.

\section{Conclusions\label{sech4:concl}}

In this Chapter, we built the grand canonical ensemble of a $d$-dimensional 
Reissner-Nordstr\"om space in a cavity, using the path integral approach. We obtained 
the partition function of the space in a cavity, by performing
the zero loop approximation to the path integral relative to the
Euclidean action, where only the term which minimizes the action
contributes to the path integral. There are two stationary points of
the action that correspond to a black hole in equilibrium with a heat
reservoir with the temperature and the electric potential fixed at the
boundary of the cavity. We have shown that the stationary point 
with lower horizon radius is unstable, while the stationary point 
with higher horizon radius is stable. We could not find analytically
the corresponding values of the event horizon radius depending on the
temperature and electric potential of the two stationary points for arbitrary dimensions. 
However, we were able to find analytical expressions for 
the event horizon radius in $d=5$, where the equation reduces to a 
quadratic polynomial.

From our analysis, there are some features of the stationary points 
in the electrically charged case that differ from the electrically uncharged case. 
First, the event horizon radius corresponding to the lowest temperature
allowed does not correspond to the photon sphere, unlike the uncharged
case. This indicates that the correspondence in the uncharged case 
is a coincidence.
Second, the larger horizon radius solution reaches the radius of
the cavity at finite temperature, unlike the uncharged case, where the
horizon radius only reaches the cavity radius at infinite temperature.

We have obtained the thermodynamics of the system, by connecting the 
partition function given by the path
integral in the zero loop approximation 
with the partition function of the grand canonical
ensemble. The grand potential of the system can be obtained in terms of 
the action in the zero loop approximation. We thus recover
the thermodynamics of the black hole corresponding to the stable
stationary point. We have shown that the system's entropy
corresponds to the Bekenstein-Hawking entropy, the pressure
corresponds to the pressure of a self-gravitating static electrically
charged spherical thin shell in equilibrium, and the thermodynamic
energy has the same expression as the expression for the quasilocal
energy.  The first law of thermodynamics with constant area is obeyed
at the stationary points of the action, as we would expect. The
stability of the stationary points is described by the heat capacity
at constant area and electric potential. If this heat capacity is
positive, then the stationary point is stable. This fits well with the
relationship between thermodynamic stability and the heat capacity.

Additionally, we made the comparison between the stable black hole solution and an
electrically charged conducting hot sphere in flat space, 
in order to see the most favorable phase of the system. 
In this case,
a configuration is more favorable than the other when its grand
potential $W$ is lower. This in turn depends on the value
of the temperature, of
the electric potential of the
reservoir, and of the radius of the
conducting sphere. Moreover, the smaller the radius of the conducting
sphere, the larger the region where the stable black hole is
favored. We also made the comparison of
the Buchdahl-Andr\'easson-Wright bound radius in $d$-dimensional
Reissner-Nordstr\"om spacetimes
with the minimum radius for which the stable black hole phase
is thermodynamically favored. We have shown that both radius do not 
coincide,
thus showing
that the connection displayed in the Schwarzschild case is not
generic, rather it is a very restricted equality holding only in
the pure gravitational situation.

\chapter{Gibbons-Hawking action for electrically 
charged black holes 
in the canonical ensemble and 
Davies' thermodynamic theory of black holes}\chaptermark{Gibbons-Hawking action 
and Davies's thermodynamic theory}
\label{ch:Daviesblackhole}

\section{Introduction}

In the previous chapter, we have analyzed the grand canonical ensemble of a 
charged black hole inside a cavity, using the Euclidean path integral approach.
Apart from this statistical treatment,
there is another thermodynamic approach based on
the use of Bekenstein-Hawking entropy~\cite{Bekenstein:1972,Bekenstein:1972b,Hawking:1975vcx}
and the first law of black hole mechanics~\cite{Bardeen:1973gs} 
to obtain the thermodynamic properties of black holes, namely the 
first law of thermodynamics. This law and its consequences 
were summarized in Davies' thermodynamic theory of black holes
~\cite{Davies:1977bgr, Davies:1989ey}. 
An important feature described by Davies was the case of an infinite discontinuity 
in the heat capacity with constant electric
charge for electric charged black holes, which was described as being similar to a second order 
phase transition. The thermodynamic analysis of charged black holes 
was also performed at the same time 
in~\cite{Hut:1977}. The characterization of this discontinuity as a second order phase 
transition was further scrutinized 
in~\cite{Sokolowski:1980uva, Kaburaki:1993ah, Katz:1993up, Parentani:1994wr}, where it was 
established that the discontinuity described a turning point or a condition of stability 
rather than a phase transition. Further works using the first law of thermodynamics 
for black hole spacetimes were done afterwards~\cite{Gibbons:2004ai, Myung:2008ze,
Maeda:2009uy,La:2010bx,Hendi:2014mna,
Clement:2019ghi,
Hajian:2021hje,
Jiang:2021pna,
Rodriguez:2021hks,
Murk:2023rwl}.

Even though the use of the first law of thermodynamics to describe the 
thermodynamics of black holes is well-motivated, there is a lack of analysis 
establishing that the construction of statistical ensembles 
using the Euclidean path integral approach 
yields the same results as just simply imposing the first law of thermodynamics 
as described in Davies' thermodynamic theory. 
Hence, in this chapter, we construct
the canonical ensemble of a charged black hole with the cavity 
at infinity in higher dimensions through the Euclidean path integral approach
with fixed temperature and electric charge. The objective is to compare the 
results from the Euclidean path integral approach, as in Gibbons and 
Hawking, with the results from Davies' thermodynamic theory.
We perform the zero loop approximation to the partition function, as in Gibbons and Hawking, 
and find two possible black hole solutions for the ensemble, 
with the larger black hole being unstable and the 
smaller black hole being stable. 
We find the thermodynamic properties of the black hole, and show that 
the Davies' thermodynamic theory for the four dimensional case, $d=4$, is in 
agreement with our results. 
We also briefly analyze the five dimensional case, $d=5$. 
We observe that the heat capacity of the black hole 
has precisely the discontinuity found by Davies~\cite{Davies:1977bgr}, and 
we show that it is in fact a turning point. 
Finally, we construct a model for charged hot flat space, 
which is described by hot flat space 
with electric charge at infinity. This allows us to 
study the phase transitions between this configuration and the 
stable black hole, which is lacking in the literature. 
We find that the charged hot flat space is always favorable 
compared to the stable black hole solution.

This chapter is presented as follows. In Sec.~\ref{sech6:Canonical1}, we construct 
the canonical ensemble 
using the partition function and perform the zero loop approximation. 
In Sec.~ \ref{sech6:thermo}, we obtain the thermodynamics of the system
from the partition function 
and we perform the analysis of phase transitions. In Sec.~\ref{sech6:Daviesd4}, we 
present the case 
$d=4$, showing that the results agree with Davies' thermodynamic theory. 
In Sec.~\ref{sech5:Daviesd5}, we present briefly the case $d=5$. 
In Sec.~\ref{sech6:Concl}, we conclude the chapter. 
The work in this chapter is based on~\cite{Tiago2024a}.

\section{\label{sech6:Canonical1}
The canonical ensemble of a charged black hole in asymptotically flat
space through the Euclidean path integral approach}\sectionmark{
    The canonical ensemble through Euclidean path integral approach
    }\thispagestyle{userightbotmark}

\subsection{The Euclidean path integral and Euclidean action for the canonical 
ensemble}

The canonical ensemble of a charged black hole in asymptotically flat space 
can be constructed through the Euclidean path integral approach, in $d$ dimensions,
with the partition function given formally by
\begin{align}\label{eqch6:partitionfunction}
        Z = \int Dg_{\alpha \beta}DA_{\gamma}\, 
        \mathrm{e}^{-I[g_{\mu\nu},A_{\sigma}]}\,\,,
\end{align}
with the Euclidean action 
\begin{align}
    I =& - \frac{1}{16\pi l_p^{d-2}}\int_M R \sqrt{g}\,d^d x 
- \frac{1}{8\pi l_p^{d-2}} \int_{\partial M} (K-K_0)\sqrt{\gamma}\, d^{d-1}x\nonumber\\
&+ \frac{(d-3)}{4\Omega_{d-2}}\int_M F_{ab}F^{ab}\sqrt{g}\,d^dx\nonumber\\
&+ \frac{(d-3)}{\Omega_{d-2}}\int_{\partial M}F^{ab}A_{a}n_b 
    \sqrt{\gamma}\,d^{d-1}x\,,
    \label{eqch6:action1}
\end{align}
where $R$ is the Ricci scalar given by first and second order
derivatives of the Euclidean metric $g_{\alpha \beta}$, 
$g$ is the determinant of $g_{\alpha \beta}$,
$K$ is the trace of the 
extrinsic curvature $K_{ab}$ of the space boundary, 
$K_0$ is the trace of the extrinsic curvature 
of the space boundary
embedded in flat Euclidean space, 
$\gamma_{ab}$
is the induced metric on the space boundary, 
$\gamma$ is the determinant of $\gamma_{ab}$, 
$\Omega_{d-2} = \frac{2\pi^{\frac{d-1}{2}}}
{\Gamma\left(\frac{d-1}{2}\right)}$
 is the surface area of the unit $(d-2)$-sphere,
$F_{\alpha \beta} = \partial_\alpha A_\beta - \partial_\beta A_\alpha$ is the Maxwell
tensor given by derivatives of the electromagnetic vector
potential $A_\alpha$, 
and $n_\beta$ is the outward unit normal vector to the space boundary.
The 
Gibbons\hskip0.02cm-\hskip-0.02cm{}Hawking-York boundary term 
involving the extrinsic curvature
must be present in the action in order
to have a well-defined variational
principle with Dirichlet boundary conditions.
The boundary term depending on the Maxwell tensor 
must be present so that the canonical 
ensemble may be prescribed, 
see \cite{Braden:1990hw}.
This can be seen from the variation of the action, as one must 
get a boundary term with the variation of flux density and not the 
variation of the Maxwell field.  
This term gives the correct identification of the action with
fixed electric flux given by the 
integral of the Maxwell tensor on a $(d-2)$-surface, 
which has the meaning of electric charge.
In the other situation, the potential 
vector 
$A_a$ must be fixed
at the boundary in order to have a well-described system, 
which means the grand canonical ensemble should be prescribed
as was done in \cite{Gibbons:1977}, see also \cite{Braden:1990hw}.

\subsection{Zero loop approximation}

\subsubsection{Euclidean Reissner-Nordstr\"om line element and 
Maxwell field}

Differently from the other chapters, here we apply the zero loop approximation 
directly in the sense of Gibbons-Hawking~\cite{Gibbons:1977}, meaning that 
the action in Eq.~\eqref{eqch6:action1} is evaluated for a space that is a 
solution to the Euclidean Einstein-Maxwell equations. 
This solution for arbitrary $d$ dimensions
with $d\geq 4$,
is described by the $d$-dimensional
Reissner-Nordstr\"om line element
\begin{align}
    ds^2 =  \left(\frac{1}{2\pi\, t_H}\right)^{\hskip -0.089 cm 2}
    \hskip -0.08cm f(r)\,
    d\tau^2 + \frac{dr^2}{f(r)} + r^2 d\Omega_{d-2}^2\,,
    \label{eqch6:ReissnerNordstrom}
\end{align}
also called 
Tangherlini
line 
element, where
the function $t_H$, the Hawking function
or Hawking temperature function, is given by
\begin{align}
t_H =
\frac{(d-3)\left(r_+^{d-3}  -
   \frac{\mu q^2}{r_+^{d-3}}\right)}
 {4\pi r_+^{d-2}}
\,,
    \label{eqch6:betahawking}
\end{align}
with $r_+$ being the horizon radius 
of the
black hole, $q$ its electric charge, the function
$f(r)$ defined by
\begin{align}
    f(r) = \left(1 - \frac{r_+^{d-3}}{r^{d-3}} \right)
    \left(1 - \frac{\mu q^2}{r_+^{d-3}r^{d-3}} \right),
    \label{eqch6:f}
\end{align}
with
\begin{align}
\mu = \frac{8\pi l_p^{d-2}}{(d-2)\Omega_{d-2}}\,,
\label{eqch6:mu}
\end{align}
and $d\Omega_{d-2}^2$ being the line element of the $(d-2)$-sphere 
with surface area $\Omega_{d-2} = \frac{2\pi^{\frac{d-1}{2}}}
{\Gamma\left(\frac{d-1}{2}\right)}$. The coordinate
range for the Euclidean time is $\tau \in \,]0,2\pi[$, the 
range for the radius coordinate is
$r\in \,]r_+,\infty[$, and the ranges of the 
angular coordinates are the usual ones.
The Maxwell electromagnetic potential field is described by 
\begin{align}
    A_\tau(y) = - \frac{iq}{2\pi (d-3)t_H }\left(
        \frac{1}{r_+^{d-3}} - \frac{1}{r^{d-3}}
     \right)\,.
    \label{eqch6:maxwellfield}
\end{align}
We now discuss the considerations used to obtain the
precise forms of the line element and
the Maxwell field
given in Eqs.~\eqref{eqch6:betahawking}-\eqref{eqch6:maxwellfield}.
First for the line element, we choose a smooth metric,
i.e., the metric cannot have conical singularities or curvature
singularities.  In order to avoid a conical singularity at the horizon
and since we have chosen $2\pi$ periodicity in the Euclidean imaginary time, 
we added the factor $\frac{1}{(2\pi t_H)^2} = 
\frac{1}{(\sqrt{f}\partial_r \sqrt{f})^{2}_{y=0}}$
to the usual $\tau \tau$
component of the Reissner-Nordstr\"om line element.  Secondly, for the
Maxwell field, we have chosen a gauge for $A_\mu$ such that only
$A_\tau$ is non-zero and we have assumed the nonexistence of magnetic
monopoles. Also, the gauge was chosen such that
$A_\mu(r_+)=0$. In this gauge, the Maxwell field in the Riemannian 
metric is tied to the
physical electric potential
given by $(d-3) \frac{2\pi i t_H}{\sqrt{f}} A_\tau$, 
which should be bounded at the horizon.

The Reissner-Nordstr\"om line element characterized
by Eqs.~\eqref{eqch6:ReissnerNordstrom}-\eqref{eqch6:mu}
has several features. The main features are the two parameters,
namely, the horizon radius
$r_+$, and the electric
charge $q$.
There is an instance where the line element 
is characterized by one parameter alone,
instead of two, which is the extremal case
\begin{align}
r_{+e}= (\mu q^2)^{\frac{1}{2d-6}} \,.
\label{eqch6:extremalr}
\end{align}
From Eq.~\eqref{eqch6:extremalr}, one sees
that for a given electric charge $q$ 
the extremal horizon radius 
$r_{+e}$ has a precise value. One can invert
Eq.~\eqref{eqch6:extremalr} so that, for a given horizon
radius $r_+$, there is an extremal electric
charge $q_e$ given by
$q_e=\sqrt{
\frac{r_+^{2d-6}}{\mu}}$.
When it is convenient, we shall
trade the horizon radius $r_+$ for
the space mass $m$ and the electric charge $q$ as
\begin{align}
    r_+ = \left(\mu m + \sqrt{\mu^2 m^2 - \mu q^2}
    \right)^{\frac{1}{d-3}}.
    \label{eqch6:r+asmandq}
\end{align}
This equation can be inverted to
give
$m = \frac{r_+^{d-3}}{2\mu} + \frac{q^2}{2r_+^{d-3}}$.
In terms of the mass, the extremal black hole
of Eq.~\eqref{eqch6:extremalr} obeys the relation
$\sqrt\mu\,m=q$, where here $q$ means the absolute
value of the electric charge.

\subsubsection{The ensemble and its solutions}

We are considering here the canonical ensemble of a charged black 
with the boundary at infinity. This boundary characterizes the 
heat reservoir with a fixed temperature $T$ and fixed electric charge 
of the whole space $Q$.
The inverse temperature at infinity, $\beta = \frac{1}{T}$, 
is determined by the Euclidean proper time at the 
boundary of the space, i.e., $\beta = 2\pi 
\left(\frac{\sqrt{f}}{2\pi t_h}\right)\sVert[2]_{r\rightarrow \infty}$. 
Using that $f(r\rightarrow \infty)=1$, one has that 
$\beta$ must be equal to the inverse
of the Hawking function $t_H$.
Now, from the path integral formalism,
$\beta$ is the fixed inverse temperature of the ensemble.
Therefore, the ensemble temperature $T$
and the Hawking temperature function $t_H(r_+,q)$
of Eq.~\eqref{eqch6:betahawking}
satisfy the relation
$
    T=t_H(r_+,q)
$.
Notice that, since the period of the Euclidean
time $\tau$ is $2\pi$, 
the factor $(2\pi t_h)^{-2}$ was introduced 
on the time-time component of the metric in order to 
have regularity, therefore one links the temperature 
function $t_h$ 
to the avoidance of a conical singularity at the horizon 
if the Einstein equations are solved. 
In addition, in this canonical ensemble,
the electric flux $\int F^{ab}dS_{ab} = -i
\frac{\Omega_{d-2}}{2}Q$ or, equivalently,
the total electric charge, with the reservoir 
at spatial infinity, is fixed
to be $Q$ so that
the electric charge of the black hole $q$ obeys
$q=Q$.
In brief, the considered canonical ensemble 
with fixed temperature $T$ and fixed electric charge $Q$
at infinity imposes the following
constraints to the possible black hole solutions,
\begin{align}
    T=t_H(r_+,Q) \,,
 \label{eqch6:Hawkbeta}
\end{align}
\begin{align}
    Q=q \,.
 \label{eqch6:QqHawk}
 \end{align}
The latter equation means that
black holes that are solutions of
this ensemble must have their electric
charge $q$ equal to the ensemble electric charge $Q$.

Inverting Eqs.~\eqref{eqch6:Hawkbeta}
and \eqref{eqch6:QqHawk},
we can see that the black hole solutions
have the generic form
\begin{align}
r_+=r_+(T,Q)\,,
 \label{eqch6:Hawkbetainverted}
\end{align}
\begin{align}
q=q(T,Q)\,,
 \label{eqch6:QqHawkinverted}
\end{align}
with this later equality having a direct solution 
$q=Q$.
Specifically, by rearranging Eq.~\eqref{eqch6:Hawkbeta} and
taking into account Eq.~\eqref{eqch6:QqHawk},
the black hole
solutions $r_+$, which are
formally represented in Eq.~\eqref{eqch6:Hawkbetainverted},
obey the condition
\begin{align}
\left(\frac{d-3}{4\pi T}\right)(r_+^{2d-6} -
\mu Q^2) - r_+^{2d-5} = 0\,.
\label{eqch6:rpinT}
\end{align}
This equation,
Eq.~\eqref{eqch6:rpinT},
is not solvable analytically
for generic $d$.
However, we can perform an analysis of 
its solutions.
The function $t_H(r_+,Q)$
in Eq.~\eqref{eqch6:betahawking}, see also Eq.~\eqref{eqch6:Hawkbeta}, 
has a maximum at 
\begin{align}
    r_{+s}= \left(\sqrt{(2d-5)\mu}\,Q\right)^\frac{1}{d-3}
    \label{eqch6:r+davies}\,,
\end{align}
which is a saddle point of the 
action for the black hole. From now onwards, $Q$
stands for the absolute value of the electric charge
$Q$ itself for convenience.
The saddle point $r_{+s}$ of the 
action of the black hole has temperature 
\begin{align}
T_s=
\frac{(d-3)^2}
{2\pi (2d-5)
(\sqrt{(2d-5)\mu}\,Q)^{\frac{1}{d-3}}}
\label{eqch6:Tdavies}\,.
\end{align}
Eliminating $Q$ in Eqs.~\eqref{eqch6:r+davies} and
\eqref{eqch6:Tdavies}, one finds $r_{+s}$ in terms of
a given temperature $T$,
$r_{+s}=\frac{(d-3)^2}{2\pi (2d-5)T}$,
or inverting, for a given $r_+$,
one finds 
$T_s=\frac{(d-3)^2}{2\pi (2d-5)r_{+}}$.
In $d=4$, the temperature $T_s$ in Eq.~\eqref{eqch6:Tdavies}
reduces to the Davies temperature,
and so, one can see Eq.~\eqref{eqch6:Tdavies} as the generalization of 
the Davies temperature to $d$ dimensions.
By inspection of Eq.~\eqref{eqch6:rpinT},
for temperatures in the
interval $0 <T \leq T_s$,
there are two solutions,
the solution
$r_{+1}(T,Q)$ and the solution $r_{+2}(T,Q)$, while 
for $T > T_s$
there are no black hole solutions.
The solution
$r_{+1}(T,Q)$ exists in the interval 
$r_{+e} <r_{+1}(T,Q)\leq r_{+s}$,
so we can summarize for the solution 1
\begin{equation}
\begin{aligned}
&r_{+1}=r_{+1}(T,Q),\quad\quad 0 <T \leq T_s,
\\
& q_1=Q,\quad\quad\quad\quad\quad\quad\,
r_{+e} <r_{+1}(T,Q)\leq r_{+s},
\end{aligned}
\label{eqch6:solutionr+1}
\end{equation}
where 
$r_{+e}$ is the
radius of the extremal black hole
given by 
$r_{+e}=r_{+1}(T\rightarrow 0,Q) = (\mu Q^2)^{\frac{1}{2d-6}}$,
see Eq.~\eqref{eqch6:extremalr},
and 
$ r_{+s}=r_{+1}(T_s,Q)$ is given in Eq.~\eqref{eqch6:r+davies}.
This solution,
$r_{+1}(T,Q)$, is an increasing monotonic
function of $T$.
The solution $r_{+2}(T,Q)$
exists in the interval 
$r_{+s}< r_{+2}(T,Q)<\infty$,
so we can summarize for the solution 2
\begin{equation}
\begin{aligned}
&r_{+2}=r_{+2}(T,Q),\quad\quad 0 <T \leq T_s,
\\
& q_2=Q,\quad\quad\quad\quad\quad\quad\,
r_{+s}< r_{+2}(T,Q)<\infty,
\end{aligned}
\label{eqch6:solutionr+2}
\end{equation}
where 
$r_{+s}=r_{+2}(T_s,Q)$, i.e., the solution 2
is bounded 
from below, and
is unbounded from above, since 
at $T \rightarrow 0$, the 
solution tends to infinity. This solution,
$r_{+2}(T,Q)$, is a decreasing monotonic
function of $T$. When the ensemble is only characterized
by the temperature $T$, with vanishing $Q$,
only the black hole solution
$r_{+2}$ survives which corresponds to the
Gibbons\hskip0.02cm-\hskip-0.02cm{}Hawking black hole 
solution.

There is however another solution which exists for all 
temperatures. This solution can be described by a limit of 
solutions in the charged matter sector. In order to 
keep a vanishing mass of space and to keep a fixed electric 
charge, one must have charged matter at infinity, at the
boundary of space. We refer to this configuration as the 
charged hot flat space, i.e. hot flat space with 
electric charged $Q$ dispersed at infinity. 
For $T>T_s$, there are no black hole solutions and one
is left with hot flat space with electric charged
$Q$
dispersed at infinity, and so the solution of the ensemble
at this temperature range can be summarized as
\begin{equation}
\begin{aligned}
&{\mathrm{charged}\;\mathrm{hot}\;\mathrm{flat}\;\mathrm{space}},\quad\quad T_s<T<\infty,
\\
& Q \;{\mathrm{dispersed}\;\mathrm{at}}\; r=\infty,\quad\quad
0\leq r<\infty.
\end{aligned}
\label{eqch6:solutionhot}
\end{equation}
Thus, the three solutions of the ensemble are displayed
in Eqs.~\eqref{eqch6:solutionr+1}-\eqref{eqch6:solutionhot}.

\subsubsection{Action of the Reissner-Nordstr\"om black hole
space and partition function}

We now evaluate the action
given in Eq.~\eqref{eqch6:action1} for the metric
in Eq.~\eqref{eqch6:ReissnerNordstrom} and for the Maxwell field in 
Eq.~\eqref{eqch6:maxwellfield}, with the
black hole solutions of the ensemble 
obeying Eq.~\eqref{eqch6:rpinT},
i.e., those formally shown in  Eq.~\eqref{eqch6:solutionr+1}
and Eq.~\eqref{eqch6:solutionr+2}.

It is useful to split the action $I$ into the gravitational action 
plus the Maxwell action, i.e. $I = I_{g\mathrm{f}} + I_{q}$, where 
\begin{align}
    & I_{g\mathrm{f}} =  -\frac{1}{16\pi l_p^{d-2}} \int_{\mathcal{M}}R\sqrt{g}d^dx 
    -\frac{1}{8\pi l_p^{d-2}}\int_{\partial M} (K-K_0) \sqrt{\gamma}
    d^{d-1}x\,,\\
    & I_{q} = \frac{(d-3)}{4\Omega_{d-2}} \int_M F_{\alpha \beta}F^{\alpha \beta}\sqrt{g} d^dx
    +  \frac{(d-3)}{\Omega_{d-2}}\int_{\partial M}F^{\alpha \beta}A_{\alpha}n_\beta 
    \sqrt{\gamma}d^{d-1}x\,\,.
\end{align}
Starting with the gravitational action, one can obtain it generally for a 
spherically symmetric metric, see Chapter~\ref{ch:Euclideanpathintegral}. 
Together with the metric form in Eq.~\eqref{eqch6:ReissnerNordstrom}, one has 
that
\begin{align}
    &I_{g\mathrm{f}} = \left(\frac{\sqrt{f}r^{d-3}}{\mu t_H}\left(1 
    - \sqrt{f}\right)\right)\sVert[3]_{r\rightarrow +\infty} 
    - \frac{\Omega_{d-2}}{4l_p^{d-2}}\left(\frac{\partial_r f r^{d-2}}
    {4\pi t_H}\right)\sVert[2]_{r\rightarrow r_H} - \frac{1}{2 t_H} \frac{q^2}{r^{d-3}_+}\,\,.
\end{align}
Regarding the action for the Maxwell field, one can simplify the 
Maxwell term as $F^{\alpha \beta}F_{\alpha \beta} = 2 F_{u\tau}F^{u\tau} = -2\frac{q^2}{r^{2d-4}}$
and the boundary term as $F^{\alpha \beta}A_{\alpha}n_\beta = \frac{2\pi i t_H q}{r^{d-2}\sqrt{f}}A_\tau$
to obtain
\begin{align}
    I_{q} = -\frac{q^2}{2t_H r_+^{d-3}} 
    + \left(\frac{q^2}{t_H \sqrt{f}}\left(\frac{1}{r_+^{d-3}} - \frac{1}{r^{d-3}} \right)\right)
    \sVert[3]_{r\rightarrow+\infty}\,\,,
\end{align}
With the action written explicitly 
in terms of the important quantities of the ensemble solution, one can now further perform 
the limits using the properties of the function $f$ and the Maxwell field $A_\tau$, in Eq.~\eqref{eqch6:f} 
and~\eqref{eqch6:maxwellfield} respectively. The gravitational action has two limits that must be 
performed. The first limit yields $\left(\sqrt{f}r^{d-3}\left(1 
- \sqrt{f}\right)\right)\sVert[1]_{r\rightarrow +\infty} = \frac{r_+^{d-3}}{2} 
+ \frac{\mu q^2}{2r_+^{d-3}}$ while the second limit yields $\left(\partial_r f r^{d-2}
\right)\sVert[2]_{r\rightarrow r_H} = 4\pi t_H r_+^{d-2}$. The action for the Maxwell 
field has one limit which yields 
$\left(\frac{q^2}{t_H \sqrt{f}}\left(\frac{1}{r_+^{d-3}} - \frac{1}{r^{d-3}} \right)\right)
\sVert[2]_{r\rightarrow+\infty} = \frac{q^2}{t_H r_+^{d-3}}$. Therefore, the full action 
is given by 
\begin{align}
I_0[T,Q]
= \frac{1}{\mu T}
    \left(\frac{r_+^{d-3}}{2} + \frac{\mu Q^2}{2r_+^{d-3}}\right) 
    - \frac{\Omega_{d-2}}{4l_p^{d-2}}r_+^{d-2},
    \label{eqch6:actioninfinity}
\end{align}
where, $T=t_H(r_+,Q)$ was used, having two black hole solutions for
$T\geq T_s$, $r_{+1}(T,Q)$ and $r_{+2}(T,Q)$, each of which gives an
expression in terms of $T$ and $Q$ that replace $r_{+}$ in Eq.~\eqref{eqch6:actioninfinity}.
Explicitly, the actions for each solution are of the form
$I_0(T,Q,r_{+1}(T,Q))$ and $I_0(T,Q,r_{+2}(T,Q))$. There is a
third solution that must be considered, corresponding to the case of
having no black hole solutions.  This case is described by hot flat
space with fixed temperature of the reservoir at infinity and with
fixed electric charge residing near the reservoir at infinity, in order
to satisfy the Gauss constraint of the electromagnetic field without
contributing to the energy content of the space. This hot flat space
in this zero loop approximation is simply classical flat space at some
temperature $T$ with no matter fields present.  The zero loop action
for classical hot flat space with electric charge at infinity is then
zero, i.e., $I_0[T,Q]=0$. The partition function $Z$ in the zero loop
approximation for the canonical ensemble is then
\begin{align}
Z=\mathrm{e}^{-I_0[T,Q]}\,,
    \label{eqch6:Zactionzeroloop}
\end{align}
with $I_0[T,Q]$ given in Eq.~\eqref{eqch6:actioninfinity}.

The partition function given in Eq.~\eqref{eqch6:Zactionzeroloop},
with the action described in Eq.~\eqref{eqch6:actioninfinity},
is valid for $d$ dimensions. 
In four dimensions, $d=4$, the partition function  
will give origin to Davies results \cite{Davies:1977bgr},
see also \cite{Hut:1977}. 
This means that
Davies' thermodynamic theory
of black holes,
in the case of electrically charged black holes, can be
seen within the canonical ensemble formalism. Here, 
the results are generalized to arbitrary $d$ dimensions, 
$d=4$ being a particular
case.

\section{Thermodynamics \label{sech6:thermo}}

\subsection{Thermodynamic quantities and properties}

We have used the
Gibbons\hskip0.02cm-\hskip-0.02cm{}Hawking Euclidean path 
integral approach
to construct the canonical ensemble
of an asymptotically flat spherically symmetric
electrically charged 
black hole space 
in arbitrary $d$ dimensions.
With the system being in
equilibrium with a heat reservoir at
infinity with temperature $T$ and electric charge $Q$,
the thermodynamics of the system can now be obtained by
considering that the partition function of the canonical ensemble is
related to the Helmholtz free energy $F$ through $Z =
\mathrm{e}^{-\frac{F}{T}}$, i.e., $F=-T\ln Z$.
From the zero loop approximation, 
Eq.~\eqref{eqch6:Zactionzeroloop},
this means $F=TI_0$.
With $I_0$ given in Eq.~\eqref{eqch6:actioninfinity}
one finds that the free energy is
$
F = \frac{1}{\mu}\left(\frac{r_+^{d-3}}{2} 
+ \frac{\mu Q^2}{2r_+^{d-3}}\right)
-\frac{\Omega_{d-2} r_+^{d-2}}{4}T$. 
Substituting $T$ for $t_H$, see Eqs.~\eqref{eqch6:betahawking}
and \eqref{eqch6:Hawkbeta}, one obtains for
the free energy the expression
\begin{align}
F(T,Q)
\hskip-0.05cm
=
\hskip-0.05cm
\frac{1}{\mu (d-2)}\hskip-0.1cm
\left(
\hskip-0.05cm
\frac{r_{+}^{d-3}}{2} 
+ (2d-5)\frac{\mu Q^2}{2r_{+}^{d-3}}
\hskip-0.05cm
\right)\hskip-0.05cm,
\label{eqch6:freeenergyinfinity}
\end{align}
where $r_+$ should be envisaged
as  $r_+=r_+(T,Q)$, since it is one of
the solutions 
$r_{+1}(T,Q)$ or
$r_{+2}(T,Q)$, given in 
Eq.~\eqref{eqch6:solutionr+1} or
Eq.~\eqref{eqch6:solutionr+2}, respectively.
Thus, the Helmholtz free energy $F$
for each solution is a function only of $T$ and $Q$,
namely, 
$F(T,Q,r_{+1}(T,Q))$ and $F(T,Q,r_{+2}(T,Q))$.

With the free energy $F$ given by Eq.~\eqref{eqch6:freeenergyinfinity},
one can obtain the thermodynamic quantities 
through its differential,
$dF = -S dT + \phi dQ$. The first component of the differential 
yields the entropy 
\begin{align}
S = \frac{A_+}{4l_p^{d-2}} \,,
\end{align}
where $A_+=\Omega_{d-2} r_+^{d-2}$ is the
area of the horizon,
and so $S$ is the Bekenstein-Hawking entropy, valid
for the two solutions 
$r_{+1}(T,Q)$ or
$r_{+2}(T,Q)$.
The second component of the differential yields the thermodynamic
electric potential 
\begin{align}
\phi = \frac{Q}{r_+^{d-3}}\,,
\end{align}
i.e., the Coulombic electric potential,
with  $r_+$ being $r_{+1}(T,Q)$ or
$r_{+2}(T,Q)$.
The thermodynamic energy, given by $E = F + TS$,
has the form 
$E = \frac{r_+^{d-3}}{2\mu} + \frac{Q^2}{2r_+^{d-3}}$,
and since  $r_+$ is $r_{+1}(T,Q)$ or
$r_{+2}(T,Q)$, there are two solutions for $E$.
This can be connected to the space mass $m$
given by
$m = \frac{r_+^{d-3}}{2\mu} + \frac{Q^2}{2r_+^{d-3}}$,
see Eq.~\eqref{eqch6:r+asmandq},
so
that here the thermodynamic energy and the black hole
mass
are equal, i.e., they obey the relation
\begin{align}
E = m\,.
\label{eqch6:energyinfinity}
\end{align}
Thus, one can write the free energy
given in Eq.~\eqref{eqch6:freeenergyinfinity} as
\begin{align}
F=m-TS\,.
\label{eqch6:freeenergyinfinityagain}
\end{align}

The energy in the
form
$E = \frac{r_+^{d-3}}{2\mu} + \frac{Q^2}{2r_+^{d-3}}$
can also be rewritten
in terms 
of the entropy and electric charge as
$E = \frac{1}{2\mu}\left(\frac{4 l_p^{d-2}S}{\Omega_{d-2}}
\right)^{\frac{d-3}{d-2}} 
+ \frac{Q^2}{2}\left(\frac{4 l_p^{d-2} S}{\Omega_{d-2}}
\right)^{\frac{3-d}{d-2}}$.
This energy function has the scaling property
$\nu^{\frac{d-3}{d-2}} E = E(\nu S, \nu^{\frac{d-3}{2(d-2)}} Q)$,
for some $\nu$, and so
through the Euler relation theorem,
one has $E= \frac{d-3}{d-2} TS + \phi Q$.
Together with Eq.~\eqref{eqch6:energyinfinity}, i.e.
$E=m$, one obtains 
\begin{align}
m = \frac{d-3}{d-2} TS + \phi Q\,,
\label{eqch6:smarrd}
\end{align}
which is the Smarr formula
for an electrically charged black hole in $d$ dimensions.
The Smarr formula is valid for the two solutions, $r_{+1}(T,Q)$ or
$r_{+2}(T,Q)$.

One can also verify that the first law
of thermodynamics,
\begin{align}
dm =  TdS + \phi dQ\,,
\label{eqch6:firstlawofbhm}
\end{align}
holds. It holds for the two solutions, $r_{+1}(T,Q)$ or
$r_{+2}(T,Q)$.
But, Eq.~\eqref{eqch6:firstlawofbhm} is also
the first law of black hole mechanics
since it involves pure black hole quantities.
This shows that
the thermodynamics that follow from the
electrically charged
canonical ensemble statistical mechanics
is equivalent to the thermodynamics that
follows from the first law of black hole mechanics.
The first law of black hole mechanics was the starting
point of Davies' analysis, while here it is a result of 
the  statistical mechanics formalism.

\subsection{Heat capacity and thermodynamic stability}

The thermodynamic stability of the system is given by the 
condition that the heat capacity
at constant electric charge must be positive, ensuring that 
the respective solution is stable. The heat capacity
at constant electric charge is
defined by 
$C_Q =\left( \frac{\partial E}{\partial T} \right)_Q$.
Since $E = m=\frac{r_+^{d-3}}{2\mu} + \frac{Q^2}{2r_+^{d-3}}$ and
$r_+=r_+(T,Q)$, one has
\begin{align}
C_Q& = \frac{1}{l_p^{d-2}}\frac{(d-2)\Omega_{d-2} r_+^{d-2}(r_+^{2d-6}-\mu Q^2)}
{4\left((2d-5)\mu Q^2 - r_{+}^{2d-6} \right)}\nonumber\\
&=
\hskip-0.1cm
\frac{m S^3  T}
{\frac{(d
\hskip-0.02cm
-
\hskip-0.03cm
3)\Omega_{\hskip-0.03cm d-2}^3}{4^5 l_p^{3d-6}\pi^2}
\hskip-0.15cm
\left[
\frac{
(\hskip-0.02cm
3d-8
\hskip-0.02cm
)
\mu^2 Q^4}
{\left(\frac{4 l_p^{d-2}S}{\Omega_{d-2}}\right)^{\hskip-0.06cm\frac{d-4}{d-2}}} 
\hskip-0.05cm
+
\hskip-0.05cm
(d
\hskip-0.02cm
-
\hskip-0.02cm
4)
\hskip-0.06cm
\left(\frac{4 l_p^{d-2} S}{\Omega_{d-2}}\right)^{\frac{3d-8}{d-2}}
\right]
\hskip-0.16cm
-
\hskip-0.11cm
T^2
\hskip-0.06cm
S^3}\hskip-0.02cm,
\label{eqch6:heatcapacityraw}
\end{align}
where in the
second equality, we have written
the heat capacity in terms of the thermodynamic 
variables $m$, $S$, $T$, and $Q$.
Note however that 
the heat capacity must be understood as a function of $\,T$ and 
$Q$, as these are the quantities controlled in the ensemble. 
This means that $r_+$ must be understood as either $r_{+1}(T,Q)$ 
or $r_{+2}(T,Q)$, as well as $m$ and $S$ must be understood as
$m=m(T,Q)$ and $S=S(T,Q)$.
The 
thermodynamic stability is satisfied if the heat capacity is 
positive. According to Eq.~\eqref{eqch6:heatcapacityraw}, the ensemble is 
stable in the range $r_{+e}\leq r_+ < r_{+s}$, where 
$r_{+e}= (\mu Q^2)^{\frac{1}{2d-6}}$ and 
$r_{+s}= \left(\sqrt{(2d-5)\mu}\,Q\right)^\frac{1}{d-3}$.
This is precisely the range 
of the solution $r_{+1}$. Therefore, one has
\begin{align}
{\mathrm{stability}\,\,\mathrm{if}}\; C_Q\geq0,\quad{\mathrm{i.e.,}} \quad
r_{+} = {r_{+1}}\,.
\label{eqch6:stability}
\end{align}
In opposition, the ensemble is unstable in the range 
$r_{+s}< r_+ < \infty$, which is the range of the solution $r_{+2}$. 
Hence, one has
\begin{align}
\hskip -0.3cm
{\mathrm{instability}\,\,\mathrm{if}}\; C_Q<0,\quad{\mathrm{i.e.,}} \quad
r_+ = r_{+2}\,.
\label{eqch6:instability}
\end{align}
So, from  Eqs.~\eqref{eqch6:stability} and
\eqref{eqch6:instability}
one has that the solution $r_{+1}$ is stable
whereas the solution $r_{+2}$ is unstable, see
Eqs.~\eqref{eqch6:solutionr+1} and \eqref{eqch6:solutionr+2}. We note
also that $r_{+1}$ is an increasing monotonic function of $T$, so that
the energy of the system increases when the temperature increases,
as it is expected from a stable system. The opposite happens to the
solution $r_{+2}$, since it is a decreasing monotonic function of $T$
and so the energy of the black hole decreases when the temperature
increases. 
From Eq.~\eqref{eqch6:stability}, one also finds that the radius
${r_{+s}}$ acts as the generalization of the Davies point for higher
dimensions. Indeed, for ${r_{+s}}$ fixed, for steady addition of
electric charge $Q$, one finds that the solution passes from an
$r_{+2}$ solution to an $r_{+1}$ solution, and eventually at the
transition, a negative $C_Q$ turns into a positive $C_Q$.
In thermodynamics, this could signal a phase transition of second
order, since the free energy $F$ and its first derivatives are
continuous, but second derivatives are discontinuous. However, this is
not the case here, we are instead in the presence of a turning point
which determines the relative scale of $r_+$ and $Q$
at which a black hole can
be in stable or metastable equilibrium
when in thermal
contact with a heat reservoir that holds $T$ and $Q$
fixed at infinity.
Indeed, in the canonical ensemble, the parameters that we can
control are $T$ and $Q$.  Maintaining $Q$ fixed, and for a given
sufficiently low $T$, there are two solutions, the stable solution
$r_{+1}(T,Q)$ and the unstable solution $r_{+2}(T,Q)$. We could try 
to start with the stable solution at low $T$ and 
devise a change of parameters $T$ and $Q$ such that $r_+$ 
was kept fixed. Eventually, we are able to reach the turning point 
and beyond it, the character of the solutions changes, i.e., the 
unstable solution $r_{+2}$ would have a fixed $r_+$, 
while the stable solution $r_{+1}$ still exists 
and would suffer a change in $r_+$. 
But any thermal perturbations would make the unstable solution $r_{+2}$
to run away from equilibrium, thus the unstable solution $r_{+2}$ is 
impossible to be maintained. And so, even for this specific change of 
parameters, with temperature up to $T_s$,
we are always in the presence of the stable solution $r_{+2}(T,Q)$, 
this being the existing solution of the ensemble at $T_s$, 
and so we should not classify 
this point as a phase transition.

Bear in mind that the thermodynamic quantities,
the first law of thermodynamics, and the Smarr formula as an
integrated first law of thermodynamics, are only
valid strictly for the stable black hole
solution $r_{+1}$, since the solution $r_{+2}$ is unstable
and does not allow a proper thermodynamic treatment.
Note also, that in the limit of
zero electric charge, $Q=0$, there is only the $r_{+2}$ solution 
corresponding to the
Gibbons\hskip0.02cm-\hskip-0.02cm{}Hawking black hole solution 
which is unstable. 
Indeed, the heat capacity in the zero electric charge case is $C =
-\frac{(d-2)\Omega_{d-2} r_+^{d-2}} {4l_p^{d-2}}$,
thus negative for all $r_+(T,Q)$.

\subsection{Favorable phases}

In a thermodynamic system, if different thermodynamic phases can take
place, we are interested to know which are the favored phases for a
given set  of parameters.  For temperature $T$ and electric charge
$Q$ fixed by the reservoir, a thermodynamic system tends to be in a
state in which its Helmholtz free energy $F$ has the lowest value.  If
a system is in a stable state but with a higher free energy $F$ than
another stable state, it is probable that the system undergoes 
a transition to the state with the lowest free
energy. Returning to the path integral calculation and the
corresponding partition function, one sees that if there are two stable
configurations, i.e., two states that minimize the action, then the
largest contribution to the partition function is given by the
configuration with the lowest action or, in thermodynamic language, with the
lowest free energy. In order to analyze these phase transitions, one 
must obtain the critical regions where the free energy is the same 
for both configurations. Generally, at these transition points,
the free energy's derivatives 
are different, signaling first order phase transitions.

In the case of a cavity within a heat reservoir at infinity
kept at $T$ and $Q$ constants,
we have seen that within the context of this chapter there are three solutions.
One is the stable black hole $r_{+1}$, Eq.~\eqref{eqch6:solutionr+1}, which
counts as a thermodynamic phase and exists for $T\leq T_s$. The other
is the unstable black hole $r_{+2}$, Eq.~\eqref{eqch6:solutionr+2}, which
also exists for $T\leq T_s$, but does not count as a thermodynamic
phase since it is unstable. The other is hot flat space with electric
charge at infinity
that exists for $T> T_s$, Eq.~\eqref{eqch6:solutionhot}. 
We have considered this phase, where there are no black holes, to be hot flat
space with electric charge dispersed at infinity, because it seems
the most natural solution, as electric charge of the same sign repels,
and eventually disperse to infinity but it can also be motivated by certain 
limits of charged matter configurations.

Thus, there are two possible phases, namely, the black hole $r_{+1}$
phase and hot flat space with electric charge at infinity.  For
$T>T_s$, only hot flat space with electric charge at infinity exists,
as seen above.  But for $T<T_s$, both $r_{+1}$ and hot flat space
with electric charge at infinity can exist. The one that is going to
dominate for $T\leq T_s$ is the one that has the lowest free energy.  Now,
the free energy of hot flat space with electric charge at infinity is
zero,
\begin{align}
F_\mathrm{ hfs}=0\,.
\label{eqch6:Fhfs}
\end{align}
The free energy of the $r_{+1}$ black hole is
always positive, $F(T,Q,r_{+1}(T,Q))>0$. This can be seen
from the on-shell expression
Eq.~\eqref{eqch6:freeenergyinfinity} which for the $r_{+1}$ solution reads
\begin{align}
F(T,Q,r_{+1}(T,Q))
\hskip-0.05cm
=
\hskip-0.05cm
\frac{1}{\mu (d-2)}\hskip-0.1cm
\left(
\hskip-0.05cm
\frac{r_{+1}^{d-3}}{2} 
+ (2d-5)\frac{\mu Q^2}{2r_{+1}^{d-3}}
\hskip-0.05cm
\right)\hskip-0.05cm.
\label{eqch6:Fr+1}
\end{align}
One finds that
Eq.~\eqref{eqch6:Fr+1}
has strictly positive 
terms. Thus, since
\begin{align}
F_\mathrm{
hfs}<F(T,Q,r_{+1}(T,Q))\,,
\label{eqch6:comparisonFs}
\end{align}
hot flat space with electric charge at
infinity is the favored phase for $T\leq T_s$.
If
the system finds itself in the black hole phase, it will
make a transition to hot flat space with electric charge
at infinity
since it has lower free energy.
We note however
that the free energy of these two phases never intersects
and so we cannot call
this a first order phase transition. An analog
to this transition is the one between supercooled
water and ice.
Moreover, hot flat space 
is the only phase for $T>T_s$.

\subsection{Interpretation}

We have deduced the thermodynamic results above starting from the path
integral approach. The
action that has entered into the path integral is the classical action,
corresponding thus to a zero loop approximation.
Although in this order of
approximation there is no mention of matter fields, which would enter
in a first loop approximation, we can try to interpret some of the
results found in zero order, in terms of wavelengths of packets of
thermal energy inside the cavity of a heat reservoir at infinity.
This is because there is a given temperature $T$ within the system,
and at a quantum level, for a given $T$, there is an
associated thermal wavelength $\lambda$, which is
$\lambda=\frac{(d-3)^2}{2\pi (2d-5)T}$.
The interpretation of the results in terms of matter 
fields is useful as we shall see now,
even if it is beyond the formalism used here.

We can start by interpreting the existence and nonexistence of the two
black hole solutions $r_{+1}$ and $r_{+2}$.
For small enough temperature $T$, and so large thermal wavelength
$\lambda$, there are two solutions for $r_+$.
The $r_+$ of the small solution is sufficiently small so that it is
smaller than $\lambda$, and so energy packets with typical wavelength
$\lambda$ are trapped in the black hole geometry and do not escape,
making the black hole a possible solution and a stable one.
The $r_+$ of the large solution is sufficiently large so that it is of
the order of $\lambda$, with $r_+$ being a bit larger,
and so energy packets with
typical wavelength $\lambda$ can escape, and backreact to turn the
black hole unstable.  Indeed, this case, with $r_+$ of the order
$\frac1T$ and so of the order of $\lambda$, corresponds to the 
black hole with the
Gibbons\hskip0.02cm-\hskip-0.02cm{}Hawking
black hole solution properties.
Now, for larger reservoir temperature $T$, the
thermal wavelength $\lambda$ gets smaller. The $r_+$ of the small
solution increases, now $r_+$ being barely smaller than
$\lambda$. The $r_+$ of the large solution
decreases,
with
$r_+$ being barely larger than $\lambda$. This latter solution
is still the one  with the
Gibbons\hskip0.02cm-\hskip-0.02cm{}Hawking 
black hole solution properties.
At a saddle or critical temperature $T_s$, the two solutions meet.
For even higher reservoir temperature
$T$, and so lower thermal wavelength
$\lambda$, there is no way to make a
black hole. The wavelength $\lambda$ is low enough that it
disperses without being able to aggregate the energy and the electric
charge in a black hole state.  In this case the electric charge
disperses to infinity, yielding hot flat space for the whole space
with the electric
charge at infinity, and so vanishing electric charge density.

We could try to interpret the favorable phases 
in terms of
wavelengths of packets of thermal energy inside the cavity, but we have not found 
a direct way to see
how the behaviour of these packets lead to hot flat space with electric charge
at infinity having always, for all parameters,
a free energy lower than the small black hole free energy.
However, it is clear what happens when one looks at the
free energy expressions.
Looking at the
original expression for the free energy of the
stable black hole, i.e.,
$
F = \frac{1}{\mu}\left(\frac{r_+^{d-3}}{2} 
+ \frac{\mu Q^2}{2r_+^{d-3}}\right)
-\frac{\Omega_{d-2} r_+^{d-2}}{4l_p^{d-2}}T$,
one sees that the entropy term which is negative has a small
contribution because $r_+$ is small, and there is the electric
charge term which goes as $\frac{Q^2}{2r_+^{d-3}}$ which
gives a large positive contribution, since $r_+$ is small,
all contributing for $F$ never being zero for any set of
parameters $T$ and $Q$.

To better understand all the issues
that we have worked out so far
and to make further progress, we have to pick up
definite dimensions. We specify the generic $d$-dimensional
results above to the case of $d=4$ and $d=5$ dimensions. 
We perform a thorough analysis for the dimension
$d=4$, while we briefly analyze the
case of $d=5$ dimensions.

\section{The case $d=4$: 
Davies' thermodynamic theory of black holes and
Davies point from the canonical ensemble\label{sech6:Daviesd4}}
\sectionmark{
    The case $d=4$: Davies' thermodynamic theory and 
    Davies point
    }\thispagestyle{userightbotmark}

\subsection{Solutions and action in $d=4$}

The dimension $d=4$ is specially interesting since it 
gives the results of Davies' thermodynamic theory of black holes
\cite{Davies:1977bgr}, see also \cite{Hut:1977}. 

We must 
start from the canonical ensemble characterized by a heat reservoir
at infinity with
temperature $T$ and electric charge $Q$
 in $d=4$. 
The black hole solutions $r_+$
of the ensemble are taken from solving Eq.~\eqref{eqch6:Hawkbeta}
together with Eq.~\eqref{eqch6:betahawking},
which in $d=4$ they yield
\begin{align}
T=t_H(r_+,Q),\quad\quad
t_H(r_+,Q)=\frac{r_+-\frac{l_p^2 Q^2}{ r_+}}
{4\pi r_+^2}\,,
\label{eqch6:Hawkbeta0d=4}
\end{align}
where again
$T$ is the temperature kept fixed at the reservoir
at infinity and 
$t_H(r_+,Q)$ is
the original Hawking function in $d=4$.
When the electric charge of the reservoir at infinity
is zero, $Q=0$, then 
$t_H(r_+,0)=\frac{1}
{4\pi r_+}$, which is the Hawking temperature of a
Schwarzschild black hole.
The electric charge $Q$ is 
the electric charge kept fixed at the reservoir
at infinity, and the black hole electric charge
$q$ 
must match it to have a consistent solution,
$q=Q$, see Eq.~\eqref{eqch6:QqHawk}.

\begin{figure}[h]
    \centering
    \includegraphics[width=0.7\linewidth]{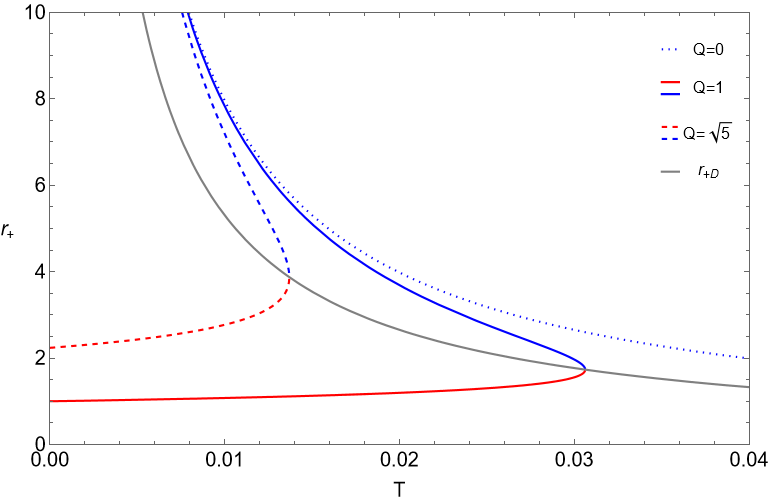}
    \caption{Plot of the two solutions $r_{+1}(T,Q)$, in red, and
    $r_{+2}(T,Q)$, in blue, of the charged black hole in the canonical
    ensemble for infinite cavity radius, as a function of $T$, for
    three values of the charge, $Q = 0$ in a dotted line, $Q = 1$ in
    filled lines, and $Q = \sqrt5$ in dashed lines, in $d=4$. The case
    $Q = 0$ is the Gibbons\hskip0.02cm-\hskip-0.02cm{}Hawking black
    hole, there is only the $r_{+1}(T,Q)$ solution, which is clearly
    unstable.  It is also plotted, in a gray line,
    the critical Davies radius as a
    function of $T$, ${r_+}_\mathrm{ D}=\frac{1}{6\pi T}$.
    \label{figch6:rtinfd4}}
    \end{figure}

To find the solutions of the canonical ensemble, one inverts 
Eq.~\eqref{eqch6:Hawkbeta0d=4}
to yield $\left( \frac{1}{4\pi T}\right)(r_+^2- l_p^2 Q^2) 
- r_+^3= 0$, which is Eq.~\eqref{eqch6:rpinT}
for $d=4$.
This equation
can be solved analytically as it is a cubic equation. 
However, we do not present the expression here. 
Alternatively, the solutions can be 
analyzed qualitatively or solved numerically. 
One can find that the function $t_H(r_+,Q)$
in Eq.~\eqref{eqch6:Hawkbeta0d=4}
has a maximum at 
$r_{+s}=\sqrt3\,Q$,
which is a saddle or critical point of the 
action of the black hole
and which it is defined as 
\begin{align}
{r_+}_\mathrm{ D}=\sqrt3\,l_p Q\,,
\label{eqch6:r+daviesd=4}
\end{align} 
since in $d=4$ it gives the Davies
black hole horizon radius.
This saddle point of the 
action is at the temperature
given by
\begin{align}
T_\mathrm{ D}=
\frac{1}
{6\sqrt{3}\,\pi l_p Q}\,,
\label{eqch6:Tdavies4d}
\end{align} 
see Eq.~\eqref{eqch6:Tdavies}, when $d=4$.
From Eqs.~\eqref{eqch6:r+daviesd=4} and
\eqref{eqch6:Tdavies4d}, one can eliminate
$Q$ to give for a given $T$,
${r_+}_\mathrm{ D}=\frac{1}{6\pi T}$,
or inverting, for a given $r_+$,
$T_\mathrm{ D}=\frac{1}{6\pi r_+}$.
The temperature given in Eq.~\eqref{eqch6:Tdavies4d}
is the Davies temperature, and it
is a result that can be extracted from \cite{Davies:1977bgr,Hut:1977}.
One finds that for temperatures $T
\leq T_\mathrm{ D}$ there are two black holes,
the solution
$r_{+1}(T,Q)$ and the solution $r_{+2}(T,Q)$, while
for $T > T_\mathrm{ D}$
there are no black hole solutions.
The solution $r_{+1}(T,Q)$ is bounded in the interval $ r_{+e}
<r_{+1}(T,Q)\leq {r_+}_\mathrm{ D}$, where $r_{+e}=r_{+1}(T\rightarrow
0,Q) = Q$ is the radius of the extremal black hole and ${r_+}_\mathrm{
D}=r_{+1}(T_\mathrm{ D},Q) = \sqrt3\, l_p Q $, so we can summarize for the
solution 1
\begin{equation}
\begin{aligned}
&r_{+1}=r_{+1}(T,Q),\quad\quad 0 <T \leq T_\mathrm{ D},
\\
& q_1=Q,\quad\quad\quad\quad\quad\quad\,
r_{+e} <r_{+1}(T,Q)\leq {r_+}_\mathrm{ D}.
\end{aligned}
\label{eqch6:solutionr+1d=4}
\end{equation}
This solution,
$r_{+1}(T,Q)$, is an increasing monotonic
function of $T$.
The solution $r_{+2}(T,Q)$ is
in the interval
${r_+}_\mathrm{ D}< r_{+2}(T,Q)<\infty$,
so we can summarize for the
solution 2
\begin{equation}
\begin{aligned}
&r_{+2}=r_{+2}(T,Q),\quad\quad 0 <T \leq T_\mathrm{ D},
\\
& q_2=Q,\quad\quad\quad\quad\quad\quad\,
{r_+}_\mathrm{ D}< r_{+2}(T,Q)<\infty,
\end{aligned}
\label{eqch6:solutionr+2d=4}
\end{equation}
where, 
at $T \rightarrow 0$, the 
solution tends to infinity there. This solution,
$r_{+2}(T,Q)$, is a decreasing monotonic
function of $T$.
When the ensemble is only characterized
by the temperature $T$, with $Q$ vanishing, $Q=0$,
only the black hole 
$r_{+2}$ survives which is the
Gibbons\hskip0.02cm-\hskip-0.02cm{}Hawking black hole 
solution.
For $T>T_\mathrm{ D}$, there are no black hole solutions and one
is left with hot flat space with electric charge
$Q$ dispersed at infinity, i.e.,
\begin{equation}
\begin{aligned}
&\mathrm{charged\; hot\;flat\;space},\quad\quad T_\mathrm{ D}<T<\infty,
\\
& Q \;\mathrm{dispersed \;at}\; r=\infty,\quad\quad
0\leq r<\infty.
\end{aligned}
\label{eqch6:solutionhot4d}
\end{equation}
We plot the two solutions $r_{+1}(T,Q)$ and $r_{+2}(T,Q)$
as functions of the temperature
in Fig.~\ref{figch6:rtinfd4} for two 
different values of the electric charge, which displays 
the features of the solutions just mentioned. 
For $T>T_\mathrm{ D}$, there are
no solutions, only hot flat space with electric charge
$Q$ at infinity.

The zero loop action 
of the canonical ensemble characterized by the
temperature $T$ and the electric charge $Q$
for $d=4$ can be
found using directly Eq.~\eqref{eqch6:actioninfinity},
i.e.,
\begin{align}
I_0[T,Q] = \frac{1}{2T}
\left(r_+
+ \frac{l_p^2 Q^2}{r_+}\right)
-\pi \frac{r_+^2}{l_p^2}
\,,
\label{eqch6:actioninfinityd=4}
\end{align}
where $\mu=l_p^2$ and $\Omega_2=4\pi$ were used.  The black hole
horizon radii $r_+$ that enter into this action are the $r_{+1}$ given
in Eq.~\eqref{eqch6:solutionr+1d=4} or the $r_{+2}$ given in
Eq.~\eqref{eqch6:solutionr+2d=4}.

\subsection{Thermodynamics in $d=4$}

With the solutions and the action of the canonical ensemble found, 
we can obtain the thermodynamics 
through the correspondence $F = T I_0$, where $F$ again is the 
Helmholtz free energy of the system.
From Eq.~\eqref{eqch6:actioninfinityd=4}, $F$ in $d=4$ is
$F =
 \frac{1}{2l_p^2}
\left(r_+
+ \frac{l_p^2 Q^2}{r_+}\right)
-T\,\pi \frac{r_+^2}{l_p^2}$,
which upon using Eq.~\eqref{eqch6:Hawkbeta0d=4}
gives
\begin{align}
F(T,Q)
=
\frac{1}{4l_p^2}
\left(
r_+ 
+  \frac{3l_p^2 Q^2}{r_{+}}
\right),
\label{eqch6:Fd=4}
\end{align}
where $r_+$ should be envisaged
as  $r_+=r_+(T,Q)$, since it is one of
the solutions 
$r_{+1}(T,Q)$ or
$r_{+2}(T,Q)$, given in
Eq.~\eqref{eqch6:solutionr+1d=4} or 
Eq.~\eqref{eqch6:solutionr+2d=4},
respectively.
From the derivatives of the free energy, 
one can obtain the entropy as 
$S = \pi \frac{r_+^2}{l_p^2}$, i.e., $S = \frac{A_+}{4l_p^2}$,
the electric potential, $\phi = \frac{Q}{r_+}$,
and 
the thermodynamic energy,
$E = \frac{1}{2l_p^2}
\left(r_+
+ \frac{l_p^2 Q^2}{r_+}\right)$, where $E = F + TS$ was used.
These expressions are valid
for both solutions, $r_{+1}$ and $r_{+2}$.
The expression for the energy is precisely the expression for
the space mass $m$, so
$E = m$. The free energy of
Eq.~\eqref{eqch6:Fd=4} is then $F=m-TS$.

Here, the Smarr formula in $d=4$ is clearly
\begin{align}
m = \frac12\, TS + \phi Q\,,
\end{align}
see Eq.~\eqref{eqch6:smarrd} for $d=4$.
Also, one has that the law
$dm =  TdS + \phi dQ$ holds, which ties 
the first law of black hole mechanics with the 
first law of thermodynamics.
The first law of black hole mechanics
is the expression from which Davies 
\cite{Davies:1977bgr} started his thermodynamic analysis,
see also \cite{Hut:1977}.
Our analysis here started from
the canonical ensemble theory and the
Euclidean path integral approach with
the action of Eq.~\eqref{eqch6:actioninfinityd=4}, which yields 
naturally the first law of thermodynamics.

The heat capacity $C_Q$ of Eq.~\eqref{eqch6:heatcapacityraw}, for $d=4$,
is given by
\begin{align}
C_Q =
\frac{1}{l_p^2}\frac
{2\pi r_+^2 \left(1-\frac{l_p^2 Q^2}{r_+^2}\right)} 
{3\frac{l_p^2 Q^2}{r_+^2}-1}=
\frac{m S^3  T}{\frac{\pi  Q^4}{4 l_p^{d-2}} -
T^2
S^3}\,,
\label{eqch6:heatcapacityraw4d}
\end{align}
where in the second equality the heat capacity was written in terms of
the thermodynamic variables $m$, $S$, $T$, and $Q$.
Note that $C_Q$ is a function of $T$ and $Q$, which are the
parameters that are controlled.
Thermodynamic stability is governed by the positivity of the heat
capacity, $C_Q\geq0$. From Eq.~\eqref{eqch6:heatcapacityraw4d}, one finds
that the range of stability is $r_{+e}\leq r_{+} < {r_+}_\mathrm{ D}$,
where $r_{+e}$ is the radius 
of the extremal black hole given by $r_{+e}=l_p Q$
and 
${r_+}_\mathrm{ D}$ is the Davies horizon radius 
given in Eq.~\eqref{eqch6:r+daviesd=4}.
This range for $r_+$
corresponds to the solution $r_{+1}$, and so one has
\begin{align}
\mathrm{ stability \,\, if}\; C_Q\geq0,\quad\mathrm{ i.e.,}\quad
r_+=r_{+1}\,.
\label{eqch6:stabilityRinfinityd=4}
\end{align}
Since  ${r_+}_\mathrm{ D}=\sqrt3 l_p\,Q$, 
Eq.~\eqref{eqch6:stabilityRinfinityd=4} is equivalent to
$l_p Q\geq\frac1{\sqrt3}r_+$, i.e., one has
$\frac1{\sqrt3}r_+\leq l_p Q\leq r_+$,
the latter term being the extremal case.
Now, the relation between the horizon radius, the
mass, and the electric charge of
the black hole is 
$r_+=l_p^2 m+\sqrt{l_p^4 m^2-l_p^2 Q^2}$, so $l_p Q\geq\frac1{\sqrt3}r_+$
is the same as $Q \leq l_p m\leq\frac{2}{\sqrt3} Q$,
which is another manner of writing the
condition for stability, and is 
the expression that can be found in \cite{Davies:1977bgr},
see also \cite{Hut:1977}. The heat capacity goes
to zero at the extremal case 
$\frac{l_p Q}{r_+}=1$.
Moreover, from Eq.~\eqref{eqch6:heatcapacityraw4d}, one finds
that the range of instability is
${r_+}_\mathrm{ D} <r_{+} <\infty$.
This range for $r_+$
corresponds to the solution $r_{+2}$, hence there is
\begin{align}
\hskip -0.3cm
\mathrm{ instability \,\,if}\; C_Q<0,\quad\mathrm{ i.e.,} \quad
r_+=r_{+2}\,.
\label{eqch6:instabilityRinfinityd=4}
\end{align}
The inequality on the horizon radius for the 
case of instability can be rewritten as $0\leq l_p Q< \frac1{\sqrt3}r_+$.
Note that
when the electric charge is zero, the heat capacity is negative
for all $r_+$,
indeed for $Q=0$ the heat capacity is 
$\frac{l_p^2 C}{r_+^2}
= -2 \pi$.
Note that $C_Q$ given in 
the second part of Eq.~\eqref{eqch6:heatcapacityraw4d}
is the same formula found in \cite{Davies:1977bgr} by 
performing in Eq.~\eqref{eqch6:heatcapacityraw4d} the redefinitions 
$S \rightarrow 8\pi S$,
$T \rightarrow \frac{1}{8\pi}T$ and 
$\frac{C_Q}{8\pi} \rightarrow C_Q$, and additionally by using Planck units.
In \cite{Hut:1977}, the conventions are yet different
from the ones we use here and from \cite{Davies:1977bgr}.

\begin{figure}[h]
\centering
\includegraphics[width=0.7\linewidth]{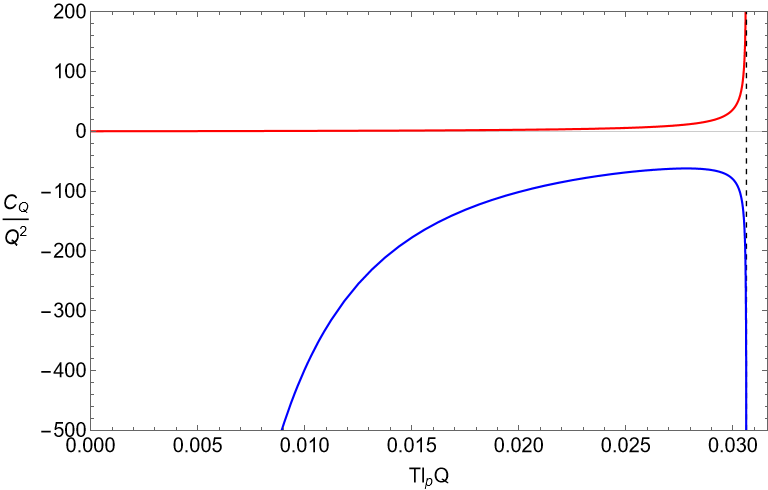}
\caption{The heat capacity $C_Q$ in $Q^2$ units,
$\frac{C_Q}{Q^2}$, is given as a function of the
temperature parameter $Tl_p Q$ in $d=4$, for the stable solution 
$r_{+1}$ in red and unstable solution $r_{+2}$ in blue. 
The heat capacity diverges for both solutions 
at the turning point $T_\mathrm{ D} l_p Q = \frac{1}{6\pi\sqrt{3}} = 0.03$, 
the latter equality being approximate.
\label{figch6:CAq4dRinfty}}
\end{figure}

The heat 
capacity $C_Q$
in units of $Q^2$, i.e.,
$\frac{C_Q}{Q^2}$, as a function
of the temperature
parameter, i.e.,  $T l_p Q$
is plotted in Fig.~\ref{figch6:CAq4dRinfty}.
For each $T l_p Q$, the heat capacity is double-valued, being 
positive for $r_{+1}$ in the red curve and negative for $r_{+2}$ in the 
blue curve. Therefore, the solution $r_{+1}$ is stable as it is 
expected from the increasing monotonic behavior of 
$r_{+1}$ with increasing temperature, while the solution 
$r_{+2}$ is unstable, having the opposite monotonic behavior. 
When $Q=0$, there is only the
$r_{+2}$ solution corresponding to the unstable 
Gibbons\hskip0.02cm-\hskip-0.02cm{}Hawking 
black hole solution. At the Davies point, corresponding to
$T_\mathrm{ D} l_p Q = \frac{1}{6\pi\sqrt{3}}$, 
the heat capacity goes to plus
infinity for the solution $r_{+1}$, and to
minus
infinity for the solution $r_{+2}$.
If, for some $T$, the configuration of the ensemble happens to be in 
the unstable $r_{+1}$ solution, then it will transition to
the stable $r_{+2}$, since any thermal perturbations make the 
solution $r_{+2}$ run away from equilibrium. 
This happens for all temperatures
up to $T_\mathrm{ D}$, where the two solutions coincide,
and for higher $T$, there are no more black hole
solutions. Thus, the point with temperature $T_\mathrm{ D}$ 
characterizes a turning
point. It was stated by Davies that such point might be 
classified as a second order phase transition. However, this 
cannot be the case for the canonical ensemble, as we discussed above, 
because only the stable solutions must be considered and the temperature 
$T_\mathrm{ D}$ signals the upper limit of existence of the stable solution. 
Another way of looking at the Davies point, through the ranges of the 
horizon radius, is that it provides the relative scale
between $r_+$ and $Q$ at which 
one has black hole stability or metastability
in the canonical ensemble with a heat reservoir at infinity.
In \cite{Davies:1977bgr}, a plot ${C_Q}\times Q$ was presented in some units
of ${C_Q}$ and of $Q$ at constant mass $m$,
whereas, here, we present the plot $\frac{C_Q}{Q^2}\times Tl_p Q$, 
where $T l_p Q$ is a temperature parameter, as $Q$ is kept
constant in the calculation of ${C_Q}$.

The analysis of the favorable thermodynamic states
for $d=4$ does not differ from the analysis for generic
$d$ given above. We summarize the analysis here for
completeness.
There are two possible phases, the stable black hole $r_{+1}$
phase and the phase with hot flat space with electric charged 
at infinity.
For $T\leq T_\mathrm{ D}$, both $r_{+1}$ and hot flat space
with electric charge at infinity can exist.
Since the free energy for hot flat space is zero
and the free energy for the black hole 
$r_{+1}$ is positive for all $T$ and $Q$,
hot flat space with electric charge at
infinity is the favored phase for $T\leq T_\mathrm{ D}$.
In this range of temperatures, 
if 
the system is in the black hole phase, it will
settle upon perturbation in 
a hot flat space with charge at infinity phase
which has lower free energy, in the same way that supercooled
water phase changes into ice.
For $T>T_\mathrm{ D}$, there are no black holes, 
hot flat space with electric charge at
infinity is the only phase.

The same interpretation in terms of wavelengths $\lambda$ of 
packets of thermal energy inside the cavity, that we gave above, can be applied 
to the specific case $d=4$. 
Note that this interpretation goes beyond the formal
results found here, since we carried out the zero loop
approximation and we do not treat quantum matter
fields. Nevertheless, it is beneficial to give an interpretation.
The essential idea is that at a given $T$ and so at a given $\lambda$,
the small black hole is smaller than $\lambda$ and the radiation is
trapped outside, while the large black hole is larger than $\lambda$
and the radiation can escape the black hole.  For sufficient high $T$,
there is too much agitation in packets of energy with small wavelength
$\lambda$, and these packets wonder undisturbed by gravity in hot flat
space with the electric charge being deposited uniformly at infinity.
In Fig.~\ref{figch6:rtinfd4}, the curve ${r_+}_\mathrm{ D}=\frac{1}{6\pi T}$
is drawn in gray, but this is the definition of $\lambda=
\frac{1}{6\pi T}$ for $d=4$. And so, Fig.~\ref{figch6:rtinfd4} describes
precisely the interpretation in terms of wavepackets given
above. Indeed, from small $T$ up to $T_\mathrm{ D}$, the gray curve is
larger than the horizon radius of the smaller black hole, while it is
smaller, although of the same order, than the horizon radius of the
larger black hole. At $T_\mathrm{ D}$, the gray curve and both solutions
meet. For larger temperatures than $T_\mathrm{ D}$, there are no black
hole solutions.

We must comment on the comparison between the approach we followed 
and the approach followed by Davies. The first law of black hole mechanics
is the expression from which Davies \cite{Davies:1977bgr} started his
analysis, see also \cite{Hut:1977}.  Our analysis here started from
the statistical mechanics canonical ensemble theory using the
Euclidean path integral approach and the action of
Eq.~\eqref{eqch6:actioninfinityd=4} rather than starting from the first
law of black hole thermodynamics.  In the Reissner-Nordstr\"om black
hole case in the canonical ensemble, as opposed to the Schwarzschild
case, there is true thermodynamics, since there are instances where
the system is thermodynamically stable.  This thermodynamic stability
of black holes for a heat reservoir at constant $T$ and $Q$ contrasts
with the thermodynamic instability of all electrically charged black
holes in a heat reservoir at constant $T$ and constant electric
potential $\phi$, i.e., Reissner-Nordstr\"om black holes in the grand
canonical ensemble.  This latter case was the case analyzed in
\cite{Gibbons:1977} using the Euclidean path integral approach for
the grand canonical ensemble, where this instability was noticed but
there was no attempt to cure the problem.  The appropriate setting
that gives a meaningful path integral and a corresponding
thermodynamics is within the electrically charged canonical ensemble
rather than the grand canonical one.

\section{The case $d=5$: A typical higher-dimensional case
\label{sech5:Daviesd5}}

\subsection{Solutions and action in $d=5$}

Here, we present the case with dimension $d=5$, as it is a
typical higher dimension, and it is the 
first possible extension of the results provided by Davies.

We must 
start from the canonical ensemble characterized by the
temperature $T$ and the electric charge $Q$
at infinity in $d=5$. 
The black hole solutions $r_+$
of the ensemble are taken from Eq.~\eqref{eqch6:Hawkbeta}
together with Eq.~\eqref{eqch6:betahawking}
which in $d=5$ give
\begin{align}
T=t_H(r_+,Q),\quad\quad
t_H(r_+,Q)=
\frac{r_+^2-\frac{4 l_p^3 Q^2}{3\pi r_+^2}}{2\pi r_+^3},
\label{eqch6:Hawkbeta0d=5}
\end{align}
where
$T$ is the temperature kept by the reservoir
at infinity and 
$t_H(r_+,Q)$ is
the Hawking function in $d=5$.
When the electric charge of the reservoir at infinity
is zero, $Q=0$, then 
$t_H(r_+,Q)=\frac{1}
{2\pi r_+}$, which is the Hawking temperature of a
Schwarzschild black hole in $d=5$.
The electric charge $Q$ is 
the electric charge kept by the reservoir
at infinity, and the black hole electric charge
$q$ 
must match it to have a consistent solution,
$q=Q$.

To find the solutions of the canonical ensemble, one inverts 
Eq.~\eqref{eqch6:Hawkbeta0d=5}
to yield
$\left( \frac{1}{2\pi T}\right)(r_+^4-\frac{4l_p^3}{3\pi} Q^2) 
- r_+^5 = 0$, which is Eq.~\eqref{eqch6:rpinT}
for $d=5$, a quintic
equation not easily  solvable analytically. 
 However, it can be 
analyzed qualitatively or solved numerically.
One finds that the function $t_H(r_+,Q)$
in Eq.~\eqref{eqch6:Hawkbeta0d=4}
has a maximum at 
\begin{align}
    r_{+s}= \left(\sqrt{\frac{20}{3\pi}}\,l_p^\frac{3}{2}Q\right)^\frac{1}{2}\,.
    \label{eqch6:r+sdaviesd=5}
\end{align} 
which is a saddle point of the 
action of the black hole,
with a corresponding temperature at the reservoir
given by
\begin{align}
T_s=
\frac{2}
{
5\pi
\left(\sqrt{\frac{20}{3\pi}}\,l_p^{\frac{3}{2}}Q\right)^{\frac12}
}
\label{eqch6:Tdaviesd=5}\,.
\end{align}
From Eqs.~\eqref{eqch6:r+sdaviesd=5} and
\eqref{eqch6:Tdaviesd=5}, one can eliminate
$Q$ to give for a given $T$,
$r_{+s}=\frac{2}{5\pi T}$,
or inverting, for a given $r_+$,
$T_s=\frac{2}{5\pi r_+}$.
One finds that for temperatures $T
\leq T_s$, there are two black hole solutions, 
the solution
$r_{+1}(T,Q)$ and the solution $r_{+2}(T,Q)$,
while for $T > T_s$
there are no black hole solutions.
The solution $r_{+1}(T,Q)$ is bounded in the interval $ r_{+e}
<r_{+1}(T,Q)\leq r_{+s}$, where $r_{+e}=r_{+1}(T\rightarrow
0,Q) = \left(\frac{2}{\sqrt{3\pi}}l_p^{\frac{3}{2}}Q\right)^{\frac{1}{2}}$ 
is the radius of the extremal black hole and
$r_{+s}=r_{+1}(T_s,Q) =
\left(\sqrt{\frac{20}{3\pi}}l_p^\frac{3}{2}Q\right)^\frac{1}{2}$,
so one can summarize 
solution 1
in the form
$r_{+1}=r_{+1}(T,Q)$,
$q_1=Q$, with
$0 <T \leq T_s$ and 
$r_{+e} <r_{+1}(T,Q)\leq r_{+s}$.
 This solution,
$r_{+1}(T,Q)$, is an increasing monotonic
function of $T$.
The solution $r_{+2}(T,Q)$ is
in the interval $ r_{+s}
<r_{+2}(T,Q)< \infty$,
so one can summarize
solution 2 in the form
$r_{+2}=r_{+2}(T,Q)$,
$q_2=Q$, with
$0 <T \leq T_s$
and
$r_{+s}< r_{+2}(T,Q)<\infty$,
where the 
solution tends to infinity at $T \rightarrow 0$. 
This solution, $r_{+2}(T,Q)$, is a decreasing monotonic
function of $T$.
When the ensemble is only characterized
by the temperature $T$, with $Q$ vanishing, $Q=0$,
only the black hole 
$r_{+2}$ survives which has the
Gibbons\hskip0.02cm-\hskip-0.02cm{}Hawking 
black hole solution 
properties.
For $T>T_s$, there are no black hole solutions and one
is left with hot flat space with electric charge
$Q$ dispersed at infinity, i.e.,
one has
charged
hot
flat
space for $T_s<T<\infty$ with $Q$
 dispersed at
$r=\infty$.

The zero loop action 
of the canonical ensemble, which is characterized by the
temperature $T$ and the electric charge $Q$
at infinity, for $d=5$ can be
found using directly Eq.~\eqref{eqch6:actioninfinity},
i.e.,
\begin{align}
I_0[T,Q] = \frac{1}{2T l_p^{3}}
\left(\frac{3\pi r_+^2}{4} 
+ \frac{l_p^3 Q^2}{r_+^2}\right)
-\frac{\pi^2 r_+^3}{2l_p^3}\,,
\label{eqch6:actioninfinityd=5}
\end{align}
where $\mu=\frac{4l_p^3}{3\pi}$
and $\Omega_3=2\pi^2$ were used.  The black hole
horizon radii $r_+$ that enter into this action are $r_{+1}$
or  $r_{+2}$.

\subsection{Thermodynamics in $d=5$}

With the solutions and the action
of the canonical ensemble found, 
we can obtain the thermodynamics 
through the correspondence $F = T I_0$, that comes from the 
zero loop approximation of the path integral, where $F$ again is the 
Helmholtz free energy of the system.
From Eq.~\eqref{eqch6:actioninfinityd=5},
$F$ in $d=5$  is
$
F = \frac{1}{2l_p^3}
\left(\frac{3\pi r_+^2}{4} 
+ \frac{l_p^3 Q^2}{r_+^2}\right)
-T\frac{\pi^2 r_+^3}{2l_p^3}\,,
$.
Substituting $T$ for $t_H$, see Eq.~\eqref{eqch6:Hawkbeta0d=5}, 
one obtains for
the free energy the expression
\begin{align}
F(T,Q)
=
\frac{\pi}{8l_p^3}
\left(
r_+^2
+ \frac{20 l_p^3 Q^2}{3\pi r_+^2}
\right),
\label{eqch6:Fd=5}
\end{align}
where $r_+$ should be envisaged
as  $r_+=r_+(T,Q)$, since it is one of
the solutions 
$r_{+1}(T,Q)$ or
$r_{+2}(T,Q)$.
Thus, the Helmholtz free energy $F$
for each solution is a function only of $T$ and $Q$,
namely, 
$F(T,Q,r_{+1}(T,Q))$ and $F(T,Q,r_{+2}(T,Q))$.
By computing the derivatives of the free energy,
one can obtain the entropy
as $S = \frac{A_+}{4l_p^3}$,
$A_+ = 2\pi^2 r_+^3$,
the 
thermodynamic
electric potential, which is $\phi = \frac{Q}{r_+^2}$,
and
the energy, which is 
$E = \frac{3\pi r_+^2}{8l_p^3} + \frac{l_p^3 Q^2}{2r_+^2}$, 
where it was used $E = F - TS$.
These expressions are valid
for both solutions, $r_{+1}$ and $r_{+2}$.
The energy has precisely the expression for
the space mass $m$, so
$E=m$.
The free energy of
Eq.~\eqref{eqch6:Fd=5} is then $F=m-TS$.

Here, in $d=5$, the Smarr formula takes the form
\begin{align}
m = \frac23\, TS + \phi Q\,.
\end{align}
Also, one has that the law
$dm =  TdS + \phi dQ$
holds. This is the first law of black hole mechanics, 
which is also the first law of black hole thermodynamics. 
And in fact, the first law of black hole thermodynamics is valid
in the electrically charged case
for the  instances where the
system is thermodynamically stable.

The heat capacity of Eq.~\eqref{eqch6:heatcapacityraw}
is in $d=5$ given by
\begin{align}
C_Q =&\frac{1}{l_p^3}
\frac
{3\pi^2  r_+^3 
\left(1-\frac{4}{3\pi}\frac{l_p^3 Q^2}{r_+^4}\right)} 
{2\left(\frac{20}{3\pi}\frac{l_p^3 Q^2}{r_+^4}-1\right)}\nonumber\\
 =&
\frac{m S^3  T}{\frac{7\pi^2}{36 l_p^3}Q^4
\left(\frac{2l_p^3 S}{\pi^2}\right)^{-\frac{1}{3}} 
+ \frac{\pi^4}{4^3}\left(\frac{2 l_p^3 S}{\pi^2}\right)^{\frac{7}{3}} -
T^2
S^3}\,,
\label{eqch6:heatcapacityraw5d}
\end{align}
where in the second equality, the heat capacity was written in terms of
the thermodynamic variables $m$, $S$, $T$, and $Q$.
The heat capacity must be seen as a function of $T$ and $Q$, with
$r_+$ being given by either the solutions $r_{+1}(T,Q)$ and $r_{+2}(T,Q)$,
or as well $m=m(T,Q)$ and $S=S(T,Q)$.
In order to have thermodynamic stability, the heat capacity must
be positive, i.e., $C_Q \geq0$, which is accomplished by the range
$r_{+e}\leq r_+\leq r_{+s}$ or in terms of electric charge
$\left(\frac{3\pi}{20}\right)^{\frac12} r_{+}^2\leq
l_p^\frac{3}{2} Q\leq\left(\frac{3\pi}{4}\right)^{\frac12} r_{+}^2$, with $r_{+s}$
given in Eq.~\eqref{eqch6:r+sdaviesd=5}. This range is precisely the one of
the solution $r_{+1}$, and so the solution $r_{+1}$ is stable.  For
the remaining range, satisfied by the solution $r_{+2}$, the heat
capacity is negative, thus the solution $r_{+2}$ is unstable.
The heat capacity, when the electric charge is zero, is negative for
all $r_+$,
given by
$\frac{l_p^3 C}{r_+^3} = -\frac{3\pi^2}{2}$.
The heat capacity has the feature that diverges for each solution
at $T_s$, which is a turning point of the two solutions.
The heat capacity goes to zero at the extremal case
$\frac{l_p^{\frac{3}{2}}Q}{r_+^2}=\left(\frac{3\pi}{4}\right)^{\frac12}$.  One can
also
infer that the solution is stable if the radius $r_+$
increases as the temperature increases,
yielding the same analysis above.

The analysis of the favorable thermodynamic states for $d=5$ follows
the same reasoning as for generic $d$.  There is the small stable
black hole phase and the hot flat space with electric charge at
infinity phase. Depending on the temperature, either the latter is favored or
it is the only phase.

An interpretation of the results in $d=5$ in terms of wavelengths
$\lambda$ of packets of thermal energy inside the cavity of a heat
reservoir at infinity also follows the analysis for generic $d$ given
above.

\section{Conclusions}
\label{sech6:Concl}

In this chapter, we have shown that the Gibbons\hskip0.02cm-\hskip-0.02cm{}Hawking
Euclidean path integral approach for electrically charged black holes
in the canonical ensemble has in its core the Davies'
thermodynamic theory of black holes. Since statistical mechanics
and its ensembles provide a deeper description of the
physics world, the results of this chapter place Davies'
thermodynamic theory on a firm basis.

To determine this connection, we computed 
the canonical partition function in the
Gibbons\hskip0.02cm-\hskip-0.02cm{}Hawking
Euclidean path integral approach for a Reissner-Nordstr\"om black hole
in $d$ dimensions.  The Euclidean action that enters into the path
integral consists of the Einstein-Hilbert-Maxwell action with the
Gibbons\hskip0.02cm-\hskip-0.02cm{}Hawking-York boundary term and an
additional Maxwell boundary term so that the canonical ensemble is
well-defined.  We have assumed that the heat reservoir resides at the
boundary of space, at infinite radius, where the temperature $T$ is
fixed as the inverse of the Euclidean proper time length at the
boundary, and also the electric charge is fixed by fixing the electric
flux at the boundary.
We then performed the zero loop approximation by giving the
expressions for the metric and the Maxwell tensor of the
Einstein-Maxwell system, obtaining the black hole solutions of the
ensemble, $r_+(T,Q)$.  We have shown that there are two solutions for
temperatures below a critical value.  The smaller black hole solution 
is stable, while the larger one is unstable.  The two solutions meet at a
saddle or critical point, given formally by $r_{+s}=r_+(T_s,Q_s)$.
Above the saddle value for the temperature, there are no black
hole solutions, only hot flat space with electric charge
dispersed at infinity.
The thermodynamics of the system follows, since the canonical
partition function connects directly to the Helmholtz free energy.
The entropy obtained from the free energy is the Bekenstein-Hawking
entropy, the electric potential is the usual Coulombic potential, and
the thermodynamic energy is the mass of the black hole. The
thermodynamic stability is controlled by the heat capacity at constant
electric charge, which must be positive for stable solutions and
negative for unstable solutions. There is a turning point 
precisely at the saddle values $T_s$ and $Q_s$.  The solution
with smaller radius is thermodynamically stable while the solution with 
larger radius is thermodynamically unstable.  
The Smarr formula relating mass, temperature, entropy,
electric potential, and electric charge follows naturally.  In
addition, the first law of thermodynamics reduces to the first law of
black hole mechanics, which, strictly speaking, is valid
only for the case of the stable solution.
We have studied the favorable phases, comparing the free energies of
the stable black hole and the hot flat space with electric
charge at infinity. We have obtained that hot flat space with electric
charge at infinity is favorable throughout the configuration
space. If, for some reason, the system finds itself in the black hole
phase, it will make a transition to
hot flat space with electric charge at infinity.
This fact is due to the 
black hole phase not being a global minimum
of the free energy $F$, the global mininum of $F$
being hot flat space with charge at infinity.
Since the free energy 
of these two phases never intersects, one
cannot call this a first order phase transition.
However, if one includes the matter sector, it may
be possible that a first order phase transition exists between black
hole and matter. We also gave an
interpretation for the solutions and their stability in terms of
wavelengths of energy packets.  By considering
the dimension $d=4$, we have shown
that Davies' thermodynamic theory of black holes follows directly from
the whole formalism presented.  Davies' starting point for the theory
was the first law of black hole thermodynamics, our starting point here was
the path integral approach with its action, and from it, we have deduced the
first law of black hole thermodynamics and the critical points found
by Davies.  The theory to the case $d=5$ was also applied. The analysis 
of this chapter
generically points towards the equivalence between the black hole
mechanics and black hole thermodynamics through the canonical
ensemble with an appropriate heat reservoir at infinity.

\chapter{Canonical ensemble of a d-dimensional
Reissner-Nordstr\"om black hole spacetime in a cavity \label{ch:canonicalblackhole}}\chaptermark{
    Canonical ensemble of a charged black hole in a cavity}\thispagestyle{userightbotmark}

\section{Introduction}

The York formalism~\cite{York:1986,Whiting:1988qr}, which uses the Euclidean 
path integral approach to quantum gravity~\cite{Gibbons:1977}, was applied 
to charged black holes to construct its grand canonical ensemble in~\cite{Braden:1990hw}, 
in the case of four dimensions.
One can also construct the canonical ensemble of charged black holes by adding 
a boundary term to the Einstein-Maxwell action, as explained in~\cite{Braden:1990hw}, 
with the analysis of the solutions being done in \cite{Lundgren:2006kt} and with 
the phase transitions between the black hole solutions having been 
done in \cite{Carlip:2003ne}, both in four dimensions. 
Also, in \cite{Lu:2010xt}, the canonical ensemble 
of charged black branes was analyzed.

As a continuity of the work presented in Chapters~\ref{ch:grandcanonicalblackhole}
and~\ref{ch:Daviesblackhole}, we construct in this chapter 
the canonical ensemble of a charged black hole inside 
a cavity for higher dimensions, with fixed 
temperature and electric charge. 
Again, we use the Euclidean path integral approach in the zero loop 
approximation to obtain the solutions of the ensemble and analyze the validity of the 
approximation. We generalize the results about the solutions of the ensemble 
for $d$ dimensions, giving analytical results for the bifurcation and meeting points. 
In sum, there are three solutions 
for electric charge lower than a critical charge, and there is only one solution for electric charge 
larger than a critical charge. Another novelty of the work 
centers on the analysis of the phase transitions between the stable black holes and 
a charged sphere with no gravity, that can model in a certain limit charged hot flat space, 
described by hot flat space with electric charge near the boundary of the cavity. 
From this analysis, there is a horizon radius at which the black hole phase starts 
to be more favorable. The comparison between this horizon radius and the 
Buchdahl-Andreásson-Wright bound~\cite{Wright:2015dma} is done, together also with the relevant horizon 
radius of the grand canonical ensemble. It is shown that both horizon radii 
do not correspond to the Buchdahl-Andreásson-Wright bound. This puts in question 
the link between matter dynamics and black hole thermodynamics for more general 
configurations than in~\cite{Andre:2021ctu}.
Further interpretation 
is given, and the particular analysis for the cases $d=4$ and $d=5$ is made.

This chapter is organized as follows. In Sec.~\ref{sech5:canonicalpathint}, we construct the 
partition function for spherically symmetric metrics with a Maxwell field. 
In Sec.~\ref{sech5:zeroloop}, we perform the zero loop approximation, 
and, we present the analysis of the solutions and their stability. 
In Sec.~\ref{sech5:thermo}, we obtain the thermodynamic quantities of the system 
from the canonical ensemble partition function. In Sec.~\ref{sech5:favorablephases}, 
we make a comparison between the solutions regarding their favorability 
and we show the presence of phase transitions. In Sec.~\ref{sech5:infinitelimit}, 
we analyze the limit of infinite cavity, recovering the results of 
Chapter~\ref{ch:Daviesblackhole}. In Sec.~\ref{sech5:BHzerofreeenergy}, 
we compare the thermodynamic radii obtained here and in Chapter~\ref{ch:grandcanonicalblackhole} 
with the Buchdahl-Andreásson-Wright bound. Finally, in Sec.~\ref{sech5:concl}, 
we present the conclusions. The work in this chapter is based on~\cite{Tiago2024b}.

\section{The canonical ensemble of a charged black hole 
in the Euclidean path integral approach\label{sech5:canonicalpathint}}\sectionmark{
    The canonical ensemble of a charged black hole
    }\thispagestyle{userightbotmark}

\subsection{The partition function}

Here, we build the canonical ensemble of a charged black hole inside cavity, in 
$d$ dimensions, using the Euclidean path integral approach 
to quantum gravity. The partition function of the system is given by
\begin{align}\label{eqch5:partitionfunction}
    Z = \int Dg_{\alpha \beta}DA_{\gamma}\, 
    \mathrm{e}^{-I[g_{\mu\nu},A_{\sigma}]}\,\,,
\end{align}
where the integral of paths must be done over periodic $g_{\mu\nu}$ and 
$A_{\sigma}$ in imaginary time, see for more details Chapter~\ref{ch:Euclideanpathintegral}. 
The Euclidean action is written in this case as
\begin{align}
    &I[g_{\mu\nu},A_{\sigma}] = - \int_{\mathcal{M}}\left(\frac{R}{16\pi l_p^{d-2}} -
    \frac{(d-3)}{4\Omega_{d-2}} F_{\alpha \beta}F^{\alpha \beta}\right)\sqrt{g} d^dx 
    \notag \\&-
   \frac{1}{8\pi l_p^{d-2}}\int_{\partial M} (K-K_0) \sqrt{\gamma}d^{d-1}x 
   + \frac{(d-3)}{\Omega_{d-2}}\int_{\partial M}F^{\alpha \beta}A_{\alpha}n_\beta 
    \sqrt{\gamma}d^{d-1}x\,,
   \label{eqch5:EucAction1}
\end{align}
where $R$ is the Ricci scalar given by derivatives and second 
derivatives of the Riemannian metric $g_{\alpha \beta}$, 
$g$ is the determinant of $g_{\alpha \beta}$, 
$K$ is the trace of the 
extrinsic curvature of the boundary of the cavity 
defined as $K_{\alpha \beta}$, 
$K_0$ is the trace of the 
extrinsic curvature of the boundary of the cavity 
embedded in flat Euclidean space, 
$\gamma$
is the determinant of
the induced metric $\gamma_{ab}$
on the boundary of the cavity, $\Omega_{d-2}$ is the surface area of the 
unit $(d-2)$-sphere,
$F_{\alpha \beta} = \partial_\alpha A_\beta - \partial_\beta A_\alpha$ 
is the Maxwell
tensor given by derivatives of the vector
potential $A_\alpha$, 
$n_\alpha$ is the outward unit normal vector to the boundary
of the cavity.

The action in Eq.~\eqref{eqch5:EucAction1} 
can be written in terms of two separate actions $I = I_{g\mathrm{f}}
+ I_{q}$, whereas
\begin{align}
    & I_{g\mathrm{f}} =  -\frac{1}{16\pi l_p^{d-2}} \int_{\mathcal{M}}R\sqrt{g}d^dx 
    -\frac{1}{8\pi l_p^{d-2}}\int_{\partial M} (K-K_0) \sqrt{\gamma}
    d^{d-1}x\,,\\
    & I_{q} = \frac{(d-3)}{4\Omega_{d-2}} \int_M F_{\alpha \beta}F^{\alpha \beta}\sqrt{g} d^dx
    +  \frac{(d-3)}{\Omega_{d-2}}\int_{\partial M}F^{\alpha \beta}A_{\alpha}n_\beta 
    \sqrt{\gamma}d^{d-1}x\,\,.
\end{align}
The action $I_{g\mathrm{f}}$ is the gravitational action with 
a zero cosmological constant, while $I_q$ is the Maxwell action 
with an additional boundary term. The boundary term depending on 
the Maxwell tensor must be included so that the canonical 
ensemble may be prescribed, see~\cite{Braden:1990hw}. 
This term allows us to fix the electric flux given by the 
integral of the Maxwell tensor on a $(d-2)$-surface, 
which has the meaning of an electric charge. This can be seen by 
performing the functional variation of the action and fixing data such that the 
variations at the boundary vanish.

\subsection{Geometry and boundary conditions}

The path integral is formally performed by summing over the Riemannian metrics 
with fixed boundary data. We choose the boundary data to be 
compatible with the data of a spherical shell with 
finite radius
embedded in the Reissner-Nordstr\"om black hole spacetime. Namely, the 
boundary of the Riemannian space describes a spherically symmetric 
heat reservoir with fixed inverse temperature, defined by the total imaginary 
proper time of the boundary, and with fixed electric flux, meaning a fixed 
electric charge. Due to the spherical symmetry of the boundary, it 
is expected that the paths having spherical symmetry contribute the 
most to the path integral. In order to simplify the analysis and towards the 
zero loop approximation, the path integral is restricted to spherical 
symmetric metrics of the form 
\begin{align}\label{eqch5:metricspherical}
    ds^2 = b(u)^2 d\tau^2 + a(u)^2 du^2 + r(u)^2 d\Omega^2_{d-2}\,\,,
\end{align}
where $b(u)$, $a(u)$ and $r(u)$ are arbitrary smooth functions of $u$, 
the coordinates have the range $\tau \in \,]0,2\pi[$ and $u\in\,]0,1[$, 
and $d\Omega^2_{d-2}$ is the $(d-2)$--sphere line element.

Moreover, the path integral also includes a sum over the possible topologies 
of the Riemannian space. Each topology, in the case of a spherically symmetric 
metric, is related to a set of regularity conditions. In the line of the 
zero loop approximation, we select the black hole sector, which 
resumes into the following regularity conditions at $u=0$
\begin{align}\label{eqch5:regularitymetric}
    & b(0) = 0\,\,,\notag\\
    & r(0) = r_+\,\,,\notag\\
    & \eval{(b'a^{-1})}_{u=0} = 1\,\,,\notag\\
    & \eval{a^{-1}(b'a^{-1})'}_{u=0} = 0\,\,,\notag\\
    & \eval{\left(\frac{r'}{a}\right)}_{u=0} = 0\,\,,
\end{align} 
where $r_+$ is the horizon radius, and also where a prime denotes 
the derivative of a function in $u$, e.g. $b'=\frac{db}{du}$.
The boundary conditions are set at the boundary of space $\partial M$,
which is assumed to be a spherical shell located at $u=1$ 
with induced metric 
\begin{align}\label{eqch5:inducedmetricboundary}
    ds_{\partial M}^2 = b(1)^2 d\tau^2 + R^2 d\Omega^2_{d-2}\,,
\end{align}
having components fixed by the inverse temperature and the radius of 
the reservoir as
\begin{align}
    &  b(1)=\frac{\beta}{2\pi}\,, \notag\\
&   r(1) = R\label{eqch5:boundarycond}\,.
\end{align}

For the electromagnetic Maxwell field, 
due to spherical symmetry, the only nonvanishing components of the 
Maxwell tensor $F_{\alpha \beta}$ are $F_{u\tau} = -F_{\tau u}$. 
Note that we are assuming the non-existence of 
magnetic monopoles. Moreover, we choose 
the gauge such that the only nonvanishing component of the 
vector potential is $A_{\tau}(u)$. Therefore, the Maxwell tensor 
$F_{\alpha \beta}$ is described 
only by the term
\begin{align}
    F_{u\tau}(u) = \frac{d A_\tau(u)}{du}\,.
    \label{eqch5:FytauAtau}
\end{align}
At $u=0$, we impose the regularity condition 
\begin{align}
    A_{\tau}(0) = 0\,,\label{eqch5:Atau0}
\end{align}
which fixes completely the gauge of the Maxwell field. 
The boundary condition at $u=1$ for the Maxwell field 
consists on a fixed electric charge. The electric charge 
can be written in terms of the electric flux 
$\int_{\substack{u=1 \\ \tau = c}} F^{\alpha \beta}dS_{\alpha \beta} =
2 i\Omega_{d-2} Q$, where $c$ is a constant, $Q$ is the electric charge 
in the cavity, $dS_{\alpha \beta} = 2 u_{[\alpha} n_{\beta]}dS$ 
is the surface element of the $y=1$ and $\tau = c$ surface, 
$u_\alpha dx^\alpha = b d\tau$, $n_\alpha dx^\alpha = a dy$, and $dS$ is 
the surface volume. For this case, the boundary condition reduces 
to
\begin{align}
    &  \Bigl(b a 
    r^{d-2}\,F^{u\tau}\Bigr) (1) =
    - i Q\,.\label{eqch5:Fcond}
\end{align}

\subsection{Action in spherical symmetry}

With the restriction to spherical symmetric metrics, the regularity 
conditions and the boundary conditions, we can simplify  
the action in the path integral. We can start with the 
gravitational action, which can be simplified into 
\begin{align}\label{eqch5:actionspherical2}
    & I_{g\mathrm{f}} =
     \left(\frac{2\pi b r^{d-3}}{\mu}
     \left(1 - \frac{r'}{a}\right)\right)\sVert[3]_{u=1}
    - \frac{\Omega_{d-2}}{4l_p^{d-2}}\left(\frac{b' r^{d-2}}{a}\right)\sVert[3]_{u=0}\notag\\
    &+ \frac{1}{8\pi l_p^{d-2}}\int_{M} a b r^{d-2}G\indices{^\tau_\tau}d^dx\,\,,
\end{align}
where 
\begin{align}\label{eqch5:mu}
    \mu = \frac{8\pi l_p^{d-2}}{(d-2)\Omega_{d-2}}\,\,,
\end{align}
and 
the Einstein tensor component $G\indices{^\tau_\tau}$ is given by
\begin{align}\label{eqch5:einsteinHam}
    G\indices{^\tau_\tau} = \frac{(d-2)}{2r'r^{d-2}}\left(r^{d-3}\left(\frac{r^{\prime2}}{a^2} 
    -1\right)\right)'\,\,.
\end{align}
Using the regularity conditions in Eq.~\eqref{eqch5:regularitymetric} and 
the boundary conditions in Eq.~\eqref{eqch5:boundarycond}, the gravitational 
action can be written as 
\begin{align}\label{eqch5:actionspherical3}
    & I_{g\mathrm{f}} =
     \left(\frac{\beta R^{d-3}}{\mu}
     \left(1 - \frac{r'}{a}\right)\right)\sVert[3]_{u=1}
    - \frac{\Omega_{d-2} r_+^{d-2}}{4l_p^{d-2}}
    &+ \frac{1}{8\pi l_p^{d-2}}\int_{M} a b r^{d-2}G\indices{^\tau_\tau}d^dx\,\,.
\end{align}
Regarding the action for the Maxwell field, one can use that 
$F^{\alpha \beta}F_{\alpha \beta} = 2 F_{u\tau}F^{u\tau} = 2\frac{A_\tau'^2}{b^2a^2}$
and also that $F^{\alpha \beta}A_{\alpha}n_\beta = - \frac{A'_\tau}{b^2 a} A_\tau$
to obtain 
\begin{align}
    I_q = - \frac{(d-3)}{2\Omega_{d-2}}\int_M \left(r^{d-2}\frac{A_\tau^{\prime 2}}{a b} 
    + 2\left(\frac{A_\tau' r^{d-2}}{ab}\right)'A_\tau \right)d^d x\,\,,
\end{align}
where the regularity condition $A_\tau(0)=0$ was used and the boundary term 
was transformed into a bulk integration term. The full action for the 
spherically symmetric metric with a Maxwell field in the canonical ensemble 
is then 
\begin{align}\label{eqch5:actionspherical5}
    &I = \left(\frac{\beta R^{d-3}}{\mu}
    \left(1 - \frac{r'}{a}\right)\right)\sVert[3]_{u=1}
   - \frac{\Omega_{d-2} r_+^{d-2}}{4l_p^{d-2}} 
   - \frac{(d-3)}{\Omega_{d-2}}\int_{M} \left(\frac{r^{d-2} A'_\tau}{b a}\right)' A_\tau d^dx
   \notag \\
   &+ \frac{1}{8\pi l_p^{d-2}}\int_{M} a b r^{d-2}
   \left(G\indices{^\tau_\tau} - 4\pi l_p^{d-2}\frac{(d-3)}{\Omega_{d-2}}
   \frac{A^{\prime2}_\tau}{b^2 a^2}\right) d^dx \,\,.
\end{align}
The statistical path integral that yields the partition function can then be 
written as
\begin{align}
    Z = \int Db Da Dr DA_\tau \mathrm{e}^{-I}\,\,,
\end{align}
with the action in Eq.~\eqref{eqch5:actionspherical5}.
For more details about the statistical ensemble through the Euclidean path integral 
approach, the
gravitational action in spherical symmetry, the regularity and boundary conditions, 
one can find them in Chapter~\ref{ch:Euclideanpathintegral}.

\section{The zero loop approximation\label{sech5:zeroloop}}

\subsection{The constrained path integral and reduced action in 
the canonical ensemble}

Given the action and the path integral for a spherically symmetric metric 
with a Maxwell field, we can proceed with the zero loop approximation 
through incremental steps. First, we impose the Hamiltonian and 
momentum constraints to the metric and the 
Gauss constraint to the Maxwell field. 
This results in a constrained path integral with a reduced 
action. We then can use the reduced action to study the validity of the 
zero loop approximation under static perturbations, which have a connection 
to the thermodynamic stability as we will show.

Starting with the constraints for the metric, 
the Hamiltonian constraint is 
$G\indices{^\tau_\tau} = 8\pi l_p^{d-2} T\indices{^\tau_\tau}$, with
$G\indices{^\tau_\tau}$ given by
Eq.~\eqref{eqch5:einsteinHam},
and
\begin{align}
    T\indices{^\tau_\tau} = \frac{(d-3)}{\Omega_{d-2}}
\frac{A'^{2}_\tau}{2a^2 b^2}\,\,,
\end{align}
where $T\indices{^\tau_\tau}$ is the time-time
component of the stress-energy tensor $T\indices{^\alpha_\beta}$. 
Thus, the Hamiltonian constraint is
\begin{align}
\frac{d-2}{ 2r' r^{d-2}}\left[r^{d-3}
\left(\frac{r'^2}{a^2} -1\right) \right]' = 
\frac{4\pi l_p^{d-2}(d-3) A'^{2}_\tau}{\Omega_{d-2}a^2 b^2}\,.
\label{eqch5:Hamconsteq}
\end{align}
The momentum constraint is trivially satisfied since 
the metric Eq.~\eqref{eqch5:metricspherical} is diagonal 
and does not depend on 
the imaginary time $\tau$.
The Gauss constraint is $\nabla_u F^{\tau u} = 0$,
which explicitly is 
\begin{align}
\left(\frac{r^{d-2}A'_\tau}{b a}\right)' =0\,,
\label{eqch5:Gaussconstr0}
\end{align}
The two constraint equations, Eqs.~\eqref{eqch5:Hamconsteq}
and \eqref{eqch5:Gaussconstr0},
are coupled, but they can be integrated in the following way.
It is better to 
start first by integrating Eq.~\eqref{eqch5:Gaussconstr0}.
Its integration yields 
\begin{align}
    A'_\tau = -i\frac{ q\, }{r^{d-2}}b a\,,
    \label{eqch5:Gaussconstr}
\end{align}
where $q$ is an integration constant. If one evaluates 
Eq.~\eqref{eqch5:Gaussconstr} at $u=1$ and uses the boundary 
condition Eq.~\eqref{eqch5:Fcond}, then one obtains that 
\begin{align}
    q=Q\,\,,\label{eqch5:qQ}
\end{align}
and so the integration constant $q$ of the Gauss constraint is 
precisely the fixed electric charge $Q$ of the ensemble. 
From this point onward, $Q$ is used to described the fixed electric 
charge. By using 
Eq.~\eqref{eqch5:Gaussconstr} and Eq.~\eqref{eqch5:qQ}, the Hamiltonian 
constraint becomes 
\begin{align}
    \frac{d-2}{2 r' r^{d-2}}\left[r^{d-3}
\left(\frac{r'^2}{a^2} -1\right) \right]' = 
- \frac{4\pi (d-3) l_p^{d-2} Q^2}{\Omega_{d-2}r^{2d-4}}\,\,,
\end{align}
which can be 
integrated to obtain
\begin{align}
\frac{r'^2}{a^2} \equiv f(r,Q,r_+) 
\,,\label{eqch5:rprimealpha0}
\end{align} 
where
\begin{align}
f(r,Q,r_+) \equiv 1 - \frac{r_+^{d-3} 
+ \frac{\mu Q^2}{r_+^{d-3}}}{r^{d-3}} + \frac{\mu Q^2}{r^{2d-6}}
\,,
\label{eqch5:rprimealpha}
\end{align} 
with $\mu$ given in Eq.~\eqref{eqch5:mu}.
We define the function 
$f$ in Eq.~\eqref{eqch5:rprimealpha}
for convenience, and the regularity 
conditions in Eq.~\eqref{eqch5:regularitymetric}  
were used to determine the integration constant $r_+$.
Although 
the second to last condition in Eq.~\eqref{eqch5:regularitymetric} 
is not used anywhere, notice for bookkeeping that, 
if $u=r$ is chosen, $r' = 1$ and $a$ diverges at 
$r=r_+$, therefore the condition
should be satisfied if
$\left(\frac{b'}{a}\right)'_{\hskip-0.1cm u=0}$
is finite. 
The function 
$A'_\tau$ in Eq.~\eqref{eqch5:Gaussconstr} is related to the 
Coulomb electric field in Lorentzian curved spacetime as
$n_\alpha E^\alpha = \frac{i A'_\tau}{b a} = \frac{Q}{r^{d-2}}$, 
where $E^\alpha$ is the electric field measured by a static observer.
It is important to write explicitly the extremal case,
i.e., when $r_{+}^{2d-6}=
\mu Q^2 $. The horizon radius for the extremal case, ${r_{+}}_e$, can be 
defined as
\begin{align}
{r_{+}}_e=\left(\mu Q^2\right)^{\frac{1}{2d-6}} 
\,,
\label{eqch5:rprimealphaextremal}
\end{align}
and function $f(r,Q,r_+)$ in Eq.~\eqref{eqch5:rprimealpha}
in the extremal case is
$f(r,Q,{r_{+}}_e) = \hskip-1mm\left( 1 
\hskip-1mm- \hskip-1mm\frac{\sqrt{\mu}Q}{r^{d-3}}\right)^2$.

The Hamiltonian, momentum, and Gauss
constraints
simplify
the action in Eq.~\eqref{eqch5:actionspherical5} considerably. 
One can see that the last term in Eq.~\eqref{eqch5:actionspherical5}
has an integrand proportional to ${G^\tau}_\tau - 8\pi{T^\tau}_\tau$
and so, applying the Hamiltonian constraint
given in Eq.~\eqref{eqch5:Hamconsteq}, this term vanishes.
Moreover, the third term in Eq.~\eqref{eqch5:actionspherical5} is 
proportional to $\left(\frac{r^{d-2}A'_\tau}{b\alpha}\right)'$
which vanishes also if the Gauss constraint
given in Eq.~\eqref{eqch5:Gaussconstr0}
is applied.
Therefore, the action Eq.~\eqref{eqch5:actionspherical5} becomes 
the reduced action $I_*$ written as
\begin{align}
    I_*[\beta,Q,R;r_+] = \frac{\beta R^{d-3}}{\mu}(1 - \sqrt{f(R,Q,r_+)}) 
    - \frac{\Omega_{d-2} r_+^{d-2}}{4l_p^{d-2}}\,,
    \label{eqch5:actioninrp}
\end{align}
which is the Euclidean action 
evaluated on the paths that obey the Hamiltonian and Gauss 
constraints, where $(r'\alpha^{-1})_{y=1}$ was substituted by the 
solution to the Hamiltonian constraint 
given in Eq.~\eqref{eqch5:rprimealpha0}.
From Eq.~\eqref{eqch5:rprimealpha},
one has that $f(r,Q,r_+)$ appearing in 
 Eq.~\eqref{eqch5:actioninrp} evaluated at the cavity
 radius $R$ is given by
\begin{align}
f(R,Q,r_+) \equiv 1 - \frac{r_+^{d-3} 
+ \frac{\mu Q^2}{r_+^{d-3}}}{R^{d-3}} + \frac{\mu Q^2}{R^{2d-6}}
\,.
\label{eqch5:Rprimealpha}
\end{align} 
The function $f(R,Q,r_+)$ for the extremal case characterized 
by Eq.~\eqref{eqch5:rprimealphaextremal}
is given by 
$f(R,Q,{r_{+}}_e)  = \left( 1 
- \frac{\sqrt{\mu}Q}{R^{d-3}}\right)^2$.

The Hamiltonian, momentum, and Gauss
constraints, 
together with the boundary conditions and the 
requirement of spherical symmetry, restrict the path integral 
considerably. The Riemannian space is determined by the functional 
$r_+$, so the path integral is the sum of spaces with all 
possible $r_+$.
Indeed, the partition function is given by the path integral
\begin{align}
Z = \int Dr_+ \mathrm{e}^{-I_*[\beta,Q,R;r_+]}\,,
\label{eqch5:partitionf1}
\end{align}
where $I_*[\beta,Q,R;r_+]$ 
is the reduced action described in Eq.~\eqref{eqch5:actioninrp}. 
There is formally another functional, the Maxwell field
$A_\tau$, but the action does not depend
explicitly on $A_\tau$, it only depends on the electric charge 
which is fixed at the cavity. This means the integration over paths 
of $A_\tau$ can be absorbed by a normalization and thus yielding 
no additional contributions to the constrained path integral.

\subsection{Stationary points of the reduced action\label{sech5:Canonical1}}

Having the constrained path integral in Eq.~\eqref{eqch5:partitionf1}, 
we can perform the zero loop approximation, which takes into 
consideration only the paths that minimize the action. The partition 
function in the zero loop approximation is given by
\begin{align}
Z[\beta, R, Q] = \mathrm{e}^{-I_{0}[\beta,R,Q]}\,,
\label{eqch5:partitionf2}
\end{align}
where
\begin{align}
I_{0}[\beta,R,Q] = I_*[\beta,R,Q; r_+[\beta,R,Q]]\,,
\label{eqch5:I0}
\end{align}
is the 
action in Eq.~\eqref{eqch5:actioninrp} evaluated at the path
that minimizes the action with 
respect to $r_+$. The function $r_+[\beta,R,Q]$ corresponds
to a black hole solution that is in thermal equilibrium with the cavity and 
it is determined by
a stationary point of the action, i.e.,
$\left(\frac{\partial I_*}{\partial r_+}\right)_{r_+=r_+[\beta,R,Q]} = 0$. 
Using Eq.~\eqref{eqch5:actioninrp}, the stationary condition reduces to 
the equation 
\begin{align}
    \beta = \iota(r_+)\,,\,
    \iota(r_+)\equiv  \frac{4\pi}{(d-3)}\frac{r_+^{d-2}}{r_+^{d-3}
    - \frac{\mu Q^2}{r_+^{d-3}}}\sqrt{f(R,Q,r_+)}\,,
\label{eqch5:beta1}
\end{align}
where $\iota(r_+)$
is the inverse temperature function, defined here for convenience.
The function $\iota$ for fixed $R$ and $Q$, which are the fixed 
quantities of the ensemble, only depends on $r_+$ alone. 
The solutions $r_+[\beta,R,Q]$ of Eq.~\eqref{eqch5:beta1}
are the stationary points or the paths that minimize
the action in Eq.~\eqref{eqch5:actioninrp},
and they are obtained from inverting 
Eq.~\eqref{eqch5:beta1}. For convenience,
we can define
a horizon radius parameter $x$ and an electric
charge parameter $y$ as
\begin{align}
x = \frac{r_+}{R}\,,\quad\quad\quad\quad
y= \frac{\mu Q^2}{R^{2d-6}}
\,.\label{eqch5:newv}
\end{align}
Rearranging
Eq.~\eqref{eqch5:beta1}, one obtains
\begin{align}
\hskip -0.20cm
(x^{2d-6}
\hskip -0.15cm
-
\hskip -0.08cm
y)^2
\hskip -0.08cm
\left(
\hskip -0.08cm
\frac{(d-3)\beta}{4\pi R}
\hskip -0.08cm\right)^2
\hskip -0.25cm
-
\hskip -0.05cm
x^{3d-7} (1
\hskip -0.08cm
 -
\hskip -0.08cm
 x^{d-3})(x^{d-3}
\hskip -0.15cm
 -
\hskip -0.08cm
y) 
\hskip -0.1cm
 =
\hskip -0.1cm
 0.\hskip -0.1cm
\label{eqch5:sols1}
\end{align}
The equation above for the horizon radius, Eq.~\eqref{eqch5:sols1},
can be reduced at most to sixth polynomial order 
for $d=5$, while for other dimensions 
the polynomial order is higher. We have not found an analytical solution
for any specific value of $d$. We note that
the non-extremal condition for the black hole
can be put in the form
\begin{align}
x_e\leq x \leq 1\,, 
\label{eqch5:extremal1}
\end{align}
where $x_e$ is the extremal $x$ related to the extremal $y$,
denoted as $y_e$, by 
\begin{align}
y_e= x_e^{2d-6}\,,
\label{eqch5:extremal2}
\end{align}
see Eq.~\eqref{eqch5:rprimealphaextremal}.

\begin{figure}[h]
\centering
\begin{subfigure}{0.5\linewidth}
\centering
\includegraphics[width=\linewidth]{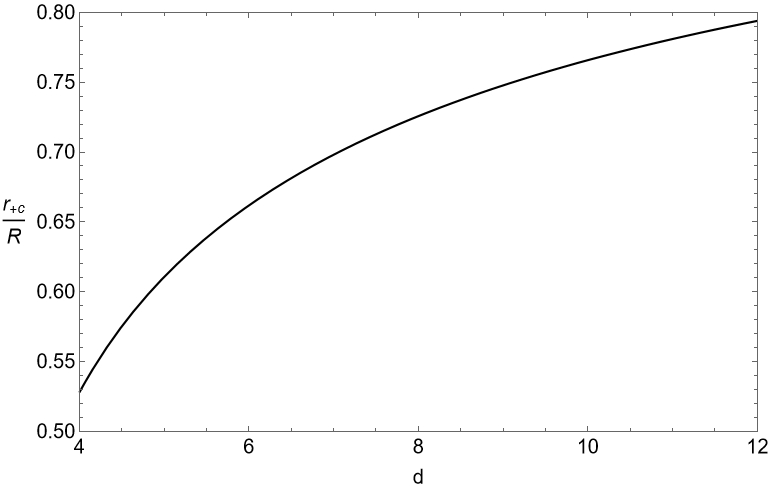}
\caption{\label{figch5:rcd}}
\end{subfigure}%
\begin{subfigure}{0.5\linewidth}
\centering
\includegraphics[width=\linewidth]{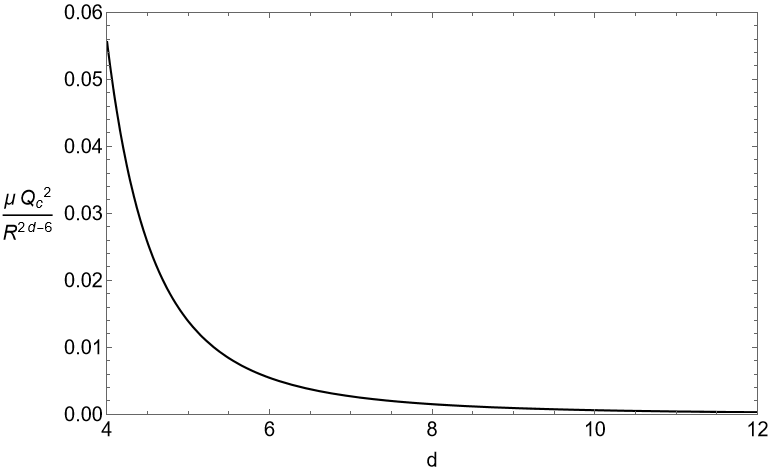}
\caption{\label{figch5:qcd}}
\end{subfigure}
\begin{subfigure}{0.5\linewidth}
\centering
\includegraphics[width=\linewidth]{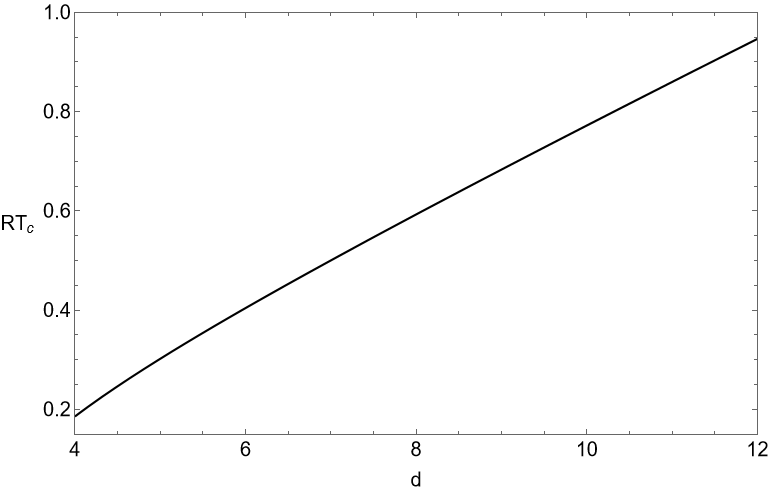}
\caption{\label{figch5:Tcd}}
\end{subfigure}
\caption{
Plots of the saddle point
$(x_s,y_s,T_s)$ of the action as functions of the
number of dimensions $d$. (a) Plot of $x_s = \frac{r_{+s}}{R}$ as a
function of $d$; (b) plot of $y_s = \frac{\mu Q_s^2}{R^{2d-6}}$ as a
function of $d$; (c) plot of $R T_s$ as a function of $d$. 
\label{figch5:saddlej}}
\end{figure}

Even though we may not
find the exact solutions for $x$, 
it is possible to obtain analytically
the limiting values for
the solutions. These limiting values are determined by
the saddle points of the action $I_*$ described as 
$\left(\frac{\partial^2 I_*}{\partial r_+^2}\right)_0 = 0$,
where the subscript $0$ 
means that the quantity inside parenthesis is evaluated at the 
stationary point. Now, 
$\left(\frac{\partial^2 I_*}{\partial r_+^2}\right)_0 = 
-\frac{\Omega_{d-2} (d-2) r_+^{d-3}}{4} \beta^{-1} 
\frac{\partial \iota}{\partial r_+}$, so
the saddle points of the action are given by
$\frac{\partial \iota}{\partial r_+}=0$
together with Eq.~\eqref{eqch5:beta1}.
This condition in terms of $x$ and $y$ yields the equation 
\begin{align}
    &\frac{d-1}{2}x^{4d-12} - (1+y)x^{3d - 9} - 3(d-3)y x^{2d-6} 
    \nonumber \\
    &+ (2d-5)y(1+y)x^{d-3} -\frac{3d-7}{2}y^2 = 0\,\,,
    \label{eqch5:critbeta}
\end{align}
which is a polynomial equation of order four 
in $x^{d-3}$ and it can be solved analytically. Its solutions are 
relevant in the qualitative behaviour of the horizon radius $x$ 
of the black hole in thermal equilibrium. For the electrically 
uncharged case $y=0$, the limiting values were
discussed in \cite{York:1986} for $d=4$,
 \cite{Andre:2020czm} for $d=5$,
and \cite{Andre:2021ctu} for generic $d$. For $0<y<y_s$, 
there are four real roots of Eq.~\eqref{eqch5:critbeta}, 
from which only two obey the non-extremal condition $0<y<x^{2d-6}$, 
and where
$y_s$ is a saddle or critical electric charge parameter to be
given below. The
two saddle points of the action, being the solutions
of interest of Eq.~\eqref{eqch5:critbeta}, are 
designated by $x_{s1} = x_{s1}(y)$ 
and $x_{s2} = x_{s2}(y)$, where $x_{s1}\leq x_{s2}$. Explicitly, 
they are given by the expressions
\begin{align}
&x_{s1}^{d-3} = \frac{1+y}{2(d-1)} + \xi- 
\frac{1}{2}\sqrt{2\eta + \frac{\zeta}{\xi} 
- 4 \xi^2}\,, \\
&x_{s2}^{d-3} = \frac{1+y}{2(d-1)} + \xi+ 
\frac{1}{2}\sqrt{2\eta + \frac{\zeta}{\xi} 
- 4 \xi^2}\,, 
\label{eqch5:xs1xs1}
\end{align}
where
\begin{equation}
\begin{aligned}
& \eta = \frac{3(1+y)^2 + 12(d-1)(d-3)y}{2(d-1)^2}\,,\\
& \zeta
\hskip-0.05cm
= 
\hskip-0.05cm
\frac{(1+y)}{(d-1)^3}
\left(y^2 - (4d^3 -24d^2 +48d - 30)y +1 \right)\,,\\
& \xi= \frac{1}{2}\sqrt{\frac{2}{3}\eta 
+ \frac{2}{3(d-1)}\frac{\sigma^2 + \sigma_0}{\sigma}}\,,\\
& \sigma = \left(\frac{\sigma_1 
+ \sqrt{\sigma_1^2 - 4\sigma_0^3}}{2}\right)^{\frac13}\,,\\
& \sigma_0 = 3(2d-5)y(1-y)^2\,,\\
& \sigma_1 = 54 (d-3)(d-2)^2(1-y)^2y^2\,\,.
\end{aligned}
\label{eqch5:defs}
\end{equation}
For the critical charge $y=y_{s}$, both saddle points merge into a single one.
The saddle point of the action at $y=y_s$ is designated by
 $x_{s}\equiv x_{s1}= x_{s2}$, 
 which is a saddle point with the feature
that the third derivative of the action 
also vanishes.
The saddle point $x_{s}\equiv x_{s1}= x_{s2}$ is given by 
\begin{align}
&x_s^{d-3} = \frac{1}{2(d-1)(2d-5)}\nonumber\\
&\times\bigg[(d-1)(3d-7)(3d^2-16d +22)\nonumber\\
& - 3\sqrt{3}(d-2)^2(d-3)\sqrt{(d-1)(3d-7)}\bigg]\,\,
\label{eqch5:xcd},
\end{align}
which occurs at $y=y_s$ given by
\begin{align}
& y_s = \frac{1}{4(d-1)(2d-5)^3(3d-7)}\nonumber\\
&\times\bigg[(d-1)(3d-7)(3d^2-16d + 22) \nonumber\\
& - 3\sqrt{3}(d-3)(d-2)^2\sqrt{(d-1)(3d-7)}\bigg]^2\,\,.
\label{eqch5:ycd}
\end{align}
We must be note that to $x_s$ corresponds an $r_{+s}$ through
$r_{+s}=x_sR$, and to 
 $y_s$ corresponds a $Q_{s}$ through
$Q_s= \frac{y_sR^{2d-6}}{\mu}$, where
the subscript
$s$ was not put in $R$ in these formulas
because, for finite $R$, one can always assume
$R$ fixed.
Putting the values given in Eqs.~\eqref{eqch5:xcd}
and \eqref{eqch5:ycd}
into Eq.~\eqref{eqch5:sols1},
one finds the temperature of the saddle point $RT_s$, 
\begin{align}
RT_s=RT_s(x_s,y_s)\,
\label{eqch5:RTs}
\end{align}
the temperature parameter at which 
$x_s$ is a solution of the black hole for $y=y_s$.
The values of $x_s$, $y_s$, and $R T_s$
are displayed for different values of $d$ 
in Fig.~\ref{figch5:saddlej}.
We can see that both $x_s$ and $RT_s$ increase as
$d$ increases, and $y_s$ decreases as $d$ increases.
For $ y_s<y<1$, there are no roots of 
Eq.~\eqref{eqch5:critbeta} that obey the non-extremal condition $0<y<x^{2d-6}$
and so there are no saddle points of the action.

Having the limiting values, 
we can now perform a qualitative analysis of the solutions for the 
horizon radius of the black hole in thermal equilibrium with the 
reservoir. For the uncharged case $y=0$, the analysis has been
discussed in \cite{York:1986} for $d=4$, \cite{Andre:2020czm} for $d=5$,
and \cite{Andre:2021ctu} for generic $d$. For $0<y<y_s$,
one can find that there are three 
solutions $x(\beta,y)$, or if one prefers $x(T,y)$,
of Eq.~\eqref{eqch5:sols1}. These three
solutions are designated by $x_1$, $x_2$,
and $x_3$. The solution $x_1$ exists in the interval of 
temperatures $0<T<T_1$
and it is bounded by $x_e<x_1(T,y)< x_{s1}(y)$, where 
the values of the solution at the bounds are 
$x_1(0,y)= x_e$,
with $x_e$ defined in Eq.~\eqref{eqch5:extremal2},
and $x_1(T_1,y) = x_{s1}(y)$, 
with $T_1$ being defined by the latter relation. 
The solution $x_2$ exists in the interval of temperatures
$T_1>  T > T_2$ and it is bounded by 
$x_{s1}(y) < x_{2}(T,y)< x_{s_2}(y)$, where the values of the 
solution at the bounds are
$x_{2}(T_1,y) = x_{s1}(y)$ 
and $x_{2}(T_2,y) = x_{s2}(y)$, with $T_2$ being defined 
by the former relation. The solution $x_{3}$ exists in the 
interval of temperatures $T_2 < T<\infty$, and 
it is bounded by $x_{s2}(y)<x_3 (T,y)< 1$, where the values of the 
solution at the bounds are 
$x_3 (T_2,y) = x_{s2}(y)$ and $x_3 (T\rightarrow \infty,y) = 1$.
As $y_s$ decreases with the increase of $d$, the 
region of existence of these solutions is 
squeezed towards lower values of the electric charge with an increase 
of $d$. For $y=y_{s}$, there are still three solutions $x_1$, $x_2$, and $x_3$, 
with the solution $x_2$ being
reduced to a point, more precisely to the 
saddle point of $\iota(r_+)$ given as $x_2 (T_s,y_s) = x_{s}$, with $T_s$ 
being defined by the latter relation. 
The bounds of $x_1$ and $x_3$ are the same as the case 
$0 < y < y_s$, except that $x_{s1}(y_s) = x_{s2}(y_s) = x_{s}$ and 
$T_s = T_1 = T_2$. For $y_s < y<1$, there is only one solution $x_4$ that 
exists for all $T$ and it is bounded by 
$x_e< x_4 (T,y) < 1$, where $x_4(0,y)
= x_e$ and $x_4 (T\rightarrow \infty,y) = 1$.

\subsection{Stability conditions\label{sech5:stabanalysis}}

To determine if the solutions are minima of the action and thus
stable, we must go beyond the zero loop approximation. This means 
we must expand the action and 
the path integral around the stationary point.
The action can be expanded as
$I_*= I_0 +
\left(\frac{\partial I_*}{\partial r_+}\right)_0 \delta r_+
+\left(\frac{\partial^2 I_*}{\partial r_+^2}\right)_0 
\delta r_+^2$,
where the subscript $0$ 
means that the quantity inside parenthesis is evaluated at the 
stationary point, $I_0= I_*(\beta,Q,R;(r_+)_0)$,
and 
$\delta r_+ = r_+ - (r_+)_0$.
Then, the partition function can be expanded as 
\begin{align}
Z = \mathrm{e}^{-I_0}\int D\delta r_+ 
\mathrm{e}^{- \left(\frac{\partial^2 I_*}{\partial r_+^2}\right)_0 
\delta r_+^2}\,.
\label{eqch5:oneloop}
\end{align}
The partition function in Eq.~\eqref{eqch5:oneloop} 
contains one loop contributions that obey the spherical symmetry 
of the geometry, the boundary conditions, and the Hamiltonian and 
Gauss constraints. For the path integral to be well-defined, 
one must have 
\begin{align}
\left(\frac{\partial^2 I_*}{\partial r_+^2}\right)_{0} > 0\,,
\label{eqch5:minimum}
\end{align}
so
that 
the stationary point is a minimum and stable, otherwise the integral 
may blow up or be continued to a complex function, 
indicating that the stationary point is not a minimum 
and it is therefore an instanton. The second derivative of the action
Eq.~\eqref{eqch5:actioninrp}
can be simplified into 
$\left(\frac{\partial^2 I_*}{\partial r_+^2}\right)_{0} = 
-\frac{\Omega_{d-2} (d-2) r_+^{d-3}}{4\beta}  
\frac{\partial \iota}{\partial r_+}$.
Thus, the stability condition reduces to 
$\frac{\partial \iota}{\partial r_+}<0$, meaning
that the solution is stable
when $\frac{r_+}{R}$ increases with a decrease in the inverse
temperature, and so with an increase in the temperature.
In terms of the variables $x$ and $y$,
see Eq.~\eqref{eqch5:newv}, the stability condition is
\begin{align}
&\frac{d-1}{2}x^{4d-12} - (1+y)x^{3d - 9} - 3(d-3)y x^{2d-6} 
\nonumber \\
&+ (2d-5)y(1+y)x^{d-3}
-\frac{3d-7}{2}y^2 > 0\,.
\label{eqch5:stabcond1}
\end{align}
The range of $x$ is $x_e<x<1$, where $x_e$ is a function of $y_e$, see
Eq.~\eqref{eqch5:extremal2}.
In the case of 
$0\leq y < y_s$, the condition of stability reduces to
two intervals in $x$, one is
$0<x<x_{s1}(y)$ and
the other is
$x_{s2}(y)<x<1$. Therefore, the solutions $x_{1}$ and $x_{3}$ 
are stable, while the solution $x_{2}$ is unstable.
Moreover, 
the points $x=x_{s1}$ and $x=x_{s2}$ are saddle points of the action
as previously stated, and so they are neutrally stable.
In the case of 
$y = y_s$, the same applies as the previous case. 
In the case 
of $y_s<y<1$, the stability condition is satisfied in the interval 
$x_e<x<1$ and so the solution $x_{4}$ is stable.

It is of interest to us to pick specific dimensions $d$.
Due to its real importance, we review the case $d=4$, and
as a typical case of higher dimension, we analyze the case
$d=5$ carefully.

\subsection{The case of $d=4$: stationary points and stability conditions}

We analyze briefly the particular case of four dimensions, $d=4$. The
original results were presented in
\cite{Carlip:2003ne,Lundgren:2006kt}, here we show that the results 
above are in agreement with the original results, and we display also 
new and interesting features for this case.

First, we should look at the qualitative behaviour of the solutions 
$x\equiv\frac{r_+}{R}$ as a
function of the temperature parameter $RT$, i.e., $x(RT)$, for the
several distinct electric charge parameter $y$ regions.  Recall that the
value of $y_s$ is important since it separates the behavior of the
solutions. From Eq.~\eqref{eqch5:ycd}, in $d=4$ it is $y_s=(\sqrt{5}
-2)^2=0.056$, the latter equality being approximate.  The solutions
can then be divided using the electric charge parameter $y$ in the
solution for the no charge case $y=0$, solutions for the charge
parameter in the region $0<y<(\sqrt{5} -2)^2$, the solution for
$y=y_s=(\sqrt{5} -2)^2$, and solutions for the charge parameter in the
region $(\sqrt{5} -2)^2<y<1$.
For $y=0$, the function $x(RT)$ describes the uncharged case and the solution
is known, it is the original York
solution \cite{York:1986}, and consists of two
solutions, here represented as $x_2$ and $x_3$.
The solution $x_{s2}$ happens
when $x_2$ and $x_3$ meet at temperature
$RT = \frac{3\sqrt{3}}{8\pi}=0.207$,
the latter equality being approximate.
For the electric charge in the range $0<y<(\sqrt{5} -2)^2$,
there are three solutions 
$x_1$, $x_2$ and $x_3$, where $x_1$ is stable,
 $x_2$ is unstable,
and $x_3$
is stable.
For very small charges, the temperature $T_1$, which is the 
temperature at which $x_{s1}$ is a solution
for the black hole at the given charge, is very high, tending to
infinite when the charge tends to zero.
For very small charges, the temperature $T_2$, which is the 
temperature at which $x_{s2}$ is a solution
for the black hole at the given charge,
is very near
the minimum temperature of the solutions of the canonical 
ensemble of the Schwarzschild black hole in four dimensions,
i.e., 
$RT = \frac{3\sqrt{3}}{8\pi}$, mentioned above.
Increasing the electric charge from small values, one has that the saddle
points $x_{s1}$ and $x_{s2}$ approach each other.
For the electric charge parameter given by
$y = (\sqrt{5} -2)^2=y_s$, the saddle points
$x_{s1}$ and $x_{s2}$ meet, and at this
electric charge, the solution $x_1$ is 
described by a curve, the solution $x_2$ is now reduced to a
point that coincides with
$x_s = x_{s1} = x_{s2}$, and the solution $x_3$ is described by another
curve.  All
solutions are stable,
more precisely, $x_1$ is stable, $x_2$ is neutrally stable, and
$x_3$ is stable.
For electric charge in
the range $ (\sqrt{5} -2)^2 < y
<1$, there is only one solution $x_4$ which
represents the union of $x_1$ and $x_3$, with 
$x_2$ having disappeared. Also, the solution $x_4$ is stable.

Second, we should look at the qualitative behaviour of the solutions 
$x\equiv\frac{r_+}{R}$ as
a function of the electric charge parameter $y\equiv\frac{\mu
Q^2}{R^2}$, with $\mu=l_p^2$ here, i.e., $x(y)$,  for the several distinct
temperature parameter $RT$ regions.  Recall that the value of $RT_s$
and the value of minimum temperature in the uncharged case 
$RT = \frac{3\sqrt{3}}{8\pi}$
are important since they separate the behavior of the solutions. In
$d=4$, the value of the temperature corresponding to $y_s$ and $x_s$
is $RT_s= 0.185$, this equality being approximate.  Thus,
the temperature parameter regions are $0<RT< 0.185$, $RT_s= 0.185$,
$0.185< RT \leq \frac{3\sqrt{3}}{8\pi}=0.207$, and
$\frac{3\sqrt{3}}{8\pi}<RT<\infty$.
For $0<RT< 0.185$, 
there are only two solutions, which are $x_1$ in the 
interval $0 < y < y_s$ and $x_4$ in the interval $y_s \leq y < 1$,
with  $y_s= (\sqrt{5} -2)^2$. 
For $RT_s= 0.185$ corresponding
to $y_s$ and $x_s$, with this equality being approximate, 
there are four solutions,
but two of them are degenerate. Indeed, there is the
$x_1$ solution, there are the $x_2$ and $x_3$ solutions
that degenerate
into a point $x_2=x_3$ at $y=y_s$, and there
is the $x_4$ solution.
For $0.185< RT \leq \frac{3\sqrt{3}}{8\pi}=0.207$,
the latter equality being approximate,
there are the four solutions 
$x_1$, $x_2$, $x_3$ and $x_4$.
The solutions $x_1$, $x_2$, and $x_3$
lie in the range $0 < y < y_s$, and
the solution $x_4$ exists only for $y_s < y <1$.
The solution $x_4$ can be seen as a continuation in $y$,
i.e., in $Q$, of the solutions $x_1$ and 
$x_3$, and so in a sense $x_4$
is the union of $x_1$ and $x_3$.
For $\frac{3\sqrt{3}}{8\pi}<RT<\infty$, 
there are also the four solutions but $x_2$ and 
$x_3$ are discontinuous.

\subsection{The case of $d=5$: stationary points and stability conditions}

\subsubsection{Behaviour of solutions and stability}

We present here the case for $d=5$ in detail, namely we explain the behaviour of 
the solutions with the aid of plots.

\begin{figure}[h]
\centering
\includegraphics[width=0.7\linewidth]{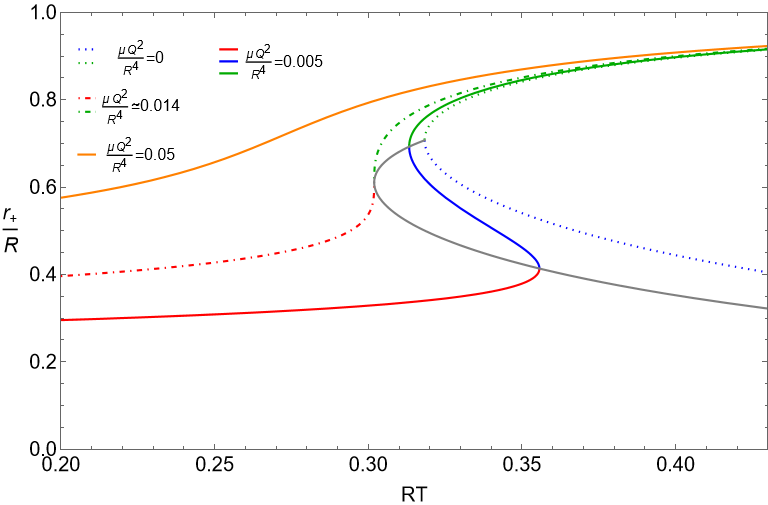}
\caption{Plots of the solutions $x\equiv\frac{r_+}{R}$ as a function
of $RT$ of the canonical ensemble in five dimensions, $d=5$, for four
values of the electric charge parameter $y\equiv\frac{\mu Q^2}{R^4}$,
with $\mu=\frac{4l_p^3}{3\pi}$ here.  The four values of the electric
charge parameter $y$ are $y=0$ in dotted lines, $y=0.005$ in full
lines, $y= \frac{(68 - 27\sqrt{6})^2}{250}=0.014$ in dot dashed lines,
the latter equality being approximate, and $y=0.05$ in an orange full
line. The solution $x_1=\frac{r_{+1}}{R}$ is represented in red,
$x_2=\frac{r_{+2}}{R}$ is represented in blue, $x_3=\frac{r_{+3}}{R}$
is represented in green, and $x_4=\frac{r_{+4}}{R}$ is represented in
orange.  The gray curve describes the trajectory of the saddle points
of the action $x_{s1}=\frac{r_{+s1}}{R}$ and
$x_{s2}=\frac{r_{+s2}}{R}$ by changing the electric charge parameter,
and it separates the regions of existence of the solutions
$x_1=\frac{r_{+1}}{R}$, $x_2=\frac{r_{+2}}{R}$, and
$x_3=\frac{r_{+3}}{R}$.
\label{figch5:rt5}
}
\end{figure}

First, we can analyze $x\equiv\frac{r_+}{R}$ as a function
of the temperature parameter $RT$, for the several regions of the
electric charge parameter $y$.  Once more, the value
of $y_s$ is important for the analysis 
since it separates the regions of different behavior for the
solutions. From Eq.~\eqref{eqch5:ycd}, in $d=5$ it is $y_s=\frac{(68 -
27\sqrt{6})^2}{250}=0.014$, the latter equality being approximate.
We can divide the analysis into the following 
regions of the electric charge parameter
$y$: the no charge case $y=0$, the electric
charge parameter in the region $0<y<\frac{(68 - 27\sqrt{6})^2}{250}$,
the specific case of the critical 
charge $y=y_s=\frac{(68 - 27\sqrt{6})^2}{250}$, and 
the electric charge parameter in the region $\frac{(68 -
27\sqrt{6})^2}{250}<y<1$.
We now describe the solutions 
$x(RT)$ for each region of $y$,
according to Fig.~\ref{figch5:rt5}, where the plots of the solutions
$x\equiv\frac{r_+}{R}$ as a function of $RT$ of the canonical ensemble
in five dimensions, $d=5$, are displayed.
%
%
An important line in such plots is the gray line,
that represents the trajectory of the saddle points $x_{s1}$ and
$x_{s2}$ of the action by varying the electric charge. This 
gray line separates
the regions where the solutions $x_1$, $x_2$, and $x_3$ can be
found. More precisely, the two saddle points $x_{s1}$ and $x_{s2}$ are
the bounds of the solution $x_2$.
For $y=0$, one has the uncharged case, which has been analyzed in 
\cite{Andre:2020czm}, and consists of two solutions, here represented as
$x_2$ and $x_3$. At the saddle point $x_{s2}$, the solutions 
$x_2$ and $x_3$ meet at temperature $RT = \frac{1}{\pi}$.
For the electric charge parameter $y$ in the region $0<y<\frac{(68 -
27\sqrt{6})^2}{250}$, which can be visualized by the $y=0.005$ case 
in the plot, there are
three solutions $x_1$, $x_2$, and $x_3$, where
again $x_1$ is stable, $x_2$
is unstable, and $x_3$ is stable, 
see below for the discussion of thermodynamic
stability.  This case is representative of small electric charges.
For very small charges, the temperature $T_1$ 
corresponding to the saddle point $x_{s1}$ 
assumes very large values and tends to infinity when the
charge tends to zero. Moreover, the temperature $T_2$,
corresponding to the saddle point $x_{s2}$ 
is close to the minimum temperature of the
solutions of the canonical ensemble of the Schwarzschild black hole in
five dimensions $RT = \frac{1}{\pi}$. Note that the
figure with the plots 
for small electric charge parameter yields a unification of York
and Davies, as the two solutions are here represented. 
More precisely, the blue and green lines
correspond to the unstable and stable black holes of York
\cite{York:1986}, respectively, and the red and blue lines correspond
to the stable and unstable black holes of Davies \cite{Davies:1977bgr},
respectively, see below for these latter black holes. Increasing the
electric charge from small values, one sees that the saddle points
$x_{s1}$ and $x_{s2}$ approach each other along the gray curve.
For the saddle electric charge $y = y_s\frac{(68 -
27\sqrt{6})^2}{250}=0.014$, with the latter equality being approximate, 
the saddle points $x_{s1}$ and $x_{s2}$ are equal as 
$x_{s1} = x_{s2} = x_{s}$. While $x_1$ and 
$x_3$ are described by a curve, the solution $x_2$ reduces to a 
point $x_2= x_s$ that connects both solutions $x_1$ and $x_3$.  
Regarding stability, $x_1$ is
stable, $x_2$ is neutrally stable, and $x_3$ is stable.
For the electric charge parameter $y$ in the region $\frac{(68 -
27\sqrt{6})^2}{250} < y <1$, which is represented in the plot 
by the case $y=0.05$, 
there is only one solution $x_4$, that is in a sense
the continuation of $x_1$ and $x_3$, with $x_2$ having disappeared. 
It must be noted that $x_4$ is a stable solution.
%

\begin{figure}[h]
    \centering
    \includegraphics[width=0.7\linewidth]{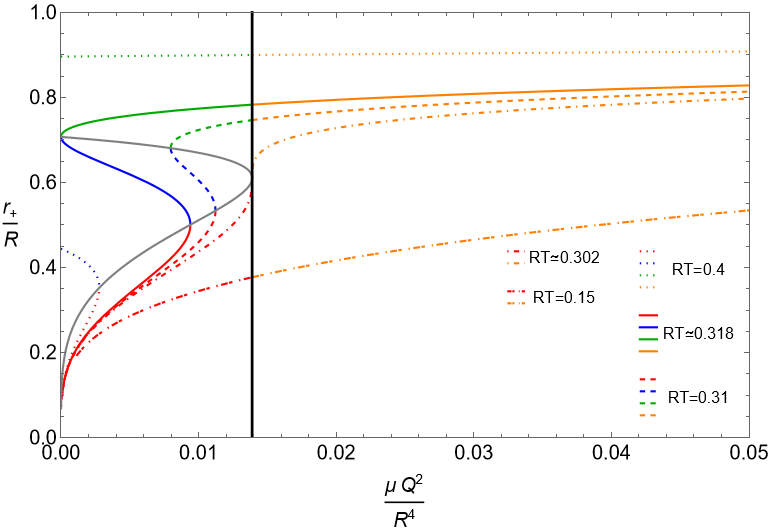}
    \caption{
    Plots of the solutions $x\equiv\frac{r_+}{R}$ as a function of
    $y\equiv\frac{\mu Q^2}{R^4}$ of the canonical ensemble in five
    dimensions, $d=5$, for five values of the temperature parameter $RT$,
    with $\mu=\frac{4}{3\pi}$.  The five values of $RT$ are $RT=0.15$ in
    double dashed lines, $RT=RT_s= 0.302$ in dot dashed lines, $RT=0.31$
    in dashed lines, $RT=\frac{1}{\pi}=0.318$, in full lines, the latter
    equality being approximate, and $RT = 0.4$ in dotted lines.  The
    solution $x_1=\frac{r_{+1}}{R}$ is represented in red,
    $x_2=\frac{r_{+2}}{R}$ is represented in blue, $x_3=\frac{r_{+3}}{R}$
    is represented in green, and $x_4=\frac{r_{+4}}{R}$ is represented in
    orange. The black line, corresponding to $y= y_s = \frac{(68 -
    27\sqrt{6})^2}{250}$, separates the solution $x_4=\frac{r_{+4}}{R}$
    from the remaining solutions.  The gray line corresponds to the
    trajectory of the saddle points of the action
    $x_{s1}=\frac{r_{+s1}}{R}$ and $x_{s2}=\frac{r_{+s2}}{R}$, which
    bounds the region where $x_2=\frac{r_{+2}}{R}$ exists.
    \label{figch5:rq5}
    }
    \end{figure}

Second, we can describe $x\equiv\frac{r_+}{R}$ as
a function of the electric charge parameter $y\equiv\frac{\mu
Q^2}{R^4}$, with $\mu=\frac{4 l_p^3}{3\pi}$ here,
for the several regions of the
temperature parameter $RT$. Here, the value of $RT_s$ and 
the value of the minimum temperature of the uncharged case 
$RT = \frac{1}{\pi}$
are important since they separate the regions of 
different behavior for the solutions. In
$d=5$, the temperature corresponding to $x_s(y_s)$ is
$RT_s= 0.302$, with this equality being approximate. The 
temperature parameter regions
$0<RT< 0.302$, $RT_s= 0.302$,
 $0.302< RT \leq \frac1\pi=0.318$, the latter equality
 being approximate, 
and $\frac1\pi<RT<\infty$ are then considered.
We describe the solution $x(y)$ here within each $RT$ region, in 
agreement with Fig.~\ref{figch5:rq5}, where the plots of the solutions
$x\equiv\frac{r_+}{R}$ as a function of $y\equiv\frac{\mu
Q^2}{R^4}$, $\mu=\frac{4}{3\pi}$,  of the canonical ensemble
in five dimensions, $d=5$, are displayed.
For the temperature parameter $RT$
in the range 
$0<RT< 0.302$, of which $RT=0.15$
is represented in the figure, 
there are only two solutions to display, which are $x_1$ in the 
interval $0 < y < y_s$, and $x_4$ in the interval $y_s \leq y < 1$,
with  $y_s= \frac{(68 - 27\sqrt{6})^2}{250}$. 
For the temperature parameter $RT$
given by $RT= RT_s= 0.302$, this equality being approximate,
one has the curves of the
$x_1$ solution and the $x_4$ solution, 
while the $x_2$ and $x_3$ solutions degenerate 
into a point $x_2=x_3$ at $y=y_s$.
For the temperatures $0.302< RT \leq \frac1\pi=0.318$,
of which  $RT = 0.31$ and $RT = \frac1\pi$ 
are represented in the figure, 
one has the solutions 
$x_1$, $x_2$ and $x_3$ lying in the range 
$0 < y < y_s$, while the solution $x_4$ 
lies in the range $y_s < y < 1$.
The figure shows explicitly that the solution $x_4$ 
is a continuation in
the electric charge parameter $y$
of the solutions $x_1$ and 
$x_3$. Note also that the gray curve in the figure
bounds the solution $x_2$.
For $\frac1\pi<RT<\infty$, which is represented by $RT=0.4$ 
in the figure,
one has also the four solutions but the segments of $x_2$ and 
$x_3$ are discontinuous.

\subsubsection{Interpretation through the thermal length}

The behaviour of the solutions merits some underlying understanding of the physics
at play, which we now give in terms of the thermal wavelength $\lambda$, 
which is proportional to the inverse of the temperature,
$\lambda=\frac1T$.  
We present the reasoning here for the plots of the
solutions $x\equiv\frac{r_+}{R}$ as a function of $RT$ of the
canonical ensemble shown in Fig.~\ref{figch5:rt5}. The solutions
are analyzed from small electric charge to 
large electric charge, and from low to high temperature $T$
with $R$ fixed. We must note that small $RT$ corresponds to low $T$ here.  

We analyze the case for a given small electric charge first.
For small $T$, the associated thermal wavelength $\lambda$ is large
and is stuck to the cavity walls, which means that if there were no
electric charge, there would be no black hole.  But since there is a
fixed electric charge, there is a small black hole with radius $r_+$
of the order of the length scale set by the charge itself. This black hole
does not form by collapse, its presence comes from
topological constraints.  The black hole is stable, small
perturbations cannot evaporate it.  For the smallest possible $T$,
$T=0$, the black hole is an extremal black hole.
For small temperature, there
is only one black hole solution which is this one.
For an intermediate $T$, as the temperature increases, one has that
the associated thermal wavelength $\lambda$ decreases. The black hole
with small $r_+$ is still there, but there is now the possibility of
forming black holes via collapse, indeed the thermal
wavelengths are no more stuck to the cavity walls and the existent
thermal energy can collapse.  One black hole that can form in this way
has radius $r_+$ of the order of $\lambda$ and is thermodynamically 
unstable since
clearly it can evaporate. The other black hole that can form in this way
has radius $r_+$ large such that $R-r_+$ is of the order of $\lambda$,
and is thermodynamically stable, the reservoir and the black hole
exchange quanta of $\lambda$ in a stabilizing way.  For intermediate
temperatures, there are thus three black hole solutions for each
temperature.
For high $T$, as the temperature increases and the associated
wavelength $\lambda$ gets even smaller.  The smallest black hole $r_+$
ceases to exist because, due to the turbulence created by the high
temperature, there is no way to maintain the electric charge
coherently at the center of the cavity.  The intermediate black hole
$r_+$ ceases to exist because the electric charge repulsion is
sufficient to halt gravitational collapse
of this black hole with
intermediate $r_+$.  The large black hole $r_+$
still exists, as it has sufficient mass to overcome the electric
repulsion and still collapses. 
For high $T$, therefore only the large black hole exists.
This is for a typical reasonably low electric charge $Q$, and
there is an interplay between the two quantities that characterize the
ensemble, namely, the temperature $T$ and the electric charge $Q$.

Second, we analyze the case of high electric charge. 
Again here, for small $T$, the associated thermal wavelength $\lambda$
is large and is stuck and cannot collapse.  But since there is a fixed
electric charge, there is a small black hole with radius $r_+$ of the
order of the length set by the charge itself, its presence comes from
topological constraints, is stable, i.e., small perturbations cannot
evaporate it. $T=0$ yields an extremal black hole.
At intermediate $T$, there is turbulence to disperse the black hole
with topological features but it is possible
to have sufficient mass to collapse the existent
thermal energy into the large black hole, with $R-r_+$ starting to be 
comparable to $\lambda$. Note that the intermediate black hole 
does not exist because the electric charge is large enough to counter 
its collapse.
For high $T$, as the temperature increases and the associated
wavelength $\lambda$ gets smaller, the large black hole $r_+$ has
sufficient mass to overcome the electric repulsion and the thermal
energy collapses, being stable.
For all temperatures, there is thus one black hole solution
only for each
temperature. It is in a sense the union of the
topological black hole with the large  collapsed black
hole as the temperature $T$
increases, the intermediate one having disappeared.
Following this
reasoning, one could also extend this interpretation to
the plots of the solutions
$x\equiv \frac{r_+}{R}$ as a function of $\frac{\mu Q^2}{R^{4}}$ in
Fig.~\ref{figch5:rq5}.

\clearpage

\section{Thermodynamics of a charged black 
hole inside a cavity in $d$ dimensions through the canonical ensemble
\label{sech5:thermo}}\sectionmark{
    Thermodynamics of a RN black hole through canonical ensemble
    }\thispagestyle{userightbotmark}

\subsection{Thermodynamic properties and stability for $d$ dimensions
\label{sech5:thermoquantities}}

\subsubsection{Thermodynamic properties}

We have constructed above the canonical ensemble for a charged black hole 
in $d$ dimensions through the Euclidean path integral approach 
subjected to the zero loop approximation. With the partition function 
calculated, we can now proceed to obtain the thermodynamic properties 
of the system. Namely, the partition function is given by 
$Z=\mathrm{e}^{-I_0[\beta,R,Q]}$, where $I_0$ is the action evaluated 
at the stationary points. Thermodynamically, the partition function 
in the canonical ensemble is also related to the 
Helmholtz free energy $F$ as $Z = \mathrm{e}^{-\beta F}$. 
Therefore, one has that the free energy is given by 
\begin{align}
    F = T\,I_0[\beta,R,Q]\,or
    \label{eqch5:freeennergygeneric}
    \end{align}
or explicitly,
\begin{align}
    F = \frac{R^{d-3}}{\mu}\left(
    1 - \sqrt{f \left(R,Q,r_+\right)}
    \right) 
    - T\frac{\Omega_{d-2} r_+^{d-2}}{4l_p^{d-2}}\,,
    \label{eqch5:freeennergy}
\end{align}
with $f(R,Q,r_+) \equiv 1 - \frac{r_+^{d-3} 
+ \frac{\mu Q^2}{r_+^{d-3}}}{R^{d-3}} + \frac{\mu Q^2}{R^{2d-6}}$,
see Eq.~\eqref{eqch5:Rprimealpha}.

The Helmholtz free energy
by definition is given in terms of the internal energy
$E$, the temperature $T$, and the entropy
$S$ by the relation
\begin{align}
F = E - TS\,,
\label{eqch5:free}
\end{align}
and it has the differential
\begin{align}
dF = - S dT - p dA + \phi dQ\,,
\label{eqch5:difffree}
\end{align}
where, in addition to the entropy 
$S$, the area 
$A$, and the electric charge
$Q$, 
there is the thermodynamic 
pressure $p$,
and the thermodynamic electric potential $\phi$. 
The thermodynamic quantities can then be obtained from the 
derivatives of the free energy $F$, more precisely, the entropy is 
$S = - \left(\frac{\partial F}{\partial T}\right)_{A,Q}$, the 
pressure is $p = - \left(\frac{\partial F}{\partial A}\right)_{T,Q}$,
and the electric potential is 
$\phi = \left(\frac{\partial F}{\partial Q}\right)_{T,A}$, where here
the subscript indicates the quantities that are fixed while 
performing the derivative. In Eq.~\eqref{eqch5:difffree}, a part of the
dependence on $T$, $A$, and 
$Q$ is implicit on the solution for the 
horizon radius $r_+ = r_+(T,A,Q)$, as it is evaluated at the 
minima of the action. To simplify the calculation of the derivatives,
one can perform the chain rule and the fact that, since
$r_+ = r_+(T,A,Q)$, the derivative of the reduced action obeys
$\left(\frac{\partial I_*}{\partial r_+}\right)_{T,R,Q} = 
 \left(\frac{\partial F}{\partial r_+}\right)_{T,R,Q}= 0$, to get 
for example
$S = - \left(\frac{\partial F}{\partial T}\right)_{A,Q} = 
-\left(\frac{\partial F}{\partial T}\right)_{R,Q,r_+} 
- \left(\frac{\partial F}{\partial r_+}\right)_{T,R,Q}
 \frac{\partial r_+}{\partial T}
= -\left(\frac{\partial F}{\partial T}\right)_{R,Q,r_+}$,
and this also holds similarly for the computation of the 
pressure and the electric potential. Therefore, the thermodynamic 
quantities can be computed as
\begin{align}
    &S = -\left(\frac{\partial F}{\partial T}\right)_{R,Q,r_+}\,\,,\,\,
    p = -\frac{1}{(d-2)\Omega_{d-2} R^{d-3}}
    \left(\frac{\partial F}{\partial R} \right)_{T,Q,r_+}\,\,,\,\,
    \phi = \left(\frac{\partial F}{\partial Q}\right)_{T,R,r_+}\,,
\end{align}
and $E = F - TS$.
The entropy is then given as
\begin{align}
S = \frac{A_+}{4l_p^{d-2}}\,,
\label{eqch5:entropy}
\end{align}
where $A_+\equiv \Omega_{d-2} r_+^{d-2}$ is
the area of the event horizon, 
and so this is the usual Bekenstein-Hawking expression
for the entropy of a black hole. The thermodynamic pressure is
\begin{align}
p = \frac{d-3}{16\pi R \sqrt{f}l_p^{d-2}}\left((1 - \sqrt{f})^2 
- \frac{\mu Q^2}{R^{2d-6}} \right)\,,
\label{eqch5:pressure}
\end{align}
the thermodynamic electric potential is
\begin{align}
\phi = \frac{Q}{\sqrt{f}}\left(\frac{1}{r_+^{d-3}} 
- \frac{1}{R^{d-3}} \right)\,,
\label{eqch5:phi}
\end{align}
and finally, from Eq.~\eqref{eqch5:free}, the
thermodynamic energy is given by
\begin{align}
E = \frac{R^{d-3}}{\mu}\left[1 - \sqrt{\left(1 - \frac{r_+^{d-3}}
{R^{d-3}} \right)
\left(1 - \frac{\mu Q^2}{r_+^{d-3}R^{d-3}}\right)} \right]\,,
\label{eqch5:energy}
\end{align}
Collecting Eqs.~\eqref{eqch5:entropy}-\eqref{eqch5:energy},
one finds that the first law of thermodynamics in the form
\begin{align}
dE = TdS-pdA+\phi dQ\,,
\label{eqch5:firstlawoft}
\end{align}
holds.
It is interesting to note,
and surely not a coincidence,
that these thermodynamic quantities are identical to the ones calculated 
for a self-gravitating charged shell, where the first law of 
thermodynamics is imposed, and, the charged shell assumes the
temperature equation 
of state of a black hole and the thermodynamic pressure equation of state 
of the cavity, see \cite{Fernandes:2022gjd}.

With the thermodynamic quantities obtained in 
Eqs.~\eqref{eqch5:entropy}-\eqref{eqch5:energy}, we can get an integrated 
first law of thermodynamics known as the Euler equation.
For that, one rewrites
the energy in Eq.~\eqref{eqch5:energy} in terms of the entropy in 
Eq.~\eqref{eqch5:entropy}, the area $A = \Omega_{d-2} R^{d-2}$, and the
electric charge $Q$ as
\begin{align}
E =& \frac{(d-2)A^{\frac{d-3}{d-2}}\Omega_{d-2}^{\frac{1}{d-2}}}
{8\pi l_p^{d-2}}\times
\nonumber\\&\left(1-\sqrt{\left(1-\left(\frac{4 S l_p^{d-2}}{A}
\right)^{\frac{d-3}{d-2}}\right)\left(1-\frac{\mu Q^2 
\Omega_{d-2}^{2\frac{d-3}{d-2}}}{(4S l_p^{d-2} A)^{\frac{d-3}{d-2}}}\right)} \right)\,
\,.\label{eqch5:energyintermsof}
\end{align}
If a 
scaling is performed on the thermodynamic quantities 
$S \rightarrow \nu S$, $A \rightarrow \nu A$ and 
$Q \rightarrow \nu^{\frac{d-3}{d-2}} Q$, then it can be verified that 
$E(\nu S, \nu A, \nu^{\frac{d-3}{d-2}}Q) = \nu^{\frac{d-3}{d-2}} 
E(S,A,Q)$. According to the Euler relation theorem, and considering 
that the differential of the energy is given by the first law of 
thermodynamics Eq.~\eqref{eqch5:firstlawoft}, the Euler equation is given 
by
\begin{align}
E = \frac{d-2}{d-3} (T S - p A) + \phi Q\,.
\label{eqch5:EulerEquation}
\end{align} 
One can furthermore differentiate Eq.~\eqref{eqch5:EulerEquation} and 
use the first law of thermodynamics to obtain
\begin{align}
\hskip -0.25cm
\frac1{d-3}\left(
{TdS - pdA}\right) + \frac{d-2}{d-3}(SdT-Adp) + Q d\phi
\hskip -0.06cm
=
\hskip -0.06cm
0\,,
\label{eqch5:gibbsduhem}
\end{align}
which is the Gibbs-Duhem relation.

\subsubsection{Thermodynamic stability and the heat capacity}

A system to be thermodynamically stable must have positive
heat capacity at constant 
area and constant electric charge $C_{A,Q}$, i.e., 
\begin{align}
C_{A,Q}\geq0\,,
\label{eqch5:caq>0}
\end{align}
where
$C_{A,Q}\equiv T\left(\frac{\partial S}{\partial T}\right)$.
We have shown in Sec.~\ref{sech5:stabanalysis} that the stability 
condition in the ensemble formalism was reduced to the condition 
$\frac{\partial \iota}{\partial r_+} < 0$.
The derivative above can be put in terms of thermodynamic variables,
and then in terms of the heat capacity.
The inverse temperature function $\iota(r_+)$
is a function of $r_+$, $R$ and $Q$. 
The variables $Q$ and $R$ are already thermodynamic variables. 
The quantity $r_+$ is also in some sense 
a thermodynamic variable since one has
that $S = \frac{\Omega_{d-2} r_+^{d-2}}{4}$.
Therefore, using $\beta=\iota(r_+)$, one has 
$\frac{\partial \iota}{\partial r_+} = 
- \frac1T \left(\frac{\partial S}{\partial r_+}\right)
\frac{1}{C_{A,Q}}$,
where the definition of the heat capacity at constant 
area and constant electric charge was used.

The heat capacity is then written as
\begin{align}
C_{A,Q}
\hskip -0.1cm
=
\hskip -0.1cm
\frac{  (d-2)  R^{d-2}f
\left(\frac{r_+^{d-3}}{R^{d-3}} 
- \frac{\mu Q^2}{R^{d-3}r_+^{d-3}}\right)
\frac{\Omega_{d-2}r_+^{d-2}}{4R^{d-2}}
}
{l_p^{d-2}
\frac{d-3}{2}
\hskip -0.1cm
\left(
\hskip -0.1cm
\frac{r_+^{d-3}}{R^{d-3}}
\hskip -0.07cm
-
\hskip -0.07cm
\frac{\mu Q^2}{r_+^{d-3}R^{d-3}}
\hskip -0.12cm
\right)^{\hskip -0.1cm 2}
\hskip -0.15cm
-
\hskip -0.1cm
f
\hskip -0.1cm
\left(
\hskip -0.1cm
\frac{r_+^{d-3}}{R^{d-3}}
\hskip -0.07cm
-
\hskip -0.07cm
(2d
\hskip -0.1cm
-
\hskip -0.1cm
5)\frac{\mu Q^2}{r_+^{d-3}R^{d-3}}
\hskip -0.1cm
\right)}.
\label{eqch5:Caq}
\end{align}
Since one has that
$C_{A,Q}\geq0$ for the system to be thermodynamically stable,
thermodynamic stability reduces to 
Eq.~\eqref{eqch5:stabcond1} after rearrangements and
definitions.
Thus,
the physical interpretation is that the stability of the solutions 
is controlled by the heat capacity at constant area and charge, as 
it should be in the canonical ensemble. This quantity is tied to 
the derivative of the inverse temperature given by 
Eq.~\eqref{eqch5:beta1} and so the condition 
reduces to the intervals given by the stationary points of 
$\iota(r_+, R,Q)$, or the saddle points of the action. Moreover, 
solutions where $r_+$ increases as $T$ increases are stable and
solutions where $r_+$ decreases as $T$ increases are unstable.

It is interesting to see what happens when one fixes
$\frac{r_+}{R}$  and changes
the electric charge parameter $\frac{\mu Q^2}{R^{2d-6}}$.
For $\frac{r_+}{R}> \left(\frac{2}{d-1}\right)^{\frac1{d-3}}$, the 
heat capacity is always positive for any electric charge, with the
limit of the bound matching the one for the uncharged black hole.
For
$0\leq
\frac{r_+}{R} \leq \left(\frac{2}{d-1}\right)^{\frac1{d-3}}$, 
the sign of the heat capacity
$C_{A,Q}$ changes according to the electric charge. $C_{A,Q}$ is
positive for sufficiently high electric charge parameter
$\frac{\mu Q^2}{R^{2d-6}}$,
and is negative for sufficiently low electric charge parameter
$\frac{\mu Q^2}{R^{2d-6}}$, the change in sign happening
at  the definite value of the charge satisfying
Eq.~\eqref{eqch5:critbeta} with
fixed $\frac{r_+}{R}$.
It is important to note that this does not indicate a phase transition since 
$\frac{r_+}{R}$ is not a thermodynamic variable controlled in the 
ensemble. At that definite value of the charge parameter,
there is rather a turning point describing the 
ratio of scales at which there is stability.

The thermodynamic variables are the temperature and
the electric charge, and therefore
the heat capacity must be
analyzed in terms of these quantities,
instead of $\frac{r_+}{R}$ and the electric charge.
For the
range of electric charges $0 < \frac{\mu Q^2}{R^{2d-6}}<
\frac{\mu Q_s}{R^{2d-6}}$, one has
three curves for the heat capacity
as a function of the temperature, one
for each solution. The heat capacity is positive for the solutions
$r_{+1}$ and $r_{+3}$, while it is negative for $r_{+2}$. The heat
capacity diverges when the solutions reach the temperatures of the
saddle points of the action, which are the turning points.  For the
critical charge parameter $\frac{\mu Q_s^2}{R^{2d-6}}$, one has two
curves for the heat capacity as a function of the temperature.  In
this
particular case, the two curves are described by the solutions
$r_{+1}$ and $r_{+3}$ and it is positive for both. Moreover, there is
a discontinuity between the two curves at $RT_s$, where the heat
capacity diverges. This point indeed does mark a second order phase
transition between $r_{+1}$ and $r_{+3}$, as both solutions are stable
and it can be seen that the free energy is continuous at $RT_s$ for
$\frac{\mu Q_s^2}{R^{2d-6}}$.  For the range $\frac{\mu
Q^2}{R^{2d-6}}>\frac{\mu Q_s^2}{R^{2d-6}}$,
there is only one curve for the heat capacity as a 
function of the temperature, corresponding to the solution $r_{+4}$
and it is always positive.

We now specify the results in this subsection for the cases 
$d=4$ and $d=5$ dimensions, supported by further comments and a figure.

\subsection{Thermodynamic properties and stability for $d=4$ dimensions}

For the case $d=4$, we can write straightfowardly the results from the 
subsection above. 
The entropy is given as
\begin{align}
    S = \pi \frac{r_+^2}{l_p^2},
\end{align}
which is the usual Hawking-Bekenstein formula $S = \frac{A_+}{4l_p^2}$, with 
$A_+=4\pi r_+^2$ being the area of the event horizon. 
The pressure is
\begin{align}
    p = \frac{1}{16\pi R l_p^2 \sqrt{f}}\left((1 - \sqrt{f})^2 
- \frac{l_p^2 Q^2}{R^{2}} \right)\,\,,
\end{align}
where it was used $\mu=l_p^2$
and
$f= 1 - \frac{r_+
+ \frac{l_p^2 Q^2}{r_+}}{R} + \frac{l_p^2 Q^2}{R^2}$.
The electric potential is
\begin{align}
    \phi = \frac{Q}{\sqrt{f}}\left(\frac{1}{r_+} 
- \frac{1}{R} \right)\,\,.
\end{align}
Finally, the mean energy is given by
\begin{align}
    E = \frac{R}{l_p^2}\left[1 - \sqrt{
\left(1 - \frac{r_+}{R} \right)
\left(1 - \frac{l_p^2 Q^2}{r_+R}\right)} \right]\,\,.
\end{align}
One can then write
the energy in terms of $S$, $A = 4\pi R^2$, and 
$Q$, i.e., $E=E(S,A,Q)$ to obtain
the Euler relation 
$E = 2(T S - p A) + \phi Q$.
The  Gibbs-Duhem 
relation is
$TdS - pdA + 2(SdT-Adp) + Qd\phi = 0$.

The heat capacity, the quantity that controls
thermodynamic stability, is
\begin{align}
C_{A,Q} = \frac{1}{l_p^2}\frac{
2R^2f
\left(\frac{r_+}{R} 
- \frac{l_p^2 Q^2}{R^2}\frac{R}{r_+} \right)
\frac{\pi r_+^2}{R^2}
}{
\frac12\left(\frac{r_+}{R} -\frac{Q^2}{R^2}\frac{R}{r_+} \right)^2 
- f\left(\frac{r_+}{R} 
- \frac{3Q^2}{R^2}\frac{R}{r_+}\right)}\,.
\label{eqch5:CAq4d}
\end{align}
One could fix $\frac{r_+}{R}$ and change
the electric charge parameter $\frac{l_p^2 Q^2}{R^2}$ in 
Eq.~\eqref{eqch5:CAq4d}.
As seen in the general $d$ case, one finds that for 
$\frac{r_+}{R} > \frac23$, the heat capacity is always 
positive, 
and for
$0\leq
\frac{r_+}{R}\leq \frac23$, the sign of the heat capacity
$C_{A,Q}$ changes depending on the electric charge, being
positive for a region of high electric charge parameter
$\frac{l_p^2 Q^2}{R^2}$,
and being negative for a region of low electric charge parameter
$\frac{l_p^2 Q^2}{R^2}$.
This does not indicate a phase transition but rather a turning
point. To see this fact and verify the true phase transitions,
one must analyze the heat capacity in terms of the fixed
quantities of the ensemble, i.e., the temperature and the electric
charge.  For the range
of charge parameters 
$0 <
\frac{l_p^2 Q^2}{R^{2}}<(\sqrt{5} - 2)^2$,
where in $d=4$ one has $\frac{l_p^2 Q_s^2}{R^{2}}=(\sqrt{5} - 2)^2$,
the heat capacity has a curve
for each solution $r_{+1}$, $r_{+2}$, and $r_{+3}$, being positive for
$r_{+1}$ and $r_{+3}$, and being negative for $r_{+2}$.  When the
solutions reach the temperatures of the saddle points of the action,
i.e., 
the turning points, the heat capacity diverges but this only indicates
conditions for stability of the ensemble,
there are no phase transitions at these points.
For the critical charge
$\frac{l_p^2 Q_s^2}{R^{2}}= (\sqrt{5} - 2)^2$, the heat capacity has two
curves as a function of the temperature, $r_{+1}$ and $r_{+3}$,
being positive for both solutions.  For this case, there is a
discontinuity between the two curves at $RT_s = 0.185$, where the heat
capacity diverges.  This point indeed signals a second order phase
transition between $r_{+1}$ and $r_{+3}$, as both solutions are stable
and it can be seen that the free energy is continuous at $RT_s= 0.185$
for $\frac{l_p^2 Q^2}{R^{2}}= (\sqrt{5} - 2)^2$.  For the range
of charge parameters
$\frac{l_p^2 Q^2}{R^{2}}>(\sqrt{5} - 2)^2$, one has that the heat
capacity of $r_{+4}$ as a function of the temperature is always
positive.
In
\cite{Carlip:2003ne,Lundgren:2006kt}
some of these results for $d=4$
are presented.

\subsection{Thermodynamic properties and stability for $d=5$ dimensions}

Here, we make the results for the case $d=5$ explicit.
The entropy is given as
\begin{align}
    S = \frac{\pi^2 r_+^3}{2l_p^3}\,\,,
\end{align}
matching the usual Bekenstein-Hawking formula $S = \frac{A_+}{4l_p^3}$, with 
$A_+=2\pi^2 r_+^3$ being the area of the event horizon. 
The pressure yields
\begin{align}
    p = \frac{2}{16\pi l_p^2 R \sqrt{f}}\left((1 - \sqrt{f})^2 
- \frac{4 l_p^3 Q^2}{3\pi R^4} \right)\,\,,
\end{align}
where it was used $\mu=\frac{4l_p^3}{3\pi}$
and
$f= 1 - \frac{r_+^2
+ \frac{4 l_p^3 Q^2}{3\pi r_+^2}}{R^2} + \frac{4l_p^3 Q^2}{3\pi R^4}$.
The electric potential yields
\begin{align}
    \phi = \frac{Q}{\sqrt{f}}\left(\frac{1}{r_+^2} 
- \frac{1}{R^2} \right)\,\,.
\end{align}
And the energy has the expression
\begin{align}
    E = \frac{3\pi R^2}{4l_p^3}\left[1 - \sqrt{\left(1 - \frac{r_+^2}
{R^2} \right)
\left(1 - \frac{4 l_p^3 Q^2}{3\pi r_+^2R^2}\right)} \right]\,\,.
\end{align}
These thermodynamic quantities are identical to the ones calculated 
for a self-gravitating charged shell, where the first law of 
thermodynamics is imposed, and the charged shell assumes the equation 
of state of the black hole, see \cite{Fernandes:2022gjd} or 
Chapter~\ref{ch:chargedselfgravitating}.
The energy can be written 
in terms of $S$, $A = 2\pi^2 R^3$, and the electric charge
$Q$, as $E=E(S,A,Q)$ to obtain
the Euler relation 
$E = \frac32(T S - p A) + \phi Q$.
The  Gibbs-Duhem 
relation yields
$\frac12\left(TdS - pdA\right) +
\frac32\left(SdT-Adp\right) + Qd\phi = 0$.

\begin{figure}[h]
\centering
\includegraphics[width=0.7\linewidth]{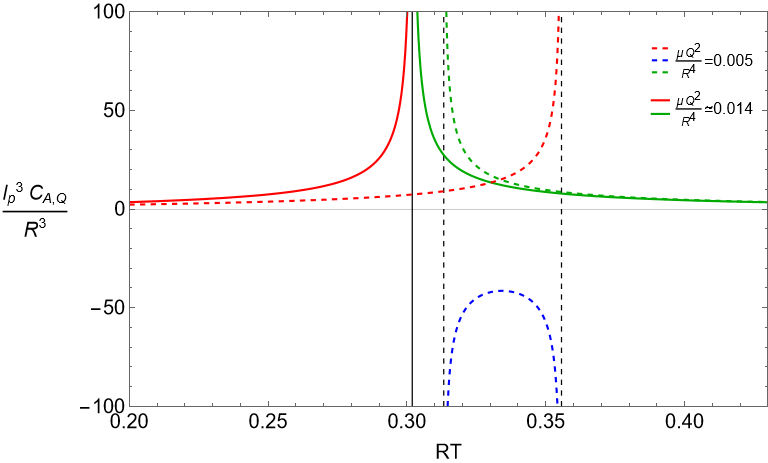}
\caption{\label{figch5:heatcapacity5d}
The heat capacity $C_{A,Q}$, namely $\frac{C_{A,Q}l_p^3}{R^3}$, 
as a function of the temperature for two values of the electric charge, 
$\frac{\mu Q^2}{R^4}=0.005$ and $\frac{\mu Q^2}{R^4}= 
\frac{\mu Q^2_s}{R^4}=0.014$ approximately, for solutions $r_{+1}$ in 
red, $r_{+2}$ in blue, and $r_{+3}$ in green. The dashed black lines 
mark the turning points of the solutions and the solid black line marks 
the second order phase transition between the stable
solutions $r_{+1}$ and $r_{+3}$.
}
\end{figure}

The heat capacity is
\begin{align}
C_{A,Q} = \frac{1}{l_p^3}\frac{
3R^3 f
\left(\frac{r_+^2}{R^2} 
- \frac{4 l_p^3 Q^2}{3\pi R^2r_+^2} \right)\frac{\pi^2r_+^3}{2R^3}
}
{
\left(
\frac{r_+^2}{R^2} -\frac{4 l_p^3 Q^2}{3\pi R^4}\frac{R^2}{r_+^2}\right)^2 
- f\left(\frac{r_+^2}{R^2} 
- \frac{20 l_p^3 Q^2}{3\pi R^4}\frac{R^2}{r_+^2}
\right)
}
\,.
\label{eq:CAq5d}
\end{align}
Regarding the behavior 
of the heat capacity with fixed $\frac{r_+}{R}$ as a function of
the electric charge parameter $\frac{l_p^3 Q^2}{R^4}$, 
one has that the heat capacity is always positive for 
$\frac{r_+}{R} > \frac{\sqrt{2}}{2}$, and the 
heat capacity changes signs for 
$0\leq \frac{r_+}{R}\leq \frac{\sqrt2}{2}$,
being positive for 
high electric charge parameter $\frac{l_p^3 Q^2}{R^4}$, and being
negative for low electric charge parameter $\frac{l_p^3 Q^2}{R^4}$.
As we already noted, to understand the turning points
and the possible phase transitions of the
solutions, one must 
analyze the behavior of the 
heat capacity through its dependence in the temperature 
and the electric charge, see Fig.~\ref{figch5:heatcapacity5d}.
For a fixed electric charge
parameter in the range $0 < \frac{\mu Q^2}{R^4}
< \frac{(68-27\sqrt{6})^2}{250}$,
where in $d=5$ one
has $\frac{\mu Q_s^2}{R^{4}}=\frac{(68-27\sqrt{6})^2}{250}$,
the heat capacity is described by three 
curves, one for each solution $r_{+1}$, $r_{+2}$, and $r_{+3}$, being 
positive for $r_{+1}$ and $r_{+3}$, and being negative for 
$r_{+2}$, see Fig.~\ref{figch5:heatcapacity5d} for the case 
$\frac{\mu Q^2}{R^4} = 0.005$. The heat capacity in this range of 
charges diverges at the turning points of the solutions, as seen 
by the dashed black lines, indicating the conditions for stability of 
the solutions and not
signalling any
phase transition. For the electric charge 
$\frac{\mu Q^2_s}{R^4}= \frac{(68-27\sqrt{6})^2}{250}$, the heat capacity 
is positive, as it is described by the curves of the solution $r_{+1}$ 
and $r_{+3}$. The heat capacity diverges at $RT_{s} =  0.302$, the 
solid black line, and here 
one in fact has a second order transition, from $r_{+1}$ to $r_{+3}$, 
as these are both stable solutions, and the free energy is continuous there.
For $\frac{\mu Q^2}{R^4} > \frac{(68-27\sqrt{6})^2}{250}$, the heat capacity is 
always positive, as it is described only by the solution $r_{+4}$.

\section{Favorable phases in the canonical ensemble of a
$d$ dimensional electrically charged
black hole in a cavity and phase transitions\label{sech5:favorablephases}}
\sectionmark{Favorable phases in the canonical ensemble}
\thispagestyle{userightbotmark}

\subsection{The black hole sector of the canonical ensemble and favorable
phases in $d$ dimensions}

We plan now to study the favorable phases of the situation at hand.
Consider first the black hole sector of the canonical ensemble and the
corresponding free energy. Since the free energy $F$ and action $I_0$ are
related by $F=\frac{I}{\beta}=T\,I_0$, the black hole free energy
$F_\mathrm{bh}$ can be taken directly from Eq.~\eqref{eqch5:freeennergy} to
be rewritten as
\begin{align}
F_\mathrm{ bh} = \frac{ R^{d-3}}{\mu}\left(1 - \sqrt{f(R,Q,r_+)}\right) -
T\frac{A_+}{4l_p^{d-2}}
\,,
\label{eqch5:freeennergybh}
\end{align}
where, in this section, the bh subscript in $F$
denotes the black hole free energy to distinguish from other
possible free energies.
Since $A_+\equiv \Omega_{d-2} r_+^{d-2}$ and
$r_+=r_+(T,R,Q)$,
the black hole
solutions have their free energies of the
form $F_\mathrm{ bh}(T,R,Q)$. For
a system characterized by the free energy,
the one that
has the lower free energy $F_\mathrm{ bh}$,  for given $R$, $T$, and $Q$,
is the one that is
thermodynamically favored. Thus, one can find the
black hole that is favored.

We have shown above that in the zero loop approximation, there are different
black hole solutions depending on the electric charge and temperature
of the reservoir, see Sec.~\ref{sech5:Canonical1}. 
For sufficiently low electric charge parameter, i.e.,
for $0\leq\frac{\mu Q^2}{R^{2d-6}}
<\frac{\mu Q_s^2}{R^{2d-6}}$,
where $Q_s$ is the saddle electric charge value,
corresponding to the saddle electric charge parameter
$y_s=\frac{\mu Q_s^2}{R^{2d-6}}$, it was seen that
there can be up to
three solutions $\frac{r_{+1}}{R}$, $\frac{r_{+2}}{R}$, and
$\frac{r_{+3}}{R}$.
The free energies $F_\mathrm{ bh}$ of these
three solutions are now commented.
The solution $\frac{r_{+1}}{R}$ has
positive free energy for all the temperatures in which
the solution exists.  The solution $\frac{r_{+2}}{R}$
has also 
positive free energy always,
but it is unstable, so this solution has no interest here. The solution
$\frac{r_{+3}}{R}$, has a temperature for each electric charge 
parameter
$\frac{\mu Q^2}{R^{2d-6}}$ at which the free energy becomes zero,
which it is defined as
$T_{F_\mathrm{ bh}=0}(Q)$
or $T_{F_\mathrm{ bh}=0}(\frac{\mu Q^2}{R^{2d-6}})$,
thus $\frac{r_{+3}}{R}$
can have
positive or negative free energy.
For the saddle charge parameter $\frac{\mu Q^2}{R^{2d-6}}
=\frac{\mu Q_s^2}{R^{2d-6}}$, the solution
$\frac{r_{+1}}{R}$ has 
positive free energy, there is a solution where
$\frac{r_{+1}}{R} =
\frac{r_{+2}}{R}=
\frac{r_{+3}}{R}$ 
 which has 
positive free energy, and the solution $\frac{r_{+3}}{R}$
has again a temperature
$T_{F_\mathrm{ bh}=0}(Q_s)$
or $T_{F_\mathrm{ bh}=0}(\frac{\mu Q_s^2}{R^{2d-6}})$, 
at which the free energy becomes zero,
thus $\frac{r_{+3}}{R}$ can have
positive or negative free energy.
For higher values of the
electric charge parameter, i.e.,
for
$\frac{\mu Q_s^2}{R^{2d-6}}<
\frac{\mu Q^2}{R^{2d-6}}<1$,
the
solution $\frac{r_{+4}}{R}$ has also a
temperature $T_{F_\mathrm{ bh}=0}(Q)$, or
$T_{F_\mathrm{ bh}=0}(\frac{\mu Q_s^2}{R^{2d-6}})$,
at which the free energy becomes zero,
thus $\frac{r_{+4}}{R}$  can have
positive or negative free energy.
The temperature $T_{F_\mathrm{ bh}=0}(Q)$
can be calculated by solving $F_\mathrm{
bh}=0$, with $F_\mathrm{ bh}$ given in Eq.~\eqref{eqch5:freeennergybh} for
either the solution $\frac{r_{+3}}{R}$
or $\frac{r_{+4}}{R}$.  One can instead put the free energy 
in terms of the mass $m$ and electric charge $Q$ through
Eq.~\eqref{eqch5:beta1} and through the relation $2\mu m = r_+^{d-3} +
\frac{\mu Q^2}{r_{+}^{d-3}}$, so that $F_\mathrm{ bh}=0$ reduces to a
quartic equation for the mass $m$ as a function of the electric
charge, see Sec~\ref{sech5:BHzerofreeenergy}. After solving it, one
can then recover the value of $r_+$ and consequently the value
$T_{F_\mathrm{ bh}=0}(\frac{\mu Q^2}{R^{2d-6}})$.  For temperatures lower
than $T_{F_\mathrm{ bh}=0}(\frac{\mu Q^2}{R^{2d-6}})$, the solutions have
positive free energy and for temperatures higher than $T_{F_\mathrm{
bh}=0}(\frac{\mu Q^2}{R^{2d-6}})$, the solutions have negative free
energy.

There is another important temperature, $T_f$, which depends on the
electric charge $Q$, i.e., on the electric charge
parameter $\frac{\mu Q^2}{R^{2d-6}}$, and
at which the favorability of one phase over the other changes.
For
the electric charge parameter within the region
 $0\leq\frac{\mu Q^2}{R^{2d-6}}
<\frac{\mu Q_s^2}{R^{2d-6}}$, 
there is a phase favorability
temperature $T_f$ at
which the solutions $\frac{r_{+1}}{R}$ and $\frac{r_{+3}}{R}$ have the
same free energy.  In other words, the solutions $\frac{r_{+1}}{R}$
and $\frac{r_{+3}}{R}$ are stable, and thus within the black hole
sector they compete between themselves to be the most favored phase.
Specifically, for temperatures lower than $T_f$, the solution
$\frac{r_{+1}}{R}$ is either more favorable than $\frac{r_{+3}}{R}$,
or is the only existing solution if the temperature is low enough.
For a temperature equal to $T_f$, the solutions $\frac{r_{+1}}{R}$ and
$\frac{r_{+3}}{R}$ are equally favorable, i.e., they coexist equally.
For temperatures higher than $T_f$, either the solution
$\frac{r_{+3}}{R}$ is more favorable than $\frac{r_{+1}}{R}$, or is
the only existing solution if the temperature is high enough.
For the electric charge parameter given by 
$\frac{\mu Q^2}{R^{2d-6}}
=\frac{\mu Q_s^2}{R^{2d-6}}$,
the temperature $T_f$
is the temperature at which
$\frac{r_{+1}}{R} =
\frac{r_{+2}}{R}=
\frac{r_{+3}}{R}$ 
and all have the same free energy, i.e.,
$\frac{r_{+1}}{R}$ and
$\frac{r_{+3}}{R}$ 
coexist. For temperatures lower than $T_f$, the solution
$\frac{r_{+1}}{R}$   
is
the only existing solution.  For temperatures higher than $T_f$, the
solution 
$\frac{r_{+3}}{R}$   
 is the only existing solution.
For the electric charge parameter within the region
$\frac{\mu Q_s^2}{R^{2d-6}}<
\frac{\mu Q^2}{R^{2d-6}}<1
$,
there is only one black hole solution, it is
$\frac{r_{+4}}{R}$. Within the black hole sector
it is surely the most favored state since
it is stable and there is no other solution.
It can 
have
positive or negative free energy.

\subsection{The hot flat space sector of the electrically
charged canonical ensemble in $d$ dimensions}

Consider now a possible electrically charged hot flat space sector, i.e.,
a cavity with nothing in it with its boundaries defined by $R$, $T$,
and $Q$, the settings of the canonical ensemble.

To have such a solution one can think in trying to decrease $r_+$ up
to zero, to a point where there is no more a black hole and
thus obtain flat
space. However, this is not possible, since there is a minimum limit
for $r_+$ given by $r_+ = {r_{+}}_e$ corresponding to the extremal black
hole. At ${r_{+}}_e$, the free energy tends to 
$F_\mathrm{bh} = \frac{Q}{\sqrt{\mu}}$, 
and it is then
impossible to decrease $r_+$ further. Regarding extremal
black holes, the only temperature that
such solutions exist is at $T=0$ and they are
considered here as it is only one point of the ensemble, although
it is a very interesting one.
It seems there is no
other immediate solution of
the action that can be a candidate for a stationary point of the
reduced action.
Thus, to emulate electrically charged hot flat space one has to go
beyond the black hole sector.
We consider, for example, a shell with radius $r_\mathrm{shell}$,
coated with the required electric charge $Q$, and with gravity
turned off, i.e., the constant of gravitation is set to zero.  The
action of the system, if we consider terms depending only on the
Maxwell field, can be calculated to give the free energy as
\begin{align}
F_\mathrm{shell} = \frac{ Q^2}{2r_{\mathrm{shell}}^{d-3}}
\left(1 -
\frac{r_{\mathrm{shell}}^{d-3}}{R^{d-3}}\right)
\,.
\label{eqch5:freeennergyshell}
\end{align}
Thus, for a
given $r_{\mathrm{shell}}$, one has
that $F_{\mathrm{shell}}$ has a given constant fixed value.  There are two
limits that we can mention.  One limit is when $r_\mathrm{shell}$ is
very small. One could see this limit as an electrically
charged central point surrounded by hot
flat space, where quantum fluctuations of the hot flat space generate
electric charge.
But this seems to lead to a
divergent free energy.
Note that the behavior mentioned for very small $r_{\mathrm{shell}}$ 
contrasts with the grand canonical ensemble
case~\cite{Fernandes:2023byx}, where $r_\mathrm{shell}=0$ corresponds to a
zero grand potential.
The other limit is when $r_\mathrm{shell}=R$ and
so the free energy is zero. This means that all the charge is
infinitesimally near the boundary of the cavity, i.e., it is at the
boundary of the cavity itself and there is hot flat space inside the
cavity.  Thus, the more interesting limit is the latter one, when
$r_{\mathrm{shell}}=R$, and the charge is gathered near the boundary of the
cavity giving $F_\mathrm{shell} = 0$.
Since, in this case, the shell emulates hot flat
space with electric charge
at the boundary, one has $F_{\mathrm{shell}} = F_{\mathrm{hfs}} =0$.
Nevertheless, it is interesting to
compare the toy model of a shell with free energy
$F_{\mathrm{shell}}$ given in Eq.~\eqref{eqch5:freeennergyshell}
for several $\frac{r_{\mathrm{shell}}}{R}$, and in
particular for $\frac{r_{\mathrm{shell}}}{R}=1$, with the black hole free energy
$F_{\mathrm{bh}}$  given in Eq.~\eqref{eqch5:freeennergybh}.

One could further
think in building an equivalent system with the constant of
gravitation turned on, such as an electrically charged
self-gravitating shell close to the boundary of the cavity. Still, it
is unclear if there is a possible conversion of this system to
a charged black hole, and vice versa,
since the two systems correspond to different
topologies and also to a different action, as here we do not consider 
the matter sector.

\subsection{First and second order phase transitions}

We are interested in studying the favorable states of
the ensemble, i.e., of an ensemble
of a cavity with fixed radius $R$, fixed
temperature $T$, and fixed electric charge $Q$, all values of these
quantities being set by the reservoir.

A thermodynamic system tends to be in a state in which its
thermodynamic potential, associated to the ensemble considered, has the
lowest value. In this case, the thermodynamic potential is the
Helmholtz free energy $F$, and so a state is favored relatively to
another if it has lower $F$ for given $R$, $T$, and $Q$. If a system
is in a stable state but with a higher free energy $F$ than another
stable state, it is probable that the system undergoes
a conversion, i.e., a phase transition, to the stable state with
the lowest free
energy. Indeed, in the calculation of the partition function by the
path integral approach, if there are two stable configurations, i.e.,
two states that minimize the action, then the largest contribution to
the partition function is given by the configuration with the lowest
action or, in thermodynamic language, with the lowest free energy.
This type of phase transitions are first order since the
free energy is continuous, but the first derivatives
are discontinuous.

In the case of the canonical ensemble of an electrically charged black
hole inside a cavity in $d$ dimensions, we must compare the free
energy between all the stable black hole solutions of the ensemble,
i.e., we have to compute $F_\mathrm{ bh}$
given in 
Eq.~\eqref{eqch5:freeennergybh}, for the possible solution $r_+(R,T,Q)$.
For any $d$ in this ensemble one can have three
solutions for the same temperature, two of them are stable.
The stable black hole with lowest $F_\mathrm{ bh}$ is the one that is
favored.
This means that considering
only the two stable black hole solutions,
one 
would then have a first order phase transition from 
$r_{+1}$ to $r_{+3}$, for the electric charge parameter
in the range 
$0 < \frac{\mu Q^2}{R^{2d-6}} <\frac{\mu Q_s^2}{R^{2d-6}}$,
and in the limit of the charge parameter with value
$\frac{\mu Q^2}{R^{2d-6}}=\frac{\mu Q^2_s}{R^{2d-6}}$,
this first order phase transition 
becomes a second order phase transition.
It is also interesting to compare the black hole solutions with 
the non-gravitating electrically
charged shell case for the same boundary data, which has
free energy given in Eq.~\eqref{eqch5:freeennergyshell}.
As we argued above, this shell is useful in 
mimicking charged
hot flat space inside the cavity. 
Depending on the value of the radius of the shell
$\frac{r_\mathrm{ shell}}{R}$, this free energy can go from infinity, when
$\frac{r_\mathrm{ shell}}{R}=0$,
to zero, when
$\frac{r_\mathrm{ shell}}{R}=1$. 
In the case of $\frac{r_\mathrm{ shell}}{R}=0$, the shell is never 
favored, while for $\frac{r_\mathrm{ shell}}{R}=1$,
i.e., the case of hot flat space with
the electric charge at the boundary, there is a region 
in which it is favored. 
In order to proceed,
it is essentially assumed a shell with 
$\frac{r_\mathrm{ shell}}{R}=1$, so that 
 $F_\mathrm{ shell}=F_\mathrm{ hfs}=0$.

Another issue that should be raised in the connection to favorable
states, although it does not come directly from the ensemble formalism
and its thermodynamics, is that there is a black hole radius $r_+$,
more precisely, there is
a ratio $\frac{r_+}{R}$, for which the thermodynamic
energy contained within $R$ is higher than the Buchdahl bound
or, in this context, the 
generalized Buchdahl bound~\cite{Wright:2015dma}.  When this
happens, that energy content should collapse into a black hole. In
this situation there is no more favorable phase considerations, the
unique phase is a black hole.  Indeed, the generalized
Buchdahl bound yields the maximum mass, or maximum
energy, that can be enclosed in a $d$-dimensional cavity with electric
charge $Q$, before the system shows up
some kind of singularity. At the bound or above, the system
most likely tends to gravitational collapse.  Since the mass of a
system is related to the gravitational radius, it also sets a bound on
the ratio $\frac{r_+}{R}$.  In this context, one should consider this
bound as yielding, for a fixed $R$, the mass $m$, or the gravitational
radius $r_+$, above which the energy within the system is sufficiently
large that the system cannot support itself gravitationally and
collapses. We can now apply this concept to the case of interest 
here.

In the Schwarzschild black hole case in $d$ dimensions it was found in
\cite{Andre:2021ctu}, that the canonical ensemble yields $F_\mathrm{ bh}=0$
when $\frac{r_+}{R}$ has the Buchdahl bound
value, $\left(\frac{r_+}{R}\right)_\mathrm{ Buch}$.
Since $R$ is fixed, one can write 
$\left(\frac{r_+}{R}\right)_\mathrm{ Buch}\equiv
\frac{r_{+\mathrm{Buch}}}{R}$ to simplify the notation.
In a $d$-dimensional Schwarzschild spacetime
one has $\frac{r_{+\mathrm{Buch}}}{R}=
\left( \frac{ 4(d-2) }
{ (d-1)^2 }
 \right)^{\frac{1}{d-3}}$.
One can infer that black hole
solutions with higher $\frac{r_+}{R}$, i.e., higher
temperatures $RT$, yield gravitational collapse. Since zero free
energy in this electrically
uncharged case, is also the free energy of hot flat space, $F_\mathrm{
hfs}=0$, one sees that in the uncharged case
one passes
directly from a situation where a hot flat
space phase is favored relatively to a black hole phase, to a situation
where the phase is a
phase where surely there is a black hole,
not merely a phase in which the black hole is favored.

Now,
in the canonical ensemble
for a black hole with electric charge,
one finds
that for $F_\mathrm{ bh}=0$ only the
bigger black hole exists, and it gives a value
for $\frac{r_{+}}{R}$ that is higher
than the 
Buchdahl bound value.
Thus, there is a definite $F_\mathrm{ bh}$ value
greater than zero where 
the 
Buchdahl value
$\frac{r_{+\mathrm{Buch}}}{R}$ 
is met. We found this  
by numerical means up to
$d=16$, but we have not found an analytical proof for all $d$.
For this definite value of $F_\mathrm{ bh}$
or lower values of it, the system
has high enough temperature and
high enough self thermodynamic
energy to undergo gravitational collapse.
When this happens there is no more
coexistence of phases, there is
only the black hole phase.
Below the saddle, or critical, charge, i.e.,
below the
electric charge parameter given
by $\frac{\mu Q_s^2}{R^{2d-6}}$,
it is the black hole solution
$\frac{r_{+3}}{R}$ that achieves 
$\frac{r_{+\mathrm{Buch}}}{R}$.
Above the saddle charge, i.e.,
above $\frac{\mu Q_s^2}{R^{2d-6}}$,
it is the black hole solution
$\frac{r_{+4}}{R}$ that achieves 
$\frac{r_{+\mathrm{Buch}}}{R}$.
In contrast,
if we considered the grand canonical ensemble with electric charge in 
Chapter~\ref{ch:grandcanonicalblackhole} or in \cite{Fernandes:2023byx},
rather than the canonical ensemble studied here, 
the point of interest would be
$W_\mathrm{ bh}=0$, where $W_\mathrm{ bh}$ is
the grand potential free energy related to
the grand canonical ensemble, and it corresponds to a
$\frac{r_+}{R}$ which is lower than the 
Buchdahl bound value.
In the grand canonical ensemble,
there is only one stable black hole. So, this means
that for
$W_\mathrm{ bh}=0$, the two phases black hole
and hot flat space coexist equally. For 
$W_\mathrm{ bh}<0$ up to some definite
negative value, then the two phases, black hole
and hot flat space, coexist but
the black hole dominates.
For the definite negative value of $W_\mathrm{ bh}$,
the radius $\frac{r_+}{R}$ is 
the
Buchdahl 
bound value $\frac{r_{+\mathrm{Buch}}}{R}$.
For even lower $W_\mathrm{ bh}$, i.e., for
higher temperature
parameter $RT$, one has $\frac{r_+}{R}$ larger than 
$\frac{r_{+\mathrm{Buch}}}{R}$ 
and 
the system collapses, or is
collapsed, there is thus no coexistence, only the black
hole phase remains.
Although numerically all three radii $\frac{r_+}{R}$, namely, the
canonical zero free
energy, the Buchdahl, and the grand canonical zero
grand potential, are very close, see
Sec.~\ref{sech5:BHzerofreeenergy}, it seems that a connection
between the ensemble stability and the mechanical stability of matter
is elusive here.
A comment is in order. The Buchdahl bound applies to a
self-gravitating mechanical system consisting of a ball of matter of
radius $R$. The system here is a thermodynamic system, with boundary data,
namely $R$, $T$, and $Q$, and contains no matter. One can argue that
in higher orders of approximation, the system contains packets of
energy and one can plausibly deduce that the system must collapse once
the Buchdahl bound is surpassed.  Be as it may, the inference made here 
comes from dynamics, not thermodynamics, and therefore is
strictly outside the followed approach.

To better understand the
issues and 
make progress, we have to pick up definite
dimensions. We now specify the generic $d$-dimensional results
to the dimensions $d=4$ and $d=5$, with 
a more thorough analysis for $d=5$.

\subsection{Full analysis in $d=4$}

For $d=4$, as for any $d$, 
this ensemble can have either one or three
black hole solutions for a given temperature.
When there are three, two of them are stable
and are of
interest in the consideration
of the most favorable phase, while the remaining
solution is unstable and is of no interest in 
the consideration of the most favorable phase.
The two that are stable
have to be compared against one another
to see which is the most favorable phase. 

We can start by comparing the free energy of 
the several black hole solutions that exist in
this ensemble between themselves. From
Eq.~\eqref{eqch5:freeennergybh}, 
in $d=4$, the black hole free energy
is
\begin{align}
F_\mathrm{ bh} = \frac{R}{l_p^2}\left(1 - \sqrt{f(R,Q,r_+)}\right) 
- T\frac{A_+}{4l_p^2}
\,,
\label{eqch5:freeennergybh4d}
\end{align}
where here one has $\frac{A_+}{4}=\pi r_+^2$,
$f(R,Q,r_+) \equiv 1 - \frac{r_+
+ \frac{l_p^2 Q^2}{r_+}}{R} + \frac{l_p^2 Q^2}{R^2}$,
where it was used $\mu=l_p^2$, and $r_+=r_+(T,R,Q)$.
In $d=4$, 
the saddle electric charge parameter value
$\frac{l_p^2 Q_s^2}{R^2} = (\sqrt{5} -2)^2
= 0.056$, 
the last equality being approximate,
separates the region with only one solution 
from the region with three solutions.

There is a first set
of general and specific comments that we must make, namely about the positivity 
of the free energy for each solution. 
For
$0\leq \frac{l_p^2 Q^2}{R^2}<\frac{Q_s^2}{R^2}$, the stable black
hole solution $\frac{r_{+1}}{R}$
has positive $F_\mathrm{ bh}$
for all the temperatures in which the solution exists.
The same happens for the solution $\frac{r_{+2}}{R}$,
but this solution is of not interest here
since it is unstable.
The other stable black
hole solution $\frac{r_{+3}}{R}$
has a temperature $T_{F_\mathrm{ bh}=0}$ depending on the
electric charge,
at which the free energy becomes zero, and
so the black
hole solution
$\frac{r_{+3}}{R}$ can have $F_\mathrm{ bh}$ positive or negative.
For the critical charge
$\frac{l_p^2 Q^2}{R^2}=\frac{l_p^2 Q_s^2}{R^2}$,
with
$\frac{l_p^2 Q_s^2}{R^2} = 0.056$ approximately, the stable black hole solution
$\frac{r_{+1}}{R}$
has positive free energy, the point
$\frac{r_{+1}}{R}=
\frac{r_{+2}}{R}=
\frac{r_{+3}}{R}
$ has positive free energy,
and the stable black hole solution $\frac{r_{+3}}{R}$
has a temperature
$T_{F_\mathrm{ bh}=0}$ at which the free
energy becomes zero.
For
$\frac{l_p^2 Q_s^2}{R^2}< \frac{l_p^2 Q^2}{R^2}<1$,
the only black hole solution is $\frac{r_{+4}}{R}$, which is stable, 
and it 
has a temperature $T_{F_\mathrm{ bh}=0}$ depending on the
electric charge,
at which the free energy becomes zero.
So, the free energy
of $\frac{r_{+4}}{R}$ can be positive or negative.
Quite generally one can calculate $T_{F_\mathrm{ bh}=0}$
by solving $F_\mathrm{ bh}=0$, with 
$F_\mathrm{ bh}$ given in Eq.~\eqref{eqch5:freeennergybh4d},
for either the solution 
$\frac{r_{+3}}{R}$ or $\frac{r_{+4}}{R}$.
The free energy can be written in terms of 
$m$ and $Q$ through Eq.~\eqref{eqch5:beta1} in $d=4$ and 
through $2 l_p^2 m = r_+ + \frac{l_p^2 Q^2}{r_{+}}$, 
allowing us to reduce $F_\mathrm{ bh}=0$ into a quartic
equation for the mass, see
Sec.~\ref{sech5:BHzerofreeenergy}.
The solutions have positive free energy for temperatures lower than 
$T_{F_\mathrm{ bh}=0}$,
and the solutions have negative free energy
for 
temperatures higher
than $T_{F_\mathrm{ bh}=0}$.

There is a second set
of general and specific comments that we must make, namely about the favorability 
between black hole solutions.
For  $0\leq\frac{l_p^2 Q^2}{R^2}
<\frac{l_p^2 Q_s^2}{R^2}$,
there is a favorability
temperature $T_f$ which depends on the
electric charge,  and at which the solutions $\frac{r_{+1}}{R}$
and
$\frac{r_{+3}}{R}$
have the same free energy.  For temperatures lower than
$T_f$, the solution $r_{+1}$ is more favorable than $\frac{r_{+3}}{R}$,
or it is the
only existing solution. For temperatures higher than $T_f$, the
solution $\frac{r_{+3}}{R}$
is more favorable than $\frac{r_{+1}}{R}$, or it is the only
existing solution.
For the critical charge
 $\frac{l_p^2 Q^2}{R^2}=\frac{l_p^2 Q_s^2}{R^2}$,
the temperature $T_f$ is the temperature at which
$\frac{r_{+1}}{R}=
\frac{r_{+2}}{R}=
\frac{r_{+3}}{R}
$
and all have the same free energy, i.e., the
stable solutions $\frac{r_{+1}}{R}$
and
$\frac{r_{+3}}{R}$ coexist.
For
$\frac{l_p^2 Q_s^2}{R^2}<\frac{l_p^2 Q^2}{R^2}<1$,
there is only one black hole solution, it is
$\frac{r_{+4}}{R}$, and, since it is stable, it is favored.
One can now consider phase transitions between the two
stable black hole solutions.
There is 
a first order phase transition from 
$r_{+1}$ to $r_{+3}$, for the electric charge
parameter in the range 
$0 < \frac{l_p^2 Q^2}{R^{2}} <\frac{l_p^2 Q^2_s}{R^{2}}$ 
and, additionally, in the limit of
the electric charge parameter with value
$\frac{l_p^2 Q^2}{R^{2}}=\frac{l_p^2 Q^2_s}{R^{2}}$, 
this first order phase transition 
turns into a second order phase transition.

We now compare, in $d=4$,
the black hole phases just discussed above with
hot flat space phase, which we emulate here by a
nonself-gravitating shell.
In $d=4$,
the free energy of the shell is
\begin{align}
F_\mathrm{ shell} = \frac{Q^2}{2r_\mathrm{ shell}}
\left(1- \frac{r_\mathrm{ shell}}{R}\right)
\,,
\label{eqch5:freeennergyshell4d}
\end{align}
where $r_\mathrm{ shell}$ is
the radius of the shell, see Eq.~\eqref{eqch5:freeennergyshell}.
So $F_\mathrm{ shell}$ depends on the
electric charge $Q$, on $r_\mathrm{ shell}$,
and on $R$, but
is a constant as a function of the
temperature $T$.
The case of
a very small shell will lead to a very high free energy due 
to the dependence on $\frac{Q}{r_\mathrm{ shell}}$,
and therefore, for this case the region of 
favorability for the shell
lies in very small values of the charge.
There are also
the cases of intermediate shell radius which
would have to be analyzed specifically.
The other limiting  
case is when the charge is near the boundary of the cavity,
with the 
free energy of this case tending to zero.
Ultimately, the black hole is favored when 
$F_\mathrm{ bh} <F_\mathrm{ shell}$, both coexist equally
when 
$F_\mathrm{ bh} =F_\mathrm{ shell}$, and
the black hole is not favored when 
$F_\mathrm{ bh} > F_\mathrm{ shell}$.
When the radius of the shell is at the cavity radius,
$\frac{r_\mathrm{ shell}}{R}=1$,
then the shell has zero free energy
and emulates hot flat space with electric charge
at the boundary.
Then, the free energy of hot flat space is
 $F_\mathrm{ shell}=F_\mathrm{ hfs}=0$. 
 The black hole is not favored when 
$F_\mathrm{ bh} >0$, both the black hole and hot flat space 
coexist equally
when 
$F_\mathrm{ bh} =0$, and
the black hole is favored when 
$F_\mathrm{ bh} <0$.
When the system finds itself in a phase that is not
favored, it will make a first order phase transition
to the favored phase.

The problem of the thermodynamic phases is even more complicated as we
have mentioned already.  When there is no electric charge, i.e., for the
Schwarzschild space in $d=4$, it was found in \cite{Andre:2021ctu} that,
in the canonical ensemble, the condition $F_\mathrm{ bh}=0$ yields a value
for $\frac{r_+}{R}$ that is equal to the generalized Buchdahl
bound~\cite{Wright:2015dma}, i.e., the limiting value
$\left(\frac{r_+}{R}\right)_\mathrm{ Buch}$ for gravitational collapse of
a self-gravitating system of energy $E$ and radius $R$.  Since
$R$ is fixed in the ensemble, one can write $\left(\frac{r_+}{R}\right)_\mathrm{
Buch}\equiv \frac{r_{+\mathrm{Buch}}}{R}$ to simplify the notation, and in
$d=4$ one has $ \frac{r_{+\mathrm{Buch}}}{R}=\frac89=0.89$, the latter
equality being approximate.  This result means that,
in the uncharged case, 
as soon as the black hole phase is favored, there is no further
coexistence with hot flat space, and the system collapses.
For nonzero electric charge there is
no more coincidence. 
Here,  to discuss this issue of favorability 
between black hole and hot flat space, we consider
the case for which the free energy of the
shell is zero, $F_\mathrm{ shell}=0$,
i.e., the case of hot flat space with electric charge
at the boundary, $\frac{r_\mathrm{ shell}}{R}=1$.  In this case,
the shell is situated at the cavity, and so $F_\mathrm{ shell}$ is the
free energy of hot flat space, $F_\mathrm{ hfs}$, which is zero.
For
nonzero electric charge $Q$, i.e., nonzero charge parameter
$\frac{l_p^2 Q^2}{R^2}$, we
find that in the canonical ensemble, the condition $F_\mathrm{ bh}=0$
yields a $\frac{r_+}{R}$ value, both for $\frac{r_{+3}}{R}$ and
$\frac{r_{+4}}{R}$, that is higher than the generalized Buchdahl bound. 
Notice that the generalized Buchdahl bound here is the limiting 
value of $\frac{r_+}{R}$ for gravitational collapse of a
self-gravitating system of energy $E$, electric charge $Q$, and radius $R$.
For an electric charge parameter
lower or equal than the saddle
value $\frac{l_p^2 Q_s^2}{R^2}$,
only the solution $\frac{r_{+3}}{R}$ can take the value of 
the Buchdahl bound, corresponding to a positive free energy and 
some temperature value $RT$. For a system with
this $RT$ or higher, then the system collapses gravitationally into a
black hole with the corresponding $\frac{r_{+3}}{R}$.  For an electric
charge higher or equal than the saddle value
$\frac{l_p^2 Q_s^2}{R^2}$, the solution
$\frac{r_{+4}}{R}$ can take the value of the 
Buchdahl bound, having a definite positive value of
$F_\mathrm{ bh}$, at some temperature parameter $RT$.
For a system with this $RT$ or higher, the system again collapses
gravitationally into a black hole with the corresponding
$\frac{r_{+4}}{R}$.
Interesting to note that in the grand canonical ensemble, where there
is only one stable black hole solution, the equation 
$W_\mathrm{ bh}=0$, $W_\mathrm{ bh}$
denoting the grand potential, yields a $\frac{r_+}{R}$ value that is
lower than the Buchdahl bound.  Thus, in this case,
when $W_\mathrm{ bh}=0$ for the system, the two phases coexist, black
hole and hot flat space. For $W_\mathrm{ bh}<0$, the black hole phase
dominates in relation to hot flat space. And 
for a certain definite negative value of $W_\mathrm{ bh}$, the
value of $\frac{r_+}{R}$ of the system is the same as the value of the
Buchdahl bound. From then on the system collapses, the only phase
being the black hole phase, and there is no coexistence of phases, see also
Sec.~\ref{sech5:BHzerofreeenergy}.

\subsection{Full analysis in $d=5$}

For $d=5$, as for any $d$, this ensemble has between one and three
black hole solutions for a given temperature.  When there are three
solutions, two of them are stable and are going to be considered here, 
while the remaining is unstable and is
of no interest in this analysis. 
The two that are stable have to be compared against
one another to see which is the most favorable phase.

\begin{figure}[h]
\centering
\includegraphics[width=0.7\linewidth]{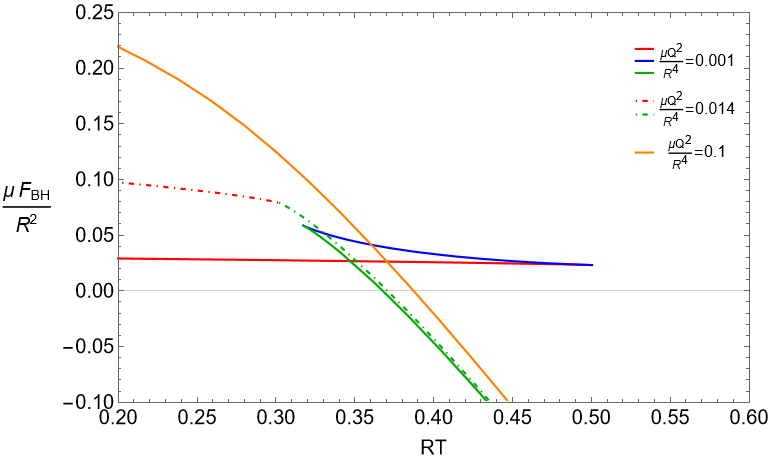}
\caption{\label{figch5:freeenergyBH}
Free energy $F_\mathrm{ bh}$ of the charged 
black hole solutions of the canonical ensemble in $d=5$,
given as a quantity with no units $\frac{\mu F_\mathrm{ bh}}{R^2}$,
as a function of the temperature parameter $RT$ for
several electric charge parameters $\frac{\mu Q^2}{R^4}$,
where $\mu=\frac{4l_p^3}{3\pi}$. 
For $\frac{\mu Q^2}{R^4} = 0.001$, the solution $r_{+1}$ is in red, 
the solution $r_{+2}$ is in blue, and the solution $r_{+3}$ is in 
green, all of them in solid lines. 
For $\frac{\mu Q^2}{R^4} = \frac{(68 - 27 \sqrt{6})^2}{250}
= 0.014$, the latter equality being approximate,
the solution $r_{+1}$ is in red 
and the solution $r_{+3}$ is in green, all of them in dashed lines.
For $\frac{\mu Q^2}{R^4} = 0.1$, the solution $r_{+4}$ is in orange,
in solid line. 
}
\end{figure}

We can start by comparing the free energy of 
the several black hole solutions that exist in
this ensemble between themselves.
In $d=5$, the black hole free energy
is
\begin{align}
F_\mathrm{ bh} = \frac{ R^2}{\mu}\left(1 - \sqrt{f(R,Q,r_+)}\right) 
- T\frac{A_+}{4l_p^3}
\,,
\label{eqch5:freeennergybh5d}
\end{align}
where here $\frac{A_+}{4}=\frac{\pi^2 r_+^3}{2}$,
$f(R,Q,r_+) \equiv 1 - \frac{r_+^2
+ \frac{\mu Q^2}{r_+^2}}{R^2} + \frac{\mu Q^2}{R^4}$,
$\mu=\frac{4 l_p^3}{3\pi}$, and $r_+=r_+(T,R,Q)$.
To help in the analysis, the free action $F_\mathrm{ bh}$ is 
plotted in Fig.~\ref{figch5:freeenergyBH}
as a function of the temperature
parameter $RT$,
for fixed electric charge parameter $\frac{\mu Q^2}{R^4}$
in $d=5$.
Recall that in $d=5$, one has
the saddle electric charge parameter value
$\frac{\mu Q_s^2}{R^4} = \frac{(68 - 27 \sqrt{6})^2}{250}
= 0.014$, the last equality being approximate.
%
%

We can make a first set
of general and specific comments directly from
Fig.~\ref{figch5:freeenergyBH}, regarding the positivity of the 
free energy for each solution. 
For
relatively low
electric charge parameter
$0\leq\frac{\mu Q^2}{R^4}<\frac{\mu Q_s^2}{R^4}$,
where $\mu=\frac{4l_p^3}{3\pi}$ in $d=5$,
the solution $\frac{r_{+1}}{R}$ 
has positive $F_\mathrm{ bh}$
for all the temperatures in which the solution exists.
The same happens for the solution $\frac{r_{+2}}{R}$, 
but this solution is of no interest here
since it is unstable.
The solution $\frac{r_{+3}}{R}$ 
has a temperature $T_{F_\mathrm{ bh}=0}$ depending on the
electric charge
at which the free energy becomes zero, and
so  $\frac{r_{+3}}{R}$  can have $F_\mathrm{ bh}$ positive or negative.
In the figure, this range of
the electric charge parameter is represented by
the case $\frac{\mu Q^2}{R^4}=0.001$.
For $\frac{\mu Q^2}{R^4}=0.001$,
one has for the  $\frac{r_{+3}}{R}$  solution that
$T_{F_\mathrm{ bh}=0}=0.367$ approximately.
For the saddle charge $\frac{\mu Q^2}{R^4}=\frac{\mu Q_s^2}{R^4}$,
with $\frac{\mu Q_s^2}{R^4} = 0.014$ approximately,
the solution $\frac{r_{+1}}{R}$ is positive, the point
$\frac{r_{+1}}{R}=
\frac{r_{+2}}{R}=
\frac{r_{+3}}{R}
$
is positive, and the
solution $\frac{r_{+3}}{R}$   
has a temperature $T_{F_\mathrm{ bh}=0}=0.37$
at which the free energy becomes zero.
For
relatively high
electric charge parameter
$\frac{\mu Q_s^2}{R^4}<\frac{\mu Q^2}{R^4}<1$,
the only solution is $\frac{r_{+4}}{R}$
and it 
has a temperature $T_{F_\mathrm{ bh}=0}$ depending on the
electric charge. So $F_\mathrm{ bh}$
of the black hole $\frac{r_{+4}}{R}$
can be positive or negative.
In the figure, this range of
$\frac{\mu Q^2}{R^4}$
is represented by
the case $\frac{\mu Q^2}{R^4}=0.1$.
For $\frac{\mu Q^2}{R^4}=0.1$,
one has that the
solution $\frac{r_{+4}}{R}$
has
$T_{F_\mathrm{ bh}=0}=0.387$ approximately.
Quite generally, one can calculate $T_{F_\mathrm{ bh}=0}$
by solving $F_\mathrm{ bh}=0$, with 
$F_\mathrm{ bh}$ given in Eq.~\eqref{eqch5:freeennergybh5d}
for either the solution 
$\frac{r_{+3}}{R}$
or $\frac{r_{+4}}{R}$.
One obtains a quartic equation for the mass $2\mu m = r_+^2
+ \frac{\mu Q^2}{r_{+}^2}$, with here $\mu=\frac{3}{4\pi}$,
as a function of the electric charge, 
see Sec.~\ref{sech5:BHzerofreeenergy}. 
For temperatures lower than 
$T_{F_\mathrm{ bh}=0}$, the solutions have positive free energy and for 
temperatures higher
than $T_{F_\mathrm{ bh}=0}$, the solutions have negative free energy.

We can make a second set
of general and specific comments directly from
Fig.~\ref{figch5:freeenergyBH}, regarding the favorability 
between black hole solutions.
For a range of low
electric charge parameter
$0\leq\frac{\mu Q^2}{R^4}< \frac{\mu Q_s^2}{R^4}$, the
solutions $\frac{r_{+1}}{R}$ and $\frac{r_{+3}}{R}$ 
have the same free energy 
at a specific temperature $T_f$, i.e., the phase
favorability temperature which
depends on $\frac{\mu Q^2}{R^4}$. 
For temperatures lower than $T_f$, the solution
$\frac{r_{+1}}{R}$ either has lower free energy than $\frac{r_{+3}}{R}$
or it is the only existing solution,
and so $\frac{r_{+1}}{R}$ is more favorable. 
For a temperature equal to $T_f$, the solutions $\frac{r_{+1}}{R}$ and
$\frac{r_{+3}}{R}$ have the same free energy and they
are equally favorable, meaning they coexist equally.
For temperatures higher than
$T_f$, the solution $\frac{r_{+3}}{R}$ either has lower free energy than 
$\frac{r_{+1}}{R}$ or it is the only existing solution, and so 
$\frac{r_{+3}}{R}$ is more favorable.
This is represented
for  $\frac{\mu Q^2}{R^4}= 0.001$ in the figure. One can see that in this
case, the favorability temperature is $RT_f=0.347$ approximately. 
Also, for $RT<0.32$, there is only the
$\frac{r_{+1}}{R}$ solution, whereas for $RT>0.50$ there is only the
$\frac{r_{+3}}{R}$ solution. The solution $\frac{r_{+2}}{R}$ is
unstable and does not enter in this analysis, however it is plotted in
the figure to show a continuity of the free energy on the three solutions.
For saddle charge
$\frac{\mu Q^2}{R^4}=\frac{\mu Q_s^2}{R^4}
= 0.014$, the latter equality being approximate,
which is shown in the figure, 
the temperature $T_f = 0.30$, approximately, 
is the temperature at which
$\frac{r_{+1}}{R}= \frac{r_{+2}}{R}= \frac{r_{+3}}{R}$, and all have
the same free energy, i.e., $\frac{r_{+1}}{R}$ and $\frac{r_{+3}}{R}$
coexist. For temperatures lower than $T_f$, the solution
$\frac{r_{+1}}{R}$ is the only existing solution.  For temperatures
higher than $T_f$, the solution $\frac{r_{+3}}{R}$ is the only existing
solution.
For the higher values of the electric charge parameter, i.e., for
$\frac{\mu Q_s^2}{R^4}<\frac{\mu Q^2}{R^4}<1$,
there is only one black hole solution
$\frac{r_{+4}}{R}$ that is stable, and so it is favorable. This is 
represented in the figure by the case $\frac{\mu Q^2}{R^4} = 0.1$.
We can now consider phase transitions between the two
stable black hole solutions.
One has a first
order phase transition from $r_{+1}$ to $r_{+3}$, for the electric
charge parameter in the range $0 < \frac{\mu Q^2}{R^{4}} <\frac{\mu
Q^2_s}{R^{4}}$.  Moreover, in the limit of the electric charge
parameter given by the value
$\frac{\mu Q^2}{R^{4}}=\frac{\mu Q^2_s}{R^{4}}$, this first
order phase transition becomes a second order phase transition. This
can be seen from Fig.~\ref{figch5:freeenergyBH}, since the intersection
point represents a first order phase transition, and at the limit of
the critical charge, this point represents a second order phase
transition.

\begin{figure}[h]
\centering
\includegraphics[width=0.7\linewidth]{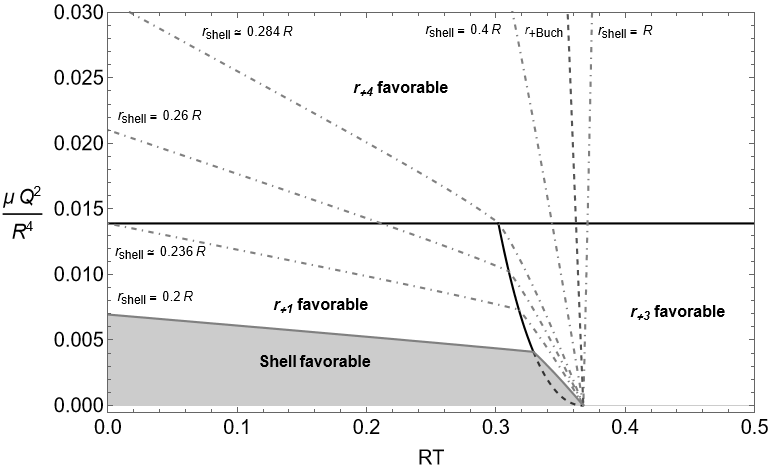}
\caption{\label{figch5:favorable}
Favorable states of the 
canonical ensemble of
an electrically 
charged black hole inside a cavity 
in $d=5$
in an electric charge  $Q$ times
temperature $T$, more precisely,
$\frac{\mu Q^2}{R^4}\times RT$ plot.
It is displayed
the region where 
the black hole
$r_{+1}$ is a
favorable phase,
the region
where the black hole
$r_{+3}$ is a favorable
phase, and the region
where the black hole
$r_{+4}$ is a favorable phase.
The delimiters of the favorable 
regions of the black hole solutions are the black lines, 
including the 
dashed line.
It is also incorporated the
solution for a nongravitating 
electrically charged shell
as a simulator for hot flat space.
The electrically charged shell with 
$\frac{r_\mathrm{ shell}}{R} = 0$ is never favored.
The electrically charged shell with 
$\frac{r_\mathrm{ shell}}{R} = 0.2$ is favored in 
the region in 
gray, this case is given as an example.
The upper delimiter of the region of favorability 
of electrically charged shells with
$\frac{r_\mathrm{ shell}}{R} = 0.236$
approximately, 
$\frac{r_\mathrm{ shell}}{R} = 0.26$, 
$\frac{r_\mathrm{ shell}}{R} = 0.284$
approximately, $\frac{r_\mathrm{ shell}}{R} = 0.4$ and 
$\frac{r_\mathrm{ shell}}{R} = 1$, which better simulates
hot flat space, are given by
the dot-dashed lines. The Buchdahl condition
line, i.e., 
$r_{+\mathrm{Buch}}$, above which there is presumably
collapse is given by a thick black dash line.}
\end{figure}

The black hole phases discussed just above
with hot flat space phase, which it is emulated by a
nonself-gravitating shell, are now compared for $d=5$,
see Fig.~\ref{figch5:favorable}.
%
%
%
%
%
%
The favorable states for each electric
charge and temperature,
and for various values of the shell radius 
can be seen in the figure. 
The free energy of the shell for the 
case $d=5$ is
\begin{align}
F_\mathrm{ shell} = \frac{Q^2}{2r_\mathrm{ shell}^2}
\left(1-\frac{r_\mathrm{ shell}^2}{R^2} \right)
\,,
\label{eqch5:freeennergyshell5d}
\end{align}
where $r_\mathrm{ shell}$ is
the radius of the shell, see Eq.~\eqref{eqch5:freeennergyshell}.
So the shell free energy $F_\mathrm{ shell}$ has a dependence on
electric charge $Q$, on $r_\mathrm{ shell}$, and on $R$, but
as a function of the
temperature $T$, the free energy is a constant.
Due to the term $\frac{Q^2}{r_\mathrm{ shell}^2}$, 
the free energy becomes divergent for 
a very small shell and fixed electric charge.
Therefore, the region of 
favorability for the very small shell
lies in very small values of the
electric charge $Q$.
There are the cases of intermediate shell radius
that are represented in the figure, namely the cases 
$\frac{r_\mathrm{ shell}}{R}={0.2, 0.236, 0.26, 0.284,0.4}$, 
with $0.236$ and $0.284$ being approximate values.
The more interesting limiting 
case is when the electric
charge is near or at the boundary of the cavity,
$\frac{r_\mathrm{ shell}}{R}=1$. The 
free energy of the shell in this limit is
zero.
The black hole solution is favored compared to the shell when $F_\mathrm{
bh} <F_\mathrm{ shell}$, while both the black hole and the shell coexist
equally when $F_\mathrm{ bh} =F_\mathrm{ shell}$, and the black hole is not
favored compared to the shell when $F_\mathrm{ bh} > F_\mathrm{ shell}$. The
gray dashed curves in the figure represent the condition $F_\mathrm{ bh}
=F_\mathrm{ shell}$ for each shell radius, delimiting the regions where
the black hole is favorable, for higher temperature, and where the
shell is favorable, for lower temperature.  When the radius of the
shell is at the cavity radius, $\frac{r_\mathrm{ shell}}{R}=1$, the free
energy of the shell becomes zero, emulating hot flat space with free
energy
$F_\mathrm{ shell}=F_\mathrm{ hfs}=0$. This is the case of hot flat space
with electric charge at the boundary.
Again, the black hole is not favored compared to hot flat space when 
$F_\mathrm{ bh} >0$, while both the black hole and hot flat space 
coexist equally when 
$F_\mathrm{ bh} =0$, and
the black hole is favored compared to hot flat space when 
$F_\mathrm{ bh} <0$. The gray dashed curve $r_\mathrm{ shell} = R$ 
in the figure corresponds to the boundary of the regions of 
favorability $F_\mathrm{ bh} =0$, and for higher temperature, the 
black hole is favorable, while for lower temperature, hot flat space 
is favorable.
If for some reason the system is in a not favored phase, 
then a first order phase transition occurs
to a favored phase.

\begin{figure}[h]
\centering 
\includegraphics[width=0.7\linewidth]{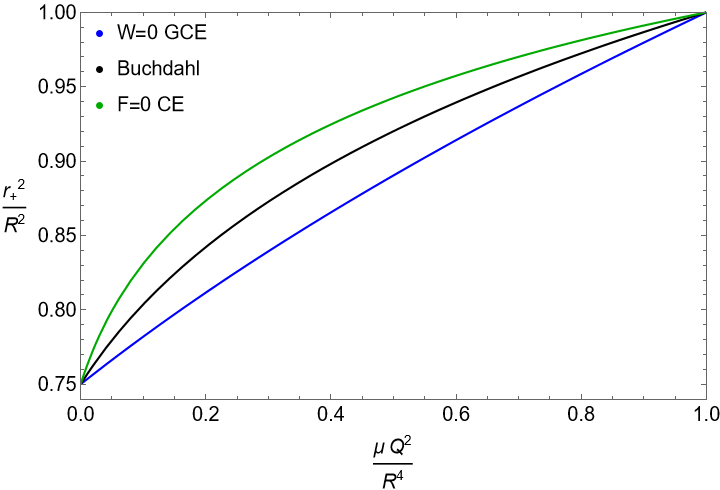}
\caption{\label{figch5:Buchdahl}
Ratio $\frac{r_+^2}{R^2}$ in terms of
the electric charge parameter
$\frac{\mu Q^2}{R^4}$, $\mu=\frac{4}{3\pi}$,
for $d=5$ for three different cases: 
given by the condition 
$F_\mathrm{ bh}=0$ in the canonical ensemble in green,
representing the stable solution $\frac{r_{+3}}{R}$; 
given by the condition $W_\mathrm{ bh} =0$ 
in the grand canonical ensemble in blue,
representing the only stable solution; 
and given by generalized Buchdahl condition in black.}
\end{figure}

The problem of the thermodynamic phases is more involved
as mentioned already above.  When there is no electric charge, 
one has Schwarzschild space in $d=5$. It was found in
\cite{Andre:2020czm,Andre:2021ctu} that, in the canonical ensemble 
of Schwarzschild space in $d=5$, the
condition $F_\mathrm{ bh}=0$ corresponds to a value for $\frac{r_+}{R}$ that is
equal to the generalized Buchdahl bound radius \cite{Wright:2015dma},
which is the value $\left(\frac{r_+}{R}\right)_\mathrm{ Buch}$ for
gravitational collapse of a self-gravitating system of energy $E$ and
radius $R$. Since $R$ is maintained fixed, it is defined
$\left(\frac{r_+}{R}\right)_\mathrm{ Buch}\equiv \frac{r_{+\mathrm{Buch}}}{R}$, 
and in $d=5$, one has $\frac{r_{+\mathrm{
Buch}}}{R}=\frac{\sqrt3}{2}=0.86$, the latter equality being
approximate. Since for $Q=0$, the free energy of hot flat space is
zero, $F_\mathrm{ hfs}=0$, meaning that there is no further coexistence with
hot flat space as soon as the black hole phase is favored, 
because the system tends to collapse.
For nonzero electric charge
parameter
$\frac{\mu Q^2}{R^4}$
there is no coincidence.
To compare the free energies,
one considers the case in which the shell
has radius equal to the cavity radius,
$\frac{r_\mathrm{ shell}}{R}=1$, and so $F_\mathrm{ shell}=0$,
meaning that the shell is a surrogate for hot flat space, i.e.,
$F_\mathrm{ shell}=F_\mathrm{ hfs}=0$,
indeed it is hot flat space
with electric charge at the boundary.
For nonzero 
$\frac{\mu Q^2}{R^4}$, one finds that in the
canonical ensemble $F_\mathrm{ bh}=0$ results in a $\frac{r_+}{R}$ value, both
for $\frac{r_{+3}}{R}$ and $\frac{r_{+4}}{R}$, 
that is higher than the generalized Buchdahl bound,
which is the value of $\frac{r_+}{R}$ for gravitational collapse
of a self-gravitating system of energy $E$, electric charge $Q$, and
radius $R$, see Fig.~\ref{figch5:Buchdahl}.  
For an electric charge parameter lower or equal than the saddle
value
$\frac{\mu Q_s^2}{R^4}$, there is a temperature $RT$ at which
the solution $\frac{r_{+3}}{R}$ can assume the 
value of the Buchdahl bound, corresponding to a positive free energy
lower than the free energy of $\frac{r_{+1}}{R}$. 
For a system with this $RT$ or
higher, the system must suffer gravitational collapse into a black hole
with the corresponding $\frac{r_{+3}}{R}$. For an electric charge
higher than the saddle
value $y_s$, there is again a temperature $RT$ at which 
$\frac{r_{+4}}{R}$ assumes the Buchdahl bound,
with positive value of $F_\mathrm{ bh}$. For a system with this $RT$ or
higher, then the system must collapse 
gravitationally into a black hole with the
corresponding $\frac{r_{+4}}{R}$.
We note that the picture in 
the grand canonical ensemble is different, as 
the equation $W_\mathrm{ bh}=0$, with $W_\mathrm{
bh}$ denoting the grand potential, results in a $\frac{r_+}{R}$ value 
for the single stable black hole, that is lower than the generalized
Buchdahl bound. One has thermodynamically that when 
the system has $W_\mathrm{ bh}=0$
the black hole phase and hot flat space phase coexist, for $W_\mathrm{
bh}<0$ the black hole phase dominates, and for a certain definite
negative value of $W_\mathrm{ bh}$ the value of $\frac{r_+}{R}$ of the
system is the same as the value of the Buchdahl bound. For larger 
temperatures, therefore the system must collapse 
gravitationally. The only phase of the system is the black hole phase and 
so there is no more coexistence, see Fig.~\ref{figch5:Buchdahl} and
Sec.~\ref{sech5:BHzerofreeenergy}.

\section{The canonical ensemble in the 
limit of infinite cavity radius: 
The Davies limit and the Rindler limit\label{sech5:infinitelimit}}
\sectionmark{The canonical ensemble in the limit 
of infinite cavity radius}\thispagestyle{userightbotmark}

\subsection{Ensemble solutions in the $R\to +\infty$ limit:
the Davies solutions and the Rindler solution}

\begin{figure}[h]
    \centering
    \includegraphics[scale=0.40]{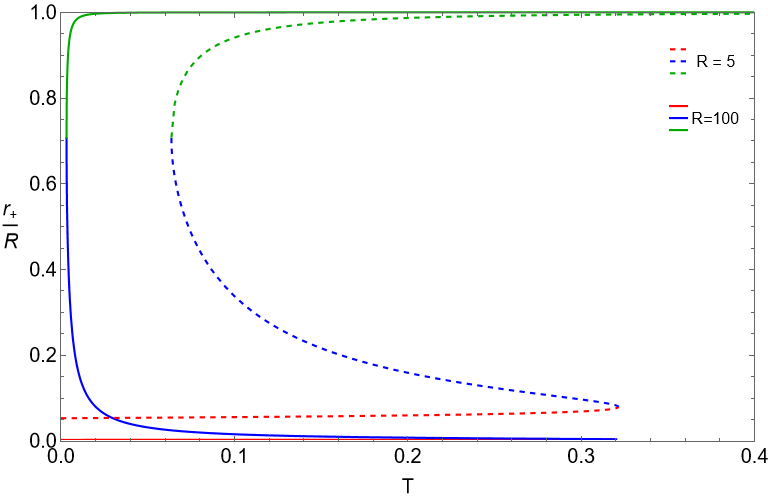}
    \caption{Plot of the solutions $r_{+1}$ in 
    red, $r_{+2}$ in blue and $r_{+3}$ in green 
    of the canonical ensemble in $d=5$ 
    as $\frac{r_+}{R}$ as a function of $T$ in Planck units, 
    for $\mu Q^2 = 0.005$, $\mu=\frac{4l_p^3}{3\pi}$,
    and for two values of $R$, $R=5$ in dashed lines, and $R=100$ 
    in filled lines. One can see the emergence of the $r_{+1}$ and 
    $r_{+2}$ solution
    limits corresponding to the Davies limit
    as they get closer to the $\frac{r_+}{R} = 0$ axis, 
    and the $r_{+3}$ solution limit corresponding to the Rindler limit 
    as it gets closer to the $\frac{r_+}{R}=1$ 
    axis.
    \label{figch5:rindlerdaviesind5}}
\end{figure}

Here, we analyze the infinite cavity radius limit, and we discuss each 
solution that arises from this limit. The importance 
of this limit is that it allows us to connect the York 
formalism with ensembles treated in \cite{Gibbons:1977} and 
to the Davies thermodynamic theory~\cite{Davies:1977bgr}.
By
performing $R\to +\infty$ limit while keeping $T$ fixed and $Q$ fixed,
three different solutions are found. One observes from
Sec.~\ref{sech5:Canonical1}, that there are three solutions for
$r_+(R,T,Q)$ if $\frac{\mu Q^2}{R^{2d-6}} < \frac{\mu Q_s^2}{R^{2d-6}}$.
By performing the
$R\to +\infty$ limit, the term $\frac{\mu Q^2}{R^{2d-6}}$ approaches
zero, and so the
solutions of the ensemble
in this limit
should correspond
to these three solutions under the
$R\to +\infty$ limit.
 For the  smallest and intermediate solutions, the
limit $R\to+\infty$ must be performed by fixing $T$ and $Q$, while
doing $\frac{r_+}{R}\to 0$.
For the largest solution, the limit $R\to
+\infty$ must be performed by fixing $T$ and $Q$, while doing
$\frac{r_+}{R}\to 1$.
The  smallest and intermediate solutions
correspond to Davies thermodynamic
solutions, while
the largest solution limit corresponds to
the Rindler solution.
These solution limits
occur for any $d$. 
In Fig.~\ref{figch5:rindlerdaviesind5},
the behavior of the three solutions in $d=5$
can be seen
for a charge $\mu Q^2 = 0.005$, $\mu=\frac{4l_p^3}{3\pi}$,
for two different $R$, $R=5$
and $R=100$, where the latter $R$
gives an idea of the $R\to\infty$ limit.
In this limit the scale $R$ is lost,
the scales set by the electric charge $Q$
and temperature $T$ at infinity are 
now the only two scales of the canonical ensemble.
We now briefly describe each solution.

The Davies solution corresponds to the smallest and intermediate
solution limits of the canonical ensemble when taking $R\to + \infty$,
with fixed $T$ and $Q$.  Thus, these are the solutions of the
electrically charged black hole in the canonical ensemble with
reservoir at infinity. This can be seen directly from the expression
of the temperature in Eq.~\eqref{eqch5:beta1}.  Since for these solutions
the behavior is $\frac{r_+}{R}\to 0$, one can maintain $r_+$ finite
during the limit $R\to +\infty$, thus obtaining the temperature
formula $T= \frac{d-3}{4\pi} \frac{r_+^{d-3} -\frac{\mu
Q^2}{r_+^{d-3}}}{r_+^{d-2}}$, which is obeyed by
the smallest and intermediate solutions.
This is precisely the Hawking temperature for
the electrically charged black hole.
From Fig.~\ref{figch5:rindlerdaviesind5}
one sees that the two
solutions tend to the axis $\frac{r_+}{R} = 0$ and  seem to get 
overlapped, which is due to the vertical axis being $\frac{r_+}{R}$. If one
regularizes the solutions through multiplying by $R$, one obtains 
the two solutions in $d$ dimensions,
which for $d=4$
are the Davies thermodynamic solutions.
Moreover, one can see that 
the solutions do not exist for all temperatures. This is because the 
two solutions only exist up to a critical temperature, the 
generalized Davies temperature,
after which there are no solutions. In the case represented  in 
Fig.~\ref{figch5:rindlerdaviesind5}
which is $d=5$, the generalized Davies temperature, i.e.,
the temperature when
$R\to\infty$,
has the expression
$T_s = \frac{4}{10\pi \left(\sqrt{5 \mu Q^2} \right)^{\frac{1}{2}}}$,
and so
for $\mu Q^2 = 0.005$ as in the
figure it yields $T_s = 0.320$, with the last equality 
being approximate.

The Rindler solution is
the largest solution limit
that can be obtained from the ensemble
by keeping $T$ and $Q$ fixed, 
while doing $R\to +\infty$ and $r_+ \to R$ in Eq.~\eqref{eqch5:beta1}.
 In 
Fig.~\ref{figch5:rindlerdaviesind5}, this solution
is the one that tends to
$\frac{r_+}{R}=1$.
The temperature dependence on the charge goes with 
$\frac{\mu Q^2}{R^{d-3}r_+^{d-3}}$, 
therefore such dependence in the limit $r_+ \to R$ and $R\to +\infty$
disappears. This happens
because the horizon radius of the black
hole tends to infinity and 
any contributions given by the charge become negligible. The 
expression for the temperature is now
the temperature of an
electrically uncharged black hole 
$T = \frac{(d-3)}{4\pi r_+ \sqrt{1 - \frac{r_+^{d-3}}{R^{d-3}}}}$.
Imposing that 
$T$ is fixed and finite leads to the condition that 
$r_+ \sqrt{1 - \frac{r_+^{d-3}}{R^{d-3}}}$ must 
tend to some constant when $R\to+\infty$ and 
$r_+ \to R$. One can show that in this limit 
the event horizon of
the black hole reduces to the Rindler horizon
and the cavity boundary is
accelerated  to yield the Unruh temperature $T$
set by the reservoir.

We now perform the analysis in full detail for 
the smallest and intermediate
solution limits 
arising from $R\to +\infty$, i.e., the Davies
solution. These are relevant since the formalism in 
this limit yields the Davies' thermodynamic theory of black holes for $d=4$.
We also analyze the largest solution limit 
arising from $R\to +\infty$, i.e., the Rindler solution.

\subsection{Infinite cavity radius and Davies' thermodynamic theory
of black holes: Canonical ensemble,  thermodynamics, and stability of
electrically charged black hole solutions in the $R\to +\infty$ limit}

\subsubsection{Action, solutions and stability for the infinite radius limit}

With the limit of infinite cavity radius for the small and intermediate 
solutions, the canonical ensemble becomes
essentially defined by the temperature $T$ and the electric
charge $Q$ at infinity. It is this $R\to +\infty$ limit
that in four dimensions
gives Davies results \cite{Davies:1977bgr}. This means that
Davies' thermodynamic theory
of black holes,
in this case of electrically charged black holes, can be
seen within the canonical ensemble formalism. Here, we present 
the results for $d$ dimensions
in the $R\to +\infty$ limit, $d=4$ being a particular
case.

In the limit of infinite radius, we must do the analysis above with care, 
since the quantities above
depend on the scale
given by the cavity radius $R$. To proceed with this limit, we must 
start from the reduced action in Eq.~\eqref{eqch5:actioninrp} and perform
the $R\to +\infty$
limit to obtain
\begin{align}
I_* = \frac{\beta}{\mu}\left(\frac{r_+^{d-3}}{2} 
+ \frac{\mu Q^2}{2r_+^{d-3}}\right)
-\frac{\Omega_{d-2} r_+^{d-2}}{4l_p^{d-2}}\,.
\label{eqch5:actioninfinity}
\end{align}
The extrema of the action occurs when
\begin{align}
\beta=\iota(r_+)\,,\quad\quad
\iota(r_+)\equiv \frac{4\pi}{(d-3)}\frac{r_+^{d-2}}{r_+^{d-3}
-\frac{\mu Q^2}{r_+^{d-3}}}\,.
\label{eqch5:Hawkbeta}
\end{align}
This is the inverse Hawking temperature of the Reissner-Nordstr\"om 
black hole measured at infinity, i.e., performing the limit of 
infinite radius into Eq.~\eqref{eqch5:beta1}.

To find the solutions of this canonical ensemble, one must invert
Eq.~\eqref{eqch5:Hawkbeta} to get $r_+(\beta,Q)$, i.e., $r_+(T,Q)$.  This
can be done by solving the following equation
\begin{align}
\left( \frac{(d-3)}{4\pi T}\right)(r_+^{2d-6}-\mu Q^2) 
- r_+^{2d-5}= 0\,,
\end{align} 
which generally is not solvable analytically
for generic $d$. However,
we can perform some qualitative analysis of the solutions.
The function $\iota(r_+)$
in Eq.~\eqref{eqch5:Hawkbeta} 
has a minimum at 
\begin{align}
r_{+s1}= \left(\sqrt{(2d-5)\mu}\,Q\right)^\frac{1}{d-3}
\label{eqch5:r+davies}\,,
\end{align}
which is a saddle point of the 
action for the black hole.
This saddle point of the 
action of the black hole has the temperature 
\begin{align}
T_{s1}=
\frac{(d-3)^2}
{2\pi (2d-5)
(\sqrt{(2d-5)\mu}Q)^{\frac{1}{d-3}}}
\label{eqch5:Tdavies}\,.
\end{align} 
In $d=4$, this $T_{s1}$ is the Davies temperature, 
and so Eq.~\eqref{eqch5:Tdavies} is the generalization of 
Davies temperature for higher dimensions.

By inspection,
one finds that for temperatures $T \leq T_{s1}$
there are two black holes, and for
$T > T_{s1}$
there are no black hole  solutions.
Indeed, for  temperatures in the
interval $0 <T \leq T_{s1}$,
there are two
solutions, the solution
$r_{+1}(T,Q)$ and the solution $r_{+2}(T,Q)$.
The solution
$r_{+1}(T,Q)$ is bounded in the interval 
$(\mu Q^2)^{\frac{1}{2d-6}}<r_{+1}(T,Q)\leq r_{+s1}$, where 
$r_{+1}(T\rightarrow 0,Q) = (\mu Q^2)^{\frac{1}{2d-6}}={r_{+}}_e$,
${r_{+}}_e$ being the radius of the extremal black hole, 
and 
$r_{+1}(T_{s1},Q) = r_{+s1}$. Moreover,
$r_{+1}(T,Q)$ is an increasing monotonic
function in $T$.
The solution $r_{+2}(T,Q)$ is bounded 
from below, i.e., $r_{+2}(T,Q) > r_{+s1}$, where 
$r_{+2}(T_{s1},Q) = r_{+s1}$, and
is unbounded from above, since 
at $T \rightarrow 0$, the 
solution $r_{+2}$ tends to infinity. Moreover,
$r_{+2}(T,Q)$ is a decreasing monotonic
function in $T$.
The action given in Eq.~\eqref{eqch5:actioninfinity}
with $r_+$ holding for $r_{+1}(T,Q)$ or 
$r_{+2}(T,Q)$ is the action in zero loop approximation
that has been found in
\cite{Tiago2024a} directly from the
Gibbons-Hawking approach, rather than from York's approach
for a given $R$ with subsequently taking the $R\to\infty$ limit,
as it is being done here.

Regarding stability, a solution is 
stable if $\frac{\partial \iota(r_+)}{\partial r_+}<0$,
as it was seen in the case of finite cavity.
This gives
\begin{align}
r_+\leq r_{+s1}\,,
\label{eqch5:stabilityRinfinity}
\end{align}
with $ r_{+s1}$ given in Eq.~\eqref{eqch5:r+davies}.
This means
that the solution is stable if the radius $r_+$ increases as the 
temperature increases. Therefore, the solution $r_{+1}$ is stable 
since it has this monotonic behavior, while the solution $r_{+2}$ 
is unstable since it has an opposite monotonic behavior.

\subsubsection{Thermodynamics in the $R\to +\infty$ limit}

With the solutions of the canonical ensemble found in the limit of 
infinite radius of the cavity,  $R\to +\infty$,
we can obtain the thermodynamics from $I_0$, i.e., 
the action in the zero loop approximation 
given in Eq.~\eqref{eqch5:actioninfinity}
evaluated at the extrema of Eq.~\eqref{eqch5:Hawkbeta}.
The thermodynamics for the system follows 
through the correspondence $F = T I_0$,  where $F$ again is the 
Helmholtz free energy of the system and thus it can be written for this 
case as
\begin{align}
F = \frac{1}{\mu}\left(\frac{r_+^{d-3}}{2} 
+ \frac{\mu Q^2}{2r_+^{d-3}}\right)
-\frac{T \Omega_{d-2} r_+^{d-2}}{4l_p^{d-2}}\,,
\label{eqch5:freeenergyinfinity}
\end{align}
where $r_+$ can be $r_{+1}(T,Q)$ or
$r_{+2}(T,Q)$.
Using the same calculation method from 
Sec.~\ref{sech5:thermoquantities}, we have that 
the entropy is 
\begin{align}
S = \frac{A_+}{4l_p^{d-2}}\,.
\end{align}
Additionally, the thermodynamic pressure $p$ is zero,
\begin{align}
p=0\,.
\end{align}
And also, the 
thermodynamic
electric potential is
\begin{align}
\phi = \frac{Q}{r_+^{d-3}}\,,
\end{align}
which is equal to the pure electric potential.
The energy, given by $E = F + TS$, can be written 
as 
$E = \frac{r_+^{d-3}}{2\mu} + \frac{Q^2}{2r_+^{d-3}}$.
But
the spacetime mass $m$ is given by
$m = \frac{r_+^{d-3}}{2\mu} + \frac{Q^2}{2r_+^{d-3}}$, see
also Sec.~\ref{sech5:BHzerofreeenergy},  so
that the thermodynamic energy and the spacetime mass
are the same in the  $R\to +\infty$ limit, i.e., 
\begin{align}
E = m\,.
\label{eqch5:energyinfinity}
\end{align}
Thus, one can write the free energy
given in Eq.~\eqref{eqch5:freeenergyinfinity} as
\begin{align}
F=m-TS\,.
\label{eqch5:freeenergyinfinityagain}
\end{align}

We must note that the expressions for the entropy,
the pressure, the thermodynamic electric potential, 
and the energy are consistent with the limit of infinite 
radius to the respective expressions in 
Sec.~\ref{sech5:thermoquantities}. Moreover, in this limit, 
the pressure $p$ vanishes, which is consistent with the 
absence of the variable $R$ in the action.

The energy in Eq.~\eqref{eqch5:energyinfinity} can be rewritten
in terms 
of the entropy and the charge as
\begin{align}
    E = \frac{1}{2\mu}\left(\frac{4S l_p^{d-2}}{\Omega_{d-2}}
\right)^{\frac{d-3}{d-2}} 
+ \frac{\mu Q^2}{2}\left(\frac{4  S l_p^{d-2}}{\Omega_{d-2}}
\right)^{\frac{3-d}{d-2}}\,\,.
\end{align}
The energy function possesses the scaling property
$\nu^{\frac{d-3}{d-2}} E = E(\nu S, \nu^{\frac{d-3}{d-2}} Q)$, which allows
the use of the Euler relation theorem to have
$E= \frac{d-3}{d-2} TS + \phi Q$,
which is the formula obtained in Sec.~\ref{sech5:thermoquantities} without 
the term $pA$. Indeed, the term $pA$ in the limit of infinite
reservoir radius 
has leading order $R^{-(d-3)}$, and so it vanishes. Since
from Eq.~\eqref{eqch5:energyinfinity}
$E=m$, one obtains 
\begin{align}
m = \frac{d-3}{d-2} TS + \phi Q\,,
\end{align}
which is the Smarr formula.

In this case the law
\begin{align}
dm =  TdS + \phi dQ\,,
\label{eqch5:fiirslawofbhm}
\end{align}
holds. This is exactly the first law of black hole mechanics.
This can be obtained from Eq.~\eqref{eqch5:firstlawoft}
in the $R\to\infty$ limit.
For $R$ finite, there is a first law of thermodynamics of the cavity
and does not correspond to the law of black hole mechanics.
For $R\to\infty$, the first law of black hole
thermodynamics and the first law of black hole mechanics
coincide into one same law, which is quite remarkable.
Moreover, in the electrically charged case, as opposed
to the Schwarzschild case, the thermodynamics of the canonical 
ensemble is valid,
since there is a region of the electric charge where the
system is thermodynamically stable.
It is from Eq.~\eqref{eqch5:fiirslawofbhm}
that Davies has started his  thermodynamic theory
of black holes for $d=4$. Here, we deduced Eq.~\eqref{eqch5:fiirslawofbhm} 
from the action in Eq.~\eqref{eqch5:actioninfinity}.

The thermodynamic
stability can be seen directly from applying the limit of 
infinite radius of the cavity in Eq.~\eqref{eqch5:Caq}, which is the 
condition for the positivity of the heat capacity. 
This condition ensures that a solution in the limit of infinite cavity 
is stable.
The heat capacity in this limit is
\begin{align}
C_{Q}& = \frac{(d-2)\Omega_{d-2} r_+^{d-2}(r_+^{2d-6}-\mu Q^2)}
{4l_p^{d-2}\left((2d-5)\mu Q^2 - r_{+}^{2d-6} \right)}\nonumber\\
&=
\frac{S^3 E T}
{\frac{(d
\hskip-0.02cm
-
\hskip-0.03cm
3)\Omega_{\hskip-0.03cm d-2}^3}{4^5 \pi^2 l_p^{3d-6}}
\hskip-0.15cm
\left[
\frac{
(\hskip-0.02cm
3d-8
\hskip-0.02cm
)
\mu^2 Q^4}
{\left(\frac{4Sl_p^{d-2}}{\Omega_{d-2}}\right)^{\hskip-0.06cm\frac{d-4}{d-2}}} 
\hskip-0.05cm
+
\hskip-0.05cm
(d
\hskip-0.02cm
-
\hskip-0.02cm
4)
\hskip-0.06cm
\left(\frac{4 S l_p^{d-2}}{\Omega_{d-2}}\right)^{\frac{3d-8}{d-2}}
\right]
\hskip-0.16cm
-
\hskip-0.11cm
T^2
\hskip-0.06cm
S^3}\hskip-0.02cm,
\label{eqch5:heatcapacityraw}
\end{align}
where the subscript $A$ in $C_{A,Q}$ has been dropped
since the evaluation is at infinity, and in the
second equality
the heat capacity was written in terms of the thermodynamic 
variables $S$, $E$, and $T$.
So there is stability if $C_Q\geq0$, i.e.,
$r_{+} \leq \left[(2d-5)\mu Q^2\right]^{\frac{1}{2d-6}}$,
which is Eq.~\eqref{eqch5:stabilityRinfinity}
together with 
Eq.~\eqref{eqch5:r+davies}.
This means that the solution $r_{+1}$ is thermodynamically
stable whereas the solution 
$r_{+2}$ is unstable. It must be noted also that $r_{+1}$ is an 
increasing monotonic function in $T$, which means the energy of the 
black hole increases of the temperature increases, as it is expected 
from a stable system. The opposite happens to the solution $r_{+2}$,
since it is a decreasing monotonic function in $T$ and so the energy
of the black hole decreases as temperature increases.

\subsubsection{Favorable phases}

There are two stable phases.
The small black hole $r_{+1}$ and hot flat space
with electric charge at infinity.
Since the  black hole $r_{+1}$ has positive
free energy and 
 hot flat space
with electric charge at infinity has
zero free energy, and systems
with lower free energy are preferred,
whenever the system finds itself
in the black hole $r_{+1}$ solution
it tends to transition to the
hot flat space 
with electric charge at infinity phase.

\subsubsection{$d=4$: Analysis leading to 
Davies' thermodynamic theory of black holes and
Davies point}

The dimension $d=4$ is specially interesting since in
the $R\to\infty$ gives the
results of Davies' thermodynamic theory of black holes
\cite{Davies:1977bgr}. In this
setting, the reservoir of temperature
$T$ and electric charge $Q$
is at infinity.

The reduced action in Eq.~\eqref{eqch5:actioninfinity} 
in $d=4$ gives
\begin{align}
I_* = \frac{\beta}{2l_p^2}
\left(r_+
+ \frac{Q^2}{r_+}\right)
-\pi \frac{r_+^2}{l_p^2}
\,,
\label{eqch5:actioninfinityd=4}
\end{align}
where $\mu=l_p^2$
and $\Omega_2=4\pi$.
The stationary points in $d=4$ occur when
\begin{align}
\beta=\iota(r_+)\,,\quad\quad \iota(r_+)\equiv
\frac{4\pi r_+^2}{r_+
-\frac{Q^2}{ r_+}}\,,
\label{eqch5:Hawkbeta0d=4}
\end{align}
corresponding to the inverse Hawking temperature 
of a charged black hole in $d=4$.

Equation~\eqref{eqch5:Hawkbeta0d=4} must be 
inverted to get the solutions 
$r_+(T,Q)$. The solutions satisfy
\begin{align}
\left( \frac{1}{4\pi T}\right)(r_+^2- Q^2) 
- r_+^3= 0\,.
\label{eqch5:r+d=infinted=4}
\end{align} 
It is possible to write analytically 
the solutions as the roots of a cubic polynomial, 
however we do not present them analytically here. 
The minimum of function $\iota(r_+)$
in Eq.~\eqref{eqch5:Hawkbeta0d=4}
occurs at 
$r_{+s1}=\sqrt3\,Q$,
being a saddle point of the 
action of the black hole.
The horizon radius of the 
saddle point is written here as 
\begin{align}
{r_+}_\mathrm{ D}=\sqrt3\,Q\,,
\label{eqch5:r+daviesd=4}
\end{align} 
as in $d=4$ it gives the Davies horizon radius.
Since $r_+=l_p^2 m+ \sqrt{l_p^4 m^2-l_p^2 Q^2}$, 
this means $l_p m=\frac{2}{\sqrt3} Q$
at the saddle point,
a result that can be found in \cite{Davies:1977bgr}. 
The temperature corresponding to the saddle point
is Eq.~\eqref{eqch5:Tdavies} in $d=4$, or explicitly
\begin{align}
T_\mathrm{ D}=
\frac{1}
{6\sqrt{3}\pi l_p Q}\,,
\label{eqch5:Tdavies4d}
\end{align} 
which is the Davies temperature, and it
is a result that can be extracted from \cite{Davies:1977bgr}.

We present a summary of the behavior of the solutions for 
$d=4$.
For $0<T \leq T_\mathrm{ D}$,
there are two
solutions, the solution
$r_{+1}(T,Q)$ and the solution $r_{+2}(T,Q)$.
The solution
$r_{+1}(T,Q)$ increases monotonically with $T$ and 
lies in the interval 
$ {r_{+}}_e
<r_{+1}(T,Q)\leq {r_+}_\mathrm{ D}$,
where 
$r_{+1}(T\rightarrow 0,Q) = {r_{+}}_e=l_p Q$
and 
$r_{+1}(T_\mathrm{ D},Q) = {r_+}_\mathrm{ D}= \sqrt3\,l_p Q
$.
The solution $r_{+2}(T,Q)$ decreases monotonically with $T$ and
lies in the interval ${r_+}_\mathrm{ D}<r_{+2}(T,Q) <\infty$, where 
$r_{+2}(T_\mathrm{ D},Q) = {r_+}_\mathrm{ D}= \sqrt3\,l_p Q$.
For $T_\mathrm{ D}<T $, there are no black hole solutions. 
Regarding stability, a solution is
stable if $\frac{\partial \iota(r_+)}{\partial r_+}\leq0$, i.e.
\begin{align}
r_+\leq {r_+}_\mathrm{ D}\,.
\label{eqch5:stabilityRinfinityd=4}
\end{align}
With ${r_+}_\mathrm{ D}$ given in Eq.~\eqref{eqch5:r+daviesd=4},
Eq.~\eqref{eqch5:stabilityRinfinityd=4} can be turned in to the 
region in the electric charge
$\frac1{\sqrt3}r_+\leq l_p Q\leq r_+$,
the latter term being simply the restriction to nonextremal case.
From Eq.~\eqref{eqch5:stabilityRinfinityd=4},
one has that the solution $r_{+1}$ is stable 
while the solution $r_{+2}$ 
is unstable.

We now summarize the results for thermodynamics in $d=4$.
The free energy of the system is $F = T I_0$, coming from the 
zero loop approximation of the path integral.
From Eq.~\eqref{eqch5:actioninfinityd=4}, the free energy is
\begin{align}
F =
 \frac{1}{2l_p^2}
\left(r_+
+ \frac{l_p^2 Q^2}{r_+}\right)
-T\,\pi \frac{r_+^2}{l_p^2}
\,.
\label{eqch5:Fd=4}
\end{align}
From the derivatives of the free energy, 
one obtains the entropy 
$S = \pi \frac{r_+^2}{l_p^2}$, i.e., $S = \frac{1}{4l_p^2} A_+$,
the thermodynamic pressure $p=0$ since there is no area dependence, 
the electric potential $\phi = \frac{Q}{r_+}$,
and the energy
$E = \frac{1}{2l_p^2}
\left(r_+
+ \frac{l_p^2 Q^2}{r_+}\right)$, from $E = F + TS$. 
Considering that this is the expression for
the spacetime mass $m$, one has
$E = m$. The free energy of
Eq.~\eqref{eqch5:Fd=4} is then $F=m-TS$.

The Smarr formula for $d=4$ is 
\begin{align}
m = \frac12\, TS + \phi Q\,.
\end{align}
Indeed, the first law of black hole mechanics 
$dm = TdS + \phi dQ$ coincides with the 
first law of thermodynamics, see above. 
The 
first law of black hole mechanics
is the expression from which Davies 
\cite{Davies:1977bgr} started his analysis.
Here, we started the analysis from
the action Eq.~\eqref{eqch5:actioninfinityd=4}
and we derived the first law from first principles.
Moreover, the system is stable thermodynamically in a 
range of values of the electric charge. 
On the other hand, 
the electrically charged case in the grand canonical
ensemble with the reservoir at infinity
is unstable. Gibbons and Hawking
through the action and the path integral
approach \cite{Gibbons:1977}
noticed this instability problem but
did not venture into the electric canonical ensemble to cure it.

The heat capacity described in Eq.~\eqref{eqch5:heatcapacityraw}, 
with $d=4$,
is given by
\begin{align}
C_Q =
\frac
{2\pi r_+^2 \left(1-\frac{Q^2}{r_+^2}\right)} 
{l_p^2\left(3\frac{Q^2}{r_+^2}-1\right)}=
\frac{S^3 E T}{\frac{\pi Q^2}{4l_p^6} -
T^2
S^3}\,,
\label{eqch5:heatcapacityraw4d}
\end{align}
where in the
second equality
the heat capacity was written in terms of the thermodynamic 
variables $S$, $E$, and $T$.
The system is thermodynamically stable if
$l_p Q\geq\frac1{\sqrt3}r_+$, i.e.,
$\frac1{\sqrt3}r_+\leq l_p Q\leq r_+$,
the latter term being the condition for nonextremal case.
The system is thermodynamically unstable if 
$0\leq l_p Q<\frac1{\sqrt3}r_+$.
This is the same result as given in 
Eq.~\eqref{eqch5:stabilityRinfinityd=4}
together with 
Eq.~\eqref{eqch5:r+daviesd=4}. The heat capacity
$C_Q$ is infinitely positive 
at the point
$l_p Q=\frac1{\sqrt3}r_+$
if one approaches it from 
higher $Q$, 
the heat capacity $C_Q$
is infinitely negative
if one approaches the point
$l_p Q=\frac1{\sqrt3}r_+$ from lower $Q$. 
The heat capacity goes to zero at the extremal case 
$l_p Q=r_+$. 
Precisely at the point $l_p Q=\frac1{\sqrt3}r_+$,
this behavior of the heat capacity was 
found in \cite{Davies:1977bgr}, and it was 
classified as being similar to a second order phase transition.
However, this
point  is a turning point rather than a
second order phase transition. This turning point
indicates the ratio of scales at which one has stability.
Indeed, when analyzing the heat capacity in terms of the 
temperature and electric charge, one has two distinctive 
curves, one for each solution, diverging at this point. 
But the unstable solution cannot be considered as a phase, 
due to its instability. The system will always remain in the 
stable configuration.
Note that the formula for $C_Q$ in 
the second line of Eq.~\eqref{eqch5:heatcapacityraw4d}
is the same formula found in \cite{Davies:1977bgr} by 
performing in Eq.~\eqref{eqch5:heatcapacityraw4d}
the redefinitions 
$S \rightarrow 8\pi S$,
$T \rightarrow \frac{1}{8\pi}T$ and 
$\frac{C_Q}{8\pi} \rightarrow C_Q$, and using Planck units.

\subsubsection{$d=5$: Analysis}

The dimension $d=5$ is a
typical higher dimension that we have been analyzing. 
We continue this trend and we present a summary for this
specific case in the $R\to +\infty$ limit.

The reduced action in Eq.~\eqref{eqch5:actioninfinity} 
in $d=5$ can be written simply as
\begin{align}
I_* = \frac{\beta}{2l_p^3}
\left(\frac{3\pi r_+^2}{4} 
+ \frac{l_p^3 Q^2}{r_+^2}\right)
-\frac{\pi^2 r_+^3}{2l_p^3}\,,
\label{eqch5:actioninfinityd=5}
\end{align}
where it was used $\mu=\frac{4 l_p^3}{3\pi}$
and $\Omega_3=2\pi^2$.
The stationary points are described by
\begin{align}
\beta=\iota(r_+)\,,
\quad\quad
\iota(r_+)\equiv
\frac{2\pi r_+^3}{r_+^2
-\frac{4 l_p^3 Q^2}{3\pi r_+^2}}.
\label{eqch5:Hawkbeta0d=5}
\end{align}
again corresponding to the inverse Hawking 
temperature of a charged black hole in $d=5$.

\begin{figure}[h]
\centering
\includegraphics[width=0.7\linewidth]{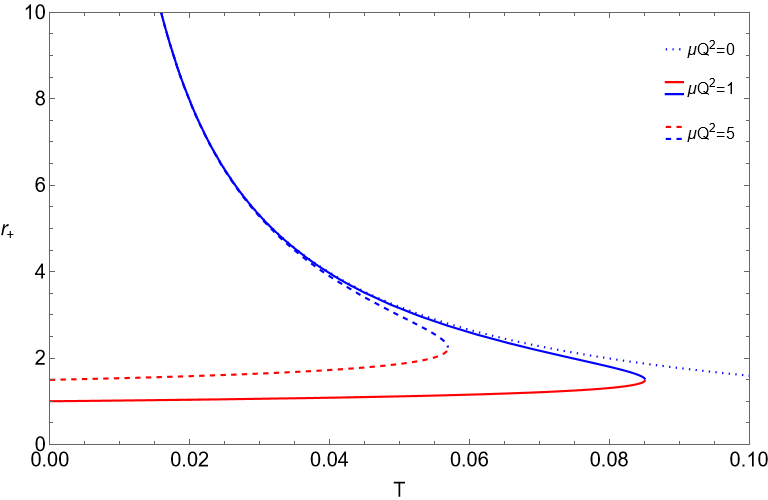}
\caption{Plot of the two solutions $r_{+1}(T,Q)$, in red, 
and $r_{+2}(T,Q)$, in blue, of the charged black hole in the 
canonical ensemble for infinite cavity radius, for two values of the 
charge, $\mu Q^2 = 1$ in filled lines, and $\mu Q^2 = 5$ in dashed 
lines, $\mu=\frac{4l_p^3}{3\pi}$, in $d=5$.
\label{figch5:rtinfd5}}
\end{figure}

The solutions are found by inverting
Eq.~\eqref{eqch5:Hawkbeta0d=5} to get $r_+(\beta,Q)$, i.e.,
$r_+(T,Q)$. 
This is the same as solving 
\begin{align}
\left( \frac{1}{2\pi T}\right)(r_+^4-\frac{4}{3\pi} l_p^3 Q^2) 
- r_+^5= 0\,,
\label{eqch5:r+s=infinted=5}
\end{align} 
which cannot be done analytically. However, it can be 
analyzed qualitatively or solved numerically,
see Fig.~\ref{figch5:rtinfd5} for this case  
of five dimensions. The function $\iota(r_+)$
in Eq.~\eqref{eqch5:Hawkbeta0d=5}
possesses a minimum at 
\begin{align}
    r_{+s1}= \left(\sqrt{\frac{20}{3\pi}}\,\,l_p^{\frac{3}{2}} 
    Q\right)^\frac{1}{2}\,,
    \label{eqch5:r+sdaviesd=5}
    \end{align} 
which corresponds to a saddle point of the 
action of the black hole.
This generalizes the Davies radius to $d=5$.
The temperature at this saddle point is
\begin{align}
T_{s1}=
\frac{4}
{
10\pi
\left(\sqrt{\frac{20}{3\pi}}\,l_p^{\frac{3}{2}}Q\right)^{\frac12}
}
\label{eqch5:Tdaviesd=5}\,.
\end{align} 
This generalizes the Davies temperature for $d=5$.

We now summarize the behavior of the solutions in $d=5$.
For temperatures $0<T \leq T_s$
there are two
solutions, the solution
$r_{+1}(T,Q)$ and the solution $r_{+2}(T,Q)$.
The solution
$r_{+1}(T,Q)$ increases monotonically with the 
temperature and
is bounded by 
${r_{+}}_e
<r_{+1}(T,Q)\leq
r_{+s}$,
where 
$r_{+1}(T\rightarrow 0,Q) 
= {r_{+}}_e=\left(\sqrt{\frac{4}{3\pi}}\,\,
l_p^{\frac{3}{2}}Q\right)^\frac{1}{2}$
is the extremal black hole, and 
$r_{+1}(T_{s1},Q) = r_{+s}=\left(\sqrt{\frac{20}{3\pi}}\,\,
l_p^{\frac{3}{2}}Q\right)^\frac{1}{2}$.
The solution $r_{+2}(T,Q)$ decreases monotonically 
with the temperature and assumes values in the interval
$r_{+s1}<r_{+2}(T,Q)<\infty$, where 
$r_{+2}(T_s,Q) = r_{+s}$.
See Fig.~\ref{figch5:rtinfd5}
for the plots of $r_{+1}$ and $r_{+2}$.
Regarding stability, a stable solution 
obeys $\frac{\partial \iota(r_+)}{\partial r_+}\leq0$.
This condition becomes
\begin{align}
r_+\leq r_{+s1}\,.
\label{eqch5:stabilityRinfinityd=5}
\end{align}
With $r_{+s1}$ given in Eq.~\eqref{eqch5:r+sdaviesd=5},
Eq.~\eqref{eqch5:stabilityRinfinityd=5} can be transformed to
$\left(\frac{3\pi}{20}\right)^{\frac12}
r_{+}^2 \leq l_p^\frac{3}{2}Q\leq\left(\frac{3\pi}{4}\right)^{\frac12}
r_{+}^2$, the latter term being 
the restriction to the nonextremal case.
From Eq.~\eqref{eqch5:stabilityRinfinityd=5},
one obtains that $r_{+1}$ is stable and that 
$r_{+2}$ is unstable.
The plots
in Fig.~\ref{figch5:rtinfd5} show the discussion above, 
namely the stable branch $r_{+1}$ and the unstable 
branch $r_{+2}$.
It is also seen clearly
that the plot of Fig.~\ref{figch5:rtinfd5}
is the limiting $R\to\infty$ case
of Fig.~\ref{figch5:rt5}. From Fig.~\ref{figch5:rt5}
one finds that when $R\to\infty$,
the solution $r_{+3}$ disappears, leaving
$r_{+1}$ and $r_{+2}$, with
$r_{+1}$ and $r_{+2}$ meeting at a maximum temperature.
Also, from Fig.~\ref{figch5:rt5}
one sees that the $r_{+2}$ and $r_{+3}$ branches
meet at a minimum temperature, and these branches are
the ones that appears in the zero charge case of York,
here slightly modified due to the existence of
an electric charge $Q$.
More specifically,  comparing Fig.~\ref{figch5:rtinfd5}
with  Fig.~\ref{figch5:rt5}, one notes
that
the red and blue lines of Fig.~\ref{figch5:rtinfd5}
are the stable and
unstable black holes of Davies, here in $d=5$,
and the red and blue lines 
of Fig.~\ref{figch5:rt5} are precisely these branches
of black holes for finite reservoir radius $R$.
The blue and green branches in 
 Fig.~\ref{figch5:rt5} correspond to
 York black holes. Thus, Fig.~\ref{figch5:rt5}
 is a unified plot of
 York and Davies black holes.
Note further 
from Fig.~\ref{figch5:rtinfd5}, that for the
electric charge going to zero,
the branch that survives in Fig.~\ref{figch5:rtinfd5}
is
the blue branch, which corresponds to the unstable black 
hole $r_{+2}$,
and the solution goes up to the point characterized
by $T=\infty$ and $r_+=0$. This branch corresponds to
the original unstable Hawking black hole, the black hole
also found in the Gibbons-Hawking 
path integral approach.

We present the summary of 
the results for the thermodynamics in $d=5$.
The free energy can be obtained 
from the zero loop approximation of the 
path integral as $F = T I_0$.
From Eq.~\eqref{eqch5:actioninfinityd=5}, 
the free energy takes the form
\begin{align}
F = \frac{1}{2l_p^3}
\left(\frac{3\pi r_+^2}{4} 
+ \frac{l_p^3Q^2}{r_+^2}\right)
-T\frac{\pi^2 r_+^3}{2l_p^3}\,.
\label{eqch5:Fd=5}
\end{align}
From its derivatives,
one obtains the entropy
as $S = \frac{A_+}{4l_p^3}$,
$A_+ = 2\pi^2 r_+^3$, the thermodynamic
pressure as $p=0$,
the 
thermodynamic
electric potential as $\phi = \frac{Q}{r_+^2}$,
and
the energy, given by $E = F - TS$, 
as 
$E = \frac{3\pi r_+^2}{8l_p^3} + \frac{Q^2}{2r_+^2}$.
Note that this is exactly the expression for
the spacetime mass $m$, so the mean energy is
$E=m$.
The free energy of
Eq.~\eqref{eqch5:Fd=5} becomes $F=m-TS$.

The Smarr formula in $d=5$ takes the form
\begin{align}
m = \frac23\, TS + \phi Q\,.
\end{align}
Also, one has that the law
$dm =  TdS + \phi dQ$
holds. And so the first law of black hole mechanics 
coincides with the first law of thermodynamics.
Also, the system is stable thermodynamically in 
a small region of the charge, so this correspondence 
is valid.

The heat capacity of Eq.~\eqref{eqch5:heatcapacityraw}
is now in $d=5$ given by
\begin{align}
C_Q =&
\frac
{3\pi^2  r_+^3 
\left(1-\frac{4}{3\pi}\frac{Q^2}{r_+^4}\right)} 
{2l_p^3\left(\frac{20}{3\pi}\frac{Q^2}{r_+^4}-1\right)}\nonumber\\
 =&
\frac{S^3 E T}{\frac{7\pi^2}{36l_p^3}Q^4
\left(\frac{2S l_p^3}{\pi^2}\right)^{-\frac{1}{3}} 
+ \frac{\pi^4}{4^3}\left(\frac{2S l_p^3}{\pi^2}\right)^{\frac{7}{3}} -
T^2
S^3}\,,
\label{eqch5:heatcapacityraw5d}
\end{align}
where in the
second equality is in terms of the thermodynamic 
variables $S$, $E$, and $T$.
One has instability if 
$0\leq l_p^\frac{3}{2}Q<\left(\frac{3\pi}{20}\right)^{\frac12}
r_{+}^2$, with $Q$ meaning its absolute modulus.
One has  thermodynamic stability if
$\left(\frac{3\pi}{20}\right)^{\frac12}
r_{+}^2\leq l_p^\frac{3}{2}Q\leq\left(\frac{3\pi}{4}\right)^{\frac12}
r_{+}^2$, the latter term being the condition for 
the nonextremal case, and
this can also be derived from 
Eq.~\eqref{eqch5:stabilityRinfinityd=5}
together with 
Eq.~\eqref{eqch5:r+sdaviesd=5}. The heat capacity
$C_Q$ is infinitely positive 
at the point
$l_p^\frac{3}{2}Q=\left(\frac{3\pi}{20}\right)^{\frac12}
r_{+}^2$, if this point is approached from 
higher $Q$, 
the heat capacity $C_Q$
is infinitely negative, if the point is approached
from lower $Q$. 
\begin{figure}[h]
\centering
\includegraphics[width=0.7\linewidth]{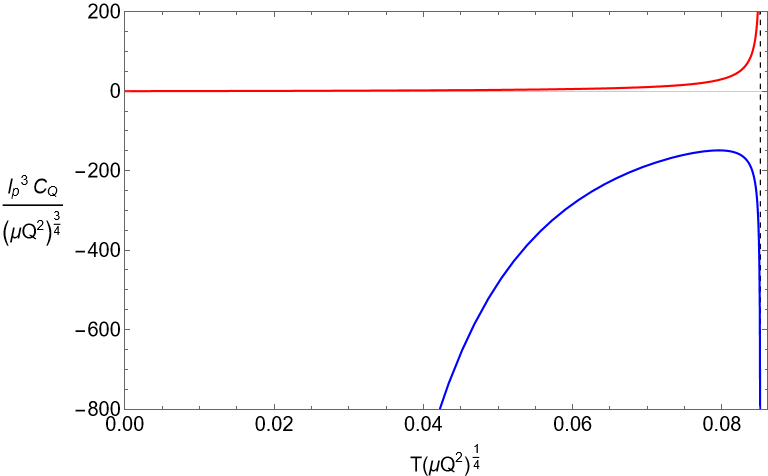}
\caption{The heat capacity $l_p^3 C_Q$ in $(\mu Q^2)^{\frac{3}{4}}$ units,
$\frac{l_p^3 C_Q}{(\mu Q^2)^{\frac{3}{4}}}$, 
is given as a function of the temperature $T (\mu Q^2)^{\frac{1}{4}}$
in $d=5$. In red, the heat capacity of $r_{+1}$ is represented, while 
in blue, the heat capacity of $r_{+2}$ is shown. 
There is a turning point at $T (\mu Q^2)^{\frac{1}{4}} =
\frac{4}{10\pi 5^{\frac{1}{4}}}$.}
\label{figch5:CAq5dRinfty}
\end{figure}
This is a turning point of the solutions, indicating 
the condition for stability. This is properly 
seen when analyzing the heat capacity with fixed 
temperature and electric charge, see Fig.~\ref{figch5:CAq5dRinfty}.
Indeed, the heat capacity is described by two curves, one for 
each solution $r_{+1}$ and $r_{+2}$, being positive for 
$r_{+1}$ and negative for $r_{+2}$. The system cannot be 
sustained in the solution $r_{+2}$ since it is unstable and 
so it can only be at the stable solution $r_{+1}$.

\subsection{Infinite cavity radius and the Rindler limit: Cavity
boundary at the Unruh temperature}

The case of the smallest and intermediate solutions was discussed above. 
We now turn to the largest solution in the limit of infinite cavity.
The largest solution in this limit can be obtained by keeping $T$ and $Q$ fixed, 
while doing $R\to +\infty$ and $r_+ \to R$ in Eq.~\eqref{eqch5:beta1}. 
The temperature dependence on the charge goes with 
$\frac{\mu Q^2}{R^{2d-6}}$, 
therefore such dependence in the limit $r_+ \to R$ and $R\to +\infty$
disappears. Intuitively, the black hole becomes very large such that 
any contributions of the charge become negligible. Then, the 
expression for the temperature reduces to the non-charged case, 
$T = (d-3)
(4\pi r_+)^{-1} (1 - \frac{r_+^{d-3}}{R^{d-3}} )^{-\frac12}$, 
but the limit still needs to be applied. The requirement that 
$T$ is fixed and so finite leads to the condition that 
$r_+ \sqrt{1 - \frac{r_+^{d-3}}{R^{d-3}}}$ must 
tend to some constant under the limit of $R\to+\infty$ and 
$r_+ \to R$. Still, it seems unclear a priori
what the system in this limit describes.

In order to understand the limit, we can first consider the 
Euclidean Schwarzschild metric
$ds^2 = R^2\frac{4 r_+^2}{R^2(d-3)^2}
(1 - \frac{r_+^{d-3}}{r^{d-3}})\,d\tau^2 + 
(1 - \frac{r_+^{d-3}}{r^{d-3}})^{-1}R^2 \,d(\frac{r}{R})^2 
+ R^2(\frac{r^2}{R^2}) d\Omega^2_{d-2}$,
where the normalization by $R$ in the line element was introduced, 
with $0\leq\tau<2\pi$ and $r_+<r \leq R$. First, we need to 
consider $r_+\to R$ in the limit of infinite cavity and only then perform
$R\to \infty$. Therefore, we must consider the near horizon 
expansion of the metric. The normalized proper radial length is given 
by
$\epsilon(r) = \frac{1}{R}
\int_{r_+}^r (1 - \frac{r_+^{d-3}}{\rho^{d-3}})^{-\frac12}
d\rho= \frac{2}{d-3} 
(\frac{R}{r_+})^{\frac{d-5}{2}}\sqrt{(\frac{r}{R})^{d-3}
- (\frac{r_+}{R})^{d-3}}$,
valid 
at the near horizon, spanning 
the interval
$0<\epsilon< \epsilon(R)$. We can 
therefore rewrite the Schwarzschild metric in this limit
as
$ds^2 = (R^2 \epsilon^2 + \mathcal{O}(\epsilon^4))d\tau^2 + R^2 d\epsilon^2 
+ (R^2 + \mathcal{O}(\epsilon^2))d\Omega^2$.
Notice however that as $r_+\to R$, the total normalized radial proper 
length $\epsilon(R)$ tends to zero. It is now that the limit 
$R\to +\infty$ is performed but such that $R\epsilon(R)$ tends to a constant,
which is defined as $\bar R$, ${\bar R}\equiv R\epsilon(R)$.
Thus, one has a new proper length
$\bar r$, defined as
\begin{align}
{\bar r}\equiv R\epsilon(r)\,,\quad\quad 0<{\bar r}<{\bar R}\,.
\label{eeqch5:barr}
\end{align}
The metric becomes in this limit 
\begin{align}
ds^2 = {\bar r}^2 d\tau^2 + d{\bar r}^2 + R^2 d\Omega^2\,,
\label{eqch5:rindlermetric}
\end{align}
i.e.,
it becomes
the two-dimensional Euclideanized Rindler metric times 
a $(d-2)$-sphere with infinite radius. The metric on the $(d-2)$-sphere 
can be normalized by choosing a specific point on the sphere and performing 
the expansion around such point, obtaining $R^2 d\Omega^2 = \sum_{i=1}^{d-2} 
(dx^i)^2$, where $x^i$ are the new coordinates. The metric then reduces to the 
$d$ dimensional Euclideanized Rindler space. The system can now be interpreted 
as follows. The event horizon of the black hole reduces to the Rindler horizon 
at ${\bar r} =0$, while the cavity boundary is located at ${\bar R}$
and it is being accelerated. The proper acceleration 
of the cavity is precisely $\frac{1}{{\bar R}}$ and the 
temperature measured at the boundary of
the cavity is $T = \frac{1}{2\pi {\bar R}}$.

We must analyze what happens to the thermodynamic quantities in this 
Rindler solution limit. First, the temperature in Eq.~\eqref{eqch5:beta1} 
is finite and equals to $T = \frac{1}{2\pi {\bar R}}$.
Since $T$ is fixed by the ensemble
this gives the solution for the cavity boundary,
namely 
\begin{align}
{\bar R} = \frac{1}{2\pi T}\,.
\label{eqch5:RrindlerfromT}
\end{align}
To be in equilibrium with the temperature $T$ of the reservoir,
the boundary itself ${\bar R}$ has to have a
Rindler acceleration that matches its Unruh temperature.
The free energy in Eq.~\eqref{eqch5:freeennergy} 
diverges negatively, $F\to-\infty$.
It diverges as $F= \frac{R^{d-3}}{\mu} 
- \frac{\Omega_{d-2}}{8\pi {\bar R}l_p^{d-2}}R^{d-2}$,
and is negative since the 
power $R^{d-2}$ is always larger than $R^{d-3}$ for 
$R\to + \infty$.
This divergence is due to the fact that the area is
divergent. Thus, it is better to work
with a specific free energy, ${\bar F}$,
a free energy per unit area, defined as
 ${\bar F}\equiv \frac{F}{\Omega_{d-2}R^{d-2}}$.
Then,
\begin{align}
{\bar F}=-\frac{1}{8\pi {\bar R}l_p^{d-2}}\,,
\label{eqch5:Frindler}
\end{align}
so it is negative.
From 
Eq.~\eqref{eqch5:entropy}, the entropy also diverges,
$S\to\infty$, it diverges as 
$S = \frac{\Omega_{d-2}R^{d-2}}{4l_p^{d-2}}$. 
Defining a specific entropy
${\bar S}\equiv \frac{S}{\Omega_{d-2}R^{d-2}}$
\begin{align}
{\bar S}=\frac{1}{4l_p^{d-2}}\,,
\label{eqch5:Srindler}
\end{align}
so it is a constant.
The thermodynamic pressure in 
Eq.~\eqref{eqch5:pressure} is  finite, which is written as
\begin{align}
{\bar p}= \frac{1}{8\pi {\bar R}l_p^{d-2}}\,, 
\label{eqch5:prindler}
\end{align}
so 
${\bar p}= \frac{T}{4 l_p^{d-2}}$.
The electric potential in Eq.~\eqref{eqch5:phi} is zero,
\begin{align}
{\bar \phi}=0\,. 
\label{eqch5:phirindler}
\end{align}
The thermodynamic energy from Eq.~\eqref{eqch5:energy}
obeys $E\to\infty$,
it diverges as $E = \frac{R^{d-3}}{\mu}$ positively.
Defining a specific energy, ${\bar E}$,
as
 ${\bar E}\equiv \frac{E}{\Omega_{d-2}R^{d-2}}$,
 one obtains
\begin{align}
{\bar E}=0\,.
\label{eqch5:Erindler}
\end{align}
The heat capacity in Eq.~\eqref{eqch5:Caq} goes
positively as
$C_{A} = \frac{(d-2)(d-3)\Omega_{d-2}}{2l_p^{d-2}}R^{d-4}
{\bar R}^2$.
So 
$
C_{A}=4\pi\frac{{\bar R}^2}{l_p^2}\;\;\mathrm{for}\,d=4$ and
$C_{A}\to\infty \;\;\mathrm{for}\,d>4$, 
i.e., for $d=4$ is finite
and depends on the temperature as $C_{A}= \frac{1}{\pi T^2 l_p^2}$,
and for
$d>4$  diverges. Since $C_A$
is positive,
this solution can then be considered stable.
Defining a specific heat capacity, ${\bar C}_A$,
as
 ${\bar C}_A\equiv \frac{C_A}{\Omega_{d-2}R^{d-2}}$
 gives
\begin{align}
{\bar C}_A=0\,.
\label{eqch5:CArindler}
\end{align}
Although this solution has divergent quantities, 
we can resort instead to mean densities or specific 
quantities, such as the specific heat, thus finding
finite thermodynamic quantities.

For the ensemble with infinite radius, we could try to analyze what is
the most preferred phase thermodynamically. However, it seems that the
two limiting solutions have different character. In the Davies
solution there is still a net electrically charge $Q$ at infinity.  In
the Rindler solution the electric charge has disappeared from the
context, so it is in fact a zero electric charge solution. Although
the starting ensembles are the same, the final ensembles in the
infinite radius limit are different. From the free energies, given
that the stable black hole in Davies solution has positive free energy
and the Rindler one has infinite negative free energy, one would
conclude that the Rindler solution is the most preferred phase. But in
fact the two solutions belong to different ensembles and cannot be
compared. As already mentioned, the Davies stable solution
tends to disperse to hot flat space with electric charge at infinity.

\section{Thermodynamic radii and 
the generalized Buchdahl radius in $d$ 
dimensions\label{sech5:BHzerofreeenergy}}
\sectionmark{Thermodynamic radii and 
the generalized Buchdahl radius}
\thispagestyle{userightbotmark}

\subsection{The uncharged case}

We analyze the thermodynamic energy or
mass to radius ratio
for the $d$-dimensional
canonical ensemble, namely, the  energy or
mass for which the black hole free
energy is zero, $F=0$.  We make the comparison between this
mass and the Buchdahl bound mass
in $d$ dimensions.

In the canonical ensemble of an uncharged spherically symmetric black
hole in $d$ dimensions~\cite{Andre:2021ctu}, which is described by the
Euclidean Schwarzschild-\hskip-0.05cm{}Tangherlini black hole space,
the canonical ensemble
is realized with a fixed temperature 
at the boundary of the cavity.
There are two black hole solutions, where the one with the largest mass is
stable and the one with the least mass is unstable.
Here one is interested in the large stable black hole.
The free energy of the ensemble also has a critical point
 at zero horizon radius, which is a
minimum, the hot flat space case.  Therefore,
we can analyze which are the favorable states in comparing the free
energies of the zero horizon radius, i.e., hot flat space, and the
stable black hole solution. The free energy of hot flat space is
zero.  The black hole solution also has zero free energy for a given
horizon radius, which is thus an important thermodynamic radius.  The
larger the temperature of the ensemble, the larger this radius, and
the lower the corresponding free energy.  Thus, we can argue that a
stable black hole is favored to hot flat space when the free energy of
the black hole is lower than the zero, which is the free energy of hot
flat space.
The radius of the black hole horizon
that yields zero free energy, i.e., $F=0$, is
$\left(\frac{r_+}{R}\right)_{F=0}=
\left(\frac{4(d-2)}{(d-1)^2}\right)^{\frac{1}{d-3}}$.
In terms of the spacetime mass $m$ this is
\begin{align}
\left(\frac{\mu m}{R^{d-3}}\right)_{F=0}= \left(
\frac{2(d-2)}{(d-1)^2}\right)\,.
    \label{eqch5:thermomassF=0}
\end{align}

The Buchdahl bound radius
marks the maximum compactness of a
spherically symmetric star before
spacetime turns singular.  The Buchdahl bound for a star
or matter configuration of
gravitational
radius $r_+$ and radius $R$ is \cite{Wright:2015dma}
$\left(\frac{r_+}{R}\right)_{\mathrm{Buch}}=
\left(\frac{4(d-2)}{(d-1)^2}\right)^{\frac{1}{d-3}}$, 
which in terms of the spacetime
mass $m$ and radius $R$ is
\begin{align}
\left(\frac{\mu m}{R^{d-3}}\right)_\mathrm{ Buch}
= \left(
\frac{2(d-2)}{(d-1)^2}\right)\,.
    \label{eqch5:buchmass}
\end{align}
It is a structural bound coming from
mechanics. Self-gravitating matter for which the
mass, or the energy, content within a radius $R$ is above the bound,
in principle collapses to a black hole.

It can be seen that both masses, or radii, although
conceptually different, have the same
expression, indeed,
$\left(\frac{\mu m}{R^{d-3}}\right)_{F=0}=
\left(\frac{\mu m}{R^{d-3}}\right)_\mathrm{ Buch}$.
Therefore, we can argue that as soon as the
black hole phase is
thermodynamically 
favorable over the hot flat space, it is actually
the only phase that exists,
the energy within the reservoir collapses to form a
black hole. This could indicate that there is a link
between black hole thermodynamics and matter mechanics.

\subsection{The charged case}

We now analyze the thermodynamic energy or mass to radius ratio
for  two ensembles, where one is
the
$d$-dimensional canonical ensemble
with electric charge that we are treating here, and
the other is the grand canonical
ensemble that was treated in Chapter~\ref{ch:grandcanonicalblackhole}, 
for which the black hole free
energies are zero, i.e., $F=0$, and $W=0$, respectively.
We make the comparison between these two energy or mass to radius ratio
and the generalized Buchdahl bound, i.e. the Buchdahl
bound in the electric charged
case in $d$ dimensions, also
called the Buchdahl-Andr\'easson-Wright
bound, see \cite{Wright:2015dma}.

In the canonical ensemble of a charged black hole inside a 
cavity in $d$
dimensions, the construction has been described throughout 
this chapter. The canonical ensemble
in this case is realized with a fixed temperature 
and fixed electric charge at the boundary of the cavity. One has in 
this case two stable black hole solutions for a charge below 
a saddle, or critical,
charge $Q_s$, and one stable black hole solution for 
a charge larger than $Q_s$. In this case,
it can be shown that the stable solution with the
largest mass for every charge can have a negative free energy, 
if the black hole has a larger mass than the one that solves this 
equation
\begin{align}
a \left(\frac{\mu m}{R^{d-3}}\right)^4 
\hskip -0.15cm
+
\hskip -0.05cm
b \left(\frac{\mu m}{R^{d-3}}\right)^3
\hskip -0.15cm
+
\hskip -0.05cm
c \left(\frac{\mu m}{R^{d-3}}\right)^2
\hskip -0.15cm
+
\hskip -0.05cm
d \left(\frac{\mu m}{R^{d-3}}\right)
\hskip -0.05cm
+
\hskip -0.05cm
e
\hskip -0.05cm
=
\hskip -0.05cm
0\,\,,
\end{align}
where
\begin{align}
&a
\hskip -0.05cm
=
\hskip -0.05cm
\left(
\hskip -0.05cm
\left(\frac{d-3}{d-2}
\hskip -0.05cm
\right)^2
\hskip -0.05cm
-
\hskip -0.05cm
4
\hskip -0.05cm
\right)^2
\hskip -0.05cm\,\,,\notag\\
&b
\hskip -0.05cm
=
\hskip -0.05cm
-4
\hskip -0.05cm
\left(
\hskip -0.05cm
4
\hskip -0.05cm
+
\hskip -0.05cm
8 y
\hskip -0.05cm
-
\hskip -0.05cm
\left(
\hskip -0.05cm
\frac{d-3}{d-2}
\hskip -0.05cm
\right)^2
\hskip -0.05cm
(
\hskip -0.05cm
3
\hskip -0.05cm
+
\hskip -0.05cm
2y
\hskip -0.05cm
)
\hskip -0.05cm
\right)\,\,,\notag\\
&c = -2\left(\frac{d-3}{d-2}\right)^4 y 
- 2 \left(\frac{d-3}{d-2}\right)^2 (y^2 - y +2)
+ 4+24(y +6y^2)\,\,\notag\\
&d = -4y\left((1+y)(1+2y)+\left(\frac{d-3}{d-2}\right)^2
(3+2y) \right)\,\,\notag\\
&e = \left(\frac{d-3}{d-2}\right)^4 y^2 + y^2 (1+y)^2 
+ 2 \left(\frac{d-3}{d-2}\right)^2 y (1+y)(2+y)\,\,,
\end{align}
with $y$ being the electric charge parameter
given by 
$y\equiv\frac{\mu Q^2}{R^{2d-6}}$,
as before. This
is a quartic equation in $\frac{\mu m}{R^{d-3}}$ and its
solution can be written
formally as
\begin{align}
\left(\frac{\mu m}{R^{d-3}}\right)_{F=0} =
g\left(d,\frac{\mu Q^2}{R^{2d-6}}\right)\,,
\label{eqch5:quartic}
\end{align}
for some calculable function $g\left(d,\frac{\mu Q^2}{R^{2d-6}}\right)$.
In the case $Q=0$, one gets
$\left(\frac{\mu m}{R^{d-3}}\right)_{F=0}= \left(
\frac{2(d-2)}{(d-1)^2}\right)^{\frac{1}{d-3}}$ as required, see
Eq.~\eqref{eqch5:thermomassF=0}.
The largest stable black hole 
with this mass has a zero
Helmholtz free energy, $F=0$.
Contrasting to the canonical ensemble of the electrically
uncharged black hole discussed above, 
the free energy 
in the electrically charged
case does not include the zero horizon radius 
case. The minimum possible horizon radius is the extremal black hole 
point ${r_{+}}_e = (\mu Q^2)^{\frac{1}{2d-6}}$, yielding a free energy 
$F_{{r_{+}}_e} = \frac{Q}{\sqrt{\mu}}$. 
To emulate hot flat space, an electrically charged
nonself-gravitating shell was used. 
The comparison was then made between the black hole 
configuration and the electrically charged 
shell with no self-gravity
at the boundary of the cavity, having 
then hot flat space inside the cavity
with the electric charge near the boundary. 
This configuration would require
us to look into the matter sector which
we have not done here. It is unclear if a
transition can occur between 
hot flat space with electric
charges near the cavity and the stable 
black holes. Nevertheless, the thermodynamic radius 
of zero free energy in the canonical ensemble is still regarded 
as an important quantity.

In the grand canonical ensemble of a charged
Reissner-Nordstr\"om  black hole inside a 
cavity for $d$ dimensions, the
construction and its thermodynamics were described
in \cite{Fernandes:2023byx}, and Chapter~\ref{ch:grandcanonicalblackhole}.
The grand canonical ensemble is realized with a fixed temperature 
and fixed electric potential
at the boundary of the cavity.
In this ensemble,  the
partition function in the zero loop approximation is given in terms of
the grand potential, or
Gibbs free energy, $W=E - TS - Q\phi$, where $E$ is the mean energy,
$T$ is the temperature, $S$ is the entropy, $Q$ is the mean charge and
$\phi$ is the electric potential. The grand potential yields $W[r_+,Q]
= \frac{R^{d-3}}{\mu}\left(1 - \sqrt{f}\right)- Q \phi - T\frac{\Omega_{d-2}
r_+^{d-2}}{4}$, with $f = \left(1 -
\frac{r_+^{d-3}}{R^{d-3}}\right) \left(1 - \frac{\mu Q^2}{(r_+
R)^{d-3}} \right)$, and the equilibrium equations that yield the black
hole solutions are $\frac1T = \frac{4\pi}{(d-3)}
\frac{r^{d-2}_+}{r^{2d-6}_+ - \mu Q^2} \sqrt{f}$ and $\phi =
\frac{Q}{\sqrt{f}}\left(\frac{1}{R^{d-3}}
-\frac{1}{r_+^{d-3}}\right)$, where the convention
for the electromagnetic coupling and electric charge was chosen so
that $Q\rightarrow \sqrt{(d-3)\Omega_{d-2}} Q$ and $\phi \rightarrow
(\sqrt{(d-3)\Omega_{d-2}})^{-1}\phi$ in the expressions in
~\cite{Fernandes:2023byx} and Chapter~\ref{ch:grandcanonicalblackhole}.
One has in 
this case up to two solutions, depending on the fixed quantities 
$T$ and $\phi$, with only one being stable.
The grand canonical
free energy of the ensemble also has a critical
point at zero horizon radius, which is a minimum, it is the
hot flat space case.
The stable black hole solution also has zero free energy for
a given horizon radius, which is thus an important thermodynamic
radius. The larger the temperature of the
ensemble, the larger this radius, and the lower the corresponding
free energy. Thus, we can argue that a stable
black hole is favored to hot flat space when the free energy
of the black hole is lower than the zero, which is
the free energy of hot flat space. The radius of the black
hole horizon that yields zero
grand potential energy, i.e., $W = 0$
is complicated to find, but 
the corresponding mass has a simple expression given by
\begin{align}
 \hskip-0.1cm
 \left(\frac{\mu m}{R^{d-3}}\right)_{W=0}\hskip-0.15cm
=&\frac{-4(d-2)^2}{(d-1)^2(d-3)^2}\hskip-0.05cm
+ \hskip-0.05cm\frac{2(d-2)((d-2)^2 + 1)}{(d-1)^2 (d-3)^2}
\notag \\
&
\times
\sqrt{1 + \frac{(d-1)^2 (d-3)^2}{4(d-2)^2}\frac{\mu Q^2}{R^{2d-6}}}\,.
\label{eqch5:thermodynamicradiusgrandcanonical}
\end{align}
Since hot flat space 
is described here by the grand potential $W[r_+,Q]$,
a possible  transition can 
occur from the charged hot flat space to the stable black hole for 
temperatures corresponding to stable black holes with higher mass 
than Eq.~\eqref{eqch5:thermodynamicradiusgrandcanonical}. 
In the case $Q=0$, one has that
$W=F$, so one
gets
$\left(\frac{\mu m}{R^{d-3}}\right)_{W=0}
=\left(\frac{\mu m}{R^{d-3}}\right)_{F=0}= \left(
\frac{2(d-2)}{(d-1)^2}\right)^{\frac{1}{d-3}}$ as required, see
Eq.~\eqref{eqch5:thermomassF=0}.

The 
Buchdahl bound was originally given
for the electrically uncharged case and in
$d=4$.
For electrically charged matter in $d$  
dimensions  one has
the generalized Buchdahl bound
that is given by \cite{Wright:2015dma}
\vfill
\begin{align}
\left(\frac{\mu m}{R^{d-3}}\right)_\mathrm{ Buch}
& = \frac{d-2}{(d-1)^2} +
\frac{1}{d-1}\frac{\mu Q^2}{R^{2d-6}}
\notag \\
&+ \frac{d-2}{(d-1)^2}\sqrt{1 + (d-1)(d-3)
\frac{\mu Q^2}{R^{2d-6}}}\,.
\end{align}
In the no charge case, $Q=0$, one
gets 
$\frac{\mu m}{R^{d-3}}= \left(
\frac{2(d-2)}{(d-1)^2}\right)^{\frac{1}{d-3}}$,
as required.

We see that the three mass to radius ratios, are
conceptually different, and now in the electrically
charged case,
have generically different expressions, indeed, 
$\left(\frac{\mu m}{R^{d-3}}\right)_{F=0}$, 
$\left(\frac{\mu m}{R^{d-3}}\right)_{W=0}$, and
$\left(\frac{\mu m}{R^{d-3}}\right)_\mathrm{ Buch}$
are not equal.
One has 
$\left(\frac{\mu m}{R^{d-3}}\right)_{F=0}\geq
\left(\frac{\mu m}{R^{d-3}}\right)_\mathrm{ Buch}\geq
\left(\frac{\mu m}{R^{d-3}}\right)_{W=0}$. This is 
an interesting result.
In the canonical ensemble,
the thermodynamic energy content within the cavity
when the black hole phase starts to be favorable, i.e.,
when $F=0$,
is higher than the 
Buchdahl bound, and so
even before the black hole is thermodynamically  favored,
collapse should occur, i.e., 
as soon as a black hole forms
there is no possibility of a thermodynamic
phase transition
to hot flat space, indeed the black hole
has been formed dynamically. 
In the grand canonical ensemble,
the energy content within the cavity
when the black hole phase starts to be favorable, i.e., 
when $W=0$, is less
than the 
Buchdahl bound, and so there
should be no collapse at this stage, indeed, collapse should only occur
when the energy content is increased
above the bound. In the grand canonical ensemble
this occurs only for some negative $W$.
Both
thermodynamic mass to radius
ratios are equal to the generalized Buchdahl
bound
when the electric 
charge is put to zero, and all the three are also
equal at the extremal point.
The plots given in Fig.~\ref{figch5:Buchdahl}
for $d=5$
help in the understanding of this behavior.
These results present a counter example to the
possible link between the black hole thermodynamics and stability 
of spherically symmetric matter. The uncharged case seems to be 
a coincidence.

\section{Conclusions\label{sech5:concl}}

We have analyzed in this chapter the canonical ensemble of a Reissner-Nordstr\"om
black hole in a cavity for arbitrary dimensions. 
We have built the canonical ensemble through the 
Euclidean path integral approach, which specifies the partition function in terms 
of a path integral involving the Euclidean action.
The Euclidean action is the usual Einstein-Hilbert-Maxwell action with the
Gibbons-Hawking-York boundary term and an additional Maxwell boundary
term so that the canonical ensemble is well-defined, all
terms having been Euclideanized.  We assumed that
the heat reservoir has a spherical boundary at finite radius $R$,
where the temperature is fixed as the inverse of the Euclidean proper
time length at the boundary, and also the electric charge is fixed by
fixing the electric flux at the boundary. We restricted the summed 
spaces in the path integral to
spherically symmetric spaces and we assumed regularity conditions 
that avoid the presence of conical and curvature singularities.

We then performed the zero loop approximation by first imposing the
Hamiltonian and the Gauss constraints, obtaining a reduced action that
depends on the fixed inverse temperature $\beta$, electric charge $Q$,
and the radius of the boundary $R$, and also depends on the radius of
the event horizon $r_+$ as a variable that is integrated through the
path integral. We found the equation for the stationary points
of the reduced action, which are the solutions $r_+[\beta,Q,R]$, and
we found also the condition of stability of the solutions.

We analyzed the existence of the solutions of the ensemble 
for arbitrary dimensions. For charges smaller than a saddle,
or critical,
electric charge, there are always three possible solutions where
the one with the smallest radius and the one with the largest radius are
stable, and the other with intermediate radius is unstable. The value
of the saddle charge and the value of the radii that bound these
solutions, which are saddle points of the reduced action, were found
analytically. For the saddle charge, the unstable solution reduces to
a point, having formally only two solutions which are stable.  For
charges larger than the saddle charge, there is only one solution, and
this solution is stable. This analysis was then applied to the four
and five dimensional cases. Regarding stability, the
solutions are stable if the radius of the event horizon increases as
the temperature increases. For this case, the condition is given in
terms of the saddle points of the reduced action.

We obtained the thermodynamics of the electrically charged black hole 
using that the partition function is related to the Helmholtz free
energy of the system in the canonical ensemble. Through the zero loop
approximation, we obtained the free energy. We retrieved the entropy, the
thermodynamic electric potential,
the thermodynamic pressure, and the
thermodynamic energy 
through the derivatives of the free energy. More precisely,
the entropy is the Bekenstein-Hawking entropy, the pressure
has the same expression of the
pressure of a self-gravitating  charged
shell with radius $R$, and the
thermodynamic electric potential is given by the usual expression. We calculated 
the mean thermodynamic energy, which can be identified with a quasilocal
energy, through the definition of
free energy. Regarding thermodynamic stability, the configurations are
stable if the heat capacity with constant charge and area is
positive. We also found the integrated first law, i.e., the Smarr formula,
and the Gibbs-Duhem relation.

We analyzed the favorable states in the canonical ensemble.  A
favorable state is a stable state of the ensemble that has the lowest
value of the free energy. In some sense, transitions can occur between
phases. Here, for an electric charge lower than the critical charge,
there are two stable black hole solutions that are in competition,
with an existing first order phase transition between them. For the
critical charge, this first order phase transition becomes a second
order phase transition.  For a charge larger than the critical charge,
there is only one stable black hole solution. In the uncharged case,
there is a stable solution and hot flat space.  Pure hot flat space
does not seem to be a solution of the canonical ensemble since the
charge is fixed. Instead, we compare the stable solutions with a
nonself-gravitating charged sphere.  This covers two limits, the case
where one has flat space with a charge at the center, which is not a
solution and is never favorable, and another case where the charge
resides near the cavity or at the cavity. In this last case, it would
act as a hot flat space with electric charge at the boundary and the
corresponding free energy vanishes.
Considering this latter case, we found a first order phase transition
between the largest black hole and hot flat space with electric charge
at the boundary.

In this chapter, regarding the canonical ensemble of a 
Reissner-Nordstr\"om
black hole in a cavity for four and higher dimensions,
there are several main achievements which can be stated:

First, we have made the construction of the canonical ensemble and the thermodynamic
analysis of all generic $d$ dimensions in a unified way. Moreover,
we presented significant cases in all the detail, namely,
the dimension $d=4$ as the most important case, and the dimension
$d=5$ as a typical higher dimensional case.

Second, in the analysis of the specific heat $C_{A,Q}$ in terms of the
temperature and the electric charge, we found the existence of a
second order phase transition between the two stable solutions for a
critical electric charge parameter
$\frac{\mu Q_s}{R^{2d-6}}$ in arbitrary
dimensions.  For lower electric charge $\frac{\mu Q}{R^{2d-6}}$, we found 
two turning points, which indicate the stability of the solutions, where the
heat capacity diverges and is double valued.  For higher charge
$\frac{\mu Q}{R^{2d-6}}$, we found that the heat capacity is always
positive.

Third, since in the canonical ensemble one can have two stable
black hole solutions, an analysis of the free energy has
enabled us to pick the black hole solution that is most favored
according to the temperature
and electric charge of the ensemble and fine the possible
first order phase transitions.  Moreover, a comparison
with the free energy of hot flat space, emulated by an electric shell
at the boundary, has revealed the thermodynamic phase that is favored.
We also argued that the Buchdahl bound is important in this
context, and the free energies for which the bound is superseded were
found, for higher free energies gravitational collapse sets in.

Fourth, we have shown that the Davies thermodynamic theory of black holes
follows from the electric charged canonical ensemble in the infinite
large reservoir limit when $d=4$.  The two ensemble solutions of lower
radii maintain, in this limit, their black hole character. One, with the
smallest radius, is the stable one, and the other with intermediate
radius is unstable.  These two solutions meet at a saddle point.  We found the
thermodynamic quantities and in particular, we found the heat
capacity at constant area and charge.  In $d=4$, the
expression of the heat capacity reduces to the expression found by
Davies.  Here, we started from the action and the path integral
approach for a reservoir at infinity and showed that the formalism
gives the first law of black hole mechanics which, of course, is also
the first law of thermodynamics for black holes.  Davies, in the $d=4$
formulation of the theory, started directly from the first law of
black hole mechanics.  These results, reached through different means,
point towards the equivalence between black hole mechanics and black
hole thermodynamics through the canonical ensemble.

Fifth, the limit of infinite radius of the
boundary of the cavity has revealed a surprise solution.
Indeed, the largest black hole solution of the ensemble, changes
character in this limit. The black hole solution
turns into a Rindler solution with the ensemble fixed temperature
being the Unruh temperature of the now accelerated boundary.

Sixth and last, we have followed the York path integral procedure, which was originally
applied to Schwarzschild black holes, throughout
this work for Reissner-Nordstr\"om black holes. We have shown that the
black hole solutions found represent the unification of York electrically
uncharged black holes and Davies electric charged black holes, in a
remarkable way. Indeed,
the two York type solutions, one larger and
stable, one smaller and unstable,
do appear, and the two Davies type solutions, the smaller and
unstable, and the even smaller and stable also do appear, in a
remarkable way.  York and Davies results follow from two different
limits of our analysis in this chapter. York results follow from taking the zero electric
charge limit. Davies results follow from taking the infinite cavity
radius limit, i.e., by putting the heat reservoir at infinity.  This
latter case can also be seen to stem from York's generic
reduced action approach with the
boundary at infinity, which in turn yields the Gibbons-Hawking path
integral formulation to black hole thermodynamics.  The
Gibbons-Hawking approach was originally applied to electrically
uncharged black holes and it was found that there was an unstable
black hole solution, the Hawking black hole, and thus no consistent
thermodynamics. It was also applied to an electrically charged black
hole in the grand canonical ensemble, and it was found a solution that
was unstable.  Had it been applied directly to electrically charged
black holes in the canonical ensemble,
one would have found that thermodynamic stable solutions exist
to vindicate the approach. We have filled this gap here.

What does remain to be understood? Here, we were interested in the
thermodynamic interaction of a black hole in a cavity with a boundary
of finite size and fixed temperature, as well as in the interaction of
the gravitational field with the electromagnetic field in such a
system.  The formalism by its very distinctive features, i.e., its
Euclidean character, applies only to the outside of a black hole event
horizon.  The black hole interior and its singularity are not
considered in the analysis.
Thus, the question about the nature of the singularity remains.  
It is expected that the
singularity is 
described by a Planck scale object, however
intricate the description might be. 
A canonical formalism for micro black holes, say of the order
of ten Planck radii, seems valid, after all Hawking
radiation, a tamed radiation at most of the scales,
if left by itself, slowly peels the singularity away.
If that radiation interacts harmoniously with the boundary
of a cavity, a thermodynamic procedure might be valid
and show how the black hole horizon and the singularity
fuse into one single describable object.

\chapter{Limits in hot spaces with negative cosmological constant 
in the canonical ensemble:
hot anti-de Sitter solution, Schwarzschild-anti 
de Sitter black hole,
Hawking-Page solution, and planar AdS black hole}
\chaptermark{Limits in hot spaces with negative $\Lambda$ in the canonical ensemble}
\label{ch:adslimits}

\section{Introduction}

In the previous chapter, we studied a charged black hole in the canonical 
ensemble. We touched briefly on the limits of infinite 
cavity which connected the solutions that characterize Davies' thermodynamic 
theory and the solutions inside a finite cavity. We detour then to the 
analysis of a specific case in which the limits connect various solutions 
existing in the literature. These are the following.
The asymptotically flat black hole solutions in thermal 
equilibrium are called the Gibbons-Hawking solutions~\cite{Gibbons:1977}.
By putting a cavity at finite radius, 
York found two solutions for the Schwarzschild black hole, which are called the 
York black hole solutions.  Moreover, Schwarzschild-AdS black holes 
in thermal equilibrium are described by the Hawking-Page solutions~\cite{Page:1982dh}. 
It is further known that very large black holes in AdS tend to the planar black hole 
solutions in AdS~\cite{Witten:1998zw}.

In this chapter, we consider the canonical ensemble of a Schwarzschild black hole in AdS 
inside a cavity, where the emphasis is to unify the aforementioned
solutions through limits in the cosmological constant and the radius of the cavity. 
These limits yield different results for each solution of the Schwarzschild AdS 
black hole, obtaining thus all the solutions mentioned above.

This chapter is organized as follows. In Sec.~
\ref{sech8:thermo}, we construct the ensemble in the zero 
loop approximation. 
In Sec.~\ref{sech8:actionandthermodynamics}, we obtain the thermodynamics 
of the system by using the partition function. In Sec.~
\ref{sech8:sols}, we analyze the solutions of the ensemble and their stability, 
with the limit of zero cosmological constant 
being trivial. In Sec.~\ref{sech8:limitR}, we consider the limit of infinite cavity 
and the limit in the cosmological constant. 
In Sec.~\ref{sech8:conc}, we present the conclusions.
This chapter is based on the ongoing work~\cite{Tiago2024bl}.

\clearpage

\section{Thermodynamics of the Schwarzschild-anti de Sitter space in
the canonical ensemble: General results for the black hole horizon
region inside a heat reservoir\label{sech8:thermo}}
\sectionmark{Thermodynamics of the 
Schwarzschild-AdS space in
the canonical ensemble}\thispagestyle{userightbotmark}

\begin{figure}[h]
    \centering
    \includegraphics*[scale=0.5]{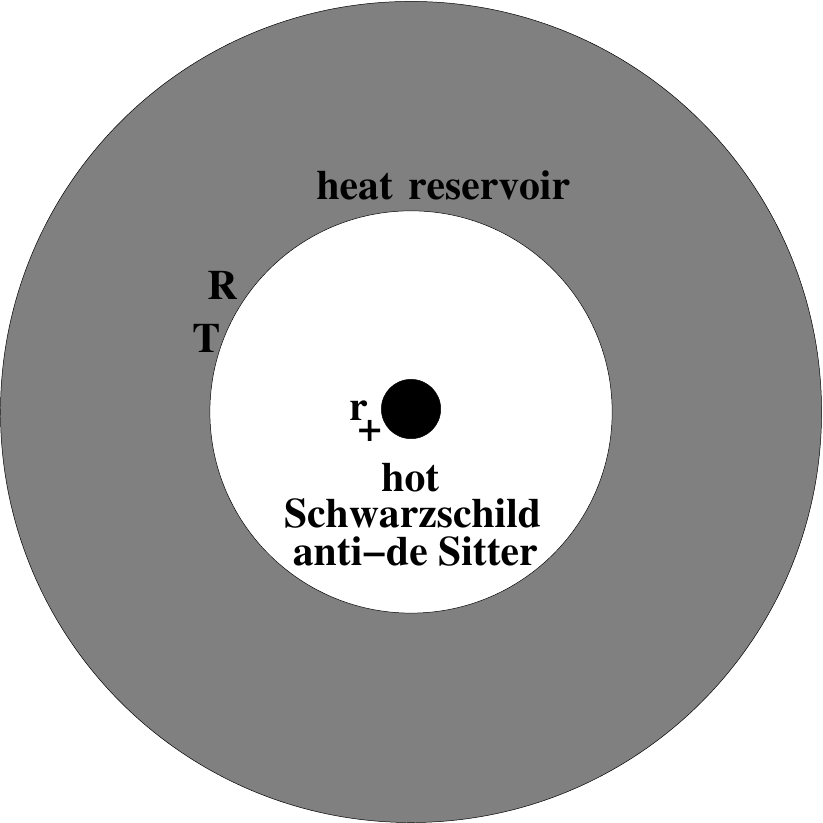}
    \caption{
    A drawing of a black hole in a cavity
    within a heat reservoir at temperature $T$
    and radius $R$ in a space with positive cosmological constant.
    Outside the black hole radius $r_+$ the geometry is a Schwarzschild-anti-de
    Sitter geometry. The Euclideanized space and its boundary have $R^2
    \times S^2$ and $S^1 \times S^2$ topologies, respectively, where the
    $S^1$ subspace with proper length $\beta=\frac1T$ is not displayed.
    }
    \label{figch8:schwantidesitterreservoir}
\end{figure}

The setup that we consider here is a space $M$ describing the case of a black hole in  
a negative cosmological constant background inside a heat reservoir, 
which is described by the boundary $\partial M$ of a cavity with radius $R$.
We then construct the canonical ensemble of this setup, where 
at the boundary we specify the data that determine
the ensemble, see Fig.~\ref{figch8:schwantidesitterreservoir}.
Namely, we fix the inverse temperature $\beta = \frac1T$ at the boundary 
with radius $R$, which we also fix. The inverse temperature $\beta$  
is given by the imaginary proper time at the boundary. Hence, we 
can construct the canonical ensemble through York formalism~\cite{York:1986}, 
see Chapter~\ref{ch:Euclideanpathintegral} for more details,
by the partition function as
\begin{align}\label{eqch8:partitionfunction}
    Z = Dg_{\alpha \beta} \mathrm{e}^{-I}\,\,,
\end{align}
where $I$ is the gravitational action given by 
\begin{align}\label{eqch8:gravaction1}
    I = -\frac{1}{16\pi l_p^{2}}\int_{M}(R - 2\Lambda)\sqrt{g}d^4x 
    - \frac{1}{8\pi l_p^{2}}\int_{\partial M}K \sqrt{\gamma} d^{3}x 
    - I_{\mathrm{AdS}}\,\,,
\end{align} 
where $R$ is the Ricci scalar, $\Lambda$ is the cosmological constant, 
$g$ is the metric determinant, 
$K = {n^\alpha}_{;\alpha}$ is trace of the extrinsic curvature of $\partial M$, 
$n^\alpha$ is the unit normal vector to $\partial M$, 
$\gamma$ is the determinant of the induced metric $\gamma_{ab}$ 
of $\partial M$ and $I_{\mathrm{AdS}}$ is the action of a reference metric 
to make $I$ finite, which is here the action of pure anti-de Sitter.
It is useful to define the anti-de Sitter or AdS length  
by $l^{2} = \frac{3}{-\Lambda}$.

We then proceed with the full zero loop approximation. The spherically 
symmetric black hole space with topology $\mathbb{R}^2\times \mathbb{S}^2$ 
that obeys the Euclidean Einstein equations with 
negative cosmological constant is 
\begin{align}\label{eqch8:lineAdS}
    ds^2 = \frac{1}{(2\pi T^\mathrm{H}_+)^2}V(r)d\tau^2 
    + \frac{dr^2}{V(R)} + r^2 d\Omega^2_2\,\,,
\end{align} 
where $\tau \in ]0,2\pi[$, $r\in ]r_+,R[$ with $r_+$ being the 
event horizon radius, $d\Omega_2^2$ being the line element of 
a unit $2$-sphere, the function $V(r)$ is
\begin{align}\label{eqch8:Vr}
    V(r) = \left(1 - \frac{r_+}{r}\right)\left(1 
    + \frac{r^2}{l^2}\left( 1 + \frac{r_+}{r} + \frac{r_+^2}{r^2}\right) \right)\,\,,
\end{align}
and
\begin{align}\label{eqch8:tbh}
    T^\mathrm{H}_+ = \frac{1 + 3\frac{r_+^2}{l^2}}{4\pi r_+}\,\,,
\end{align}
is the constant that must be added in Eq.~\eqref{eqch8:lineAdS} 
so that there is no conical singularity. The line element in 
Eq.~\eqref{eqch8:lineAdS} can be found either by 
solving the Einstein equations or by performing the Wick 
transformation to the Schwarzschild-AdS line element, 
by compactifying the imaginary time and by avoiding the conical singularity.

The black hole space is then in thermodynamic equilibrium only if the 
total imaginary proper time at the boundary of the cavity is $\beta= \frac{1
}{T}$, hence from Eq.~\eqref{eqch8:lineAdS} one has
\begin{align}\label{eqch8:thermequilibrium}
    T = \frac{T_+^\mathrm{H}}{\sqrt{V(R)}}\,\,.
\end{align}

Now, the radius of the heat reservoir $R$ sets a scale for our
problem. It is then meaningful to gauge all the length scales involved
in the problem to $R$.  Thus, the heat reservoir temperature $T$, the
cosmological constant $|\Lambda|=-\Lambda$ or the cosmological
length $l^2$,
and the black hole horizon radius $r_+$,
are gauged to
quantities without units as $RT$, $\frac{R^2}{l^2}$, and
$\frac{r_+}{R}$.  The ranges of these quantities are
important. They are: $0\leq RT<\infty$, $0\leq \frac{R^2}{l^2}<\infty$,
and $0\leq\frac{r_+}{R}\leq1$.

\section{Action, free energy, entropy, mean
energy, and heat capacity\label{sech8:actionandthermodynamics}}

The Euclidean action in Eq.~\eqref{eqch8:gravaction1} for the 
space in Eq.~\eqref{eqch8:lineAdS} is precisely the gravitational 
action with negative cosmological constant $I_{gl}$ in four dimensions, 
i.e. $I=I_{gl}$. We can easily compute it by 
substituting the line element into the action written 
in terms of the components of the spherically symmetric metric 
\begin{align}
    \label{eqch8:actionspherical2}
    &I= I_{g l} = 
    \left(\frac{\beta r}{l_p^2}\left(\sqrt{V_{\mathrm{AdS}}(r)} 
    - \sqrt{V(r)}\right)\right)\sVert[3]_{r\rightarrow R}
    - \frac{\pi}{l_p^{2}}\left(\frac{V' r^{2}}{4\pi T^\mathrm{H}_+}\right)\sVert[3]_{r=r_+}\notag\\
    &+ \frac{1}{8\pi l_p^{2}}\int_{M} \frac{r^2}{2\pi T^\mathrm{H}_+} r^{2}\left(G\indices{^\tau_\tau} 
    - \frac{3}{l^2}\right)d^4x\,\,,
\end{align}
where $V_{\mathrm{AdS}}(r) = 1 + \frac{r^2}{l^2}$ is the pure AdS redshift factor 
obtained from $V(r)$ by setting $r_+=0$,
see Chapter~\ref{ch:Euclideanpathintegral} for more details on the action. 
Now, the line element 
obeys the Einstein equation $G\indices{^\tau_\tau} 
- \frac{3}{l^2} = 0$ and so we obtain the action 
evaluated at the zero loop approximation, whose designation 
is kept as $I$, as 
\begin{align}\label{eqch8:actionzero}
    I = \beta\frac{R}{l_p^2}\left(\sqrt{V_{\mathrm{AdS}}(R)} 
    - \sqrt{V(R)}\right) - \frac{\pi r_+^2}{l_p^2}\,\,
\end{align}
where $\beta=\frac1T$ is the inverse temperature
of the ensemble, i.e., at the
boundary of the heat reservoir, and where $V(R)$ is given by 
Eq.~\eqref{eqch8:Vr} at $r=R$ as
\begin{align}
    V(R)=\left(1-\frac{r_+}{R}\right)
\left(1+\frac{ R^2}{l^2}\left(1+
\frac{r_+}{R}+\left(\frac{r_+}{R}\right)^2\right)\right)\,,
\label{eqch8:VR}
\end{align}
and also 
\begin{align}
    V_\mathrm{AdS}(R)=1+\frac{R^2}{l^2}\,.
\label{eqch8:VAdS}
\end{align}
Note that $I$ in Eq.~\eqref{eqch8:actionzero}
is $I=I(\beta,R,r_+(R,T,l),l)$. The statistical
mechanics ensemble
is characterized by $l$ which is
fixed for each space, by
$T=\frac1\beta$ and $R$
which are fixed for each ensemble,
with $r_+=r_+(R,T,l)$ solutions of 
Eq.~\eqref{eqch8:thermequilibrium}.
These $r_+=r_+(R,T,l)$ solutions 
yield the thermodynamic solutions of the
problem. Note that $l$ is also fixed.

In the zero loop approximation, the partition 
function in Eq.~\eqref{eqch8:partitionfunction} 
becomes $Z = \mathrm{e}^{-I}$. Since we are 
considering the canonical ensemble, the partition 
function is linked to the free energy $F$ of the 
system by $Z = \mathrm{e}^{-\beta F}$, defined 
by the Legendre transform of the mean energy 
as $F= E-T S$. Therefore, the action in the 
zero loop approximation is connected 
to the free energy as $I = \beta F$, 
and so the free energy is given by 
\begin{align}
    F = \frac{R}{l_p^2}\left(\sqrt{V_\mathrm{AdS}(R)}-\sqrt{V(R)}\right) -T\pi \frac{r_+^2}{l_p^2}\,.
    \label{eqch8:Ffromctionbh}
\end{align}
Now from the derivatives of $F$, we are able to 
obtain the thermodynamic properties of the 
canonical ensemble of a Schwarzschild-AdS black hole 
inside a cavity. Namely, the entropy can be given by 
$S = -\left(\frac{\partial F}{\partial T}\right)_{R}$, 
where the subscript means the quantity that is kept 
fixed while performing the derivative, obtaining thus
\begin{align}
    \label{eqch8:Sentropyschwarzschildmain}
  S = \pi \frac{r_+^2}{l_p^2}.
\end{align} 
The thermodynamic pressure can also be obtained, 
through the derivative 
$8\pi R p = - \left(\frac{\partial F}{\partial R}\right)_{T}$, 
as 
\begin{align}
    p = \frac{1}{8\pi R l_p^2}\left(\frac{1 + 2\frac{R^2}{l^2} - \frac{r_+}{2R}(1 + \frac{r_+^2}{l^2})
    }{\sqrt{V(R)}} - \frac{1 + 2\frac{R^2}{l^2}}{\sqrt{V_\mathrm{AdS}(R)}}\right)\,\,.
\end{align}
Finally, the mean energy can be obtained from the 
Legendre transformation $E = F + TS$ as 
\begin{align}\label{eqch8:energy}
    E =  \frac{R}{l_p^2}\left(\sqrt{V_\mathrm{AdS}(R)}-\sqrt{V(R)}\right).
\end{align}

Regarding thermodynamic stability, the quantity that gives information 
about stability is the heat capacity at constant area $A = 4\pi R^2$, 
$C_A$. The heat capacity is given by 
$C_A=\left(\frac{\partial E}{\partial T}\right)_{\hskip -0.15cm A}$, 
which by using Eq.~\eqref{eqch8:energy} together with the solutions of 
the ensemble $r_+(T,R)$ given by solving Eq.~\eqref{eqch8:thermequilibrium},
we obtain
\begin{align}\label{eqch8:heatcapacity}
    C_A = \frac{4\pi r_+^2 (1 + 3\frac{r_+^2}{l^2})V(R)}
    {l_p^2\left(\frac{r_+}{R}\left(1 + 3\frac{r_+^2}{l^2} \right)^2 
    + 2 V(R)\left(3\frac{r_+^2}{l^2} - 1\right)\right)}\,\,,
\end{align}
which can also be obtained from the second derivative of 
the free energy $F$. 
Now, one has a thermodynamically stable system if
\begin{eqnarray}
    \label{eqch8:CAstabilitymain}
    C_A \geq 0\,\,,
\end{eqnarray}
which is verified if the denominator in Eq.~\eqref{eqch8:heatcapacity} 
is positive. Another alternative to understand the condition Eq.~\eqref{eqch8:CAstabilitymain}
is by relating it to the derivative of the solution, 
as one has $C_A = - \frac{R}{2l_p^2 \sqrt{V(R)}}\frac{\partial V(R)}
{\partial r_+}\frac{\partial r_+}{\partial T}$. Since the derivative 
$\frac{\partial V(R)}
{\partial r_+}$ is negative, the condition for stability in Eq.~\eqref{eqch8:CAstabilitymain} 
is satisfied if $\frac{\partial r_+}{\partial T}$ is positive, i.e. the solution that 
obeys  
$\frac{\partial r_+}{\partial T}> 0$ is stable. We must comment on the type of stability 
considered here. The stability condition in Eq.~\eqref{eqch8:CAstabilitymain} 
is the stability condition of the canonical ensemble with fixed area and fixed 
temperature only. This is different from intrinsic thermodynamic stability, which requires
further conditions on the concavity of the free energy. Here, we are only interested 
on the thermodynamic stability of the ensemble.

\section{Thermodynamic solutions of
Schwarzschild-anti-de Sitter black holes in the canonical ensemble
\label{sech8:sols}}\sectionmark{
Schwarzschild-AdS black hole solutions in the canonical ensemble}
\thispagestyle{userightbotmark}

\subsection{Temperature equation}

The solutions of the event horizon radius can be found by the 
condition of temperature equilibrium at the boundary of the cavity. 
This condition is written in Eq.~\eqref{eqch8:thermequilibrium}, which 
is translated by setting the local temperature according to the 
Tolman formula at the boundary to be the fixed temperature $T$. 
We can indeed write Eq.~\eqref{eqch8:thermequilibrium} explicitly 
as 
\begin{align}\label{eqch8:thermoeq}
    4\pi T = \frac{1}{r_+}\frac{1 + 3\frac{R^2}{l^2} \frac{r_+^2}{R^2}}
    {\sqrt{1 - \frac{r_+}{R}}\sqrt{1 + \frac{R^2}{l^2}\left(1 + \frac{r_+}{R}
    + \frac{r_+^2}{R^2}\right)}}\,\,.
\end{align}
The strategy is then to invert 
Eq.~\eqref{eqch8:thermoeq} in order to find the solutions 
$r_+$ in function of the fixed parameters of the ensemble $T$ and $R$, 
and also in function of the AdS length $l$. Depending on the 
parameters $(T,R,l)$, there can be no solution, 
one solution, or two solutions for $r_+$. 
When there are two solutions, these are denoted by 
\begin{align}
    &r_{+1} = r_{+1}(R,l,T)\,\,,\\
    &r_{+2} = r_{+2}(R,l,T)\,\,,
\end{align} 
with $r_{+1}\leq r_{+2}$. For a given $T$ and $R$, the existence of 
two roots $r_{+1}$ and
$r_{+2}$ is similar to the case of the Schwarzschild space \cite{York:1986},
with here having additionally the parameter $l$.

We now proceed with the analysis of the solutions for the cases, 
$0\leq\frac{R^2}{l^2}<\infty$ and $0<RT<\infty$. 
In general there are no analytical solutions.
However, special 
attention in this paper is given to two limiting cases, 
for small cosmological constant 
 $0\leq \frac{R^2}{l^2}\ll 1$, and in the 
 very high temperature $RT\to \infty$, where 
 analytical solutions can be found.

 \subsection{Solutions in two limiting cases}

\subsubsection{Solutions for
small cosmological constant, $\frac{R^2}{l^2}\ll1$}

For very small $\frac{R^2}{l^2}$, 
$\frac{R^2}{l^2}\ll 1$, we find
from Eq.~\eqref{eqch8:thermoeq} that
there are no black hole solutions 
for
\begin{equation}
R T<\frac{\sqrt{27}}{8\pi}
\left( 1+\frac{5}{18}\,\frac{R^2}{l^2}\right)
\,,\quad\quad\quad\frac{R^2}{l^2}\ll 1\,,
\label{eqch8:nosolutionslimit}
\end{equation}
and there
are two black hole solutions for
\begin{equation}
RT
\geq \frac{\sqrt{27}}{8\pi} \left( 1+\frac{5}{18}\,
\frac{R^2}{l^2}\right)\,,\quad\quad\quad
\frac{R^2}{l^2}\ll 1
\,.
\label{eqch8:solutionslimit}
\end{equation}
One of the two solutions is the small black
hole $r_{+1}(R,l,T)$,
and the other solution is the large black hole
$r_{+2}(R,l,T)$.
The two solutions merge
into one sole solution when 
the equality sign in
Eq.~\eqref{eqch8:solutionslimit}
holds. In this case, the coincident
double solution has horizon radius
given by
\begin{equation}
\frac{r_{+1}}{R}=\frac{r_{+2}}{R}=\frac23
\left(1-\frac{17}{27}\frac{R^2}{l^2}\right)\,,
\quad \frac{R^2}{l^2}\ll1\,.
\label{eqch8:r+solutionslimit}
\end{equation}
For zero cosmological constant,
$|\Lambda| R^2=0$, i.e.,
$\frac{R^2}{l^2}=0$,
we have a pure Schwarzschild black
hole and we recover York's result of 
$RT
\geq \frac{\sqrt{27}}{8\pi}$ to have black
hole solutions, the solutions merge
with 
$\frac{r_{+1}}{R}=\frac{r_{+2}}{R}=\frac23$.

One could work out in the regime $\frac{R^2}{l^2}\ll1$
the action $I$, the thermodynamic energy $E$, the
entropy $S$, and the heat capacity $C_A$, given through
Eqs.~\eqref{eqch8:actionzero}
to \eqref{eqch8:heatcapacity}.  Apart from the entropy
expression $S=4\pi r_+^2$, valid for each of the two black hole
solutions, the calculation of the other quantities is not practical
and they are not particularly illuminating.  
However, an instance where all
quantities can be worked out, in particular the heat capacity $C_A$
with a simple expression
is the high temperature limit, which we turn now.

\subsubsection{Solution in the high temperature limit,  $RT$
high}

For the range of values of the cosmological constant considered in
this section, $0\leq\frac{R^2}{l^2}<\infty$, we can find solutions
in the limit of $RT$ goes to infinity, see \eqref{eqch8:thermoeq}. 
Since $R$ is the quantity that we consider as the gauge, 
$RT$ going to infinity is the same in this context as 
$T$ going to infinity.

For a given $T$, there are two black hole solutions, the small black
hole solution $r_{+1}$ and the large black hole solution $r_{+2}$.
We set
the heat reservoir temperature 
$T$ fixed but very high, in the sense that $T\rightarrow
\infty$. From Eq.~\eqref{eqch8:thermequilibrium} there are two possibilities.  
Either 
$T^\mathrm{ H}_+\rightarrow \infty$ which corresponds to the small black
hole solution having a very small $r_{+1}$,
or $V(R)\rightarrow 0$ which corresponds to
the large black hole solution $r_{+2}$
approaching the reservoir radius. 

The first solution 
for a very high
heat reservoir temperature,
$T\to\infty$, 
corresponds to the limit 
$T^\mathrm{ H}_+\rightarrow \infty$, which from 
Eq.~\eqref{eqch8:tbh} means that 
$r_+=r_{+1}\rightarrow 0$. In this limit,
we have
\begin{equation}
T^\mathrm{ H}_{+1}= \frac{1}{4\pi r_{+1}}\,,
\label{eqch8:TH+1}
\end{equation}
where the equality sign is valid
within the approximation taken.
From Eq.~\eqref{eqch8:thermoeq}, one
can find the leading order behaviour of the small black hole
solution $r_{+1}$ as 
\begin{equation}
\frac{r_{+1}}{R}=
\frac{1}{4\pi RT\sqrt{1+\frac{R^2}{l^2}}}\,,
\label{eqch8:r+1Tinfinite}
\end{equation}
where the equality sign is valid within the approximation
taken. 
The expression inside the square root of Eqs.~\eqref{eqch8:r+1Tinfinite} is
clearly positive. As a by-product, one can also find
the black hole mass
$2l_p^2 m=r_+
   +\frac{r_+^3}{l^3}$ that in this limit one has
$m_1l_p^2 =\frac{r_{+1}}{2}$.
One could work out in this order, i.e., $T\to\infty$,
the action $I$, the energy $E$, the
entropy $S$, and the heat capacity $C_A$, given
through Eqs.~\eqref{eqch8:actionzero}
to \eqref{eqch8:heatcapacity}.
The most interesting quantity is
the heat capacity $C_A$,
which yields the criterion for thermodynamic stability,
indeed when $C_A<0$ the solution is thermodynamically
unstable, when $C_A\geq0$ the solution is thermodynamically
stable. From 
$C_A=\left(\frac{\partial E}{\partial T}\right)_A$,
one finds from Eq.~\eqref{eqch8:energy} that
$
C_A=\frac{1}{2l_p^2\sqrt{V(R)}}
\left(\frac{\partial r_{+1}}{\partial T}\right)_R$,
which upon using Eq.~\eqref{eqch8:r+1Tinfinite} yields
\begin{equation}
C_{A_{+1}}=-\frac{1}{8\pi l_p^2 T^2\left(1+\frac{R^2}{l^2}\right)^\frac{3}{2}}<0\,,
\end{equation}
so that $C_A$ for the small black hole
$r_{+1}$
is negative. The heat capacity $C_{A_{+1}}$ can also be computed through
Eq.~\eqref{eqch8:heatcapacity} with this limit applied.
The small black hole $r_{+1}$
solution is thus
unstable. Note that actually, the black hole should be surrounded by quantum
fields, with their backreaction on the metric being 
neglected here. However, if
$T^\mathrm{ H}\rightarrow \infty$, the energy density
and other components of the renormalized stress-energy tensor
should diverge. To avoid
this, we restrict $r_{+1}$ in the sense that it has
to be larger than the Planck length scale
$l_p$, i.e., $r_{+1}> l_p$.

The second solution for a very high
heat reservoir temperature,
$T\to\infty$
has $V(R)\rightarrow 0$.
It is clear from Eqs.~\eqref{eqch8:thermequilibrium} 
and \eqref{eqch8:VR}
that the condition $V(R)\rightarrow 0$, implies,
for the whole range
$0\leq\frac{R^2}{l^2}<\infty$, that
$r_{+2}$ should be near the cavity radius, i.e. 
$r_{+2} = R$ minus corrections.
Now, from Eq.~\eqref{eqch8:tbh}, one has in
this limit 
\begin{equation}
T^\mathrm{ H}_{+2}= \frac{1+\frac{R^2}{l^2}}{4\pi R}\,,
\end{equation}
where the equality sign is valid within the approximation
taken. In first order, one can
perform a Taylor expansion, and write
$V(R)=\left(\frac{dV}{dr}\right)\sVert[2]_{r=r_{+2}}
\left(R-r_{+2}\right)$ plus higher order terms.
Since $\left(\frac{dV}{dr}\right)\sVert[2]_{r=r_{+2}}=
4\pi T^\mathrm{ H}_{+2}$, one can write
$V(R)= 4\pi
T^\mathrm{ H}_{+2}\left(R-r_{+2}\right)$.
Using Eq.~\eqref{eqch8:thermequilibrium}, 
or Eq.~\eqref{eqch8:thermoeq},
one has
\begin{equation}
\frac{r_{+2}}{R}= 1-\frac{1+3\frac{R^2}{l^2}}{(4\pi RT)^2}\,,
\label{eqch8:r+2highT}
\end{equation}
where the equality is valid within the approximation taken.
As a by-product, the ADM mass can be found through
$2ml_p^2=r_+ +\frac{r_+^3}{l^3}$, 
which in this limit
becomes
$m_2l_p^2=\frac{R}{2}\left[1+\frac{R^2}{l^2}
-\frac{(1+3\frac{R^2}{l^2})^2}{(4\pi RT)^2}\right]$.
One could work out in this order, i.e., $T\to\infty$,
the action $I$, the energy $E$, the
entropy $S$, and the heat capacity $C_A$, given
through Eqs.~\eqref{eqch8:actionzero} to \eqref{eqch8:heatcapacity}.
Again, the most interesting quantity is 
the heat capacity $C_A$.
For the
heat capacity $C_A$, given by
$C_A=\left(\frac{\partial E}{\partial T}\right)_A$,
one finds from Eq.~\eqref{eqch8:energy} that
$C_A=\frac{1}{l_p^2 \sqrt{V(R)}} 
\left(\frac{\partial m_2}{\partial T}\right)_R$,
where it
was used the expression $V(R)=1-\frac{2m_2}{R} +
\frac{R^2}{l^2}$.  Thus, using the expression for $m_2$
just found above, one has $C_A=\frac{1}{l_p^2 \sqrt{V(R)}} \frac12\frac{1+
3\frac{R^2}{l^2}}
{16\pi^2 T^3 R}$ and since $\sqrt{V(R)}=
\frac{1+3\frac{R^2}{l^2}}{4\pi RT}$ it gives
\begin{equation}
C_{A_{+2}}=\frac{1+3\frac{R^2}{l^2}}{4\pi l_p^2 T^2} >0\,,
\label{eqch8:CAr+2}
\end{equation}
so that $C_{A+2}$ is small and positive.  The large black hole $r_{+2}$
solution is thus stable.

\subsection{Full spectrum of the
Schwarzschild-anti de Sitter
thermodynamic black hole solutions
and diagrams}
\label{sch8:diagramsandanalysis}

\subsubsection{Preliminaries}

We now display the solutions in figures accompanied by a qualitative 
analysis. The figures are important to understand the
thermodynamic solutions
of the Schwarzschild-anti-de Sitter horizons
in a cavity. There are two different figures. 
The first figure contains the curves $\frac{r_+}{R}$, 
that are solution of the thermodynamic equilibrium,
for a fixed value of $4\pi RT$,
as a function of $\frac{R}{l}$,
see Fig.~\ref{figch8:figureyofx}.   The second and third figures
contain the curves $\frac{r_+}{R}$, for a
fixed value of $\frac{R^2}{l^2}$, 
as a function of $4\pi RT$, see
Fig.~\ref{figch8:yofw}, and as a 
function of $4\pi l T$, see Fig.~\ref{figch8:yofbarw}. 
We discuss the physical
interpretation and present the mathematical analysis of the solutions
afterwards.

We can use the variable $T$, $RT$, or $l T$ to 
perform the analysis of the solutions. The difference between 
them relates to the different limiting cases one wishes to 
analyze. York maintains $R$ fixed and $T$ fixed independently,
so $RT$ fixed may be good in certain circumstances,
but when one fixes a parameter independently
of the other, it may be better to use $l T$. Then, $T$ fixed
is the same as $l T$ fixed.
In terms of $RT$, which is a good choice for
$R<\infty$, and for $R\to\infty$ with $T\to0$, one has from
Eq.~\eqref{eqch8:thermoeq} that 
\begin{equation}
4\pi RT=
\frac{1}{\frac{r_+}{R}}
\frac{1+3\frac{r_+^2}{R^2}\frac{R^2}{l^2}}
{
\sqrt{1-\frac{r_+}{R}}\,
\sqrt{1+\frac{R^2}{l^2}
\left(
1+
\frac{r_+}{R} +
\frac{r_+^2}{R^2}
\right)
}
}\,.
\label{eqch8:tbh2}
\end{equation}
In terms of $l T$, which is a good choice for
$T$ fixed and to consider the
limit $R\to\infty$,
one has from Eq.~\eqref{eqch8:thermoeq} that
\begin{equation}
4\pi l T=
\frac{1}{\frac{r_+}{R}\frac{R}{l}}
\frac{1+3\frac{r_+^2}{R^2}\frac{R^2}{l^2}}
{
\sqrt{1-\frac{r_+}{R}}\,
\sqrt{1+\frac{R^2}{l^2}
\left(
1+
\frac{r_+}{R} +
\frac{r_+^2}{R^2}
\right)
}
}\,.
\label{eqch8:tbh5}
\end{equation}

We can now analyze for each region of parameters the solutions
of thermodynamic equilibrium.

\subsubsection{Solutions and behaviour display for the Schwarzschild-anti-de
Sitter thermodynamic black hole solutions with $RT$ fixed}

\begin{figure}[h]
\centering
\includegraphics[width=.7\columnwidth]{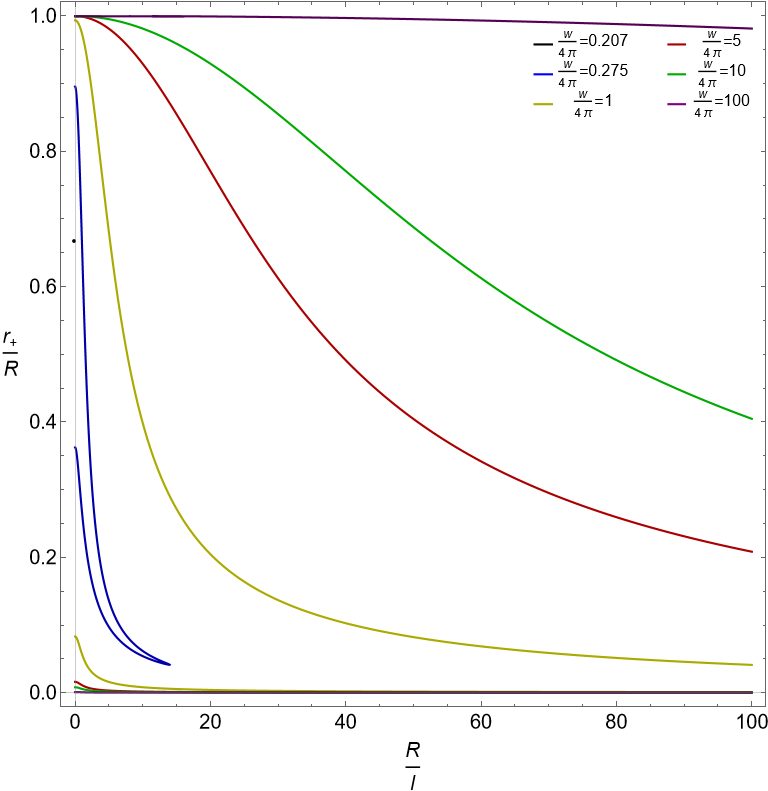}
\caption{
Plots of $\frac{r_+}{R}$ as a function of
$\frac{R}{l}$ for six different values of $4\pi RT$:
$4\pi RT = \frac{\sqrt{27}}{2}=2.60$ with $RT = \frac{\sqrt{27}}{8\pi}=0.207$
as a black dot, $4\pi RT =3.46$ with $RT = 0.275$ as a blue curve, 
$4\pi RT = 12.57$ with $RT=1$ as a yellow curve, $4\pi RT = 62.83$
with $RT= 10$ as a red curve, $4\pi RT = 125.7$ with $RT= 10$ as a 
green curve, and $4\pi RT =1250$ with $RT = 100$ as a purple curve. 
}
\label{figch8:figureyofx}
\end{figure}

The first figure is shown in Fig.~\ref{figch8:figureyofx}, describing
how the black hole horizon
radii $\frac{r_+}{R}$ behave in relation to
$\frac{R}{l}$ for each $4\pi RT$.

The case $4\pi RT=\frac{\sqrt{27}}{2}=2.598$, i.e.,
$RT=\frac{\sqrt{27}}{8\pi}=0.207$, the equalities
in decimal notation being
approximate,
is the first solution displayed in Fig.~\ref{figch8:figureyofx}
by a black dot. This solution corresponds to the one with zero
cosmological constant, $\frac{R}{l}=0$, i.e.,
$\sqrt{\Lambda} R=0$, which is the pure Schwarzschild
solution found first by
York. 
This solution is a coincident horizon solution with
$\frac{r_{+1}}{R}=\frac{r_{+2}}{R}=\frac23=0.667$, the equality
in decimal notation being
approximate. For other larger $\frac{R}{l}$
there are no solutions.

The case $4\pi RT=3.456$,
the equality
in decimal notation being
approximate,
i.e.,
$RT=0.275$, is displayed by a blue curve in 
Fig.~\ref{figch8:figureyofx}.
There are two solutions in this case, and when
$\frac{R}{l}=14$, approximately, then
$\frac{r_{+1}}{R}=\frac{r_{+2}}{R}=0.04$, approximately.
For other larger $\frac{R}{l}$, there are no solutions.

The case $4\pi RT=12.57$,
the equality
in decimal notation being
approximate
i.e.,
$RT=1$, is displayed by a yellow curve in Fig.~\ref{figch8:figureyofx}.
There are two solutions that exist through every $\frac{R}{l}$, 
that only meet at $\frac{R}{l}\rightarrow +\infty$.


The same behaviour of the case $4\pi RT=12.57$ 
occurs to the cases $4\pi RT=62.83$ with $RT=5$, 
$4\pi RT = 125.66$ with $RT=10$ and 
$4\pi RT = 1256.6$ with $RT=100$. They are 
respectively displayed by a curve in red, green and 
purple in Fig.~\ref{figch8:figureyofx}. Again there are 
two solutions, one that starts near $\frac{r_+}{R}=1$ and 
decreases towards zero for increasing $\frac{R}{l}$, 
and another that starts near $\frac{r_+}{R}=0$ and decreases 
towards zero. The solutions never meet.   

%

%

%

\subsubsection{Solutions and behaviour display for the Schwarzschild-anti-de Sitter
thermodynamic  black hole solutions with $\frac{R}{l}$
fixed}

\begin{figure}[h]
\centering
  \includegraphics[width=.7\columnwidth]{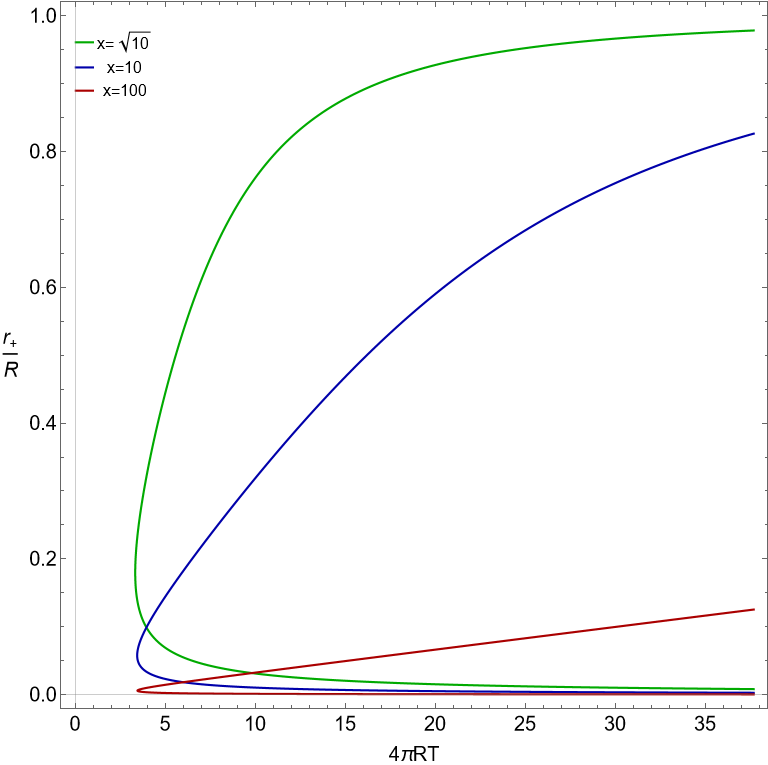}
\caption{
Plots of $\frac{r_+}{R}$ as a function of $4\pi RT$ for three different
values of $\frac{R}{l}$: $\frac{R}{l} = \sqrt{10}$ as the 
green curve, $\frac{R}{l} = 10$ as the blue curve and 
$\frac{R}{l} = 100$ as the red curve.
}
\label{figch8:yofw}
\end{figure}

\begin{figure}[h]
    \centering
      \includegraphics[width=.7\columnwidth]{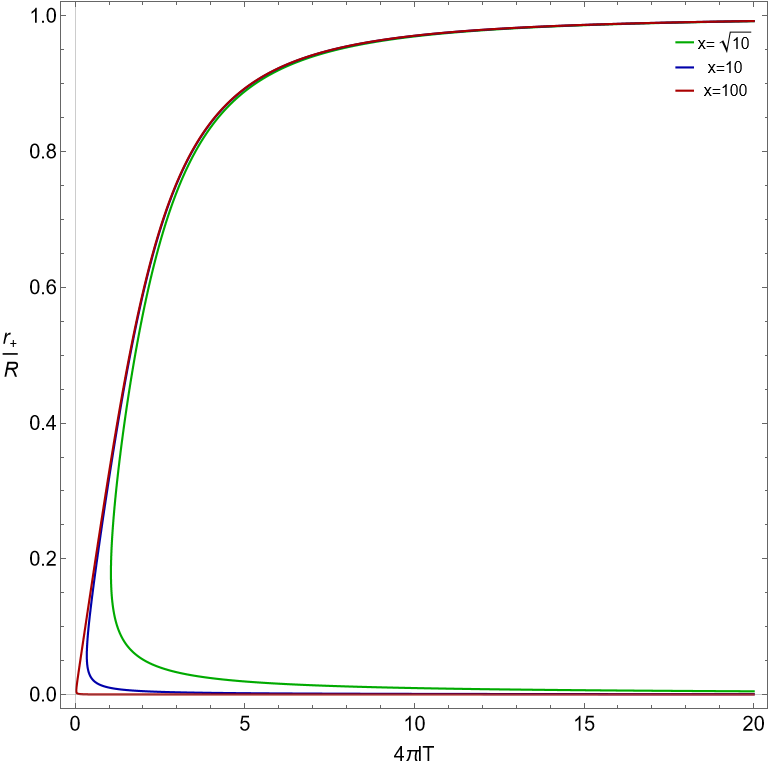}
    \caption{
    Plots of $\frac{r_+}{R}$ as a function of $4\pi l T$ for three different
    values of $\frac{R}{l}$:  $\frac{R}{l} = \sqrt{10}$ as the 
    green curve, $\frac{R}{l} = 10$ as the blue curve and 
    $\frac{R}{l} = 100$ as the red curve.
    }
    \label{figch8:yofbarw}
    \end{figure}

We display a snapshot for each $\frac{R}{l}$ of how the black hole horizon
radii $\frac{r_{+}}{R}$ behave in relation to $T$, in 
Figs.~\ref{figch8:yofw} and~\ref{figch8:yofbarw}.
Specifically, Fig.~\ref{figch8:yofw} shows the behaviour in $4\pi RT$,
while Fig.~\ref{figch8:yofbarw} 
shows the behaviour in $4\pi l T$.

With respect to the curves $\frac{r_+}{R}$ in function of 
$4\pi RT$, the cases 
$\frac{R}{l} = \sqrt{10}$, $\frac{R}{l} = 10$ and 
$\frac{R}{l}=100$ are displayed as green, blue and red 
curves, respectively in Fig.~\ref{figch8:yofw}. For the 
three cases, there are no black hole solutions for 
$4\pi RT < 3.339$ for the green curve, $4\pi RT < 3.448$
for the blue curve and $4\pi RT < 3.463$ for the 
red curve, where the numerics are approximate. 
For larger temperatures, there are always two solutions. 
The solutions start from a bifurcating point where both solutions 
coincide and the small black hole decreases towards zero 
while the large black hole increases towards $\frac{r_+}{R}=1$, 
for increasing temperature. It is important to note the 
similarities in behaviour of the solutions with the York's case, i.e. 
$\frac{R}{l}=0$ or zero cosmological constant.

With respect to the curves $\frac{r_+}{R}$ in function of 
$4\pi lT$, the cases 
$\frac{R}{l} = \sqrt{10}$, $\frac{R}{l} = 10$ and 
$\frac{R}{l}=100$ are displayed as green, blue and red 
curves, respectively in Fig.~\ref{figch8:yofbarw}. For the 
three cases, there are no black hole solutions for 
$4\pi lT < 1.056$ for the green curve, $4\pi lT < 0.344$
for the blue curve and $4\pi RT < 0.034$ for the 
red curve, where the numerics are approximate. 
For larger temperatures, there are always two solutions. 
The solutions start from a bifurcating point where both solutions 
coincide and the small black hole decreases towards zero 
while the large black hole increases towards $\frac{r_+}{R}=1$, 
for increasing temperature. While these cases can be described as 
a rescale of the previous cases, the plot in function of 
$4\pi l T$ shows information about the limit of large 
cosmological constant, which we shall explore further below.

\subsection{Physical analysis of the solutions}

We now make additional qualitative comments to
the plots that have been displayed
in Figs.~\ref{figch8:figureyofx}-\ref{figch8:yofbarw}.

One striking feature, that can be deduced from the plots, is that the
space of black hole horizon radius solutions is enlarged as the
reservoir temperature $T$, or rather $4\pi RT$, is increased. In fact, for
very low temperatures there are no solutions for any $\Lambda$, or
rather, for
any $\Lambda R^2$.  At the temperature $4\pi RT=\frac{\sqrt{27}}{2}=2.598$
there is
only one solution, the pure Schwarzschild solution
with zero cosmological constant, and it is the coincident solution.
For higher $4\pi RT$ there are two
solutions, one large, one small,
up to a value of $\Lambda R^2$. This value grows rapidly
with increasing temperature. Also, with growing temperature,
the large and small black holes tend to radius 1 and radius 0, for small 
$\Lambda R^2$

With the help of
Figs.~\ref{figch8:figureyofx}-\ref{figch8:yofbarw}, we can
give a qualitative explanation for the reason of why
black hole solutions with some nonzero
cosmological constant appear only for
ever higher temperatures $RT$.
The temperature of the reservoir defines a
thermal length scale $\lambda=\frac1T$
for the system. There is also
another length scale, the reservoir
radius $R$, and the cosmological length
$l=\frac3{\sqrt{|\Lambda|}}$.
Thus, we can start from
$|\Lambda|=0$, so that the cosmological length scale
$l=\frac3{\sqrt{|\Lambda|}}$ is infinite, 
$l=\infty$.
In this case there is no coupling of
this length scale with the other two,
$\lambda=\frac1T$ and $R$.
In this situation, we see that
for low $T$, or high $\lambda$, one has $\lambda\gg R$. Since
the thermal wavelength is very large compared to the
reservoir radius $R$, then this wavelength
is stuck to the reservoir
and the corresponding energy
cannot collapse to form a black hole
in any circumstances.
When $T$ is sufficiently increased, i.e., 
$RT=\frac{R}{\lambda}$ is larger than approximately $0.2$, 
the wavelength $\lambda$ is
sufficiently small,
and the corresponding thermal energy 
can travel freely inside the reservoir
and can collapse,
so that formation of black holes is possible.  The value
$RT=\frac{R}{\lambda}=\frac{\sqrt{27}}{8\pi}=0.206$, with the 
last equality approximate,
divides no black hole from two black hole
solutions. The existence of two black hole
solutions for a given temperature $T$, i.e., a
given thermal wavelength $\lambda$ can also be explained.
The small black hole form with an $r_+$ of the order of
$\lambda$, and is unstable as the energy packets with 
length $\lambda$ that escape
from the black hole cannot be scattered back in enough time
to maintain $r_+$ stable.
The large black hole forms with an $R-r_+$
of the order of $\lambda$ so the black hole
and the reservoir exchange energy in a stable manner,
as  the energy packets with length $\lambda$ that escape 
from the black hole are scattered back in enough time
to maintain $r_+$ stable.
Now, we do the analysis 
for the case of finite cosmological constant, i.e. $l$
finite.  For low enough $l$ but slightly larger than $R$, the space
inside the reservoir shrinks, due to the
negative cosmological constant, and so in some way this inner space
has less proper length along the radius. Although the reservoir
radius related to its area is still $R$, the radial length related to
the volume is small, and so the volume is also small.  This means that
energy packets with same $\lambda$, same temperature, relatively to
the cases with infinite $l$, cannot yet travel freely inside the cavity
and cannot form black holes.
As the temperature increases, $\lambda$ decreases and one can have two
black holes down to some finite $l$ which gives the two coincident
solutions. The value of $r_+$ for this finite $l$ case decreases to
smaller values as $\frac{R}{l}$ increases.  This can be understood
as well.  As the temperature increases more, $\lambda$ decreases more,
and one can have a higher $\sqrt{\Lambda} R=3\frac{R}{l}$ for the
coincident solution. But higher $\sqrt{\Lambda} R$ means the space 
is further shrank and so the coincident solution has a value
$r_{+1}=r_{+2}$ small. There is however a certain finite temperature  
at which the coincident solution only occurs for infinite 
cosmological constant, and the same happens for larger temperatures.
We still haven't understood why this occurs. 

This physical interpretation holds either for fixing $RT$ or fixing
$\frac{R}{l}$ as the existence or not of black hole solution is an
interplay between $R$, $T$, and $l$, as we described.

\subsection{Mathematical analysis of the solutions}

\subsubsection{Nomenclature}

We now obtain through a mathematical
analysis some important features displayed in the
plots above, Figs.~\ref{figch8:figureyofx},~\ref{figch8:yofw} 
and~\ref{figch8:yofbarw}. The important equation to analyze here 
is Eq.~\eqref{eqch8:tbh2}. The
natural variables without units are
$\frac{R}{l}$ and
$\frac{r_+}{R}$.
To shorten the notation, we define the
variables $x$ and $y$ as
\begin{equation}
x\equiv\frac{R}{l}\,,
\label{eqch8:xdef}
\end{equation}
\begin{equation}
y\equiv\frac{r_+}{R}\,,
\label{eqch8:ydef}
\end{equation}
with the range of the variables being $0\leq x<\infty$, and
$0\leq y\leq 1$. Furthermore, 
the variable $w$ is additionally defined as 
\begin{equation}
w\equiv4\pi RT\,.
\label{eqch8:zdef}
\end{equation}
Then, with these definitions, Eq.~\eqref{eqch8:tbh2} becomes
\begin{equation}
w=
\frac{1+3x^2y^2}{y\sqrt{1-y}
\sqrt{1+x^2(1+y+y^2)}}\,.
\label{eqch8:zofxy}
\end{equation}
There are solutions for
$w_0\leq w<\infty$, where for convenience,
$w_0\equiv\frac{\sqrt{27}}{2}= 2.598$ is defined, with equality being 
approximate.
Now, for a fixed temperature $T$, or fixed $w$, one
has $dw=0$, and so
$\frac{dy}{dx}=-\frac{\frac{\partial w}{\partial x}}
{\frac{\partial w}{\partial y}}$.
After some calculations, one can obtain
that $\frac{\partial y}{\partial x}$ at constant $w$ is
\begin{equation}
\frac{\partial y}{\partial x}=\frac{2xy(1-y)Q(x,y)}{3R(x,y)}\,,
\label{eqch8:dydx}
\end{equation}
where
\begin{equation}
Q(x,y)\equiv(1+y+y^2)(1-3x^2y^2)-6y^2\,,
\label{eqch8:Qy}
\end{equation}
and
\begin{equation}
R(x,y) \equiv  
- 2( 1  -
3 x^2y^2)( 1 - y)[
1 + 
x^2( 1  + y
+ y^2 )]+
( 1+
3 x^2
y^2)^2y.
\label{eqch8:Ry}
\end{equation}
In addition, we need in the analysis
$\frac{\partial w}{\partial y}$ at constant $x$.
One can obtain from Eq.~\eqref{eqch8:zofxy} that
\begin{equation}
\frac{\partial w}{\partial y}=
\frac{R(x,y)}{2y^2(1-y)^{3/2}
\Bigl[1+x^2(1+y+y^2)\Bigr]^{3/2}}\,.
\label{eqch8:wy}
\end{equation}

We must recall that for each
$\frac{R}{l}$
there are two solutions, $r_{+1}$,
the small solution,
and, $r_{+2}$, the large solution,
which change as $RT$ is changed,
i.e., for each $x$, there are $y_1$
and $y_2$, which change as $w$ is changed.
To summarize, the ranges of $x$, $y$, $w$
for Eq.~\eqref{eqch8:zofxy}
can be written explicitly, 
\begin{equation}
0\leq x<\infty,\,\quad\quad
0\leq y\leq 1,\quad\quad
w_0\leq w<\infty\,,
\label{eqch8:w0old}
\end{equation}
with $w_0\equiv\frac{\sqrt{27}}{2} = 2.598$.

\subsubsection{Analysis}

With $w$ fixed, we first analyze the coincident solutions $y_c=\frac{r_{+c}}{R}$, 
where $y_c = y_1=y_2$, which are crucial for the 
analysis. We then analyze the solutions $y_1$ and $y_2$.

\begin{figure}
    \centering
    \subfloat[\label{figch8:lll}]{
        \includegraphics[width=0.4\columnwidth]{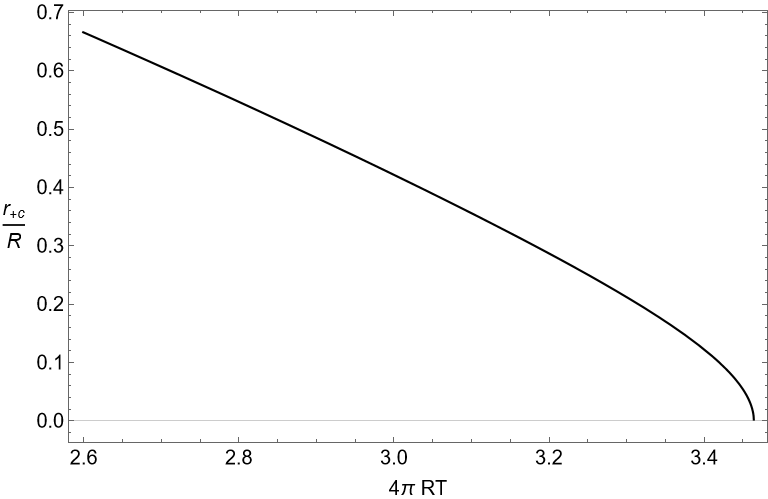}}
    \subfloat[\label{figch8:llll}]{
        \includegraphics[width=0.4\columnwidth]{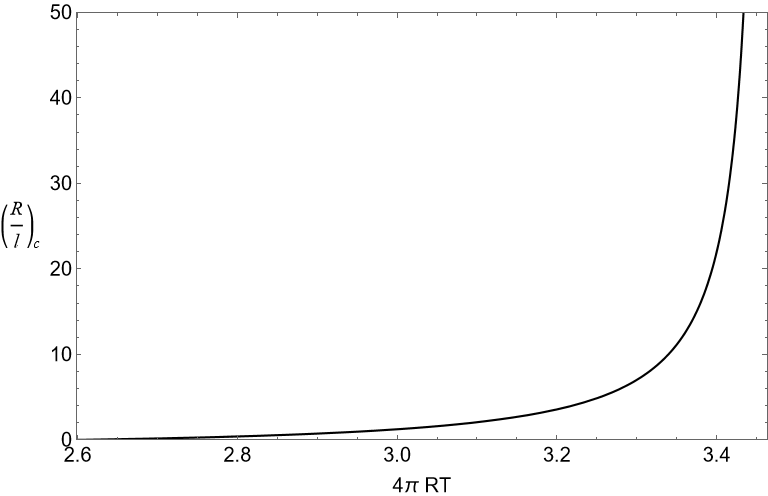}}
    \caption{\label{figch8:coincident} Plot of the 
    coincident solution $y_c$ in (a) and $x_c$ in (b)  
    in function of $w$. 
    For $y_c = \frac{2}{3}$, $x_c=0$ and $x_c$ is then increased towards 
    infinity, yielding $y_c = 0$ at $w=w_1 = 2\sqrt{3}$.}
\end{figure}

With $w$ fixed, we can look at the point $x=x_c$ where
$y_1=y_2=y_c$. Unfortunately, we have not found a closed analytic solution. 
Nevertheless, we are able to obtain certain features.
The point $x=x_c$ occurs when
$\frac{\partial y}{\partial x}=\infty$,
see Eq.~\eqref{eqch8:dydx}, i.e, ${R}=0$. So, the
coincident solution satisfies the equation
$2(1-3 x^2y^2)(1-y)[1+
x^2(1+y+y^2)]-(1+3x^2y^2)^2y=0$ together 
with Eq.~\eqref{eqch8:zofxy}. Since both equations 
are quadratic in $x$, one can subtract one by the other 
and obtain a formula for $x_c(y_c,w)$. One can further 
insert this relation into Eq.~\eqref{eqch8:zofxy} 
and obtain the equation for $y_c(w)$ 
\begin{align}\label{eqch8:eqyc}
    w^4 y^6_c + (2w^2 - 8) w^2 y_c^3 -12 w^2 y^2_c 
    + 48 - 4 w^2 = 0\,\,, 
\end{align}
which is a sixth order polynomial equation and cannot be solved numerically.
The expression for $x_c$ however can be simplified further 
using Eq.~\eqref{eqch8:eqyc}, yielding
\begin{align}\label{eqch8:eqxc}
    x_c^2 = \frac{w^2 (12 - 16 y_c - w^2 y_c^4)}{12(12 - w^2)}\,\,.
\end{align}
We note that there is an important value of 
$w$, which we denote here by $w_1 = 2\sqrt{3} = 3.4641$ approximately. 
The pole in Eq.~\eqref{eqch8:eqxc} shows that the coincident solution 
$y_c$ at $x_c$ does not exist for $w > w_1$, since the numerator 
is positive for the solution $y_c$. In Fig.~\ref{figch8:coincident},
we present the plot of the coincident solution $y_c(w)$, 
where $x=0$ when $y_c = \frac{2}{3}$ and $x$ tends to infinity 
when $y_c$ tends to zero. As seen analytically, this last case happens 
when $w=w_1$. 

To summarize, between
$w_0<w<w_1$, there is a $y_c$
in the range $0\leq y \leq \frac23$
and a $x_c$
where both solutions $y_1$ and $y_2$ coincide.
In the limit $w=w_1$, the coincident solution 
becomes $y_c=0$ with $x_c$ going to infinity. 
For the range $w> w_1$, there is no coincident solution. 

Considering the first solution $y_1$ from Eq.~\eqref{eqch8:zofxy},
we find that $y_1$ obeys an equation of the type 
\begin{align}
y_1\sqrt{1-y_1}\sqrt{1+x^2(1+y_1+y_1^2)}
w=1+3x^2y_1^2\,.
\label{eqch8:y1}
\end{align}
for a given fixed $w$
in the range
$w_0\leq w\leq\infty$, $w_0=\frac{\sqrt{27}}{2}$. 
An important
property are the points where $\frac{\partial y_1}{\partial x}=0$.
From Eq.~\eqref{eqch8:dydx}, this happens
when $x=0$ or $Q(x,y_1)=0$.
The point $x=0$ can be 
point of a minimum, a saddle, or a maximum
of $y_1$ depending on the value of $w$. 
The point given by the root $Q(x,y_1)=0$
corresponds to a minimum of $y_1$
when it exists. The condition $Q(x,y_1)=0$ can be reduced 
from Eq.~\eqref{eqch8:Qy} to the condition
$x^2=\frac{1+y_1-5y_1^2}{3 y_1^2(1+y_1+y_1^2)}$,
where $x=x(y_1,w)$ according to Eq.~\eqref{eqch8:zofxy}.
This latter equation has solutions for $x\geq0$.
When $x=0$, one can find
that the solution is $y_1=\frac{1+\sqrt{21}}{10}$, which putting back
into Eq.~\eqref{eqch8:zofxy}, yields $w=\frac{10\sqrt{10}}
{(1+\sqrt{21})\sqrt{9-\sqrt{21}}}\equiv w_*$, where $w_*$ was defined and
it has the value $w_*=2.695$ approximately.
For this temperature, there is thus a minimum of $y_1$
provided by $Q(x,y_1)=0$.
For $w> w_1$, there are no solutions of  $Q(x,y_1)=0$. This can be 
seen by understanding that there is no coincident solution 
for this range and the solution $y_1$ always decreases towards 
zero. For $w_*<w<w_1$, there are solutions
of $Q(x,y_1)=0$ for points $x>0$, with the minimum of 
$y_1$ decreasing for larger $w$ and $x$ also increasing with 
larger $w$.
On the other hand for $w<w_*$, there are also no solutions
of $Q(x,y_1)=0$.

Now, we are able to describe the solution $y_1$ in function of $x$ with 
a fixed $w$.
For $w_0<w<w_*$, the solution at $x=0$
gives a zero derivative of $y_1$.
We then conclude that for 
$w_0<w<w_*$, the solution $y_1$ at $x=0$ is a minimum of $y_1$, i.e.,
$y_1$ starts from some value and then increases towards $y_c$.
For $w=w_*$, the solution $y_1$ at $x=0$ is a saddle point of $y_1$ but 
still it is the lowest value of $y_1$, i.e. $y_1$ starts from $x=0$ at some 
value and increases towards $y_c$.
For $w_*<w<w_1$, the solution $y_1$ at $x=0$ is a maximum of $y_1$, 
and $y_1$ then decreases towards the minimum given by 
$Q(x,y_1)=0$, and afterwards increases towards the coincident solution 
$y_c$. For $w\geq w_1$, the solution $y_1$ at $x=0$ continues 
to be a maximum of $y_1$ 
and the solution decreases and tends to zero for larger $x$.


We now analyze the case of $y_2$, the larger solution,
in function of $x$ with fixed $w$. From Eq.~\eqref{eqch8:zofxy}, one has
\begin{align}
y_2\sqrt{1-y_2}\sqrt{1+x^2(1+y_2+y_2^2)}\,&w=1+3x^2y_2^2\,,
\label{eqch8:y2new}
\end{align}
for a given fixed $w$
in the range
$w_0\leq w\leq\infty$, $w_0=\frac{\sqrt{27}}{2}$. 
An interesting property is the existence of the 
point $\frac{dy_2}{dx}=0$. 
For the solution $y_2$, the zero derivative
occurs when $x=0$, which corresponds to the only maximum of $y_2$.
One has that 
$y_2$ starts
at $x=0$ at a maximum, and then decreases for all
$x$ increasing. For $w_0 < w < w_1$, the solution $y_2$ 
decreases until it reaches the coincident solution $y_c$ at 
some $x=x_c$. For $w_1<w$, the solution $y_2$ decreases 
and tends to zero as $x$ tends to infinity.
As $w$ increases from $w_0$, the maximum of $y_2$,
which is at $x=0$, increases. When $w\to\infty$ this
maximum $y_2\to1$ for all $x$.

The solutions $y_1$ and $y_2$ in function of $w$ with fixed 
$x$ always follow the same pattern. For a certain 
value of $w$, the coincident solution $y_1=y_2=y_c$ appears 
and for increasing $w$, the solution $y_1$ decreases and tends 
to zero while $y_2$ increases and tends to $y=1$. This happens 
for any value of $0<x<\infty$.




\section{The planar AdS black hole and the Hawking-Page black hole
solutions: Taking the boundary to infinity, $R\to \infty$\label{sech8:limitR}}
\sectionmark{The planar AdS and the Hawking-Page black hole
solutions}\thispagestyle{userightbotmark}

\subsection{Preliminary analysis}

We now consider the analysis of the solutions 
when the boundary goes to infinity $R\to\infty$.
Due to the scaling property of the equations, we chose the temperature 
parameter as $RT$, but such parameter is 
not convenient for the limit $R\to\infty$ with $T$ fixed
for any $T\geq0$. It turns out that we have to separate
the cases $T>0$ fixed, and $T=0$ fixed in a correct manner. 
When $T>0$, the limit to $R\to\infty$ is direct,  
the two black hole solutions
$r_{+1}$ and $r_{+2}$ in this limit have a certain
characteristic behavior. 
Indeed, the small unstable solution $r_{+1}$
behaves as
$r_{+1}= \frac{l}{4 \pi RT}$, and so
goes to zero $r_{+1}=0$, therefore  
$\frac{r_{+1}}{R}=0$.
The large solution $r_{+2}$ behaves as 
$r_+ =c R$ for some $c>0$,
so $\frac{r_{+2}}{R}=c$ and this is the stable solution.
The small unstable solution 
then tends to hot anti-de Sitter, while the large stable solution 
tends to a planar black hole,
as we will see.
When $T=0$, the limit to $R\to\infty$ has to be taken
with care. Indeed, the $R\to\infty$ in this case is such
that $RT$ should be finite, and so $T\to0$ must be done in
a definite manner. One can therefore parametrize 
$T=\frac{l}{R}T_*$
for some finite $T_*$.
This case gives the two solutions of Hawking-Page,
namely the unstable solution $r_{+1}$ and the stable solution $r_{+2}$.

To see what type of geometry we obtain when performing the limit,
we can write the Schwarzschild-AdS line element
given in Eqs.~\eqref{eqch8:lineAdS}
explicitly as 
\begin{align}
ds^2=&\frac{1}{(2\pi T^\mathrm{ H}_+)^2}
\left(1-\frac{r_+}{r}
+
\frac{r^2}{l^2}\left(1-\left(\frac{r_+}{r}\right)^3\right)\right)
d\tau^2
+\nonumber\\
&\frac{dr^2}{
1-\frac{r_+}{r}
+
\frac{r^2}{l^2}\left(1-\left(\frac{r_+}{r}\right)^3\right)}
+r^2d\Omega^2\,, \nonumber\\
& r_+\leq r\leq R\,,
\label{eqch8:schwadsforlimits}
\end{align}
where 
$0\leq \tau<2\pi$, with
$\frac{1}{T^\mathrm{ H}_+}=\frac{4\pi r_+}{1+3\frac{r_+^2}{l^2}}$,
$T^\mathrm{ H}_+=
\frac{1}{4\pi r_+} \left(1+3\frac{r_+^2}{l^2}\right)$.
The space with the metric above 
is in thermal equilibrium at temperature $T$,
the temperature of a reservoir placed at $R$, 
given by the imaginary proper time at the boundary.
In agreement with the Tolman formula, the local temperature 
is then given by the $\tau \tau$ component of the 
metric. The relation between the temperature 
$T$ and the event horizon radius in thermodynamic equilibrium 
is Eq.~\eqref{eqch8:thermequilibrium}, which can be put in the form
\begin{align}
4\pi RT=& \frac{1}{ \frac{r_+}{R}} \frac{
1+3\left(\frac{R}{l}\right)^2 \left(\frac{r_+}{R}\right)^2} {
\sqrt{1-\frac{r_+}{R}}\,
\sqrt{\left(1+\frac{R^2}{l^2}\left(1+
\frac{r_+}{R}+\left(\frac{r_+}{R}\right)^2\right)\right)} }\,,
\nonumber\\
&0\leq T<\infty\,,\quad\quad r_+\leq R<\infty\,,
\label{eqch8:tbhforlimits}
\end{align}
so the reservoir temperature is fixed for each situation
with $T\geq0$. 
With Eqs.~\eqref{eqch8:schwadsforlimits} and \eqref{eqch8:tbhforlimits}
we can now see the solutions that arise when $R\to\infty$
in the case the temperature of the reservoir is nonzero
$T>0$ and in the case the temperature of the reservoir is zero
$T=0$.

\subsection{First limit: The planar AdS black hole solutions.
Taking constant $T$ with $T\geq0$ first, and
performing after the $R\to\infty$ limit}

\subsubsection{The planar AdS black hole solutions}

For a given $R$, there can be up to two black hole solutions, 
if $T$ is greater than a certain value. These are
the small unstable solution $r_{+1}$
and the large solution $r_{+2}$. For $T$ less
than this value, there is no black hole solution, 
but one can choose a different topology sector to obtain 
hot AdS space, i.e. pure AdS space with a temperature inside 
the cavity, which we regard here as a solution.
For $T>0$ and $R\to\infty$, one can
show that the two black hole solutions still exist. 
Moreover, the small solution tends to $r_{+1}=0$ 
and so degenerates to the hot AdS space in a sense, 
while the solution $r_{+2}$ becomes a 
planar black hole in AdS.

Regarding the $r_{+1}$ solution, 
for $R\to\infty$, the leading order expansion yields 
$\frac{r_{+1}}{R}=\frac{1}{4\pi RT}$ and thus
$r_{+1}=\frac{1}{4\pi T}$ since $T$ is a finite number. Therefore,
we conclude that this is the Hawking small black hole
or the Gibbons-Hawking small black hole. The leading order 
expansion gives precisely the relation of the Hawking temperature 
at spatial infinity.

Regarding the case of $r_{+2}$, the large and stable solution,
for $T\geq0$, and $R\to\infty$, Eq.~\eqref{eqch8:tbhforlimits}
yields
$4\pi l T=3\frac{\frac{r_+}{R} }{\sqrt{1-\frac{r_+}{R}}
\sqrt{1+\frac{r_+}{R}+(\frac{r_+}{R})^2}}$ with
$T\geq0$ and $R=\infty$, where it was assumed that 
$r_+=c R$, with $c$ being some number. Note that we can 
do this since this is the behaviour of $r_{+2}$. 
Then, this suggests
the following
coordinate change and definitions
\begin{align}
&{\bar\tau}=\tau\,,\quad
{\bar r}=c\frac{r}{R}l\,,
\nonumber \\
&{\bar r}_+=c\frac{r_+}{R}l\,,\quad\bar R=cl\,,
\quad {\bar l}= \,l\,,
\label{eqch8:changecoordplanar}
\end{align}
with $c$ a real number, $c>0$.
Note that the new reservoir coordinate
radius is $\bar R=cl$.
Then, we find
$d s^2=\frac{1}{(2\pi {\bar T}^\mathrm{ H}_+)^2}\frac{1}{{\bar l}^2}
\left({\bar r}^2-\frac{{\bar r}_+^3}{{\bar r}}\right)
d\bar\tau^2+
\frac{ {\bar l}^2d{\bar r}^2}{
{\bar r}^2-\frac{{\bar r}_+^3}{{\bar r}}
}
+\frac{{\bar r}^2}{{\bar l}^2}(
\frac{R^2}{c^2}
d\Omega^2)$
where $0<{\bar\tau}<2\pi$, and now 
$\frac{1}{ {\bar T}^\mathrm{ H}_+}=
\frac{4\pi {\bar l}^2}
{3{\bar r}_+}$.
We note that the metric is blowing because it is covering the
whole sphere with its radius $R$ tending to infinity. This can be
cured by precisely selecting a very small section of the sphere, which
locally is flat. Choose a precise point with coordinates
$\theta=\theta_0$, $\phi=\phi_0$, and expand the
spherical metric around those points with arbitrarily small
$\Delta\theta$ and $\Delta\phi$
but such that $\bar x =
\frac{R}{c}\Delta\theta$ and $\bar
y = \frac{R}{c}
\sin\theta_0\Delta\phi$. Then the
metric around such patch is
\begin{align}
d s^2=&\frac{1}{(2\pi {\bar T}^\mathrm{ H}_+)^2}\frac{1}{{\bar l}^2}
\left({\bar r}^2-\frac{{\bar r}_+^3}{{\bar r}}\right)
d\bar\tau^2+
\frac{ {\bar l}^2d{\bar r}^2}{
{\bar r}^2-\frac{{\bar r}_+^3}{{\bar r}}
}\nonumber\\
&+\frac{{\bar r}^2}{{\bar l}^2}
(d{\bar x}^2+d{\bar y}^2),\quad\quad {\bar r_+}\leq {\bar r}\leq{\bar R}\,,
\label{eqch8:planarbh}
\end{align}
with 
$0\leq\bar\tau <2\pi$, ${\bar r_+}\leq {\bar R}<\infty$,
$-\infty< \bar x <\infty$, $-\infty< \bar y <\infty$,
which is the planar black hole line element.
The Hawking temperature is 
${\bar T}^\mathrm{ H}_+=\frac{3{\bar r}_+}
{4\pi {\bar l}^2}$.
One can verify
that there is no conical singularity
in the $\bar\tau\times \bar r$ plane 
at $\bar r={\bar r}_+$. The condition is
$\sqrt{
\frac{\partial_{\bar r}g_{\tau\tau}}{g_{{\bar r}{\bar r}}}
}=1$.
If we write the metric as 
$d s^2=\frac{1}{{\bar l}^2}
(\frac{2{\bar l}}{3{\bar r}_+})^2
\left({\bar r}^2-\frac{{\bar r}_+^3}{{\bar r}}\right)
d\bar\tau^2+
\frac{ {\bar l}^2d{\bar r}^2}{
{\bar r}^2-\frac{{\bar r}_+^3}{{\bar r}}
}$, with $0<\bar\tau<2\pi$, we can see
that 
$\sqrt{\frac{\partial_{\bar r} g_{\tau\tau}}{g_{{\bar r}{\bar r}}}}
=
\left(
\frac{1}{3{\bar r}_+}
(2{\bar r}+\frac{{\bar r}_+^3}{{\bar r}^2})
({\bar r}^2-\frac{{\bar r}_+^3}{{\bar r}}
)^{-\frac12}
({\bar r}^2-\frac{{\bar r}_+^3}{{\bar r}}
)^{\hskip -0.02cm \frac12}\right)_{{\bar r}_+}
=1$, as it should.

The temperature $T$ of the reservoir
is the inverse of the Euclidean
time length given by
\begin{align}
T=\frac{3{\bar r}_+}{4\pi{\bar l}
\sqrt{{\bar R}^2-\frac{{\bar r}_+^3}{{\bar R}}} },
\label{eqch8:temperatureplanarbh}
\end{align}
where ${\bar R}$ is the new coordinate radial
position of the reservoir, given by $ {\bar R}=cl$
for some $c$, and is precisely
the equation 
$4\pi l T=\frac{3\,\frac{r_+}{R} }{\sqrt{1-\frac{r_+}{R}}
\sqrt{1+\frac{r_+}{R}+(\frac{r_+}{R})^2
}
}$ with
$T>0$ and $R=\infty$, from which the analysis was started
but with new definitions.
Now, Eq.~\eqref{eqch8:temperatureplanarbh} can be put
in the form
$4\pi{\bar l}T=\frac{3\frac{{\bar r}_+}{\bar R}}{
\sqrt{1-\frac{{\bar r}_+^3}{{\bar R}^3}} }$,
which shows that
$\frac{{\bar r}_+}{\bar R}$ is a function of
${\bar l}T$ alone, i.e.
$\frac{{\bar r}_+}{\bar R}=\frac{{\bar r}_+}{\bar R}
({\bar l}T)$. From
Eq.~\eqref{eqch8:temperatureplanarbh}, when
it exists, there is only one solution for
$\frac{{\bar r}_+}{\bar R}
({\bar l}T)$ as expected from the
limit we took. Moreover, there is
always a solution for any $T$, contrarily
to the spherical case where for $T$ below a certain value
there are not solutions.



\subsubsection{The case of zero cosmological constant $\Lambda=0$,
i.e., $l=\infty$:
The Gibbons-Hawking small black hole and 
Rindler boundary at infinity with flat space inside}

It is interesting to consider 
the case of zero cosmological constant $\Lambda=0$,
i.e., $l=\infty$. The small unstable black hole 
$r_{+1}$ in this limit reduces to the 
Schwarzschild solution given by Gibbons and Hawking
in~\cite{Gibbons:1977}. We have seen that 
the large stable solution $r_{+2}$ 
in the infinite cavity 
limit gives a planar black hole solution. Going further 
with the limit of zero cosmological constant, 
we must proceed with care. 
Indeed, from Eq.~\eqref{eqch8:planarbh}, 
we can deduce that the space
becomes now Rindler space with the boundary at infinity receding
with appropriate temperature $T$.

The limit that allows one to obtain the Rindler 
space from the planar black hole solution can be seen as 
follows. Having
the planar black hole line element, Eq.~\eqref{eqch8:planarbh},
written as
$ds^2 = \left(\frac{2 l^2}{3\bar{r}_+}\right)^2
\frac{1}{l^2 \bar{r}}
  (\bar{r}^3 - \bar{r}_+^3)d\tau^2 + \frac{l^2 \bar{r}
  d\bar{r}^2}{\bar{r}^3 - \bar{r}_+^3} + \frac{\bar{r}^2}{l^2}
  \frac{l^2}{\bar{R}^2}(dx^2 + dy^2)$, one can employ the limit
$\bar{r}_+ \rightarrow \bar{R}$ and $l\rightarrow +\infty$, but such
that $l \sqrt{\bar{R}^3 - \bar{r}_+^3}$ is finite. In order to do
this, one can evaluate the proper radial length as
$\tilde{r}(\bar{r}) = l\int_{r_+}^r \frac{\sqrt{\rho}
d\rho}{\sqrt{\rho^3 - \bar{r}_+^3}}$, which gives
\begin{align}
\tilde{r}(\bar{r})= \frac{2 l}{3
\bar{r}_+^{\frac{3}{2}}}\sqrt{\bar{r}^3 - \bar{r}_+^3}\,,
\label{eqch8:tilder}
\end{align}
which is valid for very small $\bar{r} - \bar{r}_+$. 
The planar black hole metric then becomes
$ds^2 = \frac{\bar{r}_+}{\bar{r}}\tilde{r}^2
d\tau^2 + d\tilde{r}^2 +
\frac{\bar{r}^2}{\bar{R}^2}(dx^2 + dy^2)$.
Consider that from Eq.~\eqref{eqch8:tilder}, one has
$\bar{r}^3 = \left(\frac{3 r_+^{3/2} \tilde{r}}{2 l}\right)^2
+ \bar{r}_+^3$
and also that $\bar{r}_+\rightarrow \bar{R}$, then
$l\rightarrow
+\infty$ implies $\frac{\bar{r}_+}{\bar{r}} = 1 -
\mathcal{O}\left(\frac{1}{l}\right)$
and $\frac{\bar{r}^2}{\bar{R}^2} = 1 -
\mathcal{O}\left(\frac{1}{l}\right)$.
Therefore,
we find that the line element in this limit $l\to\infty$ is
\begin{align}
ds^2 = \tilde{r}^2 d\tau^2 + d\tilde{r}^2 + dx^2 + dy^2\,,
\label{eqch8:rindler}
\end{align}
which is the Rindler line element
in the $\tau\times r$ plane times a flat plane.
Note that the Rindler horizon is at $\tilde{r}=\tilde{r}_+=0$,
which corresponds to the old black hole horizon 
$\bar{r}=\bar{r}_+$, see Eq.~\eqref{eqch8:tilder}. Thus, the situation is
the following after the limit. 
There is a reservoir at $\tilde R$ at temperature
$T$ accelerating away with acceleration $a=\frac{1}{2\pi T}$,
with $T$ corresponding to the Unruh temperature.

\subsection{Second limit: The Hawking-Page spherical black hole
solutions. Taking the $T\to0$ limit, and concomitantly
taking $R\to\infty$, with constant $RT$}

\subsubsection{The Hawking-Page spherical black holes}

When $R\to\infty$ is taken first,
we see from Eq.~\eqref{eqch8:tbhforlimits}
that the Hawking-Page solutions can be recovered by 
performing the limit $T\to0$
such that $RT=\mathrm{ constant}$.
We therefore have $r_+$ finite, although $\frac{r_+}{R}=0$ since
we are taking $R\to\infty$. 

The limit can be seen with more care. From 
$4\pi RT= \frac{1}{ \frac{r_+}{R}} \frac{
1+3\left(\frac{R}{l}\right)^2 \left(\frac{r_+}{R}\right)^2} {
\sqrt{1-\frac{r_+}{R}}\,
\sqrt{\left(1+ \frac{R^2}{l^2}\left(1+
\frac{r_+}{R}+\left(\frac{r_+}{R}\right)^2\right)\right)} }$,
see Eq.~\eqref{eqch8:tbhforlimits},
the limit $R\to\infty$
gives
$4\pi RT= \frac{1+\left(\frac{r_+}{l}\right)^2}{ r_+}l$.
The idea is to define a new conformal temperature such that 
\begin{equation}
T=T_*\frac{l}{R}\,,
\quad\quad T\to0,\;R\to\infty,
\label{eqch8:tlimitads}
\end{equation}
and so the thermodynamic equilibrium equation becomes
\begin{equation}
T_*=\frac{1}{4\pi}\frac{1+
3\left(\frac{r_+}{l}\right)^2}{ r_+}
\,.
\label{eqch8:tbhawkingpage}
\end{equation}
This is the limit $T\to0$ and $R\to\infty$.
Note that $T_*$ is essentially 
$T^\mathrm{ H}_+$ of Eq.~\eqref{eqch8:schwadsforlimits}.
The equation in Eq.~\eqref{eqch8:tbhawkingpage} 
yields the two Hawking-Page $r_+$ solutions. 
One is the solution 
\begin{equation}
\frac{r_{+1}}{l}=
\frac{2\pi}{3} l T-\sqrt{
\left(\frac{2\pi}{3}
l T_*\right)^2-\frac13}
\,,
\label{eqch8:smallhawkingpage}
\end{equation}
which is the small solution and it is unstable.
The other solution is 
\begin{equation}
\frac{r_{+2}}{l}=
\frac{2\pi}{3} l T+\sqrt{
\left(\frac{2\pi}{3}
l T_*\right)^2-\frac13}
\,,
\label{eqch8:largehawkingpage}
\end{equation}
which is the large solution and it is stable.
These two black hole solutions exist 
for temperatures obeying 
$T_*\geq\frac{\sqrt3}{2\pi l}$.
When there is equality 
$T_*=\frac{\sqrt3}{2\pi l}$, the two solutions
merge into one given by 
$\frac{r_{+1}}{l}=\frac{r_{+2}}{l}=
\frac{2\pi}{3} l T$.
When 
$T_*<\frac{\sqrt3}{2\pi l}$, i.e., for
low enough temperatures,
there are no black hole solutions, one is
in the presence
of pure hot AdS space, also called classical
hot space.
Thus, the Hawking-Page solutions inherit from
the $R$ finite solutions the same properties.

\subsubsection{The case of zero cosmological constant $\Lambda=0$,
i.e., $l=\infty$:
The Gibbons-Hawking black hole
and hot flat planar space}

In the limit of infinite cavity, while keeping $RT$ constant, 
we have the two Hawking-Page solutions. It is also interesting 
to proceed with the limit of zero cosmological constant 
in these two solutions. When $l=\infty$, 
the solution $r_{+1}$ becomes the Gibbons-Hawking
unstable Schwarzschild black hole solution, 
while the solution $r_{+2}$ becomes the Rindler solution, 
but now the cavity resides at infinity.

Taking the limit 
 $l\to\infty$ with $T_*$ constant, one gets from Eq.~\eqref{eqch8:smallhawkingpage}
that $r_{+1}=\frac{1}{4\pi T_*}$ which is the 
Gibbons-Hawking black hole solution. At spatial infinity, the 
temperature is $T_*$. The thermal energy of this solution is equal to
its mass $E=m$.
The heat capacity is negative $C=-2\pi
r_+^2$, therefore the solution is unstable.

Also, from Eq.~\eqref{eqch8:largehawkingpage},
we find
$r_{+2}=\frac{4\pi T_*}{3}l^2$, i.e.,
$r_{+2}=\infty$ when $l\to\infty$. 
It may seem that there is no way to understand this 
limit. However, we can look into the line element 
and see the consequences of doing $l\to \infty$. 
From Eqs.~\eqref{eqch8:lineAdS} and~\eqref{eqch8:Vr}, the line 
element is 
\begin{align}
    &ds^2 = \left(\frac{2r_+}{1+3\frac{r_+^2}{l^2}}\right)^2V(r)d\tau^2 
    + \frac{dr^2}{V(r)} + r^2 d\Omega^2\,\,,\notag\\
    &V(r) = 1 + \frac{r^2}{l^2} - \left(1 + \frac{r_+^2}{l^2}\right)\frac{r_+}{r}\,\,.
\end{align}
The idea is now to substitute $r_{+2}=\frac{4\pi T_*}{3}l^2$ and perform 
the coordinate transformation $r = \frac{4\pi T_*}{3}l^2 \bar{x}$, with $\bar{x}\in ]1,+\infty[$. 
Due to the limit $l\to\infty$, the function $V(r)$ has leading 
order terms $V(r)\rightarrow \left(\frac{4}{3}\pi T_*\right)^2 l^2
\left(\bar{x}^2 - \frac{1}{\bar{x}}\right)$. Moreover, the $2$-sphere line element 
$r^2 d\Omega^2$ becomes $\left(\frac{4\pi T_*}{3}\right)^2 \bar{x}^2 l^4 d\Omega^2$. Similar to the case 
of the planar black hole, we can regularize the $2$-sphere line element 
by considering very small angles around a specific point 
$(\theta_0,\phi_0)$ such that we have new coordinates 
$dy= \frac{4\pi T_*}{3} l^2 \bar{x} d\theta$ and $dz =\frac{4\pi T_*}{3} l^2 \bar{x} \sin(\theta_0)d\phi$. 
Hence, the leading 
order line element in the limit $l\to\infty$ becomes 
\begin{align}\label{eqch8:planarlinfinity}
    ds^2 = l^2 \left(\frac{4}{9}\left(\bar{x}^2 - \frac{1}{\bar{x}} \right)d\tau^2 
    + \frac{d\bar{x}^2}{\bar{x}^2 - \frac{1}{\bar{x}}}\right) + dy^2 + dz^2\,\,,
\end{align}
which has the form of the hot planar black hole geometry. Note that the appearance 
of this geometry here is not surprising. The length $l$ can be regarded 
as the radius of a natural cavity in AdS and we already have seen that 
the limit of infinite cavity in AdS gives precisely the hot planar black 
hole. But here, we still 
must perform the limit $l \to \infty$ in Eq.~\eqref{eqch8:planarlinfinity}.
In some sense, performing the limit $l\to \infty$ to the $r_{+2}$ solution of Hawking and 
Page is the same as performing the limit $l\to \infty$ to the hot planar 
black hole, which arises from the limit of infinite cavity. 

From Eq.~\eqref{eqch8:planarlinfinity}, we can see that the limit 
$l\to \infty$ gives an infinite line element without any further 
considerations. That means only that all the points outside the 
neighbourhood of $\bar{x}=1$ are at infinite distance from points 
at $\bar{x}=1$. In order to regularize the metric, we must thus expand in the 
neighbourhood of $\bar{x}=1$. The proper radius length 
$\epsilon = l \int_1^{\bar{x}} \frac{\sqrt{\bar{x}}d\bar{x}}{\sqrt{\bar{x}^3 - 1}}$ is given at 
leading order by $\epsilon = l \frac{2}{\sqrt{3}}\sqrt{\bar{x}-1}$. 
We now must perform the limit $l\to \infty$ with $\sqrt{\bar{x}-1}$ 
being very small, such that $\epsilon$ is finite. Then, 
$ l^2 \bar{x}^3 - \frac{l^2}{\bar{x}} \rightarrow \frac{9}{4}\epsilon^2$. 
The line element of the space in the limit of $l \to \infty$ becomes 
\begin{align}\label{eqch8:rindleragain}
    ds^2 = \epsilon^2 d\tau^2 + d\epsilon^2 + dy^2 + dz^2\,\,,
\end{align}
with $\epsilon \in ]0,+\infty[$. This is again the Rindler metric but now 
the boundary of space is at infinity. Indeed, the inverse temperature 
at the boundary of Rindler is infinite, i.e. the temperature is zero, which 
agrees with the limit $l \to \infty$ in $T_* = \frac{3 r_+}{4\pi l^2}$ while 
keeping $\frac{r_+}{l}$ finite. One can further make 
a coordinate transformation to obtain the Euclidean flat 
space
\begin{align}
&ds^2 = dq^2 + dw^2 + dy^2 + dz^2\,,
\label{eqch8:rindler2flatflat}
\end{align}
where $q = \epsilon \cos(\tau)$ and $w = \epsilon\sin(\tau)$.
One can think of this space as the hot flat planar 
space which has topology $R^2\times R^2=R^4$.
This solution cannot be found from the
original Gibbons-Hawking action because it has
different boundary conditions and the class of spherically 
symmetric metrics chosen does not cover this solution. 
Indeed, the choice of writing the metric in terms of the 
compactified imaginary time means that 
the hot flat planar space can only be achieved by 
finding the Rindler space first. However, the Rindler 
space is not spherically symmetric. This can be seen by 
transforming Eq.~\eqref{eqch8:rindleragain} into 
spherical coordinates, having then 
$ds^2 = r^2 \cos^2(\theta)d\tau^2 + dr^2 + r^2 d\Omega^2_2$.

While the Rindler metric obtained and the hot flat planar space 
are related by a coordinate transformation, we note that the 
physical situation described here is the one of an accelerated 
observer at infinity. This is so because the local temperature is 
defined by the length along orbits of the imaginary proper time, which 
in this case correspond in the physical space to the trajectories of 
constant accelerated observers. In other terms, the temperature 
is measured by constant accelerated observers and the fixed temperature 
of space corresponds to the temperature measured by the 
constant accelerated observer at infinity. And so, the Rindler 
metric describes explicitly the physical situation, although being 
equivalent to hot flat planar space.

A property of these limits in this subsection
is that procedures somehow commute. 
For the case of the small black hole solution, both 
procedures give an endpoint described by the Gibbons-Hawking 
solution.  
Regarding the large solution, 
if one starts from the planar solution and takes zero
cosmological constant, one obtains the
planar Rindler solution with the reservoir accelerated. 
If one instead starts with the large Hawking-Page solution and takes zero
cosmological constant, the solution also becomes the
Rindler solution with some coordinates accelerated.

\subsection{The limits visualized}

The figures displayed in Figs.~\ref{figch8:yofw}
and~\ref{figch8:yofbarw} are helpful to visualize the limits 
described in this section. Indeed, in Fig.~\ref{figch8:yofw}, 
where the solutions are plotted in function of $w=4\pi RT$, 
one can see that for large $\frac{R}{l}$ the distance between 
the solutions starts to narrow. It was checked in fact that 
for larger and larger $\frac{R}{l}$, the solutions tend 
to merge towards zero. These are the two Hawking-Page solutions 
which can be recovered if one rescales $r_+$ by $l$ instead of $R$. 

On the other hand, in Fig.~\ref{figch8:yofbarw}, where the 
solutions are plotted in function of $4\pi l T$, the two 
solutions have different behaviours for large $\frac{R}{l}$. 
The small black hole solution seems to tend towards zero 
while the large black hole solution tends to a smooth curve. 
The smooth curve corresponds to the solution of the planar AdS 
black hole. An interesting point is $l T=0$ in the limit of 
infinite $\frac{R}{l}$. The planar solution at $l T=0$ has 
$r_+=0$, meaning that the location of event horizon's 
plane is pulled towards an infinite proper length, i.e. to infinity. 
But at $l T=0$, there are also the Hawking-Page solutions as the limits 
above show.

\section{Conclusions}
\label{sech8:conc}

In this chapter, we have analyzed the canonical ensemble of a Schwarzschild-AdS 
black hole inside a cavity, with particular 
focus on the limits of the horizon radius solutions that are 
in thermodynamic equilibrium with the cavity.

We have shown that, with $\Lambda R^2\ll1$, York's
solution for pure Schwarzschild is automatically incorporated when
$\Lambda R^2=0$, appearing first for $RT=\frac{\sqrt{27}}{8\pi}$, with
a coincident black hole horizon radius $r_{+1}=r_{+2}=\frac23 R$.  For
higher $\Lambda R^2$, the coincident black hole horizon radius gets
decreased values for some higher $RT$. The value of $RT$ for the coincident 
black hole solution saturates to a particular value 
$RT = \frac{2\sqrt{3}}{4\pi}$ for infinite $\Lambda R^2$ and it has 
zero event horizon radius. We gave a heuristic understanding of
this behavior.  Changing the values of $\Lambda R^2$
and $RT$, we obtain either two thermodynamics solutions, one which is 
a small solution, $r_{+1}$,
and one which is large, $r_{+2}$. 
The solution $r_{+1}$ is thermodynamically unstable, while the
solution $r_{+2}$ is stable.

We have shown that for $|\Lambda| R^2\to\infty$,
unexpected solutions also arise.
There are two different classes of solutions in this limit.
One class is obtained by keeping a constant finite $T$ 
and by performing the limit of $R\to \infty$. 
For this class, the small unstable black
hole solution  $r_{+1}$ disappears, whereas the 
 large stable black
hole solution $r_{+2}$ turns into a planar black hole.
The second class is obtained by making the limit $R\to \infty$ but 
also by putting the temperature to zero, such that $RT$ is constant.
For this class, the two black hole solutions yield the Hawking-Page
spherical solutions in AdS.
The $|\Lambda|=0$ case in this limit was also considered, 
where the small black hole solution $r_{+1}$ becomes the 
Gibbons-Hawking solution and the large black hole solution 
$r_{+2}$ becomes a Rindler space with accelerated boundary.

Our work in this chapter establishes the connection between the existing 
solutions in the literature in a unifying way through 
the limits performed. It would be interesting to expand 
this analysis to a larger family of ensembles with more 
parameters.

\chapter{The canonical ensemble
of a self-gravitating
matter thin shell
in AdS}
\label{ch:thinshellAdS}

\section{Introduction}

With the previous chapters being based on configurations with black holes 
either with a Maxwell field or a negative cosmological constant, we now turn 
our attention towards spacetimes containing self-gravitating matter.

As discussed in the previous chapter, when considering 
asymptotically anti-de Sitter (AdS) spacetimes, it was shown~\cite{Page:1982dh} 
that for Schwarzschild-AdS there would be two black hole solutions, with the 
largest being stable. Hence, asymptotically AdS spacetimes stabilize thermodynamically 
black hole configurations as the negative cosmological constant makes the 
spacetime being described as a box. Moreover, it was found in~\cite{Page:1982dh} 
the existence of a phase transition between the hot thermal AdS, i.e. 
pure AdS containing nonself-gravitating gravitons, and the stable black hole, 
the so called Hawking-Page phase transition. Other ensembles in asymptotically AdS were 
also further studied, see~\cite{Peca:1998cs,Chamblin:1999hg}. 

The application of the formalism to self-gravitating matter is of great interest 
to explore the effects of thermodynamics in curved spacetime and uncover also 
the connection between thermodynamics and gravity. The inclusion of matter shells 
as simple descriptions of matter surrounding a black hole 
has been done in the construction of 
ensembles with curved space~\cite{Martinez:1989hn}, 
where it was shown that the total entropy of 
the system is the sum of matter entropy with the black hole entropy. A more 
thorough analysis was done in~\cite{Lemos:2023yiz}, while keeping the radius of 
the shell fixed.

In this chapter, we consider the canonical ensemble of 
matter shell in asymptotically AdS space, using the Euclidean path integral 
approach. The objective is to analyze the phase transitions between 
a black hole and a self-gravitating configuration which may mimic hot thermal AdS. 
We consider a matter action which is approximated by a fluid description, which is 
motivated from the path integral over the self-gravitating matter fields. We impose the 
zero loop approximation, and analyze the equilibrium and stability conditions describing 
this approximation. 
The thermodynamic quantities of the system are then obtained from the 
partition function. A characteristic of the system is that the 
condition corresponding to the mechanical stability of the shell is not 
accessible by thermodynamics.
We choose a specific equation of state, corresponding 
to a matter gas with mass, or alternatively, to a graviton gas restricted to a 
thin shell. We find four solutions for the shell, with only one being fully 
stable. We analyze the favorability between the thin shell and the black hole solutions 
and we find the phase transition between the two phases, which possesses a behaviour
analogous to the Hawking-Page phase transition.

This chapter is organized as follows. In Sec.~\ref{sech7:pathintegral}, 
we construct the canonical ensemble of matter thin shell in AdS. In Sec.~
\ref{sech7:zeroloop}, we apply the zero loop approximation, and 
we obtain the 
equilibrium and the stability conditions an arbitrary 
equation of state. In Sec.~
\ref{sech7:thermodynamics}, we obtain the thermodynamics of the thin shell in AdS
from the partition function. In Sec.~\ref{sech7:unspecifiedeos}, 
we choose a specific equation of state and we study the solutions of the ensemble. 
In Sec.~\ref{sech7:tsxbh}, we compare the AdS black hole solutions to the matter 
thin shell solutions, and we find a phase transition. In Sec.~\ref{sech7:concl}, 
we present the conclusions. This chapter is based on~\cite{Tiago2024bk}.

\section{Canonical ensemble of a self-gravitating matter thin shell in
asymptotically AdS space
\label{sech7:pathintegral}}\sectionmark{Canonical ensemble 
of a self-gravitating matter thin shell in AdS}\thispagestyle{userightbotmark}

\subsection{The partition function}

The canonical ensemble of a four dimensional 
curved space with negative cosmological constant 
and with matter fields can be 
given by $Z = \int Dg_{\alpha \beta}
D\psi\, \mathrm{ e}^{-I[g_{\mu \nu}, \psi]}$, where $g_{\alpha \beta}$ 
represents the Euclidean metric, $\psi$ describes the matter
fields, and $I$ is the Euclidean action. Due to the difficulties in 
performing the full path integral, we perform here
the zero loop approximation of the path integral, but we do it
in steps. We assume that the path integral over
the matter fields can be put inside the path integral over metrics in the sense
of $Z = \int Dg_{\alpha \beta}\, \mathrm{ e}^{-I_{g l}}
\int D\psi \,\mathrm{ e}^{-I_\psi}$, where $I_{gl}=I_{gl}[g_{\mu \nu}]$
is the Euclidean gravitational action with negative cosmological constant
and $I_\psi=I_\psi[g_{\mu \nu},\psi]$ is the
Euclidean matter action of any field $\psi$. We assume minimal
coupling between the field $\psi$ and the metric $g_{\alpha \beta}$.
While for the general case one cannot perform the path integral on
matter, for the case of a matter thin shell in spherical symmetry one
can perform the path integral exactly, if the action is quadratic in
the field. This is because the metric components are seen as constants in
the action of the matter thin shell and the path integral becomes an 
integration over gaussian functions, yielding
$\int D\psi
\,\mathrm{ e}^{-I_\psi[g_{\mu \nu},\psi]}=
\mathrm{ e}^{-I_\mathrm{ m}[g_{\mu \nu}]}$,
where $I_\mathrm{m}$ is an effective matter action. Therefore, the 
partition function considered here is
\begin{align}
    Z = \int Dg_{\alpha \beta}\, \mathrm{e}^{-I_{gl} - I_\mathrm{m}}\,,
\label{eqch7:partitionfunctiongeneric}
\end{align}
Since a matter thin 
shell, denoted by $\mathcal{C}$, is considered, the asymptotically AdS space $M$ 
is split into two spaces $M_1$ and $M_2$. The outer boundary of $M$ 
is represented as $\partial M$. The gravitational action is then given 
by 
\begin{align}\label{eqch7:gravaction}
    I_{gl} =& - \frac{1}{16\pi l_\mathrm{ p}^2}\int_{M\setminus\{\mathcal{C}\}} 
   \left(R + \frac{6}{l^2}\right)\sqrt{g} d^4x 
   + \int_{\mathcal{C}}  \frac{[K]}{8\pi l_\mathrm{ p}^2} 
    \sqrt{\gamma}d^3y 
   \notag \\
   &- \frac{1}{8\pi l_\mathrm{ p}^2}\int_{\partial M} K \sqrt{\gamma}d^3y 
   - I_{\mathrm{AdS}}
   \,\,,
\end{align}
where
$l_\mathrm{p}$ is the Planck length, 
$R$ is the Ricci scalar, $g$ is the metric determinant, 
$l = \sqrt{-\frac{3}{\Lambda}}$ is defined as the AdS length,
with $\Lambda$ 
being the negative
cosmological constant,
 $\gamma_{ab}$ is the induced 
metric from the space metric $g_{\alpha \beta}$
on the hypersurface in analysis,
$\gamma$ is the
determinant of $\gamma_{ab}$,
$K_{ab}$ is the extrinsic curvature 
of the hypersurface in analysis, with 
trace $K$ given by $K= {n^\alpha}_{;\alpha}$,
$n^\alpha$ being the normal vector to the  hypersurface in analysis,
i.e., either $\mathcal{C}$ or $\partial M$,
the bracket $[K] = \eval{K}_{M_2}- \eval{K}_{M_1}$ 
means the difference  between 
$K$ evaluated at $M_2$ and $K$ evaluated at $M_1$,
and $I_{\mathrm{AdS}}$ is the action of pure AdS 
which is the reference space with negative cosmological constant.
In relation to the matter part of the
Euclidean action,
the Lagrangian density is taken as
the matter free energy per
unit area $\mathcal{F}_\mathrm{ m}$.
This stems from the fact 
the canonical ensemble is being considered
which is connected to the thermodynamic Helmholtz free energy.
Then, 
\begin{align}\label{eqch7:mataction}
    I_\mathrm{ m}= \int_{\mathcal{C}} \mathcal{F}_\mathrm{ m}[\gamma_{ab}] \sqrt{\gamma}d^3x\,\,,
\end{align}
where $\mathcal{F}_\mathrm{ m}$
is a functional of the induced metric $\gamma_{ab}$
on the shell, with $\gamma$ being the 
determinant of $h_{ab}$. 
The Euclidean action of the system
$I=I_{gl}+I_\mathrm{ m}$ is then given 
by  
\begin{align}
   I = &- \frac{1}{16\pi l_\mathrm{ p}^2}\int_{M\setminus\{\mathcal{C}\}} 
   \left(R + \frac{6}{l^2}\right)\sqrt{g} d^4x 
   + \int_{\mathcal{C}} \left( \frac{[K]}{8\pi l_\mathrm{ p}^2} 
   + \mathcal{F}_\mathrm{ m}[\gamma_{ab}]\right)\sqrt{\gamma}d^3y \notag\\
   & - \frac{1}{8\pi l_\mathrm{ p}^2}\int_{\partial M} 
   K \sqrt{\gamma}d^3y - I_\mathrm{AdS}\,,
   \label{eqch7:action}
\end{align}
with all quantities having been properly defined.

\subsection{Geometry and boundary conditions}

In the analysis, we only consider paths which are spherically
symmetric. 
We also assume that the spaces are static.
The metric for the space $M_1$ is written as 
\begin{align}
   & ds^2_{M_1} = 
   b_1^2(u) \frac{b^2_2(u_\mathrm{ m})}{b_1^2(u_\mathrm{ m})}d\tau^2 
   + a_1^2(u)dy^2 + r(u)^2 d\Omega^2\,,\quad\quad 0\leq u<u_\mathrm{ m}\,.
\label{eqch7:metricM1}
\end{align}
For the thin shell $\mathcal{C}$, the induced metric is written as
\begin{align}
   ds^2_{\mathcal{C}} = b^2_2(u_\mathrm{ m})d\tau^2 
   + r^2(u_\mathrm{ m})d\Omega^2\,,\quad\quad
   \quad\quad\quad\quad\quad\quad u=u_\mathrm{ m}\,.
   \label{eqch7:inducedmetricA}
\end{align}
For space $M_2$, the metric is 
\begin{align}
   & ds^2_{M_2} = b^2_2(u) d\tau^2
    + a_2^2(u)dy^2 + r^2(u)d\Omega^2
   \,,\quad\quad\quad\quad\quad\quad u_\mathrm{ m}< u<1\,.
\label{eqch7:metricM2}
\end{align}
Here $b_1$, $b_2$, $a_1$, $a_2$, and $r$ are functions of the
coordinate $u$. The
Euclidean time coordinate $\tau$ is chosen to be an angular
coordinate in the interval $0<\tau <2\pi$ on $M$, the
radial coordinate $u$
takes values as above, $d\Omega^2=d\theta^2+\sin^2\theta \,d\phi^2$ is
the line element of the $2$-sphere with surface area $\Omega = 4\pi$,
and the coordinates $\theta$ and $\phi$ are the usual spherical
coordinates.  The points at the thin shell are located at $u=u_\mathrm{
m}$ and $u_\mathrm{ m}$ is exhibited as a label that is fixed, while the
radius of the shell $r(u_\mathrm{ m})$ depends on the arbitrary function
$r(u)$.

We must further impose regularity conditions and boundary conditions 
on the metrics that are being summed in the path integral. 
In the region $M_1$, the interior
region, we impose regularity conditions 
at $u=0$ corresponding to flat conditions at the origin, i.e.,
\begin{align}
   r|_{y=0}= 0\,,\quad\quad\
 b_1|_{y=0} \,\,\mathrm{ finite\,\, and \,\,positive}\,,\notag\\
   \eval{\frac{r'}{a_1}}_{y=0} = 1\,,\,\quad
   \eval{\frac{1}{r'}\left(\frac{r'}{a_1}\right)'}_{y=0} = 0\,,
\quad
\eval{\frac{b_1'}{a_1}}_{y=0} = 0 \,,
   \label{eqch7:regularity}
\end{align}
where a primed quantity means derivative
with respect to $u$, e.g.,
$r' = \frac{dr}{du}$.

In the region $M_2$, the exterior region,
we impose that the space behaves as asymptotically 
AdS, when $u\rightarrow 1$. As seen in Chapter~\ref{ch:Euclideanpathintegral}, 
the AdS boundary conditions are summarized as 
\begin{align}
    \eval{\frac{b_2(u)}{r(u)}}_{u\to1} = 
      \frac{\bar{\beta}}{2\pi l}\,,\notag\\
    \eval{\frac{a_2(u) r(u)}{r'(u)}}_{u\to1} = l
    \,,
    \label{eqch7:boundary}
\end{align}
where the parameter $\bar{\beta}$
is defined to be the fixed quantity of the ensemble. In some sense,
the parameter $\bar{\beta}$ is proportional to the total proper length of
the conformal boundary with induced metric 
$(\frac{l^2}{r(y)^2}ds^2)\sVert[1]_{u\rightarrow 1}$, with conformal 
factor $\frac{l}{r(y)}$, 
and the $\bar{\beta}$ is identified to the
inverse of the local temperature $\bar{T}$
of the conformal boundary,
such that $\bar{\beta} = \frac{1}{\bar{T}}$. It must be pointed out that fixing the inverse
temperature in this conformal boundary as $\bar{\beta}$ is a
choice of the formalism. Here, the choice coincides with the usual
Euclidean proper time approach formalism, e.g., the way 
the temperature is defined at infinity for a black hole yields the same as
the Hawking-Page definition. One could have chosen a different 
conformal transformation as long as the asymptotic AdS behaviour 
is imposed. This indeed leads to possible different choices 
of the fixed inverse temperature $\bar{\beta}$. However, one can view 
the different fixed temperatures as being
related to the choice of conformal observer that measures the 
temperature. In order to obtain a nonsingular conformal metric, 
the conformal transformation must behave asymptotically as
$\frac{c}{r(y)}$, with $c$ being a constant. This leads to a 
$\bar{\beta}$ only differing by a constant multiplication factor 
which can be thought of as a change of scale for the temperature 
and energy. The physical results do not alter from such choice.

\subsection{Matter free energy and stress-energy tensor}

The matter Lagrangian density can be identified to the thermodynamic 
Helmholtz free energy density since we are
dealing with the canonical ensemble.
The Helmholtz free energy potential $F$  is described by 
$F=E-TS$, where $E$ is the thermodynamic 
energy, $S$ the entropy, and $T$ the temperature of the reservoir.
We must then analyze the associated density quantities, and so the 
free energy density $\mathcal{F_\mathrm{ m}}$ can be written as
\begin{align}
\mathcal{F_\mathrm{ m}}[\gamma_{ab}] = \epsilon_\mathrm{ m}[\gamma_{ab}] 
- T_\mathrm{ m}[\gamma_{ab}] s_\mathrm{ m}[\gamma_{ab}]
  \,,
   \label{eqch7:matterfe}
\end{align}
where $\epsilon_\mathrm{ m}$ is the total energy density of the matter,
$T_\mathrm{ m}$ is the local temperature of the shell, and 
$s_\mathrm{ m}$ is the entropy density of the matter. All these quantities are 
functionals of the induced metric $\gamma_{ab}$.
Since from Eq.~\eqref{eqch7:inducedmetricA} one has that $\gamma_{ab}$
depends on two arbitrary quantities that are seen as constants at the 
shell, 
$b_2(y_\mathrm{ m})$ and $r(y_\mathrm{ m})$,
the matter free energy density 
$\mathcal{F_\mathrm{ m}}[\gamma_{ab}]$ depends only locally
on these two quantities, and so a
dependence on derivatives of $\gamma_{ab}$ is ruled out.

The radius $\alpha$ 
of the shell is defined as
\begin{align}
   \alpha = r(u_\mathrm{ m})\,.
   \label{eqch7:numberdensityshell1}
\end{align}
Although in order to keep a consistent nomenclature 
we should have defined the radius of the shell
as $r_\mathrm{ m}=r(u_\mathrm{ m})$,
it is preferred to stick with
$\alpha = r(u_\mathrm{ m})$ to not overcrowd the symbols ahead.
We can define a local temperature at some point $u$
as $T(u) = \frac{1}{2\pi b_2(u)}$. So the local temperature of 
the shell is 
\begin{align}
   T_\mathrm{ m} = \frac{1}{2\pi b_2(u_\mathrm{ m})}\,.
   \label{eqch7:tempshell1}
\end{align}
The rationale for this definition comes from continuity, since in the
canonical ensemble one fixes the Euclidean proper time length at the
boundary and assigns to it the meaning of an inverse temperature. 
One must keep in mind however that this definition does not
give information about the specific expression of the temperature
since $b_2(u_\mathrm{ m})$ is arbitrary.  

The free energy per unit area
$\mathcal{F}_\mathrm{ m}[\gamma_{ab}]=\mathcal{F}_\mathrm{ m}[b_2(u_\mathrm{ m}),
r(u_\mathrm{ m})]$
can then be put in the form
\begin{align}
   &\mathcal{F}_\mathrm{ m}[h_{ab}] = \mathcal{F}_\mathrm{{m}}
   [\alpha,T_\mathrm{ m}]\,,
   \label{eqch7:freeenergyShell1}
\end{align}
upon using Eq.~\eqref{eqch7:numberdensityshell1} and \eqref{eqch7:tempshell1}.
Now, we assume the first law to describe the matter energy  density 
as
$d \epsilon_\mathrm{ m} = T_\mathrm{ m} ds_\mathrm{ m} -
2(\epsilon_\mathrm{ m}
-T_\mathrm{ m}s_\mathrm{ m}
+ p_\mathrm{ m})\frac{d\alpha}{\alpha}$,
where $p_\mathrm{ m}$ 
is the matter tangential pressure at the shell.
Thus, from Eq.~\eqref{eqch7:matterfe}, the free energy density has
the differential
\begin{align}
d \mathcal{F}_\mathrm{ m} = - s_\mathrm{ m} dT_\mathrm{ m} -
2(\mathcal{F}_\mathrm{ m} + p_\mathrm{ m})
\frac{d\alpha}{\alpha}
\,.
\label{eqch7:difffreeenergy}
\end{align}
With the known differential of the free energy density regarding its
dependence on the metric components, one can compute the
surface stress-energy tensor $S^{ab}$
as the functional derivative, $S^{ab}
= - \frac{2}{\sqrt{\gamma}} \frac{\delta (\sqrt{\gamma}
\mathcal{F_\mathrm{ m}})[\gamma_{ab}]}{\delta \gamma_{ab}}$.
From Eq.~\eqref{eqch7:inducedmetricA},
one has $h_{\tau\tau}=b^2_2(u_\mathrm{ m})$
and $h_{\theta\theta}=\frac{h_{\phi\phi}}{\sin^2(\theta)}=r^2(u_\mathrm{
m})$.
Then, from
 $\alpha = r(u_\mathrm{ m})$ and 
$T_\mathrm{ m} = \frac{1}{2\pi b_2(u_\mathrm{ m})}$,
see Eqs.~\eqref{eqch7:numberdensityshell1}
and \eqref{eqch7:tempshell1}, one finds that
the variation yields
 ${S^{\tau}}_{\tau} =
- \mathcal{F}_\mathrm{ m} + T_\mathrm{ m}
\frac{\partial \mathcal{F}_\mathrm{ m}}{\partial T_\mathrm{ m}}$
and 
 ${S^{\theta}}_{\theta} = {S^{\phi}}_{\phi} =
-\frac12\alpha\frac{\partial \mathcal{F}_\mathrm{ m}}{\partial \alpha} -
\mathcal{F}_\mathrm{ m}
$, where the partial derivatives are done keeping the
hidden variable constant,
and $\delta \gamma= \gamma \gamma^{ab} \delta \gamma_{ab}$ has been used. 
Some care is needed while performing the variational
derivative to obtain ${S^{\theta}}_{\theta}$ and ${S^{\phi}}_{\phi}$,
as one
must calculate $d\alpha$ in $d\mathcal{F}_m$ as 
$d\alpha = d\left(\sqrt[4]{\frac{\gamma_{\theta\theta} 
\gamma_{\phi \phi}}{\sin^2\theta}}\right)$.
From Eqs.~\eqref{eqch7:matterfe} and \eqref{eqch7:difffreeenergy},
one has  $\epsilon_\mathrm{ m} = \mathcal{F}_\mathrm{ m} - T_\mathrm{ m}
\frac{\partial \mathcal{F}_\mathrm{ m}}{\partial T_\mathrm{ m}}$ and $p_\mathrm{ m} =
-\frac12\alpha\frac{\partial \mathcal{F}_\mathrm{ m}}{\partial \alpha} -
\mathcal{F}_\mathrm{ m}$.
Thus, the stress-energy tensor ${S^a}_{b}$ has components
\begin{align}
{S^{\tau}}_{\tau} = - \epsilon_\mathrm{ m}\,,\quad\quad
{S^{\theta}}_{\theta} = {S^{\phi}}_{\phi} = p_\mathrm{
m}\,.
\label{eqch7:stress-energytensor}
\end{align}
The fluid is thus isotropic, more specifically, it is
a perfect fluid. 
Note that $\epsilon_\mathrm{ m}=\epsilon_\mathrm{ m}(\alpha,T_\mathrm{ m})$
and $p_\mathrm{ m}=p_\mathrm{ m}(\alpha,T_\mathrm{ m})$.
The rest mass $m$ of the shell is important for the analysis below and 
it is defined as
\begin{align}
m= 4\pi \alpha^2
\epsilon_\mathrm{ m}\,.
\label{eqch7:massshell}
\end{align}
Since $\epsilon_\mathrm{ m}=\epsilon_\mathrm{ m}(\alpha,T_\mathrm{ m})$, one has
$m=m(\alpha,T_\mathrm{ m})$. The dependence of the thermodynamic
quantities in $\alpha$ and $T_\mathrm{ m}$ is helpful when one makes
variations of the action on the metric components to find the
Hamiltonian constraint, however it is also helpful  to
invert the first law
of thermodynamics to get $ds_\mathrm{ m} = \frac{1}{T_\mathrm{
m}}d\epsilon_\mathrm{ m} + 2(\epsilon_\mathrm{ m} - T_\mathrm{ m} s_\mathrm{ m} +
p_\mathrm{ m}) \frac{d\alpha}{\alpha}$. One can integrate over the area
to obtain the first law of thermodynamics in the form, 
\begin{align}
T_\mathrm{ m}dS_\mathrm{ m} = dm+
p_\mathrm{ m} dA_\mathrm{ m},
\label{eqch7:1stlawintegral}
\end{align}
where
\begin{align}
A_\mathrm{ m}= 4\pi \alpha^2,
\label{eqch7:areaAm}
\end{align}
\begin{align}
S_\mathrm{ m}= s_\mathrm{ m} A_\mathrm{ m},
\label{eqch7:entropySm}
\end{align}
are the area of the shell and 
the entropy of the matter in the shell, respectively.
Written likes this, also
the quantities $S_\mathrm{ m}$, $T_\mathrm{ m}$, and $p_\mathrm{ m}$ become
functions of $m$ and $\alpha$. The
dependencies used below shall be explicitly indicated.

\subsection{Euclidean action in spherical symmetry}

Since only spherically symmetric metrics are considered, we can write 
the action explicitly in terms of its components. The gravitational action 
with negative cosmological constant, written in Eq.~\eqref{eqch7:gravaction}, 
for a $C^0$ metric is
\begin{align}\label{eqch7:actionnegativecosmological2}
   &I_{gl} = 
   \left(\frac{2\pi b_2 r }{l_p^2}\left(\left(\frac{r'}{a_2}\right)_{\mathrm{AdS}} 
   - \frac{r'}{a_2}\right)\right)\sVert[3]_{u\rightarrow1}
   -\frac{\pi}{l_p^{2}}\left(\frac{b'_1 b_2(u_m) r^{2}}
   {a_1 b_1(u_m)}\right)\sVert[3]_{u=0} \notag\\
   & + \frac{1}{8\pi l_p^{2}}\int_{M_1} a_1 b_1 \frac{b_2(u_m)}{b_1(u_m)} 
   r^{2}\left(G\indices{_1^\tau_\tau} 
   - \frac{3}{l^2}\right) d^4x \notag\\
   &+ \frac{1}{8\pi l_p^{2}}\int_{M_2} a_2 b_2 r^{2}\left(G\indices{_2^\tau_\tau} 
   - \frac{3}{l^2}\right)d^4x\notag\\
   & - \frac{1}{8\pi l_p^{2}}\int_{\mathcal{C}} ([K\indices{^\tau_\tau}] 
   - [K])\sqrt{\gamma}d^{3}x\,\,,
\end{align}  
where $I_\mathrm{AdS} = 
-\left(\frac{2\pi b_2 r }{l_p^2}\left(\frac{r'}{a_2}\right)_{\mathrm{AdS}}\right)
\sVert[3]_{u\rightarrow1}$ with $\left(\frac{r'}{a_2}\right)_{\mathrm{AdS}}$ being the 
redshift factor of AdS, see Chapter~\ref{ch:Euclideanpathintegral},  the Einstein 
tensor component $G\indices{_1^\tau_\tau}$ and $G\indices{_2^\tau_\tau}$ are given by 
\begin{align}\label{eqch7:Einsteintensor}
   &G\indices{_1^\tau_\tau} = \frac{1}{r'r^{2}}\left(r\left(\frac{r^{\prime2}}{a_1^2} 
    -1\right)\right)'\,\,.\notag\\
   &G\indices{_2^\tau_\tau} = \frac{1}{r'r^{2}}\left(r\left(\frac{r^{\prime2}}{a_2^2} 
   -1\right)\right)'\,\,,
\end{align}
and with the terms depending on the extrinsic curvature being given as 
\begin{align}\label{eqch7:hamiltoniandelta}
   [K] - [K\indices{^\tau_\tau}] = 
   \frac{2}{r}\eval{\left(\frac{r'}{a_2} 
   - \frac{r'}{a_1}\right)}_{u=u_m}\,\,.
\end{align}
We now can use the regularity and boundary conditions in Eqs.~\eqref{eqch7:regularity} 
and~\eqref{eqch7:boundary}, respectively, to further simplify the gravitational 
action as 
\begin{align}\label{eqch7:actionnegativecosmological23}
   &I_{gl} = 
   \left(\frac{\bar{\beta} r^2 }{l_p^2l}\left(\left(\frac{r'}{a}\right)_{\mathrm{AdS}} 
   - \frac{r'}{a_2}\right)\right)\sVert[3]_{u\rightarrow1}\notag\\
   & + \frac{1}{8\pi l_p^{2}}\int_{M_1} a_1 b_1 \frac{b_2(u_m)}{b_1(u_m)} 
   r^{2}\left(G\indices{_1^\tau_\tau} 
   - \frac{3}{l^2}\right) d^4x \notag\\
   &+ \frac{1}{8\pi l_p^{2}}\int_{M_2} a_2 b_2 r^{2}\left(G\indices{_2^\tau_\tau} 
   - \frac{3}{l^2}\right)d^4x\notag\\
   & - \frac{1}{8\pi l_p^{2}}\int_{\mathcal{C}} ([K\indices{^\tau_\tau}] 
   - [K])\sqrt{\gamma}d^{3}x\,\,,
\end{align}  
Finally, we must look towards the thin shell matter action in Eq.~\eqref{eqch7:mataction}. 
Through the definition of the matter free energy $\mathcal{F}_{\mathrm{m}} = 
\epsilon_\mathrm{m} - T_\mathrm{m} s_\mathrm{m}$, one 
can rewrite the matter action as 
\begin{align}
   I_m = \int_{\mathcal{C}}\epsilon_\mathrm{m} \sqrt{\gamma}d^3x - S_\mathrm{m}\,\,,
\end{align} 
where it was used that $T_m= \frac{1}{2\pi b_2(u_m)}$, and $S_\mathrm{m} = 4\pi \alpha^2 s_\mathrm{m}$. 
The full action can then be written as 
\begin{align}\label{eqch7:actionnegativecosmological3}
   &I = 
   \left(\frac{\bar{\beta} r^2 }{l_p^2l}\left(\left(\frac{r'}{a}\right)_{\mathrm{AdS}} 
   - \frac{r'}{a_2}\right)\right)\sVert[3]_{u\rightarrow1} - S_\mathrm{m}\notag\\
   & + \frac{1}{8\pi l_p^{2}}\int_{M_1} a_1 b_1 \frac{b_2(u_m)}{b_1(u_m)} 
   r^{2}\left(G\indices{_1^\tau_\tau} 
   - \frac{3}{l^2}\right) d^4x \notag\\
   &+ \frac{1}{8\pi l_p^{2}}\int_{M_2} a_2 b_2 r^{2}\left(G\indices{_2^\tau_\tau} 
   - \frac{3}{l^2}\right)d^4x\notag\\
   & - \frac{1}{8\pi l_p^{2}}\int_{\mathcal{C}} ([K\indices{^\tau_\tau}] 
   - [K] - 8\pi l_p^2 \epsilon_\mathrm{m})\sqrt{\gamma}d^{3}x\,\,.
\end{align}  
Further details can be found on Chapter~\ref{ch:Euclideanpathintegral} on the 
construction of the path integral, on the regularity conditions,
on the boundary conditions, and on the expression of the action 
for spherically symmetric spaces.

\section{The zero loop approximation\label{sech7:zeroloop}}

\subsection{The constrained path integral and reduced action}

Having the action in spherical symmetry, we now proceed with the 
zero loop approximation. We make this approximation in steps, starting by 
imposing
the Hamiltonian and momentum constraint equations, so
the path integral is along the constraint paths. We don't apply the 
zero loop approximation straight away since we want
to perform a stability analysis, or rather to see the validity of the 
zero loop approximation. Only afterwards
we perform the full zero-loop approximation. We start with the
Hamiltonian constraint, consisting of one equation for each region
$M_1$ and $M_2$, and a junction condition on the matter shell $\mathcal{C}$.  
We analyze then the momentum constraint.

The Hamiltonian constraint 
in the regions $M_1$ and $M_2$ makes use of
the Einstein tensor component
${G^{\tau}}_{\tau}$, in Eq.~\eqref{eqch7:Einsteintensor}
for each $M_1$ and $M_2$.
The Hamiltonian constraint is the 
Einstein equation ${G^{\tau}}_{\tau} = \frac{3}{l^2}$, which 
can be integrated in both spaces $M_1$ and $M_2$ to yield
\begin{align}
   & \left(\frac{r'}{a_1}\right)^2  = 
   1 + \frac{r^2}{l^2}\equiv f_1(r)\,,
   \label{eqch7:a1constraint}\\
   &  \left(\frac{r'}{a_2}\right)^2  = 
   1 + \frac{r^2}{l^2} 
   - \frac{\tilde{r}_+ + \frac{\tilde{r}_+^3}{l^2}}{r}
   \equiv f_2(r,\tilde{r}_+)\,,
   \label{eqch7:a2constraint}
\end{align}
where the regularity condition $\frac{r'}{a_1}\sVert[1]_{y=0} = 1$ in
Eq.~\eqref{eqch7:regularity} was used, $\tilde{r}_+$ is the
gravitational radius of the system and it is featured as an
integration constant obeying $\tilde{r}_+ < \alpha$, and
the functions $f_1(r)$ and $f_2(r,\tilde{r}_+)$ have been defined for
convenience. Due to the order of the differential equation in the
Hamiltonian constraint equation, the regularity condition
$\frac{1}{r'}\left(\frac{r'}{a_1}\right)'\sVert[2]_{y=0} = 0$ in
Eq.~\eqref{eqch7:regularity} was not used but it is naturally
satisfied.  The same thing happens for the function $\frac{r'}{a_2}$
which obeys naturally the boundary condition Eq.~\eqref{eqch7:boundary}.

The Hamiltonian constraint in the hypersurface $\mathcal{C}$ is described by the
junction condition $[{K^{\tau}}_{\tau}] - [K] = - 8\pi l_\mathrm{ p}^2
{S^{\tau}}_{\tau}$, where ${S^{\tau}}_{\tau}$ is the $\tau\tau$
component of the surface stress-energy tensor. 
The extrinsic curvature term $[K\indices{^{\tau}_{\tau}}] - [K]$ 
is given by Eq.~\eqref{eqch7:hamiltoniandelta}.
The surface stress-energy tensor is the functional derivative
$S^{ab}$, with $S\indices{^{\tau}_\tau}=-\epsilon_\mathrm{ m}$, see
Eq.~\eqref{eqch7:stress-energytensor}.
Then, for the mass $m= 4\pi \alpha^2 \epsilon_\mathrm{ m}$,
Eq.~\eqref{eqch7:massshell}, one finds that the Hamiltonian constraint at the
shell is
\begin{align}
   m = \frac{\alpha}{l_\mathrm{ p}^2}(\sqrt{f_1(\alpha)} -
   \sqrt{f_2(\alpha,\tilde{r}_+)})
   \label{eqch7:junctioncondhamiltonian}
   \,,
\end{align}
with $f_1(\alpha)=\eval{\left(\frac{r'}{a_1}\right)^2}_\alpha
=1 + \frac{\alpha^2}{l^2}$ and
$f_2(\alpha,\tilde{r}_+)=
\eval{\left(\frac{r'}{a_2}\right)^2}_\alpha
=1 + \frac{\alpha^2}{l^2} 
   - \frac{\tilde{r}_+ + \frac{\tilde{r}_+^3}{l^2}}{\alpha}$,
   see Eqs.~\eqref{eqch7:a1constraint}
and \eqref{eqch7:a2constraint}, respectively.
While the dependence of $m$ in the metric components 
is described by $m = m(\alpha,T_\mathrm{ m})$,
one can invert in order to $T_\mathrm{ m}$
and get $T_\mathrm{ m} = T_\mathrm{ m}(m,\alpha)$. Using now the 
junction condition Eq.~\eqref{eqch7:junctioncondhamiltonian}, 
one obtains the temperature of the shell as a function of $\tilde{r}_+$ 
and $\alpha$ as $T_\mathrm{ m} = T_\mathrm{
m}(m(\tilde{r}_+,\alpha),\alpha)$, as long as the equation of
state $T_\mathrm{ m}(m,\alpha)$ is provided. 

In relation to the momentum constraints, due to the spherical symmetry
of the metrics in Eqs.~\eqref{eqch7:metricM1} and~\eqref{eqch7:metricM2} and
the symmetry on translations in $\tau$, the momentum constraints 
in the regions $M_1$ and $M_2$
are satisfied a priori. Moreover, the momentum constraint at the shell is
satisfied since the matter shell stress tensor is diagonal as
$\mathcal{F}_\mathrm{ m}$ is a functional only of $b_2$ and $\alpha$.

Imposing the Hamiltonian constraints in both spaces $M_1$ and $M_2$, 
together with the junction condition at the shell, the bulk terms 
in the action Eq.~\eqref{eqch7:actionnegativecosmological3} vanish. 
The term that remains to be calculated is the limit 
$\left(r^2\left(\frac{r'}{a}\right)_{\mathrm{AdS}} - r^2\frac{r'}{a_2} \right)
\sVert[2]_{u\rightarrow 1}$. Through the Hamiltonian 
constraints, one has that $\left(\frac{r'}{a}\right)_{\mathrm{AdS}} = \sqrt{f_1(r)}$ 
since it is the redshift factor of pure $AdS$
and  $\frac{r'}{a_2} = \sqrt{f_2(r,\tilde{r}_+)}$. Hence, the limit 
yields 
\begin{align}
   \left(r^2\sqrt{f_1(r)} - r^2\sqrt{f_2(r,\tilde{r}_+)}\right)
\sVert[2]_{u\rightarrow 1} = \frac{l}{2}
(\tilde{r}_+ + \frac{\tilde{r}_+^3}{l^2})\,\,.
\end{align}
The action in Eq.~\eqref{eqch7:actionnegativecosmological3} then becomes 
the reduced action 
\begin{align}
   I_*[\bar{\beta};\tilde{r}_+,\alpha] 
   = \frac{\bar{\beta}}{2 l_\mathrm{ p}^2}\left(\tilde{r}_+ +
   \frac{\tilde{r}_+^3}{l^2}\right) 
   - S_\mathrm{ m}(m(\tilde{r}_+,\alpha),\alpha)
   \,,\label{eqch7:reducedaction1}
\end{align}
where $m(\tilde{r}_+,\alpha)$ is given by the right-hand side of
Eq.~\eqref{eqch7:junctioncondhamiltonian}. The partition function
of Eq.~\eqref{eqch7:partitionfunctiongeneric}
with its path integral reduces thus to the 
following expression 
\begin{align}
   Z[\bar{\beta}] = \int D\tilde{r}_+ D\alpha 
   \,\mathrm{ e}^{- I_*[\bar{\beta}; \tilde{r}_+,\alpha]}\,,
   \label{eqch7:reducedactionpathintegral}
\end{align}
as the sum over different metrics with spherical symmetry reduces to
the sum over metrics with different $\tilde{r}_+$ and different
$\alpha$. For clarification, the integration over $\alpha$ arises due
to the sum over metric
functions $r(y)$. Although the Hamiltonian constraint
ensures that the metric in the bulk has the same form for any arbitrary 
function $r(y)$, through a coordinate transformation $r=r(y)$,
the value $\alpha = r(y_\mathrm{ m})$ that
separates the regions $M_1$ and $M_2$ depends on the specific function
$r(y)$, and so one must sum over the possible values of $\alpha$.

\subsection{
The zero-loop approximation from the reduced action 
and stationary conditions}

With the reduced action of the system being given by
Eq.~\eqref{eqch7:reducedaction1},
we now minimize it to find
the action in the zero-loop 
approximation. To find
the minimum of the action, we need
to find its stationary conditions which
are given by
\begin{align}
\frac{\partial I_*[\bar{\beta};\tilde{r}_+,\alpha]}{\partial \alpha}= 0   \,,
\label{eqch7:stationcond1}
\end{align}
\begin{align}
\frac{\partial I_*[\bar{\beta};\tilde{r}_+,\alpha]}{\partial \tilde{r}_+}= 0\,.
\label{eqch7:stationcond2}
\end{align}
The stationary conditions given in Eqs.~\eqref{eqch7:stationcond1}
and~\eqref{eqch7:stationcond2} can be understood as the remaining
Einstein equations whose solutions minimize the action.  Since the
reduced action is essentially the Einstein-Hilbert action together
with the matter action, having the Hamiltonian constraint being
imposed, the minimization of the action in relation to $\alpha$ and to
$\tilde{r}_+$ is equivalent to the minimization in relation to the
metric components $g_{\theta \theta}$ and to $g_{uu}$. And so, these
conditions yield the Dirac delta terms of $G_{\theta \theta}=8 \pi
T_{\theta \theta}$, and the equation $G_{uu} = 0$, where $G_{\theta
\theta}$ and $G_{uu}$ are the corresponding components of the Einstein
tensor, and $T_{\theta \theta}$ is the corresponding component of the
stress-energy tensor.
In order to further develop the derivatives of
Eqs.~\eqref{eqch7:stationcond1} and \eqref{eqch7:stationcond2},
one has to find  $\frac{\partial
S_\mathrm{ m}}{\partial \alpha}$ and $\frac{\partial S_\mathrm{ m}}{\partial \tilde{r}_+}$.
From the first law of thermodynamics
given in Eq.~\eqref{eqch7:1stlawintegral}, the matter entropy has
the differential form $dS_\mathrm{ m} = 
\frac{dm(\tilde{r}_+,\alpha)}{T_\mathrm{ m}} +
\frac{p_\mathrm{ m}}{T_\mathrm{ m}}dA_\mathrm{ m}$,
and so to find the derivatives of $S_\mathrm{ m}$
one has to find $\frac{\partial A_\mathrm{ m}}{\partial \alpha}$,
$\frac{\partial A_\mathrm{ m}}{\partial \tilde{r}_+}$,
$\frac{\partial m}{\partial \alpha}$, and
$\frac{\partial m}{\partial \tilde{r}_+}$.
From the expression of $A_\mathrm{ m}$,
Eq.~\eqref{eqch7:areaAm}, one has
\begin{align}
\frac{\partial A_\mathrm{ m}}{\partial \alpha} = 8\pi
\alpha, \quad\quad\quad \frac{\partial A_\mathrm{ m}}{\partial \tilde{r}_+} = 0\,.
\label{eqch7:derivarea}
\end{align}
From the expression of $m(\tilde{r}_+,\alpha)$,
Eq.~\eqref{eqch7:junctioncondhamiltonian}, one has
\begin{align}
&\frac{\partial m}{\partial \alpha} \equiv - 8\pi \alpha p_\mathrm{ g}\,,
\quad\quad\quad 
\frac{\partial m}{\partial \tilde{r}_+} = \frac{1 +
\frac{3\tilde{r}_+^2}{l^2}} {2 l_\mathrm{ p}^2 \sqrt{f_2(\alpha, \tilde{r}_+)}}\,,
\nonumber
\\
&
p_\mathrm{ g} \equiv \frac{1}{8\pi \alpha l_\mathrm{ p}^2}
   \left(\frac{1 + 2 \frac{\alpha^2}{l^2}
   - \frac{\tilde{r}_+ + \frac{\tilde{r}_+^3}{l^2}}{2\alpha}}
   {\sqrt{f_2(\alpha, \tilde{r}_+)}} - 
   \frac{1 + 2\frac{\alpha^2}{l^2}}{\sqrt{f_1(\alpha)}}\right)
\,,
\label{eqch7:pg}
\end{align}
where $p_\mathrm{ g}$ is defined as the gravitational pressure.
Using then the chain rule on the first law of thermodynamics,
one gets the following 
derivatives for the matter entropy
\begin{align}
\frac{\partial
   S_\mathrm{ m}}{\partial \alpha} = \frac{8\pi \alpha}{T_\mathrm{ m}}(p_\mathrm{ m} -
   p_\mathrm{ g}), \quad\quad\quad 
    \frac{\partial S_\mathrm{ m}}{\partial \tilde{r}_+} = \frac{1 + 3
   \frac{\tilde{r}_+^2}{l^2}} {2 l_\mathrm{ p}^2 T_\mathrm{ m}
   \sqrt{f_2(\alpha,\tilde{r}_+)}}\,.
   \label{eqch7:derTSm}
\end{align}
Since the differential of the entropy has been
recast in terms of  $\alpha$ and $\tilde{r}_+$, the temperature of the
shell and the pressure of the shell also have that dependence as
$T_\mathrm{ m}= T_\mathrm{ m}(m(\alpha,\tilde{r}_+),\alpha) =T_\mathrm{
m}(\alpha,\tilde{r}_+)$ and $p_\mathrm{ m}=p_\mathrm{
m}(m(\tilde{r}_+,\alpha),\alpha) = p_\mathrm{ m}(\alpha,\tilde{r}_+)$.
From now on, we abbreviate this dependence to avoid cluttering,
however, the dependence must be assumed.

Then, using the reduced action given in Eq.~\eqref{eqch7:reducedaction1}
and the derivatives of the matter entropy in Eq.~\eqref{eqch7:derTSm}, we
find the stationary conditions.
The stationary condition of 
Eq.~\eqref{eqch7:stationcond1} yields 
\begin{align}
p_\mathrm{ g} = p_\mathrm{ m}\,.
\label{eqch7:statio1}
\end{align}
This equation gives the condition for mechanical
equilibrium of the shell. We call 
Eq.~\eqref{eqch7:statio1} as the balance of pressure
equation.
The stationary condition of 
Eq.~\eqref{eqch7:stationcond2}  yields 
\begin{align}
\bar{\beta} = \frac{1}{T_\mathrm{ m} \sqrt{f_2(\alpha, \tilde{r}_+)}}\,.
   \label{eqch7:statio2}
\end{align}
This equation gives the condition for thermodynamic
equilibrium of the shell. We call
Eq.~\eqref{eqch7:statio2} as
the balance of temperature equation.
Thus, these two 
equations above give, as expected, the conditions for equilibrium.

One can
verify that Eq.~\eqref{eqch7:statio1} only depends on $\tilde{r}_+$ and 
$\alpha$, which means the solutions to this equation can 
be expressed as
\begin{align}
   \alpha=\alpha(\tilde{r}_+)\,.
   \label{eqch7:statio1final}
\end{align}
Then, one can input such solutions into
Eq.~\eqref{eqch7:statio2} to obtain an equation only dependent on
$\tilde{r}_+$ and $\bar{\beta}$, which can be solved by a function
$\tilde{r}_+(\bar{\beta})$, i.e.,
\begin{align}
   \bar{\beta} = \frac{\iota(\tilde{r}_+)}
{1 + 3\frac{\tilde{r}_+^2}{l^2}}\,,\quad\quad
\mathrm{ implying}
\quad\quad
\tilde{r}_+=\tilde{r}_+(\bar{\beta})\,,
\label{eqch7:statio2final}
\end{align} 
where $\iota(\tilde{r}_+)$
is a function of $\tilde{r}_+$ that appears for
convenience and is given for the thin shell by
\begin{align}
\iota(\tilde{r}_+)=\frac{1 + 3\frac{\tilde{r}_+^2}{l^2}}
   {T_\mathrm{ m}(\tilde{r}_+,\alpha(\tilde{r}_+))
   \sqrt{f_2(\alpha(\tilde{r}_+))}}.
\label{eqch7:iota}
\end{align}
In comparison, for the Hawking-Page black hole one has
$\iota(\tilde{r}_+)=4\pi r_+$. Of course, the expression
of Eq.~\eqref{eqch7:iota} means we are dealing with a thin shell.

Since, from Eq.~\eqref{eqch7:statio1final}, one has
$\alpha=\alpha(\tilde{r}_+)$,
the reduced action $I_*[\bar{\beta};\alpha,\tilde{r}_+]$
of Eq.~\eqref{eqch7:reducedaction1} under the mechanical stationary 
condition
can be written as an effective reduced action
of the form
$I_*[\bar{\beta};\tilde{r}_+]$.
This in turn implies that
the partition function 
given in Eq.~\eqref{eqch7:reducedactionpathintegral}
is now 
 $Z[\bar{\beta}] = \int D\tilde{r}_+
\,\mathrm{ e}^{- I_*[\bar{\beta}; \tilde{r}_+]}$, 
as the zero-loop approximation 
in the path integral over $\alpha$ as been done, i.e.,
using the reduced
action evaluated at the  stationary point
provided by Eq.~\eqref{eqch7:statio1}, or what amounts
to the same thing, by Eq.~\eqref{eqch7:statio1final}.
It is interesting to note that this behavior can be deduced from the
structure of the reduced action in Eq.~\eqref{eqch7:reducedaction1}
together with Eq.~\eqref{eqch7:reducedactionpathintegral}. In fact, the 
path integral over $\alpha$ in the partition function, 
$\int D\alpha \,\mathrm{ e}^{S_\mathrm{ m}}$, corresponds to
the partition function of the 
microcanonical ensemble.
Therefore, this
indicates that the canonical ensemble of the full system can be
described by an effective reduced action determined by the
microcanonical ensemble of a hot
self-gravitating matter thin shell with
fixed $\tilde{r}_+$, $I_*[\bar{\beta};\tilde{r}_+,\alpha(\tilde{r}_+)]=
I_*[\bar{\beta};\tilde{r}_+]$, while the solutions $\alpha(\tilde{r}_+)$ are 
but a consequence of performing the zero-loop approximation 
on the path integral over $\alpha$, i.e., of performing the zero
loop approximation
on the microcanonical ensemble.
Given Eqs.~\eqref{eqch7:reducedaction1}, \eqref{eqch7:statio1final},
and \eqref{eqch7:statio2final}, 
the solutions $\tilde{r}_+(\bar{\beta})$ of the
canonical ensemble give the path that extremizes the reduced action.
Having $\tilde{r}_+(\bar{\beta})$, one finds
$\alpha=\alpha(\tilde{r}_+(\bar{\beta}))$,
and then the action $I_*[\bar{\beta};\tilde{r}_+(\bar{\beta})]$,
which is the action of the stationary points.
This action is the zero-loop approximation action $I_0(\bar{\beta})$.
Indeed, 
\begin{align}
I_{0}[\bar{\beta}] \equiv
I_*[\bar{\beta};\tilde{r}_+(\bar{\beta})],
\label{eqch7:expandaction}
\end{align}
i.e., the zero-loop action
$I_{0}[\bar{\beta}]$ is found by evaluating the reduced
action around its stationary points with $\alpha(\tilde{r}_+(\bar{\beta}))$
and $\tilde{r}_+(\bar{\beta})$ being found from 
the stationary conditions.
The partition function of the canonical ensemble can then be obtained as
\begin{align}
    Z(\bar{\beta})=\mathrm{ e}^{- I_{0}[\bar{\beta}]} \,,
   \label{eqch7:zerolooppartitionfunction}
\end{align}
in the zero
loop approximation, and the thermodynamic properties of the system can 
be extracted.

\subsection{The stability criteria from the reduced action of a
hot self-gravitating thin shell in asymptotically AdS space}

Going a step further within this formalism,
we can apply the zero-loop approximation of the path integral in
Eq.~\eqref{eqch7:reducedactionpathintegral},
and go one order up to first order approximation
by evaluating the reduced
action around its stationary points up until second order and
write
\begin{align}
   I_*[\bar{\beta}; \tilde{r}_+,\alpha] = I_{0}[\bar{\beta}]
   + \sum_{ij} H_{ij} \delta r^i \delta r^j\,,
   \label{eqch7:expandaction2}
\end{align}
where $I_{0}[\bar{\beta}] = 
I_*[\bar{\beta};\tilde{r}_+(\bar{\beta}),\alpha(\bar{\beta})]$
is the reduced action evaluated at the stationary points
given in Eq.~\eqref{eqch7:expandaction},
with 
$\tilde{r}_+(\bar{\beta})$ and $\alpha(\bar{\beta})$
being found from 
the stationary conditions of $I_*$,
and $H_{ij} = \eval{\frac{\partial^2 I_*}
{\partial r^i \partial r^j}}_0$ is the Hessian of the
reduced action $I_*$ evaluated at the 
stationary points, 
with the parameters $r^i = (\alpha,\tilde{r}_+)$, with $i = 1,2$. 
The partition function can then 
be written in the saddle point approximation as
\begin{align}
   Z[\bar{\beta}] = \mathrm{ e}^{- I_{0}[\bar{\beta}]} \int D r^i 
   \mathrm{ e}^{- \sum_{jk} H_{jk} \delta r^j \delta r^k}
   \label{eqch7:Z1}\,,
\end{align} 
where the first and second factors are the zero and first-loop
contributions. 
Although we only consider the zero-loop contribution,
i.e., the zero-loop approximation, we also take into account
the first-loop contribution in the sense that 
it gives some information about the stability of the approximation. For the
path integral to converge and so for the formalism to be stable, the
Hessian $H_{ij}$ must be positive definite, i.e.,
the stationary points must correspond to a local minimum
of the reduced action.

The components of the Hessian are
\begin{align}
   & H_{\alpha\alpha} = \frac{8\pi \alpha}{T_\mathrm{ m}}
   \left(\left(\frac{\partial p_\mathrm{ g}}{\partial \alpha}\right)_{\tilde{r}_+} 
   - \left(\frac{\partial p_\mathrm{ m}}{\partial \alpha}\right)_\mathrm{ m} 
   + 8 \pi \alpha p_\mathrm{ m}
\left(\frac{\partial p_\mathrm{ m}}{\partial m} \right)_\alpha \right)\,
   \,,\label{eqch7:I*alphaalpha}
\end{align}
\begin{align}
   & H_{\alpha\,\tilde{r}_+} = 
   \left(\frac{1 +  \frac{3\tilde{r}_+^2}{l^2}}
   {2 T_\mathrm{ m}\sqrt{f_2} l_\mathrm{ p}^2}\right)
   \left(
   \frac{\frac{\alpha}{l^2} + \frac{\tilde{r}_+
   +
   \frac{\tilde{r}_+^3}{l^2}}{2\alpha^2}}
   {f_2} - 8\pi \alpha
   \left(\frac{\partial p_\mathrm{ m}}{\partial m}\right)_{\alpha}\right)
   \,,\label{eqch7:I*rpalpha}
\end{align}
\begin{align}
    H_{\tilde{r}_+\tilde{r}_+} = 
   \left(\frac{1 + \frac{3\tilde{r}_+^2}{l^2}}
   {2 T_\mathrm{ m} \sqrt{f_2} l_\mathrm{ p}}\right)^2
   \left[\frac{1}{l_\mathrm{ p}^2}\left(\frac{\partial T_\mathrm{ m}}{\partial m}\right)_\alpha 
   - \frac{T_\mathrm{ m}}{\alpha \sqrt{f_2}}\right]\,,
   \label{eqch7:I*rprp}
\end{align}
where
\begin{align}\label{eqch7:derpressure}
   \left(\frac{\partial p_\mathrm{ g}}{\partial \alpha}\right)_{\tilde{r}_+} = 
   \frac{1}{8\pi \alpha^2 l_\mathrm{ p}^2}
   \left(\frac{3\frac{\tilde{r}_+ +
   \frac{\tilde{r}_+^3}{l^2}}{2\alpha}
   \left(1 - \frac{\alpha^2}{l^2}\right) - 1 
   - 3\frac{\left(\tilde{r}_+ +
   \frac{\tilde{r}_+^3}{l^2}\right)^2}{4\alpha^2}}{f_2^{3/2}}
   + \frac{1}{f_1^{3/2}}\right)\,\,,
\end{align}
and since here the function depends on three variables, 
the variable that is kept constant while performing the partial derivative 
is written in the subscript of the parenthesis of the partial derivative. We choose  
the sufficient conditions for the positive definiteness of the Hessian to be
\begin{align}
    H_{\alpha \alpha} > 0 \,,
    \label{eqch7:stab1}
\end{align}
\begin{align}
    H_{\tilde{r}_+ \tilde{r}_+} 
   - \frac{H_{ \alpha\,\tilde{r}_+}^2}{H_{\alpha \alpha}} > 0\,.
\label{eqch7:stab2}
\end{align}
Applying Eqs.~\eqref{eqch7:I*alphaalpha}-\eqref{eqch7:I*rprp}
to Eqs.~\eqref{eqch7:stab1} and \eqref{eqch7:stab2}, and including
the marginal case, one has
\begin{align}
 \left(\frac{\partial p_\mathrm{ g}}{\partial \alpha}\right)_{\tilde{r}_+} 
   - \left(\frac{\partial p_\mathrm{ m}}{\partial \alpha}\right)_\mathrm{ m} 
   + 8 \pi \alpha p_\mathrm{ m}
\left(\frac{\partial p_\mathrm{ m}}{\partial m} \right)_\alpha \geq0\,,
    \label{eqch7:stab1final}
\end{align}
\begin{align}
\frac{d
\tilde{r}_+}{d \bar{T}}\geq0\,,
    \label{eqch7:stab2final}
\end{align}
respectively.
Indeed,  one can obtain the derivative of the 
solution $\tilde{r}_+(\bar{\beta})$ 
by applying the derivative of $\bar{\beta}$ to 
Eqs.~\eqref{eqch7:statio1} and~\eqref{eqch7:statio2}, obtaining 
\begin{align}
   \frac{d \tilde{r}_+}{d \bar{\beta}} = - \frac{1}{2 l_\mathrm{ p}^2}\left( 1 + 3
\frac{\tilde{r}_+^2}{l^2}\right)
\left(H_{\tilde{r}_+ \tilde{r}_+}
- \frac{H^2_{\tilde{r}_+ \alpha}}
{H_{\alpha \alpha}} \right)^{\hskip-0.15cm -1}\,\,.
\end{align}
And so Eq.~\eqref{eqch7:stab2} implies that $
\frac{d
\tilde{r}_+}{d \bar{\beta}}<0$, or in terms of temperature
$\frac{d
\tilde{r}_+}{d \bar{T}}>0$, leading to Eq.~\eqref{eqch7:stab2final},
when one includes the marginal case.
Regarding the meaning of these stability conditions, one can verify
that Eq.~\eqref{eqch7:stab1final} is precisely the mechanical stability
condition for a static shell in AdS with constant $\tilde{r}_+$.
Regarding the other condition in Eq.~\eqref{eqch7:stab2final}, in some
sense it is a thermal stability condition, which shall be seen in the 
thermodynamic analysis.

We must comment about the quantity $\frac{d \alpha}{d \bar{\beta}}$.
One can also obtain the derivative of the 
solution $\alpha(\bar{\beta})$ 
by applying the derivative of $\bar{\beta}$ to 
Eqs.~\eqref{eqch7:statio1} and~\eqref{eqch7:statio2}, obtaining
\begin{align}
   \frac{d \alpha}{d \bar{\beta}} = -
\frac{H_{\tilde{r}_+ \alpha}}{H_{\alpha \alpha}} \frac{d\tilde{r}_+}{d\bar{\beta}}\,\,,
\end{align}
i.e., 
$\frac{d \alpha}{d \bar{T}} = -
\frac{H_{\tilde{r}_+ \alpha}}{H_{\alpha \alpha}} \frac{d\tilde{r}_+}{d\bar{T}}$.
Thus, if mechanical stability holds, $H_{\alpha \alpha} > 0$,
Eq.~\eqref{eqch7:stab1},
then the radius of
the shell $\alpha$ decreases with ensemble
temperature if $H_{\tilde{r}_+ \alpha}>0$, and
increases if $H_{\tilde{r}_+ \alpha}<0$. The sign
of $H_{\tilde{r}_+ \alpha}$ depends on the particular
shell one is studying.

\section{Thermodynamics of the hot self-gravitating thin shell 
in the zero-loop approximation\label{sech7:thermodynamics}}\sectionmark{Thermodynamics 
of the hot self-gravitating thin shell}
\thispagestyle{userightbotmark}

In the
statistical mechanics formalism of the 
canonical ensemble, the partition 
function is given by the free energy $F$ as 
$Z= \mathrm{ e}^{-\bar{\beta} F}$, while the zero-loop approximation 
gives a partition function $Z= \mathrm{ e}^{-I_0}$.
By connecting both, one has
$F = \bar{T} I_0$, where $\bar{T}$ is the temperature
of the system,  $\bar{T}=\frac{1}{\bar{\beta}}$.
Then, from
Eq.~\eqref{eqch7:reducedaction1}, the free energy is
\begin{equation}
F =\frac{\tilde{r}_+ \left( l^2 + \tilde{r}_+^2
\right)}{2l^2 l_\mathrm{ p}^2} - \bar{T} S_\mathrm{ m}\,,
\label{eqch7:freeenergy}
\end{equation}
with $\tilde{r}_+$
given by
the solution
$\tilde{r}_+=\tilde{r}_+(\bar{T})$
of Eq.~\eqref{eqch7:statio2final},
and $\alpha$ given by
the solution
$\alpha=\alpha(\tilde{r}_+(\bar{T}))$ of Eq.~\eqref{eqch7:statio1final}
together
with Eq.~\eqref{eqch7:statio2final}.

We can now obtain the thermodynamic quantities
for the system from the derivatives 
of the free energy. In terms of the thermodynamic
energy $E$, the temperature $\bar{T}$, and the entropy $S_\mathrm{ m}$, the
free energy of a system 
and its 
differential are given by 
\begin{align}
F= E - \bar{T} S\,,\quad\quad\quad dF = - S d\bar{T}\,,
\label{eqch7:F1}
\end{align}
respectively.
From the Eqs.~\eqref{eqch7:freeenergy} and \eqref{eqch7:F1},
we obtain 
that the entropy of the system is 
\begin{align}
  S = S_\mathrm{ m}\,,
  \label{eqch7:entropy}
\end{align}
and the mean energy is
\begin{equation} 
E = \frac{1}{2 l_\mathrm{ p}^2}
\tilde{r}_+ \left( 1 + \frac{\tilde{r}_+^2}{l^2} \right)\,,
\label{eqch7:energy}
\end{equation}
with $\tilde{r}_+=\tilde{r}_+(\bar{T})$.
One can identify the Schwarzschild-AdS mass $M$ 
as the right-hand side of Eq.~\eqref{eqch7:energy}, i.e.,  
$E = M$. 

Regarding thermodynamic stability,  
we must verify if the heat capacity
$C$
is positive. If 
\begin{equation}
C \geq0,
\label{eqch7:heatcapacity>0}
\end{equation}
the system is thermodynamically stable,
where the limiting case was included, otherwise it
is unstable.
The heat capacity is defined as
$C =  \frac{d E}{d \bar{T}}$.
Using Eq.~\eqref{eqch7:energy} together with 
Eq.~\eqref{eqch7:statio2final},
we find 
\begin{equation}
C = \frac{1}{l_\mathrm{ p}^2}\frac{ \left( 1 +
\frac{3\tilde{r}_+^2}{l^2} \right) \iota^2(\tilde{r}_+) }{
\frac{12\tilde{r}_+
\iota(\tilde{r}_+)}{l^2} - 2 \left( 1 +
\frac{3\tilde{r}_+^2}{l^2} \right) 
\frac{\partial \iota(\tilde{r}_+)}{\partial \tilde{r}_+} }\,,
\label{eqch7:heatcapacity}
\end{equation}
where $\iota(\tilde{r}_+)=\frac{1 + 3\frac{\tilde{r}_+^2}{l^2}}
{T_\mathrm{ m}(\tilde{r}_+,\alpha(\tilde{r}_+))
\sqrt{f_2(\alpha(\tilde{r}_+))}}$, see 
Eq.~\eqref{eqch7:iota}, and $\tilde{r}_+=\tilde{r}_+(\bar{T})$.
One can see that the heat capacity is positive if
\begin{align}
6\frac{\tilde{r}_+}{l^2} - \iota'(\tilde{r}_+) \bar{T}\geq 0\,,
\label{eqch7:stabheta} 
\end{align}
where the limiting case was included.
Using Eq.~\eqref{eqch7:statio2final}, we find that Eq.~\eqref{eqch7:stabheta}
is equivalent to $\frac{d\tilde{r}_+}{d \bar{T}}\geq0$ which is
Eq.~\eqref{eqch7:stab2final} of the ensemble theory.  One can now see that the
remaining stability condition Eq.~\eqref{eqch7:stab1} is not present or
cannot be accessed by the thermodynamics of the system.
It is moreover interesting to see that the
thermodynamic stability of the canonical ensemble given by
Eqs.~\eqref{eqch7:heatcapacity>0}-\eqref{eqch7:stabheta} is
also given by the saddle point
approximation of the effective reduced
action  $I_*[\bar{\beta};\tilde{r}_+]$, only dependent in the parameter
$\tilde{r}_+$.  This may be due to the fact that
$\tilde{r}_+$ is associated to the quasilocal energy and so the
effective reduced
action $I_*[\bar{\beta};\tilde{r}_+]$
plays the role of the appropriate generalized free
energy that when minimized yields indeed the thermodynamic equilibrium
and stability of the canonical ensemble. It is important to note also
that the thermodynamic stability of the canonical ensemble is
different from the intrinsic stability of the system in the sense of
Callen, as intrinsic stability requires more conditions on the
concavity of the free energy.

\section{Specific case of matter thin shell with 
barotropic equation of state
\label{sech7:unspecifiedeos}}

In order to proceed with the analysis of the
canonical ensemble of a self-gravitating 
matter thin shell, we must now
give the equations of state for the matter
in the shell. Here, we give an
equation of state for the pressure
in the form of a barotropic equation, i.e.,
\begin{align}
  p_\mathrm{ m}(m,\alpha) = \frac13\frac{m}{4\pi\alpha^2}\,.
  \label{eqch7:eospressure}
\end{align}
We choose the equation of state for
the temperature of the matter as
\begin{align}
  T_\mathrm{ m} = \frac{4}{3 C_0} \frac{m^{\frac{1}{4}}}
  {\left( 4\pi \alpha^2\right)^{\frac{1}{4}}}\,,
   \label{eqch7:tempspecific}
\end{align}
where $C_0$ is a constant with units.
Then, integrating the first law of thermodynamics yields
that the matter
entropy has the equation 
\begin{align}
  S_\mathrm{ m} = C_0 m^{\frac{3}{4}} (4\pi \alpha^2)^{\frac{1}{4}}\,.
  \label{eqch7:entropyspecific}
\end{align}
A more general equation for $p_\mathrm{ m}(m,\alpha)$ in
Eq.~\eqref{eqch7:eospressure}
could be chosen,
e.g., $p_\mathrm{ m}(m,\alpha) = \lambda\frac{m}{4\pi\alpha^2}$,
where $\lambda$ is a constant,
with $\lambda = \frac{1}{2}$ corresponding to the barotropic 
equation of state of a two-dimensional ultrarelativistic 
gas.
The
mechanical stability condition, Eq.~\eqref{eqch7:stab1final},
requires that  $\lambda< \frac{1}{2}$.
Since the dominant energy condition requires 
$-1<\lambda<1$, a reasonable range for $\lambda$ is 
$0 <\lambda< \frac{1}{2}$, as we require the pressure to be 
positive.
If this general expression
for the pressure is integrated, using from the first law of thermodynamics
that $p_\mathrm{ m} = - 
\left(\frac{\partial m}{\partial A_\mathrm{ m}}\right)_{S_\mathrm{ m}}$, the
partial derivative in relation the
area $A_\mathrm{ m}$ being defined at constant matter entropy,
one would obtain the expression 
for $S_\mathrm{ m}$ in the form
$S_\mathrm{ m}(m,\alpha) = S_\mathrm{ m}(m(4\pi \alpha^2)^{\lambda})$, i.e.,
$S_\mathrm{ m}$ is an arbitrary function of $m(4\pi \alpha^2)^{\lambda}$.
Here, we choose a power-law expression for the entropy
of the form 
$S_\mathrm{ m}(m,\alpha) = C_0 m^{\delta} (4\pi \alpha^2)^{\delta \lambda}$,
with $C_0$ being a constant and $\delta$ a number.
Then, the temperature
equation of state
could be deduced using the first law of thermodynamics,
i.e., 
$\frac1{T_\mathrm{ m}}= 
\left(\frac{\partial S_\mathrm{ m}}{\partial m}\right)_{A_\mathrm{ m}}$
giving $T_m =
\frac{m^{1-\delta}}{\delta C_0 (4\pi \alpha^2)^{\delta \lambda}}$.
We could instead
have picked up this equation of state for 
$T_m$ from the necessity of the equality of the second order
cross derivatives to have an exact $S_\mathrm{ m}$, and then
integrate the first law to find $S_\mathrm{ m}$ itself.
We narrow further the analysis to a
specific $\lambda$ and $\delta$, using 
Eq.~\eqref{eqch7:eospressure}-\eqref{eqch7:entropyspecific}.
In particular, we choose $\lambda = \frac{1}{3}$, with the 
pressure given in Eq.~\eqref{eqch7:eospressure}.
We also choose
$\delta = \frac{3}{4}$, so that the
temperature and the entropy have
equations given in 
Eq.~\eqref{eqch7:tempspecific} and
Eq.~\eqref{eqch7:entropyspecific}, respectively.
For numerical purposes, it is best to write the pressure as 
$l_\mathrm{p}^2 l
p_\mathrm{ m}(m,\alpha) =
\frac13 
\frac{m\frac{l_\mathrm{ p}^2}{l}}{4\pi\frac{\alpha^2}{l^2}}
$, the temperature as
$T_\mathrm{ m} = \frac1l \frac{4}{3c_0}
\left(\frac{l}{l_\mathrm{ c}}\right)^\frac14
\left(\frac{l}{l_\mathrm{ p}}\right)^\frac12
\left(\frac{m l_\mathrm{ p}^2}{l}\right)^{\frac{1}{4}} \left( 4\pi
\frac{\alpha^2}{l^2}\right)^{-\frac{1}{4}}$, and
the 
entropy as $S_\mathrm{ m} = c_0 \left(\frac{l_\mathrm{ c}}{l}\right)^\frac14
\left(\frac{l}{l_\mathrm{ p}}\right)^\frac32
\left(\frac{m \,l_\mathrm{ p}^2}{l}\right)^{\frac{3}{4}}
\left(4\pi \frac{\alpha^2}{l^2}\right)^{\frac{1}{4}}$, i.e., $C_0 =
c_0 l_\mathrm{ c}^{\frac{1}{4}}$, where $c_0$ has no units
and $l_\mathrm{ c}$ can be
understood as the Compton wavelength associated to the rest mass of
the constituents of the shell.
The motivation for the matter equations of state given above is
both physical and mathematical.  Physically, the equations of state
resemble the equations of state of a radiation gas. Namely, the
equation of state for the pressure $p_\mathrm{ m}(m,\alpha)$ can be
thought of as the equation of state of a three-dimensional radiation
gas confined in a very thin shell of small width $l_c$. It can also be
thought of as a two-dimensional gas of a fundamental field with some
Compton wavelength $l_c$.
Mathematically, it allows for an analytical
treatment of the balance of the pressure which facilitates the search
for the solutions of the shell and the analysis of its stability.

\begin{figure}[h]
   \centering
   \includegraphics[width=0.60\textwidth]{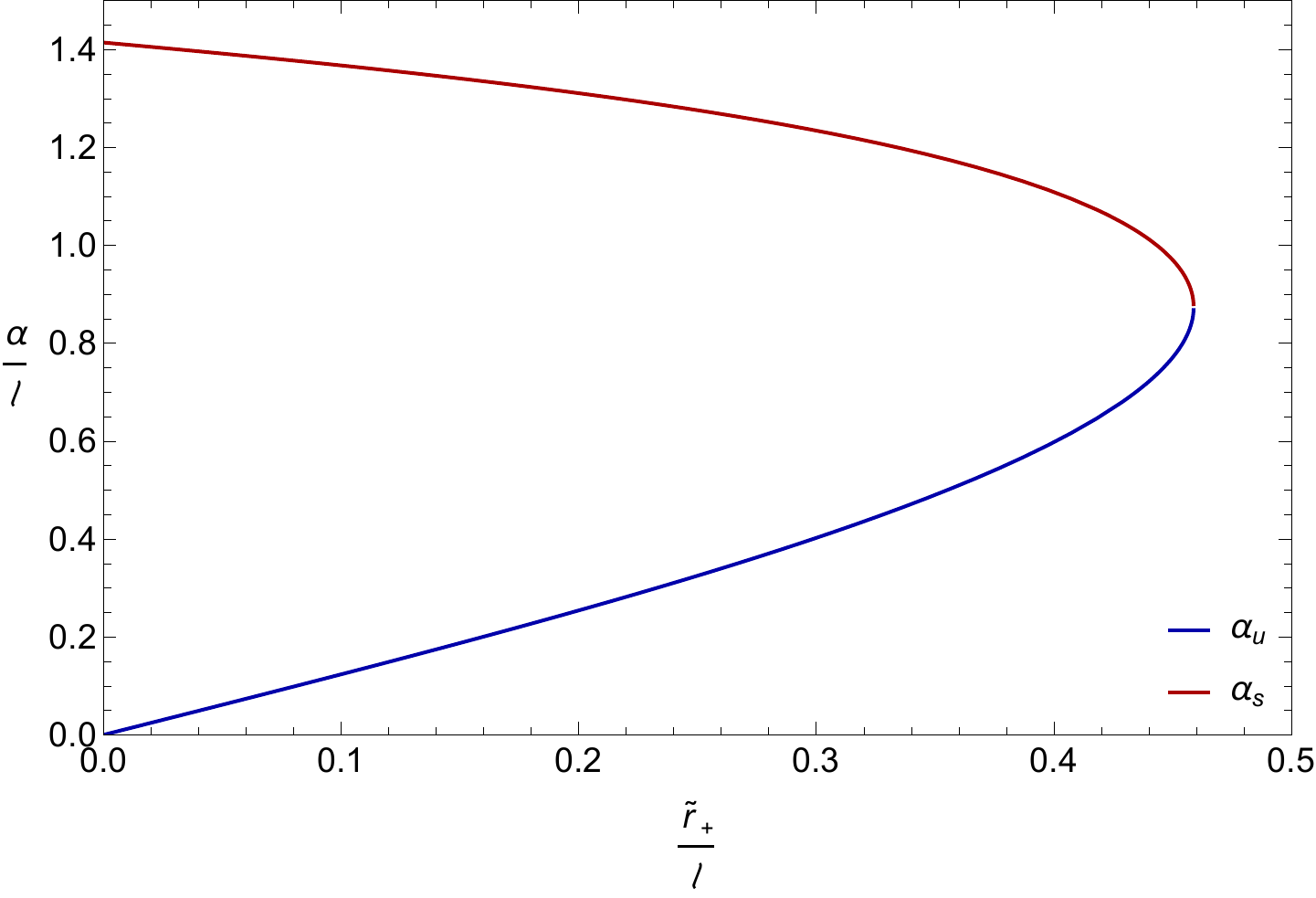}
   \caption{\label{figch7:alphainrp}Solutions of the balance of pressure 
   $\frac{\alpha_\mathrm{ u}}{l}$ and $\frac{\alpha_\mathrm{ s}}{l}$
   as function of $\frac{\tilde{r}_+}{l}$.}
\end{figure}

With the equations of state described, we can solve numerically 
Eq.~\eqref{eqch7:statio2}, to obtain two solutions for the
radius $\alpha$ of the shell which is
written as $\alpha_\mathrm{ u}(\tilde{r}_+)$
and $\alpha_\mathrm{ s}(\tilde{r}_+)$,
see Fig.~\ref{figch7:alphainrp}, the meaning of the subscripts u and s
will turn up shortly.
The solution
$\alpha_\mathrm{ u}(\tilde{r}_+)$ 
is monotonically increasing, 
while $\alpha_\mathrm{ s}(\tilde{r}_+)$ 
is monotonically decreasing, until both meet 
a common point. By evaluating the stability condition 
in Eq.~\eqref{eqch7:stab1final},
it turns out that the solution
$\alpha_\mathrm{ u}(\tilde{r}_+)$  is 
mechanically unstable, while
$\alpha_\mathrm{ s}(\tilde{r}_+)$  is mechanically stable, hence the 
nature of the subscripts for the solutions.
This mechanical
behavior of the thin shell
is rather like the radius-mass behavior of a white dwarf
or a neutron star. The two solutions translate into two possible radii
for the shell for a given energy, one which is large and another which
is small.  The small radius solution has very high pressure and is
unstable, the large radius solution has low pressure and is stable.
Physically, this is similar to the two solutions appearing in models
of astrophysical objects, such as white dwarfs, neutron stars
with polytropic-type equations of state, one solution being unstable
while the other being stable.

\begin{figure}[h]
   \centering
   \includegraphics[width=0.60\textwidth]{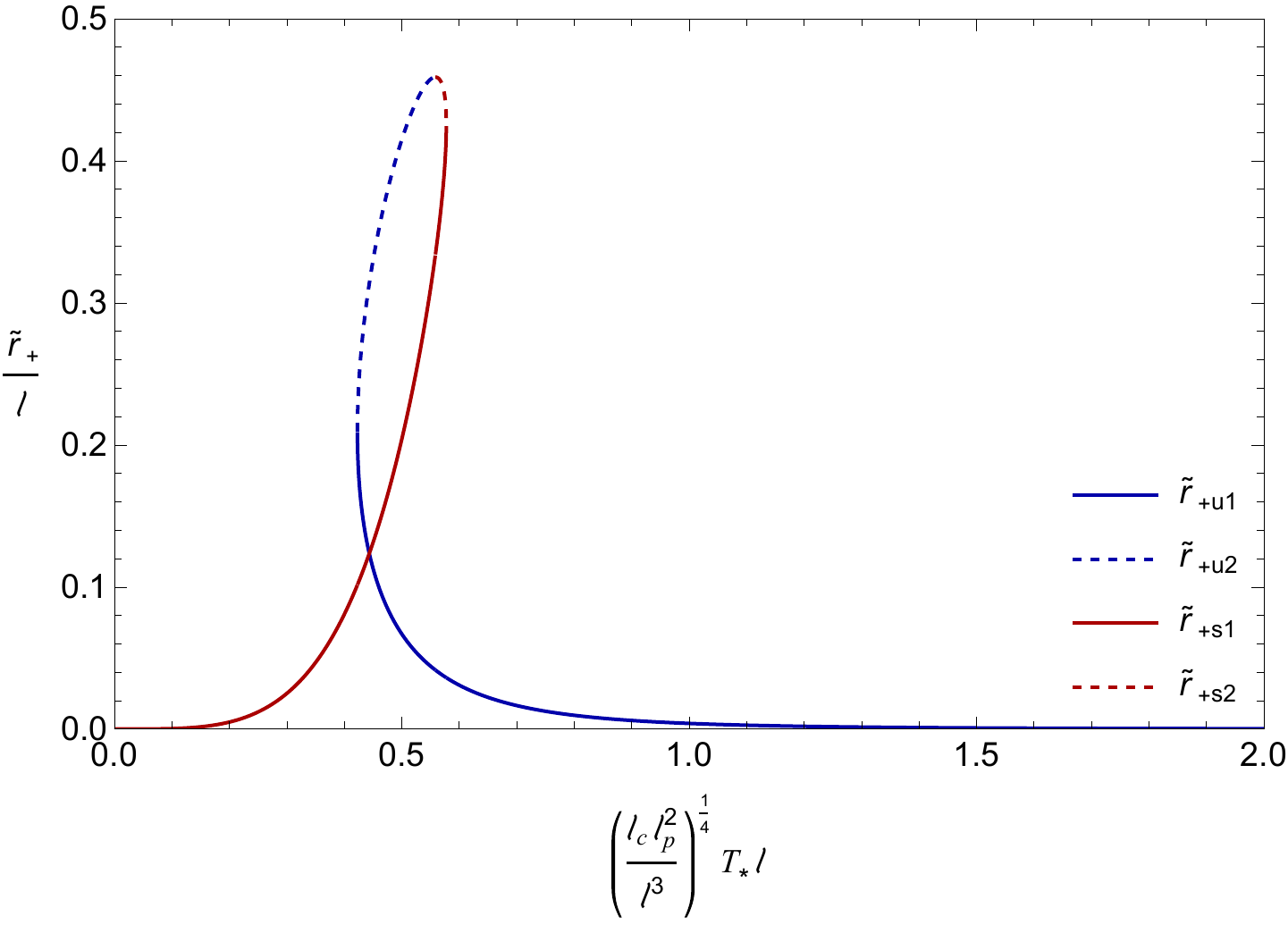}
   \caption{\label{figch7:rpintstar} Solutions of the ensemble 
$\frac{\tilde{r}_{+\mathrm{ u}1}}{l}$,
$\frac{\tilde{r}_{+\mathrm{ u}2}}{l}$,
$\frac{\tilde{r}_{+\mathrm{ s}1}}{l}$, and 
$\frac{\tilde{r}_{+\mathrm{ s}2}}{l}$,
as functions of
$\bar{T}l
\left(\frac{l_\mathrm{ c}}{l}\right)^\frac14
\left(\frac{l_\mathrm{ p}}{l}\right)^\frac12$.
Both
$\frac{\tilde{r}_{+\mathrm{ u}1}}{l}$ and
$\frac{\tilde{r}_{+\mathrm{ u}2}}{l}$
have shell radius $\alpha_\mathrm{ u}$,
while both
$\frac{\tilde{r}_{+\mathrm{ s}1}}{l}$ and
$\frac{\tilde{r}_{+\mathrm{ s}2}}{l}$
have shell radius $\alpha_\mathrm{ s}$.}
\end{figure}

Knowing the solutions
$\alpha_\mathrm{ u}(\tilde{r}_+)$ and
$\alpha_\mathrm{ s}(\tilde{r}_+)$
for the radius of the shell, we can put them
into Eq.~\eqref{eqch7:statio2} in order to obtain the solutions
for the gravitational
radius $\tilde{r}_+(\bar{T})$. We find that there are
four solutions in total,
two solutions
$\tilde{r}_{+\mathrm{ u}1}(\bar{T})$ and
$\tilde{r}_{+\mathrm{ u}2}(\bar{T})$  with
shell radius
$\alpha_\mathrm{ u}$, i.e.,
$\alpha_\mathrm{ u} (\tilde{r}_{+\mathrm{ u}1}(\bar{T}))$ and
$\alpha_\mathrm{ u} (\tilde{r}_{+\mathrm{ u}2}(\bar{T}))$, respectively,
and other two solutions
$\tilde{r}_{+\mathrm{ s}1}(\bar{T})$ and
$\tilde{r}_{+\mathrm{ s}2}(\bar{T})$  with shell radius
$\alpha_\mathrm{ s}$, i.e., with
$\alpha_\mathrm{ s}(\tilde{r}_{+\mathrm{ s}1}(\bar{T}))$
and
$\alpha_\mathrm{ s}(\tilde{r}_{+\mathrm{ s}2}(\bar{T}))$,
respectively.
These solutions are
shown in Fig.~\ref{figch7:rpintstar}. Regarding stability, the solutions
are thermodynamically stable if the gravitational radius increases
with the temperature $\bar{T}$.
From the figure,
$\tilde{r}_{+\mathrm{ u}1}(\bar{T})$ 
and $\tilde{r}_{+\mathrm{ s}2}(\bar{T})$
are thermodynamically unstable, and 
$\tilde{r}_{+\mathrm{ u}2}(\bar{T})$
and $\tilde{r}_{+\mathrm{ s}1}(\bar{T})$  are
thermodynamically stable.

\begin{figure}[h]
\centering
\includegraphics[width=0.60\textwidth]{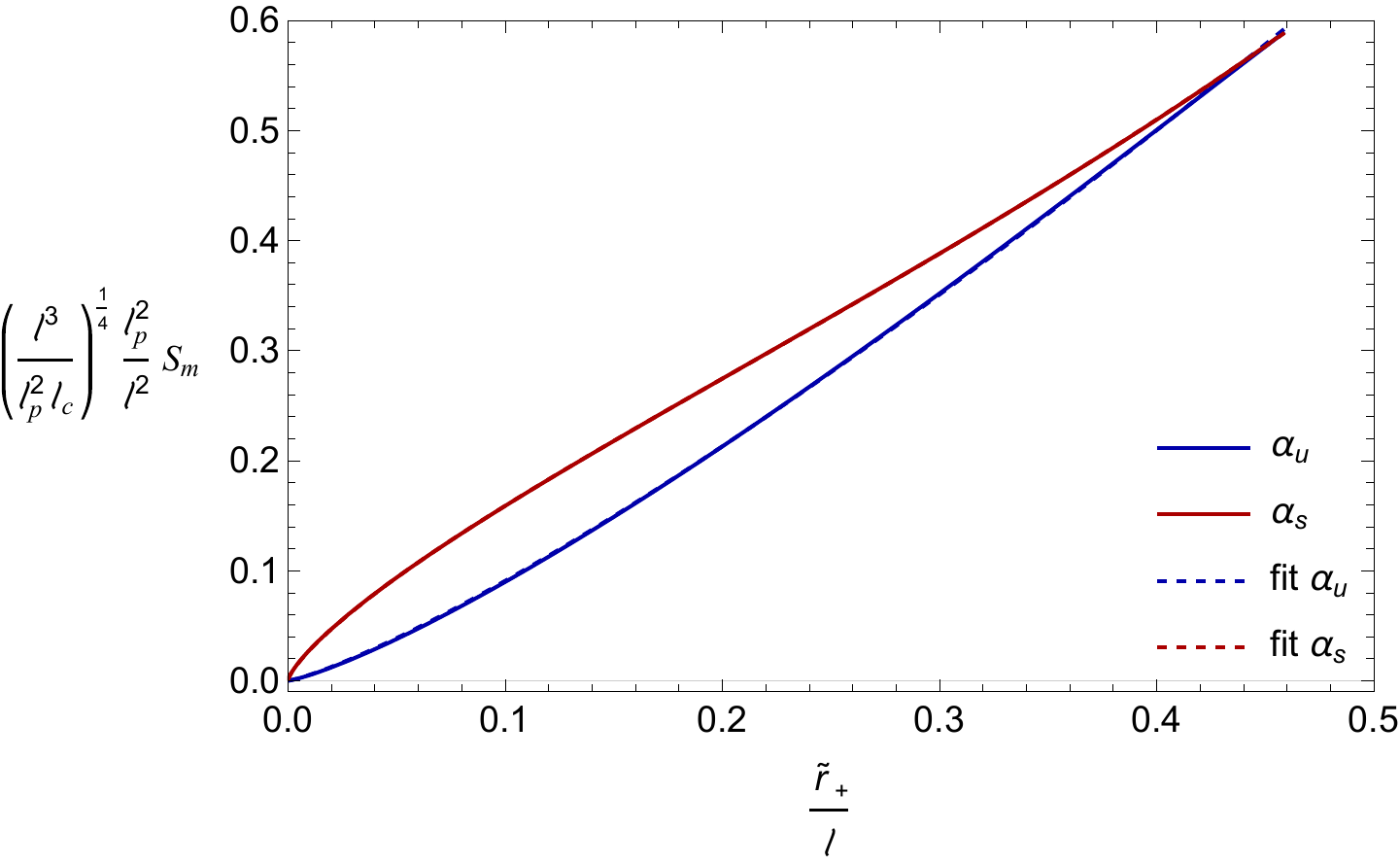}
\caption{\label{figch7:entropy} Matter entropy $\left(\frac{l^3}{l_\mathrm{ p}^2
l_\mathrm{ c}}\right)^{\frac{1}{4}} \frac{l_\mathrm{ p}^2}{l^2}S_\mathrm{ m}$ in function of the
gravitational radius $\frac{\tilde{r}_+}{l}$ for the two shell radius
solutions $\alpha_\mathrm{ u}(\tilde{r}_+)$
and $\alpha_\mathrm{ s}(\tilde{r}_+)$.  A fit
was performed for each branch, with $\left(\frac{l^3}{l_\mathrm{ p}^2
l_\mathrm{ c}}\right)^{\frac{1}{4}} \frac{l_\mathrm{ p}^2}{l^2}S_\mathrm{ m}=1.54662
(\frac{\tilde{r}_+}{l})^{1.2323}$ for the case of
$\alpha_\mathrm{ u}(\tilde{r}_+)$ and $\left(\frac{l^3}{l_\mathrm{ p}^2
l_\mathrm{ c}}\right)^{\frac{1}{4}} \frac{l_\mathrm{ p}^2}{l^2}S_\mathrm{
m}=0.898912(\frac{\tilde{r}_+}{l})^{0.755675}
+0.867397(\frac{\tilde{r}_+}{l})^{2.91424}$ for
$\alpha_\mathrm{ s}(\tilde{r}_+)$, with respective coefficients
of determination
$R^2 = 0.999992$ and $R^2=1$, with this last equality being
approximate.}
\end{figure}

We now analyze the entropy $S_\mathrm{ m}$, see Fig.~\ref{figch7:entropy}.
We performed a polynomial fit to the matter entropy in order to understand 
its leading power of $\tilde{r}_+$.
The fit for $S_\mathrm{ m}$
given in Eq.~\eqref{eqch7:entropyspecific} for the solution $\alpha_\mathrm{
u}$ as a function of $\tilde{r}_{+\mathrm{ u}}$ is described by the
function $\left(\frac{l^3}{l_\mathrm{ p}^2 l_\mathrm{ c}}\right)^{\frac{1}{4}}
\frac{l_\mathrm{ p}^2}{l^2}S_\mathrm{ m}=1.54662 (\frac{\tilde{r}_{+\mathrm{
u}}}{l})^{1.2323}$ with a coefficient of determination $R^2=0.999992$.
The fit of $S_\mathrm{ m}$ given in Eq.~\eqref{eqch7:entropyspecific} for the
solution $\alpha_\mathrm{ s}$ as a function of $\tilde{r}_{+\mathrm{ s}}$ is
described by the function $\left(\frac{l^3}{l_\mathrm{ p}^2
l_\mathrm{ c}}\right)^{\frac{1}{4}} \frac{l_\mathrm{ p}^2}{l^2}S_\mathrm{
m}=0.898912(\frac{\tilde{r}_{+\mathrm{ s}}}{l})^{0.755675}
+0.867397(\frac{\tilde{r}_{+\mathrm{ s}}}{l})^{2.91424}$ with a
coefficient of determination $R^2=1$, with this equality being
approximate. We attempted another fit of $S_\mathrm{ m}$ for the solution $\alpha_\mathrm{
s}$ with just one power, giving $\left(\frac{l^3}{l_\mathrm{ p}^2
l_\mathrm{ c}}\right)^{\frac{1}{4}} \frac{l_\mathrm{ p}^2}{l^2}S_\mathrm{ m}=1.12374
(\frac{\tilde{r}_{+\mathrm{ s}}}{l})^{0.867504}$, with $R^2 = 0.999607$,
however the differences of the fit are visible in the plot and 
we considered instead the fit with two powers.
The fits we obtained are very close to the numerical results for the
$S_\mathrm{ m}$, which is surprising and one could wonder if there might
be an analytic solution.  But in order to obtain the expression of
$S_\mathrm{ m}$, one needs to solve a quintic polynomial equation and we
were not able to find an analytic solution.  A feature that the fits
do not capture is the fact that $S_\mathrm{ m}$ is only defined in the
interval $0<\frac{\tilde{r}_+}{l}<
0.4589$ with this last number being approximate.

\begin{figure}[h]
\centering
\includegraphics[width=0.60\textwidth]{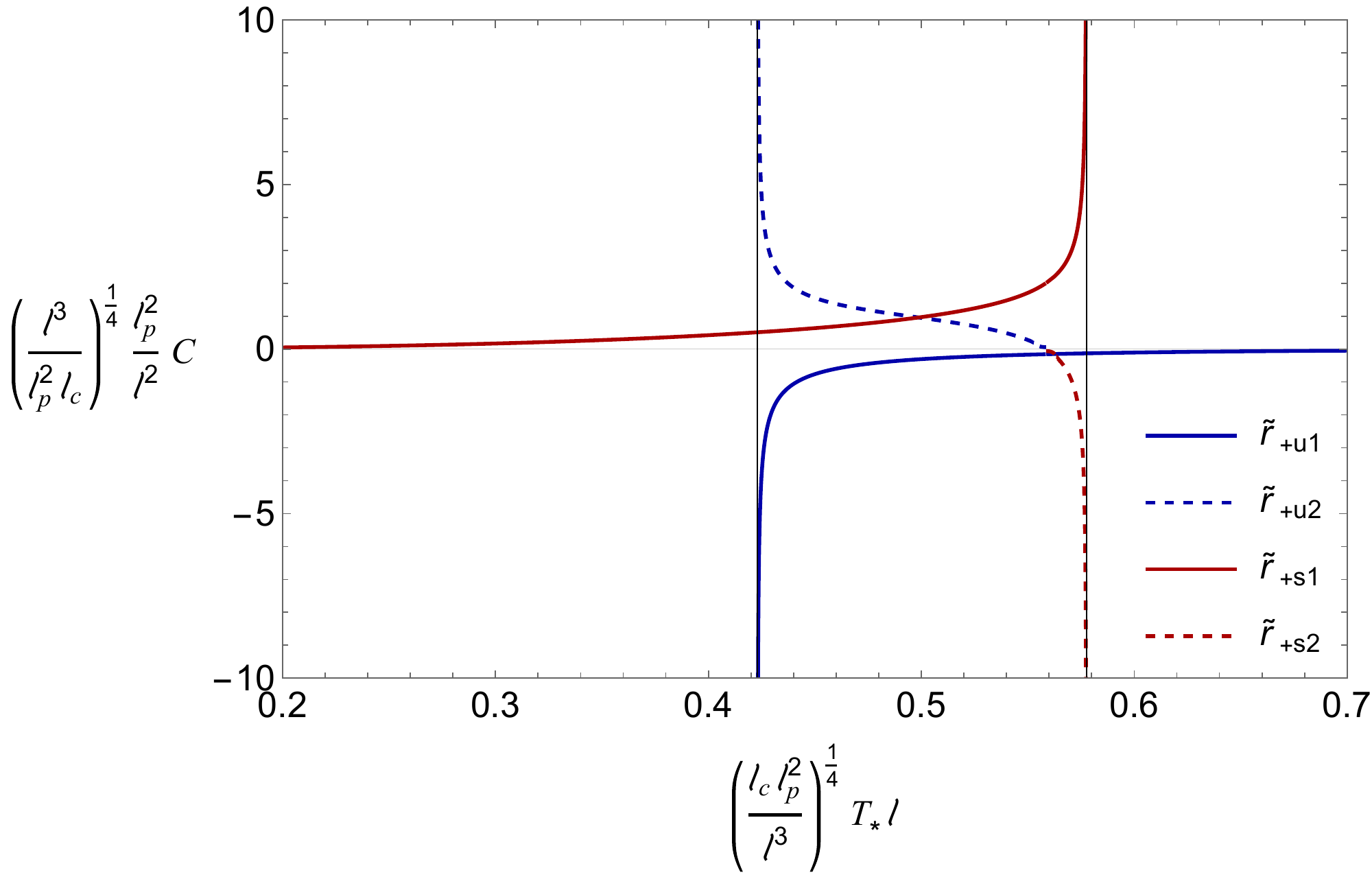}
\caption{\label{figch7:Cintstar} Adimensional heat capacity for the
solutions
$\tilde{r}_{+\mathrm{ u}1}$,
$\tilde{r}_{+\mathrm{ u}2}$,
$\tilde{r}_{+\mathrm{ s}1}$, and
$\tilde{r}_{+\mathrm{ s}2}$
as functions of $\bar{T}l
\left(\frac{l_\mathrm{ c}}{l}\right)^\frac14
\left(\frac{l_\mathrm{ p}}{l}\right)^\frac12$,
where the solutions $\alpha_\mathrm{ u}$
and $\alpha_\mathrm{ s}$ are also assumed.
The solutions
$\tilde{r}_{+\mathrm{ u}1}$ and
$\tilde{r}_{+\mathrm{ s}2}$ are thermodynamically unstable, while
$\tilde{r}_{+\mathrm{ u}2}$ and
$\tilde{r}_{+\mathrm{ s}1}$ are thermodynamically stable.}
\end{figure}

Another equivalent indicator of
thermal stability is given by the positivity of the heat capacity,
see Fig.~\ref{figch7:Cintstar}.  Yet, it must be noticed that
$\tilde{r}_{+\mathrm{ u}2}(\bar{T})$ has a shell radius
$\alpha_\mathrm{ u}(\tilde{r}_{+\mathrm{ u}2}(\bar{T}))$,
which is mechanically unstable. Therefore, the only fully stable
solution is $\tilde{r}_{+\mathrm{ s}1}(\bar{T})$ with shell radius
$\alpha_\mathrm{ s}(\tilde{r}_{+\mathrm{ s}1}(\bar{T}))$.

\section{Hot thin shell versus black hole in AdS}
\label{sech7:tsxbh}

\subsection{The black hole}


We now present the relevant quantities of the canonical 
ensemble of a Schwarzschild-AdS black hole for completeness and 
also because it is important in the analysis of phase transitions regarding the 
hot thin shell. In order to obtain the reduced 
action and then the zero-loop
action, we can carry on in a similar way the 
calculations above but now with black hole boundary conditions rather than
thin shell ones, or we can simply use the Hawking-Page results.

The
action of the Hawking-Page black hole solutions is
\begin{align}
I_\mathrm{ bh} = 
\frac{1}{2 l_\mathrm{ p}^2\bar{T}}
\left( r_+ + \frac{r_+^3}{l^2}
\right)  - \frac{\pi r_+^2}{l_\mathrm{ p}^2}
\,,
\label{eqch7:actionbh}
\end{align}
with the radius $r_+$ being a function of $\bar{T}$.
Indeed here, the gravitational radius $r_+$, which
is also a horizon radius, is
given by the equation
$\bar{\beta} \equiv\frac{1}{\bar{T}}= \frac{\iota(r_+)}
{1 + 3\frac{r_+^2}{l^2}}$, see also Eq.~\eqref{eqch7:statio2final} with
$\iota(r_+)=4\pi r_+$.

\begin{figure}[h]
\centering
\includegraphics[width=0.60\textwidth]{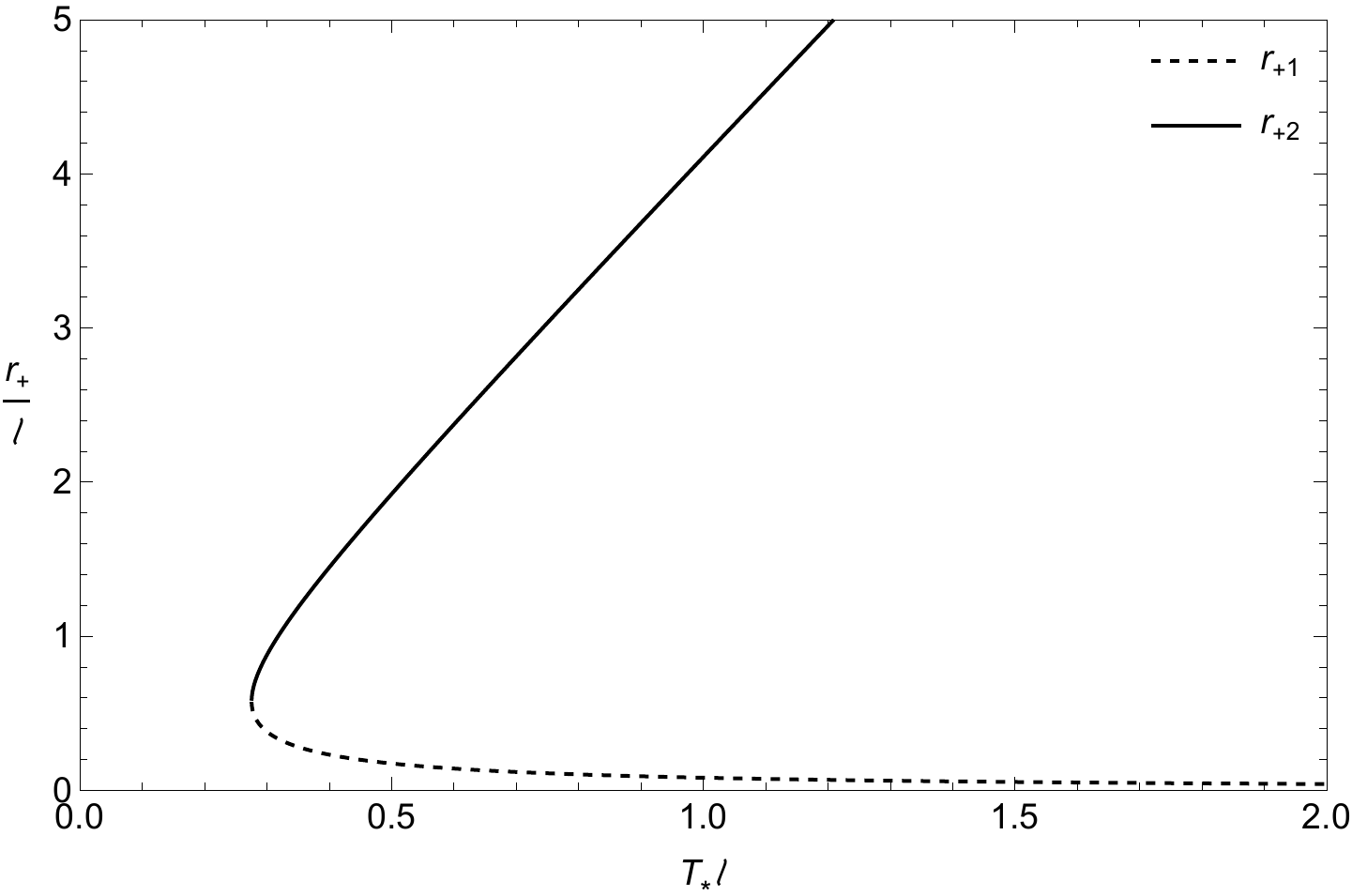}
\caption{\label{figch7:r+bh} Solutions of the ensemble 
$\frac{r_{+1}}{l}$
and 
$\frac{r_{+2}}{l}$
for the black hole in asymptotically AdS.}
\end{figure}

Therefore, we have to solve  for $r_+$ the equation
$\bar{T}=\frac{1 + 3\frac{r_+^2}{l^2}}{4\pi r_+}$.
The solutions
$\frac{r_+}{l}$ as a function of $\bar{T}l$ are given
by
\begin{align}
\frac{r_+}{l} = 
\frac{2\pi l \bar{T}}{3} \pm \frac{1}{3}\sqrt{(2\pi l \bar{T})^2 - 3}.
\label{eqch7:radiusbh}
\end{align}
Thus, for $l \bar{T}\geq\frac{\sqrt3}{2\pi}$, there are two black
hole solutions, $r_{+1}(\bar{T})$, the solution with
the minus sign, which is thermodynamically unstable, and $r_{+2}(\bar{T})$,
the solution with the plus sign, 
which is stable. 
When equality holds, $l \bar{T} = \frac{\sqrt3}{2\pi}$, one has a degenerate
solution,
 $r_{+1}(\bar{T})=r_{+2}(\bar{T})=\frac{1}{\sqrt{3}}$. For $l \bar{T}<\frac{\sqrt3}{2\pi}$
there are no solutions, see Fig.~\ref{figch7:r+bh}. 
The numerical value of this 
critical temperature is
$l \bar{T}=\frac{\sqrt3}{2\pi}=0.276$ approximately, with 
the corresponding horizon radius $\frac{r_+}{l}=\frac{1}{\sqrt{3}}=0.577$.

\begin{figure}[h]
\centering
\includegraphics[width=0.60\textwidth]{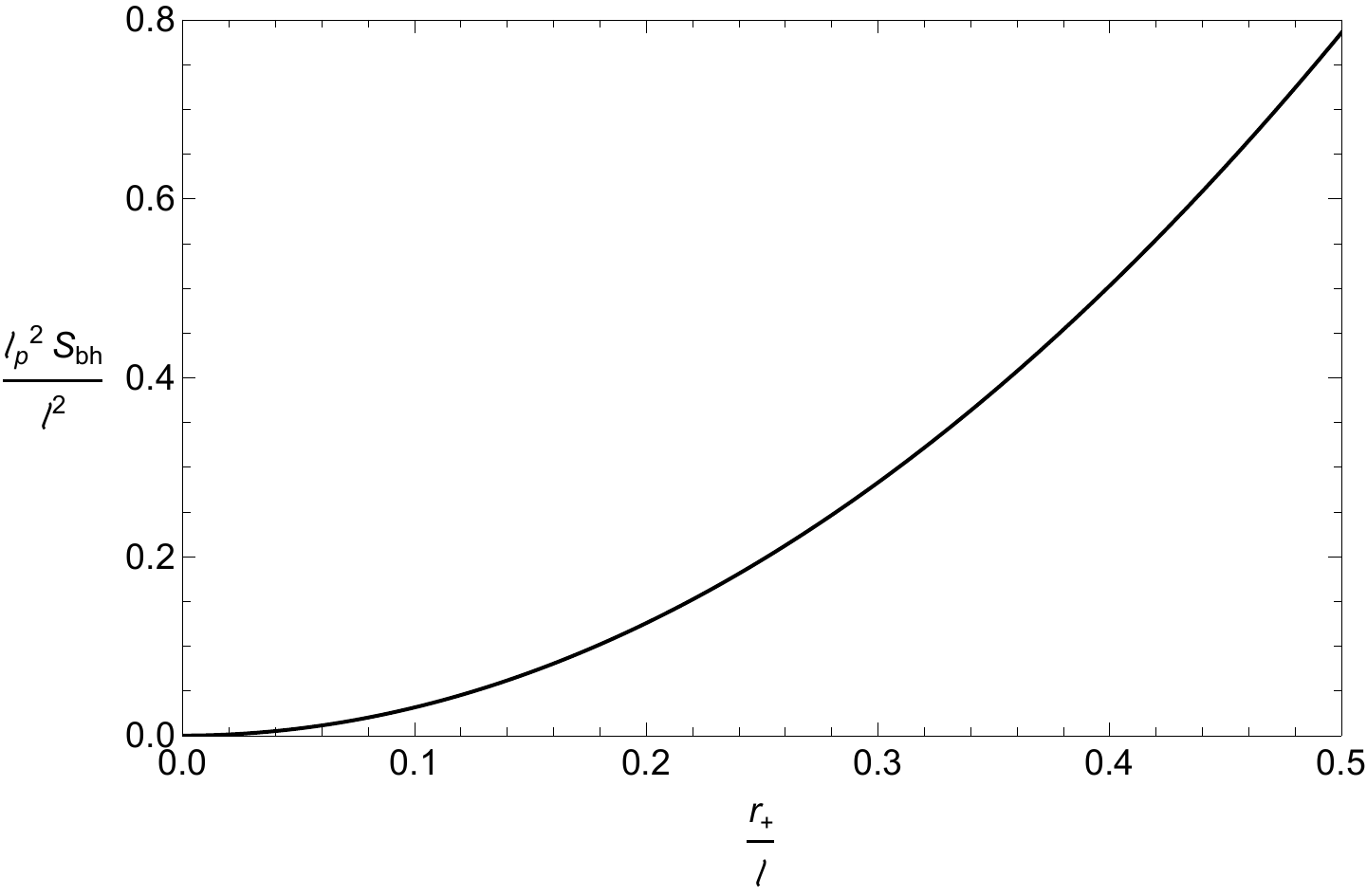}
\caption{\label{figch7:entropybh} Black hole entropy 
$\frac{l^2_\mathrm{ p}}{l^2}S_\mathrm{ bh}$ as a function of 
the horizon radius $\frac{r_+}{l}$,
which stands either for $\frac{r_{+1}}{l}$
or for $\frac{r_{+2}}{l}$.}
\end{figure}

The black hole entropy can be obtained from the action as
\begin{align}
S_\mathrm{ bh}=\frac{\pi r_+^2}{l_\mathrm{ p}^2}\,,
\label{eqch7:Sbh}
\end{align}
i.e., the Bekenstein-Hawking entropy, with
$r_+$ standing for $r_{+1}$ or $r_{+2}$. 
The entropy describes the usual parabola, see
Fig.~\eqref{figch7:entropybh}.

\begin{figure}[h]
\centering
\includegraphics[width=0.60\textwidth]{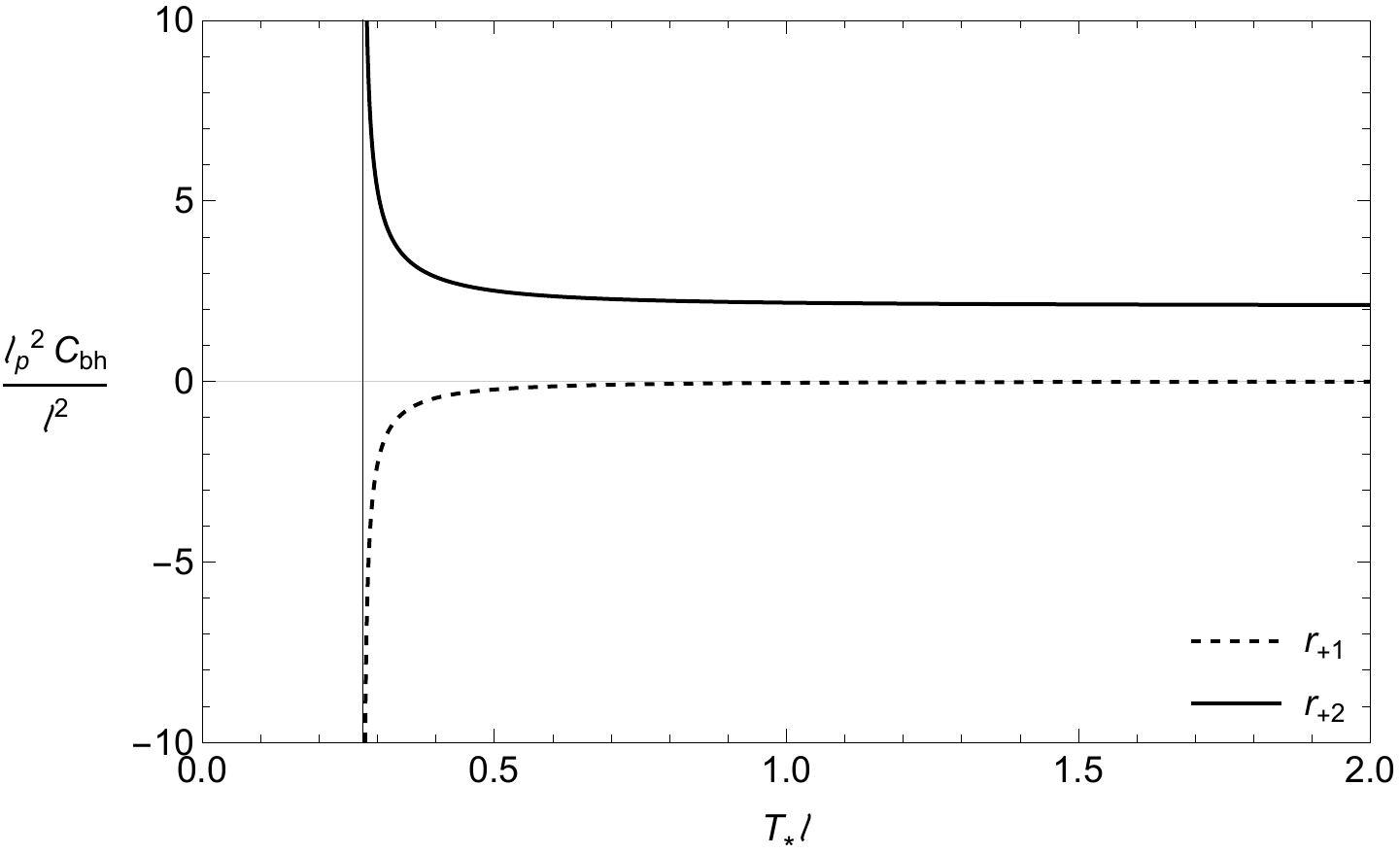}
\caption{\label{figch7:Cbh} Adimensional heat capacity $\frac{l_\mathrm{ p}^2
C_\mathrm{ bh}}{l^2}$ of the black hole solutions $r_{+1}$ and $r_{+2}$ in
function of $\bar{T} l$.}
\end{figure}

The heat capacity for the Schwarzschild-AdS black hole is 
\begin{align}
C_\mathrm{ bh}=\frac{2\pi r_+^2 \left(1 + 3\frac{\tilde{r}_+^2}{l^2}\right)}
{3\left(\frac{r_+^2}{l^2} -
\frac{1}{3}\right)}\,, \label{eqch7:Cbh}
\end{align}
for each solution $r_{+1}(\bar{T})$ and $r_{+2}(\bar{T})$, see
Fig.~\ref{figch7:Cbh}.  The heat capacity is positive for $\frac{r_+}{l}
> \frac{\sqrt{3}}{3}$, and so $r_{+1}$ is thermodynamic unstable and
$r_{+2}$ is thermodynamic stable.

We must add that pure hot AdS and
black hole in AdS compete to be the prominent
thermodynamic phase, with the phase that has the minimum action
being the one that is favored. Pure hot AdS has zero action,
$I_\mathrm{ PAdS}=0$, so if $I_\mathrm{ bh}>0$ then AdS is favored,
if $I_\mathrm{ bh}=0$ the two phases coexist equally,
and if $I_\mathrm{ bh}<0$ then the black hole is favored. 
From the black hole action, Eq.~\eqref{eqch7:actionbh}, 
one finds that
as one increases $l \bar{T}$ from zero, there is a
first order phase transition
from thermal AdS to the stable black hole state, the transition
happening at $l \bar{T}=\frac{1}{\pi}=0.318$, the latter
number being approximate, to the stable black 
hole with horizon radius $\frac{{r}_{+2}}{l}=1$.

\subsection{Hot thin shell versus black hole and favorable states}

\subsubsection{Gravitational radii, entropies, and heat capacities}

We are interested in the comparison between the properties obtained for
the hot
thin shell in AdS and the properties of the black hole in AdS.
More specifically, we can compare the gravitational radii of each thin
shell with the gravitational radius of the black hole, examine their
entropies, and analyze their heat capacities.

First, we compare the two possible gravitational
radii of each thin shell and the gravitational radii of the black
hole.
For the mechanically unstable hot thin shell, which is given by the shell
radius $\alpha_\mathrm{ u}$, we have found that there are two branches for
the gravitational radius. One of the branches, $\tilde{r}_{+\mathrm{ u}1}$
is thermodynamically unstable, while the other branch
$\tilde{r}_{+\mathrm{ u}2}$ is thermodynamically stable.  As well, for the
Hawking-Page black hole, the horizon radius $r_+$ has one branch
$r_{+1}$ which is thermodynamically unstable, and another branch
$r_{+2}$ which is thermodynamically stable.  It is clear from
Fig.~\ref{figch7:alphainrp} and Fig.~\ref{figch7:r+bh} that the two
solutions for the gravitational radius with thin shell radius
$\alpha_\mathrm{ u}$, which is mechanically unstable, share similarities
with the two solutions for the horizon radius of the black hole. In
detail, the thermodynamically unstable branch of the thin shell
solution, $\tilde{r}_{+\mathrm{ u}1}$, follows the same behavior as the
thermodynamically unstable black hole solution, $r_{+1}$. As well, the
thermodynamically stable branch of the thin shell solution,
$\tilde{r}_{+\mathrm{ u}2}$, follows the same behavior as the
thermodynamically stable black hole solution, $r_{+2}$.  These
similarities could perhaps be expected since the mechanically unstable
shell is bound to collapse into a black hole and can be understood as
a black hole precursor.
For the mechanically stable hot thin shell, which is given by the shell
radius $\alpha_\mathrm{ s}$, we have found that there are also two
branches for the gravitational radius. One of the branches,
$\tilde{r}_{+\mathrm{ s}1}$ is thermodynamically unstable, while the other
branch $\tilde{r}_{+\mathrm{ s}2}$ is thermodynamically stable.
From Fig.~\ref{figch7:alphainrp} and Fig.~\ref{figch7:r+bh}, it can be 
seen that the
two solutions for the gravitational radius with thin shell radius
$\alpha_\mathrm{ s}$, which is mechanically stable, share no similarities
with the two solutions for the horizon radius of the black hole.
However, the solutions $\tilde{r}_{+\mathrm{ s}1}$ 
and $\tilde{r}_{+\mathrm{ s}2}$ appear to have similarities with the
behavior of the Davies black hole solutions, which correspond to an
electrically charged black hole \cite{Tiago2024a} in the
canonical ensemble. These black hole solutions have a stable branch
that start at zero temperature with a horizon radius given by the
electric charge and then the horizon radius increases with the
temperature up until a maximum temperature.  The same happens with the
mechanically and thermodynamically stable hot
matter thin shell, starting
at zero temperature with zero gravitational radius instead of a
non-zero value. This behavior is expected from a solution of hot
self-gravitating matter that models hot AdS space
with radiation at the same order of approximation.

Second, we now compare the matter thin shell entropy with the black hole
entropy. Both entropies depend on their own gravitational radius,
which in the black hole case is also a horizon radius.
For the mechanical unstable and stable thin shells, we have seen that
the matter entropy $S_\mathrm{ m}$
can be described approximately by a power law.
For the unstable shell $\alpha_\mathrm{ u}$, we found that $S_\mathrm{ m} = \xi
\tilde{r}_{+}^{\;1.2323}$, for some $\xi$.
For the stable shell $\alpha_\mathrm{ s}$, we
found that $S_\mathrm{ m} = \xi \tilde{r}_{+}^{\;0.867504}$, for some
other $\xi$.
For the black hole, the entropy is also given by a power law $S_\mathrm{
bh}= \chi r_+^2$, for some $\chi$.
Both the mechanical unstable shell $\alpha_\mathrm{ u}$ and
the black hole have exponents $\gamma $ satisfying $\gamma >1$,
while the stable
shell has an exponent $\gamma <1$. This behavior reinforces the similarity
properties between the mechanical unstable shell and the black hole,
as advocated above. To understand this, we can resort
to a Bekenstein argument \cite{Bekenstein:1972b}.
It is stated in it, without making calculations, that one should
expect that black holes have an entropy with an exponent $\gamma $
obeying $\gamma >1$. 
Suppose that they had an exponent $\gamma <1$, then 
one would have that two isolated black holes that will merge into one
should have a final mass lower than the sum of the initial
mass, part of the initial mass being lost in
gravitational radiation. Concomitantly
the entropy of the final black hole
should be greater than the sum of the
entropies of the initial black hole, to have the second
law of thermodynamics obeyed.
The two conditions cannot be met simultaneously when $\gamma <1$.
Suppose for these purposes that the black hole
entropy is proportional to a power of the gravitational radius,
$S_\mathrm{ bh}= \chi r_+^\gamma$,
for some $\chi$ and $\gamma$.
For black hole $a$ and black hole $b$ merging into 
a third black hole $c$,
one has from the first condition $r_{+c}< r_{+a} + r_{+b}$ and from 
the second condition $r_{+c}^\gamma > r_{+a}^\gamma + r_{+b}^\gamma$, i.e., 
one has the range for the horizon radius of the black hole $c$
obeying
$(r_{+a}^\gamma + r_{+b}^\gamma)^{\frac{1}{\gamma}} < r_{+c} < 
r_{+a} + r_{+b}$.
This inequality can only be fulfilled for $\gamma>1$. 
From thermodynamic arguments then it was chosen $\gamma=2$,
the correct value for black holes.
The point here is that
the unstable shell $\alpha_\mathrm{ u}$ has an exponent
$\gamma=1.2323$ that is indeed greater than one, and
following the arguments above it is black hole like, i.e.,
the shell behaves as the black hole that it can originate
from gravitational collapse.
The 
stable shell $\alpha_\mathrm{ s}$ has an exponent
$\gamma=0.867504$ that is less than one, and
thus it does not behave as a black hole that it could
originate upon collapse.
Another remark
in relation to the entropy
is that for the black hole $S_\mathrm{ bh}= \pi
\frac{l^2}{l_\mathrm{ p}^2}\left(\frac{r_+}{l}\right)^2$,
the exponent is $\gamma=2$,
and it grows much faster than for the matter for large gravitational
radius, as expected, since it is known that black holes have the
maximum entropy.  However, for small gravitational radius, the matter
entropy $S_\mathrm{ m}$ is larger than the black
hole entropy, so this might indicate that there are no
stable
black holes for small gravitational radius, see Fig.~\ref{figch7:r+bh}
noting that $r_{+1}$ is unstable.

Third, we can compare the heat capacities for the
hot thin
shell and black hole, $C$ and $C_\mathrm{ bh}$, respectively, 
with the help of Figs.~\ref{figch7:Cintstar} and \ref{figch7:Cbh}.
The heat capacity $C$ as a function of the temperature for the
mechanically unstable shell $\alpha_\mathrm{ u}(\tilde{r}_+)$ behaves in
the same manner as the heat capacity of the black hole $C_\mathrm{ bh}$ as
a function of temperature. There are parts that are thermodynamically
unstable, $\tilde{r}_{+\mathrm{ u}1}$ for the shell and $r_{+1}$ for the
black hole, and parts that are thermodynamically stable
$\tilde{r}_{+\mathrm{ u}2}$ for the shell, and $r_{+2}$ for the black
hole.
The heat capacity $C$ as a function of the temperature for the
mechanically stable shell $\alpha_\mathrm{ u}$ behaves differently from the
heat capacity of the black hole $C_\mathrm{ bh}$ as a function of
temperature. However, it shows similarities to the heat capacity of
the electrically charged black hole in the canonical
ensemble~\cite{Tiago2024a}.  In particular, the
thermodynamically stable $\tilde{r}_{+\mathrm{ s}1}$ branch of the heat
capacity is similar to the heat capacity of the stable branch of
an electrically charged black hole.  These similarities
displayed here for
the heat capacity are the same as the similarities that we found
above when
comparing the gravitational radii, and indeed they come from the fact
that the heat capacity is related to the first derivative in
temperature of the gravitational radius, and so the similarities from
the gravitational radius solutions are carried into the heat capacity.

\subsubsection{Favorable states: Comparison between hot thin shell,
black hole, and pure hot AdS thermodynamic states}

We now make the identification of the favorable states of the ensemble, i.e.,
given a fixed temperature, we analyze if the
hot thin shell is favored in relation to the black hole
or if the contrary happens.

In order to do this,
to identify the favorable states at a fixed temperature 
of the ensemble, we must compare the action of the stable hot
self-gravitating matter thin shell with the action of the stable black
hole solution of Schwarzschild-AdS. This is so because the sector with
a self-gravitating matter shell may compete with the black hole sector
and the pure hot AdS sector in the path integral.
From thermodynamics, it is known that the preferred
configuration is the one with the least free energy.
This also means the one with the least
free action, since $I_0= \bar{\beta} F$.
This can be seen because the partition function is $Z=\mathrm{ e}^{-I_0}$,
and thus the configuration with less $I_0$ is the more probable one.
Now, if at a certain temperature, the
configuration with the least action changes, then this marks a first order
phase transition as the action is continuous but not differentiable
there.  In~\cite{Page:1982dh}, hot thermal AdS, i.e.,
hot AdS in one-loop approximation, and the stable
black hole solution were discussed, where it was discovered the
Hawking-Page phase transition from hot thermal AdS to the stable black
hole. Here, the stable self-gravitating matter thin shell can be 
understood as one possible description of hot AdS with thermal 
self-gravitating matter, i.e., hot curved AdS.

To help in the comparison of the actions, 
we can write the action of the matter thin shell
given in Eq.~\eqref{eqch7:reducedactionpathintegral} as
$\frac{l_\mathrm{ p}^2}{l^2}I_0 =
\frac{\frac{\tilde{r}_+}{l} \left( 1 + \frac{\tilde{r}_+^2}{l^2}
\right)}{2 \bar{T}l} -
c_0\left( \frac{l_\mathrm{ p}^2l_\mathrm{ c}}{l^3} \right)^\frac14
\left(\frac{m \,l_\mathrm{ p}^2}{l}\right)^{\frac{3}{4}}
\left(4\pi \frac{\alpha^2}{l^2}\right)^{\frac{1}{4}}$,
with $\tilde{r}_+$
given by
the stable solution
$\tilde{r}_+=\tilde{r}_+(\bar{T})$
of Eq.~\eqref{eqch7:statio2final}, i.e.,
$\tilde{r}_{+\mathrm{ s}2}$,
and $\alpha$ given by 
the stable solution
$\alpha=\alpha(\tilde{r}_+(\bar{T}))$ of Eq.~\eqref{eqch7:statio1final}
together
with Eq.~\eqref{eqch7:statio2final}, i.e.,
$\alpha_\mathrm{ s}=\alpha_\mathrm{ s}(\tilde{r}_{+\mathrm{ s}2}(\bar{T}))$,
and the action has been set without units.  It is useful to define
$z=\left(\frac{l^3}{l_\mathrm{ p}^2l_\mathrm{ c}} \right)^\frac14$ as the
parameter without units
that establishes the relevant scale ratios between the
Planck length, AdS length, and the Compton length in the case of the
hot matter thin shell with the chosen equations of state above.
Then, we can write the adimensional action of the matter thin shell as
\begin{align}
&I_0=\frac1z {\bar
I}_0\left(\frac{\bar{T} l}{z}\right),
\quad\quad\quad\quad\quad\quad
z\equiv\left(\frac{l^3}{l_\mathrm{ p}^2l_\mathrm{ c}} \right)^\frac14
\,,
\nonumber
\\
&{\bar I}_0\left(\frac{\bar{T}
l}{z}\right)\equiv \frac{l^2}{l_\mathrm{ p}^2}
\frac{\frac{\tilde{r}_+}{l} \left( 1 +
\frac{\tilde{r}_+^2}{l^2} \right)}{2 \frac{\bar{T}l}{z}} -
\frac{l^2}{l_\mathrm{ p}^2} c_0 \left(\frac{m
\,l_\mathrm{ p}^2}{l}\right)^{\frac{3}{4}}
\left(4\pi \frac{\alpha^2}{l^2}\right)^{\frac{1}{4}}\,.
\label{eqch7:Inounits}
\end{align}
This property of the thin shell action
is also useful for numerical purposes, as one can compare
actions 
with a given parameter $z$.
The behavior of the action of the
matter thin shell, Eq.~\eqref{eqch7:Inounits}, evaluated at the 
solutions of the shell, can be summarized as follows. 
The action starts at zero at $l\bar{T}=0$ and decreases for increasing 
$l\bar{T}$. After a final temperature $lT_{*f}$, 
the stable thin shell solution ceases to exist, which can be
interpreted as
the matter 
having larger thermal agitation than the permitted to
have a shell, implying that the shell can collapse to a black
hole or disperse to infinity.
This $lT_{*f}$ depends on the scale ratio parameter
$z=\left( \frac{l^3}{l_\mathrm{ p}^2l_\mathrm{ c}} \right)^\frac14$,
which itself depends
on the natural gravitational scale ratio $\frac{l}{l_\mathrm{ p}}$ and on the
matter scale ratio $\frac{l}{l_\mathrm{ c}}$.

In relation to the action of the stable black hole, one has
the Hawking-Page action given by
\begin{align}
\frac{l_\mathrm{ p}^2}{l^2}I_\mathrm{ bh} = 
\frac{\frac{r_+}{l} \left( 1 + \frac{r_+^2}{l^2}
\right)}{2 \bar{T}l}  - 
\frac{\pi r_+^2}{l^2}
\,,
\label{eqch7:actionbhrepeat}
\end{align}
with $r_+$ being the stable black hole
solution, i.e.,
$r_{+2}$, 
given
as
a function of $l\bar{T}$
by
$\frac{\tilde{r}_+}{l} = 
\frac{2\pi l \bar{T}}{3} + \frac{1}{3}\sqrt{(2\pi l \bar{T})^2 - 3}$.
We have seen that the stable solution only exists
for $l \bar{T}\geq\frac{\sqrt3}{2\pi} =0.276$, with last equality 
being approximate.

\begin{figure}[h]
\centering
\includegraphics[width=0.60\textwidth]{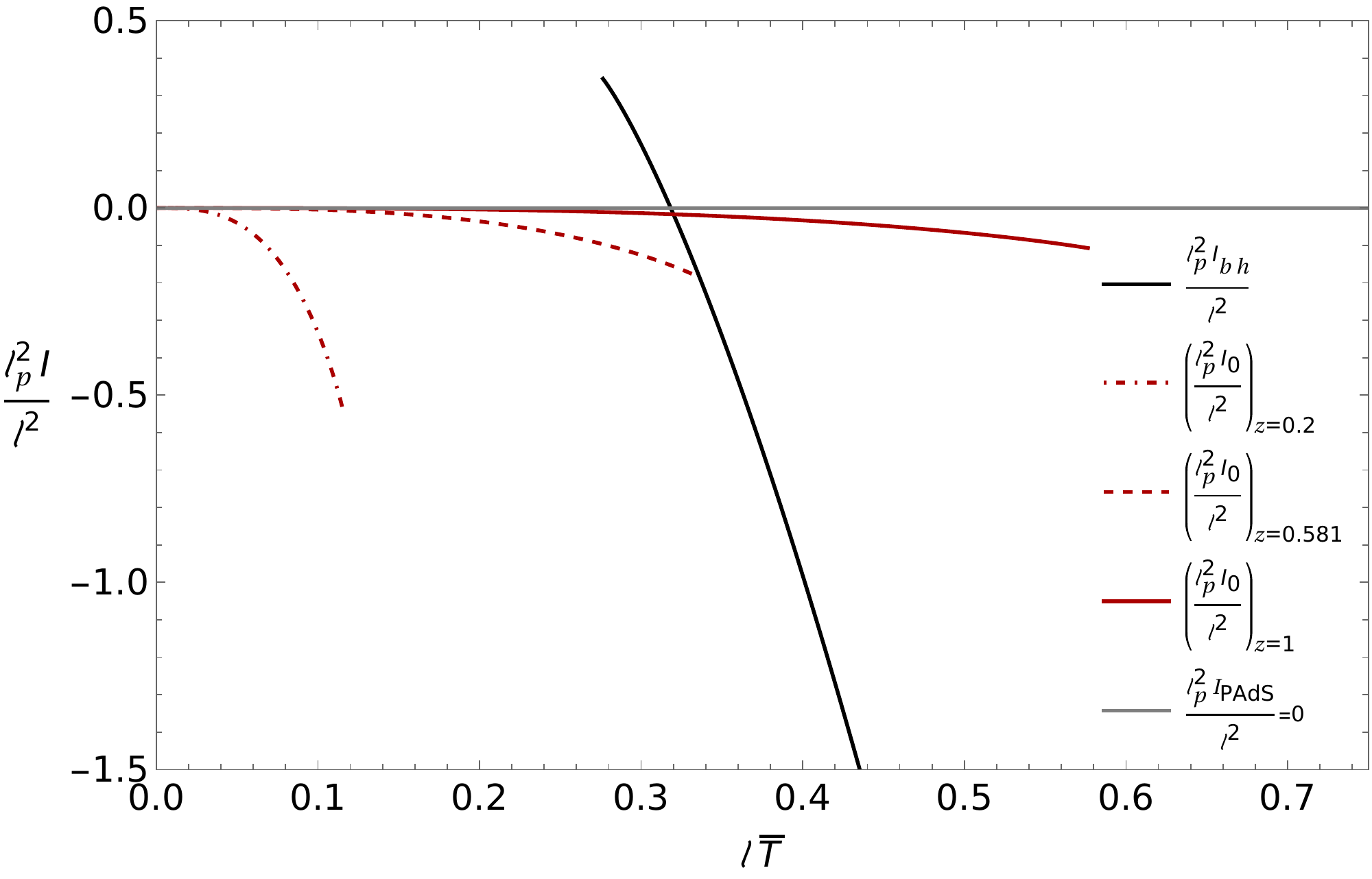}
\caption{\label{figch7:actionintstar}
Plot of the actions
$I_\mathrm{ bh}$, $I_0$, and 
$I_\mathrm{ PAdS}$
as functions of the temperature $l \bar{T}$. The
solution that has lower action between stable black
hole, hot shell, and pure hot AdS
is the one that is favored.  It is chosen
$z=\left( \frac{l}{l_\mathrm{ c}}\right)^\frac{1}{4} \left(\frac{l}{l_\mathrm{
p}}\right)^\frac{1}{2} = 0.2,\, 0.581,\,1$ to compare the actions.
For $z=0.2$ the hot shell ceases to exist at temperature 
$l T_{*\mathrm{ f}} = 0.115$, for
$z=0.581$ at temperature 
$l T_{*\mathrm{ f}} = 0.335$, and for 
$z=1$ at temperature 
$l T_{*\mathrm{ f}}=0.577$.
}
\end{figure}

Now, we must compare the matter action with the black hole action for
each possible parameter $z$ and $c_0$, and also with
pure hot AdS space
characterized by $I_\mathrm{ PAdS} = 0$.  Here, we made the choice $c_0=1$ 
since $c_0$ can be in some sense absorbed by the parameter $z$.
Then, the comparison of the actions can be made only on $z$.  The plot of the
actions is shown in Fig.~\ref{figch7:actionintstar}, where for the matter
shell three values of $z$ are considered,
namely, $z =0.1$, $z=0.581$, and $z=1$.  The
action of matter thin shell is zero at $l\bar{T}=0$ 
and decreases to negative values for increasing temperature, until the
final temperature $lT_{*\mathrm{ f}}$ is reached with a corresponding
minimum negative action. 
Note that the maximum temperature depends on the parameter 
$z$ as $l T_{*\mathrm{ f}} = 0.577 z$, where $0.577$ is approximate.
Above this temperature $l T_{*\mathrm{ f}}$ the shell stops
to exist and probably collapses.
With respect to the black hole case, the action $I_\mathrm{ bh}$ 
only starts to exist for $l \bar{T}\geq \frac{\sqrt3}{2\pi}=0.276$,
and decreases with increasing
temperature. The black hole  action is positive in the 
range $\frac{\sqrt3}{2\pi} \leq l T_{*} \leq 0.318$
where $0.318$ is approximate, is zero at $lT_{*} = 0.318$,
and is negative for $l \bar{T}>0.318$.
The point of intersection between the two actions is given by the
equality $I_\mathrm{ bh}(l T_{*}) = \frac{1}{z} \bar{I}_0(\frac{l
T_{*}}{z})$ for each $z$. For example, in the case of $z=1$, one has
that the actions intersect at $l \bar{T} = 0.320$, and so as one increases
the temperature around this point, there is a first order phase
transition from the hot
matter thin shell to the stable black hole.  This
is analogous to the case of the Hawking-Page phase transition, where
the matter is treated in a one-loop approximation, rather than in
zero loop.
It can be found that the intersection between the matter thin shell
action and the black hole action only happens for a range of $z$.  As
one decreases $z$, the maximum temperature of the thin shell also
decreases, while the black hole action is unaltered. And so, there
must be a minimum value of $z$ for which the intersection occurs. The
minimum value can be found by considering that the two actions
intersect exactly at the maximum temperature of the shell,
i.e., $I_\mathrm{ bh}( 0.577 z) = \frac{1}{z} \bar{I}_0( 0.577)$, which
numerically can be solved and gives $z= 0.581$ approximately and the
first order phase transition occurs for this case at $l \bar{T}=0.336$,
approximately. As a consequence, the action of the matter thin shell
intersects the action of the black hole only if
$0.581\leq z <\infty$, with
first number being approximate. If $0< z < 0.581$, the matter shell
solution ceases to exist before it intersects the curve of the action
of the black hole. Therefore, there is only a first order phase
transition from the matter thin shell to the black hole when
$0.581\leq z <\infty$.

We must comment about the phase of pure hot AdS space,
i.e., AdS space in zero loop,
in comparison with the hot thin shell and
black hole configurations. 
For $0.581<z<\infty$, the matter thin shell action always intersects
the black hole action and it is always negative, and so
the pure hot AdS space is always the least favorable.
For $0.547 \leq z < 0.581$, with $0.547$ being approximate,
the matter thin shell does not intersect the black hole
action. However, when the matter thin shell ceases to exist at the
maximum temperature, there is a black hole solution with negative
action and so both the thin shell and the black hole solutions are
still more favorable than pure
hot AdS space. For the range $0\leq z \leq
0.547$, there is an interval of temperatures, between the maximum
temperature of the shell and the temperature at which the black hole
solution has zero action, where pure hot AdS space is favorable.

It is worth stressing that
the parameter $z$ can be restricted from validity arguments of the
zero-loop approximation. The zero-loop approximation should be valid
for the cases $l \gg l_\mathrm{ p}$ but with $l$ not that large and since
$\alpha$ is comparable to $l$, one must have $l \gg l_\mathrm{ c}$ also so that
matter at the thin shell can be judged thermodynamic.
Moreover, all scales must be much greater than the Planck scale
$l_\mathrm{ p}$, as the zero-loop approximation is being used here.  Therefore,
one must have $l \gg l_\mathrm{ c} \gg l_\mathrm{ p}$, which means a large
value of $z=\left( \frac{l^3}{l_\mathrm{ p}^2l_\mathrm{ c}}
\right)^\frac14$. In this regime, the first order phase transition can
always occur, and both the matter thin shell and the black hole
solutions are more favorable than pure hot AdS space $I_\mathrm{ PAdS}=0$.
This strengthens the interpretation that the matter thin shell with
the chosen equation of state models hot AdS space with
self-gravitating radiation matter at low temperatures.

\subsubsection{Favorable states: Comparison between thin shell, black
hole, and hot thermal AdS thermodynamic states}

We have seen how the action
of a stable self-gravitating matter system in AdS,
which is a realization of hot curved AdS, compares with
the action of the stable AdS black hole solution,
and the action of pure AdS, describing classical
AdS space devoid of any matter.

It is also interesting to substitute pure AdS for hot thermal AdS,
i.e., AdS space with
nonself-gravitating radiation obtained from the one-loop approximation, 
and compare with the action for black
hole and the thin shell. 
The action $I_\mathrm{ TAdS}$ for hot thermal AdS is given by 
\begin{align}
I_\mathrm{ TAdS}=-\frac{\pi^4 (l \bar{T})^3}{45}\,,
\label{eqch7:actionhotthermalads}
\end{align}
where hot thermal AdS is assumed to be made of
particles, each with 
effective number of spin states equal to two, such as gravitons do.
We can then compare the first
order phase transition treated above
with the Hawking-Page phase
transition, which is a transition between hot thermal AdS
with action given in  Eq.~\eqref{eqch7:actionhotthermalads}
and the black hole
action given in Eq.~\eqref{eqch7:actionbh} \cite{Page:1982dh}.
Moreover, when the temperature
of the radiation is sufficiently high, 
the radiation forms a singularity of the Buchdahl type
in the center and then it presumably collapses
to a black hole. We find that 
this maximum Buchdahl radiation temperature
for which radiation ceases to exist is given by 
$lT_\mathrm{ Buch}=0.4234\left(\frac{2\pi^2}{30}\right)^{-\frac14}
\left(\frac{l}{l_\mathrm{ p}}\right)^{\frac12}$
\cite{Page:1982dh}, see also \cite{Page:1985em}.

\begin{figure}[h]
\centering
\includegraphics[width=0.60\textwidth]{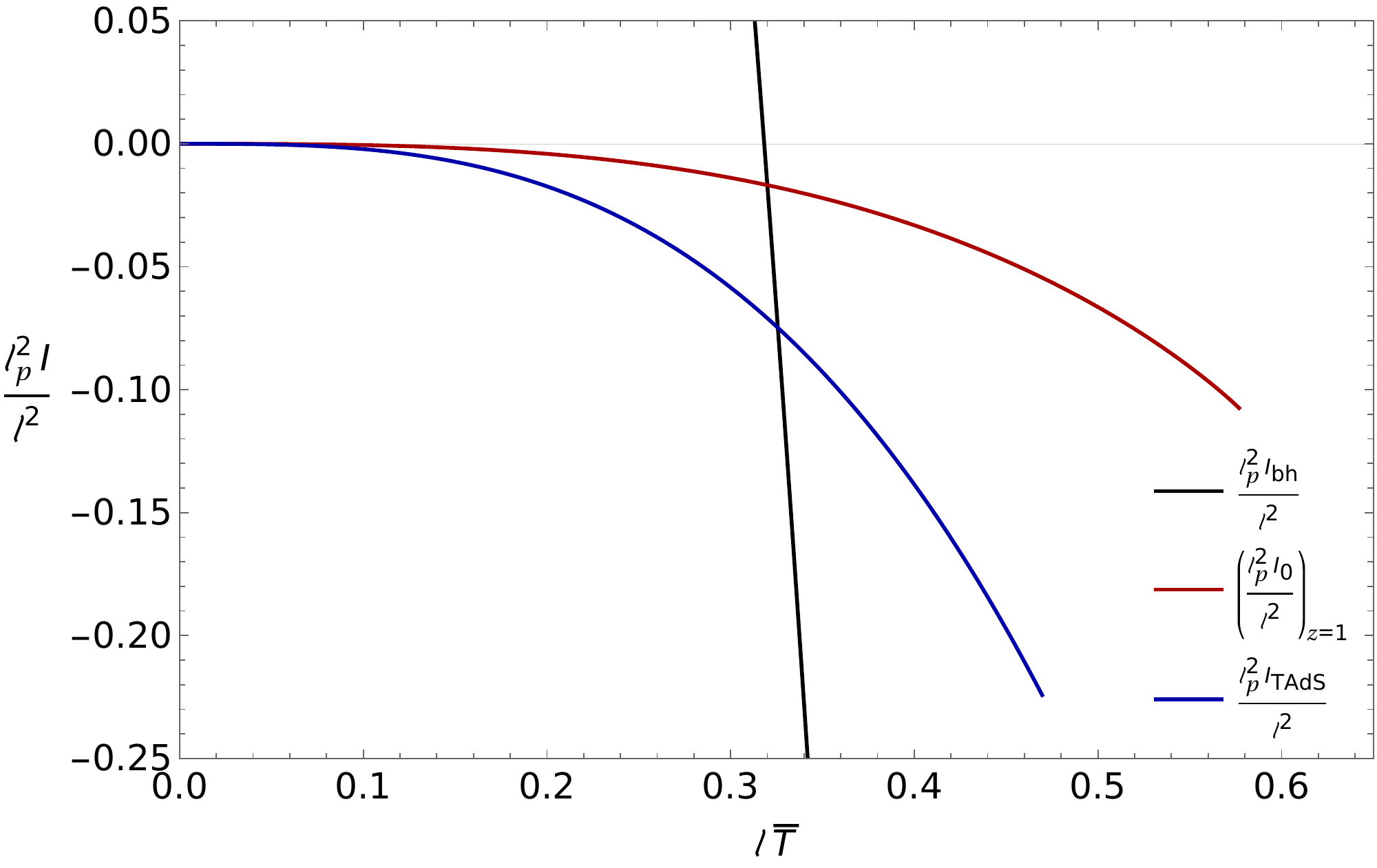}
\caption{\label{figch7:actionradiation}
Plot of the actions
$I_\mathrm{ bh}$, $I_0$, and 
$I_\mathrm{ TAdS}$ as functions of the temperature $l \bar{T}$. The
solution that has lower action between  stable black
hole, hot shell, and thermal hot AdS,
i.e., AdS with nonself-gravitating radiation,
is the one that is favored.
It is chosen $z=\left( \frac{l}{l_\mathrm{ c}}\right)^\frac{1}{4} 
\left(\frac{l}{l_\mathrm{ p}}\right)^\frac{1}{2} = 1$ and 
$\frac{l_p}{l} = 1$ to compare the actions.
For $z=1$, the hot shell ceases to exist at temperature 
$l T_{*\mathrm{ f}} = 0.577$. 
Thermal hot AdS ceases to exist at temperature
$lT_\mathrm{ Buch}=0.4701$.}
\end{figure}

In Fig.~\ref{figch7:actionradiation}, we plot the actions for
the stable black hole, the
hot matter thin shell, and hot thermal AdS.
In the figure, we consider the parameters
$z=\left( \frac{l}{l_\mathrm{ c}}\right)^\frac{1}{4} 
\left(\frac{l}{l_\mathrm{ p}}\right)^\frac{1}{2} = 1$ and 
$\frac{l_p}{l} = 1$ to compare the actions.
First, we can comment on the transitions
between stable black hole and hot thermal AdS thermodynamic states
and, second, we can compare the results for the
hot matter thin shell and hot thermal AdS.
In relation to the first point, we
have seen that there are no black holes, and therefore
no stable black hole, in the range of temperatures
$0\leq l\bar{T}<\frac{\sqrt3}{2\pi}=0.276$, the last number
being approximate.
From 
$\frac{\sqrt3}{2\pi}\leq l\bar{T}<0.325$, hot thermal AdS
is favored in relation to a black hole state.
At $l\bar{T}=0.325$ hot thermal AdS and black hole coexist equally,
and this is the temperature at which a first order
phase transition occurs. 
This is the Hawking-Page phase transition.
For
$0.325<l\bar{T}<lT_\mathrm{ Buch}$, where $lT_\mathrm{ Buch}=0.4701$, the black
hole is favored over hot thermal AdS, meaning that is more probable to
find the system in a black hole state.  For $lT_\mathrm{
Buch}<l\bar{T}<\infty$, the system is in a collapsed black hole state in
AdS, meaning that at these temperatures it is not possible to find the
system in a hot thermal AdS state.
In relation to the second point, we can see from the figure 
that the first order phase transition between the matter thin
shell and the black hole has the same behavior as the Hawking-Page
phase transition between hot thermal AdS and the black hole.
However,
it seems that  hot thermal AdS, i.e., AdS with nonself-gravitating
radiation, is more favorable than the matter thin shell, with the
differences being very small as one increases $z$ and $\frac{l}{l_p}$.
Since the action for hot thermal AdS does not include gravitation, as
it corresponds to nonself-gravitating radiation, it is clear that the
hot thin shell should mimic self-gravitating radiation due to its
similar behavior around the phase transition.
In addition, they share the feature of a maximum temperature,
$l T_{*\mathrm{ f}}= 0.577$
for the hot thin shell and
$lT_\mathrm{ Buch}=0.4701$
for hot thermal AdS, the values being for $z=1$ and $l=l_p$.

\section{Conclusions}
\label{sech7:concl}

In this chapter, we studied the canonical ensemble of a hot
self-gravitating matter thin shell in AdS by finding the partition
function of the system via the Euclidean path integral approach to
quantum gravity. Our study was restricted to spherically symmetric
metrics and we established the boundary conditions. Imposing the
Hamiltonian constraint, we obtained the reduced action of the matter
thin shell in AdS, and from the reduced action, we obtained 
the stationary and stability conditions.  There
are two equations for the stationary condition, i.e., the balance of
pressure and the balance of temperature, and two stability conditions,
i.e., the mechanical stability condition and the thermodynamic
stability condition. The fact that one can obtain the two stability
conditions shows the power of the reduced action in the Euclidean path
integral approach, in that it gives not only information about
thermodynamics but also of mechanics.

We have shown that for the case of the matter thin shell in AdS, one can
obtain the canonical ensemble of the thin shell by establishing an
effective reduced action only dependent on the gravitational radius of
the thin shell. This eases the analysis of the canonical ensemble and
further shows that one can build an effective reduced action dependent
only on the gravitational radius for this case, hiding the description
of matter in the form of the effective entropy. We established here 
the link between the effective entropy and the specific description of
the shell with an equation of state, in the zero-loop approximation.

The thermodynamics of the system follows directly from zero-loop
approximation consisting of the reduced action evaluated at the
stationary points, which is equivalent to finding the action for the
specific solution of Einstein equation.  In this approximation, we
obtained directly the relevant thermodynamic quantities, namely, the
mean free energy, the entropy, the mean energy, and the heat capacity.
We found there is a correspondence between thermodynamic stability in the
ensemble theory and positive heat capacity in the derived
thermodynamics, as it should.  On the other hand, within
thermodynamics itself, we cannot determine mechanical stability by
varying the thermodynamic quantity fixed at the conformal boundary,
i.e., the temperature. This fact seems to be a consequence of applying
the zero-loop approximation to the internal degree of freedom, the
radius of the shell. It also means that the zero-loop approximation of
the effective reduced action yields the expected thermodynamic
stability condition, since it can be seen as a generalized free energy
function.

We introduced an equation of state for the matter 
and we obtained the solutions of the
canonical ensemble for the matter thin shell. We found that there are in total four
solutions, with only one of them being stable both mechanically and
thermodynamically. We compared the action of the stable
matter shell solution with the Hawking-Page AdS black hole stable
solution and we verified the existence of a first order phase transition 
in a physically reasonable range of scale lengths.  We have shown that
this first order phase transition follows an analogous behaviour to the
Hawking-Page phase transition, and that the hot matter thin shell can
mimic self-gravitating radiation.

It will be interesting to uncover the sector with a black hole and a
shell together to fully understand the space of configurations and
respective phase transitions. Moreover, it would be interesting to
explore additional fixed parameters of the ensemble, such as the
chemical potential, to understand if one is able to access the
mechanical stability condition from varying these fixed parameters, in
the sense that they are needed for thermodynamic stability. This line
of research motivates Chapter~\ref{ch:Chatelier}.

\chapter{Thermodynamic ensembles of a black hole 
and a self-gravitating matter thin shell with
a fixed chemical potential: equilibrium, stability 
and Le Chatelier-Braun principle}
\chaptermark{Ensembles of a black hole 
and a self-gravitating matter thin shell}
\label{ch:Chatelier}

\section{Introduction}

The study of the canonical ensemble including self-gravitating matter 
thin shell
and a black hole was first done in~\cite{Martinez:1989hn}, using the 
Euclidean path integral approach to quantum gravity~\cite{Gibbons:1977}. 
The analysis of this system was further deepened in~\cite{Lemos:2023yiz}, 
by keeping the radius of the shell fixed. In Chapter~\ref{ch:thinshellAdS}, 
we considered the canonical ensemble of a matter thin shell in anti-de Sitter 
with a chosen equation of state, revealing that the mechanical stability 
of the shell appeared as a condition for the validity of the zero loop approximation 
but it was not needed for the thermodynamic stability. 

In this chapter, we progress further in the study of self-gravitating 
matter thin shells to study how the formalism handles the stability of 
composed systems. In that regard, 
we construct the grand canonical ensemble of a self-gravitating matter thin 
shell with a black hole inside, with the system surrounded by 
a finite cavity. We calculate the partition function through the Euclidean 
path integral approach in the zero loop approximation.
We introduce the chemical potential in the description of 
matter in order to understand its implications to thermodynamic stability 
and the system at hand. For convenience, 
the partition function is written in terms of the another partition function 
for an ensemble with cavity at infinity, with fixed energy $E$ and fixed logarithm 
of the fugacity $\beta\mu$, which we call here the $(E,\beta\mu)$ ensemble. 
Note that this ensemble is a modification of the microcanonical ensemble. 
We show that the Le Chatelier-Braun principle follows 
from the validity of the zero loop approximation and also that the conditions for 
the validity of the zero loop approximation, including the mechanical stability 
of the shell, must be considered to infer thermodynamic stability, due to the 
presence of the chemical potential.

This chapter is organized as follows. In Sec.~
\ref{sech9:pathintegral}, we construct the partition function of the ensemble, 
where we obtain that the partition function of the grand canonical ensemble can 
be described in terms of another partition function related to 
the $(E,\beta\mu)$ ensemble. 
In Sec.~\ref{sech9:chemicalzeroloop}, we perform the zero loop approximation to 
the partition function of the $(E,\beta\mu)$ ensemble. 
In Sec.~\ref{sech8:grandcan}, 
the zero loop approximation is performed to 
the partition function of the grand canonical ensemble. In Sec.~\ref{sech9:thermomicro},
we obtain the thermodynamics of the system composed by a black hole with a thin shell surrounding, 
from the $(E,\beta\mu)$ ensemble. In Sec.~\ref{sech9:thermograndcan},
we obtain the thermodynamics from the grand canonical ensemble. 
In Sec.~\ref{sech9:martinez}, we consider the Martinez equation of state and 
we obtain another fundamental equation of state. In Sec.~\ref{sech9:hessian}, 
we display the Hessian of the relevant actions. In Sec.~\ref{sech9:stabshellmech},
we review the mechanical stability of a thin shell surrounding a black hole.
In Sec.~\ref{sech9:conc}, we present the conclusions.

\section{The grand canonical ensemble and the $(E,\beta\mu)$ ensemble 
through the path integral approach\label{sech9:pathintegral}}\sectionmark{Ensembles 
through the path integral approach}
\thispagestyle{userightbotmark}

\subsection{The grand canonical statistical partition function}

The construction of the grand canonical ensemble of a curved space $M$
with matter can be done with the Euclidean path integral approach 
to quantum gravity, with the partition function given by 
\begin{align}\label{eqch9:partition}
    Z = Dg_{\alpha \beta}D\psi \mathrm{e}^{-I[g_{\gamma \nu},\psi]}\,\,,
\end{align}
where $g_{\alpha \beta}$ is the Euclidean metric, $\psi$ represents 
a matter field, $I[g_{\gamma \nu},\psi]$ is the Euclidean 
action, and $Dg_{\alpha \beta}$ and $D\psi$ are the integration measures 
for $g_{\alpha \beta}$ and $\psi$, respectively. In the canonical ensemble,
the integration is done over periodic fields $g_{\alpha \beta}$ and $\psi$, 
if $\psi$ is bosonic. However, this condition can suffer modifications 
according to the type of ensemble one considers.

In the case of this chapter, we are interested in including and fixing the chemical 
potential of the matter field, hence the ensemble we are considering is the 
grand canonical ensemble. While we explained the Euclidean path integral approach 
in Chapter~\ref{ch:Euclideanpathintegral}, here it 
is important to explain how we can introduce the chemical potential in the 
action $I[g_{\gamma \nu},\psi]$. We can first trace back the partition 
function in the formal representation $Z = \Trace(\mathrm{e}^{-\beta H})$ 
for the canonical ensemble, where $\beta$ is the fixed inverse 
temperature defined
by $\beta = \int_{\partial M} (g^{\tau \tau})^{-1/2} d\tau$, with $\tau$ being 
the imaginary Euclidean time having period $2\pi$. 
In order to consider the grand canonical ensemble, 
one must modify the partition function as 
$Z = \Trace(\mathrm{e}^{-\beta H + \beta \mu N})$, where $\mu$ is the fixed 
chemical potential and $N$ is the mean particle number. The operator $\beta H$
can be defined by the mean Euclidean Hamiltonian as 
$\beta H = \int_M \mathfrak{h} \,d^3x d\tau$,
where $\mathfrak{h}$ is the Hamiltonian tensor density with respect to 
$g_{\alpha \beta}$ of the Euclideanized space, i.e. $\mathfrak{h}$ 
transforms as the determinant of the metric $\sqrt{g}$. Depending on the Hamiltonian of the 
field, one can find the functional version of the particle 
number as $N(\tau) = \int \mathfrak{n} \,d^3x$ for a slice of constant $\tau$, 
where $\mathfrak{n}$ is the 
particle number tensor density that transforms like the determinant of 
the induced metric of hypersurfaces with constant $\tau$. 
From here, it seems non-trivial to include such operator in the trace without 
assumptions, since it is an integral dependent on the slice of constant $\tau$. A simple 
way to avoid such dependence is to use $N = \frac{1}{2\pi} \int N(\tau) d\tau$, that is 
the mean particle number over the slices of constant $\tau$. In principle, if the 
particle number is conserved over the slices, then $N=N(\tau)$, giving the right result.
One could then build the term $\beta \mu N$ 
as $\beta \mu N = \int (g^{\tau \tau})^{-1/2} \mu(x)\mathfrak{n}\, d^3x d\tau$, 
where $\mu(x)$ is a scalar and it is the local chemical potential defined by
$\frac{\beta \mu}{2\pi} = (g^{\tau \tau})^{-1/2} \mu(\tau,x)$. The introduction of 
this local chemical potential allows to construct a covariant integral. The full trace 
can then be transformed into the Euclidean path integral in Eq.~\eqref{eqch9:partition}, where 
$I[g_{\alpha \beta},\psi]$ is the Euclidean action of the metric space 
and matter fields with a modification 
that depends on the local chemical potential $\mu(x)$. 
For complex matter fields with a kinetic part which 
is quadratic, such modification can be obtained by a simpler manner. One can consider 
the typical matter field lagrangian but with a transformation of the field, for example 
$\psi = \mathrm{ e}^{-\frac{\beta \mu}{2\pi}\tau} \hat{\psi}$ and $\psi^\dagger = 
\mathrm{ e}^{\frac{\beta \mu}{2\pi} \tau} \hat{\psi}^\dagger$
for complex bosonic fields, where $\dagger$ means complex conjugate. This gives a modified 
lagrangian for the fields $\hat{\psi}$ and $\hat{\psi}^\dagger$ which include the 
chemical potential. The periodic conditions 
are satisfied not by $\psi$ and $\psi^\dagger$ but for $\hat{\psi}$ and $\hat{\psi}^\dagger$. 
This agrees with the fact that $\mu$, although being a constant of the ensemble, is 
a scalar density with respect to a one dimensional metric. The fact we are fixing $\mu$ 
means that we are choosing a specific foliation of space in 
hypersurfaces of constant $\tau$. Nevertheless, we can proceed assuming the identity 
$\frac{\beta \mu}{2\pi} = (g^{\tau \tau})^{-1/2} \mu(\tau,x)$.
It is also convenient to define 
a notion of local temperature as 
$\frac{1}{T(\tau,x)} = 2\pi (g^{\tau \tau})^{-1/2}$, although it must be emphasized 
that only $\beta$ and $\beta \mu$ have definite meanings as they are fixed in the ensemble. 
The result $\beta \mu = \mu(\tau,x)/T(\tau,x)$ then comes naturally.

In this paper, we consider a spherically symmetric 
cavity with a matter thin shell inside together with a black hole. 
The partition function for the system is assumed to be given by 
$Z = \int Dg_{\alpha \beta} \,\mathrm{e}^{-I_g[g_{\alpha \beta}]} 
\int D\psi \,\mathrm{e}^{-I_m[g_{\alpha \beta},\psi]}$, where 
the gravitational action $I_g$ is given by the Einstein Hilbert action plus 
the Gibbons-Hawking-York term, and the matter action $I_m$ depends on the 
type of matter considered. In general, it is not possible to obtain a closed 
form for the matter path integral, even more it is not possible 
to guarantee its convergence after the techniques of regularization and 
renormalization. Yet, we can assume that such path integral, if it is 
convergent, yields a general expression 
$\mathrm{e}^{-\int \mathcal{W}[g_{\alpha \beta}] \sqrt{g}d^4 x} = 
\int D\psi \, \mathrm{e}^{-I_m[g_{\alpha \beta},\psi]}$, 
where $\mathcal{W}$ is the 
matter grand potential density. For the case of a matter thin shell
inside a spherically symmetric cavity, the induced metric in the thin shell 
is described by the metric of a $2$-sphere with constant radius plus the metric of 
a ring with constant radius parametrized by Euclidean imaginary time. 
Therefore, the path integral in principle can be integrated. If we take 
$\gamma_{ab}$ to be the induced metric on the shell, we can assume the 
matter grand potential being described by the functional 
$\mathcal{W} = \mathcal{F}(T_m,n_m) - \mu_m n_m$, 
where $T_m$ is the local temperature at the shell defined as 
$\frac{1}{T_m} = 2\pi (\sqrt{\gamma^{\tau \tau}})^{-1}$, 
$n_m$ is the particle number density defined as 
$n_m = (\sqrt{\gamma}\sqrt{\gamma^{\tau \tau}})^{-1}\mathfrak{n}_m$
with $\mathfrak{n}_m$ being the particle number tensor density, and 
$\mu_m$ is the local chemical potential conjugate to the particle number, 
where $\frac{\mu_m}{T_m}$ must be constant.

\subsection{Grand canonical action for a black hole and a matter thin shell}

Here, we consider spherically symmetric 
configurations which are stationary and so these have a connection to a 
physical spacetime, 
in particular a black hole with a 
matter thin shell in equilibrium inside a cavity. 
The Euclideanized spacetime 
$M$ is split into two parts, $M_1$ 
and $M_2$, by the thin shell described by a hypersurface 
$\mathcal{C}$. The space $M_1$ designates the Euclidean space 
with boundary $\mathcal{C}$ only, while $M_2$ designates the 
Euclidean space with two disjoint boundaries $\mathcal{C}$ 
and $\partial M$, with the latter being the boundary of the cavity. 
In some sense, $M_1$ is the inner Euclidean 
space while $M_2$ is the outer Euclidean space.

The action for the system inside the cavity is given by
\begin{align}
    & I = - \int_{M\setminus\{\mathcal{C}\}}\frac{1}{16\pi l_p^2} R 
    \sqrt{g}d^4 x\,\,\nonumber\\
    & + \int_\mathcal{C} \left(\frac{[K]}{8 \pi l_p^2} 
    + \mathcal{F}(T_m,n_m,\sqrt{\gamma}) - \mu_m n_m\right)\sqrt{\gamma}d^{3}s\nonumber\\
    & - \frac{1}{8\pi l_p^2}
    \int_{\partial M} (K - K_0)\sqrt{\gamma}d^{3}s\,\,,
    \label{eqch9:actionEHS}
\end{align}
where $R$ is the Ricci tensor, $g_{\alpha \beta}$ is the Euclidean metric,   
$K = r^\alpha_{;\alpha}$ is the trace of the extrinsic curvature with $r_a$
being the outward unit normal to the considered hypersurface, $\gamma_{ab}$ 
represents both the induced metric on $\mathcal{C}$ 
and the induced metric on $\partial M$ with determinant $\gamma$, depending on the 
context,
written in 
a chosen coordinate system $s^i = (\tau, \theta, \phi)$,
$\mathcal{F}$ is the free energy density of the thin shell, 
$ \frac{1}{T_m} = 2\pi (\sqrt{\gamma^{\tau \tau}})^{-1}$ is the local 
inverse temperature at the shell, $n_m$ is the 
particle number density of the 2-surface of the thin shell which is a 
functional $n_m = (\sqrt{\gamma^{\tau \tau}} \sqrt{\gamma})^{-1}\mathfrak{n}_m$
with $\mathfrak{n}_m$ being the particle number scalar density,
$\mu_m$ is the local chemical potential at the 
shell, and $K_0$ 
is the extrinsic curvature of the 
hypersurface considered embedded in flat space. 
Also, $[K]$ means the difference $\eval{K}_{M_2}- \eval{K}_{M_1}$ evaluated 
at the hypersurface, where $K$ can be any tensor. 
For convenience, in connection to Chapter~\ref{ch:Euclideanpathintegral}, 
one can split the action $I = I_{g\mathrm{f}} + I_m$ 
into the gravitational action $I_{g\mathrm{f}}$ 
expressed by 
\begin{align}
    & I_{g\mathrm{f}} = - \int_{M\setminus\{\mathcal{C}\}}\frac{1}{16\pi l_p^2} R 
    \sqrt{g}d^4 x\,\,\nonumber\\
    & + \int_\mathcal{C} \left(\frac{[K]}{8 \pi l_p^2}\right)\sqrt{\gamma}d^{3}s\nonumber\\
    & - \frac{1}{8\pi l_p^2}
    \int_{\partial M} (K - K_0)\sqrt{\gamma}d^{3}s\,\,,
    \label{eqch9:actiongrav}
\end{align}
and the matter action $I_m$ with the expression
\begin{align}
    I_m = \int_\mathcal{C} \left(\mathcal{F}(T_m,n_m,\sqrt{\gamma}) 
    - \mu_m n_m\right)\sqrt{\gamma}d^{3}s\,\,.
\end{align}

\subsection{Geometry and the matter thin shell description}

We fix the geometry of the boundary of the cavity $\partial M$ as a 
spherically symmetric hypersurface, with topology $\mathbb{S}^1\times \mathbb{S}^{2}$. 
Because of this fixing, we assume that spherically symmetric spaces contribute the 
most for the path integral due to a spherically symmetric space $M$. The 
metrics considered on $M_1$ are
\begin{align}
    ds^2_{M_1} = b_1^2(u)\frac{b_2^2(u_m)}{b_1^2(u_m)}d\tau^2 + 
    a_1^2(u)du^2 + r(u)^2 d\Omega^2\,\,,\label{eqch9:metricM1}
\end{align}
and the metrics considered on $M_2$ are 
\begin{align}
    ds^2_{M_2} = b_2^2(u) d\tau^2 + a_2^2(u)du^2 + r^2(u)d\Omega^2\,\,,
    \label{eqch9:metricM2}
\end{align}
where $b_1$, $b_2$, $a_1$, $a_2$, and $r$ are functions of $u$ only, 
$d\Omega^2$ is the line element of the $2$-sphere, $\mathbb{S}^{2}$, 
and also the coordinates are chosen so that $\tau \in \,]0,2\pi[$ on $M$, 
$u \in\, ]0,y_m[$ on $M_1$ and $u\in\, ]y_m,1[$ on $M_2$, and $\theta$ and 
$\phi$ are the spherical coordinates on $M$.

The hypersurface $\mathcal{C}$ is described by the condition $u=u_m$, 
with induced metric 
\begin{align}
    ds^2_{\mathcal{C}} = b^2_{2}(u_m) d\tau^2 + \alpha^2 d\Omega^2\,\,,
    \label{eqch9:metricA}
\end{align}
where $\alpha$ is defined as $r(u_m)\equiv \alpha$, i.e. the radius of the 
matter thin shell. Notice that the choices for the metrics on $M_1$ and 
on $M_2$ ensure that the metric is continuous on $\mathcal{C}$, i.e. the metric 
on $M$ is $C^0$. Relatively 
to the quantities at the shell, we assume that matter is
in equilibrium. These considerations allow us to describe the 
differential of the free energy as 
$d\mathcal{F} = - s_m dT_m + \mu_m dn_m - \frac{\chi}{\sqrt{\gamma}}d\sqrt{\gamma}$, 
with the 
quantity $\chi$ defined as $\chi = \epsilon_m + p_m 
- s_m T_m - \mu n_m$, with $s_m$ being the entropy per area 
and $\epsilon_m$ being the energy 
density. The quantity $\chi$ is present to include the possibility 
of having degrees of homogeneity different from unity, such is the case 
for black hole like equations of state, see Chapter~\ref{ch:chargedselfgravitating}. 
For a degree of homogeneity of one, the quantity $\chi$ 
should be zero. Moreover, the matter shell obeys 
several equations of state, i.e. an expression for 
$\mathcal{F}(T_m,n_m,\sqrt{\gamma})$ must be known apriori as it is derived 
from the matter path integral. Notice that by integrating 
$\mathcal{F}(T_m,n_m,\sqrt{\gamma})$ along the shell, 
one obtains the mean free energy which obeys the typical 
thermodynamic differential.

The hypersurface $\partial M$ describing the boundary of the cavity is 
given by the condition $u=1$, with induced metric
\begin{align}
    ds^2_{\partial M} = b_2^2(1)d\tau^2 + R^2 d\Omega^2\,\,,
    \label{eqch9:metricB}
\end{align}
where $R$ is defined as $r(1) = R$.

\subsection{Grand canonical boundary conditions}

We must impose boundary conditions to select the topology of the spaces that
are summed on the path integral and to establish the quantities that are 
fixed on the ensemble.

At $u=0$, we impose black hole regularity conditions which are summarized by
\begin{align}
    & b(0)=0 \,\,,\label{eqch9:boundcondb0}\\
    & \eval{\frac{b_1'}{a_1}}_{u=0}\frac{b_2(u_m)}{b_1(u_m)} = 1\,\,,
    \label{eqch9:boundcondbprime0}\\
    & \eval{\frac{1}{a_1}\left(\frac{b_1'}{a_1}\right)'}_{u=0}
    \frac{b_2(u_m)}{b_1(u_m)} = 0\,\,,
    \label{eqch9:boundcondbprimeprime0}\\
    & r(0) = r_+ \,\,,\label{eqch9:boundcondrp}\\
    & \eval{\frac{r'}{a_1}}_{u=0} = 0\,\,,\label{eqch9:boundcondr0}
\end{align}
where a primed quantity means derivative over $u$, 
i.e. $b'_1 = \frac{d b_1}{du}$, see Chapter~\ref{ch:Euclideanpathintegral} 
for more details.

At $u=1$, the boundary conditions are specific to the fixed quantities of 
the ensemble. In this case, we fix the geometry of $\partial M$, having
a topology $\mathbb{S}^1 \times \mathbb{S}^{2}$, with the metric components
\begin{align}
    & r(1) = R\,\,,\label{eqch9:Rat1}\\
    & 2\pi b_2(1) = \beta \,\,,\label{eqch9:bat1}
\end{align}
i.e. the radius of the boundary of the cavity is fixed to be $R$ and the 
Euclidean time length corresponds to the inverse temperature as 
$\int (\gamma^{\tau\tau})^{-1/2} d\tau  = \beta = T^{-1}$.
Finally, we also fix the chemical potential $\mu$ at $\partial M$ 
which obeys the relation 
\begin{align}
    \beta \mu = \frac{\mu_m}{T_m}\,,\label{eqch9:boundchemical}
\end{align}
where $\beta \mu$ can be understood as the logarithm of the fugacity, 
and we shall consider it instead of 
$\mu$ for convenience.

\subsection{Constraint equations}

In order to simplify the path integral, we perform the zero loop 
approximation. An intermediate step for this approximation, which also 
avoids metrics whose action is arbitrarily negative, is to impose the 
constraint equations that are obeyed by the stationary points of the 
action. In some sense, one is integrating over metrics that are physically 
relevant but do not necessarily obey the evolution equations. In this case, 
the constraint equations consist on the Hamiltonian and momentum constraints 
for the Euclidean space and the Gauss constraint to the Maxwell field. 
We here impose these constraints in $M_1$, $M_2$ and $\mathcal{C}$.
Notice that the momentum constraints are apriori satisfied since we have 
a static spacetime with matter.

The Hamiltonian constraint for spaces $M_1$ and $M_2$ are given by 
${G^{\tau}}_{\tau} = 8\pi l_p^2 {T^{\tau}}_{\tau}$, where ${G^{\tau}}_{\tau}$
is the $\tau\tau$-component of the Einstein tensor given in this case by
\begin{align}
    & {G^{\tau}}_{\tau}|_{M_1} = \frac{2}{2r' r^{2}}
    \left(r\left[\left(\frac{r'}{a_1}\right)^2 -1\right] \right)'\,\,,
    \label{eqch9:einsteintensorM1}\\
    & {G^{\tau}}_{\tau}|_{M_2} = \frac{2}{2r' r^{2}}
    \left(r\left[\left(\frac{r'}{a_2}\right)^2 -1\right] \right)'\,\,,
    \label{eqch9:einsteintensorM2}
\end{align}
for $M_1$ and $M_2$, respectively, and ${T^{a}}_{b} 
= 0$ since one has vacuum space. 
Therefore, the Hamiltonian constraints given for $M_1$ and $M_2$ 
are respectively
\begin{align}
    &\frac{2}{r' r^{2}}
    \left(r\left[\left(\frac{r'}{a_1}\right)^2 -1\right] \right)'
    = 0,
    \label{eqch9:hamiltonianconstrM1}\\
    & \frac{2}{r' r^{2}}
    \left(r\left[\left(\frac{r'}{a_2}\right)^2 -1\right] \right)'
    = 0\,\,.
    \label{eqch9:hamiltonianconstrM2}
\end{align}
For $\mathcal{C}$, one has the terms of the Hamiltonian constraint depending
on a Dirac delta positioned at the shell, which leads to the junction 
condition $[{K^{\tau}}_{\tau}] - h^\tau_{\tau}[K] 
= - 8\pi l_p^2 {S^{\tau}}_{\tau}$, where $S^{\tau}_{\tau} = -\epsilon_m$ 
is the $\tau\tau$ component
of the surface stress-energy tensor of the shell. Notice that this stress-energy 
tensor is diagonal, with the other diagonal components being 
${S^{\theta}}_{\theta} = {S^{\phi}}_{\phi} 
= p_m$, i.e.
the tangential pressure. This stress-energy 
tensor is the same as if one considered the variational principle of a 
perfect fluid, see~\cite{Brown:1992kc}. In our case, it comes from the fact that 
the term $\mu_\mathrm{m} n_\mathrm{m} \sqrt{g}$ gives $\frac{\beta\mu}{2\pi} 
\mathfrak{n}_\mathrm{m}$ and so does not depend on the metric, while the 
variation of $\mathcal{F}$ in order to the metric is 
\begin{align}
    \delta \mathcal{F} = - s_\mathrm{m}\delta \left( \frac{\sqrt{\gamma^{\tau \tau}}}{2\pi}\right)
- \frac{\epsilon_{\mathrm{m}} + p_\mathrm{m} - T_\mathrm{m} s_\mathrm{m}}{\sqrt{\gamma^{\tau \tau} \gamma}}
\delta\left(\sqrt{\gamma^{\tau\tau}\gamma} \right)\,\,.
\end{align}
It is useful to explicitly express 
the extrinsic curvatures for constant $u$ hypersurfaces as
\begin{align}
    & K_{ab}dx^a dx^b|_{M_1} = \frac{b'_1 b_1}{a_1} 
    \left(\frac{b_2^2(u_m)}{b_1^2(u_m)}\right)d\tau^2 
    + \frac{r'}{a_1 r}d\Omega^2\,\,,\label{eqch9:extrinsicM1}\\
    & K_{ab}dx^a dx^b|_{M_2} = \frac{b'_2 b_2}{a_2} d\tau^2 
    + \frac{r'}{a_2 r}d\Omega^2\,\,,
    \label{eqch9:extrinsicM2}
\end{align}
with
\begin{align}
    & K|_{M_1} = \frac{b_1'}{a_1 b_1} + 2\frac{r'}{a_1 r}\,\,,
    \label{eqch9:extrinsicscalarM1}\\
    & K|_{M_2} = \frac{b_2'}{a_2 b_2} + 2\frac{r'}{a_2 r}\,\,,
    \label{eqch9:extrinsicscalarM2}
\end{align}
for $M_1$ and $M_2$ respectively. And so the junction condition from the 
Hamiltonian constraint is 
\begin{align}
    \frac{\alpha}{l_p^2} \left[\frac{r'}{a_1}-\frac{r'}{a_2}\right] 
    = m \,\,,\label{eqch9:junctioncond}
\end{align} 
where the mass of the shell $m = 4\pi \alpha^{2}\epsilon_m$ has been defined.

We can then integrate the Hamiltonian constraints in 
Eqs.~\eqref{eqch9:hamiltonianconstrM1} and~\eqref{eqch9:hamiltonianconstrM2}. 
In particular, the Hamiltonian 
constraints are satisfied with the following expressions for $a_1$ and $a_2$
\begin{align}
    &\left(\frac{r'}{a_1}\right)^2 \equiv f(r_+,r) \equiv f_{1}(r) = 1 
    - \frac{r_+}{r}
    \,\,,\label{eqch9:f1}\\
    &\left(\frac{r'}{a_2}\right)^2 \equiv f(\tilde{r}_+,r) \equiv 
    f_{2}(r) = 1 - \frac{\tilde{r}_+}{r}\,\,,\label{eqch9:f2}
\end{align}  
where the boundary condition Eq.~\eqref{eqch9:boundcondrp} was used to find the 
integration constant of $f_{1}$, which corresponds to the horizon radius 
given by $r_+$, and the 
integration constant of $f_{2}$ was parametrized with the total gravitational radius of the 
system given by $\Tilde{r}_+$. The function $f_{1}$ and $f_2$
are actually the same function $f_{1} = f_{2} = f$ if the arguments are the same, but we make 
the distinction here to treat $f_{1}(r)$ as $f$ parametrized by $r_+$, 
while $f_{2}(r)$ is treated as $f$ parametrized by $\Tilde{r}_+$, i.e. to avoid bloating.

Notice that the mass of the shell $m$ is determined by the total gravitational 
radius $\tilde{r}_+$, the horizon radius $r_+$ and the shell radius $\alpha$, 
through the junction condition in Eq.~\eqref{eqch9:junctioncond}, i.e. 
\begin{align}
    m = m(\tilde{r}_+,r_+,\alpha) =\frac{\alpha}{l_p^2}\left[\sqrt{f_{1}(\alpha)}
    -\sqrt{f_2(\alpha)}\right] \,\,.\label{eqch9:m=rtilde}
\end{align} 

From now on, we abbreviate the dependence of $\tilde{r}_+$, 
$r_+$ and $\alpha$ throughout the 
chapter, except when explicitly stated otherwise.

\subsection{Grand canonical reduced action}

With the boundary conditions established, the geometry chosen and with the 
constraint equations in mind, we can express the action in 
Eq.~\eqref{eqch9:actionEHS} for the metrics obeying the constraints. 
We can split the action as the sum $I = I_{g\mathrm{f}} + I_m$, where 
$I_{g\mathrm{f}}$ is the gravitational action which in spherical symmetry 
gives
\begin{align}\label{eqch9:actionspherical}
    &I_{g\mathrm{f}} = 
    \left(\frac{2\pi b_2 r }{l_p^2}
    \left(1 - \frac{r'}{a_2}\right)\right)\sVert[3]_{u\rightarrow1}
    -\frac{\Omega}{4l_p^{2}}\left(\frac{b'_1 b_2(u_\mathrm{m}) 
    r^{2}}{a_1 b_1(u_\mathrm{m})}\right)\sVert[3]_{u=0} \notag\\
    & + \frac{1}{8\pi l_p^{2}}\int_{M_1} a_1 b_1 \frac{b_2(u_\mathrm{m})}{b_1(u_\mathrm{m})} 
    r^{2}G\indices{_1^\tau_\tau} d^4x
    + \frac{1}{8\pi l_p^{2}}\int_{M_2} a_2 b_2 r^{2}G\indices{_2^\tau_\tau} d^4x\notag\\
    & - \frac{1}{8\pi l_p^{2}}\int_{\mathcal{C}} 
    ([K\indices{^\tau_\tau}] - [K])\sqrt{\gamma}d^{3}s\,\,,
\end{align}
see Chapter~\ref{ch:Euclideanpathintegral} for more details, 
and the matter action can be written using the property 
$\mathcal{F} = \epsilon_m - T_m s_m$ as
\begin{align}
    I_m = \int_\mathcal{C} \left(\epsilon_m - T_m s_m,
    - \mu_m n_m\right)\sqrt{\gamma}d^{3}s\,\,.
\end{align}
Using the properties of the local temperature and the 
local chemical potential, i.e. $2\pi T_m = (\sqrt{\gamma^{\tau \tau}})^{-1}$
 and $\frac{\mu_m}{T_m} = \beta \mu$, one can further reduce the matter 
action as 
\begin{align}
    I_m = \int_\mathcal{C} \epsilon_m \sqrt{\gamma}d^{3}s
     -S_m - \beta \mu N_m\,\,,
\end{align}
where it was defined $S_m = 4\pi \alpha^2 s_m$ and $N_m = 4\pi \alpha^2 n_m$.
Putting now together the actions, one gets 
\begin{align}
    &I = 
    \left(\frac{2\pi b_2 r }{l_p^2}
    \left(1 - \frac{r'}{a_2}\right)\right)\sVert[3]_{u\rightarrow1}
    -\frac{\Omega}{4l_p^{2}}\left(\frac{b'_1 b_2(u_\mathrm{m}) 
    r^{2}}{a_1 b_1(u_\mathrm{m})}\right)\sVert[3]_{u=0} - S_m - \beta \mu N_m\notag\\
    & + \frac{1}{8\pi l_p^{2}}\int_{M_1} a_1 b_1 \frac{b_2(u_\mathrm{m})}{b_1(u_\mathrm{m})} 
    r^{2}G\indices{_1^\tau_\tau} d^4x
    + \frac{1}{8\pi l_p^{2}}\int_{M_2} a_2 b_2 r^{2}G\indices{_2^\tau_\tau} d^4x\notag\\
    & - \frac{1}{8\pi l_p^{2}}\int_{\mathcal{C}} 
    ([K\indices{^\tau_\tau}] - [K] - \epsilon_m)\sqrt{\gamma}d^{3}s\,\,.
\end{align}
By applying the boundary conditions in Eqs.~\eqref{eqch9:boundcondb0}
-\eqref{eqch9:boundchemical} and the Hamiltonian constraints with the junction 
condition, one obtains finally the expression for the reduced action as 
\begin{align}
    &I^*(R,T,\beta\mu; \tilde{r}_+,r_+,\alpha)
    = \beta \frac{R}{l_p^2} \left(1 
    - \sqrt{f_{2}(R)}\right)\notag\\ &
    - \frac{\pi r_+^{2}}{l_p^2} - S_m - \beta \mu N_m
    \,\,,\label{eqch9:reducedaction1}
\end{align}
where $S_m + \beta \mu N_m$ must be a function of $m$ given 
by Eq.~\eqref{eqch9:m=rtilde}, the area of the shell 
$A(\alpha)=4\pi \alpha^{2}$, the chemical potential over temperature 
$\beta \mu$, and, the variables to the left of $;$ are fixed.
The dependence of the matter terms
can be seen by inverting the first law of thermodynamics 
applied to $m(S_m,A(\alpha),N_m)$, 
i.e. $dm = T_m dS_m - p_m dA(\alpha) + \mu_m dN_m$, 
to get the function $S_m(m, A(\alpha),N_m)$ and then add 
$\beta \mu N_m$. The differential is then $d(S_m + \beta \mu N_m)
= \frac{dm}{T_m} + \frac{p_m}{T_m}dA(\alpha) 
+ N_m d(\beta \mu)$, and the reduced action is then fully determined 
by giving equations of state that describe this differential plus the 
expression for $m$ in Eq.~\eqref{eqch9:m=rtilde}. 

For convenience, we define the function $\mathcal{S}$ by the quantity
\begin{align}
    &\mathcal{S}(\beta \mu;\tilde{r}_+,r_+,\alpha) = 
   \frac{\pi r_+^2}{l_p^2} + S_m(m(\tilde{r}_+,r_+,\alpha),A(\alpha),\beta \mu) 
    \notag \\
    &+ \beta \mu N_m(m(\tilde{r}_+,r_+,\alpha),A(\alpha),\beta\mu)
    \,\,.\label{eqch9:legendreentropy} 
\end{align}

\subsection{The constrained path integral for the grand canonical ensemble}

The constrained path integral over configurations obeying the boundary conditions 
above becomes now 
\begin{align}
    Z = \int D\bm{\omega}\, \mathrm{e}^{-I^*(\bm{z};\bm{\omega})}\,\,,
    \label{eqch9:constrainedpathintegral}
\end{align}
where $I^*$ is the reduced action depending on the vector  
$\bm{z} = (R,T,\beta \mu)$ whose components are the fixed parameters
$z^i$ with $i \in\, {1,2,3}$, which correspond to a fixed radius of the 
cavity, a fixed temperature and a fixed 
logarithm of the fugacity at the cavity, 
but it also depends on the vector of variables 
integrated over the path integral,
$\bm{\omega}$, with components $\omega^i$ corresponding to
$\omega^i = (\Tilde{r}_+, r_+,\alpha)$. 
We can see how the constraint equations reduce the path integral in this way. 
Initially, the variables to be integrated on the path integral are 
$b_1$, $b_2$, $a_1$, $a_2$ and $r$, on the space of 
physical metrics. The Hamiltonian constraints
imply that the dependence on $b_1$ and $b_2$ disappear from the action, 
in particular the junction condition removes the dependence of $b_2(u_m)$
from the action. Due to these types of constraints, the functions $r'/a_1$ and 
$r'/a_2$ are functionals of the variable $r_+$ and $\tilde{r}_+$, 
respectively. This means there can be a change of integration element
$Da_1 Da_2$ to $D\tilde{r}_+ Dr_+$. 
Moreover $r'/a_1$ and $r'/a_2$ are functions of $r$ alone. This means one can 
invert the function $r(u)$ to perform an arbitrary change of coordinates 
$u = u(r)$, which will not change the metric except on the location of the 
shell $\alpha = r(u_m)$. Therefore, the integration element $Dr$ becomes
$D\alpha$. Notice here that the Jacobian of these 
transformations on the integration element were not considered 
since we are only interested on 
the zero loop approximation.  
An equation of state for $S_m + \beta \mu N_m$ in function of $m$ given by 
Eq.~\eqref{eqch9:m=rtilde}, $A$ and 
$\mu_m/T_m = \beta \mu$ is still required to determine the partition function.

\subsection{The partition function of the 
$(E,\beta\mu)$ ensemble and 
its relation to the grand canonical ensemble}

Interestingly, it is possible to rewrite the
constrained path integral in Eq.~\eqref{eqch9:constrainedpathintegral} 
in the following way 
\begin{align}
    Z = \int D \tilde{r}_+ \mathrm{e}^{-\beta \frac{R}{l_p^2} \left(1 
    - \sqrt{f_{2}(R)}\right)} Z_{\mathcal{S}}(\beta\mu; \tilde{r}_+)
    \,\,,\label{eqch9:constrainedpathintegral1}
\end{align}
where the functional $Z_{\mathcal{S}}$ is 
\begin{align}
    Z_{\mathcal{S}}(\beta\mu; \tilde{r}_+) = \int D\hat{\bm{\omega}}\, 
    \mathrm{e}^{\mathcal{S}(\beta \mu;\tilde{r}_+,\hat{\bm{\omega}})}\,\,,
\end{align}
where $\hat{\bm{\omega}}$ is defined as the vector with components 
$\hat{\omega}^A = (r_+ , \alpha)$, with indices $A \in \{2,3\}$. 
Therefore, the partition function of the grand canonical ensemble 
is given by a Laplace-like transform~\cite{Brown:1992bq} of 
the functional $Z_{\mathcal{S}}$. 
If we did not consider the chemical potential, then $Z_{\mathcal{S}}$
would describe the microcanonical ensemble
of the black hole with a 
self-gravitating matter shell, as it is the path integral with the 
action without the gravitational boundary term. 
However, with the chemical potential, 
the functional $Z_{\mathcal{S}}$ does not fit into the partition functions 
of the usual ensembles, as it describes a partition function of the 
black hole and self-gravitating shell with fixed $\tilde{r}_+$, or fixed
energy, and $\beta \mu$ fixed. As already stated, for simplicity, 
we call this ensemble the $(E,\beta \mu)$ ensemble of a black hole with a self-gravitating shell. 
It would be interesting to obtain the partition function 
$Z_\mathcal{S}$ from first principles as we did for the grand canonical 
ensemble with a finite cavity. 
However, the calculations are similar as the grand canonical ensemble.
One would have to consider the action without the Gibbons-Hawking-York boundary 
term, which is consistent with fixing the quasilocal energy at the cavity rather than 
the temperature of the cavity in the boundary conditions. Notice that having a 
cavity at finite or infinite radius in the $(E,\beta\mu)$ ensemble 
is equivalent to 
fixing either the quasilocal energy at the cavity at a radius $R$ 
or the gravitational radius 
$\tilde{r}_+$. Since here we fix $\tilde{r}_+$, we can consider $Z_{\mathcal{S}}$
as the partition function of the $(E,\beta\mu)$ ensemble with a cavity 
at infinity, indeed $Z_{\mathcal{S}}$ does not depend 
on $R$.

Assuming that we could determine $Z_{\mathcal{S}}$, either by 
performing the path integral or the zero loop approximation, we can define 
the function $\tilde{S}$ as 
\begin{align}
    \mathrm{e}^{\tilde{\mathcal{S}}(\beta\mu; \tilde{r}_+)} 
    = \int D\hat{\bm{\omega}}\, 
    \mathrm{e}^{\mathcal{S}(\beta \mu;\tilde{r}_+,\hat{\omega})} = 
    Z_{\mathcal{S}}\,\,,\label{eqch9:tildeS}
\end{align}
and the grand canonical partition function can be given by 
\begin{align}
    Z = \int D \tilde{r}_+ \mathrm{e}^{- \tilde{I}(\bm{z};\tilde{r}_+)} 
    \,\,,\label{eqch9:constrainedpathintegral2}
\end{align}
with 
\begin{align}
    \tilde{I}(\bm{z};\tilde{r}_+) = \beta R \left(1 - \sqrt{f_2(R)}\right)
    - \tilde{\mathcal{S}}(\beta\mu;\tilde{r}_+)\,\,,
    \label{eqch9:effectiveaction}
\end{align}
being the effective action of the grand canonical ensemble. The result of 
Eq.~\eqref{eqch9:constrainedpathintegral2} together with Eq.~\eqref{eqch9:tildeS}
allows for a better understanding of the full zero loop approximation
applied to Eq.~\eqref{eqch9:constrainedpathintegral}, as we shall see below.
Moreover, Eq.~\eqref{eqch9:constrainedpathintegral2} means that the 
system of a black hole and self-gravitating matter thin shell can 
be described by an effective action, and the freedom of choosing 
the equations of state for the shell turns into some freedom on 
the expression of the function $\tilde{\mathcal{S}}$.

\section{$(E,\beta\mu)$ ensemble in the zero loop approximation
\label{sech9:chemicalzeroloop}}

\subsection{Expansion around the stationary points}

Here, we treat the zero loop approximation applied to the path integral in 
Eq.~\eqref{eqch9:tildeS}, i.e. to the partition 
function of the black hole plus a thin shell with fixed 
$\tilde{r}_+$ and $\beta\mu$. This means that the function $\mathcal{S}$ 
must be expanded around its stationary points $\hat{\omega}_0^A = 
(r_+(\beta\mu;\tilde{r}_+), \alpha(\beta\mu; \tilde{r}_+))$ defined by 
$\frac{\partial \mathcal{S}}{\partial \hat{\omega}^A}|_{\hat{\bm{\omega}}= 
\hat{\bm{\omega}}_0} = 0$.
The path integral in Eq.~\eqref{eqch9:tildeS} can be expanded up to second order as 
\begin{align}
    Z_{\mathcal{S}} = \mathrm{e}^{\mathcal{S}(\beta\mu;\tilde{r}_+, 
    \hat{\bm{\omega}}_0)} 
    \int D\delta\hat{\bm{\omega}} \mathrm{e}^{- \hat{H}_{\hat{\omega}^A \hat{\omega}^B} 
    \delta\hat{\omega}^A 
    \delta\hat{\omega}^B}\,\,,
    \label{eqch9:expansionchemicalpartition}
\end{align}
with $\hat{H}_{\hat{\omega}^A \hat{\omega}^B}$ 
being the negative of the second derivatives of $\mathcal{S}$ 
as $\hat{H}_{\hat{\omega}^A \hat{\omega}^B} = 
-\frac{\partial \mathcal{S}}{\partial \hat{\omega}^A \partial 
\hat{\omega}^B}$ evaluated at $\bm{\omega} = \bm{\omega}_0$ 
and $\delta \hat{\omega}^A = \hat{\omega}^A  - \hat{\omega}^A_0$. 
In order for the zero loop approximation to be well-defined, the path integral 
in Eq.~\eqref{eqch9:expansionchemicalpartition} should be convergent. This means that 
the matrix $\hat{H}_{\hat{\omega}^A \hat{\omega}^B}$ 
must be positive definite and so the stationary points 
must be a maximum of the function $\mathcal{S}$. By truncating the expansion at zeroth 
order, we obtain $Z_{\mathcal{S}} = \mathrm{e}^{\mathcal{S}(\beta\mu;\tilde{r}_+, 
    \hat{\bm{\omega}}_0)}$
and so, from the definition of the function $\tilde{\mathcal{S}}$ in 
Eq.~\eqref{eqch9:tildeS}, we have
\begin{align}\label{eqch9:zeroloopchemicalentropy}
    \tilde{\mathcal{S}}(\beta\mu;\tilde{r}_+) =  \mathcal{S}(\beta\mu;\tilde{r}_+, 
    \hat{\bm{\omega}}_0)\,\,.
\end{align}

\subsection{Stationary equations}

The stationary conditions follow from finding the stationary points of 
the function $\mathcal{S}$, with fixed $\tilde{r}_+$ and $\beta \mu$.
And so the stationary points $\hat{\bm{\omega}}_0 = (r_+(\hat{\bm{z}}),
\alpha(\hat{\bm{z}}))$, with $\hat{\bm{z}} = (\tilde{r}_+,\beta \mu)$, 
are such that 
$(\frac{\partial \mathcal{S}}{\partial \hat{\omega}^A})
|_{\hat{\bm{\omega}} = \hat{\bm{\omega}}_0}=0$. The derivatives of 
  $\mathcal{S}$ are 
\begin{align}\label{eqch9:stationaritycondinternal}
    & \frac{\partial \mathcal{S}}{\partial r_+} = 
    \left(1 - \frac{T(r_+,\alpha)}{T_m}\right)\frac{2\pi r_+}{l_p^2}\,\,,\notag \\
    & \frac{\partial \mathcal{S}}{\partial \alpha} = \frac{2 \alpha }{T_m}
    \left(4\pi p_m - 4\pi p(\alpha) \right)\,\,,
\end{align}
where the following definitions for the temperature 
and pressure functions, for simplicity, were used
\begin{align}
    &T(r_+,\alpha) = 
    \frac{1}{4\pi r_+ \sqrt{f(r_+;\alpha)}}\,\,,
    \label{eqch9:temperaturefunction}\\
    &4\pi p(\alpha) = \frac{1}{4 \alpha l_p^2}
    \left(\frac{1 + f_2(\alpha)}{\sqrt{f_2(\alpha)}} 
    - \frac{1 + f_1(\alpha)}{\sqrt{f_1(\alpha)}}\right)
    .\label{eqch9:pressurefunction}
\end{align}
Then, the stationary conditions become 
\begin{align}
    & T_m = T(r_+,\alpha) \,\,,\label{eqch9:statior1}\\
    & 4\pi p_m =4\pi p(\alpha) \,\,\label{eqch9:statioalpha}.
\end{align}
Therefore, the stationary conditions imply that the temperature of 
the shell must be at the temperature given by the Tolman formula 
and that there must be an equilibrium of pressures at the shell.

\subsection{Stability conditions and their relation to the behaviour of the 
solutions
\label{sech9:stabinternal}}

The zero loop approximation in the context 
of the $(E,\beta\mu)$ ensemble is valid if the stationary points 
obtained 
from solving Eqs.~\eqref{eqch9:statior1} and~\eqref{eqch9:statioalpha}
are local maxima of $\mathcal{S}$. To such stationary points, we designate 
them as stable solutions. As we have seen, this only happens 
if the matrix $\hat{H}_{\hat{\omega}^A \hat{\omega}^B}$ 
is positive definite. Since the matrix is $2\times2$,
we can use Sylvester's criterion to obtain the two sufficient conditions for stability
as
\begin{align}
    & \hat{H}_{\alpha \alpha} > 0\,\,,\label{eqch9:stabinternalcond1}\\ 
    & \frac{|\hat{H}|}{\hat{H}_{\alpha \alpha}} > 0 \,\,,\label{eqch9:stabinternalcond2}
\end{align}
where $|\hat{H}|$ is the determinant of the matrix 
$\hat{H}_{\hat{\omega}^A \hat{\omega}^B}$. The components of 
the hessian $\hat{H}_{\hat{\omega}^A \hat{\omega}^B}$ 
are presented in Sec.~\ref{sech9:hessian}. Namely, the component 
$\hat{H}_{\alpha \alpha}$ is related to the mechanical stability of the 
shell. Indeed, $\hat{H}_{\alpha \alpha}$ corresponds 
to the derivative of the difference of pressures, and the 
condition in Eq.~\eqref{eqch9:stabinternalcond1} is precisely 
the condition that must be obeyed for the shell to be mechanically stable, 
as we show in Sec.~\ref{sech9:stabshellmech}.
This is quite interesting as the validity of the 
zero loop approximation through the path integral approach 
gives precisely the mechanical stability condition of the shell.

We can write the stability conditions in a different way which may help to 
understand the behaviour of the solutions under stability. The system in 
Eqs.~\eqref{eqch9:stabinternalcond1} and~\eqref{eqch9:stabinternalcond2} does not give an 
explicit connection of the stability conditions with the behaviour of the solutions. 
However, we can establish this connection by considering the following. The 
stationary solutions $\hat{\omega}^A_{0}$ are described by 
Eqs.~\eqref{eqch9:statior1} and~\eqref{eqch9:statioalpha}. 
One can now perform on Eqs.~\eqref{eqch9:statior1} and~\eqref{eqch9:statioalpha}
the total derivative in $\hat{z}^A = (\tilde{r}_+,\beta \mu)$, which are the quantities 
that are fixed in the $(E,\beta\mu)$ ensemble. 
One can obtain the following relations
$\hat{\xi}_{\hat{z}^C \hat{\omega}^A} + \hat{H}_{\hat{\omega}^A \hat{\omega}^B} 
\frac{\partial \hat{\omega}^B_0}{\partial \hat{z}^C} = 0$, where 
$\hat{\xi}_{\hat{z}^C \hat{\omega}^A} = 
-\eval{\frac{\partial^2 \mathcal{S}}{\partial \hat{\omega}^A \partial \hat{z}^C}}
_{\hat{\bm{\omega}} = \hat{\bm{\omega}}_0}$. 
And so, these relations can be inverted to 
yield the derivatives of the solutions of the $(E,\beta\mu)$ ensemble as
\begin{align}
    & \frac{\partial \hat{\omega}^A_0}{\partial \hat{z}^C} 
    = - (\hat{H}^{-1})^{\hat{\omega}^A \hat{\omega}^B} 
    \hat{\xi}_{\hat{z}^A \hat{\omega}^B}\,\,,\label{eqch9:solinternalderivatives1}
\end{align}
Now, we can build a matrix $\hat{\mathcal{H}}_{\hat{z}^D \hat{z}^C} 
= - \hat{\xi}_{\hat{z}^D\hat{\omega}^A} \frac{\partial \hat{\omega}^A_0}{\partial \hat{z}^C}$, 
which is related to the inverse of the hessian by Eq.~\eqref{eqch9:solinternalderivatives1},
or explicitly $\hat{\mathcal{H}}_{\hat{z}^D \hat{z}^C} 
= \hat{\xi}_{\hat{z}^D \hat{\omega}^A}(\hat{H}^{-1})^{\hat{\omega}^A \hat{\omega}^B} 
    \hat{\xi}_{\hat{z}^C \hat{\omega}^B}$. 
The vectors $\hat{\xi}_{\hat{z}^A \hat{\omega}^A}$ are 
\begin{align}
    & \hat{\xi}_{\tilde{r}_+ \hat{\omega}^A} = - \frac{\partial}{\partial \hat{\omega}^A}
    \left(\frac{T(\tilde{r}_+,\alpha)}{T_m}\frac{2\pi \tilde{r}_+}{l_p^2}\right)\,\,,
    \label{eqch9:hatxitilderp}\\
    & \hat{\xi}_{\beta \mu \hat{\omega}^A} = - \frac{\partial N_m}{\partial \hat{\omega}^A}\,\,.
    \label{eqch9:hatxibetamu}
\end{align}
Therefore, we can write the matrix $\mathcal{H}_{\hat{z}^C \hat{z}^D}$ as
\begin{align}
    \mathcal{H} = \begin{pmatrix}
        \eval{\frac{\partial}{\partial |\tilde{r}_+}
        \left( \frac{T(\tilde{r}_+,\alpha)}{T_m}\frac{2\pi \tilde{r}_+}{l_p^2}\right)}
        _{\hat{\bm{\omega}}= \hat{\bm{\omega}}_0}
        & \eval{\frac{\partial N_m}{\partial |\tilde{r}_+}}_{\hat{\bm{\omega}}= \hat{\bm{\omega}}_0} \\
        \eval{\frac{\partial N_m}{\partial |\tilde{r}_+}}_{\hat{\bm{\omega}}= \hat{\bm{\omega}}_0} 
        & \eval{\frac{\partial N_m}{\partial |\beta \mu}}_{\hat{\bm{\omega}}= \hat{\bm{\omega}}_0}
    \end{pmatrix}\,\,,\label{eqch9:mathcalH}
\end{align}
where $\frac{\partial}{\partial |\hat{z}^C}
 = \frac{\partial \hat{\omega}^A_0}{\partial \hat{z}^C} 
\frac{\partial }{\partial \hat{\omega}^A}$
represents the partial derivative over the implicit dependence of 
$\hat{z}^C$, and also $\frac{\partial N_m}{\partial |\tilde{r}_+} = 
\frac{\partial }{\partial |\beta \mu}
\left(\frac{T(\tilde{r}_+,\alpha)}{T_m}2\pi \tilde{r}_+\right)$ 
due to the hessian being symmetric.
The stability conditions stipulate that the matrix 
$\hat{H}_{\hat{\omega}^A \hat{\omega}^B}$ must be positive definite, which means that 
$(\hat{H}^{-1})^{\hat{\omega}^A \hat{\omega}^B}$ must also be positive definite. 
Since the vectors $\hat{\xi}_{\hat{z}^C \hat{\omega}^A}$ for each $\hat{z}^C$ 
are in principle independent, they
can be represented as a nonsingular matrix, the matrix 
$\mathcal{H}_{\hat{z}^A \hat{z}^B}$ can be seen as 
$\mathcal{H}_{\hat{z}^A \hat{z}^B} = \hat{\xi}_{\hat{z}^A \hat{\omega}^C} 
(\hat{H}^{-1})^{\hat{\omega}^C \hat{\omega}^D} \hat{\xi}^T_{\hat{\omega}^D \hat{z}^B}$, 
where $\hat{\xi}^T_{\hat{\omega}^D \hat{z}^B}$ 
are the transpose components of the matrix $ \hat{\xi}_{\hat{z}^B \hat{\omega}^D}$. 
Therefore, $\mathcal{H}_{\hat{z}^A \hat{z}^B}$ must be positive definite if and only 
if $\hat{H}_{\hat{\omega}^A \hat{\omega}^B}$ is positive definite. This statement 
is related to the thermodynamics of the system, which we discuss below.

\section{Grand canonical ensemble in the zero loop approximation
\label{sech8:grandcan}}

\subsection{Grand canonical path integral expansion 
around the stationary points}

We now proceed with the zero loop approximation of the statistical path 
integral for the grand canonical ensemble 
in Eq.~\eqref{eqch9:constrainedpathintegral}, with the reduced 
action given by Eq.~\eqref{eqch9:reducedaction1} and with the relations given 
by Eq.~\eqref{eqch9:m=rtilde}.
We then perform the zero loop approximation by expanding the 
reduced action around its stationary points. These stationary points are the solutions  
$\bm{\omega}_0 = (\tilde{r}_+(\bm{z}),r_+(\bm{z}),
\alpha(\bm{z}))$ such that
$\eval{\frac{\partial I^*}{\partial \omega^i}}_{\bm{\omega}=\bm{\omega}_0}=0$.
The path integral around the stationary point can then be rewritten as 
\begin{align}
    Z = \mathrm{e}^{-I_0(\bm{z})}\int D\delta\bm{\omega}\,
    \mathrm{e}^{-(H_{\omega^i \omega^j})\delta\omega^i \delta\omega^j}\,\,,
    \label{eqch9:firstloop}
\end{align} 
where $I_0(\bm{z}) = I^*(\bm{z}; \bm{\omega}_0)$ 
is the reduced action evaluated at the stationary 
point, $H_{\omega^i \omega^j} = \eval{\frac{\partial^2 I^*}{\partial \omega^i \partial \omega^j}}
_{\bm{\omega}=\bm{\omega}_0}$ is the hessian 
of the reduced action over the variables $\bm{\omega}$, evaluated at the 
stationary point, and 
$\delta\bm{\omega}$ is the difference vector of the variables to the 
solutions of the stationary points, i.e. 
$\delta \omega^i = \omega^i - \omega^i_0$. We adopt the notation that 
the partial derivatives are done while keeping the variables of the 
definition of the function constant. 
The zero loop approximation is 
then valid if the hessian is positive definite i.e. for solutions that are 
minima of the action. If the solutions of the ensemble are minima, then the 
solutions are stable, otherwise they are maxima and unstable or saddle points
and so marginally 
stable. The partition function for the stable solutions is then 
$Z=\mathrm{e}^{-I_0(\bm{z})}$.

However, with the zero loop approximation done to the partition function of the 
$(E,\beta\mu)$ ensemble, it is better to envision the zero 
loop approximation of the path integral describing the grand canonical ensemble through the 
identity in Eq.~\eqref{eqch9:constrainedpathintegral1}. We can apply the 
zero loop approximation in parts, starting by the functional $Z_{\mathcal{S}}$, 
which was obtained in Sec.~\ref{sech9:chemicalzeroloop}, and the partition 
function of the grand canonical ensemble becomes
\begin{align}
    &Z =   \int D \tilde{r}_+ \left[ \mathrm{e}^{-\tilde{I}(\bm{z};\tilde{r}_+)} 
    \int D\delta\hat{\bm{\omega}} 
    \mathrm{e}^{- \hat{H}_{\hat{\omega}^A \hat{\omega}^B} 
    \delta\hat{\omega}^A 
    \delta\hat{\omega}^B}\right]\,\,,\label{eqch9:constrainedpathintegral3}
\end{align}
where the effective action $\tilde{I}(\bm{z},\tilde{r}_+)$ is given by 
Eq.~\eqref{eqch9:effectiveaction} and the function $\tilde{\mathcal{S}}$ is provided 
by Eq.~\eqref{eqch9:zeroloopchemicalentropy}, i.e. it is
determined in this case through the zero loop approximation of the path integral 
describing the
$(E,\beta\mu)$ ensemble of the black hole and the self-gravitating 
thin shell, $\tilde{\mathcal{S}} = 
\mathcal{S}(\beta\mu;\tilde{r}_+,\hat{\bm{\omega}}_0)$. 
Note that this assignment is done by deprecating the 
path integral over the fluctuations of the parameters $\hat{\omega}^A$, 
since we are interested in the zero loop. Now we can 
proceed with the expansion over the stationary points of the 
effective action, $\frac{\partial \tilde{I}}{\partial \tilde{r}_+}|_{\tilde{r}_+
=\tilde{r}_+(\bm{z})} = 0$, 
obtaining 
\begin{align}
    &Z = \mathrm{e}^{-I_0(\bm{z})} \int D\delta\tilde{r}_+ 
    \left[\mathrm{e}^{- \tilde{H}_{\tilde{r}_+ \tilde{r}_+} 
    \delta \tilde{r}_+ \delta \tilde{r}_+} \right.\notag\\
    & \times\left.\int D\delta \hat{\bm{\omega}}  
    \mathrm{e}^{- \hat{H}_{\hat{\omega}^A \hat{\omega}^B} 
    \delta\hat{\omega}^A 
    \delta\hat{\omega}^B}\right]\,\,,
    \label{eqch9:constrainedpathintegral4}
\end{align} 
where $\tilde{H}_{\tilde{r}_+ \tilde{r}_+} = 
\frac{\partial^2 \tilde{I}}{\partial \tilde{r}_+^2}|_{\tilde{r}_+=\tilde{r}_+(\bm{z})}$, 
and $I_0(\bm{z}) = \tilde{I}(\bm{z};\tilde{r}_+(\bm{z}))$. 
In connection with the expansion in Eq.~\eqref{eqch9:firstloop}, the 
zeroth order action is the same, i.e. 
\begin{align}
    I_0(\bm{z}) = \tilde{I}(\bm{z};\tilde{r}_+(\bm{z})) = 
I^*(\bm{z};\tilde{r}_+(\bm{z}),
\hat{\bm{\omega}}_0|_{\tilde{r}_+ = \tilde{r}_+(\bm{z})})\,\,,
\end{align}
and the stationary points in Eq.~\eqref{eqch9:firstloop} are the same, i.e. 
\begin{align}
    \omega_0^i = (\tilde{r}_+(\bm{z}), 
    \hat{\omega}_0^A|_{\tilde{r}_+ = \tilde{r}_+(\bm{z})})
    \,\,. 
\end{align}
Considering the second order perturbations of the action in the two 
expansions, i.e. in Eqs.~\eqref{eqch9:firstloop} and~\eqref{eqch9:constrainedpathintegral4},
we can prove the equivalence between the two by using the transformation 
$\delta \omega^A = \delta \hat{\omega}^A + \frac{\partial \hat{\omega}^A_0}{\partial \tilde{r}_+} 
\delta \tilde{r}_+$, with 
$\frac{\partial \hat{\omega}^A_0}{\partial \tilde{r}_+} =  -H_{\tilde{r}_+ \hat{\omega}^B} 
(\hat{H}^{-1})^{\hat{\omega}^A \hat{\omega}^B}|_{\tilde{r}_+=\tilde{r}_+(\bm{z})}$, 
see Sec.~\ref{sech9:hessian} for the expression of the Hessians. Note that the 
Hessians in this case behave as tensors since the first derivatives of the 
respective actions vanish due to the stationary conditions.
Therefore, the conditions for the validity of the zero loop approximation in both 
expansions are equivalent. However, the expansion in 
Eq.~\eqref{eqch9:constrainedpathintegral4} gives a more clear interpretation 
of the zero loop approximation of the path integral describing the grand canonical ensemble and the 
meaning of the conditions for its validity. It is thus convenient to work 
with Eqs.~\eqref{eqch9:constrainedpathintegral3} 
and~\eqref{eqch9:constrainedpathintegral4} and the effective 
action given in Eq.~\eqref{eqch9:effectiveaction}, i.e. 
$\tilde{I}(\bm{z};\tilde{r}_+) = 
\beta R\left(1 - \sqrt{f_2(R)}\right) 
- \tilde{\mathcal{S}}\left(\beta \mu; \tilde{r}_+\right)$, 
with the identification 
in Eq.~\eqref{eqch9:zeroloopchemicalentropy}, 
i.e. $\tilde{\mathcal{S}}(\beta\mu;\tilde{r}_+) =  
\mathcal{S}(\beta\mu;
\tilde{r}_+,\hat{\bm{\omega}}_0)$.

\subsection{Stationary equation}

The stationary equation describing the minimum of the effective 
action is determined by $\frac{\partial \tilde{I}}{\partial \tilde{r}_+}|_{\tilde{r}_+
=\tilde{r}_+(\bm{z})} = 0$, with the minimum 
$\tilde{r}_+ = \tilde{r}_+(\bm{z})$. Knowing the expression of the 
effective action in Eq.~\eqref{eqch9:effectiveaction}, we obtain the 
stationary equation as
\begin{align}
    \beta = B(\beta \mu; \tilde{r}_+)\sqrt{f_2(R)}\,\,,
\label{eqch9:stationtildes}
\end{align}
where $B(\beta \mu; \tilde{r}_+) = 2\frac{\partial \tilde{\mathcal{S}}}{\partial \tilde{r}_+}$. 
Using the fact that 
$\tilde{\mathcal{S}}(\beta\mu;\tilde{r}_+) =  
\mathcal{S}(\beta\mu;
\tilde{r}_+,\hat{\bm{\omega}}_0)$ for the black hole and thin shell in the 
zero loop approximation, the function $B(\beta\mu; \tilde{r}_+)$ can be written 
in terms of the quantities of the black hole and shell evaluated at the 
stationary points of the $(E,\beta\mu)$ ensemble, yielding
\begin{align}
    & B(\beta \mu;\tilde{r}_+) = 
    \frac{1}{\sqrt{f_2(\alpha)} T_m(m(\tilde{r}_+,\hat{\omega}_0),A(\alpha(\hat{\bm{z}})),
    \beta \mu)}\,\,.
    \label{eqch9:B}
\end{align}
It is also interesting to consider the number of particles 
$\frac{\partial \tilde{\mathcal{S}}}{\partial \beta \mu} = 
\tilde{N}(\beta \mu ;\tilde{r}_+)$ which for the case of the black hole and 
thin shell is $\tilde{N}(\beta \mu ;\tilde{r}_+) = 
N_m(m(\tilde{r}_+,\hat{\omega}_0),A(\alpha(\hat{\bm{z}})),\beta\mu)$.

\subsection{Stability condition}

For the validity of the 
zero loop approximation of the path integral describing the grand canonical ensemble, 
we must require that $\tilde{H}_{\tilde{r}_+ \tilde{r}_+} 
= \eval{\frac{\partial^2 \tilde{I}}{\partial \tilde{r}_+^2}}
_{\tilde{r}_+ = \tilde{r}_+(\bm{z})} > 0$, which reduces to the 
condition
\begin{align}
    \tilde{H}_{\tilde{r}_+ \tilde{r}_+} 
    = \eval{\left(\frac{B}{2 f_2(R) R} - \frac{\partial B}{\partial \tilde{r}_+}\right)}
    _{\tilde{r}_+=\tilde{r}_+(\bm{z})}
    \geq 0\,\,.
    \label{eqch9:stabtilde}
\end{align}
This condition can be tied to the behaviour of the solution 
$\tilde{r}_+=\tilde{r}_+(\bm{z})$. Indeed, by using Eq.~\eqref{eqch9:stationtildes}, 
the derivative of the 
solution $\tilde{r}_+(\bm{z})$ is given by 
\begin{align}\label{eqch9:derTtilde}
    \frac{\partial \tilde{r}_+}{\partial T} = \eval{\frac{ 2 R B^2 f_2^{\frac{3}{2}}(R)}
    {B - 2 f_2(R) R \frac{\partial B}{\partial \tilde{r}_+}}}_{\tilde{r}_+=\tilde{r}_+(\bm{z})}\,\,,
\end{align}
where the partial derivative in $T$ is done by keeping $R$ and $\beta\mu$ 
constant. And so the stability condition in Eq.~\eqref{eqch9:stabtilde} 
leads to the condition that
    $\frac{\partial \tilde{r}_+}{\partial T} > 0$,
the gravitational radius must increase with the temperature of the ensemble.
It is also convenient to write the other derivative of the solution from applying 
the derivative over $\beta\mu$ on the stationary condition, giving
\begin{align}\label{eqch9:derbetamutilde}
    \frac{\partial \tilde{r}_+}{\partial \beta\mu} = 
    \frac{T}{B} \frac{\partial \tilde{r}_+}{\partial T}
    \eval{\frac{\partial \tilde{N}}{\partial \tilde{r}_+}}_{\tilde{r}_+ = \tilde{r}_+(\bm{z})}
    \,\,.
\end{align}
Here, the sign of the derivative $\frac{\partial \tilde{r}_+}{\partial \beta\mu}$ depends on 
the sign of $\frac{\partial \tilde{N}}{\partial \tilde{r}_+}$, and ultimately depends on 
the choice of equation of state for the shell.

Note however that the stability conditions of the 
$(E,\beta\mu)$ ensemble, Eqs.~\eqref{eqch9:stabinternalcond1} 
and~\eqref{eqch9:stabinternalcond2}, must be satisfied simultaneously with
Eq.~\eqref{eqch9:stabtilde}, yielding precisely the positive definiteness 
condition of $H_{\omega^i \omega^j}$ in Eq.~\eqref{eqch9:firstloop}. 
The reason for these stability conditions is the 
identification in Eq.~\eqref{eqch9:zeroloopchemicalentropy}, which comes 
from applying the zero loop approximation to $\mathcal{S}$. 
If one did not perform the zero loop approximation to $\mathcal{S}$ 
but performed the zero loop approximation on the effective action, 
one would only have the stability condition in Eq.~\eqref{eqch9:stabtilde}.
This means that the stability condition in Eq.~\eqref{eqch9:stabtilde} is the 
one inherent to the grand canonical ensemble.

\section{Thermodynamics of a self-gravitating 
matter thin shell and a black hole in the $(E,\beta\mu)$ ensemble with cavity 
at infinity\label{sech9:thermomicro}}\sectionmark{Thermodynamics 
in the $(E,\beta\mu)$ ensemble}\thispagestyle{userightbotmark}

\subsection{The $(E,\beta\mu)$ ensemble from statistical mechanics}

The general idea to build the $(E,\beta\mu)$ ensemble 
from statistical arguments is to start by constructing the partition function of 
system with a number of discrete states that exchanges particles with the 
reservoir. For that, we can make use of the microcanonical ensemble of 
the system $A_s$ and the reservoir $A_r$. The system plus the reservoir 
only exchange the number of particles but such that the total number 
of particles is conserved $N^{(0)} = N_s + N_r$, where $N_s$ is the 
number of particles of system $A_s$ and $N_r$ is the total number of 
particles of the reservoir. If the system finds itself in just one 
state with $N_s$ particles, the reservoir will find itself with 
possible $\Omega'(N^{(0)}-N_s)$ number of states with the 
number of particles $N_r = N^{(0)}-N_r$. Meaning that the probability 
of the system to be at exactly one state with the number of particles 
$N_r$ is $P_r = c \Omega'(N^{(0)}-N_r)$,
which comes from the postulate of equal probability between states and 
$c$ is a normalization constant to be determined by the sum of probabilities 
being unity. Since $A_{r}$ is a reservoir, the number of particles 
$N_s$ must be much smaller than $N_r$. This means one can expand 
$\Omega'(N^{(0)}-N_s)$ as $\ln(\Omega'(N^{(0)} - N_s)) = 
\ln(\Omega'(N^{(0)})) - \partial_{N^{(0)}}\ln(\Omega'(N^{(0)})) N_s$. 
With the definition of $\beta \mu$ being $\beta \mu = 
- \partial_{N^{(0)}}\ln(\Omega'(N^{(0)}))$, one gets the probability 
$P_s = \frac{1}{Z_{\mathcal{S}}} \mathrm{e}^{\beta \mu N_s}$, where 
$Z_{\mathcal{S}}$ is the partition function of the $(E,\beta\mu)$ 
ensemble. Since $Z_{\mathcal{S}}$ is determined by normalization of the 
probability, one obtains
\begin{align}
    Z_{\mathcal{S}} = \sum_{N_s} \mathrm{e}^{S_s + \beta\mu N_s}\,\,,
\end{align}
where $\sum_{N_r}$ is done over the possible number of particles of 
the system, $S_s$ is the entropy of the system with number of particles 
$N_s$ correspondent to the logarithm of the number of states of the system 
with $N_s$ particles. Now $Z_{\mathcal{S}}$ is a function of the energy of 
the system $E$ and $\beta\mu$ as $Z_{\mathcal{S}}(E,\beta\mu)$.
From the definition of the inverse temperature as 
$\beta_{C} = \partial_{E}\ln(Z_{\mathcal{S}}(E,\beta\mu))$, together with the fact 
that the mean number of particles is given by 
$N_{C} = \partial_{\beta\mu}\ln(Z_{\mathcal{S}}(E,\beta\mu))$, the differential 
of the logarithm of the partition function is 
\begin{align}
    d \ln(Z_{\mathcal{S}}) = \beta_{C} dE - N_{C} d\beta\mu\,\,.
\end{align} 
But, from the first law of thermodynamics, one has that 
$d(S_{C} + \beta\mu N_{C}) = \beta_{C} dE - N_{C} d\beta\mu$, where 
$S_{C}$ is the entropy of the system. Therefore, the partition function can 
be related to the thermodynamic quantity $S_{C} + \beta\mu N_{C}$ as
\begin{align}
    Z_{\mathcal{S}} = \mathrm{e}^{S_{C} + \beta\mu N_{C}}\,\,,
\end{align}
as a function of the energy and $\beta \mu$. 

Such ensemble is not used frequently as it seems difficult to 
realize a reservoir that only exchanges particles but not energy. However, for our 
purposes, it is convenient to consider it as a step towards the 
grand canonical ensemble. The arguments to obtain the partition function 
are based from Reif's book, but we adapted them here to the number of particles.

\subsection{Connection between the action and thermodynamics}

The $(E,\beta\mu)$ partition function with
fixed total gravitational radius $\tilde{r}_+$ and a fixed $\beta \mu$
should be described by $Z_{\mathcal{S}} = \mathrm{e}^{\tilde{\mathcal{S}}
(\beta\mu;\tilde{r}_+)}$, 
with $\tilde{\mathcal{S}}(\beta\mu;\tilde{r}_+) = 
\mathcal{S}(\beta\mu; \tilde{r}_+,\hat{\bm{\omega}}_0)$. 
From statistical mechanics, the partition function of a system with constant energy 
$\tilde{r}_+/2$ and constant $\beta \mu$ is given by 
$Z = \mathrm{e}^{S_{C} + \beta\mu N_{C}}$,
where $S_{C}$ and $N_{C}$ are the 
entropy and the mean particle number, respectively, of the 
$(E,\beta\mu)$ ensemble, with the subscript $C$ standing for chemical. 
By connecting the two partition functions, 
we obtain that  
\begin{align}
    S_{C} + \beta \mu N_{C} = 
    \mathcal{S}(\beta \mu; \tilde{r}_+,\hat{\bm{\omega}}_0)\,\,.
    \label{eqch9:legendreentropythermodynamics}
\end{align}
From here, we can compute the relevant thermodynamic quantities of the 
$(E,\beta\mu)$ ensemble.

\subsection{Entropy, temperature and particle number}

From the thermodynamic quantity $S_{C} + \beta \mu N_{C}$, one has the differential
\begin{align}
    d(S_{C} + \beta \mu N_{C}) = \frac{1}{T_{C}} d\left(\frac{\tilde{r}_+}{2}\right) 
    + N_{C} d\beta\mu \,\,,\label{eqch9:differentiallegendrethermo} 
\end{align}
where the derivatives of the Legendre transform of the entropy are 
given by $\frac{1}{T_{C}} = 2\frac{\partial (S_{C} + \beta\mu N_{C})}{\partial \tilde{r}_+} $
and $N_{C} = \frac{\partial (S_{C} + \beta \mu N_{C})}{\partial \beta \mu}$. 
From the expression of $\mathcal{S}$, 
we have then that the temperature of the $(E,\beta\mu)$ ensemble is 
\begin{align}
    & T_{C} = T_m(m(\tilde{r}_+,\hat{\bm{\omega}}_0), 
    A(\alpha(\hat{\bm{z}})),\beta \mu)\sqrt{f_2(\alpha(\hat{\bm{z}}))}
    \,\,,\label{eqch9:temperaturemixedensemble}
\end{align}
while the particle number is 
\begin{align}
    &N_{C} = N_m(m(\tilde{r}_+,\hat{\bm{\omega}}_0), 
    A(\alpha(\hat{\bm{z}})),\beta \mu) \,\,.\label{eqch9:particlemixedensemble}
\end{align}
Finally, we can compute the entropy of the system as 
$S_{C}=\mathcal{S}(\beta \mu; \tilde{r}_+,\hat{\bm{\omega}}_0) - \beta \mu N$, 
yielding 
\begin{align}
    S_{C} = \pi r_+^2(\hat{\bm{z}}) + S_m(m(\tilde{r}_+,\hat{\bm{\omega}}_0), 
    A(\alpha(\hat{\bm{z}})),\beta \mu)\,\,.
    \label{eqch9:totalentropymixedensemble}
\end{align}

\subsection{Thermodynamic stability of the $(E,\beta\mu)$ 
ensemble with the reservoir}

In order to analyze the thermodynamic stability of the $(E,\beta\mu)$ 
ensemble, we 
must use the total entropy functional of the system plus the reservoir 
at infinity, which only fixes $\bar{\beta \mu}$ of the system. 
This functional 
is $\bar{\mathcal{S}} = S_{C} + \bar{\beta \mu}N_{C}$, whose variation
represents the variation of the total entropy of the system and the reservoir together, 
as one has 
$d\bar{\mathcal{S}} = dS_{C} + dS_{CM\,r}$, with the variation of the entropy of the 
reservoir being $dS_{CM\,r} = -\bar{\beta \mu}dN_{CM\,r}$. Since the variation on the 
particle number of the reservoir is $dN_{CM\,r} = -dN_{C}$ due to number 
particle conservation, and 
since the variation is done with 
fixed energy, then one has $d\bar{\mathcal{S}} = (\bar{\beta \mu} - \beta\mu)dN_{C}$. 
Now, the total entropy must be at its maximum, 
leading to $d\bar{\mathcal{S}} = 0$, i.e. 
$\beta\mu = \bar{\beta\mu}$,
and leading to the stability condition 
\begin{align}
\frac{\partial N_{C}}{\partial \beta \mu} > 0\,\,, \label{eqch9:stabilitymixedensemble} 
\end{align}
where $N_{C}$ is given by 
Eq.~\eqref{eqch9:particlemixedensemble}.

We can now establish the connection of the stability or validity 
of the zero loop approximation with the thermodynamic stability of the ensemble. 
The stability condition in Eq.~\eqref{eqch9:stabilitymixedensemble} can be expanded into 
$\eval{\frac{\partial N_m}{\partial \beta \mu}}_{\hat{\bm{\omega}} = \hat{\bm{\omega}}_0} 
+ \frac{\partial \hat{\omega}^A_0}{\partial \beta \mu} 
\eval{\frac{\partial N_m}{\partial \omega^A}}_{\hat{\bm{\omega}} = 
\hat{\bm{\omega}}_0} > 0$.
From the matrix in Eq.~\eqref{eqch9:mathcalH}, one of the conditions for the 
positive definiteness 
of the matrix $\mathcal{H}$ is $\eval{\frac{\partial N_m}{\partial |\beta \mu}}
_{\hat{\bm{\omega}}= \hat{\bm{\omega}}_0}
=\frac{\partial \hat{\omega}^A_0}{\partial \beta \mu} 
\eval{\frac{\partial N_m}{\partial \omega^A}}_{\hat{\bm{\omega}} = 
\hat{\bm{\omega}}_0}
>0$, which is not enough to ensure Eq.~\eqref{eqch9:stabilitymixedensemble}. 
We must then consider the positivity 
of $\eval{\frac{\partial N_m}{\partial \beta \mu}}_{\hat{\bm{\omega}}
=\hat{\bm{\omega}}_0}$, which depends heavily on the 
choice of the equations of state for the shell as 
$dS_m = \frac{1}{T_m}dm + \frac{p_m}{T_m}d(4\pi \alpha^2) - \beta \mu dN_m$. 
Moreover, its intrinsic stability must require that $\frac{\partial N_m}{\partial \beta \mu} > 0$. 
Therefore, if one chooses a thermodynamically intrinsically 
stable shell, the maximization of $\mathcal{S}$ indicates that the ensemble is 
thermodynamically stable, as Eq.~\eqref{eqch9:stabilitymixedensemble} is satisfied.

It is also interesting to explore the relation between the thermodynamic 
stability in Eq.~\eqref{eqch9:stabilitymixedensemble} and the mechanical stability of 
the shell in Eq.~\eqref{eqch9:stabinternalcond1}. From the relation $\hat{\mathcal{H}}_{\hat{z}^D \hat{z}^C} 
= \hat{\xi}_{\hat{z}^D \hat{\omega}^A}(\hat{H}^{-1})^{\hat{\omega}^A \hat{\omega}^B} 
\hat{\xi}_{\hat{z}^C \hat{\omega}^B}$ in Sec.~\ref{sech9:stabinternal}, 
one obtains explicitly 
\begin{align}
    & \frac{\partial N_{C}}{\partial \beta \mu} = \frac{\hat{H}_{\alpha \alpha}}{|\hat{H}|}
    \eval{\left(\frac{\partial N_m}{\partial r_+} - \frac{\partial N_m}{\partial \alpha} \frac{H_{r_+ \alpha}}{H_{\alpha \alpha}}
    \right)^2}_{\hat{\bm{\omega}} = 
    \hat{\bm{\omega}}_0} + \frac{1}{H_{\alpha \alpha}} \eval{\left(\frac{\partial N_m}{\partial \alpha}\right)^2}
    _{\hat{\bm{\omega}} = 
    \hat{\bm{\omega}}_0}
    + \eval{\frac{\partial N_m}{\partial \beta\mu}}_{\hat{\bm{\omega}} = 
    \hat{\bm{\omega}}_0}\,\,,
\end{align}
and so mechanical stability is not sufficient to guarantee 
thermodynamic stability, however to have thermodynamic stability 
one needs mechanical stability. This effect is due to the ensemble we are 
considering with $\beta \mu$ fixed.

The conditions for the stability of the zero loop approximation 
seem to be more restrictive than the thermodynamic stability condition of the 
$(E,\beta\mu)$ ensemble. One must remember that the 
system describes actually two subsystems in equilibrium. Therefore, we 
must analyze the thermodynamics of the $(E,\beta\mu)$ ensemble 
as the interaction of two 
systems at constant total energy and constant $\beta \mu$.

\subsection{The $(E,\beta\mu)$ ensemble 
describing two systems in equilibrium
and Le Chatelier-Braun principle}

\subsubsection{The equilibrium of the two systems plus the reservoir and 
the recovery of the 
thermodynamic quantities of the black hole inside a cavity}

In order to treat the ensemble as two systems in equilibrium rather than a system as 
a whole, we must expand the entropy functional in terms of variables of both systems.
We can choose the entropy of the black hole, $S_{bh} = \pi r_+^2$ and the area of the 
cavity $4\pi \alpha^2$. The total energy is held constant, which means that the energy 
of the shell and the energy of the black hole inside the cavity obey a certain relation, 
which is precisely $m = m(\tilde{r}_+,r_+,\alpha)$. 
Then, the differential of the functional $\bar{\mathcal{S}}$
is given by
\begin{align}
    &d\bar{\mathcal{S}} = \left(1 - \frac{T_{bh}}{T_m}\right)dS_{bh} 
    + \frac{1}{T_m}\left(p_{m} 
    - p_{bh}\right)d(4\pi \alpha^2) + (\Bar{\beta \mu} - \beta \mu) dN_m \,\,,
    \label{eqch9:functionalentropytwosystems}
\end{align}
where $T_{bh}$ and $p_{bh}$ are the temperature and the mean pressure of the black hole 
inside a cavity. But the two terms are precisely the differential of $\mathcal{S}$ in 
Eq.~\eqref{eqch9:stationaritycondinternal}, meaning that $T_{bh} = T(r_+,\alpha)$ and 
$p_{bh} = p(\alpha)$. One thus recovers the thermodynamic quantities of the black hole 
inside a cavity of radius $\alpha$. 

From the principle that the functional must be a maximum, this means that 
the first derivatives must vanish, i.e. $T_{bh} = T_m$ and $p_m = p(\alpha)$, which are 
exactly the equilibrium conditions found by the zero loop approximation, 
and also one has $\beta \mu = \bar{\beta \mu}$.

\subsubsection{Le Chatelier-Braun principle}

Since the functional $\bar{\mathcal{S}}$ 
must be a maximum, and with the vanishing 
first derivatives, we must also consider the condition that the hessian of the 
functional must be negative definite, i.e. $d^2\bar{\mathcal{S}} < 0$. 
This precisely yields 
that the matrix
\begin{align}
    &d^2\bar{\mathcal{S}} = \begin{pmatrix}
        -\Lambda^T \cdot\hat{H}\cdot \Lambda & 0_{2} \\
        0_{2}^T &  - \eval{\frac{\partial N_m}{\partial \beta \mu}}_{\hat{\bm{\omega}}
        = \hat{\bm{\omega}}_0} 
    \end{pmatrix}\,\,,\label{eqch9:superfunctionalhessianinternal}\\
    & \Lambda = \begin{pmatrix}
        \eval{(\frac{\partial S_{bh}}{\partial r_+})^{-1}}_{\hat{\bm{\omega}}
        = \hat{\bm{\omega}}_0} & 0 \\
        0 & \eval{(\frac{\partial (4\pi \alpha^2)}{\partial \alpha})^{-1}}_{\hat{\bm{\omega}}
        = \hat{\bm{\omega}}_0}
    \end{pmatrix} \,\,,\label{eqch9:lambdainternal}
\end{align}
must be negative definite,
where $0_{2}$ is the two-dimensional zero vector and $0_2^T$ its transpose. 
This means that precisely $\hat{H}$ must be positive definite and 
that the shell must be intrinsically 
stable as $\frac{\partial N_m}{\partial \beta \mu}> 0$.

The thermodynamic meaning of the positive definiteness 
of $\hat{H}$ is exactly the Le Chatelier-Braun principle of the two subsystems 
in equilibrium, i.e. the black hole inside a cavity made by the 
matter shell and the matter shell itself. Indeed, 
one can use the variable of the black hole $S_{bh}$, which has a conjugate 
variable $\left(1 - \frac{T(r_+,\alpha)}{T_m} \right)$, being zero if the 
black hole is in equilibrium with the cavity. The other variable $\alpha$, 
which is a variable of 
matter shell, has a conjugate variable $\frac{1}{T_m}(p_m - p(\alpha))$, 
which is zero if the black hole inside the cavity and the matter shell are 
in mechanical equilibrium. Now, if one assumes that the black hole seizes to be in 
thermal equilibrium, the black hole entropy increases 
and its conjugate variable varies as
$\Delta\left(1 - \frac{T(r_+,\alpha)}{T_m} \right) = \eval{\left((\frac{\partial S_{bh}}{\partial r_+})^{-1}
\frac{\partial}{\partial r_+}\left(1 - \frac{T(r_+,\alpha)}{T_m} \right)\right)}_{\hat{\bm{\omega}}
= \hat{\bm{\omega}}_0} \Delta S_{bh}$, while maintaining 
$4\pi \alpha^2$ constant. These variations result in a violation also of the equilibrium 
condition $\frac{1}{T_m}(p_m - p(\alpha))$. As equilibrium is restored, one has 
a variation of the equilibrium condition 
$\Delta\left(1 - \frac{T(r_+,\alpha(\tilde{r}_+,\beta\mu,r_+))}{T_m} \right)
=  \left((\frac{\partial S_{bh}}{\partial r_+})^{-1}
\frac{\partial}{\partial r_+}\left(1 - \frac{T(r_+,\alpha(r_+,\tilde{r}_+,\beta \mu))}{T_m} \right)
\right)_{\hat{\bm{\omega}}= \hat{\bm{\omega}}_0}
 \Delta S_{bh}$, where $p_m = p(\alpha)$ is assumed, yielding solutions to 
 $\alpha = \alpha(\tilde{r}_+,\beta\mu, r_+)$.
One can write this last derivative as 
\begin{align}
    & \eval{\left(\left(\frac{\partial S_{bh}}{\partial r_+}\right)^{-1}\frac{\partial}{\partial r_+}
    \left(1 - \frac{T(r_+,\alpha(r_+,\tilde{r}_+,\beta \mu))}{T_m} \right)\right)}_
    {\hat{\bm{\omega}}=\hat{\bm{\omega}}_0}
     \nonumber \\
    &= 
    \eval{\left(\left(\frac{\partial S_{bh}}{\partial r_+}\right)^{-1}\frac{\partial}{\partial \alpha}
    \left(1 - \frac{T(r_+,\alpha)}{T_m}\right)\right)}_{\hat{\bm{\omega}}= \hat{\bm{\omega}}_0}
    - \eval{\frac{ T_m \left(\frac{\partial}{\partial \alpha}\left(1 - \frac{T(r_+,\alpha)}{T_m}\right)\right)^2}
    {8\pi \alpha \frac{\partial}{\partial \alpha}(p_m - p(\alpha))}}_{\hat{\bm{\omega}}= \hat{\bm{\omega}}_0}\,\,,
    \label{eqch9:Chatelierresponse}
\end{align}
where it was used 
\begin{align}
\frac{\partial \alpha(\tilde{r}_+, \beta \mu,r_+)}{\partial r_+} 
= - \frac{2\pi r_+ T_m 
\partial_{\alpha}\left(1 - \frac{T(r_+,\alpha)}{T_m}\right)}
{\partial_{\alpha}\left(p_m - p_{\alpha}\right)8\pi \alpha}\,\,.
\label{eqch9:derivativeChatelier}
\end{align}
Now, we can identify 
the right-hand side of Eq.~\eqref{eqch9:Chatelierresponse} as being proportional to 
$|\hat{H}|/\hat{H}_{\alpha \alpha}$, which is positive by the stability condition of the 
minimization of the reduced action. Therefore, we obtain that
\begin{align}
    & \eval{\frac{\partial}{\partial S_{bh}}\left(1 - \frac{T(r_+,\alpha)}{T_m}\right)}_{\hat{\bm{\omega}}
    = \hat{\bm{\omega}}_0} 
    \nonumber >
    \eval{\frac{\partial}{\partial S_{bh}}\left(1 - \frac{T(r_+,\alpha(r_+,\tilde{r}_+,\beta \mu))}{T_m} \right)}
    _{\hat{\bm{\omega}} = \hat{\bm{\omega}}_0}> 0\,\,,\label{eqch9:ChatelierPrinciple}
\end{align}
This is precisely the Le Chatelier-Braun principle for the black hole inside a 
cavity and the shell, arising from the stability of the zero loop approximation.

Curiously, the positive definiteness of $\hat{H}$ implies that 
$\mathcal{H}$ is positive definite as we have shown in Sec.~\ref{sech9:stabinternal}. 
The positive definiteness of $\hat{H}$ represents 
the Le Chatelier-Braun principle as we have shown above, 
but the meaning of the positive definiteness of 
$\mathcal{H}$ seems to be still illusive. Indeed, one of the conditions of 
positive definiteness of $\mathcal{H}$ contributes to the thermodynamic stability of the 
ensemble, however another condition remains.
This condition may yield a statement in the sense of Le Chatelier-Braun principle, since it 
only involves implicit derivatives of $\tilde{r}_+$ and $\beta \mu$. Indeed, one 
can rewrite the matrix as 
\begin{align}
&\mathcal{H} = \mathcal{H}_t - \mathcal{H}_p\,\,,\\
& \mathcal{H}_t = \begin{pmatrix}
    \frac{\partial }{\partial \tilde{r}_+}\left(\frac{1}{2T_{C}}\right) & 
    \frac{\partial N_{C}}{\partial \tilde{r}_+} \\
    \frac{\partial N_{C}}{\partial \tilde{r}_+} & \frac{\partial N_{C}}{\partial \beta\mu}
\end{pmatrix}\,\,,\\
& \mathcal{H}_p = \begin{pmatrix}
    \eval{\frac{\partial}{\partial \tilde{r}_+}
    \left(\frac{1}{2T_m \sqrt{f_2(\alpha)}}\right)}_{\hat{\bm{\omega}}=
    \hat{\bm{\omega}}_0} & 
    \eval{\frac{\partial N_m}{\partial \tilde{r}_+}}_{\hat{\bm{\omega}}=
    \hat{\bm{\omega}}_0} \\
    \eval{\frac{\partial N_m}{\partial \tilde{r}_+}}_{\hat{\bm{\omega}}=
    \hat{\bm{\omega}}_0} & 
    \eval{\frac{\partial N_m}{\partial \beta \mu}}_{\hat{\bm{\omega}}=
    \hat{\bm{\omega}}_0}
\end{pmatrix}\,\,,\label{eqch9:matrixmathcalHsep}
\end{align}
where $\mathcal{H}_t$ has components with total derivatives at the solutions 
of the inner system and $\mathcal{H}_p$ are the partial derivatives. 
Then, the positive definiteness of the matrix $\mathcal{H}$ seems to indicate 
that, in some sense, when more energy and $\beta\mu$ are available to the system 
with the system remaining in the previous state, the two systems respond in order to
have 
\begin{align}
    & \frac{\partial}{\partial \tilde{r}_+}\left(\frac{1}{2T_{C}}\right) > 
    \eval{\frac{\partial}{\partial \tilde{r}_+}\left(\frac{1}{2T_m \sqrt{f_2(\alpha)}}\right)}
    _{\hat{\bm{\omega}}= \hat{\bm{\omega}}_0} \,\,,\label{eqch9:entropychat1}
    + \eval{\frac{\left(\frac{\partial N_{C}}{\partial \tilde{r}_+} - \frac{\partial N_m}{\partial \tilde{r}_+}\right)^2}
    {\left(\frac{\partial N_{C}}{\partial \beta\mu} - \frac{\partial N_m}{\partial \beta\mu}\right)}}_
    {\hat{\bm{\omega}}= \hat{\bm{\omega}}_0}\,\,,\\
    & \frac{\partial N_{C}}{\partial \beta \mu} >  \eval{\frac{\partial N_m}{\partial \beta \mu}}
    _{\hat{\bm{\omega}}= \hat{\bm{\omega}}_0} \,\,.\label{eqch9:entropychat2}
\end{align}
For the first 
condition, the difference of the inverse temperature is further 
increased by the response of the system. For the second condition, 
the difference in the mean number of particles is also further increased 
by the response of the system, and it leads to the thermodynamic stability of the 
$(E,\beta\mu)$ ensemble if the shell has $\frac{\partial N_m}{\partial \beta \mu}> 0$, 
as seen previously.

\section{Thermodynamics in the grand canonical ensemble of a black hole and a 
self-gravitating matter thin shell inside a cavity\label{sech9:thermograndcan}}
\sectionmark{Thermodynamics in the grand canonical ensemble}\thispagestyle{userightbotmark}

\subsection{The grand potential of a black hole and a self-gravitating matter thin shell 
inside a cavity}

We now obtain the thermodynamic properties from the grand canonical ensemble.
The grand canonical ensemble of the outer system inside a cavity is described by the 
partition function $Z = \mathrm{e}^{- \beta W}$, where $W$ is the 
grand potential of the outer system inside a cavity. From the path integral approach 
in the zero loop approximation, 
the partition function is related to the reduced action evaluated at the solutions 
of the ensemble, i.e. $Z = \mathrm{e}^{- \tilde{I}_0}$, where $\tilde{I}_0 = 
\tilde{I}(\bm{z}; \tilde{r}_+(\bm{z}))$, with $\tilde{I}$ 
given in Eq.~\eqref{eqch9:effectiveaction} and $\tilde{r}_+=\tilde{r}_+(\bm{z})$
being the solution to Eq.~\eqref{eqch9:stationtildes}. By combining the two expressions 
for the partition function, one obtains the relation
\begin{align}
    W(T,A(R),\beta\mu) = T \tilde{I}_0(\bm{z}) 
    \,\,,
    \label{eqch9:tildegrandpotential1}  
\end{align} 
where $A(R) = 4\pi R^2$. Therefore, the expression for the grand potential can be written 
explicitly as 
\begin{align}
    &W = R(1 - \sqrt{f(\tilde{r}_+(\bm{z}),R)}) - T 
    \tilde{\mathcal{S}}(\beta \mu; \tilde{r}_+(\bm{z}))\,\,,
    \label{eqch9:tildegrandpotential2}
\end{align}
where $\tilde{\mathcal{S}}(\beta \mu; \tilde{r}_+)$ is the functional given 
by the zero loop approximation of the path integral describing $(E,\beta\mu)$ ensemble in 
Eq.~\eqref{eqch9:zeroloopchemicalentropy}, i.e. $\tilde{\mathcal{S}}(\beta\mu;\tilde{r}_+) 
=  \mathcal{S}(\beta\mu;\tilde{r}_+,\hat{\bm{\omega}}_0)$
and $\hat{\bm{\omega}}_0$ 
are the solutions to the system in Eqs.~\eqref{eqch9:statior1} and~\eqref{eqch9:statioalpha}. 

The grand potential is described thermodynamically as the Legendre transform of the energy 
as 
\begin{align}
    W = E - T S - \mu N \,\,,\label{eqch9:legendregrandpot}
\end{align}
where the energy is written in terms of $E=E(S,A(R),N)$, and where 
$T$ is the temperature, $S$ is the total entropy, $\mu$ is the chemical potential and 
$N$ is the mean particle number. 
From here and from the derivatives of the grand potential, we can obtain the 
thermodynamic 
mean quantities.

\subsection{Mean energy, entropy and mean particle number}

The differential of the grand potential can be written as 
$W = W(T,A(R), \beta \mu)$ is
\begin{align}
    dW = - (S + N \beta \mu) dT - p dA(R) - T N d(\beta \mu)\,\,,\label{eqch9:diffgrandpot}
\end{align}
where the derivatives of the grand potential can then be read out 
as $S + N \beta \mu = \frac{\partial W}{\partial T}$, 
$p= - \frac{\partial W}{\partial A(R)}$ and 
$T N = - \frac{\partial W}{\partial \beta\mu}$. Now, due to the 
fact that the stationary points obey the condition 
$\eval{\frac{\partial \tilde{I}}{\partial \tilde{r}_+}}_{\tilde{r}_+ = \tilde{r}_+(\bm{z})} 
= 0$, 
we can perform the derivatives of the grand potential
using the chain rule to obtain thermodynamic quantities of the total gravitational 
system as
$S + N \beta \mu = - \eval{\frac{\partial (T\tilde{I})}{\partial T}}_{\tilde{r}_+= \tilde{r}_+(\bm{z})} $,
$p= -\eval{\frac{\partial (T\tilde{I})}{\partial A(R)}}_{\tilde{r}_+= \tilde{r}_+(\bm{z})}$, and
$T N = - \eval{\frac{\partial (T\tilde{I})}{\partial \beta \mu}}_{\tilde{r}_+=\tilde{r}_+(\bm{z})}$.
Therefore, we obtain that $S + N\beta \mu = 
\tilde{\mathcal{S}}(\beta \mu; \tilde{r}_+(\bm{z}))$,  
and that $T N = T \tilde{N}(\beta\mu,\tilde{r}_+(\bm{z}))$. 
In terms of the quantities for the black hole and the shell, the 
entropy is
\begin{align}
    S =  \pi r_+^2(\bm{z}) + S_m(m(\bm{\omega}_0),A(\alpha(\bm{z})),\beta\mu)\,\,,
    \label{eqch9:entropygrandcan1}
\end{align}
the mean particle number is
\begin{align}
    N = N_m(m(\bm{\omega}_0),A(\alpha(\bm{z})),\beta \mu) \,\,,
    \label{eqch9:meanparticlegrandcan1}
\end{align}
and the mean pressure is
\begin{align}
    p = \frac{1}{16\pi R \sqrt{f_2(R)}}\left(1 - \sqrt{f_2(R)}\right)^2\,\,.
    \label{eqch9:pressuregrandcan1}
\end{align}
Finally, from Eqs.~\eqref{eqch9:entropygrandcan1}-\eqref{eqch9:pressuregrandcan1} and 
Eq.~\eqref{eqch9:legendregrandpot}, the mean energy can be computed to be 
\begin{align}
 E = R(1 - \sqrt{f_2(R)})\,\,.\label{eqch9:meanenergygrandcan}
\end{align}

\subsection{Thermodynamic stability of the grand canonical ensemble with 
the reservoir}

In order to study the thermodynamic stability of the grand canonical 
ensemble, we need to consider the total entropy of the ensemble plus the 
reservoir. A small difference in this entropy is given 
by $dS + dS_{res}$, where $dS_{res} = \frac{1}{\bar{T}}dE_{res} 
- \bar{\beta \mu} dN_{res}$
and $\bar{T}$ and $\bar{\beta \mu}$ are fixed quantities of the ensemble. 
Since one has $dE_{res} = -dE$ and $dN_{res} = - dN$ due to conservation of 
energy and particle number, then the difference in the sum of entropies 
becomes $dS - \frac{1}{\bar{T}}dE + \bar{\beta \mu}dN = - \frac{d\bar{W}}{\bar{T}}$,
where $\bar{W}$ is the grand potential functional given by
\begin{align}
    \bar{W}(T,\beta\mu) = E - \bar{T}S - \bar{T} \bar{\beta \mu} N\,\,,
    \label{eqch9:functionalgrandpot}
\end{align} 
where the dependence on $R$ is omitted since the quantity must be constant.
To have stability, the sum of entropies must be a maximum for the 
equilibrium configurations. Since the difference of the sum of entropies 
is $- \frac{d\bar{W}}{\bar{T}}$, then this means that the grand potential 
functional should be at a minimum in stable equilibrium configurations.
The differential of the grand potential functional is given by 
$d\bar{W} = (T - \bar{T})dS + (T \beta\mu - \bar{T}\bar{\beta \mu})dN$, and at 
equilibrium it must vanish, yielding the equilibrium conditions 
$T = \bar{T}$ and $\beta \mu = \bar{\beta \mu}$. Now the condition 
of the stable configurations being at a minimum of the grand potential 
functional translates into having a positive definite hessian of 
the grand potential functional, i.e.
\begin{align}
    d^2 \bar{W} = \begin{pmatrix}
    \frac{\partial(S + \beta \mu N)}{\partial T} & T \frac{\partial N}{\partial T}\\
    T \frac{\partial N}{\partial T} & T \frac{\partial N}{\partial \beta \mu}
    \end{pmatrix} \,\,,\label{eqch9:d2Wbar}
\end{align}
must be positive definite. In terms of thermodynamic coefficients, 
the stability conditions can be related to the condition $C_{A,N} > 0$
and $\frac{\partial N}{\partial \beta\mu} > 0$, where $C_{A,N}$ is 
the heat capacity at constant area and particle number 
$C_{A,N} = T\left( \frac{\partial S}{\partial T}\right)_{N,A}$ given by
\begin{align}
    C_{A,N} = T\frac{\partial(S + \beta \mu N)}{\partial T} 
    - T^2\left(\frac{\partial N}{\partial \beta\mu} \right)^{-1}
    \left(\frac{\partial N}{\partial T}\right)^2 > 0\,\,.
\end{align} 

In order to connect the thermodynamic 
stability conditions in Eq.~\eqref{eqch9:d2Wbar} with the stability conditions 
of the zero loop approximation regarding the effective action, we  can rewrite the 
components of Eq.~\eqref{eqch9:d2Wbar} as
\begin{align}
    & \frac{\partial(S + \beta \mu N)}{\partial T} = \frac{\partial \tilde{r}_+}{\partial T}
    \eval{\frac{\partial \tilde{S}}{\partial\tilde{r}_+}}
    _{\tilde{r}_+=\tilde{r}_+(\bm{z})}
    \,\,,\\
    & T\frac{\partial N}{\partial T} = T \frac{\partial \tilde{r}_+}{\partial T} 
    \eval{\frac{\partial \tilde{N}}{\partial \tilde{r}_+}}
    _{\tilde{r}_+=\tilde{r}_+(\bm{z})}\,\,,\\
    & T \frac{\partial N}{\partial \beta\mu} = T \eval{\left( 
     \frac{\partial \tilde{N}}{\partial \beta\mu} + \frac{\partial \tilde{N}}{\partial \beta\mu}   
    \frac{\partial \tilde{r}_+}{\partial \beta\mu}\right)}
    _{\tilde{r}_+=\tilde{r}_+(\bm{z})}\,\,.
\end{align}
The thermodynamic stability conditions for the grand canonical ensemble 
then simplify, using the relation 
$\frac{\partial \tilde{r}_+}{\partial \beta\mu} = \frac{T}{B}\frac{\partial \tilde{r}_+}{\partial T}$,
as 
\begin{align}
    & \frac{\partial \tilde{r}_+}{\partial T} > 0\,\,,\label{eqch9:thermograndcan1}\\
    & \eval{\frac{\partial \tilde{N}}{\partial \beta\mu}}_{\tilde{r}_+=\tilde{r}_+(\bm{z})} 
    > 0\,\,.\label{eqch9:thermograndcan2}
\end{align}
The first condition, Eq.~\eqref{eqch9:thermograndcan1}, 
is exactly the same as the stability condition in Eq.~\eqref{eqch9:stabtilde}
of the zero loop approximation. The second condition, Eq.~\eqref{eqch9:thermograndcan2},
is precisely the condition of thermodynamic stability of the $(E,\beta\mu)$ 
ensemble.

Therefore, there is thermodynamic stability if the zero loop approximation 
of effective action is valid and moreover, if the 
zero loop approximation of the path integral describing the $(E,\beta\mu)$ 
ensemble is valid. This relation is captured 
because of the fixed parameter $\beta\mu$ which is also the intrinsic 
parameter of the shell since $\beta\mu$ is constant throughout the space. 
This does not happen for example in \cite{Tiago2024bk}, where thermodynamic
stability is completely disconnected from 
mechanical stability. 
We must note however that the connection between mechanical stability and 
thermodynamic stability is thin, because the condition is a sum of 
the mechanical condition with another term involved in the Le Chatelier-Braun 
principle.  Additionally, the shell must have 
$\frac{\partial N_m}{\partial \beta\mu}>0$, 
which depends on the choice of the equation of state for the shell.

\section{Fundamental equations of state\label{sech9:martinez}}

\subsection{The Martinez pressure equation of state}

Here, we evaluate the possibility of giving the pressure equation 
of state from general relativity, as was done in \cite{Martinez:1996ni}. The 
differential of the functional $\mathcal{S}$ for the shell can be written as 
\begin{align}
    d\mathcal{S} = \frac{1}{T_m}dm + \frac{p_m}{T_m} dA + N_m 
    d\beta\mu \,\,.
\end{align}
Now, the energy of the shell is given by Eq.~\eqref{eqch9:m=rtilde} and 
the pressure is assumed to be given by the equation of state 
$p_m = p(\alpha)$, where $p(\alpha)$ is given in Eq.~\eqref{eqch9:pressurefunction}.
But for this to be true, $p(\alpha)$ must be a function of $m$, $\alpha$ and 
$\beta\mu$. This was true for the case of a shell only, since 
only $\tilde{r}_+$ appeared and so the dependence on $m$ or $\tilde{r}_+$ is 
equivalent through a transformation of variables. However, for the case 
of a thin shell with a black hole inside, one also has the dependence on $r_+$. 
From Eq.~\eqref{eqch9:m=rtilde}, the function $p(\alpha)$ must then be a function 
$P(\sqrt{f_2} - \sqrt{f_1},\alpha,\beta\mu)$ for it to be valid as an equation of 
state. The pressure equation of state can be rewritten as 
being proportional to $m \left(\frac{1}{\sqrt{f_1 f_2}} - 1 \right)$. Taking 
$\sqrt{f_1}$ and $\sqrt{f_2}$ as independent variables, it can be seen that 
$\sqrt{f_1}\sqrt{f_2}$ can never be written as $\sqrt{f_1} - \sqrt{f_2}$. 
Hence, the equilibrium of pressures obtained from the Einstein equations, 
or more specifically the junction conditions, cannot be used as an equation 
of state. 

We can, however, extract the equation of state from a self-gravitating 
matter thin shell and impose here. From~\cite{Martinez:1996ni}, 
the pressure equation of state is 
\begin{align}
    p_m = \frac{l_p^2 m^2}{16\pi \alpha^3\left(1 - \frac{l_p^2 m}{\alpha}\right)}\,\,.
\end{align}
Note that $p_m$ only depends here on the mass of the shell $m$ and the area 
of the shell $A = 4\pi \alpha^2$. 
Using the integrability conditions, we can further obtain that the 
inverse temperature must satisfy
\begin{align}
    \frac{1}{T_m} = \left(1 - \frac{l_p^2 m}{\alpha} \right)
    g\left[m(2 - \frac{l_p^2 m}{\alpha}),\beta\mu\right]\,\,,
\end{align}
where $g$ is some function of $m(2 - \frac{l_p^2 m}{\alpha})$ and $\beta\mu$. 
The functional $\mathcal{S}$ can be described as
\begin{align}
    \mathcal{S} = \frac{1}{2}\int^{m\left(2 - \frac{l_p^2 m}{\alpha}\right)}_0 g(s,\beta\mu)ds\,\,,
\end{align}
which depends on the choice of the function $g$. Preliminary analysis of this 
equation of state with the choice $b(x,\beta\mu) \propto x^{\frac{3}{2}}$ 
indicate that there are no stable shell solutions with a black 
hole inside. Rather, we should consider $\mathcal{S}$ evaluated at the 
limits of the parameter space, i.e. when there is no black hole or when the 
black hole meets the shell, or when the black hole sits inside shell but the 
gravitational radius of the system meets the shell. The largest value of 
$\mathcal{S}$ for these cases seems to vary with the fixed parameters 
of the ensemble and with the chosen function $g$.

\subsection{A fundamental equation of state for the shell with a black hole inside}

There is however a fundamental equation of state for the configuration of a 
shell in equilibrium with a black hole inside that we briefly explore here. This 
equation of state must be seen from the equilibrium equations for the pressure 
and temperature. Namely, one has 
\begin{align}
    &T_m^{-1} = 4\pi r_+ \sqrt{f_1(\alpha)}\,\,,\\
    &p_m = \frac{1}{16\pi \alpha l_p^2}
    \left(\frac{1 + f_2(\alpha)}{\sqrt{f_2(\alpha)}} 
    - \frac{1 + f_1(\alpha)}{\sqrt{f_1(\alpha)}}\right)\,\,,
\end{align}
with the expression for the shell mass $m = \alpha l_p^{-2}\left(\sqrt{f_1(\alpha)} 
- \sqrt{f_2(\alpha)}\right)$. The idea is to substitute $f_2(\alpha)$ 
by $m$, $\alpha$ and $f_1(\alpha)$ as 
\begin{align}
    \sqrt{f_2(\alpha)} = \sqrt{f_1(\alpha)}-\frac{l_p^2 m}{\alpha}\,\,.
\end{align}
The two equilibrium equations become 
\begin{align}\label{eqch9:equilibEOS}
    &\frac{\partial \mathcal{S}}{\partial m} = 4\pi \alpha (1 - f_1(\alpha)) \sqrt{f_1(\alpha)}\,\,,\notag\\
    &\frac{\frac{\partial \mathcal{S}}{\partial A}}{\frac{\partial \mathcal{S}}{\partial m}} 
    = \frac{m}{16\pi \alpha^2}\left[\frac{1}{\sqrt{f_1(\alpha)}(\sqrt{f_1(\alpha)} 
    - \frac{l_p^2 m}{\alpha})} -1\right]\,\,,
\end{align}
where it was used $\frac{1}{T_m} = \frac{\partial \mathcal{S}}{\partial m}$ and 
$\frac{p_m}{T_m} = \frac{\partial \mathcal{S}}{\partial A}$. The question we 
are trying to answer here is if there is an equation of state such that 
one can eliminate the dimensions of the system in Eq.~\eqref{eqch9:equilibEOS}. 
The answer seems to be positive. Indeed, we can choose an equation of state 
of the form 
\begin{align}
    \mathcal{S} = \frac{\alpha^2}{l_p^2}\varphi\left(\frac{l_p^2 m}{\alpha}\right)\,\,,
\end{align}
where $\varphi(\frac{l_p^2 m}{\alpha})$ is a function to be determined 
in terms of $\frac{l_p^2 m}{\alpha}$. Putting this equation of state into 
Eq.~\eqref{eqch9:equilibEOS}, one obtains
\begin{align}\label{eqch9:EOSeqs}
    & \varphi' = 4\pi (1 - f_{1})\sqrt{f_1}\,\,,\notag\\
    & \frac{4\varphi}{\varphi'} = \frac{l_p^2 m}{\alpha}\left(\frac{1}{\sqrt{f_1}(\sqrt{f_1} 
    - \frac{l_p^2 m}{\alpha})} + 1\right)\,\,,
\end{align}
where the dependence on $\alpha$ was dropped and $\varphi'$ is the derivative on the 
argument of $\varphi$. Notice now that Eq.~\eqref{eqch9:EOSeqs} can be further simplified 
by solving the first equation in Eq.~\eqref{eqch9:EOSeqs} to obtain $f_1 = f_1(\varphi')$ 
and substitute it into the second equation to obtain a differential equation 
for $\varphi(\frac{l_p^2 m}{\alpha})$ as
\begin{align}
    \frac{4\varphi}{\varphi'} = \frac{l_p^2 m}{\alpha}\left(\frac{1}{\sqrt{f_1(\varphi')}(\sqrt{f_1(\varphi')} 
    - \frac{l_p^2 m}{\alpha})} + 1\right)\,\,.
\end{align}
One can have multiple solutions for $\varphi$, since both equations need to be inverted 
to finally obtain an expression for $\varphi'$ in function of $\frac{l_p^2 m}{\alpha}$ 
and $\varphi$.
If there is indeed such a $\varphi$ that solves the differential equation, 
then both equilibrium equations are satisfied 
if and only if the first equation in Eq.~\eqref{eqch9:EOSeqs} is satisfied. 
By using $\frac{l_p^2 m}{\alpha} = \sqrt{f_1} - \sqrt{f_2}$, this means 
that the solution is some $\sqrt{f_1}$ in function of $\sqrt{f_2}$, i.e. $\frac{r_+}{\alpha}$ 
in function of $\frac{\tilde{r}_+}{\alpha}$. With fixed $\tilde{r}_+$, one is then free to 
pick an $\alpha$ such that the solution is still valid, obtaining a value of $r_+$ for each 
$\alpha$. For each fixed $\tilde{r}_+$, there is then a non-countable collection of 
solutions described by a curve in the $\alpha\times r_+$ plane.
This accomplishes the same functionality of the Martinez' equation of state, 
however it is much more involved for this case. 
Unfortunately, for some solutions we analyzed numerically, 
these solutions seem to be minima of $\mathcal{S}$.

\section{Hessians related to the actions\label{sech9:hessian}}

In order to analyze stability, we need to evaluate the hessian of the 
reduced action on the stationary points. In this section, the six 
components of the hessian for the case of a black hole inside a 
self-gravitating matter thin shell in a cavity are presented. The 
stationary points are given by solving the simultaneous vanishing of the 
components of the gradient of the reduced action.

The components of the hessian respective to at least one derivative on 
$\tilde{r}_+$, i.e. $H_{\tilde{r}_+ \bm{\omega}^A}$, are 
\begin{align}
    &H_{\tilde{r}_+ \tilde{r}_+} = 
    \frac{1}{4 l_p^2 f_2(R)}
    \Bigg[ \frac{1}{l_p^2 T_m^2}\frac{\partial T_m}{\partial m}\frac{f_2(R)}{f_2(\alpha)} 
    - \frac{1}{(f_2(\alpha))^{\frac{3}{2}}T_m}\left(\frac{1}{\alpha} - \frac{1}{R}\right)
     \Bigg]\,\,,\label{eqch9:apprtildertildeaction}\\
    &H_{\tilde{r}_+ r_+} = - \frac{1}{4 l_p^4 f_2(\alpha) T_m^2} \frac{\partial T_m}{\partial m}
    \,\,,\label{eqch9:apprtildeq1action}\\
    &H_{\tilde{r}_+ \alpha} = 
    \frac{4\pi \alpha}{l_p^2}
    \Bigg[ \left(\frac{\partial T_m}{\partial A(\alpha)} - p(\alpha)\frac{\partial T_m}{\partial m}\right)
    \frac{1}{T_m^2 \sqrt{f_2(\alpha)}} 
    + \frac{\tilde{r}_+}{16\pi \alpha^3 (f_2(\alpha))^{\frac{3}{2}} T_m}\Bigg]\,\,.
    \label{eqch9:appdertildealphaaction}
\end{align}
The components to at least a derivative in $r_+$ are
\begin{align}
    &\hat{H}_{r_+ r_+} = H_{r_+ r_+} = \frac{4\pi^2 r_+^2}{l_p^2}\Bigg[ 
    \frac{1 - 3f_1(\alpha)}{4\pi r_+^2 f_1(\alpha)}
    + \frac{1}{l_p^2}\frac{\partial T_m}{\partial m}\Bigg]\,,\label{eqch9:appderrprpaction}\\
    &\hat{H}_{r_+ \alpha} = H_{r_+ \alpha} = - \frac{16\pi^2 r_+ \alpha}{l_p^2}\Bigg[ 
    \frac{r_+}{16\pi \alpha^3 f_1(\alpha)} 
    + \frac{1}{T_m}\left(\frac{\partial T_m}{\partial A(\alpha)}
    - \frac{\partial T_m}{\partial m}\, p(\alpha) \right)\Bigg]\,\,,
    \label{eqch9:appderrpalphaaction}
\end{align}
And finally the last component of the hessian is
\begin{align}
    &\hat{H}_{\alpha \alpha} = H_{\alpha \alpha} = \frac{64 \pi^2 \alpha^2}{T_m}
    \Bigg[\frac{1}{8\pi \alpha}\frac{\partial p(\alpha)}{\partial \alpha} 
    - \frac{\partial p_m}{\partial A(\alpha)} + p(\alpha)\frac{\partial p_m}{\partial m}
    \Bigg]\,,\label{eqch9:deralphaalphaaction}
\end{align}
 where $\frac{\partial p(\alpha)}{\partial \alpha}$ is given by 
\begin{align}\label{eqch9:appderpderalpha}
    \frac{\partial p(\alpha)}{\partial \alpha} = \frac{1}{16\pi l_p^2}
    \Bigg( \frac{3f_1^2 +  1}{2\alpha^2 f_1^{\frac{3}{2}}}
    -\frac{3f_2^2 +  1}{2\alpha^2 f_2^{\frac{3}{2}}} \Bigg)\,\,.
\end{align}
The hessian of the effective action in Eq.~\eqref{eqch9:effectiveaction} is
\begin{align}
    \tilde{H}_{\tilde{r}_+ \tilde{r}_+} 
    = \eval{\left(\frac{B}{2 f_2(R) R} - \frac{\partial B}{\partial \tilde{r}_+}\right)}
    _{\tilde{r}_+=\tilde{r}_+(\bm{z})}\,\,,
    \label{eqch9:appstabtilde}
\end{align}
which by using the transformation $\delta \omega^A = \delta \hat{\omega}^A 
+ \frac{\partial \hat{\omega}^A_0}{\partial \tilde{r}_+} \delta \tilde{r}_+$, with 
$\eval{\frac{\partial \hat{\omega}^A_0}{\partial \tilde{r}_+}}_{\tilde{r}_+=\tilde{r}_+(\bm{z})} 
= -H_{\tilde{r}_+ \hat{\omega}^B} 
(\hat{H}^{-1})^{\hat{\omega}^A \hat{\omega}^B}|_{\tilde{r}_+=\tilde{r}_+(\bm{z})}$ and 
the consistency relations of the inverse matrix $(H^{-1})^{\omega^i \omega^j}$,
one obtains the relation between $\tilde{H}_{\tilde{r}_+ \tilde{r}_+}$ with 
$H_{\omega^i \omega^j}$ as
\begin{align}
    \tilde{H}_{\tilde{r}_+ \tilde{r}_+} = 
    \frac{|H|}{|H|_{\{\tilde{r}_+\},
    \{\tilde{r}_+\}}}\,\,.
\end{align}

\section{Mechanical stability of a shell around a black hole
\label{sech9:stabshellmech}}

We relate here the derivative of the difference of the pressures with the 
mechanical stability condition of a shell with a black hole inside. 
Following~\cite{Brady:1991np}, the equations of motion for a shell are
\begin{align}
    & \ddot{r} = \frac{8\pi r k_2 k_1}{l_p m}\left[p_m + \frac{m}{8\pi r^2} 
    - \frac{\tilde{r}_+ k_1 - r_+ k_2}{l_p^2 r^2 k_1 k_2} \right]\,\,,\\
    & m = \frac{r}{l_p^2} \left(k_1 - k_2\right)\,\,.\\
    & k_1 = \sqrt{f_1 +\dot{r}^2} \,\,,\,\,k_2 = \sqrt{f_2 + \dot{r}^2}\,\,.
\end{align}
For a shell in equilibrium at radius $r=\alpha$, 
it is required that $\dot{r}=\ddot{r}=0$, which 
gives the shell pressure equilibrium equation $p_m = p(\alpha)$, with $p(\alpha)$ 
described in Eq.~\eqref{eqch9:pressurefunction}. For small perturbation in the 
radius, one has $r= \alpha + \delta r$, where the equation for the perturbations 
is given by 
\begin{align}
    & \delta\ddot{r} = \frac{8\pi \alpha f_2 f_1}{l_p m}\frac{\partial}{\partial\alpha}
    \left[p_m - p(\alpha)\right]\delta r\,\,.
\end{align} 
To have a mechanically stable shell, $\delta r$ must have an oscillatory 
motion and not an exponential one. And so mechanical stability means 
$\frac{\partial}{\partial\alpha}\left(p(\alpha) - p_m\right) > 0$, i.e. 
$H_{\alpha \alpha} > 0$.

\section{Conclusions\label{sech9:conc}}

In this chapter, we constructed the grand canonical ensemble of 
a thin shell with a black hole inside a cavity and also the 
$(E,\beta\mu)$ ensemble of a black hole and a thin shell, which is 
an ensemble with fixed energy and fixed chemical potential. 
We construct the $(E,\beta\mu)$ 
ensemble because it gives a better 
understanding and serves as a first step towards the 
construction of the grand canonical ensemble. We apply the zero loop 
approximation, leading to equilibrium equations and stability conditions 
for the validity of the approximation. To obtain the solutions of the 
ensembles, we still need to make a choice of equations of state.

In this chapter, we have shown the power of the Euclidean path integral approach 
in the construction of the ensembles of self-gravitating systems. 
In the zero loop approximation, the formalism provides the 
equilibrium equations and the stability conditions. These last conditions 
include the mechanical stability of the shell and also lead to the 
Le Chatelier-Braun principle. In connection to the thermodynamics of the 
system in the grand canonical ensemble and the $(E,\beta\mu)$ 
ensemble, the stability conditions lead to thermodynamic stability, but 
the stability conditions are more restrictive. This means one cannot 
infer that the stability conditions are satisfied if there is thermodynamic 
stability. This may be due to the nature of the zero loop approximation, 
since the stability conditions are tied to the expansion of the 
integral over the minima of the action. Indeed, if one was able to 
obtain the path integral in a convergent way, the stability conditions 
would not exist. As we have seen in Chapter~\ref{ch:thinshellAdS}, 
the mechanical stability is disconnected from thermodynamic stability of the canonical 
ensemble of a shell in AdS, but this is because the chemical potential was not 
included. Since the chemical potential times the inverse temperature 
is constant throughout the space, 
in some way, we have some limited access to the properties of the thin shell 
in the ensemble.

The task of finding an equation of state such that yields stable solutions 
for a thin shell in equilibrium with a black hole still remains. While the 
fundamental equation of state from Martinez~\cite{Martinez:1996ni} does 
not give solutions of a shell with a black hole inside, we found another fundamental equation 
of state. However, it seems that the solutions are unstable. 
It was argued in~\cite{Brady:1991np} that the configuration of a shell with a black hole 
inside would always be unstable. However, a linear equation of state was used for the 
thin shell and there were no thermodynamic considerations in the analysis. A more thorough 
study of stability is still needed to ascertain the families of equations of state 
that could yield a stable configuration.

\chapter{Concluding Remarks}\label{ch:Conclusion}

In this thesis, we explored the thermodynamics of curved spacetimes and self-gravitating 
matter via two methods, by imposing the first law of thermodynamics and 
by constructing statistical ensembles through the Euclidean path integral 
approach in quantum gravity. There are however a few caveats to the 
analysis we took here and there are also future research lines. 

In the first Part, composed only by the Chapter~\ref{ch:chargedselfgravitating}, we considered
an electrically charged self-gravitating matter thin shell 
and we imposed the first law of thermodynamics to such shell. Furthermore, we imposed the Martinez 
fundamental pressure equation of state, allowing the shell to be in 
mechanical equilibrium for every radius of the shell. We chose the temperature equation of 
state so that it allowed the black hole limit, in order to recover 
black hole thermodynamics. We further analyzed the intrinsic thermodynamic stability. 
We showed that the shell can be put at the brink of becoming a black hole and 
also that the shell becomes marginally stable when doing so. The Martinez equation of 
state gives special characteristics to the shell.
However, what is the extent of the existence of such fundamental equation of state
for an isolated thin shell? 
It is not clear if the fundamental pressure equation of state is specific to 
general relativity and to matter thin shells. A further study must be done 
for alternative theories of gravity, where the junction conditions are different. 
For example, for $f(R,T)$ theories, i.e. theories with a lagrangian which is dependent 
on the Ricci scalar and the trace of the stress energy tensor, it was shown 
in~\cite{Rosa:2021teg} that 
a self-gravitating thin shell must be at a specific radius 
for mechanical equilibrium, but however its thermodynamic implications still need to be explored.

In the second Part, we used the Euclidean path integral approach to compute the 
partition function of several curved spacetime configurations. 
In Chapter~\ref{ch:Euclideanpathintegral}, we reviewed
the formalism for the case of spherically symmetric metrics, which 
established a basis for the rest of the Chapters. Additionally, in 
all Chapters of this Part, we considered the zero loop approximation, where in some cases the 
Hamiltonian and momentum constraints were imposed to the Euclidean action 
to obtain a reduced action. While the choice of topology 
for the path integral has been motivated by the topology of the boundary 
of the space, it would be of interest to extend the study in a more thorough way 
to axisymmetric complex spaces. 
Kerr-Newmann complexified spaces~\cite{Brown:1990fk} were considered already, but  
a deeper study of the reduced action for these spaces has not been done yet. Furthermore, 
an issue that is transversal to this Part is that we analyzed the validity of the zero loop 
approximation using the reduced action, which already 
assumes the field constraints. A further study regarding the equivalence between 
the existence of negative modes of Riemannian or pseudo-Riemannian spaces and the 
analysis of the reduced action must still be done.

In particular, in Chapters~\ref{ch:grandcanonicalblackhole},
~\ref{ch:Daviesblackhole} and~\ref{ch:canonicalblackhole}, we considered ensembles of Reissner-
Nordstr\"om black holes in arbitrary dimensions, in the zero loop approximation. 
Namely, in Chapter~\ref{ch:grandcanonicalblackhole}, we considered the grand canonical ensemble 
inside a cavity. In Chapter~\ref{ch:Daviesblackhole}, 
we considered the canonical ensemble of a charged black hole inside a cavity with infinite radius.
And in Chapter~\ref{ch:canonicalblackhole}, we considered the canonical ensemble of a charged black hole
inside a finite cavity. Note that in Chapter~\ref{ch:Daviesblackhole}, we have shown explicitly that the results 
from imposing the first law of thermodynamics agree with the Euclidean path integral 
approach. An innovation that we introduced was the modelling of a configuration corresponding 
to hot flat space in each ensemble. In the grand canonical ensemble, we modelled the configuration 
by a charged sphere with no gravity in the limit of very small radius, 
which allowed the existence of an electric potential difference. In some sense, 
it described hot flat space with an electric potential difference. In the 
canonical ensemble, we modelled the configuration by a charged shell with no gravity 
in the limit that the shell approached the boundary of space. In some sense, it 
described hot flat space with electric charge at the boundary of the cavity. These 
configurations yielded a zero action in their respective ensembles. We were able to
study the phase transitions between the charged black hole and these configurations, 
something that was missing in the literature. However, these configurations are heavily 
simplified. In order to further improve the analysis in this thesis, the study 
of the matter section with electric charge is required, which may give a better description 
of these configurations.

In Chapter~\ref{ch:adslimits}, we considered the limits of the solutions of the 
zero loop approximation 
of Schwarzschild-AdS black holes inside a cavity. Indeed, we have shown 
that these 
limits unify the existing black hole solutions described in~\cite{Gibbons:1977},
~\cite{York:1986}, the planar AdS black hole solution and the 
Rindler solution.

In Chapter~\ref{ch:thinshellAdS}, we studied the canonical ensemble for a 
matter thin shell in anti-de Sitter (AdS). We gave an equation of state 
for the shell  that is similar to an equation of state of a graviton 
gas trapped in a shell. Such equation of state allowed the existence of a stable solution 
for the shell. 
While one could choose another equation of state, our main goal was to analyze 
the phase transitions between the matter thin shell and the black hole, where we 
have shown that the phase transition is similar to the Hawking-Page phase transition. 
It is expected that such phase transition occurs between self-gravitating matter 
and black holes in AdS and also in asymptotically flat, but further analysis must be 
done. An interesting avenue is to consider self-gravitating fluids in the formalism 
to model exactly a gas of gravitons and photons with backreaction, which we 
started to study but did not include it in the thesis since it is very preliminary.

In Chapter~\ref{ch:Chatelier}, we have constructed the grand canonical ensemble of 
a self-gravitating 
matter thin shell with a black hole inside, all within a cavity, 
including the chemical potential of the shell. We have shown that the 
analysis of the stationary points of the reduced action yields precisely the 
Le Chatelier-Braun principle between the thin shell and the black hole, and 
also yields thermodynamic stability of the full system. Related to the 
work in Chapter~\ref{ch:chargedselfgravitating}, the Martinez equation of state 
does not seem to give a stable solution for a shell with a black hole inside and only 
the limiting cases of a thin shell alone or a black hole alone are permitted. 
There seems to be another fundamental equation state different from Martinez equation 
of state allowing a non-countable number of equilibrium configurations, however 
they seem unstable thermodynamically. Further research is needed to understand if 
there is always a fundamental equation of state for shells in equilibrium with other 
systems.

We conclude by stating that main message of the thesis. 
The formalism of the Euclidean path integral 
approach is quite powerful in describing the thermodynamics of curved spacetimes, including 
self-gravitating matter. However, there are still many angles that need to be 
explored, not only on the limitations of the formalism but also regarding the 
connection to other semiclassical descriptions.


\cleardoublepage
\defbibheading{bibintoc}[\bibname]{%
  \phantomsection
  \manualmark
  \markboth{\spacedlowsmallcaps{#1}}{\spacedlowsmallcaps{#1}}%
  \addtocontents{toc}{\protect\vspace{\beforebibskip}}%
  \addcontentsline{toc}{chapter}{\tocEntry{#1}}%
  \chapter*{#1}%
}
\printbibliography[heading=bibintoc]


\end{document}